\pdfoutput=1 
\documentclass[12pt,pdftex,fleqn,twoside,a4paper]{DissTB}
\usepackage[pdftex,hypertexnames=false,pdfpagelabels,plainpages=false]{hyperref}

\usepackage[utf-8]{inputenc}
\usepackage{a4,rotating,ngerman,enumerate,placeins}
\usepackage{graphicx,epsfig,color,stmaryrd}
\usepackage{mdwtab,floatflt,fancyhdr}
\usepackage{subfig}
\usepackage{longtable}
\usepackage[mathscr]{euscript}
\usepackage[intlimits,sumlimits]{amsmath}
\usepackage{amssymb,amsfonts,amsopn,amsbsy,bbm,mathrsfs}

\numberwithin{equation}{chapter}
\numberwithin{table}{chapter}
\numberwithin{figure}{chapter}

\newlength{\wordwidth}
\newlength{\quadwidth}
\newlength{\stepb}
\newcommand{\xtilde}[2][XX]{\text{$#2$\settowidth{\wordwidth}{$#2$}\settowidth{\quadwidth}{#1}\setlength{\stepb}{0.5\wordwidth}\hspace{-\stepb}\hspace{-0.5\quadwidth}$\widetilde{\phantom{#1}}$\hspace{\stepb}\hspace{-0.5\quadwidth}\hspace{-0.2\quadwidth}}}

\makeatletter
  \renewcommand{\p@subtable}{\thetable}
\makeatother

\makeatletter
  \renewcommand{\p@subfigure}{\thefigure}
\makeatother

\makeatletter
  \renewcommand{\@pnumwidth}{2.2em}
\makeatother

\newcommand{\helpkbar}{\phantom{\big\vert}\!}

\newcommand{\Ell}{\mathcal{L}}                 

\newcommand{\vph}{\varphi}
\newcommand{\vth}{\vartheta}

\newcommand{\vep}{\varepsilon}
\newcommand{\vrh}{\varrho}

\renewcommand{\u}[1]{\,{\rm #1}}               

\DeclareMathOperator{\Imag}{Im}   
\DeclareMathOperator{\Real}{Re}   
\DeclareMathOperator{\Res}{Res}   

\DeclareMathOperator{\diag}{diag}

\DeclareMathOperator{\const}{const}    

\DeclareMathOperator{\Spur}{Sp}

\newcommand{\e}{\mathrm{e}}

\newcommand{\del}{\ensuremath{\partial}}
\renewcommand{\d}{\ensuremath{\u d}}    
\newcommand{\dpa}{\ensuremath{\partial}}

\newcommand{\dd}[2]{\ensuremath{\frac{\d #1}{\d #2}}}            
\newcommand{\ddp}[2]{\ensuremath{\frac{\dpa #1}{\dpa #2}}}       

\newcommand{\ten}[1]{\ensuremath{\cdot 10^{#1}}}                

\renewcommand{\v}[1]{\ensuremath{\boldsymbol{#1}}}  

\newcommand{\mc}[1]{\ensuremath{\mathcal{#1}}}      
\newcommand{\mb}[1]{\ensuremath{\mathbb{#1}}}       
\newcommand{\mbm}[1]{\ensuremath{\mathbbm{#1}}}     
\newcommand{\ms}[1]{\ensuremath{\mathscr{#1}}}      

\newcommand{\I}{{\rm i}}                            

\newcommand{\um}{\ensuremath{\mathbbm1}}                            

\newcommand{\OO}{\mathcal{O}}                       

\newcommand{\klr}[1]{\left(#1\right)}               
\newcommand{\kle}[1]{\left[#1\right]}               
\newcommand{\klg}[1]{\left\{#1\right\}}             
\newcommand{\abs}[1]{\left|#1\right|}               

\newcommand{\an}{\Big\vert}                         

\newcommand{\nab}{\v\nabla}                         

\newcommand{\subt}[1]{_{\text{#1}}}                 
\newcommand{\supt}[1]{^{\text{#1}}}                 

\newcommand{\unl}[1]{\underline{#1}}                                     

\newcommand{\verw}[1]{\ensuremath{\left<#1\right>}}                       

\newcommand{\bra}[1]{\ensuremath{\langle#1|}}                             
\newcommand{\ket}[1]{\ensuremath{|#1\rangle}}                             
\newcommand{\bracket}[2]{\ensuremath{\langle#1|#2\rangle}}                
\newcommand{\matelem}[3]{\bra{#1}#2\ket{#3}}                              

\newcommand{\HC}{^\dag}                                                   

\newcommand{\rbra}[1]{\ensuremath{(#1|}}      
\newcommand{\rket}[1]{\ensuremath{|#1)}}                             
\newcommand{\rbracket}[2]{\ensuremath{(#1|#2)}}                
\newcommand{\rmatelem}[3]{\rbra{#1}#2\rket{#3}}                              

\newcommand{\lbra}[2][XX]{\xtilde[#1]{\bra{#2}}}
\newcommand{\lket}[2][XX]{\xtilde[#1]{\ket{#2}}}

\newcommand{\lrbra}[2][XX]{\xtilde[#1]{\rbra{#2}}}

\newcommand{\lrbracket}[3][XX]{\ensuremath{\xtilde[#1]{(#2|}#3})}

\newcommand{\lrmatelem}[3]{\lrbra{#1}#2\rket{#3}}

\newcommand{\Pro}{\unl{\mbm P}}

\newcommand{\umc}[1]{\unl{\mc{#1}}}

\newcommand{\uop}[1]{\unl{\ms{#1}}}

\newcommand{\vmc}[1]{\ensuremath{\v{\mc{#1}}}}

\newcommand{\vu}[1]{\unl{\v{#1}}}

\newcommand{\uk}{\hat{\umc K}}                     
\newcommand{\uell}{\underline{\hat\Ell}}           

\newcommand{\diffuz}[2][z]{\left\llbracket#2\right\rrbracket^{#1}_{u}}
\newcommand{\diffuza}[3][z]{\left\llbracket#2\right\rrbracket^{#1}_{#3}}

\newcommand{\PV}{\subt{PV}}
\newcommand{\PVU}{\supt{PV}}

\newcommand{\LambShift}{\ensuremath{\mc S}}
\newcommand{\FineStructure}{\ensuremath{\Delta}}
\newcommand{\HyperFineSplitting}{\ensuremath{\mc A}}

\newcommand{\mycell}[2]{\parbox{#1}{\vspace{1mm}\centering\baselineskip=18pt #2\vspace{1mm}}}

\newcommand{\tfxxx}{\tfrac52}
\newcommand{\tfxx}{\tfrac32}
\newcommand{\tfx}{\tfrac12}

\newcommand{\wigjb}[6]{
  \ensuremath{\begin{Bmatrix} #1 & #2 & #3\\ #4 & #5 & #6\end{Bmatrix}}} 

\newcounter{dummy}
\begin{document}
\bibliographystyle{disstbg}
\pagestyle{empty}
\begin{center}
\LARGE\sffamily
{\bfseries Inaugural-Dissertation\\[5mm]
\large
zur\\[1mm]
Erlangung der Doktorwürde\\[1mm]
der\\[1mm]
Naturwissenschaftlich-Mathematischen\\[1mm]
Gesamtfakultät\\[1mm]
der Ruprecht-Karls-Universität\\[1mm]
Heidelberg\\[1mm]
}
\vfill\normalsize
vorgelegt von\\[4mm]
{\large\bfseries Dipl.-Phys. Timo Friedhelm Bergmann}\\[4mm]
aus Kassel\\[15mm]
Tag der mündlichen Prüfung: 19.07.2006
\cleardoublepage\rmfamily

{\LARGE\sffamily\phantom{x}

\vspace{4cm}
Theorie des
longitudinalen Atomstrahl-Spinechos\\[5mm] 
und\\[5mm] 
paritätsverletzende Berry-Phasen
in Atomen\\[5mm]
}

\vfill\large
\begin{tabular}{ll}
{\bf Gutachter:}\qquad & Prof. Dr. Otto Nachtmann\\ 
 & Priv.-Doz. Maarten DeKieviet, PhD.
\end{tabular}
\end{center}

\begin{small}
\cleardoublepage\noindent
{\bf Theorie des longitudinalen Atomstrahl-Spinechos und paritätsverletzende Berry-Phasen in Atomen}\\

In dieser Arbeit entwickeln wir eine nichtrelativistische Theorie zur
quantenmechanischen Beschreibung longitudinaler Atomstrahl-Spinecho-Experimente, 
bei denen ein Strahl neutraler Atome eine Anordnung statischer elektrischer und magnetischer Felder durchquert. 
Der Gesamtzustand des Atoms ist die Lösung der Schrödinger-Gleichung mit matrixwertigem Potential und
kann als Superposition lokaler (atomarer) Eigenzustände der Potentialmatrix geschrieben werden.
Die orts- und zeitabhängigen Amplitudenfunktionen jedes Eigenzustands repräsentieren
die atomaren Wellenpakete und können mit der von uns aufgestellten Master-Formel
in einer Reihenentwicklung berechnet werden.
Die nullte Ordnung dieser Reihenentwicklung beschreibt den adiabatischen Grenzfall,
in höheren Ordnungen werden Mischungen der Zustände und der zugeordneten Amplitudenfunktionen berücksichtigt.
Wir geben eine Anleitung zur theoretischen Beschreibung eines longitudinalen Atomstrahl-Spinecho-Experiments
und des sogenannten Fahrplanmodells an, bei dem die Schwerpunktsbewegung der Wellenpakete
in den angelegten Feldern grafisch dargestellt wird. Die Schwerpunktsbewegung der Wellenpakete wird dabei
durch die unterschiedliche Aufspaltung der lokalen Eigenwerte der Potentialmatrix bestimmt.
Wir studieren als Beispiel für eine Anwendung der Theorie paritätsverletzende, geometrische (Berry-)Phasen in Atomen.
In diesem Zusammenhang führen wir geometrische Flussdichten ein, mit denen man für spezielle
Feldkonfigurationen die geometrischen Phasen direkt anhand eines Vektordiagramms veranschaulichen kann. 
Am Beispiel einer bestimmten Feldkonfiguration zeigen wir die Existenz einer paritätsverletzenden geometrischen 
Phase.\\[2cm]
{\bf Theory of Longitudinal Atomic Beam Spin Echo and Parity Violating Berry-Phases in Atoms}\\

We present a nonrelativistic theory for the quantum mechanical description of longitudinal atomic beam spin echo
experiments, where a beam of neutral atoms is subjected to static electric and magnetic fields.
The atomic wave function is the solution of a matrix-valued Schrödinger equation and can be written
as superposition of local (atomic) eigenstates of the potential matrix. The position- and time-dependent
amplitude function of each eigenstate represents an atomic wave packet and can be calculated in a
series expansion with a master formula that we derive.
The zeroth order of this series expansion describes the adiabatic limit, whereas the higher order contributions
contain the mixing of the eigenstates and the corresponding amplitude functions. We give a tutorial for the theoretical
description of longitudinal atomic beam spin echo experiments and for the so-called Fahrplan model, 
which is a visualisation tool for the propagation of wave packets of different atomic eigenstates. 
As an example for the application of our theory, we study parity violating geometric (Berry-)phases. In this context, 
we define geometric flux densities, which for certain field configurations can be used to illustrate geometric phases 
in a vector diagram. Considering an example with a specific field configuration, we prove the existence of a 
parity violating geometric phase.
\end{small}

\cleardoublepage
\begin{flushright}
\phantom{|}
\Large\em\vspace{10cm}
Für Dajana und Elena.
\end{flushright}

\fancypagestyle{plain}{
\fancyhead{}
\fancyfoot{}
\renewcommand{\headrulewidth}{0pt}
\renewcommand{\footrulewidth}{.5pt}
\fancyfoot[EL,OR]{\bfseries\thepage}
}

\cleardoublepage
\pagestyle{fancy}
\renewcommand{\headheight}{15pt}
\fancyhead{}
\fancyfoot{}
\fancyhead[ER]{\slshape Inhaltsverzeichnis}
\fancyhead[OL]{\slshape Inhaltsverzeichnis}
\fancyhead[EL,OR]{\bfseries\thepage}
\fancyfoot[EL,OR]{ {\phantom{\AA{}}} }
\renewcommand{\headrulewidth}{.5pt}
\renewcommand{\footrulewidth}{.5pt}
\renewcommand{\thepage}{\roman{page}}
\setcounter{page}{1}
\refstepcounter{dummy}
\addcontentsline{toc}{chapter}{Inhaltsverzeichnis}
\tableofcontents

\chapter*{Notation und Konventionen}
\fancyhead[ER]{\slshape Notation und Konventionen}
\fancyhead[OL]{\slshape Notation und Konventionen}
\addcontentsline{toc}{chapter}{Notation und Konventionen}

In der gesamten Arbeiten werden natürliche Einheiten verwendet, d.h. es ist $\hbar = c = 1$. Die eigentlich
magnetische Flussdichte genannte Größe $\vmc B$ wollen wir salopp als magnetische Feldstärke bezeichen. Sie ist
nicht zu verwechseln mit $\v H = \vmc B/\mu_0 - \v M$ (SI-Einheiten, $\v M$ ist die Magnetisierung).
Desweiteren wird die folgende Notation vereinbart:\\[5mm]

\noindent
\begin{longtable}{lp{1cm}p{12cm}}
$\v L,\v S,\v J,\v I,\v F$ && Operatoren für den Bahndrehimpuls des Elektrons ($\v L$), den Elektronspin ($\v S$),
den Gesamtdrehimpuls des Elektrons ($\v J = \v L + \v S$), den Kernspin ($\v I$) und den atomaren Gesamtdrehimpuls
($\v F = \v J + \v I$).\\
$\ket{n L_J,F,F_3}$ && Atomarer Gesamtdrehimpulszustand ohne Berücksichtigung der Hyperfeinwechselwirkung. Eigenzustand 
der Drehimpulsoperatoren $\v L^2$, $\v S^2$, $\v J^2$, $\v I^2$, $\v F^2$ sowie $F_3$. Die im Atom unveränderlichen 
Quantenzahlen $S$ und $I$ für den Spin des Elektrons und des Kerns werden in der Notation weggelassen. Die Hauptquantenzahl
$n$ soll daran erinnern, dass in der Ortsdarstellung dieser Zustände die (nichtrelativistischen) Wellenfunkionen für
Wasserstoff, $\Psi_{n,L,L_3}(r,\vth,\vph) = R_{n,L}(r)Y_{L,L_3}(\vth,\vph)$, zu verwenden sind.\\
$\uop A, \uop B, \uop C,\ldots $ && Matrizen im Unterraum der atomaren Zustände mit fester Hauptquantenzahl $n$, 
dargestellt in der Basis $\klg{\ket{n L_J,F,F_3}}$ der Gesamtdrehimpulszustände.\\
$\uop M(Z)$ && Die vom Ort $Z$ abhängige (nichthermitesche) Massenmatrix, d.h. die Darstellung des atomaren
Hamiltonoperators im Unterraum zu fester Hauptquantenzahl $n$.\\
$\alpha=1,\ldots,N$ && Index für die Nummerierung der Eigenzustände von $\uop M(Z)$. Die Dimension
des jeweils betrachteten Unterraums zu fester Hauptquantenzahl wird mit $N$ bezeichnet.\\
$\rket{\alpha(Z)}$ && Rechter Eigenzustand mit Index $\alpha$ der Massenmatrix $\uop M(Z)$ zum komplexen 
Eigenwert $E_\alpha(Z)$. Auch als atomarer Eigenzustand oder innere atomarer Zustand bezeichnet.\\
$\lrbra{\alpha(Z)}$ && Linker Eigenzustand von $\uop M(Z)$ zum komplexen Eigenwert $E_\alpha(Z)$.\\
$\Pro_\alpha(Z)$ && Quasiprojektor $\Pro_\alpha(Z)=\rket{\alpha(Z)}\lrbra{\alpha(Z)}$ des inneren atomaren Zustands mit 
Index $\alpha$.\\
$\uop A_{\beta\alpha}(Z)$ &&
Matrixelemente der Matrix $\uop A$ bzgl. der lokalen Eigenzustände von $\uop M(Z)$.
Es ist z.B. $\uop A_{\beta\alpha}(Z) = \lrbra{\beta(Z)}\uop A\rket{\alpha(Z)}$.\\
$\Psi_\alpha(Z,t)$ && Orts- und zeitabhängige Wellenfunktion des atomaren Zustands mit Index $\alpha$.\\
$\phi_\alpha(Z,t)$ && Phasenwinkel des in $\Psi_\alpha(Z,t)$ enthaltenen Phasenfaktors.\\ 
$A_\alpha(Z,t)$ && Amplitudenanteil in $\Psi_\alpha(Z,t)$.\\ 
$\vmc E, \vmc B$ && Elektrische und magnetische Feldstärke.\\
$\del_x$ && Symbol für die partielle Ableitung nach $x$, $\del_x = \del/\del x$.\\
$(\hat{\ms O}f)(Z,t)$ && Bei der Anwendung eines Operators $\hat{\ms O}$ auf eine Funktion $f(Z,t)$ entstehende Funktion 
$(\hat{\ms O}f)(Z,t)$.\\
$\hat L,\ \umc{\hat{\mc L}}$ && Differentialoperatoren (skalar, matrixwertig).\\
$\hat K,\ \umc{\hat{\mc K}}$ && Integraloperatoren (skalar, matrixwertig).\\
$d$ && Symbol für die äußere Ableitung.\\
$d f$ && Äußere Ableitung einer Funktion $f$ (Differentialform, vollständiges Differential einer Funktion).\\
d$x$ && Symbol für ein Differential (z.B. innerhalb eines Integrals oder einer Differentialform).\\
$\oint_{\mc C}$ && Integral über eine geschlossene Kurve $\mc C$.\\
$\del\mc F$ && Rand einer Fläche $\mc F$.\\
lABSE && Abkürzung für {\em longitudinal Atomic Beam Spin Echo}.\\
PC  && Abkürzung für {\em Parity Conserving}.\\
PV  && Abkürzung für {\em Parity Violating}.\\
APV && Abkürzung für {\em Atomic Parity Violation}.
\end{longtable}

\cleardoublepage
\parindent0pt
\parskip10pt

\sloppy
\cleardoublepage

\newcounter{totalpage}
\setcounter{totalpage}{\value{page}}
\setcounter{page}{1}

\pagestyle{fancy}
\fancyhead{}
\fancyfoot{}

\renewcommand{\chaptermark}[1]{\markboth{Kapitel \thechapter. #1}{}}
\renewcommand{\sectionmark}[1]{\markright{#1}}
\fancyhead[ER]{\slshape\leftmark}
\fancyhead[OL]{\slshape\rightmark}
\fancyhead[EL,OR]{\bfseries\thepage}
\fancyfoot[EL,OR]{ \phantom{\AA{}} }
\renewcommand{\headrulewidth}{.5pt}
\renewcommand{\footrulewidth}{.5pt}
\renewcommand{\thefootnote}{\arabic{footnote}}
\newcommand{\overref}[2]{\overset{\text{\footnotesize #1}}{#2}}


\chapter{Einleitung}\label{s1:Einleitung}

\setcounter{page}{1}
\renewcommand{\thepage}{\arabic{page}}

\section{Historischer Rückblick}

Das Phänomen der Interferenz, die nichts anderes ist als die Superposition von zwei oder mehr Wellen, 
ist - ob nun bewusst oder unbewusst - jedem Kind bekannt, das einmal zwei Steine gleichzeitig
ins Wasser geworfen hat. Die Amplituden der beteiligten Wellen addieren sich dabei auf und man kann Effekte
wie Auslöschung und Verstärkung an verschiedenen Punkten der interferierenden Wellen beobachten.

Dass man mit der Interferenz auch hochpräzise, experimentelle Messungen machen kann, ist in der Physik schon
seit sehr langer Zeit bekannt. So nutzte z.B. der Michelson-Morley-Versuch \cite{Mic1881,MiMo1887}, 
der Ende des 19. Jahrhunderts die Existenz des Äthers zeigen sollte die Wellennatur des Lichts
und die damit verbundenen Interferenzeffekte aus. Dieser Versuch
basierte auf einem Lichtstrahl, der in zwei zueinander senkrechte Strahlen aufgespalten wurde, 
die dann nach Reflexion wieder miteinander vereinigt und zur Interferenz gebracht wurden.
Der Grundgedanke des Michelson-Morley-Experiments war, dass der sogenannte Ätherwind 
zu einer Laufzeitdifferenz und somit zu einem Unterschied der relativen Phase der beiden Lichtstrahlen führt,
so dass - abhängig von der Ausrichtung des Versuchsaufbaus - relativ zum Ätherwind Interferenzeffekte
zu beobachten sind. Die Messgenauigkeit ist dabei im wesentlichen durch
die halbe Wellenlänge des verwendeten Lichts gegeben und liegt somit im Bereich von einigen hundert Nanometern.
Der Versuch brachte jedoch nicht das damals gewünschte Ergebnis sondern lieferte im Gegenteil
einen sehr eindrucksvollen Beweis dafür, dass es den Äther nicht gibt.

Der entscheidende Schritt auf dem Weg zur Interferenz von Materiewellen war die 1924 von Louis Victor, Duc de Broglie
in seiner Dissertation \cite{deBr25} eingeführte Beziehung $\lambda = h/(m v)$, 
die einem Teilchen der Masse $m$ mit einem Impuls $mv$ eine Wellenlänge $\lambda$ zuordnet 
($h$ ist das Plancksche Wirkungsquantum).
Durch diese Beziehung wurde der bereits von den Eigenschaften des Lichts bekannte Welle-Teilchen-Dualismus nun auch auf 
massenbehaftete Teilchen übertragen. 
Die Wellennatur von Atomen wurde wenige Jahre später von Estermann und Stern \cite{EsSt30} 
durch den Nachweis von Beugung an einem Kristallgitter experimentell bestätigt, doch das erste Interferenzexperiment
mit Materiewellen (Elektronen) folgte erst in den 50er Jahren von J. Marton {\em et al.} \cite{Mar52,MaSiSu54}.
Interferometrie mit Neutronen folgte in den 60er \cite{MaSp62} und 70er \cite{RaTrBo74} Jahren, bis dann erstmals 
im Jahre 1991 ein Interferometer für Atome \cite{Pri91} vorgestellt wurde. Der wesentliche Vorteil der Interferometrie
mit Atomen gegenüber der Interferometrie mit Elektronen oder Neutronen ist ihre wesentlich komplexere, innere Struktur,
durch die sich eine Vielzahl von theoretischen und experimentellen Möglichkeiten eröffnet.

Die Atominterferometrie hat sich seit der damaligen Zeit weiterentwickelt 
und findet heute zahlreiche Anwendungen in der Atomoptik. Für einen Überblick über das Gebiet sei
auf die Artikel \cite{Pri01, Ber97} und die darin zitierten Referenzen verwiesen.
Die vorliegende
Arbeit beschäftigt sich mit der theoretischen Beschreibung eines speziellen Atominterferometers, das
in Heidelberg von Prof. D. Dubbers und PD M. DeKieviet, PhD., und Mitarbeitern \cite{ASSE1} 
im Jahre 1995 zunächst für Helium
und später auch für Wasserstoff und Deuterium \cite{DissAR, DeHaRe06}
entwickelt wurde und auf der Messung atomarer Spinechosignale\footnote{
Der Zusammenhang zwischen dem Spinecho, das man aus der Kernspinresonanz kennt, und
dem hier angewendeten Strahlspinecho wird z.B. in \cite{DissAR}, Kapitel 1 ausführlich diskutiert.}
basiert. 
Hierbei wird ein sogenanntes Spinechofeld angelegt, das aus zwei hintereinander geschalteten, in Strahlrichtung
orientierten, antiparallelen Magnetfeldern besteht. Die Energieniveaus der verschiedenen atomaren
Gesamtdrehimpulszustände spalten im Magnetfeld unterschiedlich auf, was zu einem Auseinanderdriften 
der zugeordneten Wellenpakete führt. Im zweiten, antiparallelen Magnetfeld kann durch geschickte
Wahl der Feldstärke ein erneutes Zusammenlaufen der Wellenpakete verursacht werden, so dass es
zu einem Überlapp kommt. Durch einen geeigneten Detektor kann man dann die Oszillation der relativen
Phase der beteiligten Wellenpakete in Abhängigkeit des angelegten Spinechofeldes 
als Interferenzsignal (Spinechosignal) messen. 

Wie in \cite{ASSE1} gezeigt wurde, ist das Heidelberger Atomstrahl-Spinecho-Experiment hochsensitiv auf 
Änderungen der relativen Phase der atomaren Wellenpakete 
und ist somit ein interessantes Werkzeug für die Untersuchung einer Reihe von Phänomenen,
z.B. in der Oberflächenphysik \cite{ASSE-App1}. Eine weitere Anwendung, die uns in der vorliegenden Arbeit
als Motivation für die Entwicklung einer Theorie zur Beschreibung von Atomstrahl-Spinecho-Experimenten dient, 
ist die Messung paritätsverletzender Effekte in den leichtesten Atomen - Wasserstoff
und Deuterium - bei denen man theoretische Rechnungen bis zu großer Genauigkeit durchführen kann.

Bevor wir diesen Zusammenhang weiter diskutieren, wollen wir zunächst einen kurzen Überblick über die historische
Entwicklung der Untersuchung von Paritätsverletzung in Atomen geben. Der erste Hinweis auf eine Verletzung der
Parität in der Natur war das sogenannte $\theta$-$\tau$-Rätsel. Mit diesen Buchstaben bezeichnete man zwei Teilchen
mit gleicher Masse und gleicher Ladung, die aber in zwei bzw. drei Pionen zerfallen, also in Endzustände
mit unterschiedlicher Parität. Lee und Yang \cite{LeYa56} zogen 1956 daher erstmals die Verletzung
der Spiegelsymmetrie in der schwachen Wechselwirkung in Erwägung und betrachteten in diesem Zusammenhang 
$\theta$ und $\tau$ als ein und dasselbe Teilchen. Heute ist die P-Verletzung in
der vereinheitlichten Theorie der elektroschwachen Wechselwirkung ein fester Bestandteil des
Standardmodells \cite{Wei67, Sal68, Gla70}.

In einem Atom ist das Elektron an den positiven Atomkern hauptsächlich durch die elektromagnetische Wechselwirkung
gebunden, die durch Austausch virtueller Photonen vermittelt wird. Die schwache Wechselwirkung trägt zu dieser
Bindung nur zu einem kleinen Teil durch den sogenannten neutralen schwachen Strom zwischen Elektron und den Quarks im 
Atomkern bei, der durch den Austausch virtueller, neutraler $Z$-Bosonen vermittelt wird, die eine Masse von
etwa $90\u{GeV}$ haben.

Lange vor der experimentellen Entdeckung des neutralen schwachen Stroms\footnote{Für einen historischen Überblick
über die Entdeckung des neutralen schwachen Stroms verweisen wir auf \cite{Hai04}.}
untersuchte Ya. B. Zel'dovich \cite{Zel59} im Jahre 1959 den Effekt eines hypothetischen
aber mit dem heutigen neutralen schwachen Strom vergleichbaren Beitrag zur Wechselwirkung zwischen Elektron und
Atomkern. Er stufte die damit verbundenen Phänomene allerdings als unbeobachtbar ein\footnote{
Eine gute Darstellung der historischen Entwicklung der
Untersuchung der Paritätsverletzung in Atomen findet sich in \cite{Bud99}.}. 
Erst viele Jahre später, nachdem der experimentelle Nachweis des neutralen schwachen Stroms erbracht
und das Standardmodell der Elementarteilchen aufgestellt war,
zeigten M.A. und C. Bouchiat \cite{Bou74} im Jahre 1974, dass der Beitrag der schwachen Wechselwirkung zur Bindung
des Elektrons an den Atomkern ungefähr mit der dritten Potenz der Kernladungszahl $Z$ skaliert.
Dies war der Beginn der intensiven theoretischen und experimentellen Untersuchung der Paritätsverletzung in
Atomen (APV\footnote{In der Literatur wird häufig die Abkürzung APV für {\em Atomic Parity Violation} verwendet.}).

Bereits Ende der 70er Jahre wurde der experimentelle Nachweis der Paritätsverletzung in Atomen
von zwei unabhängigen Gruppen an Wismut (Bi, $Z=83$) und Thallium (Tl, $Z=81$) erbracht \cite{BaZo79,Con79} und bis heute
werden Experimente auf diesem Gebiet aufgrund der bereits erwähnten Skalierung mit der
Kernladungszahl ausschließlich an schweren Atomen durchgeführt. Die Hauptmotivation für die Untersuchung
der APV ist auch heute noch die Messung der schwachen Kernladung $Q_W$, aus der man
den Weinberg-Winkel $\vth_W$ ableiten kann. Das besondere dabei ist, dass der Zugang, den die APV
für die Bestimmung des Weinberg-Winkels liefert, komplementär zum Zugang der Hochenergiephysik ist.
Wir verweisen für eine ausführliche Diskussion dieser Thematik auf einen Übersichtsartikel 
der Bouchiats \cite{Bou97} und für einen aktuellen Überblick über den derzeitigen Stand der 
Untersuchung der APV auf \cite{Bou05}.

\section{Motivation}

Wir interessieren uns in dieser Arbeit für Paritätsverletzung in leichten Atomen, insbesondere Wasserstoff und
Deuterium. Wie oben bereits erwähnt, kann man bei diesen einfachen atomaren Systemen numerische
Rechnungen bis zu hoher Genauigkeit durchführen. Die experimentellen Methoden werden immer empfindlicher,
wie das Beispiel des Atomstrahl-Spinechos zeigt, so dass nach unserer Meinung die berechtigte Hoffnung besteht,
auch bei diesen leichten Atomen P-verletzende Effekte zu messen. Neben der oben diskutierten Motivation
der Messung der schwachen Kernladung kann ein Vergleich der theoretischen und experimentellen Resultate
eine sehr genaue Bestätigung des Standardmodells bei niedrigen Energien liefern, aber auch Licht in
das Dunkel der sogenannten Spinkrise \cite{Ash88,Ash89} bringen. 
Mit diesem Begriff wird die auch heute noch bestehende
Unsicherheit \cite{AsFl06} über die genaue Zusammensetzung des Proton- bzw. Neutronspins bezeichnet, 
der nur zu einem kleinen Teil von den drei Valenzquarks getragen wird. Für eine ausführlichere Diskussion
dieses Zusammenhangs verweisen wir auf \cite{BoBrNa95}.

Bereits 1983 wurden von Prof. O. Nachtmann und W. Bernreuther \cite{BeNa83} paritätsverletzende Rotationen
der Polarisation von Wasserstoff in elektrischen Feldern theoretisch untersucht, in den 90er Jahren wurden diese
Ideen weiter ausgebaut \cite{BoBrNa95,DissTG,BrGaNa99,GaNa00}.
Es wurde eine Methode zur resonanten Verstärkung der P-verletzenden Polarisations-Rotationen
in elektrischen Feldern vorgeschlagen \cite{BoBrNa95}, P-verletzende Energieverschiebungen und geometrische Phasen
\cite{DissTG,BrGaNa99} studiert, sowie P-verletzende Effekte in atomarem Dysprosium bei beinahe entarteten
Energieniveaus untersucht \cite{GaNa00}.

Im Rahmen der dieser Arbeit vorangegangenen Diplomarbeit \cite{DiplTB} untersuchten wir
abermals P-verletzende Polarisations-Rotationen in elektrischen Feldern, 
damals am Beispiel von Deuterium. Wir zeigten die Existenz P-erhaltender Polarisations-Rotationen
von Deuterium, einem wesentlich stärkeren Effekt als die P-verletzenden Rotationen, der aus diesem Grund
sehr geeignet wäre für ein erstes, vorbereitendes Atomstrahl-Spinecho-Experiment auf dem Weg zur Messung P-verletzender
Effekte in leichten Atomen.

Aus der direkten Nachbarschaft zur Atomstrahl-Spinecho-Gruppe um PD M. DeKieviet, PhD., und der vielversprechenden
Messgenauigkeit der Apparatur entstand die Motivation, die theoretische Untersuchung der P-verletzenden Effekte bei
leichten Atomen in eine quantenmechanische Beschreibung eines Atomstrahl-Spinecho-Experiments einzubetten. In allen
vorherigen Arbeiten wurden Atome in Ruhe in zeitabhängigen Potentialen betrachtet, bei denen man sich auf 
eine quantenmechanische Beschreibung im Hilbertraum der atomaren Zustände beschränken konnte. Bei einem 
Atomstrahl-Spinecho-Experiment dagegen muss man die Atome darüberhinaus als orts- und zeitabhängige Wellenpakete
beschreiben, die die vorgegebene Feldkonfiguration mit einer festgelegten Geschwindigkeitsverteilung
durchqueren.

Im Rahmen dieser Arbeit beschränken wir uns auf die Betrachtung des sogenannten longitudinalen Atomstrahl-Spinechos,
bei dem es nur eine Strahlachse gibt und die Wellenpakete entlang dieser Achse auseinanderdriften und wieder
zusammengeführt werden. Im Gegensatz dazu gibt es noch die Möglichkeit des transversalen Atomstrahl-Spinechos, bei
dem die Strahlen räumlich getrennt und wieder zusammengeführt werden. Wir beschränken uns weiterhin nur
auf zeitunabhängige elektrische und magnetische Felder. 

\section{Gliederung der Arbeit}

Die vorliegende Arbeit liefert eine ausführliche Herleitung einer Theorie des longitudinalen
Atomstrahlstrahl-Spinechos. Obwohl diese Theorie insbesondere für neutrale, wasserstoffähnliche, metastabile
Atome in statischen elektrischen und magnetischen Feldern entwickelt wurde, sollte eine Übertragung auf
andere Systeme (mit zeitunabhängigem Matrixpotential) ohne größere Schwierigkeiten möglich sein.

In Kapitel \ref{s2:Vorbetrachtungen} gehen wir erste Schritte auf dem Weg zur Beschreibung eines 
Atomstrahl-Spinecho-Experiments, indem wir zwei einfache Näherungslösungen berechnen, die auf ähnlichen Ansätzen
basieren. Wir diskutieren die physikalischen Eigenschaften der beiden Lösungen.

In Kapitel \ref{s3:Anwendungen} betrachten wir einige Anwendungen der physikalisch sinnvolleren Näherungslösung,
die auf der Superposition ebener WKB-Wellen basiert. Wir untersuchen Reflexion und Transmission an einer
Potentialstufe und am Potentialwall und geben in Abschnitt \ref{s3:Fahrplan} bereits eine theoretische
Beschreibung des sogenannten Fahrplanmodells \cite{DeHaRe06} im adiabatischen Grenzfall an.
In diesem Zusammenhang berechnen wir bereits ein erstes Spinechosignal.

In Kapitel \ref{s4:Formalismus} entwickeln wir eine Methode zur Berechnung der Lösung der 
Schrödinger-Gleichung mit reellem, skalaren Potential in Form einer
Reihenentwicklung nach Potenzen eines Integraloperators $\hat K$. Mit Hilfe dieser Methode sind wir dann in der
Lage, die Näherungslösungen aus Kapitel \ref{s2:Vorbetrachtungen} quantitativ zu analysieren und zu vergleichen.
Wir untersuchen desweiteren die Korrekturen erster Ordnung zur WKB-Näherungslösung und geben ein Verfahren zur
Berücksichtigung der Dispersion bereits ab nullter Ordnung der Entwicklung an.

In Kapitel \ref{s5:Zerfall} betrachten wir komplexe, skalare Potentiale, bei denen das Wellenpaket mit fortschreitender 
Zeit zerfällt. Wir demonstrieren die Fähigkeit des Formalismus aus Kapitel \ref{s4:Formalismus} 
zur Selbstkorrektur des ursprünglich angesetzten Phasenfaktors.

In Kapitel \ref{s6:Formalismus} übertragen wir alle bis dahin gesammelten Erkenntnisse 
auf den Fall matrixwertiger Potentiale und gelangen so zu einer Theorie für die 
Beschreibung longitudinaler Atomstrahl-Spinecho-Experimente. Wir führen eine Erweiterung des Fahrplanmodells
auf den nichtadiabatischen Fall ein und geben die allgemeine Vorgehensweise für die Berechnung
eines Atomstrahl-Spinecho-Signals an.

In Kapitel \ref{s7:Berry} stellen wir dann die Verbindung zur Paritätsverletzung in Atomen her
und betrachten als Beispiel paritätsverletzende geometrische Phasen. Wir entwickeln eine grafische Methode
zur Veranschaulichung geometrischer Phasen, die im Spezialfall eines konstanten magnetischen (elektrischen)
Feldes und veränderlicher elektrischer (magnetischer) Feldstärke angewendet werden kann. Schließlich zeigen
wir die Existenz P-verletzender geometrischer Phasen für eine spezielle Feldkonfiguration.

Kapitel \ref{s8:Summary} schließt den Hauptteil der Arbeit mit einer Zusammenfassung 
und einem Ausblick auf zukünftige Arbeiten ab.

In Anhang \ref{sA:QMBasics} werden einige quantenmechanische Grundlagen angegeben, die bei den Rechnungen
der vorliegenden Arbeit verwendet wurden. Nach einer Zusammenfassung der für die Beschreibung metastabiler Atome
relevanten Wigner-Weisskopf-Methode in Abschnitt \ref{sA:WWF} folgt in Abschnitt \ref{sA:ZweiTeilchen} 
eine grundlegende Diskussion des quantenmechanischen Zwei-Teilchen-Problems, die zur Aufstellung der
matrixwertigen Schrödinger-Gleichung führt. Daraufhin geben wir in Abschnitt \ref{sA:StoeRe} Grundlagen
für die Störungsrechnung mit nichthermiteschen Massenmatrizen an, die wir in Abschnitt \ref{sA:OrtsablMat}
zur Berechnung lokaler Matrixdarstellungen von Ortsableitungsoperatoren einsetzen. Abschnitt \ref{sA:Parity}
schließt Anhang \ref{sA:QMBasics} mit einer Hilfsrechnung für die Paritätstransformation der in Kapitel \ref{s7:Berry}
behandelten geometrischen Flussdichten ab.

In Anhang \ref{sB:MundEWP} werden alle Grundlagen für die numerische Behandlung von metastabilem Wasserstoff und Deuterium
in elektrischen und magnetischen Feldern inklusive der P-verletzenden Beiträge zur Massenmatrix angegeben.
Dort finden sich zahlreiche Tabellen mit den Matrixdarstellungen der Operatoren für das magnetische Moment
und das Dipolmoment von Wasserstoff und Deuterium, sowie die zugehörigen freien Massenmatrizen.

Anhang \ref{sC:WFT} behandelt die sogenannte Gabor-Transformation, die eine Methode zur Superposition von Funktionen
mit lokalisierten Wellenpaketen ist. Eine vereinfachte Version dieser Transformation wird in Abschnitt \ref{s4:Dispersion}
für das Verfahren zur besseren Berücksichtigung der Dispersion der Wellenpakete in der Reihenentwicklung der Lösung der 
Schrödinger-Gleichung verwendet.

\chapter{Vorbetrachtungen}\label{s2:Vorbetrachtungen}

\newcommand{\tfracMk}{\tfrac M{\overset{ }{\bar k}}}

\section{Longitudinales Atomstrahl-Spinecho}\label{s2:LABSE}

\begin{figure}[h!]
\centering
\subfloat[]{\label{f2:lABSE}\includegraphics[width=12cm]{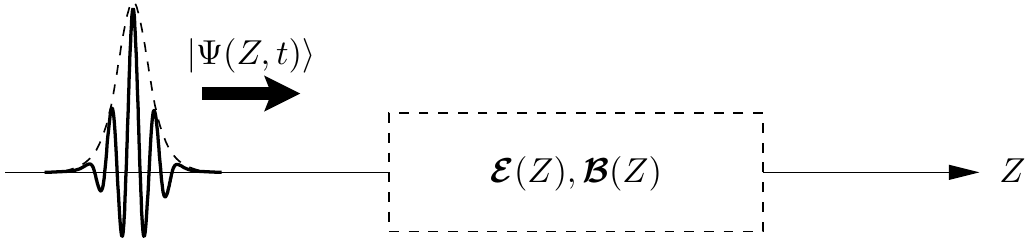}}\\
\subfloat[]{\label{f2:Spinechofeld}\includegraphics[width=12cm]{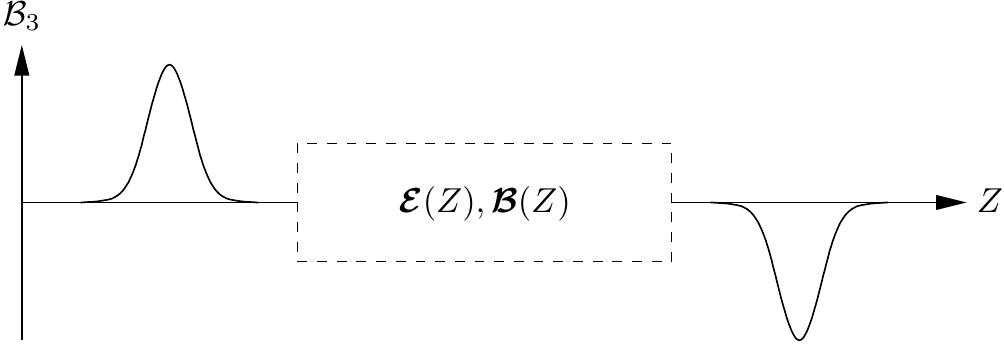}}
\caption[Schematischer Aufbau eines longitudinalen Atomstrahl-Spinecho-Experiments.]{
Schematischer Aufbau eines longitudinalen Atomstrahl-Spinecho-Experiments. Erläuterungen siehe Text.}
\label{f2:lABSE_Schema}
\end{figure}

In diesem Kapitel wollen wir erste Schritte auf dem Weg zur theoretischen Beschreibung eines longitudinalen
Atomstrahl-Spinecho-Experiments gehen, bei dem ein Atom eine Anordnung statischer elektrischer und magnetischer Felder durchqueren soll, siehe Abb. \subref*{f2:lABSE}.
Bereits diese sehr oberflächlichen Beschreibung des in der vorliegenden Arbeit betrachteten Spezialfalls des
lABSE\footnote{Die Abkürzung ABSE für den englischen Ausdruck {\em Atomic Beam Spin Echo} ist üblich. 
Da wir hier nur das longitudinale ABSE untersuchen, verwenden wir in diesem Zusammenhang die Abkürzung lABSE.} 
zeigt, was eine entsprechende Theorie leisten muss, nämlich die quantenmechanische Beschreibung 
des Atoms als Wellenpaket unter dem Einfluss der durch die elektrischen und magnetischen Felder erzeugten Potentiale.

Wie in der Einleitung bereits erklärt, befindet sich die gewünschte Feldkonfiguration beim lABSE in der Regel 
zwischen den sogenannten Spinecho-Feldern, siehe Abb. \subref*{f2:Spinechofeld}. 
Dies sind einfach zwei antiparallele Magnetfelder in Strahlrichtung, die dafür sorgen, 
dass die Wellenpakete aufgrund unterschiedlicher Potentiale im ersten Magnetfeld 
zunächst auseinander driften und im zweiten Feld wieder zusammengeführt werden, so dass es durch den Überlapp je zweier
Teilwellenpakete zu beobachtbaren Interferenzmustern kommt. In Abschnitt \ref{s3:Fahrplan} werden wir all dies an
einem einfachen Beispiel genauer erläutern. Zuvor jedoch wollen wir im folgenden Abschnitt \ref{s2:Charakteristiken}
einen ersten einfachen Ansatz für die Beschreibung eines Atoms in einem matrixwertigen Potential machen.
Die Unzulänglichkeiten, die sich bei diesem Ansatz offenbaren werden, machen eine genauere Untersuchung
von zunächst eindimensionalen Wellenpaketen in skalaren Potentialen notwendig, die in Abschnitt
\ref{s2:WKB} begonnen wird und erst in Kapitel \ref{s6:Formalismus} wieder auf den Fall matrixwertiger
Potentiale ausgedehnt wird.

\section{Die Charakteristiken-Lösung}\label{s2:Charakteristiken}

\subsection{Eine Näherungslösung der Schrödinger-Gleichung}\label{s2:Bewegungsgleichung}

Beginnen wir mit einer einfachen, qualitativen Überlegung: Der quantenmechanische
Gesamtzustand $\ket{\Psi(Z,t)}$ des Atoms ist eine Superposition innerer atomarer Zustände\footnote{Die 
genauere Spezifizierung dieser inneren atomaren Zustände wollen wir auf
einen späteren Zeitpunkt verschieben. In diesem einführenden Abschnitt \ref{s2:Charakteristiken} 
wollen wir uns zunächst ein grobes Bild der zu beschreibenden Situation machen. Der ungeduldige Leser
sei bereits auf Anhang \ref{sA:ZweiTeilchen} verwiesen, wo entsprechende Grundlagen ausführlich behandelt werden.}.
Zusätzlich muss das Atom im Ortsraum als zeitabhängige Wellenfunktion beschrieben werden, also
von der Koordinate $Z$ und der Zeit $t$ abhängen. Insgesamt kann also das Atom als ein
verallgemeinerter Spinor geschrieben werden, dessen Komponenten die Wellenfunktionen der inneren, atomaren Zustände
sind. Das elektrische und das magnetische Feld führen zur Wechselwirkung der Komponenten-Wellenfunktionen,
d.h. das durch die Felder verursachte, ortsabhängige Potential kann in Form einer Matrix $\uop M(Z)$
geschrieben werden.

Wir können also eine matrixwertige Schrödinger-Gleichung für den atomaren Gesamtzustand in der Form
\begin{align}\label{e2:SG}
  \klr{-\frac1{2M}\del_Z^2 + \uop M(Z)}\ket{\Psi(Z,t)} = \I\del_t\ket{\Psi(Z,t)}\ ,
\end{align}
ansetzen. Wir separieren im atomaren Gesamtzustand eine ebene Welle mit mittlerer Wellenzahl $\bar k$
und Energie $\bar\omega$ ab und schreiben
\begin{align}\label{e2:AnsatzWF}
  \ket{\Psi(Z,t)} = \e^{\I(\bar kZ-\bar\omega t)}\ket{A(Z,t)}\ .
\end{align}
Nun nehmen wir an, die Potentialmatrix $\uop M(Z)$ variiere nur langsam mit dem Ort und die zugrundeliegenden
Felder seien schwach genug, um nur geringe Unterschiede in den Potentialen der einzelnen atomaren Zustände
zu verursachen. Ferner nehmen wir an, wir betrachten nur sehr breite Wellenpakete im Ortsraum. 
Unter diesen Voraussetzungen wird auch der Amplituden-Spinor $\ket{A(Z,t)}$ im Ort nur langsam variieren
und wir können seine zweite Ortsableitung in der Schrödinger-Gleichung vernachlässigen.

Wir erhalten dann nach Einsetzen von (\ref{e2:AnsatzWF}) in Gl. (\ref{e2:SG})
\begin{align}\label{e2:SG.2}
\begin{split}
(\bar\omega + \I\del_t)\ket{A(Z,t)} &= 
\kle{-\frac1{2M}\klr{-\bar k^2+2\I \bar k\del_Z + \del_Z^2}+ \uop M(Z)}\ket{A(Z,t)}\\
&\approx \kle{-\frac1{2M}\klr{-\bar k^2+2\I \bar k\del_Z}+ \uop M(Z)}\ket{A(Z,t)}
\end{split}
\end{align}
Nun setzen wir $\bar\omega = \bar k^2/2M$ und erhalten damit die in der hier gemachten Näherung gültige 
Schrödinger-Gleichung
\begin{align}\label{e2:SG.3}
  \klr{\I\frac{\bar k}M\del_Z + \I\del_t - \uop M(Z)}\ket{A(Z,t)} = 0
\end{align}
für den Amplituden-Spinor $\ket{A(Z,t)}$.

Eine Standard-Methode zum Lösen dieser einfachen partiellen Differentialgleichung (DGL) ist die Methode der Charakteristiken
(eine gute Darstellung findet sich z.B. in \cite{MaWa70}, Kap. 8.2). Hierbei führt man eine Variablentransformation durch,
die die partielle DGL zweier Variabler in eine gewöhnliche DGL einer Variablen überführt. Wir definieren hier
\begin{align}\label{e2:Transformation}
  \begin{split}
    \eta  &:= \frac12\klr{t + \frac M{\bar k}Z}\ ,\\
    \xi &:= \frac12\klr{t - \frac M{\bar k}Z}\ .
  \end{split}
\end{align}
Damit gilt die Rücktransformation
\begin{align}\label{e2:Ruecktransform.}
  \begin{split}
    t &= \eta + \xi\ ,\\
    Z &= \frac{\bar k}M(\eta - \xi)\ ,
  \end{split}
\end{align}
und die Differentialoperatoren lauten
\begin{align}\label{e2:Ableitungen}
  \del_Z &= \ddp{\eta}{Z}\del_\eta + \ddp{\xi}Z\del_\xi 
     = \frac m{2\bar k}(\del_\eta - \del_\xi)\ ,\\
  \del_t &= \ddp{\eta}{t}\del_\eta + \ddp{\xi}t\del_\xi =
  \frac12(\del_\eta + \del_\xi)\ ,
\end{align}
und somit können wir Gl. (\ref{e2:SG.3}) umschreiben in
\begin{align}\label{e2:SG.4}
  \big[\I\del_\eta - \uop M(Z(\eta,\xi))\big]\ket{A(Z(\eta,\xi),t(\eta,\xi))} = 0
\end{align}
Mit der Definition $\ket{B(\eta,\xi)} := \ket{A(Z(\eta,\xi),t(\eta,\xi))}$
erhalten wir die Gleichung
\begin{align}\label{e2:SG.5}
  \big[\I\del_\eta - \uop
    M(Z(\eta,\xi))\big]\ket{B(\eta,\xi)} = 0\ ,
\end{align}
deren Lösung
\begin{align*}
  \ket{B(\eta,\xi)} = \mc P\exp\klg{-\I\int_{\eta_0}^\eta\d\eta'\ \uop
    M(Z(\eta',\xi))}\ket{B(\eta_0,\xi)}\ ,\qquad (\eta\geq\eta_0)
\end{align*}
lautet. Die Einschränkung $\eta>\eta_0$ ist wichtig,
da die Pfadordnung, die hier bzgl. der Integrationsvariable $\eta'$
zu nehmen ist, ansonsten die Operatoren $\uop M(Z(\eta',\xi))$ in die
verkehrte Reihenfolge bringen würde\footnote{Der Pfad wird stets von $\eta_0\to\eta$
durchlaufen, die Pfadordnung würde für $\eta<\eta_0$ die Operatoren jedoch
so anordnen, also würde der Pfad von $\eta\to\eta_0$ durchlaufen werden.\label{fn:1}}.
Dieses Problem umgeht man mit einer
Parametrisierung des Weges von $\eta_0$ nach $\eta$:
\begin{align}\label{e2:Parametrisierung}
  \eta'(\vrh) = (1-\vrh)\eta_0 + \vrh\eta,\qquad (0\leq \vrh\leq 1)\ .
\end{align}
Dann wird nämlich
\begin{align}\label{e2:Def.Eta.Rho}
  \d\eta' = (\eta-\eta_0)\d\vrh
\end{align}
und die Lösung kann nun für beliebige Werte von $\eta$ geschrieben werden als
\begin{align}\label{e2:C.Lsg.tmp}
  \ket{B(\eta,\xi)} = \mc P\exp\klg{-\I\int_{0}^1\d\vrh\ (\eta-\eta_0)
    \uop M(Z(\eta'(\vrh),\xi))}\ket{B(\eta_0,\xi)}
\end{align}
Dabei steht $\mc P$ für die Pfadordnung/Zeitordnung bzgl. des
Parameters $\vrh$.  Im Argument der Matrix $\uop M(Z(\eta'(\vrh),\xi))$ 
ist $Z$ gemäß der Transformationsvorschrift (\ref{e2:Ruecktransform.}) 
und $\eta'(\vrh)$ aus (\ref{e2:Def.Eta.Rho}) einzusetzen.
Betrachtet man Gl. (\ref{e2:C.Lsg.tmp}) genauer, so erkennt man, dass der Operator
\begin{align}\label{e2:Konnektor}
\umc V(\eta,\eta_0;\xi)\ket{B(\eta_0,\xi)} := \mc P\exp\klg{-\I\int_{0}^1\d\vrh\ (\eta-\eta_0)
\uop M(Z(\eta'(\vrh),\xi))}
\end{align}
den Zustand $\ket{B(\eta_0,\xi)}$ mit dem Zustand $\ket{B(\eta,\xi)}$ verbindet, und zwar entlang
einer Linie mit konstantem $\xi$. Der Operator $\umc V(\eta,\eta_0;\xi)\ket{B(\eta_0,\xi)}$ ist
ein Konnektor (siehe z.B. \cite{NaQCD96}, Gl. (2.23 f.)) mit den folgenden Eigenschaften:
\begin{itemize}
\item {\bf Neutrales Element:} Der Konnektor entlang eines geschlossenen Pfades ist
  gleich dem Einheitsoperator:
  \begin{align}\label{e2:Konn.Eins}
    \umc V(\eta,\eta;\xi) = \unl\um\ .
  \end{align}
\item {\bf Inverser Konnektor:} Den inversen Konnektor erhält man durch
  Vertauschen von Anfangs- und Endpunkt und dem rückwärts
  durchlaufenen Pfad, d.h. hier: bei festem $\xi$ sind die
  Integrationsgrenzen zu vertauschen:
  \begin{align}\label{e2:Konn.Invers}
    \umc V(\eta_2,\eta_1;\xi)\umc V(\eta_1,\eta_2;\xi) = \unl\um\ .
  \end{align}
\item {\bf Produkt von Konnektoren:} Das Produkt zweier Konnektoren
  für zwei verbundene Pfade entspricht dem Konnektor über den
  gesamten Pfad, d.h. hier bei festem $\xi$ gilt:
  \begin{align}\label{e2:Konn.Produkt}
    \umc V(\eta_3,\eta_2;\xi)\umc V(\eta_2,\eta_1;\xi) = \umc
    V(\eta_3,\eta_1;\xi)
  \end{align}
\end{itemize}

\subsection{Eigenschaften der Charakteristiken-Lösung}\label{s2:Anfangsbedingungen}

\subsubsection{Anfangsbedingungen}

Ein Problem der Lösung aus Gl. (\ref{e2:C.Lsg.tmp}) sind die Anfangsbedingungen. Ausgehend von
einem Amplituden-Spinor $\ket{A(Z,t=t_0)}$ zur Zeit $t_0$ interessiert man sich für den Amplituden-Spinor
zu einer Zeit $t>t_0$. Das $\eta$-$\xi$-Koordinatensystem, in dem die Lösung (\ref{e2:C.Lsg.tmp}) lebt,
ist aber relativ zum $Z$-$t$-Koordinatensystem gedreht (da sowohl $\eta$ als auch $\xi$ lineare Funktionen in $Z$
und $t$ sind). Um eine anschauliche Lösung zu erhalten, müssen wir Gl. (\ref{e2:C.Lsg.tmp}) in die ursprünglichen
Koordinaten $(Z,t)$ zurücktransformieren. Dazu verwenden wir die Eigenschaften der eben eingeführten Konnektoren.

Mit Hilfe der Transformationsvorschrift (\ref{e2:Transformation}) identifizieren wir
\begin{align}\label{e2:A.B.t0}
  \ket{A(Z,t_0)} =\ket{B(\eta(Z,t_0),\xi(Z,t_0))} = \ket{B(\tfrac12(t_0+\tfrac
  M{\bar k}Z),\tfrac12(t_0-\tfracMk Z))}\ .
\end{align}
Nun setzen wir auf der rechten Seite die Lösung (\ref{e2:C.Lsg.tmp}) ein und erhalten
für beliebiges $\eta_0$
\begin{align}\label{e2:A.B.t0.2}
  \ket{A(Z,t_0)} &= \umc V(\tfrac12(t_0+\tfrac
  M{\bar k}Z),\eta_0;\xi)\ket{B(\eta_0,\xi)}
\end{align}
mit
\begin{align}\label{e2:xi.t0}
  \xi &= \tfrac12(t_0-\tfracMk Z)\ .
\end{align}
Wir können nun die Lösung zu einer späteren Zeit $t$ durch erneutes Anwenden eines Konnektors
erhalten. Dabei müssen wir allerdings beachten, dass der Konnektor nur Punkte entlang Linien mit konstantem
$\xi$ verbindet. Gemäß
\begin{align}\label{e2:xi.t}
  \xi = \tfrac12(t-\tfracMk Z') \overref{(\ref{e2:xi.t0})}=
  \tfrac12(t_0-\tfracMk Z)
\end{align}
gelangen wir also im $Z$-$t$-Koordinatensystem auch zu einem anderen Ort $Z'$. Anhand von
Gl. (\ref{e2:xi.t}) können wir in Verbindung mit der Definition von $\xi(Z,t)$ und $\eta(Z,t)$
aus Gl. (\ref{e2:Transformation}) den Wert von $\eta$ bestimmen, für den wir den Konnektor
benötigen. Es folgt
\begin{align}\label{e2:Amplitude.1}
  \ket{A(Z',t)} = \ket{B(\eta(Z',t),\xi(Z',t))} = \umc V(\tfrac12(t +
  \tfracMk Z'),\eta_0;\xi)\ket{B(\eta_0,\xi)}\ .
\end{align}
Um die Verbindung zwischen dem Anfangswert $\ket{A(Z,t_0)}$ und $\ket{A(Z',t)}$
herzustellen, lösen wir (\ref{e2:A.B.t0.2}) nach $\ket{B(\eta_0,\xi)}$ auf und
setzen das Ergebnis in (\ref{e2:Amplitude.1}) ein. Es folgt
\begin{align}\label{e2:Amplitude.2}
  \ket{A(Z',t)} = \umc V(\tfrac12(t+\tfracMk Z'),\eta_0;\xi) \umc
  V^{-1}(\tfrac12(t_0+\tfracMk Z),\eta_0;\xi)\ket{A(Z,t_0)}\ .
\end{align}
Mit Hilfe der Relationen (\ref{e2:Konn.Invers}) und (\ref{e2:Konn.Produkt}) 
\begin{align}
  \ket{A(Z',t)} = \umc V(\tfrac12(t+\tfrac
  M{\bar k}Z'),\tfrac12(t_0+\tfracMk Z);\xi)\ket{A(Z,t_0)}\ .
\end{align}
Die beiden Werte $Z$ und $Z'$ hängen über Gl. (\ref{e2:xi.t})
voneinander ab. Wir ersetzen daher
\begin{align}
  Z = Z' - \tfrac{\bar k}M(t-t_0)
\end{align}
und erhalten
\begin{align}\label{e2:C.Amplitude.0}
  \ket{A(Z',t)} = \umc V(\tfrac12(t+\tfrac
  M{\bar k}Z'),t_0-\tfrac12(t-\tfracMk Z');\xi)
  \ket{A(Z'-\tfrac{\bar k}M (t-t_0),t_0)}\ .
\end{align}
Setzen wir die explizite Darstellung des Konnektors aus (\ref{e2:C.Lsg.tmp}) ein, so folgt
schließlich nach Umbenennung von $Z'$ in $Z$
\begin{align}\label{e2:C.Amplitude.tmp}
  \ket{A(Z,t)} &= \mc P\exp\klg{-\I\int_{0}^{1} \d\vrh\ (t-t_0)\uop M(
    Z(\eta'(\vrh),\xi))}\ket{A(Z-\tfrac{\bar k}M(t-t_0),t_0)}
\end{align}
mit $\eta'(\vrh)$ aus (\ref{e2:Parametrisierung}), sowie $\eta$ und $\xi$ aus
(\ref{e2:Transformation}).

Wir können im pfadgeordneten Integral die Substitution $Z' = \tfrac{\bar k}M(\eta'-\xi)$ durchführen
(siehe (\ref{e2:Ruecktransform.})), was wegen $\xi=\const$ zu 
$\d\eta' = \tfracMk \d Z'$ führt. Wegen (\ref{e2:Parametrisierung})
und (\ref{e2:C.Amplitude.0}) folgt damit $\d\eta' = (t-t_0)\d\vrh = \tfrac M{\bar
  k}\d Z'$ und somit $Z'(\vrh) = \tfracMk (t-t_0)\vrh + Z'_0$.
Mit $Z'(1) = Z$ und $Z'(0) = Z'_0 = Z - \tfrac{\bar k}M(t-t_0)$ erhalten wir schließlich
\begin{align}\label{e2:Lsg.C}\hspace{-1cm}
  \boxed{\ket{A(Z,t)} = \mc P\exp\klg{-\I\frac{M}{\bar k}\int_{Z-\tfrac{\bar k}M
        (t-t_0)}^{Z} \d Z'\ \uop M(Z')}
    \ket{\tilde\vph\klr{Z-\frac{\bar k}{M}(t-t_0)}}\ ,\quad(t\geq t_0)}\ .
\end{align}
Wir haben hier eine Umbenennung
$\ket{\tilde\vph\klr{Z-\frac{\bar k}{M}(t-t_0)}}=\ket{A(Z-\tfrac{\bar k}M(t-t_0),t_0)}$
durchgeführt, um Vergleiche mit den Ergebnissen der Rechnungen in den folgenden Kapiteln zu erleichtern. 
Die Einschränkung $t\geq t_0$ ist wieder notwendig, damit
die Pfadordnung das richtige Ergebnis liefert (siehe Fußnote \ref{fn:1}
auf Seite \pageref{fn:1}).

\subsubsection{Grafische Veranschaulichung der Charakteristiken-Lösung}

\begin{figure}[!Hb]
\centering
\includegraphics[width=14cm]{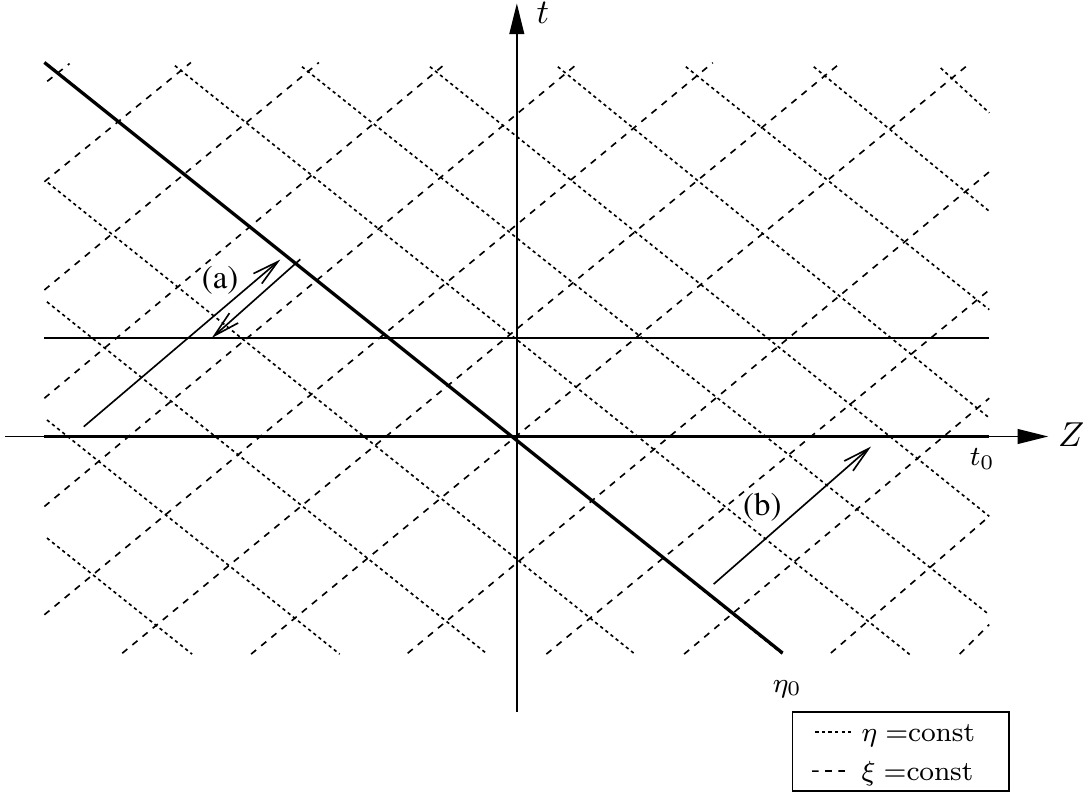}
\caption[Anfangswertproblem der Charakteristiken-Lösung]{
Geometrische Veranschaulichung der Problematik des Anfangswertproblems.
Die Bewegungsgleichung für das Wellenpaket wurde in den neuen Koordinaten $(\eta,\xi)$
gelöst. Die gestrichelten Linien sind die Linien mit jeweils konstantem $\eta$
bzw. $\xi$.}
\label{f2:Visualisierung} 
\end{figure}

Wir wollen nun die eben gemachte Rechnung und Vorgehensweise grafisch
veranschaulichen. Zunächst betrachten wir Gl. (\ref{e2:A.B.t0.2}). In der
grafischen Interpretation entspricht diese Gleichung der Verbindung
eines Punktes $(\eta_0,\xi)$ mit $(z(\eta,\xi),t_0(\eta,\xi))$
über den Konnektor, angedeutet durch den Pfeil (b) in Abb.
\ref{f2:Visualisierung}. Der Wert $\ket{B(\eta_0,\xi)}$ wird durch
(\ref{e2:A.B.t0.2}) auf den Wert $\ket{A(Z,t_0)}$ abgebildet.

Stellen wir die Gleichung nach $\ket{B(\eta_0,\xi)}$ um, so kehrt der
Pfeil seine Richtung um, d.h. es wird der Anfangswert $\ket{A(Z,t_0)}$ mit
dem Wert $\ket{B(\eta_0,\xi)}$ verbunden. Da wir aber an $\ket{A(Z',t)}$ zu
einer festen Zeit $t$ interessiert sind, muss der Wert
$\ket{B(\eta_0,\xi)}$ erneut mit $\ket{A(Z',t)}$ verbunden werden.  Dies ist mit
den beiden Pfeilen (a) auf der linken Seite von Abb. \ref{f2:Visualisierung}
angedeutet und entspricht Gl. (\ref{e2:Amplitude.2}).

Die beiden Konnektoren, die je zwei Funktionswerte entlang desselben
Pfades verbinden, lassen sich schließlich zu einem einzigen Konnektor
zusammenfassen, so dass am Ende nur die Verbindung von $\ket{A(Z,t_0)}$ mit
$\ket{A(Z',t)}$ verbleibt.
\FloatBarrier

\subsubsection{Direkte Implementation der Anfangsbedingungen}

Motiviert von dieser grafischen Veranschaulichung wollen wir nun versuchen, den
direkten Weg von $\ket{A(Z,t_0)}$ zu $\ket{A(Z',t)}$ gehen. Dazu setzen wir bei
der Differentialgleichung (\ref{e2:SG.4}) an, deren Lösung (mit korrekter
Parametrisierung des Pfades, s. Gl. (\ref{e2:Parametrisierung}))
\begin{align}\label{e2:Amplitude.3}
  \begin{split}
    \ket{A(Z(\eta,\xi),t(\eta,\xi))} &= \mc P\exp\klg{-\I\int_0^1 
      \d\vrh\ (\eta-\eta_0)\uop M(Z(\eta'(\vrh),\xi))}\\
     & \times \ket{A(Z(\eta_0,\xi),t(\eta_0,\xi))}
  \end{split}
\end{align}
lautet. Gemäß Abb. \ref{f2:Visualisierung} wird der Wert des Amplituden-Spinors an einer Stelle
$(Z_0,t_0)$ entlang der Linie mit festem $\xi$ mit dem Wert $\ket{A(Z,t)}$
über das pfadgeordnete Integral (Konnektor) verbunden. Interessiert
man sich nun für den Wert $\ket{A(Z,t)}$, so legt dies
eindeutig den Wert von $\xi$ fest, s. Gl. (\ref{e2:Transformation}).  Bei
gegebener Anfangszeit $t_0$ kann man damit auf die ursprüngliche
Stelle $Z_0$ schließen, mit dem der Punkt $(Z,t)$ verbunden ist. Es
gilt
\begin{align}\label{e2:xi}
&\xi \overref{(\ref{e2:Transformation})}= \tfrac12\klr{t-\tfrac M{\overset{ }{\bar k}}Z} =
  \tfrac12\klr{t_0-\tfracMk Z_0} \intertext{und somit}\label{e2:z(t-t0,z0)}
  &Z(t-t_0,Z_0) = \tfrac{\bar k}M(t-t_0) + Z_0\ ,\qquad Z_0(t-t_0,Z) = Z
  - \tfrac{\bar k}M(t-t_0)\ .
\end{align}
Im Konnektor werden die Anfangs- und Endwerte $\eta_0,\eta$ für die
betrachteten Punkte $(Z_0,t_0)$, $(Z,t)$ benötigt. Es gilt nach
(\ref{e2:Transformation}):
\begin{align}\label{e2:Eta0.Eta}
  \begin{split}
    \eta_0 &= \tfrac12\klr{t_0+\tfracMk Z_0} 
      = \tfrac12\klr{2t_0 - t +\tfracMk Z} = t_0 - \xi,\\
    \eta &= \tfrac12\klr{t+\tfracMk Z},\qquad \eta-\eta_0 = t-t_0\ .
  \end{split}
\end{align}
Nach (\ref{e2:Ruecktransform.}) gilt außerdem
\begin{align}
  Z(\eta_0,\xi) = \frac{\bar k}M(\eta_0 - \xi)
\overref{(\ref{e2:xi}),(\ref{e2:Eta0.Eta})}= Z_0,\quad t(\eta_0,\xi) =
  \eta_0 + \xi = t_0\ .
\end{align}
Damit lautet die Lösung (\ref{e2:Amplitude.3})
\begin{align}\label{e2:C.Amplitude}
  \begin{split}
    \ket{A(Z,t)} &= \mc P\exp\klg{-\I\int_{0}^{1} \d\vrh\ 
      (t-t_0)\uop M(\tfrac{\bar k}M(\eta'(\vrh) - \xi))}
    \ket{A(Z_0,t_0)}\ .
  \end{split}
\end{align}
Dieses Ergebnis ist, nach Einsetzen von $Z_0$ aus (\ref{e2:z(t-t0,z0)}), völlig
identisch mit (\ref{e2:C.Amplitude.tmp}) und somit äquivalent zu (\ref{e2:Lsg.C}). 
Wir haben es hier jedoch ohne die Verwendung der Zwischenfunktion
$\ket{B(\eta_0,\xi)}$ erhalten. 

\subsubsection{Abschließende Bemerkungen zur Charakteristiken-Lösung}

Eine interessante Eigenschaft der Lösung (\ref{e2:C.Amplitude}) sieht man im adiabatischen Grenzfall. 
Dann ist die Potentialmatrix $\uop M(Z)$ diagonal, die Pfadordnung entfällt und es treten keinerlei Mischungen 
zwischen Wellenpaketen unterschiedlicher Zustände auf. 
Jedes Wellenpaket bekommt dann gemäß der Lösung einen dynamischen Phasenfaktor. 
Da die Wellenpakete nur von einer einzigen, orts- und zeitabhängigen Funktion 
$Z_0(Z,t) = Z-\tfrac{\bar k}M(t-t_0)$ abhängen, gibt es keine Dispersion.
Desweiteren ist die Geschwindigkeit der Wellenpakete mit $\bar v = \bar k/M$ auch in den Potentialen stets konstant.

Offenbar hat also die Verwendung des sehr einfachen Ansatzes (\ref{e2:AnsatzWF}) 
mit dem Phasenwinkel einer ebenen Welle in Verbindung mit der
Vernachlässigung der zweiten Ortsableitung des Amplituden-Spinors zu einer sehr eingeschränkt gültigen
Näherungslösung geführt. Dies soll uns als Motivation dienen, nach besseren Lösungsverfahren der matrixwertigen
Schrödinger-Gleichung zu suchen. Insbesondere wollen wir im folgenden Abschnitt \ref{s2:WKB}
zunächst eine verbesserte Näherung für ein eindimensionales Wellenpaket in einem reellen Potential untersuchen. 
Erst nachdem wir diese Näherungslösung eingehend studiert haben, kommen wir in Kap. \ref{s5:Zerfall}
zunächst zu komplexen, skalaren Potentialen und in Kap. \ref{s6:Formalismus} dann schließlich wieder zurück
zu matrixwertigen Potentialen.

\section{Die WKB-Lösung}\label{s2:WKB}

Ab sofort betrachten wir nur stationäre, reelle, skalare Potentiale die von einer Ortskoordinate $z$ abhängen. 
Erst in Kapitel \ref{s5:Zerfall} werden wir zu komplex-skalaren Potentialen und in
Kapitel \ref{s6:Formalismus} dann schließlich zu allgemeinen Matrixpotentialen übergehen.
Die im vorangegangenen Abschnitt \ref{s2:Charakteristiken} verwendeten Spinoren 
$\ket{\Psi(Z,t)}$ und $\ket{A(Z,t)}$ werden nun also
durch skalare Funktion $\Psi(z,t)$ und $A(z,t)$ ersetzt. Da die quantenmechanischen 
Überlegungen der folgenden Kapitel allgemeinerer Natur sind und sich nicht mehr direkt auf die
Beschreibung eines Atoms beziehen, schreiben wir
anstelle der Gesamtmasse $M$ des Atoms und seiner Schwerpunktskoordinate $Z$ nun allgemeiner $m$ für die Masse
des durch die Wellenfunktion repräsentierten Teilchens und $z$ für die Ortskoordinate.

\subsection{Eine verbesserte Näherungslösung der Schrödinger-Gleichung}\label{s2:WKB.Lsg}

Macht man in der ursprünglichen Schrödinger-Gleichung (mit reellem, skalaren Potential $V(z)$),
\begin{align}\label{e2:s.SG.1}
\klr{\del_z^2 - 2mV(z) + 2m\I\del_t}\Psi(z,t) = 0\ ,
\end{align}
den allgemeinen Ansatz
\begin{align}\label{e2:Allg.AnsatzWF}
\Psi(z,t) = \e^{\I\phi(z,t)}A(z,t)\ ,
\end{align}
so erhält man nach Einsetzen in die Schrödinger-Gleichung
\begin{align}\label{e2:s.SG.2}
\begin{split}
  0 &= (\I \del_z^2\phi(z,t) - (\del_z\phi(z,t))^2 - 2mV(z) - 2m\del_t\phi(z,t))A(z,t)\\ 
    &+ 2\I(\del_z\phi(z,t))(\del_z A(z,t)) + \del_z^2A(z,t) + 2m\I\del_t A(z,t)\ .
\end{split}
\end{align}
Vernachlässigt man nun generell alle zweiten Ortsableitungen, d.h.
\begin{align}\label{e2:s.SG.3}
\del_z^2 \phi(z,t) \approx 0,\qquad \del_z^2A(z,t) \approx 0\ ,
\end{align}
und betrachtet weiterhin nur Lösungen, die der Forderung
\begin{align}\label{e2:PW.Bedingung}
(\del_z\phi(z,t))^2 + 2mV(z) + 2m\del_t\phi(z,t) = 0
\end{align}
genügen, so muss für die Amplitudenfunktion $A(z,t)$ die Gleichung
\begin{align}\label{e2:s.SG.4}
\frac1m\klr{\del_z\phi(z,t)}\klr{\del_z A(z,t)} + \del_t A(z,t) = 0
\end{align}
gelten. 

Die Gleichung (\ref{e2:PW.Bedingung}) für den Phasenwinkel $\phi(z,t)$ wird gelöst von
\begin{align}\label{e2:Lsg.PW}
\phi_±(z,t) = -\frac{\bar k^2}{2m}t ±\int^z_{z_0}\d z'\ \sqrt{\bar k^2 - 2mV(z')}\ ,
\end{align}
wobei $\bar k$ die mittlere Wellenzahl des betrachteten Teilchens ist. Die Gesamtenergie des
Teilchens, die im potentialfreien Fall identisch mit der kinetischen Energie ist, lautet demzufolge
\begin{align}\label{e2:Eges}
E\subt{ges} = \frac{\bar k^2}{2m}\ .
\end{align}
Wir wählen hier die Lösung $\phi_+(z,t)$, die dem Phasenwinkel einer bei
fortschreitender Zeit in positive $z$-Richtung laufenden ebenen Welle entspricht. Im Folgenden
sei also
\begin{align}\label{e2:PW.WKB}
\phi(z,t) = -\frac{\bar k^2}{2m}t + \int^z_{z_0}\d z'\ \sqrt{\bar k^2 - 2mV(z')}\ .
\end{align}
Setzen wir dies in die Gl. (\ref{e2:s.SG.4}) für die Amplitudenfunktion ein, so folgt
\begin{align}\label{e2:s.SG.5}
\frac{\sqrt{\helpkbar\bar k^2-2mV(z)}}m\del_z A(z,t) +\del_t A(z,t) = 0\ ,
\end{align}
was von jeder Funktion $A(z,t)=\vph(\zeta(z)-\tau(t))$ erfüllt wird, sofern man
die neuen Koordinaten $\zeta(z)$ und $\tau(t)$ definiert als
\begin{align}\label{e2:Def.Zeta.Tau}
\boxed{
\zeta(z):=\int_{z_0}^z\d z'\ \frac{\bar k}{\sqrt{\phantom{\big\vert}\!\bar k^2-2mV(z')}}\ ,\qquad\tau(t) := \frac{\bar k}mt
}\ .
\end{align}
Insgesamt lautet die Wellenfunktion also nach Einsetzen aller Größen
\begin{align}\label{e2:Lsg.Alternativ}
\Psi(z,t) = \exp\klg{-\I\frac{\bar k^2}{2m}t + \I\int^z_{z_0}\d z'
\ \sqrt{\bar k^2 - 2mV(z')}}\vph(\zeta(z)-\tau(t))\ .
\end{align}

Der im Phasenwinkel von Gl. (\ref{e2:Lsg.Alternativ}) enthaltene Anteil
\begin{align}\label{e2:SWKB}
S_{\bar k}(z) := \int_{z_0}^z\d z'\ \sqrt{\bar k^2-2mV(z')}
\end{align}
entspricht dem Phasenwinkel der WKB-Näherung. Die WKB-Methode (siehe z.B. \cite{Messiah}, Bd. 1, Kap. 6.2) 
liefert eine semiklassische Näherungslösung der Schrödinger-Gleichung in dem Sinne, 
dass $\hbar$ als kleiner Parameter betrachtet wird, nach dem die Wellenfunktion entwickelt wird 
und Beiträge der Ordnung $\OO(\hbar^2)$ vernachlässigt werden. Für eine stationäre, ebene Welle
mit Wellenzahl $k$ erhält man dann die Lösung
\begin{align}\label{e2:WKBW}
\psi_k\supt{WKB}(z) = A_k\supt{WKB}(z)\exp\klg{\I S_k(z)}\ ,
\end{align}
mit $S_k(z)$ wie in (\ref{e2:SWKB}) und
\begin{align}\label{e2:Amplitude.WKB}
A_k\supt{WKB}(z) = \frac{C}{\sqrt{\del_z S_k(z)}}\ .
\end{align}
Es zeigt sich, dass diese Lösung genau dann eine gute Näherung ist, wenn
\begin{align}\label{e2:WKB-Naeherung}
\frac{|m \del_z V(z)|}{\abs{(k^2-2m V(z))}^{3/2}} \ll 1
\end{align}
erfüllt ist, d.h. die Änderung des Potentials auf der Skala der lokalen Wellenlänge muss
vernachlässigbar klein sein.

In einem Atomstrahl-Spinecho-Experiment wie wir es in der vorliegenden Arbeit betrachten wollen,
muss diese Näherung nicht unbedingt erfüllt sein, insbesondere wenn nicht-adiabatische Übergänge
berücksichtigt werden müssen. Zunächst wollen wir aber annehmen, die WKB-Näherung sei erfüllt und
Korrekturen zur Näherungslösung später untersuchen. 

Eine Näherung, die bei lABSE-Experimenten in der Regel sehr gut erfüllt ist, 
ist die, dass das Potential $V(z)$ sehr viel kleiner
als die Gesamtenergie (\ref{e2:Eges}) des Teilchens ist (siehe \cite{DissAR}, Abschnitt 3.3.1, S. 53).
Tatsächlich gilt etwa
\begin{align}\label{e2:ABSE.kin.E.}
\frac{V(z)}{E\subt{ges}} = \frac{2mV(z)}{\bar k^2} \approx 10^{-5}\ .
\end{align}
In diesem Fall wird sich die lokale Wellenzahl, definiert durch
\begin{align}
k(z) := \del_zS_{\bar k}(z) = \sqrt{\bar k^2 - 2mV(z)}
\end{align}
nur wenig ändern und man kann für die WKB-Amplitude
\begin{align}\label{e2:AWKB.app}
A_k\supt{WKB}(z) \approx 1
\end{align}
annehmen. Aus den WKB-Wellen (\ref{e2:WKBW}) mit Amplitude $A_k\supt{WKB}(z) \approx 1$ wollen wir nun
ein Wellenpaket superponieren. Es sei also
\begin{align}\label{e2:WP.WKB.FT}
\Psi(z,t) = \int\d k\ \tilde\Psi(k-\bar k)\psi_k(z,t)\ ,
\end{align}
wobei die Amplitudenfunktion $\tilde\Psi(k-\bar k)$ bei Null ein stark ausgeprägtes Maximum haben
soll, d.h. $\bar k$ ist der mittlere Impuls des freien Wellenpakets.

Mit der Näherung (\ref{e2:AWKB.app}) und der Substitution
$k' = k-\bar k$ folgt weiter
\begin{align}\label{e2:WP.WKB.FT.approx}
\Psi(z,t) = \int\d k'\ \tilde\Psi(k')\ \exp\klg{-\I\frac{(\bar k + k')^2}{2m}t
+ \I\int^z_{z_0}\d z'\ \sqrt{(\bar k+k')^2 - 2mV(z')}}\ .
\end{align}
In der Entwicklung 
\begin{align}\label{e2:kbar+k.approx}
(\bar k+k')^2 = \bar k^2 + 2k'\bar k + {k'}^2 \approx \bar k^2 + 2k'\bar k\ .
\end{align} 
ist die Vernachlässigung von ${k'}^2$ gerechtfertigt durch die geforderte, starke Lokalisierung von $\tilde\Psi(k)$ um
Null, d.h. $\abs{k'/\bar k}\ll1$. Wir können also schreiben
\begin{align}\label{e2:WP.WKB.FT.approx.2}
\begin{split}
\Psi(z,t) &= \int\d k'\ \tilde\Psi(k')
\ \exp\klg{-\I\frac{\bar k^2}{2m}t - \I\frac{k'\bar k}mt}\\
&\times \exp\klg{\I\int^z_{z_0}\d z'\ \sqrt{\bar k^2 - 2mV(z') + 2k'\bar k}}
\end{split}
\end{align}
und erhalten mit
\begin{align}\label{e2:lokWZ.approx}
\begin{split}
\sqrt{\bar k^2 - 2mV(z') + 2k'\bar k} &= 
\sqrt{\bar k^2 - 2mV(z')}\klr{1 + \frac{2k'\bar k}{\bar k^2-2mV(z')}}^{1/2} \\
&\approx \sqrt{\bar k^2-2mV(z')}\klr{1 + \frac{k'\bar k}{\bar k^2-2mV(z')}}
\end{split}
\end{align}
die Wellenfunktion
\begin{align}\label{e2:WP.WKB.FT.approx.3}
\begin{split}
\Psi(z,t) &= \int\d k'\ \tilde\Psi(k')
\ \exp\klg{-\I\frac{\bar k^2}{2m}t - \I\frac{k'\bar k}mt}\\
&\times \exp\klg{\I\int^z_{z_0}\d z'\ \sqrt{\bar k^2 - 2mV(z')}  
+ \I \frac{k'\bar k}m\int^z_{z_0}\d z'\ \frac m{\sqrt{\helpkbar\bar k^2 - 2mV(z')}}} \ .
\end{split}
\end{align}
Wir können nun den von $k'$ unabhängigen Anteil aus dem
Integral herausziehen und erhalten schließlich
\begin{align}\label{e2:WP.WKB.FT.approx.4}
\begin{split}
\Psi(z,t) &= \exp\klg{-\I\frac{\bar k^2}{2m}t 
+ \I\int^z_{z_0}\d z'\ \sqrt{\bar k^2 - 2mV(z')}}\\
&\times \int\d k'\ \tilde\Psi(k')
\ \exp\klg{- \I\frac{k'\bar k}m
\klr{t -\int^z_{z_0}\d z'\ \frac m{\sqrt{\helpkbar\bar k^2 - 2mV(z')}}}} \ .
\end{split}
\end{align}
Der Phasenfaktor vor dem Integral ist bereits identisch mit dem Phasenfaktor aus (\ref{e2:Lsg.Alternativ}).
Das Integral selbst ist eine Fouriertransformation der Funktion $\tilde\Psi(k')$ an der Stelle
\begin{align}
\zeta(z) - \tau(t) = \frac{\bar k}m \klr{\int^z_{z_0}\d z'\ \frac m{\sqrt{\helpkbar\bar k^2 - 2mV(z')}}-t}\ ,
\end{align}
in der wir die bereits in (\ref{e2:Def.Zeta.Tau}) definierten Funktionen $\zeta(z)$ und $\tau(t)$
wiedererkennen. Die Fouriertransformierte von $\tilde\Psi(k')$ wollen wir mit dem Buchstaben $\vph$ bezeichnen, d.h.
wir erhalten hier die Amplitudenfunktion
\begin{align}\label{e2:DefAWP}
\boxed{\vph(\zeta(z)-\tau(t)) := \int\d k'\ \tilde\Psi(k') \e^{\I k'(\zeta(z)-\tau(t))}}
\end{align}
und für die gesamte Wellenfunktion den Ausdruck
\begin{align}\label{e2:Lsg.WKB}
\boxed{
\Psi(z,t) = \exp\klg{-\I\frac{\bar k^2}{2m}t 
+ \I\int^z_{z_0}\d z'\ \sqrt{\bar k^2 - 2mV(z')}}
\vph(\zeta(z)-\tau(t))}\ .
\end{align}
Dieses Ergebnis stimmt mit Gl. (\ref{e2:Lsg.Alternativ}) überein und zeigt somit, dass die Bezeichnung dieser
Wellenfunktion als WKB-Lösung gerechtfertigt ist, denn wir haben hier WKB-Wellen superponiert. Wir wollen 
zum Abschluss dieses Abschnitts noch einmal alle Näherungen auflisten, die wir bei der Herleitung der Lösung (\ref{e2:Lsg.WKB}) gemacht haben:
\begin{enumerate}[W1)]
\item WKB-Näherung (\ref{e2:WKB-Naeherung}), \label{WKB.WKB}
\item $E\subt{ges} = \tfrac{\bar k^2}{2m}\gg V(z)$, d.h. $A_k\supt{WKB}(z) \approx 1$, \label{WKB.E}
\item $\bar k \gg \Delta k$, wobei $\Delta k$ die Breite der Fouriertransformierten $\tilde\Psi(k)$ der Amplitudenfunktion
$\vph(\zeta)$ im Impulsraum sein soll. \label{WKB.Breit}
\end{enumerate}

\subsection{Eigenschaften der WKB-Lösung}\label{s2:WKB-Eigenschaften}

\subsubsection{Anfangsbedingungen}

Die Anfangsbedingungen für die Wellenfunktion (\ref{e2:Lsg.WKB}) sind leicht zu implementieren. Zur Zeit $t=0$
wird nämlich
\begin{align}\label{e2:Psi.t0}
\Psi(z,t=0) = \exp\klg{\I\int^z_{z_0}\d z'\ \sqrt{\bar k^2 - 2mV(z')}}\vph(\zeta(z))\ .
\end{align}
Wir können o.B.d.A. stets davon ausgehen, dass die Amplitudenfunktion $\vph(\zeta)$ um $\zeta = 0$ zentriert ist.
Somit ist nach Definition von $\zeta(z)$ die Amplitudenfunktion $\vph(\zeta(z))$ im Ortsraum um die in $\zeta(z)$
vorkommende Integrationsuntergrenze $z_0$ zentriert. Wählt man $z_0$ so, dass das Anfangswellenpaket vollständig
im potentialfreien Bereich lokalisiert ist\footnote{
Diese Annahme wird uns im weiteren Verlauf der Arbeit noch sehr häufig begegnen.
}, so kann man, wegen $k(z)=\bar k$ für $\vph(\zeta(z))\neq 0$,
vereinfachend $\zeta(z)$ durch $z-z_0$ ersetzen und erhält
\begin{align}\label{e2:Psi.t0.2}
\Psi(z,t=0) = \exp\klg{\I\bar k\,(z-z_0)}\vph(z-z_0)\ .
\end{align}
Hier kann man nun leicht die gewünschte Einhüllende $\vph(z-z_0)$ des Anfangswellenpakets im Ortsraum festlegen.

\subsubsection{Gaußsches Wellenpaket}

\begin{figure}[!ht]
\centering
\includegraphics{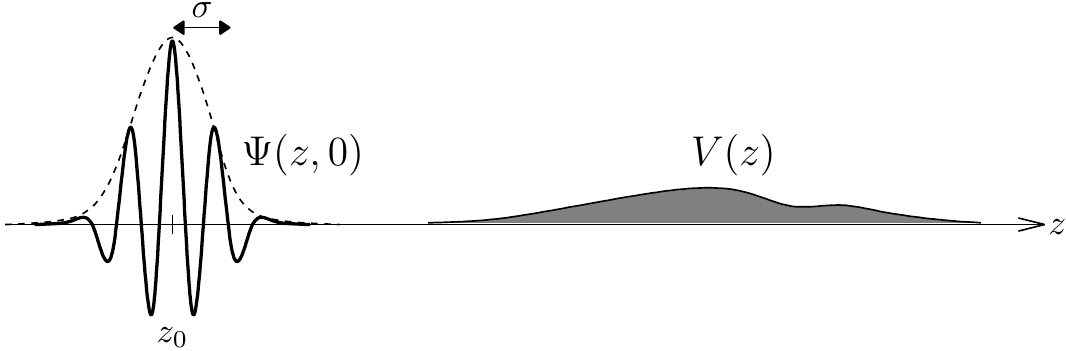}
\caption[Gaußsches Anfangswellenpaket]{Das Gaußsche Wellenpaket zur Zeit $t=0$.}
\label{f3:Gausspaket}
\end{figure}

Wählen wir z.B. als Anfangsbedingung ein um $z_0$ zentriertes Wellenpaket mit Wellenzahl $\bar k$, d.h.
\begin{align}\label{e2:GaussWP}
\Psi(z,t_0 = 0) = \mc N(\sigma)\exp\klg{\I \bar k (z-z_0) -\tfrac{(z-z_0)^2}{2\sigma^2}}
\end{align}
mit
\begin{align}\label{e2:NormierungGauss}
\mc N(\sigma) = (\sigma\sqrt\pi)^{-1/2}\ ,
\end{align}
so ist die Einhüllende gemäß (\ref{e2:Psi.t0.2}) gegeben durch
\begin{align}\label{e2:GaussEnvelope}
\vph(z-z_0) = (\sigma\sqrt\pi)^{-1/2}\exp\klg{-\tfrac{(z-z_0)^2}{2\sigma^2}}\ .
\end{align}
Die Breite\footnote{Wir richten uns bei der Definition des Gaußschen Wellenpakets nach
\cite{Cohen}, Kap. 1.11.1. Als Breite bezeichnen wir nach den Konventionen von
\cite{Cohen} die Breite von $\abs{\Psi(z,t)}^2$ bzgl. des $1/\e$-fachen des Maximums.} 
$2\sigma$ und $z_0$ seien so gewählt, dass das Potential $V(z)$ über die gesamte 
Ausdehnung des Wellenpakets als Null angesehen werden kann, siehe Abb. \ref{f3:Gausspaket}.
\FloatBarrier

Unter diesen Voraussetzungen gilt nach (\ref{e2:DefAWP})
\begin{align}
\begin{split}
\vph(z-z_0) &= \mc N(\sigma)\exp\klg{\-\tfrac{(z-z_0)^2}{2\sigma^2}}
= \int\d k'\ \tilde\Psi(k') \exp\klg{\I k' (z-z_0)}
\end{split}
\end{align}
und daher mit $z' = z-z_0$
\begin{align}\label{e2:Gauss.FT}
\begin{split}
\tilde\Psi(k') = \frac{\mc N(\sigma)}{2\pi}\int\d z'
\ \exp\klg{-\frac{{z'}^2}{2\sigma^2}}\exp\klg{-\I k' z'} 
= \frac\sigma{\sqrt{2\pi}} N(\sigma) \exp\klg{-\frac{\sigma^2}2{k'}^2}\ .
\end{split}
\end{align}
Die Breite des Gaußschen Wellenpaketes im Impulsraum ist
also $2/\sigma$ (vgl. $2\sigma$ im Ortsraum), d.h. für
$k' = \pm \tilde\sigma := \pm 1/\sigma$ ist die Amplitude im Impulsraum auf $1/\sqrt e$
abgefallen. Wir verlangen gemäß der Näherung (\ref{e2:kbar+k.approx})
\begin{align}
(\bar k+k')^2 = \bar k^2 + 2k'\bar k + {k'}^2 = \bar k^2\kle{1 + \frac{2k'}{\bar k} +
\klr{\frac{k'}{\bar k}}^2}
\approx  \bar k^2\kle{1 + \frac{2k'}{\bar k}}\ ,
\end{align}
also
\begin{align}\label{e2:kbar.k.approx.1}
2\abs{\frac{k'}{\bar k}}\gg \abs{\frac{k'}{\bar k}}^2
\end{align}
für alle $k'$, bei denen $\tilde\Psi(k')$ signifikant von Null verschieden ist.
Dies impliziert für die Breite $\tilde\sigma = 1/\sigma$ im Impulsraum
\begin{align}\label{e2:kbar.k.approx.2}
2\gg \frac{\tilde\sigma}{\bar k} = \frac1{\sigma\bar
k}\qquad\Rightarrow\qquad\sigma\gg\frac1{2\bar k} = \frac{\bar \lambda}{4\pi}
\end{align}

Anschaulich heißt dies, dass die Breite des Wellenpakets
viel größer sein muss, als die de-Broglie-Wellenlänge der
Atome im Strahl. Geht man von einem Strahl atomaren Wasserstoffs
mit mittlerer Geschwindigkeit $\bar v=3500\u m/\u s$ aus\footnote{M. DeKieviet, private Diskussion.}, so
ergibt sich eine de-Broglie-Wellenlänge von etwa $\bar\lambda = 0,11\u {nm}$ und
folglich muss für die Breite des Wellenpakets in etwa $\sigma\gg 0,01\u {nm} = 0,1 $\AA\ gelten.

\subsubsection{Dispersionsfreiheit und Schwerpunktsbewegung}

Man sieht Gl. (\ref{e2:Lsg.WKB}) direkt an, dass die WKB-Lösung dispersionsfrei ist. Die Einhüllende
$\vph(\zeta(z)-\tau(t))$ hängt effektiv nur von einer einzigen Koordinate ab, die in der WKB-Lösung
eine Funktion von Ort und Zeit ist. Mit dem Preis der Dispersionsfreiheit erkauft man sich aber
eine sehr einfach zu verstehende Schwerpunktsbewegung\footnote{Der Ausdruck Schwerpunktsbewegung für ein
dispersionsfreies Wellenpaket ist eigentlich nicht nötig, da ja das gesamte Wellenpaket mit gleicher 
Geschwindigkeit propagiert. Man kann aber davon ausgehen, dass zumindest im adiabatischen Grenzfall auch der Schwerpunkt
eines dispergierenden Wellenpakets in guter Näherung die hier berechnete Bewegung erfährt.}.
Hat $\vph(\zeta-\tau)$ nämlich ein ausgeprägtes Maximum an der Stelle $\hat\zeta$, so bewegt
dieses sich gemäß 
\begin{align}\label{e2:Maximum}
\hat\zeta = \zeta(\hat z(t)) - \tau(t)
= \int_{z_0}^{\hat z(t)}\d z'\ 
\frac{\bar k}{\sqrt{\phantom{\big\vert}\!\bar k^2-2mV(z')}} - \frac{\bar k}mt= \const
\end{align}
fort, wobei $\hat z(t)$ der Ort des Maximums sei. Die Zeitableitung
dieser Gleichung, multipliziert mit $m/\bar k$,
\begin{align}
0 = \frac m{\sqrt{\phantom{\big\vert}\!\bar k^2-2mV(z(t))}}\dd{\hat z(t)}t - 1\ ,
\end{align}
führt auf eine Energie-Erhaltungsgleichung der Form
\begin{align}\label{e2:Schwerpunkt}
\frac12 m \klr{\dd{\hat z(t)}t}^2 + V(\hat z(t))= \frac{\bar k^2}{2m}\ .
\end{align}
Das Maximum des Wellenpakets bewegt sich also wie ein klassisches
Teilchen mit der Gesamtenergie $\frac{\bar k^2}{2m}$ durch das Potential. Dies spiegelt
auch den semiklassischen Charakter der WKB-Lösung wider.

Im Vergleich zur Charakteristiken-Lösung (\ref{e2:Lsg.C}) stellt dies bereits eine enorme
Verbesserung dar. Wie am Ende von Abschnitt \ref{s2:Anfangsbedingungen} bereits angesprochen,
bewegt sich der Schwerpunkt des Wellenpakets der Charakteristiken-Lösung (im Fall eines skalaren Potentials)
nämlich auch in Potentialen stets mit konstanter Geschwindigkeit $\bar v= \bar k/m$. Diese Geschwindigkeit
wird vom WKB-Wellenpaket nur im potentialfreien Raum eingenommen.

Die physikalisch sehr gut motivierte WKB-Lösung soll uns als Ausgangspunkt für die weitere Untersuchung
der Theorie des longitudinalen Atomstrahl-Spinechos dienen. Im folgenden Kapitel werden wir einige Anwendung
der WKB-Lösung studieren. Neben der Potentialstufe und dem Potentialwall werden wir bereits ein einfaches
lABSE-Experiment beschreiben und das sogenannte Fahrplanmodell einführen.

\chapter{Anwendungen der WKB-Lösung}\label{s3:Anwendungen}

\section{Streuung an der Potentialstufe}\label{s3:Potentialstufe}

Wir wollen nun einige Anwendungen der WKB-Lösung (\ref{e2:Lsg.WKB}) studieren. Wir beginnen
mit der Untersuchung des Verhaltens eines WKB-Wellenpakets an einer Potentialstufe, wobei wir
immer von den Näherungen (W1)-(W3) auf Seite \pageref{WKB.WKB} ausgehen.

\subsection{Rechnung für ebene WKB-Wellen}\label{s3:PS:WKBW}

\begin{figure}[!ht]
  \centering
  \includegraphics{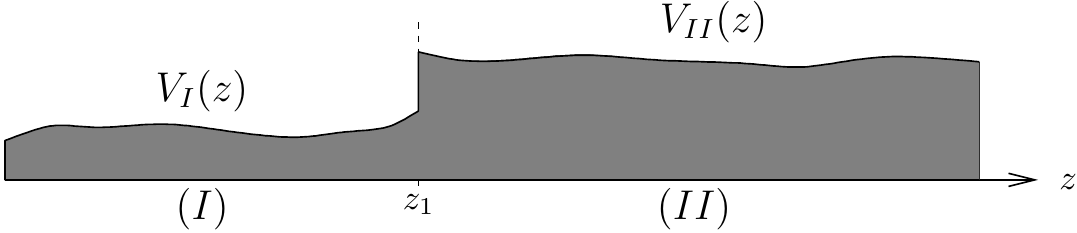}
  \caption{Allgemeines Potential mit einem Sprung bei $z=z_1$.}
  \label{f3:Potentialstufe}
\end{figure}

Wir betrachten nun eine Potentialstufe der Form
\begin{align}\label{e3:Potentialstufe}
  V(z) = \Theta(z_1-z) V_{I}(z)+\Theta(z-z_1)V_{II}(z)
\end{align}
mit einem Sprung bei $z=z_1$, siehe Abb \ref{f3:Potentialstufe}.

Um das Verhalten des WKB-Wellenpakets (\ref{e2:Lsg.WKB}) an der Potentialstufe
zu verstehen, müssen wir zunächst das Verhalten der ebenen WKB-Wellen (\ref{e2:WKBW}), wieder
in der Näherung (W2), $A\supt{WKB}_k(z)\approx 1$, untersuchen. Dann superponieren wir wie
im letzten Abschnitt die Lösungen mit festem $k$ erneut zu einem Wellenpaket.

Den Bereich, in dem  $V(z)=V_I(z)$ ist bezeichnen wir mit $I$, den Bereich mit
$V(z) = V_{II}(z)$ bezeichnen wir mit $II$. Wir verwenden im Folgenden aus Gründen
der Übersichtlichkeit die Abkürzungen
\begin{align}\label{e3:DefU}
U_I(z) = 2mV_I(z)\ ,\quad U_{II}(z) = 2mV_{II}(z)\ .
\end{align}
Im Bereich $I$ setzt sich die Lösung
aus einer einlaufenden und einer reflektierten Welle zusammen.
Mit der Näherung (W3), siehe Gl. (\ref{e2:kbar+k.approx}), folgt
\begin{align}\label{e3:PS:Psi.k.I}
  \begin{split}
    \psi_{k',I}(z,t) &\approx \exp\klg{-\I\frac{\bar k^2}{2m}t - \I\frac{k'\bar k}m t}\\
    &\times \Bigg[\exp\klg{\I\int_{z_0}^z\d z'\ \sqrt{\helpkbar\bar k^2-U_I(z')+2\bar kk'}}\\
    &\qquad+ R(k')\exp\klg{-\I\int_{z_0}^z\d z'\ \sqrt{\helpkbar\bar k^2-U_I(z')+2\bar kk'}}\Bigg]\ ,
  \end{split}
\end{align}
wobei wir gemäß (\ref{e2:kbar+k.approx}) quadratische Beiträge in $k'$
vernachlässigt haben. Die Entwicklung der Wurzeln verschieben
wir auf später, da die Rechnung dadurch nur unübersichtlich würde.
Der reflektierte Anteil erhält vor den Integralen ein zusätzliches Minuszeichen, da es sich
hierbei um eine nach links laufende Welle handelt.

Im Bereich $II$ gibt es nur einen transmittierten Anteil. Hier erhalten wir
\begin{align}
  \begin{split}
    \psi_{k',II}(z,t) &= S(k')\exp\klg{-\I\frac{\bar k^2}{2m}t - \I\frac{k'\bar k}m t}\\
    &\times \exp\klg{\I\int_{z_1}^z\d z'\ \sqrt{\helpkbar\bar k^2-U_{II}(z')+2\bar kk'}}\ .
  \end{split}
\end{align}
Die Amplitude $S(k')$ des transmittierten Anteils der Welle
ist nicht zu verwechseln mit dem WKB-Anteil des Phasenwinkels $S_k(z)$, definiert in (\ref{e2:SWKB}).
Die Untergrenze der Integration im Bereich $II$ wählen wir als $z_1$, so 
dass sich die Integration nur über $U_{II}$ erstreckt. Gäbe es noch weitere 
Unstetigkeitsstellen im Potential, so wären die Integrationsgrenzen 
auf ähnliche Weise zu wählen, so dass niemals über eine Unstetigkeitsstelle
integriert wird.

Die im Allgemeinen komplexwertigen Amplituden $R(k')$ und $S(k')$ des reflektierten bzw. transmittierten Anteils
der Welle berechnen wir aus den Forderungen der Stetigkeit der Wellenfunktion
und ihrer ersten Ableitung an der Potentialstufe bei $z=z_1$. 

\textbf{Stetigkeit der Wellenfunktion:} $\psi_{k',I}(z_1,t) \overset!= \psi_{k',II}(z_1,t)$
\begin{align}
  \begin{split}
    S(k') &= \exp\klg{\I\int_{z_0}^{z_1}\d z'\ \sqrt{\helpkbar\bar k^2-U_I(z')+2\bar kk'}}\\ 
      &\quad+ R(k')\exp\klg{-\I\int_{z_0}^{z_1}\d z'\ \sqrt{\helpkbar\bar k^2-U_I(z')+2\bar kk'}} 
  \end{split}
\end{align}
\textbf{Stetigkeit der ersten Ableitung der Wellenfunktion:} $\del_z\psi_{k',I}(z_1,t) \overset!=
\del_z\psi_{k',II}(z_1,t)$
\begin{align}
  \begin{split}
    S(k')\sqrt{\frac{\helpkbar\bar k^2-U_{II}(z_1)+2\bar kk'}{\helpkbar\bar k^2 - U_{I}(z_1)+2\bar kk'}} &= 
      \Bigg[\exp\klg{\I\int_{z_0}^{z_1}\d z'\ \sqrt{\helpkbar\bar k^2-U_I(z')+2\bar kk'}}\\
    &\qquad- R(k')\exp\klg{-\I\int_{z_0}^{z_1}\d z'\ \sqrt{\helpkbar\bar k^2-U_I(z')+2\bar kk'}}
    \Bigg]
  \end{split}
\end{align}
Hieraus erhält man {\small
\begin{align}\label{e3:PS:R}
  \begin{split}
    R(k') &= \frac{\sqrt{\helpkbar\bar k^2 - U_{I}(z_1)+2\bar kk'}-\sqrt{\helpkbar\bar k^2-U_{II}(z_1)+2\bar kk'}}
            {\sqrt{\helpkbar\bar k^2 - U_{I}(z_1)+2\bar kk'}+\sqrt{\helpkbar\bar k^2-U_{II}(z_1)+2\bar kk'}}\\
      &\quad\times \exp\klg{2\I\int_{z_0}^{z_1}\d z'\ \sqrt{\helpkbar\bar k^2 -
            U_{I}(z')+2\bar kk'}}\\
        &=: \hat R(k')\exp\klg{2\I\int_{z_0}^{z_1}\d z'\ \sqrt{\helpkbar\bar k^2 -
            U_{I}(z')+2\bar kk'}}\ ,
  \end{split}
\end{align}
\begin{align}\label{e3:PS:S}
  \begin{split}
    S(k') &= \frac{2\sqrt{\helpkbar\bar k^2 - U_{I}(z_1)+2\bar kk'}}
              {\sqrt{\helpkbar\bar k^2 - U_{I}(z_1)+2\bar kk'}+\sqrt{\helpkbar\bar k^2 - U_{II}(z_1)+2\bar kk'}}\\
      &\quad\times \exp\klg{\I\int_{z_0}^{z_1}\d z'\ \sqrt{\helpkbar\bar k^2 -
              U_{I}(z_1)+2\bar kk'}}\\
          &=: \hat S(k')\exp\klg{\I\int_{z_0}^{z_1}\d z'\ \sqrt{\helpkbar\bar k^2 -
              U_{I}(z_1)+2\bar kk'}}\ ,
  \end{split}
\end{align}}

wobei wir die reellen Amplituden $\hat R(k')$ und $\hat S(k')$ definiert
haben. Man kann sich leicht davon überzeugen, dass für $R(k')$ und $S(k')$ die Beziehung
\begin{align}\label{e3:PS:Stromerhaltung}
\begin{split}  
1 &= \abs{R(k')}^2 + \sqrt\frac{\bar k^2 - U_{II}(z_1)+2\bar kk'}{\bar k^2 - U_{I}(z_1)+2\bar kk'}\abs{S(k')}^2\\
&=|\hat R(k')|^2 + \sqrt\frac{\bar k^2 - U_{II}(z_1)+2\bar kk'}{\bar k^2 - U_{I}(z_1)+2\bar kk'}|\hat S(k')|^2\ ,
\end{split}
\end{align}
erfüllt ist. Stellt man sich die ebene Welle als Teilchenstrahl vor, so bedeutet (\ref{e3:PS:Stromerhaltung}),
dass die Summe der Wahrscheinlichkeiten für die Reflexion und Transmission des Teilchens gleich Eins ist, d.h.
ein Teilchen im Strahl wird am Potentialwall stets entweder reflektiert oder transmittiert.
In diesem Zusammenhang bezeichnet man $|\hat R(k')|^2$ als Reflexionskoeffizient und
\begin{align}\label{e3:PS:T}
  T(k') = \sqrt\frac{\bar k^2 - U_{II}(z_1)+2\bar kk'}{\bar k^2 - U_{I}(z_1)+2\bar kk'}|\hat S(k')|^2
\end{align}
als Transmissionskoeffizienten.

\subsection{Anwendungen der Näherung der WKB-Lösung}

Die Näherung (W3) von S. \pageref{WKB.Breit} bedeutet, dass wir nur breite Wellenpakete $\Psi({}z,t)$
im Ortsraum betrachten wollen, deren Fouriertransformierte $\tilde\Psi(k')$ um eine 
sehr große Wellenzahl $\bar k$ sehr stark lokalisiert ist, d.h. wir setzen $\abs{k'/\bar k} \ll1$ für alle signifikant 
von Null verschiedenen Werte von $\tilde\Psi(k')$ voraus. Wir betrachten weiterhin gemäß Näherung (W2) 
nur sehr kleine Potentiale, verglichen mit der kinetischen Energie (W2), 
d.h. es gilt weiter $U(z)/\bar k^2\ll 1$. Insgesamt wird also
\begin{align}
\abs{\frac{U(z)}{\bar k^2} - 2\frac{k'}{\bar k}}\ll 1
\end{align}
für alle Orte $z$ gelten.
Wir können damit die in den Ausdrücken der Reflexions- und Transmissionsamplituden auftretenden
Wurzeln wie folgt entwickeln:
\begin{align}\label{e3:Wurzel.approx}
\begin{split}  
\sqrt{\bar k^2 - U(z)+2\bar kk'} &= \bar k\sqrt{1 - \klr{\frac{U(z)}{\bar k^2} - 2\frac{k'}{\bar k}}}\\
&\approx \bar k\klr{1-\frac12\klr{\frac{U(z)}{\bar k^2} - 2\frac{k'}{\bar k}}}\\
&= \bar k(1 - \vep(z)) + k'\ .
\end{split}
\end{align}
Hier haben wir die durch
\begin{align}
\vep(z) := \frac{U(z)}{2\bar k^2}
\end{align}
definierte Hilfsgröße eingeführt, für die stets $\abs{\vep(z)}\ll 1$ gilt.

Wir werden nun diese Entwicklung in den in  (\ref{e3:PS:R}) und (\ref{e3:PS:S}) definierten Amplituden 
$\hat R(k')$ und $\hat S(k')$ anwenden, die Wurzeln in den Phasenfaktoren aber zunächst unberührt
lassen und erst bei der Superposition der ebenen Wellen entwickeln.
Für $\hat R(k')$ folgt
\begin{align}
  \hat R(k') \approx \frac{\vep_{II}(z_1) - \vep_{I}(z_1)}{2 + 2k'/\bar k- \vep_I(z_1) -
\vep_{II}(z_1)}\ .
\end{align}
Das reflektierte Wellenpaket ergibt sich im Wesentlichen als Fouriertransformation des Produkts
$\hat R(k')\tilde\Psi(k')$. Am Beispiel des Gaußschen Wellenpakets wollen wir uns davon überzeugen,
dass bei den hier gemachten Näherungen der Anteil von $\hat R(k')$ bei der Superposition des
Wellenpakets als konstant im Vergleich zu $\tilde\Psi(k')$ angesehen werden kann.
Setzt man nämlich gemäß (\ref{e2:Gauss.FT}) $\tilde\Psi(k') = \sigma N(\sigma)\exp(-\sigma^2{k'}^2/2)$,
wobei nach (\ref{e2:kbar.k.approx.2}) $\sigma^2\gg\bar k^{-2}$ erfüllt ist, so führt die Entwicklung
der Amplitude $\hat R(k')$ nach der kleinen Größe $k'/\bar k$,
\begin{align}
  \hat R(k')f(k') \propto \klr{1 - \frac{k'}{\bar k} + \OO((k'/\bar
k)^2)}\exp\klg{-\sigma^2{k'}^2/2}
\end{align}
auf das folgende Ergebnis:
Das Argument $\sigma^2{k'}^2/2$ der Gauß-Funktion wächst betragsmäßig im Vergleich zu $k'/\bar k$
sehr schnell an. Dies bedeutet, die Gauß-Funktion schneidet für $\abs{k'}>0$ die Entwicklung 
von $\hat R(k')$ nach $k'/\bar k$ schnell ab. Auf der anderen Seite geht für $k'/\bar k\to0$
die Gaußfunktion gegen Eins und $\hat R(k')$ kann durch $\hat R(0)$ approximiert werden.
Somit kann die Abhängigkeit von $k'$ in $\hat R(k')$ gegenüber der Abhängigkeit von $k'$
in $\tilde\Psi(k')$ in guter Näherung ganz vernachlässigt werden und wir können uns im Folgenden
auf die Betrachtung von $\hat R(0)$ beschränken.

Für $\hat S(k')$ gilt die analoge Überlegung. Die Entwicklung der Wurzeln ergibt
\begin{align}
  \hat S(k') \approx \frac{1-\vep_{I}(z_1) + k'/\bar k}{1 + k'/\bar k -
\tfrac12(\vep_I(z_1)+\vep_{II}(z_1))}\ ,
\end{align}
entwickeln wir nun weiter nach $k'/\bar k$, so folgt
\begin{align}
  \hat S(k')f(k') \propto \klr{1 + C\frac{k'}{\bar k} + \OO((k'/\bar
k)^2)}\exp\klg{-\sigma^2{k'}^2/2}
\end{align}
mit einer Konstante $C\approx 1$. Mit der gleichen Argumentation wie für $\hat R(k')$
können wir uns im Folgenden also auch auf die Betrachtung von $\hat S(0)$ beschränken.

Wir definieren daher
\begin{align}
  \begin{split}\label{e3:PS:R.approx}
    \hat R_0 &:= \hat R(0) = \frac{\sqrt{\helpkbar\bar k^2 - U_{I}(z_1)}-\sqrt{\helpkbar\bar k^2-U_{II}(z_1)}}
            {\sqrt{\helpkbar\bar k^2 - U_{I}(z_1)}+\sqrt{\helpkbar\bar k^2-U_{II}(z_1)}} \approx \hat R(k')\ ,
  \end{split}\\\label{e3:PS:S.approx}
  \begin{split}
    \hat S_0 &:= \hat S(0) = \frac{2\sqrt{\helpkbar\bar k^2 - U_{I}(z_1)}}
              {\sqrt{\helpkbar\bar k^2 - U_{I}(z_1)}+\sqrt{\helpkbar\bar k^2 - U_{II}(z_1)}} \approx \hat S(k')
  \end{split}
\end{align}
und man sieht sofort, dass auch für die approximierten Amplituden
\begin{align}\label{e3:PS:R+S=1}
  1 + \hat R_0 = \hat S_0
\end{align}
erfüllt ist. Entwickeln wir nun die Wurzeln in $\hat R_0$ und $\hat S_0$ gemäß der Näherung (W2) der
kleinen Potentiale bis zur Ordnung $U/\bar k^2$, so folgt weiter
\begin{align}\label{e3:PS:R.approx.2}
    \hat R_0 &\approx \frac{U_{II}(z_1)-U_{I}(z_1)}{4\bar k^2}\ ,\\ \label{e3:PS:S.approx.2} 
    \hat S_0 &\approx \klr{1 + \frac{U_{II}(z_1)-U_I(z_1)}{4\bar k^2}}
\end{align}
und somit
\begin{align}
  |\hat R_0|^2 \approx |R(k')|^2 \approx 0\ ,\qquad |\hat S_0|^2 \approx |S(k')|^2 \approx \klr{1 +
\frac{U_{II}(z_1)-U_I(z_1)}{2\bar k^2}}
\end{align}
In unserer Näherung sollte also der reflektierte Anteil nicht relevant
sein. Die Gleichung (\ref{e3:PS:Stromerhaltung}) für die Erhaltung des Wahrscheinlichkeitsstroms gilt 
auch für die Näherungen $\hat R_0$ und $\hat S_0$, wenn man überall $k'=0$ setzt und den Vorfaktor
entwickelt:
\begin{align}\label{e3:PS:T.approx}
  \begin{split}
    1 &\overset!\approx |\hat R_0|^2 + \sqrt{\frac{1-U_{II}(z_1)/\bar k^2}{1-U_I(z_1)/\bar
k^2}}|\hat S_0|^2\\
     &\approx 0 + \klr{1 - \frac{U_{II}(z_1)-U_I(z_1)}{2\bar k^2}}\klr{1 +
\frac{U_{II}(z_1)-U_I(z_1)}{2\bar k^2}}\\
     &\approx 1 \approx \hat T_0
  \end{split}
\end{align}
Hiermit haben wir auch gezeigt, dass der Transmissionskoeffizient $\hat T_0$ (der ja dem eben
berechneten Ausdruck wegen $|\hat R_0|^2\approx 0$ entspricht) wie zu erwarten gleich Eins ist.

\subsection{Superposition des Wellenpakets}

Mit den Ergebnissen des letzten Abschnitt wollen wir nun das Wellenpaket als Superposition
der ebenen Wellen berechnen. Neben den reellen Amplituden $\hat R_0$ und $\hat S_0$
müssen wir aber noch die in $R(k')$ und $S(k')$ vorkommenden Phasenfaktoren berücksichtigen,
die mit den Phasenfaktoren der ebenen Wellen zusammengefasst werden müssen.

Die zu superponierenden Wellen in den Bereichen $I$ und $II$ lauten
{\small
\begin{align}\label{e3:Welle.I}\hspace{-5mm}
\begin{split}
\psi_{k',I}(z,t) &\approx \exp\klg{-\I\frac{\bar k^2}{2m}t - \I\frac{k'\bar k}m t}
\Bigg[\exp\klg{\I\int_{z_0}^z\d z'\ \sqrt{\helpkbar\bar k^2-U_I(z')+2\bar kk'}}\\
&+ \hat R_0\exp\klg{2\I\int_{z_0}^{z_1}\d z'\ \sqrt{\helpkbar\bar k^2 -
U_{I}(z')+2\bar kk'}-\I\int_{z_0}^z\d z'\ \sqrt{\helpkbar\bar k^2-U_I(z')+2\bar kk'}}\Bigg]\ ,\\
\psi_{k',II}(z,t) &\approx \hat S_0\exp\klg{-\I\frac{\bar k^2}{2m}t - \I\frac{k'\bar k}m t}\\
&\times \exp\klg{\I\int_{z_0}^{z_1}\d z'\ \sqrt{\helpkbar\bar k^2 -
U_{I}(z_1)+2\bar kk'} + \I\int_{z_1}^z\d z'\ \sqrt{\helpkbar\bar k^2-U_{II}(z')+2\bar kk'}}\ .
\end{split}
\end{align}
}

Die weitere Rechnung verläuft analog wie in Abschnitt \ref{s2:WKB.Lsg}, Gl. (\ref{e2:WP.WKB.FT} ff.), d.h.
nach Entwicklung der Wurzeln wie in (\ref{e2:lokWZ.approx}) ziehen wir die von $k'$ unabhängigen Teile
des Phasenfaktors vor das Fourier-Integral und erhalten nach einiger Rechnung schließlich im Bereich $I$
das Ergebnis
\begin{align}\label{e3:PS:Psi.I}
  \begin{split}
    \Psi_I(z,t) &= \exp\klg{-\I\frac{\bar k^2}{2m}t + \I\int_{z_0}^{z}\d z'\ \sqrt{\bar
k^2-U_I(z')}}
       \vph(\zeta_I(z)-\tau(t))\\
       &+\exp\klg{-\I\frac{\bar k^2}{2m}t + 2\I\int_{z_0}^{z_1}\d z'\ \sqrt{\bar k^2-U_I(z')}
         - \I\int_{z_0}^{z}\d z'\ \sqrt{\bar k^2-U_I(z')}}\\
       &\times \vph\subt{ref}(2\zeta_I(z_1)-\zeta_I(z)-\tau(t))\ .
  \end{split}
\end{align}
Hier ist $\vph(\zeta(z)-\tau(t))$ wieder definiert wie in (\ref{e2:DefAWP}) und 
$\zeta_I(z)$ und $\tau(t)$ wie in (\ref{e2:Def.Zeta.Tau}) (mit dem Potential $V(z) = V_I(z) = U_I(z)/(2m)$)
Die Einhüllende der reflektierten Welle $\vph\subt{ref}$ ist definiert durch
\begin{align}
  \vph\subt{ref}(\zeta-\tau) := \hat R_0\vph(\zeta-\tau)\ .
\end{align}
Wie am Argument von $\vph\subt{ref}$ in (\ref{e3:PS:Psi.I}) zu erkennen ist, bewegt sich aufgrund des
gleichen Vorzeichens bei $\zeta_I(z)$ und $\tau(t)$ der Schwerpunkt des reflektierten Wellenpakets
genau in entgegengesetzte Richtung wie der Schwerpunkt des einlaufenden Pakets.

Im Bereich $II$ ($z_1\leq z$) hat die Wellenfunktion folgende Gestalt:
\begin{align}\label{e3:PS:Psi.II}
  \begin{split}
    \Psi_{II}(z,t) &= \exp\klg{-\I\frac{\bar k^2}{2m}t 
      +\I\int_{z_0}^{z_1}\d z'\ \sqrt{\bar k^2-U_I(z')} + \I\int_{z_1}^z\d z'\ \sqrt{\bar
k^2-U_{II}(z')}}\\
    &\times \vph\subt{trans.}(\zeta_I(z_1)+\zeta_{II}(z)-\tau(t))
  \end{split}
\end{align}
mit 
\begin{align}\label{e3:PS:DefZeta.II}
  \zeta_{II}(z) &:= \int_{z_1}^{z}\d z'\ \frac{\bar k}{\sqrt{\bar
  k^2-U_{II}(z')}}\ ,\\ \label{e3:PS:WP.out}
\vph\subt{trans.}(\zeta-\tau) &= \hat S_0(\zeta-\tau)\ .
\end{align}
Den Wellenfunktionen $\Psi_{I,II}(z,t)$ sieht man sofort an, dass $\Psi_I(z_1,t)=\Psi_{II}(z_1,t)$
erfüllt ist, da man dann die Gleichung (\ref{e3:PS:R+S=1}) erhält.

\section{Streuung am Potentialwall}\label{s3:Potentialwall}

\subsection{Rechnung für ebene WKB-Wellen}\label{s3:PW:WKBW}

\begin{figure}[!ht]
  \centering
  \includegraphics{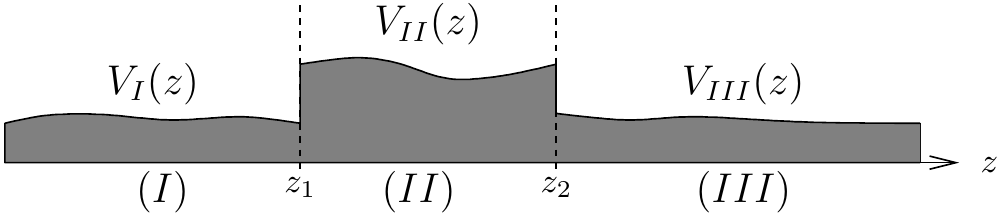}
  \caption{Allgemeines Potential mit einem Wall zwischen $z=z_1$ und $z=z_2$.}
  \label{f3:Potentialwall}
\end{figure}

Wir gehen nun einen Schritt weiter und betrachten einen
Potentialwall (siehe Abb. \ref{f3:Potentialwall}) der Form
\begin{align}\label{e3:Potentialwall}
  V(z) = 
  \begin{cases}
    V_I(z)     & z < z_1\\
    V_{II}(z)  & z_1\leq z\leq z_2\\
    V_{III}(z) & z_2 < z
  \end{cases}
\end{align}
Die Näherungen (W1)-(W3) des letzten Abschnitts seien auch hier erfüllt.

Wir werden die Rechnung, die vom Prinzip her völlig analog zur Rechnung
an der Stufe ist, hier nicht in voller Länge aufschreiben und uns stattdessen
auf die Angabe von Resultaten beschränken. Um die Formeln möglichst übersichtlich
zu gestalten, müssen wir erst einige Abkürzungen einführen. Es sei
\begin{align}\label{e3:PW:DefW}
  {\mc W}_N(z) &:= \sqrt{(\bar k + k')^2 - U_N(z)}\qquad (N=I,II,III)\\ \label{e3:PW:DefE}
  \Upsilon_N(z_b,z_a) &:= \exp\klg{\I\int_{z_a}^{z_b}\d z' {\mc W}_N(z')}
\end{align}
Wie üblich wenden wir in den Wurzeln ${\mc W}_N$ stets die Näherung (W2) an, d.h. wir vernachlässigen
quadratische Beiträge in $k'$,
\begin{align}\label{e3:PW:W.approx}
  {\mc W}_N(z) \approx \sqrt{\bar k^2 - U_N(z) + 2\bar kk'}\ .
\end{align}
Aus der Definition von $\Upsilon_N(z_b,z_a)$ erhalten wir die folgenden Eigenschaften:
\begin{subequations}\label{e3:PW:E.props}
\begin{align}
  \Upsilon_N(z_b,z_a) &= \Upsilon_N^{-1}(z_a,z_b)\ ,\\
  \Upsilon_N(z,z) &= 1\ .
\end{align}
\end{subequations}
Mit diesen Abkürzungen lauten die Wellenfunktionen in den einzelnen
Bereichen analog zu (\ref{e3:PS:Psi.k.I}):
\begin{align}\label{e3:PW:Psi.k.I-III}
  \begin{split}
    \psi_{k',I}(z,t) &= e^{-\I\klr{\bar k^2 + 2\bar k k'}
    t/(2m)}\klr{\Upsilon_{I}(z,z_0) + R(k')\Upsilon_{I}(z_0,z)}\ ,\\
    \psi_{k',II}(z,t) &= e^{-\I\klr{\bar k^2 + 2\bar k k'}
    t/(2m)}\klr{A(k')\Upsilon_{II}(z,z_1) + B(k')\Upsilon_{II}(z_1,z)}\ ,\\
    \psi_{k',III}(z,t) &= e^{-\I\klr{\bar k^2 + 2\bar k k'}
    t/(2m)} S(k')\Upsilon_{III}(z,z_2)\ .
  \end{split}
\end{align}

Das zu lösende lineare Gleichungssystem, das aus der Stetigkeit der 
Wellenfunktion und ihrer Ableitung an den Punkten $z_{1,2}$ folgt,
lautet mit diesen Bezeichnungen
\begin{subequations}\label{e3:PW:LGS}
\begin{align}
  \Upsilon_I(z_1,z_0) + R(k')\,\Upsilon_I(z_0,z_1) &= A(k') + B(k')\\
  A(k')\,\Upsilon_{II}(z_2,z_1) + B(k')\,\Upsilon_{II}(z_1,z_2) &= S(k')\\[1mm]
  \Upsilon_I(z_1,z_0) - R(k')\,\Upsilon_I(z_0,z_1) &= \frac{{\mc W}_{II}(z_1)}{{\mc W}_I(z_1)}(A(k') - B(k'))\\
  A(k')\, \Upsilon_{II}(z_2,z_1) - B(k')\, \Upsilon_{II}(z_1,z_2) &= \frac{{\mc W}_{III}(z_2)}{{\mc W}_{II}(z_2)}S(k')
\end{align}
\end{subequations}
Hierbei sind $A(k')$ und $B(k')$ die Koeffizienten für die in positive bzw. negative
$z$-Richtung laufenden Anteile der Wellenfunktion im Bereich II. 
Das LGS (\ref{e3:PW:LGS}) besteht aus vier Gleichungen für die vier
unbekannten Koeffizienten $A(k'),\,B(k'),\,R(k'),\,S(k')$. Arbeitet man sich durch die Rechnungen,
so erhält man schließlich

{\footnotesize
\begin{align}\label{e3:PW:R}
 \hspace{-10mm}  R(k') &= \Upsilon_I^2(z_1,z_0)\frac{\Upsilon_{II}(z_2,z_1)
({\mc W}_I+{\mc W}_{II})\an_{z_1}({\mc W}_{II}-{\mc W}_{III})\an_{z_2}
    + \Upsilon_{II}^{-1}(z_2,z_1)({\mc W}_I-{\mc W}_{II})\an_{z_1}({\mc W}_{II}+{\mc W}_{III})\an_{z_2}}{
    \Upsilon_{II}(z_2,z_1) ({\mc W}_I-{\mc W}_{II})\an_{z_1}({\mc W}_{II}-{\mc W}_{III})\an_{z_2}
    + \Upsilon_{II}^{-1}(z_2,z_1)({\mc W}_I+{\mc W}_{II})\an_{z_1}({\mc W}_{II}+{\mc W}_{III})\an_{z_2}}\\ \label{e3:PW:S}
\hspace{-10mm} S(k') &= \frac{4\Upsilon_{I}(z_1,z_0) {\mc W}_I(z_1){\mc W}_{II}(z_2)}{
    \Upsilon_{II}(z_2,z_1) ({\mc W}_I-{\mc W}_{II})\an_{z_1}({\mc W}_{II}-{\mc W}_{III})\an_{z_2}
    + \Upsilon_{II}^{-1}(z_2,z_1)({\mc W}_I+{\mc W}_{II})\an_{z_1}({\mc W}_{II}+{\mc W}_{III})\an_{z_2}}
\end{align}
}

Setzt man in diesen Ausdrücken $U_I = U_{III} = 0$ und $U_{II} = U = \const$,
so erhält man schließlich die aus den Büchern über Quantenmechanik
(z.B. \cite{Messiah}, Bd.1, S. 93) bekannte Formel für den Transmissionskoeffizienten
\begin{align}
  T(k') = \abs{S(k')}^2 = \frac{4\bar k^2(\bar k^2-U)}
    {4\bar k^2(\bar k^2-U) + \sin^2\klr{\sqrt{\bar k^2-U}\ (z_2-z_1)}U^2}\ .
\end{align}
Somit haben wir die Richtigkeit der allgemeinen Formeln (\ref{e3:PW:R}), (\ref{e3:PW:S})
für die Koeffizienten $R(k')$ und $S(k')$ in diesem Fall überprüft.

\subsection{Anwendungen der Näherung der WKB-Lösung}

In den Ausdrücken für $R(k')$ und $S(k')$ betrachten wir nun wieder die
gleichen Näherungen wie bei der Potentialstufe, d.h. wir verwenden
Näherungen $\hat R_0 = \hat R(k'=0)$ und $\hat S_0=\hat S(k'=0)$
für die Amplituden der Koeffizienten.
In den Phasenfaktoren $\Upsilon_N$ behalten wir analog
zur Potentialstufe die $2\bar k k'$-Terme unter den Wurzeln ${\mc W}_N$ bei.

Es gilt dann für im Rahmen der Näherungen (W2) und (W3) von Seite \pageref{WKB.Breit} für die Wurzel ${\mc W}_N$:
\begin{align}
  {\mc W}_N(z) \approx \sqrt{\bar k^2-U_N(z)} \approx \bar k\klr{1-\frac{U_N(z)}{2\bar k^2}}\ .
\end{align}

Betrachten wir zunächst den Zähler von $S(k')$ aus (\ref{e3:PW:S}), so ist
\begin{align}
  4\Upsilon_I(z_1,z_0){\mc W}_I(z_1){\mc W}_{II}(z_2) \approx 4\Upsilon_I(z_1,z_0)\klr{\bar k^2 - U_I(z_1) - U_{II}(z_2)}\ .
\end{align}
Für die Produkte der Wurzeln ${\mc W}_N$ gilt weiter
{\small
\begin{align}
\begin{split}
  ({\mc W}_I - {\mc W}_{II})\an_{z_1}({\mc W}_{II}-{\mc W}_{III})\an_{z_2} &= {\mc W}_I(z_1){\mc W}_{II}(z_2) +
{\mc W}_{II}(z_1){\mc W}_{III}(z_2)\\
  &-{\mc W}_I(z_1){\mc W}_{III}(z_2) - {\mc W}_{II}(z_1){\mc W}_{II}(z_2)\\
  &\approx -U_I(z_1)-U_{II}(z_2)-U_{II}(z_1)-U_{III}(z_2)\\
  &\quad+U_I(z_1)+U_{III}(z_2)+U_{II}(z_1)+U_{II}(z_2)\\
  &= 0
\end{split}
\end{align}
\begin{align}
\begin{split}
  ({\mc W}_I + {\mc W}_{II})\an_{z_1}({\mc W}_{II}+{\mc W}_{III})\an_{z_2} &= {\mc W}_I(z_1){\mc W}_{II}(z_2) +
{\mc W}_{II}(z_1){\mc W}_{III}(z_2)\\
  &+{\mc W}_I(z_1){\mc W}_{III}(z_2) + {\mc W}_{II}(z_1){\mc W}_{II}(z_2)\\
  &\approx -U_I(z_1)-U_{II}(z_2)-U_{II}(z_1)-U_{III}(z_2)\\
  &\quad-U_I(z_1)-U_{III}(z_2)-U_{II}(z_1)-U_{II}(z_2)\\
  &= 4\bar k^2-2\kle{U_I(z_1)+U_{II}(z_1) + U_{II}(z_2) + U_{III}(z_2)}
\end{split}
\end{align}
\begin{align}
\begin{split}
  ({\mc W}_I + {\mc W}_{II})\an_{z_1}({\mc W}_{II}-{\mc W}_{III})\an_{z_2} &= {\mc W}_I(z_1){\mc W}_{II}(z_2) -
{\mc W}_{II}(z_1){\mc W}_{III}(z_2)\\
  &-{\mc W}_I(z_1){\mc W}_{III}(z_2) + {\mc W}_{II}(z_1){\mc W}_{II}(z_2)\\
  &\approx 4\bar k^2-U_I(z_1)-U_{II}(z_2)+U_{II}(z_1)+U_{III}(z_2)\\
  &\quad+U_I(z_1)+U_{III}(z_2)-U_{II}(z_1)-U_{II}(z_2)\\
  &= -2\kle{U_{II}(z_2) - U_{III}(z_2)}
\end{split}
\end{align}
\begin{align}
\begin{split}
  ({\mc W}_I - {\mc W}_{II})\an_{z_1}({\mc W}_{II}+{\mc W}_{III})\an_{z_2} &= {\mc W}_I(z_1){\mc W}_{II}(z_2) -
{\mc W}_{II}(z_1){\mc W}_{III}(z_2)\\
  &+{\mc W}_I(z_1){\mc W}_{III}(z_2) - {\mc W}_{II}(z_1){\mc W}_{II}(z_2)\\
  &\approx -U_I(z_1)-U_{II}(z_2)+U_{II}(z_1)+U_{III}(z_2)\\
  &\quad-U_I(z_1)-U_{III}(z_2)+U_{II}(z_1)+U_{II}(z_2)\\
  &= -2\kle{U_I(z_1) - U_{II}(z_1)}
\end{split}
\end{align}
}

Somit wird $S(k')$ zu
\begin{align}\label{e3:PW:S.approx}
  \begin{split}
    S(k') &\approx \frac{\Upsilon_{II}(z_2,z_1)\Upsilon_I(z_1,z_0)\kle{\bar
        k^2-U_I(z_1)-U_{II}(z_2)}} {\bar k^2-\tfrac12\kle{U_I(z_1)+U_{II}(z_1)
        + U_{II}(z_2) + U_{III}(z_2)}}\\
    &=: \Upsilon_{II}(z_2,z_1)\Upsilon_I(z_1,z_0)\hat S_0\ ,
\end{split}
\end{align}
und setzt sich somit (im Falle von reellen Potentialen) wieder aus einem Phasenfaktor
\begin{align}
  \begin{split}
    \Upsilon_{II}(z_2,z_1)\Upsilon_I(z_1,z_0) &\approx \exp\Bigg\{\I\int_{z_0}^{z_1}\d z'\ 
    \sqrt{\bar k^2-U_I(z')+2\bar k k'}\\ &\qquad\!\!\quad+ \I\int_{z_1}^{z_2}\d z'\ \sqrt{\bar
      k^2-U_{II}(z')+2\bar k k'}\Bigg\}
  \end{split}
\end{align}
und einer reellen, von $k'$ unabhängigen Amplitude $\hat S_0$ zusammen. Der Reflexionskoeffizient
$R(k')$ wird zu
\begin{align}\label{e3:PW:R.approx}
  \begin{split}
  R(k') &\approx -\Upsilon_{II}(z_2,z_1)\Upsilon_I^2(z_1,z_0)\\ 
&\times \frac{\Upsilon_{II}(z_2,z_1)\kle{U_{II}(z_2)-U_{III}(z_2)}+\Upsilon^{-1}_{II}(z_2,z_1)\kle{U_{I}(z_1)-U_{II
}
(z_1)}}
  {2\bar k^2-\kle{U_I(z_1)+U_{II}(z_1) + U_{II}(z_2) + U_{III}(z_2)}}\ .
  \end{split}
\end{align}
Bei reellen Potentialen setzt sich der Reflexionskoeffizient für den Potentialwall also 
aus der Summe zweier Ausdrücke zusammen,
die jeweils die Gestalt eines Phasenfaktors multipliziert mit einer Amplitude
haben. Des Weiteren sind die 
Größenverhältnisse von $R(k')$ und $S(k')$ wie bei der Potentialstufe.

Die Berechnung der Wellenpakete gestaltet sich nun analog wie bei der
Potentialstufe, d.h. wir spalten die Koeffizienten $R(k')$ und $S(k')$
jeweils auf in Amplitude und Phasenfaktor.
Die Phasenfaktoren ändern dann die Argumente der reflektierten
bzw. transmittierten Einhüllenden $\vph\subt{ref/trans.}$
ab und liefern außerdem zusätzliche, konstante Phasenfaktoren. 

Es ergibt sich
\begin{align}
	R(k') &\approx \exp\klg{2\I\phi_I(k')+2\I\phi_{II}(k')}\hat R^{(II)}_0
				+ \exp\klg{2\I\phi_I(k')}\hat R^{(I)}_0\\
	S(k') &\approx \exp\klg{\I\phi_I(k')+\I\phi_{II}(k')}\hat S_0
\end{align}
mit
\begin{subequations}
	\begin{align}
		\phi_{I}(k') &:= \int_{z_0}^{z_1}\d z'\ \sqrt{\bar k^2-U_I(z')+2\bar k k'}\ ,\\
		\phi_{II}(k') &:= \int_{z_1}^{z_2}\d z'\ \sqrt{\bar k^2-U_{II}(z')+2\bar k k'}\ ,\\
		\hat R^{(I)}_0 &:= \frac{U_{II}(z_1)-U_{I}(z_1)}{2\bar k^2-(U_I(z_1)+U_{II}(z_1)+U_{II}(z_2)+U_{III}(z_2))}\ ,\\
		\hat R^{(II)}_0 &:= \frac{U_{III}(z_2)-U_{II}(z_2)}{2\bar k^2-(U_I(z_1)+U_{II}(z_1)+U_{II}(z_2)+U_{III}(z_2))}\ ,\\
		\hat S_0 &:= \frac{2(\bar k^2-U_I(z_1)-U_{II}(z_2))}{2\bar k^2-(U_I(z_1)+U_{II}(z_1)+U_{II}(z_2)+U_{III}(z_2))}\ .
	\end{align}
\end{subequations}

Für die Wellenpakete in den Bereichen $I$ und $III$ folgt dann nach einiger Rechnung
{\small
\begin{subequations}
\begin{align}\hspace{-5mm}
	\begin{split}
		\Psi_I(z,t) &= \exp\klg{-\I\frac{\bar k^2}{2m}t + \I\int_{z_0}^z\d z'\ \sqrt{
			\bar k^2-U_I(z')}}\vph(\zeta_I(z)-\tau(t))\\
			&+ \exp\klg{-\I\frac{\bar k^2}{2m}t + 2\I\phi_I(0) - \I\int_{z_0}^z\d z'\ \sqrt{
				\bar k^2-U_I(z')}}\vph^{(I)}\subt{ref}(z,t)\\
			&+ \exp\klg{-\I\frac{\bar k^2}{2m}t + 2\I\kle{\phi_I(0)+\phi_{II}(0)} - \I\int_{z_0}^z\d z'\ \sqrt{
\bar k^2-U_I(z')}}\vph^{(II)}\subt{ref}(z,t)\ ,
	\end{split}
\end{align}
\begin{align}
	\vph^{(I)}\subt{ref}(z,t)  &:= \hat R^{(I)}_0\vph(2\zeta_I(z_1)-\zeta_I(z)-\tau(t))\ ,\\[1mm]
	\vph^{(II)}\subt{ref}(z,t) &:= \hat R^{(II)}_0\vph(2\zeta_I(z_1)+2\zeta_{II}(z_2)-\zeta_I(z)-\tau(t))
\end{align}
\end{subequations}
}

und
{\small 
\begin{subequations}\label{e3:PW:Psi.III}
\begin{align}
	\begin{split}
		\Psi_{III}(z,t) &= \exp\klg{-\I\frac{\bar k^2}{2m}t + \I\phi_I(0)+\I\phi_{II}(0) 
			+ \I\int_{z_2}^z\d z'\ \sqrt{\bar k^2-U_{III}(z')}}\\
			&\quad\times \vph\subt{trans.}(z,t)\ ,
	\end{split}\\[2mm]
	\vph\subt{trans.}(z,t) &:= \hat S_0\vph(\zeta_I(z_1)+\zeta_{II}(z_2)+\zeta_{III}(z)-\tau(t))\ .
\end{align}
\end{subequations}
}

Die Definitionen von $\zeta_{II,III}(z)$ sind analog wie in (\ref{e3:PS:DefZeta.II}), mit dem Ort des
jeweils vorangegangenen Potentialsprungs als Untergrenze der Integration. Eine Verallgemeinerung der Formeln 
auf beliebig viele Potentialsprünge ist mit den hier dargestellten Ergebnissen auf einfache Weise möglich,
sofern man sich nur für das einlaufende (und insgesamt reflektierte) und das auslaufende Wellenpaket interessiert.

\section{Anwendung auf ein einfaches, longitudinales Atomstrahl-Spinecho-Ex\-peri\-ment}\label{s3:Fahrplan}

Wir wollen nun an einem sehr einfachen Beispiel demonstrieren, dass wir im Prinzip bereits
jetzt in der Lage sind, das Fahrplanmodell so, wie es in \cite{DissAR}, Abschnitt 3.3.3, S. 55ff. dargestellt ist,
zu reproduzieren. Dazu ist es allerdings nötig, bereits die inneren atomaren Zustände zu berücksichtigen,
die wir bisher aber noch nicht eingeführt haben. Wir wollen deshalb im nächsten Abschnitt
\ref{s3:FahrplanBasics} einige Grundlagen zur Beschreibung eines Atoms mit mehreren inneren Zuständen
angeben, ohne jedoch zu sehr ins Detail zu gehen. In Kap. \ref{s6:Formalismus} greifen wir dieses Thema
erneut auf und werden dann auch die theoretischen Hintergründe genauer erarbeiten. Um mit der
Notation in Kap. \ref{s6:Formalismus} übereinzustimmen, werden wir bei allen Formeln, die
im Zusammenhang mit der Betrachtung eines Atoms stehen, wieder Großbuchstaben für die Ortskoordinate
$Z$ (Schwerpunktskoordinate des Atoms) und die Masse $M$ (Gesamtmasse des Atoms) verwenden.

In Abschnitt \ref{s3:fruehereArbeiten} wollen wir einen kurzen Rückblick auf frühere Arbeiten wagen,
die sich mit Atomen in elektrischen Feldern im Zusammenhang mit der Paritätsverletzung beschäftigt haben
(siehe z.B. \cite{BoBrNa95}). Das Thema der Paritätsverletzung in Atomen, für die wir ja die
Theorie des longitudinalen Atomstrahl-Spinechos eigentlich entwickeln, greifen wir in der vorliegenden
Arbeit aber erst in Kap. \ref{s7:Berry} wieder auf, wo wir P-verletzende geometrische Phasen studieren werden.

In Abschnitt \ref{s3:Fahrplanmodell} kommen wir endlich zur Einführung des Fahrplanmodells und
der Beschreibung des lABSE-Beispielexperiments.

\subsection{Grundlagen zur Beschreibung eines Atoms mit mehreren inneren Zuständen}\label{s3:FahrplanBasics}

Der atomare Gesamtzustand $\ket{\Psi(z,t)}$ ist im Allgemeinen eine Superposition
mehrerer, innerer atomarer Eigenzustände, die sich durch gewissen Quantenzahlen 
(z.B. Hauptquantenzahl, Gesamtdrehimpuls usw.) unterscheiden. Für die in diesem Abschnitt
angestrebte Beschreibung des Fahrplanmodells genügt eine oberflächliche Betrachtung
des Atoms. Wir wollen daher einfach die inneren atomaren Zustände kennzeichnen durch
\begin{align}
\rket{\alpha}\ ,\quad \alpha=1,\ldots,N\ .
\end{align}
Der Index $\alpha$ nummeriert die insgesamt $N$ atomaren Zustände in geeigneter Weise
durch. Die Diracsche Schreibweise mit runden Klammern verwenden wir generell zur Kennzeichnung
eines inneren atomaren Zustands.

Zur Zeit $t=0$ befinde sich das Atom am Ort $Z_0$ (im feldfreien Raum) in einem inneren Gesamtzustand
\begin{align}\label{e3:Initial.Ket}
\ket{\chi} = \sum_{\alpha=1}^N \hat\chi_\alpha\rket{\alpha(Z_0)}\ ,\qquad \sum_{\alpha=1}^N \abs{\hat\chi_\alpha}^2 = 1\ ,
\end{align}
was z.B. eine vorgegebene Polarisation sein könnte. Es liegt nun nahe, den gesamten
atomaren Anfangszustand als ein Produkt von Wellenfunktion und innerem Gesamtzustand
zu schreiben, d.h.
\begin{align}\label{e3:Initial.Atom}
\ket{\Psi(Z,t=0)} = \Psi(Z,t=0)\ket{\chi}\ .
\end{align}
So bekommt schließlich jeder einzelne, innere atomare Zustand $\rket{\alpha(Z_0)}$ ein eigenes Wellenpaket 
mit einer durch die Schrödinger-Gleichung bestimmten Zeitentwicklung, die im Allgemeinen für jedes Wellenpaket
unterschiedlich ist. Die atomaren Eigenzustände  $\rket{\alpha(Z)}$
an einem Ort $Z\neq Z_0$ sind bei angelegten äußeren Feldern Superpositionen der freien Zustände 
$\rket{\alpha(Z_0)}$, den atomaren Gesamtzustand am Ort $Z$ kann man aber dennoch zu jeder Zeit $t$ 
als Superposition der lokalen atomaren Eigenzustände schreiben, d.h.
\begin{align}\label{e3:Initial}
\ket{\Psi(Z,t)} = \sum_\alpha\Psi_\alpha(Z,t)\rket{\alpha(Z)}\ .
\end{align}

In diesem Abschnitt wollen wir nur den adiabatischen Grenzfall betrachten, in dem sich die
Zusammensetzung der inneren atomaren Zustände in den Potentialen so langsam ändert, dass die Amplituden
$\Psi_\alpha(Z,t)$ nicht miteinander mischen und somit unabhängig 
voneinander sind. Bei vorausgesetzter Gültigkeit der Näherungen (W1)-(W3) können wir dann z.B. die WKB-Wellenpakete
als Grundlage für die $\Psi_\alpha(Z,t)$ wählen, d.h.
\begin{align}\label{e3:WP.WKB}
	\Psi_\alpha(Z,t) &= \hat\chi_\alpha\e^{\I\phi_\alpha(Z,t)}\vph(\zeta_\alpha(Z)-\tau(t))
\end{align}
mit
\begin{align}\label{e3:WKB-PW}
	\phi_\alpha(Z,t) &:= 
    -\frac{\bar k^2}{2M}t + \int_{Z_0}^Z\d Z'\ \sqrt{\bar k^2-2M V_\alpha(Z')}\ , 
\end{align}
und
\begin{align}\label{e3:Zeta}
	\zeta_{\alpha}(Z) &= \int_{Z_0}^Z\d Z'\ \frac{\bar k}{\sqrt{\helpkbar\bar k^2-2M V_\alpha(Z')}}\ ,
\end{align}
denn die lokalen Zustände $\rket{\alpha(Z)}$ werden im Allgemeinen 
unterschiedlichen Potentialen $V_\alpha(Z)$ unterliegen.
In Kap. \ref{s6:Formalismus} werden wir die Definition von $\zeta_\alpha(Z)$ noch etwas genauer untersuchen
und besser motivieren. Für den Augenblick soll uns die Tatsache ausreichen, dass durch die Definition
(\ref{e3:Zeta}) jedes WKB-Wellenpaket eine individuelle Schwerpunktsbewegung erfährt, abhängig vom jeweiligen
Zustands-Potential $V_\alpha(Z)$. Sind die Potentiale Null\footnote{
In Kapitel \ref{s6:Formalismus} werden wir sehen, dass anstelle der Potentiale $V_\alpha(Z)$ eigentlich
die Potentialdifferenz $\Delta V_\alpha(Z) = V_\alpha(Z)-V_\alpha(Z_0)$ zu verwenden ist, wenn man von
gewissen Voraussetzungen ausgeht. Wir wollen in diesem Kapitel nicht weiter auf diese subtilen Feinheiten
der Theorie eingehen.}, so haben alle Wellenpakete die gleiche, 
mittlere Geschwindigkeit $\bar k/m$.

\subsection{Zusammenhang mit früheren Arbeiten}\label{s3:fruehereArbeiten}

In bisherigen Arbeiten zum Thema Paritätsverletzung in leichten Atomen
(siehe z.B. \cite{BeNa83, BoBrNa95, DissTG, BrGaNa99, DiplTB}) wurden nur Atome in Ruhe betrachtet, 
die sich in einem zeitabhängigen Potential befinden. Auf diese Weise spart man sich die Beschreibung
des Atoms als Wellenfunktion. Um zu verstehen, wie man prinzipiell P-Verletzung mit einem lABSE-Experiment nachweisen
könnte, kommt man aber nicht um die Beschreibung der Atome im Ortsraum als Wellenfunktion herum.

Im Ruhesystem des Atoms sollte unsere Beschreibung aber äquivalent zur früheren Betrachtungsweise sein
und zu den gleichen Resultaten führen. Wir wollen uns in diesem Abschnitt ansatzweise von der Richtigkeit
dieser Behauptung überzeugen.

In früheren Arbeiten wurde das Atom durch den rein zeitabhängigen Zustandsvektor $\ket t$ (hier schreiben wir $\ket{\Psi(t)}$) repräsentiert, 
der sich gemäß der SG (s. \cite{BoBrNa95}, Gl. (3.9))
\begin{align}\label{e3:SG.ext}
\I\dd{}{t}\ket{\Psi(t)} = H\ket{\Psi(t)}
\end{align}
zeitlich entwickelte. Der Hamilton-Operator $H$ des Atoms (inkl. P-Verletzung und
Zerfall aus angeregten Niveaus) wurde in einem Unterraum zu fester Hauptquantenzahl 
$n$ als Matrix $\uop M$ repräsentiert und die Zeitentwicklung von $\ket{\Psi(t)}$
konnte geschrieben werden als (s. \cite{BoBrNa95}, Gl. (3.11))
\begin{align}\label{e3:Lsg.ext}
  \ket{\Psi(t)} = \e^{-\I\uop M t}\ket{\Psi(0)}\ ,
\end{align}
Im Bereich eines konstanten Potentials zwischen zwei Zeiten $t_1<t_2$
kann das Eigenwertproblem für $\uop M$ gelöst werden und man erhält
dann die Zeitentwicklung
\begin{align}\label{e3:EZ.ext}
  \ket{\Psi_E(t_2)} = \e^{-\I E (t_2-t_1)}\ket{\Psi_E(t_1)} = \e^{-\I E \Delta t}\ket{\Psi_E(t_1)}\ ,
\end{align}
für einen Eigenzustand von $\uop M$ zur Energie $E$.

\begin{figure}
	\centerline{\includegraphics[width=15cm]{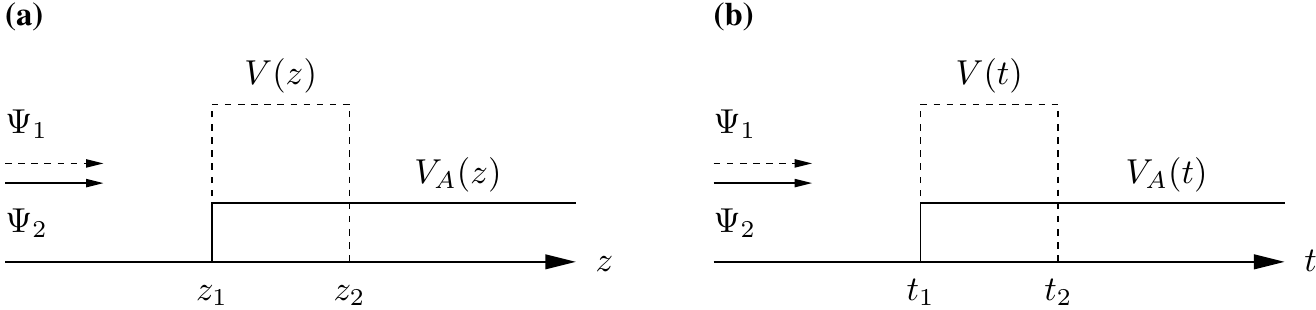}}
\caption[Zum Zusammenhang mit früheren Arbeiten.]{
Ein Beispiel zur Verdeutlichung der Äquivalenz der Beschreibungen aus früheren Arbeiten (b)
		und aus dieser Arbeit (a).}\label{f3:Zusammenhang}
\end{figure} 
Ein einfaches Beispiel soll den Zusammenhang der Rechnungen in dieser Arbeit und den
Rechnungen vorhergehender Arbeiten demonstrieren. Wir betrachten einfach eine kohärente
Superposition zweier Zustände, die wir mit $\Psi_1$ und $\Psi_2$ bezeichnen wollen.
Der Zustand $\Psi_1$ durchquert einen Potentialwall der Höhe $V$, der Zustand $\Psi_2$,
den wir als Analysatorzustand verwenden wollen, passiere eine Potentialstufe der
Höhe $V_A$. In Abb. \ref{f3:Zusammenhang}a ist die Situation dargestellt, die wir in der vorliegenden Arbeit betrachten:
Die Zustände sind mit Wellenpaketen verknüpft, die die stationären, ortsabhängigen 
Potentiale durchlaufen. In Abb. \ref{f3:Zusammenhang}b ist die Situation aus früheren Arbeiten
gezeigt, wo das betrachtete Atom in Ruhe ist und die Potentiale zeitabhängig sind.

Wir beginnen bei der Situation aus Abb. \ref{f3:Zusammenhang}b, d.h. wir gehen von einem
Anfangszustand\footnote{Die in diesem Abschnitt betrachteten Zustände sind lediglich Beispiele ohne
wirkliche physikalische Bedeutung.} zur Zeit $t=0$ aus,
\begin{align}
	\ket{\Psi(t=0)} = \tfrac1{\sqrt2}\klr{\ket{\Psi_1} + \ket{\Psi_2}}\ .
\end{align}
Wir entwickeln diesen Zustand nun nach der Zeit, wobei wir, wie in Abb. \ref{f3:Zusammenhang}b dargestellt,
der Einfachheit halber von einem verschwindenden Potential vor der Zeit $t_1$ für beide Zustände
ausgehen. Nach (\ref{e3:EZ.ext}) lautet der Zustand zu einer Zeit $t\geq t_2$ dann
\begin{align}
	\ket{\Psi(t)} = \tfrac1{\sqrt2}\klr{\e^{-\I V (t_2-t_1)}\ket{\Psi_{1}} + \e^{-\I V_A(t-t_1)}\ket{\Psi_{2}}}\ .
\end{align}
Wir können zu dieser Zeit z.B. eine Messung des Anteils des Anfangszustands
im Endzustand durchführen, d.h. wir betrachten die Wahrscheinlichkeitsdichte
\begin{align}\label{e3:Density.2}
	\begin{split}
		\rho(t) &= \abs{\bracket{\Psi(t=0)}{\Psi(t)}}^2 = \tfrac12\abs{\e^{-\I V (t_2-t_1)}+\e^{-\I V_A (t-t_1)}}^2\\
			&= \tfrac14\abs{1+\e^{-\I (V_A-V) (t_2-t_1) - \I V_A (t-t_2)}}^2\ .
	\end{split}
\end{align}
Wie man sieht, wird die Wahrscheinlichkeitsdichte mit der Zeit oszillieren. Die unterschiedlichen Potentiale
der beiden Zustände sorgen für einen zusätzlichen Phasenschub. Wären beide Potentiale identisch gewesen, würde
die Wahrscheinlichkeitsdichte einfach den Wert Eins annehmen.

Nun betrachten wir die in dieser Arbeit vorliegende Situation, wobei wir für den Moment nur die relative Phase
berücksichtigen wollen, die Schwerpunktsbewegung wird im nächsten Abschnitt ausführlich betrachtet.
Wir setzen daher in den Formeln für das WKB-Wellenpaket stets die Einhüllende zu $\vph(\zeta) \equiv 1$.
Nach den Formeln für die transmittierten Anteile der Wellenfunktionen bei der Potentialstufe (\ref{e3:PS:Psi.II})
und beim Potentialwall (\ref{e3:PW:Psi.III}) haben die Phasenfaktoren der beteiligten Zustände an einem Ort $z\geq z_2$
die Gestalt
\begin{align}
	\begin{split}
		\ket{\Psi(z,t)} &\sim \exp\klg{-\I\frac{\bar k^2}{2m}t}\\
		  &\times \Bigg( 
			   \exp\klg{\I\bar k(z_1-z_0)+\I\sqrt{\bar k^2 - 2m V}\,(z_2-z_1)+\I\bar k(z-z_2)}\ket{\Psi_1}\\
			&\quad\quad \exp\klg{\I\bar k(z_1-z_0)+\I\sqrt{\bar k^2 - 2m V_A}\,(z-z_1)}\ket{\Psi_2}\Bigg)
	\end{split}
\end{align}
Wir haben die Einhüllende $\vph(\zeta)$ hier einfach vernachlässigt bzw. gleich Eins gesetzt.
Nun projizieren wir wieder auf den Anfangszustand, der an einem Anfangsort $z_0$ die Form
\begin{align}
	\ket{\Psi(z_0,t=0)} = \tfrac1{\sqrt2}\klr{\ket{\Psi_1}+\ket{\Psi_2}}
\end{align}
haben soll. Wir erhalten dann (unter der Annahme, das die $\ket{\Psi_1}$, $\ket{\Psi_2}$ orthonormal zueinander sind)
analog wie oben die Wahrscheinlichkeitsdichte
\begin{align}
	\begin{split}
		\rho(z) &= \abs{\bracket{\Psi(t=0)}{\Psi(t)}}^2\\
			&= \tfrac14\abs{1+\e^{-\I (\sqrt{\bar k^2 - 2m V}-\sqrt{\bar k^2 - 2m V_A})(z_2-z_1) 
				- \I(\bar k - \sqrt{\bar k^2 - 2m V_A})\,(z-z_2)}}^2\ .
	\end{split}
\end{align}
Wie man sieht, ist in diesem speziellen, einfachen Fall keine Zeitabhängigkeit mehr
in der Wahrscheinlichkeitsdichte vorhanden. Der relative Phasenwinkel kann für sehr kleine Potentiale
entwickelt werden und lautet dann
\begin{align}
\begin{split}
	&\ (\sqrt{\bar k^2 - 2m V}-\sqrt{\bar k^2 - 2m V_A})(z_2-z_1)+(\bar k - \sqrt{\bar k^2 - 2m V_A})\,(z-z_2)\\
	\approx&\ (V_A-V)\frac{\bar k}m(z_2-z_1) + V_A\frac{\bar k}m(z-z_2)
\end{split}
\end{align}
Dies kann direkt mit dem relativen Phasenwinkel aus (\ref{e3:Density.2}) verglichen werden, d.h. mit
\begin{align}
	(V_A-V)(t_2-t_1) + V_A(t-t_2)\ .
\end{align}
Im ortsabhängigen Fall wird also lediglich die Zeitdifferenz $t_2-t_1$ durch die mittlere Geschwindigkeit
$\bar k/m$ multipliziert mit der Breite des Potentialwalls ersetzt, d.h. durch die
(mittlere) Durchflugszeit. 

Dieses einfache Beispiel zeigt also, das die bisher aufgestellte Theorie bereits in der Lage ist, die Ergebnisse
der früheren Arbeiten zu reproduzieren und in dem hier betrachteten Spezialfall auch äquivalent zur damaligen 
Beschreibung ist. Durch die Berücksichtigung von orts- und zeitabhängigen Wellenfunktionen in Verbindung 
mit den inneren atomaren Zuständen werden wir aber am Ende von Kapitel \ref{s6:Formalismus} in der Lage sein, 
wesentlich komplexere, physikalische Vorgänge als in früheren Arbeiten theoretisch zu beschreiben. 
Diese Komplexität hat natürlich ihren Preis und wird sich, wie wir später noch sehen
werden, auch in den Formeln niederschlagen.

\subsection{Das Fahrplanmodell}\label{s3:Fahrplanmodell}

Das Fahrplanmodell (siehe \cite{DissAR}, Kap. 3.3.3)\footnote{
Dem aufmerksamen Leser mag aufgefallen sein, dass die Nummer des Kapitels \glqq Fahrplanmodell\grqq~der
zitierten, experimentellen Dissertation \cite{DissAR} exakt 
mit der Nummer dieses Abschnitts übereinstimmt, der ebenfalls das Fahrplanmodell behandelt.
Ich versichere, dass dies reiner Zufall ist und nicht beabsichtigt war, lade den Leser aber herzlich
zum Nachdenken über die mystische Bedeutung dieses Zufalls ein.
} 
liefert eine anschauliche, grafische Darstellung der Schwerpunktsbewegung der
Wellenpakete, die zu verschiedenen atomaren Eigenzuständen
gehören und gibt einen einfachen Überblick über die Lage der Interferenzpunkte und
über die beteiligten Zustände.

Eine saubere theoretische Beschreibung des Fahrplanmodell gab es bisher nicht.
In diesem Abschnitt wollen wir demonstrieren dass wir bereits jetzt, mit den
bisher aufgestellen Formeln, in der Lage sind, das Fahrplanmodell theoretisch
zu beschreiben. Wir sind darüberhinaus in der Lage, die zugrundeliegenden physikalischen
Vorgänge zu verstehen und Grenzen des Fahrplanmodells aufzuzeigen.

\begin{floatingfigure}[r]{6cm}
\centerline{\includegraphics[width=6cm]{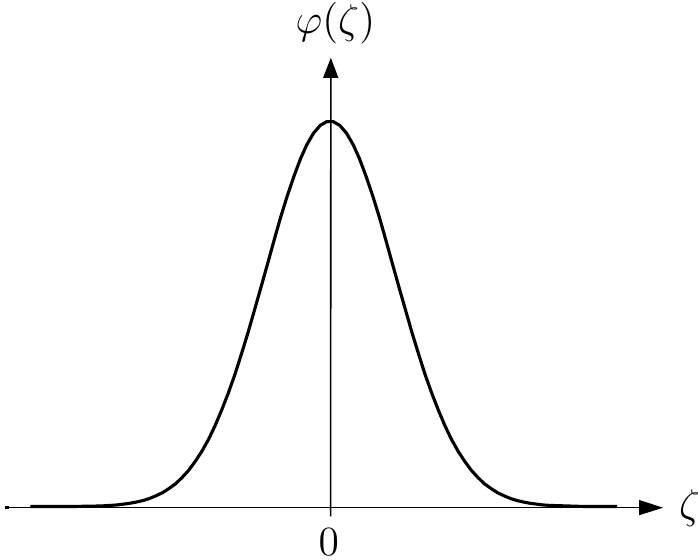}}
\caption{Ein um $\hat\zeta=0$ zentriertes Gauß-Paket.}\label{f3:GaussUmNull}
\end{floatingfigure}
Wir wollen uns in diesem Abschnitt eng an das Beispiel des Wasserstoffspinechos halten,
das auch ausführlich in \cite{DissAR}, Kap. 3, diskutiert ist. Auf diese Weise können wir unsere
Resultate stets direkt mit den Ergebnissen aus \cite{DissAR} vergleichen.

Wir betrachten der Einfachheit halber ein Gaußsches Wellenpaket, dessen Einhüllende
$\vph(\zeta)$ um den Punkt $\hat\zeta = 0$ zentriert sei, siehe Abb. \ref{f3:GaussUmNull}. 
Die Schwerpunktsbewegung des zu einem bestimmten, atomaren Zustand $\rket{\alpha(Z)}$ gehörenden WKB-Wellenpakets
(\ref{e3:WP.WKB}) ergibt sich, wie bei der Diskussion am Ende von Abschnitt \ref{s2:WKB-Eigenschaften},
aus der Kombination der der Funktionen $\zeta_\alpha(Z)$ aus (\ref{e3:Zeta}) und $\tau(t)$
aus (\ref{e2:Def.Zeta.Tau}) im Argument der Einhüllenden und der Forderung
\pagebreak
\begin{align}\label{e3:SP.Bew.allg.}
\zeta_\alpha(\hat Z_\alpha(t))-\tau(t) = \int_{Z_0}^{\hat Z_\alpha(t)}\d Z'\ \frac{\bar k}{\sqrt{\helpkbar\bar
  k^2-2M V_\alpha(Z')}} - \tau(t) \overset!= \hat\zeta = 0\ .
\end{align}
Die zeitabhängige Bewegung der Schwerpunktskoordinate $\hat Z_\alpha(t)$ ist durch diese Gleichung implizit
festgelegt und entspricht der Bewegung eines klassischen Teilchens durch das Potential $V_\alpha(Z)$.

In dem in \cite{DissAR} definierten Fahrplanmodell wird nun nach der 
Position des Schwerpunkts (hier gleich dem Maximum) des $\alpha$-ten Wellenpakets zu einer Zeit $t$ gefragt,
bei der der Schwerpunkt eines freien Wellenpakets am Ort $Z$ liegt.
Für ein freies Wellenpaket gilt nach Voraussetzung $V(Z)\equiv 0$, also ist die freie Schwerpunktsbewegung
nach (\ref{e3:SP.Bew.allg.}) einfach durch
\begin{align}\label{e3:SP.frei}
\hat Z\subt{frei}(t) = Z_0 + \tau(t)
\end{align}
gegeben. Befindet sich das freie Wellenpaket am Ort $\hat Z\subt{frei}(t) = Z_0 + \tau(t)$, so ist die
Koordinate $\zeta_\alpha(\hat Z_\alpha(t))$ des Wellenpakets mit Index $\alpha$ zur gleichen Zeit $t$ gegeben durch
\begin{align}\label{e3:SP.zeta.t}
0 = \zeta_\alpha(\hat Z_\alpha(t)) - \tau(t) = \zeta_\alpha(\hat Z_\alpha(t)) - (\hat Z\subt{frei}(t) - Z_0)\ .
\end{align}
Wir können die Zeitabhängigkeit eliminieren, indem wir $Z = \hat Z\subt{frei}(t)$ und somit 
nach Gl. (\ref{e3:SP.frei}) $t=M(Z-Z_0)/\bar k$ setzen und
\begin{align}\label{e3:SP.zeta.Z}
\zeta_\alpha(\hat Z_\alpha(t))\big\vert_{t=M(Z-Z_0)/\bar k} = Z - Z_0
\end{align}
erhalten. Durch diese Gleichung ist implizit die Abhängigkeit des Schwerpunkt $\hat Z_\alpha$
des Wellenpakets mit Index $\alpha$ vom Ort $Z$ auf der Strahlachse definiert, der 
laut Rechnung dem Ort des Schwerpunkts eines freien Wellenpakets entspricht.

Die Berechnung von $\hat Z_\alpha$ als Funktion von $Z$ 
durch Lösen der Gleichung (\ref{e3:SP.zeta.Z}) ist sehr umständlich.
Einfacher ist die direkte Betrachtung der Koordinatenfunktion $\zeta_\alpha(Z)$ am Ort $Z$. Für das freie
Wellenpaket gilt
\begin{align}\label{e3:zeta.frei}
\zeta\subt{frei}(Z) = Z-Z_0
\end{align}
und wir können z.B. nach dem Wert der Funktionen
\begin{align}\label{e3:Delta.Zeta}
\Delta\zeta_\alpha(Z) := \zeta_\alpha(Z) - \zeta\subt{frei}(Z) = \zeta_\alpha(Z) - (Z-Z_0)
\end{align}
fragen. Die physikalische Bedeutung dieser Größe sieht man nach Multiplikation mit der
inversen, mittleren Geschwindigkeit $\bar v^{-1} = M/\bar k$. Dann folgt nämlich
nach Einsetzen der Definition (\ref{e3:Zeta}) von $\zeta_\alpha(Z)$
\begin{align}\label{e3:temp.T}
\frac{M}{\bar k}\Delta\zeta_\alpha(Z) = \int_{Z_0}^Z\d Z'\ \frac{M}{\sqrt{\helpkbar\bar
k^2-2M V_\alpha(Z')}} - T\subt{frei}(Z)\ ,
\end{align}
wobei $T\subt{frei}(Z)$ die Zeit ist, die ein freies Wellenpaket mit der Geschwindigkeit $\bar v$
braucht, um von $Z_0$ nach $Z$ zu gelangen. Im gleichen Sinne kann das in (\ref{e3:temp.T})
auftretende Integral interpretiert werden. Der Integrand ist das Inverse der lokalen
Geschwindigkeit
\begin{align}\label{e3:lok.v}
v_\alpha(Z) := \frac{\sqrt{\helpkbar\bar
k^2-2M V_\alpha(Z)}}{M}
\end{align}
und demzufolge ist das Ortsintegral einfach die Gesamtzeit $T_\alpha(Z)$, die der Schwerpunkt des $\alpha$-ten
Wellenpakets braucht, um durch das Potential $V_\alpha(Z')$ von $Z_0$ nach $Z$ zu gelangen. Der gesamte Ausdruck
(\ref{e3:temp.T}) ist also nichts weiter als die Zeitdifferenz $\Delta T_\alpha(Z)$
zwischen dem Eintreffen der Schwerpunkte des $\alpha$-ten und des freien Wellenpakets 
am Ort $Z$, wenn beide gleichzeitig bei $Z_0$ gestartet sind. Wir definieren also
\begin{align}\label{e3:Delta.T}
\begin{split}
\Delta T_\alpha(Z) =&~\frac{M}{\bar k}\Delta\zeta_\alpha(Z)\\
:=&~T_\alpha(Z) - T\subt{frei}(Z) =
\int_{Z_0}^Z\d Z'\ \frac{M}{\sqrt{\helpkbar\bar k^2-2M V_\alpha(Z')}} - \frac{M}{\bar k}(Z-Z_0)\ .
\end{split}
\end{align}

Die Größe $\Delta T_\alpha(Z)$ ist viel leichter zugänglich als der tatsächliche Ort $\hat Z_\alpha(Z)$
des Schwerpunkts, für den man die Koordinate $\zeta_\alpha(Z)$ erst invertieren müsste. Ein viel klareres
und auch einfacher zu berechnendes Bild der Schwerpunktsbewegung bekommt man tatsächlich durch die
Verwendung von $\Delta T_\alpha(Z)$. Hierfür muss man nur die Koordinatenfunktionen $\zeta_\alpha(Z)$
kennen und numerisch berechnen, was man aber für die Berechnung der Wellenfunktion ohnehin hätte
tun müssen. Wir bekommen bei der Beschreibung eines lABSE-Experiments den Fahrplan also sozusagen geschenkt.
Auf der anderen Seite kann man den Fahrplan leicht berechnen, ohne die Wellenfunktionen zu kennen,
d.h. man kann auf einfache Weise Fahrpläne für verschiedene Feldkonfigurationen erstellen und erst
dann entscheiden, für welche Konfiguration man eine komplette theoretische Beschreibung des atomaren Zustands
und schließlich des lABSE-Experiments liefert.

Nun wollen wir zum Wasserstoffspinecho zurück kommen. Im Grundzustand gibt es
vier atomare Eigenzustände: drei mit Gesamtdrehimpuls $F=1$ ($F_3=±1,0$)
und einen mit $F=F_3=0$. Diese koppeln in einem Magnetfeld zu
Energieeigenzuständen, die wir wieder mit $\rket\alpha$ $(\alpha=1,2,3,4)$ bezeichnen
wollen\footnote{Natürlich hängt die Zusammensetzung dieser Zustände vom lokalen Wert des
Magnetfelds und somit vom Ort $Z$ ab. Wir werden aber für den Rest dieses Kapitels
diese Abhängigkeit nicht mehr explizit in der Form $\rket{\alpha(Z)}$ anschreiben,
da dies für das Verständnis des Fahrplanmodells nicht notwendig ist und um die Notation
so kurz und übersichtlich wie möglich zu halten.}.
Das Eigenwertproblem für Wasserstoff im Grundzustand in einem
magnetischen Feld wurde gelöst und ist im Anhang \ref{sB:H1} dargestellt.
In Tabelle \ref{tB:MEWEV} kann man die eben angesprochenen Eigenzustände im Magnetfeld ablesen,
ebenso wie die jeweiligen Eigenenergien. 

Die grafische Darstellung der Eigenenergien in Abhängigkeit des angelegten magnetischen Felds $\mc B = \mc B_3$
in $Z$-Richtung kann man im Breit-Rabi-Diagramm, Abb. \ref{f3:BR-H1}, ablesen.
Wie man sieht, haben die Zustände $\rket1$ und $\rket3$ eine lineare Energieabhängigkeit
vom Magnetfeld, der Zustand $\rket2$ (und auch $\rket4$) besitzt dagegen eine quadratische Energieabhängigkeit.
\begin{floatingfigure}[r]{7cm}
\includegraphics[width=7cm]{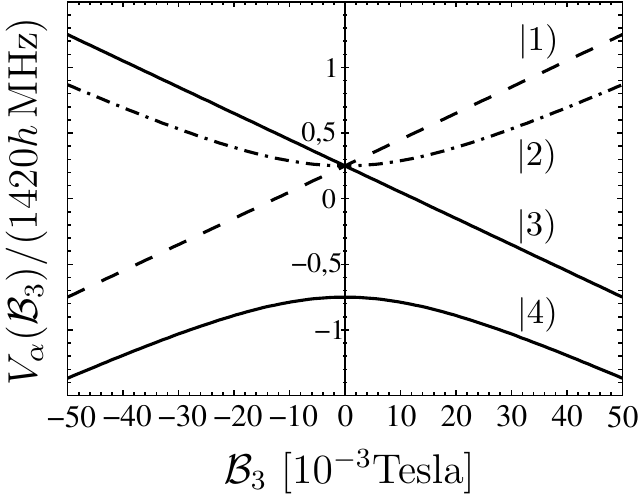}
\caption{Breit-Rabi-Diagramm für Wasserstoff im $(n=1)$-Unterraum}
\label{f3:BR-H1}
\end{floatingfigure}

Wir betrachten nun einen Zustand im Magnetfeld mit Gesamtdrehimpuls Eins in $x$-Richtung,
d.h. einen Eigenzustand zum Operator $F_x$ mit Eigenwert $F_x=1$, sowie zum Operator $\v F^2$ mit Eigenwert $F(F+1) = 2$:
\begin{align}\label{e3:Def.1x}
  \ket{1_x}:=\tfrac12\rket1 + \tfrac1{\sqrt2}\rket2 + \tfrac12\rket3\ .
\end{align}
In diesem Fall ist es günstig, die Energien der Zustände $\rket{1}$, $\rket2$ und $\rket3$
im Limes $\mc B\to0$ als Nullpunkt zu wählen. Weiterhin betrachten wir
nur sehr geringe Magnetfelder\footnote{M. DeKieviet, private Diskussion.},
die etwa bei 1-4\% des kritischen Magnetfelds\footnote{
Im Zusammenhang mit der Aufspaltung der atomaren Energieniveaus in einem angelegten Magnetfeld unterscheidet man
zwischen dem Zeeman-Bereich und dem Paschen-Back-Bereich. Der Zeeman-Bereich liegt für kleine Magnetfelder
vor, wo die Kopplung des Bahndrehimpulses des Elektrons und seines Spin zum Gesamtdrehimpuls 
stärker ist als die Kopplung beider Drehimpulse an das angelegte Magnetfeld. 
Bei starken Magnetfeldern überwiegt dagegen die Kopplung der Einzeldrehimpulse des Elektrons an das Magnetfeld
und man spricht vom Paschen-Back-Bereich. Im Breit-Rabi-Diagramm \ref{f3:BR-H1} 
erkennt man den Übergang zwischen beiden Bereichen z.B. an dem Übergang der Energie des Zustands $\rket{2}$
von einer quadratischen zu einer linearen Feldabhängigkeit. Als kritisches Magnetfeld bezeichnen wir die 
(ungefähre) Grenze zwischen Zeeman- und Paschen-Back-Bereich.} von $\mc B\subt{kr}=50\u {mT}$ liegen.
In diesem Bereich gilt näherungsweise (siehe Tab. \ref{tB:MEWEV})
\begin{align}\label{e3:EW-B-H1.1+3}
  E_{1,3}(\mc B) = ±\tfrac12 g\mu_B\mc B,\qquad E_2(\mc B) \approx
  \frac{\klr{\tfrac12 g \mu_B \mc B}^2}{2A}\ ,
\end{align}
wobei $g\approx 2$ der Landé-Faktor des Elektrons, $\mu_B$ das Bohrsche Magneton
und $A=1420 h \u {MHz} \approx 6\ten{-6}\u {eV}$ die Hyperfeinaufspaltung des
Wasserstoffs ist.

\pagebreak
Wählen wir nun $\mc B=2\u {mT}$, so erhalten wir die Energien
\begin{align}\label{e3:FP-EW.typisch}
  E_{1,3}(2\u {mT}) \approx ±1\ten{-7}\u {eV},\qquad E_2(2\u {mT}) \approx
  1\ten{-9}\u {eV}\ .
\end{align}
Weiterhin betrachten wir einen Strahl von Wasserstoffatomen mit einer mittleren Geschwindigkeit von
$\bar v = 2880\u {m/s}$, d.h. kinetischer Energie $E\subt{kin}=0,025\u {eV}$. Diese Parameter
sind so gewählt, dass die resultierenden Diagramme einen guten Überblick über die
physikalischen Vorgänge liefern. Ob sie sich für wirklich für Experimente eignen, spielt also
im Moment keine Rolle für uns.

Die Näherung für kleine Potentiale ist mit den eben angegebenen Parametern 
sehr gut erfüllt, z.B. gilt $E_1/E\subt{kin} = 4\ten{-6}$. In einem solchen Strahl haben die Atome eine
Wellenzahl von $\bar k=3,4\ten{10}/\u m$ und eine Wellenlänge 
$\bar\lambda = 1,8\ten{-10}\u m$, d.h. für eine Breite von $\sigma=10^{-7}\u m=
1800\bar\lambda \gg \bar\lambda/(4\pi)$ stellt das WKB-Wellenpaket nach (\ref{e2:kbar.k.approx.2})
eine gute Näherung dar.

Nun wählen wir eine Magnetfeldkonfiguration, die aus zwei antiparallel gerichteten Feldern
bestehen soll. Beide Felder wollen wir als stückweise konstant wählen
und in Strahlrichtung orientieren. Das erste Feld erzeuge also einen
Potentialwall im Bereich $-2\u m\leq-1\u m$, das zweite Feld wollen wir für
$z\geq 1\u m$ bis ins Unendliche ausgedehnt wählen (zur besseren
Demonstration der Kreuzungspunkte). 

\begin{figure}[hp]
  \centering
\subfloat[]{\label{f3:FPPotBsp.a}\includegraphics[width=12cm]{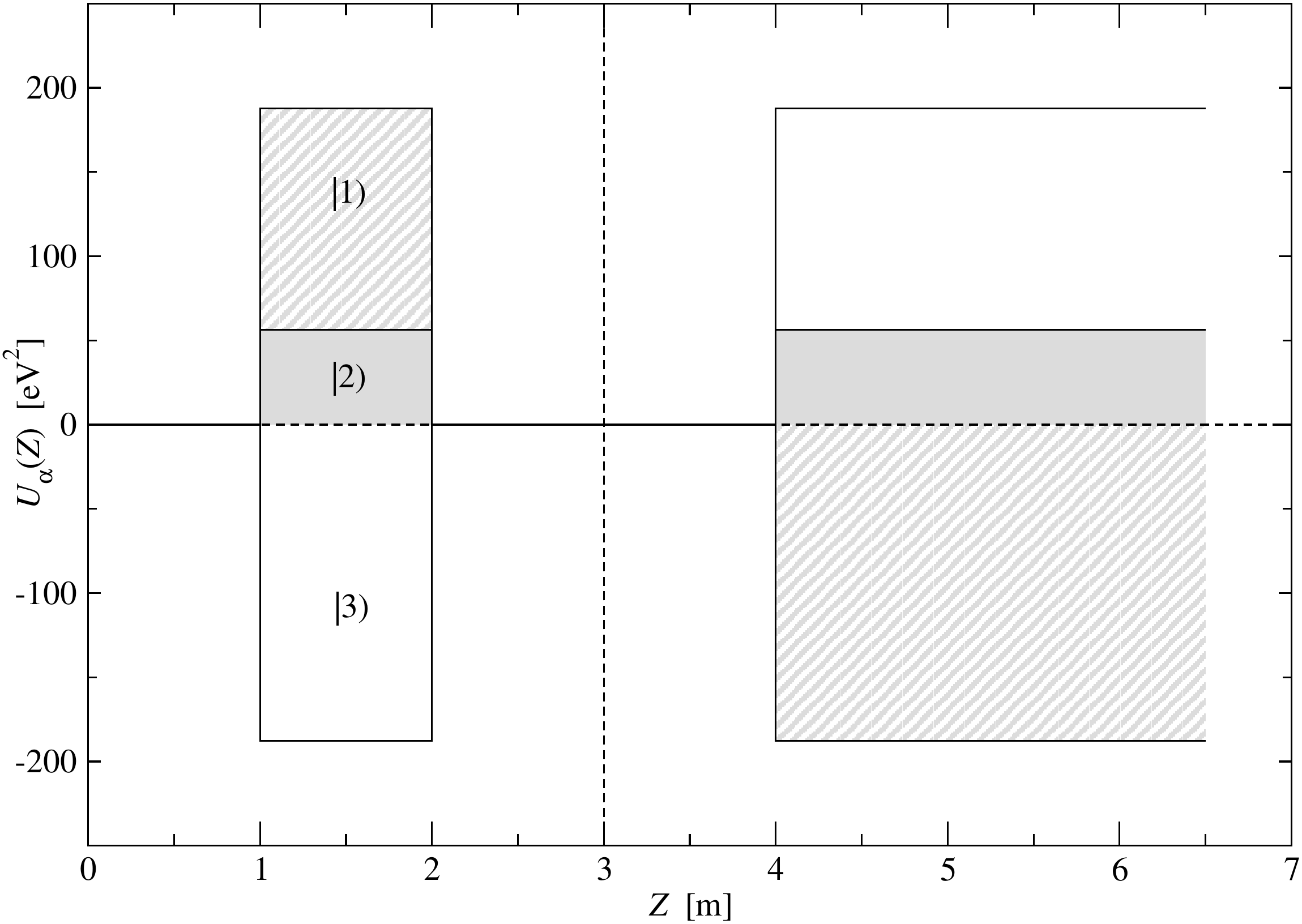}}\\[2mm]
  \subfloat[]{\label{f3:FPPotBsp.b}\includegraphics[width=12.1cm]{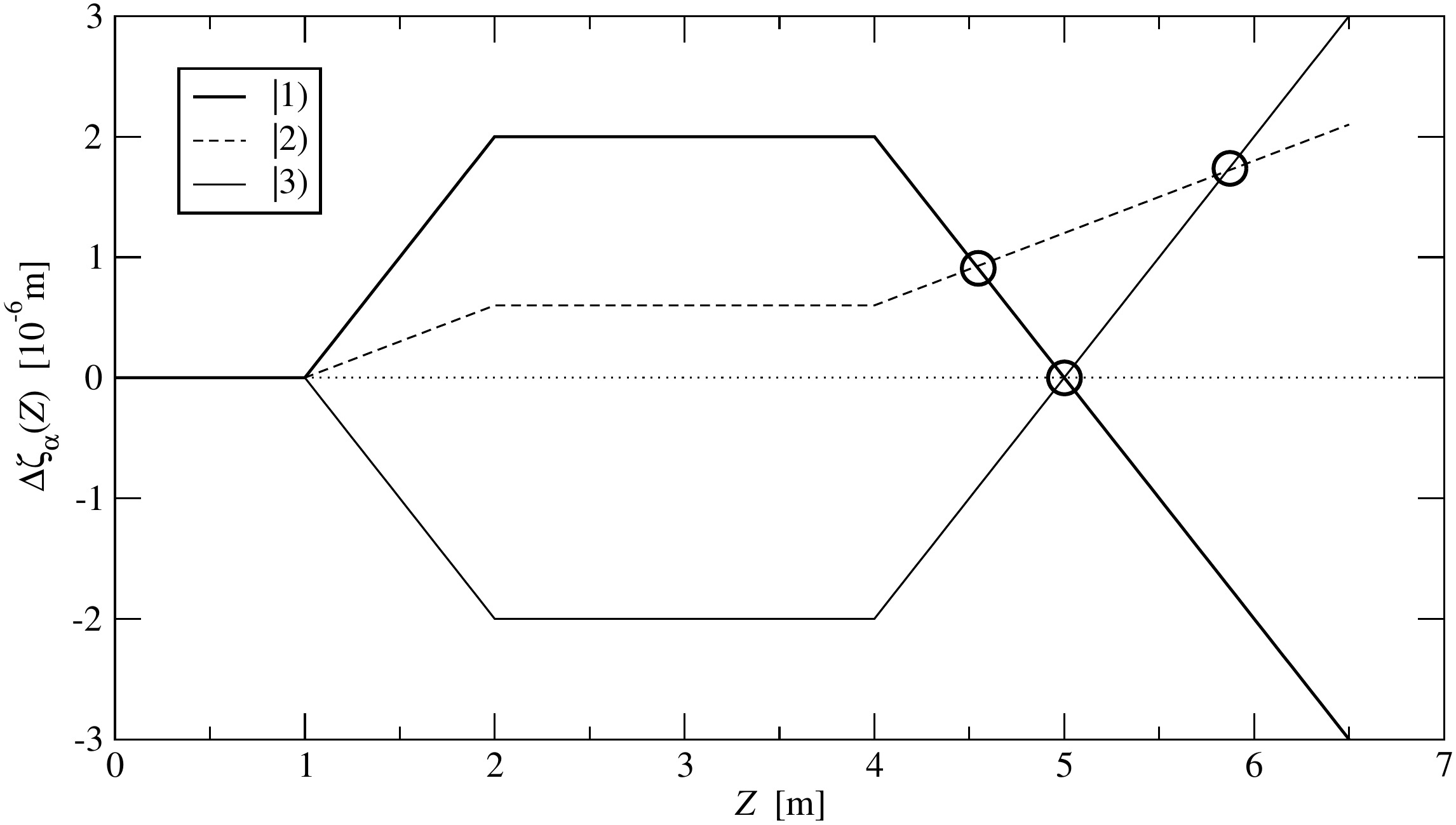}}
  \caption[Die Potentiale für den polarisierten Zustand $\ket{1_x}$]{
    Die Potentiale $U_\alpha(z)=2m V_\alpha(z)$ für die Komponenten des polarisierten Zustands
    $\ket{1_x}$, dargestellt in \subref*{f3:FPPotBsp.a}. Das Potential $U_2(z)$ ist dabei
    um einen Faktor 30 ggü. dem realistischen Potential für Wasserstoff in einem
    Magnetfeld mit $\mc B = 2\u {mT}$ skaliert. In \subref*{f3:FPPotBsp.b} ist der Fahrplan mit der daraus resultierenden
    Vorhersage für die Kreuzungspunkte der Schwerpunkte der einzelnen Wellenpakete gezeigt.}
  \label{f3:FPPotBsp}
\end{figure}

An dieser Stelle sei ein wichtiges Detail angemerkt: Da laut
Gl. (\ref{e3:FP-EW.typisch}), gemäß dem Breit-Rabi-Diagramm, 
die Energie $E_2$ um den Faktor 100 kleiner als
$E_{1,3}$ ist, bewegt sich der Zustand $\rket2$ praktisch frei
durch die Felder. Für die Demonstration, die wir hier
anstreben, erweist sich dieser Umstand als sehr ungeeignet, weshalb wir bis
auf weiteres die Energie $E_2$ um einen Faktor 30 verstärkt haben.
Die resultierenden Potentiale $U_\alpha(Z) = 2M V_\alpha(Z)$ können
Abb. \subref*{f3:FPPotBsp.a} entnommen werden.

In Abb. \subref*{f3:FPPotBsp.b} ist der Fahrplan für diese Potentiale dargestellt, d.h. wir
haben hier die in (\ref{e3:Delta.Zeta}) definierten Differenzen $\Delta\zeta_\alpha(Z)$
für die jeweiligen Wellenpakete zum freien Wellenpaket gegen die Ortskoordinate $Z$ aufgetragen.
Es sei daran erinnert, dass diese Abstände die durch die Geschwindigkeit des freien Teilchens
dividierten Zeitdifferenzen $\Delta T_\alpha(Z)$ aus (\ref{e3:Delta.T}) sind.

Im ersten Magnetfeld haben die Zustände $\rket1$ und $\rket2$ positive Potentiale und
\glqq verspäten\grqq~sich gegenüber dem freien Wellenpaket, d.h. im Fahrplan tritt eine
positive Abweichung auf. Zustand $\rket3$ hat ein negatives Potential, so dass sein Wellenpaket
einen Vorsprung ggü. dem freien Wellenpaket bekommt. Im zweiten Magnetfeld dreht sich die
Situation für die Zustände $\rket1$ und $\rket3$ gerade um, während Zustand $\rket2$ das
gleiche Potential wie im ersten Feld spürt. Die sich hieraus ergebenden Kreuzungspunkte sind
in Abb. \subref*{f3:FPPotBsp.b} eingekreist. Würde man im zweiten Magnetfeld einen Detektor platzieren,
so würde man an diesen Stellen Interferenz beobachten können, da je zwei Wellenpakete gleichzeitig
auf den Detektor treffen und daher einen nichtverschwindenden Überlapp haben. Bei Verschiebung
des Detektors würde man die ortsabhängige Oszillation der relativen Phasen der beteiligten Wellenpakete
sehen, sowie eine Änderung der Amplitude dieser Oszillation, die durch die Änderung des Überlapps
entsteht.

Diese physikalische Überlegungen wollen wir im nächsten Abschnitt quantitativ mit den bisherigen Methoden
untersuchen und das Kapitel mit der Berechnung des lABSE-Interferenz-Signals abschließen.

\FloatBarrier
\subsection{Die Berechnung des Interferenzsignals}

Der atomare Gesamtzustand am Ort $Z$ lautet gemäß (\ref{e3:Initial}) und (\ref{e3:WP.WKB}) mit dem
Polarisationszustand (\ref{e3:Def.1x})
\begin{align}\label{e3:PsiDetektor}
\begin{split}
	\ket{\Psi(Z,t)} &= \tfrac12\e^{\I\phi_{1}(Z,t)}\vph(\zeta_1(Z)-\tau(t))\rket1\\
		&+ \tfrac1{\sqrt2}\e^{\I\phi_{2}(Z,t)}\vph(\zeta_2(Z)-\tau(t))\rket2\\
		&+ \tfrac12\e^{\I\phi_{3}(Z,t)}\vph(\zeta_3(Z)-\tau(t))\rket3
\end{split}
\end{align}
mit den WKB-Phasenwinkeln aus (\ref{e3:WKB-PW}).
Nun sei am Ort $Z$ ein Detektor platziert, der den Anteil der ursprünglichen Polarisation
im Zustand (\ref{e3:PsiDetektor}) misst, also den Zustand $\ket{\Psi(Z,t)}$ auf $\ket{1_x}$
projiziert. Die Wellenfunktion, die der Detektor messen wird, lautet damit
\begin{align}
\begin{split}
\Psi_{1x}(Z,t) &= \bracket{1_x}{\Psi(Z,t)}\\
&= \tfrac14\e^{\I\phi_{1}(Z,t)}\vph(\zeta_1(Z)-\tau(t))\\
&+ \tfrac12\e^{\I\phi_{2}(Z,t)}\vph(\zeta_2(Z)-\tau(t))\\
&+ \tfrac14\e^{\I\phi_{3}(Z,t)}\vph(\zeta_3(Z)-\tau(t))
\end{split}
\end{align}
und die Wahrscheinlichkeitsdichte
\begin{align}\label{e3:Special.Density}
\begin{split}\hspace*{-2mm}
\rho_{1x}(Z,t) &= \abs{\Psi_{1x}(Z,t)}^2\\ 
&= \frac1{16}\Big|\vph(\zeta_1(Z)-\tau(t))
+2\e^{\I\phi_{2,1}(Z)}\vph(\zeta_2(Z)-\tau(t))+\e^{\I\phi_{3,1}(Z)}\vph(\zeta_3(Z)-\tau(t))\Big|^2\ .
\end{split}
\end{align}
Dabei haben wir einen globalen Phasenfaktor $\exp\klg{\I\phi_1(Z,t)}/4$ aus dem Betragsquadrat herausgezogen
und die Phasenwinkeldifferenzen
\begin{align}\label{e3:PWDiff}
\begin{split}
\phi_{\beta,\alpha}(Z) &= \phi_\beta(Z,t) - \phi_\alpha(Z,t) \\
&= \int_{Z_0}^Z\d Z'\ \klr{\sqrt{\bar k^2-2M V_\beta(Z')}-\sqrt{\bar k^2-2M V_\alpha(Z')}}\ .
\end{split}
\end{align}
eingeführt.

Ein lABSE-Experiment misst stets den zeitlich integrierten Fluss von Atomen.
Dies entspricht dem Zählen von Atomen, die vom Detektor gemessen werden und sich somit (in unserem Beispiel)
im Zustand $\ket{1_x}$ befinden, während eines festgelegten Zeitintervalls. 
Bezogen auf die Gesamtzahl von Atomen, die am Detektor während dieses Zeitintervalls angekommen sind, 
ergibt sich dann die Wahrscheinlichkeit $P_{1x}(Z)$ dafür, 
dass sich ein Atom am Detektorort im Zustand $\ket{1_x}$ befindet.
In dem eindimensionalen Fall, den wir beim lABSE stets betrachten, ist der Fluss $\Phi_{1x}(Z)$
stets identisch mit der Wahrscheinlichkeitsstromdichte $j_{1x}(Z,t)$, für die
wir in diesem Abschnitt die Näherung
\begin{align}\label{e3:Fluss}
\Phi_{1x}(Z,t) := j_{1x}(Z,t) \approx \frac{\bar k}M \rho_{1x}(Z,t)\ .
\end{align}
machen. Da wir uns den Detektor im Magnetfeld platziert denken, haben die Wellenpakete eigentlich unterschiedliche
lokale Geschwindigkeiten. Für die sehr kleinen Magnetfelder, die wir in diesem Abschnitt betrachten, sollte
(\ref{e3:Fluss}) aber eine sehr gute Näherung darstellen. Eine korrekte Definition der Wahrscheinlichkeitsstromdichte
wird in Kap. \ref{s6:Formalismus} gegeben.

Das lABSE-Signal, dass wir nun berechnen wollen lautet also
\begin{align}\label{e3:intFluss}
P_{1x}(Z) = \int\d t\ \Phi_{1x}(Z,t) = \frac{\bar k}M\int\d t\ \rho_{1x}(Z,t)\ .
\end{align}

An dieser Stelle wollen wir noch eine Anmerkung zum Gesamtfluss der Atome machen. Dieser ergibt sich
in der Näherung aus (\ref{e3:Fluss}) zu
\begin{align}
\Phi\subt{ges}(Z,t) &:= \frac{\bar k}M\bracket{\Psi(Z,t)}{\Psi(Z,t)} 
= \frac{\bar k}M\sum_{\alpha=1,2,3}\hat\chi_\alpha^2\vph(\zeta_\alpha(Z)-\tau(t))^2\ .
\end{align}
Die Wahrscheinlichkeit, dass das Atom am Detektor ankommt ist dann gegeben durch
\begin{align}
\begin{split}
P\subt{ges}(Z) &= \int_{-\infty}^\infty\d t\ \frac{\bar k}M
   \sum_{\alpha=1,2,3}\hat\chi_\alpha^2\vph(\zeta_\alpha(Z)-\tau(t))^2\\
&= \sum_{\alpha=1,2,3}\hat\chi_\alpha^2\int_{-\infty}^\infty\d \tau\ \vph(\zeta_\alpha(Z)-\tau)^2\\
&= \sum_{\alpha=1,2,3}\hat\chi_\alpha^2\\
&= 1
\end{split}
\end{align}
und ist wie zu erwarten gleich Eins.

Verwenden wir nun die Dichte $\rho_{1x}(Z,t)$ aus (\ref{e3:Special.Density}) und berechnen (\ref{e3:intFluss}),
so erhalten wir nach Substitution $\frac{\bar k}m t\to \tau$ und anschließender Translation/Spiegelung
$\tau\to \zeta_1(Z)-\tau$
\begin{multline}
	P_{1x}(Z) = \frac1{16}\int_{-\infty}^\infty\d\tau\ \Big|\vph(\tau)+2\e^{\I\phi_{2,1}(Z)}\vph(\zeta_2(Z)-\zeta_1(Z)+\tau)\\
		+ \e^{\I\phi_{3,1}(Z)}\vph(\zeta_3(Z)-\zeta_1(Z)+\tau)\Big|^2\ .
\end{multline}
Nach Ausmultiplikation des Betragsquadrats erhalten wir weiter
\begin{align}
	\begin{split}\hspace*{-4mm}
		P_{1x}(Z) = \frac1{16}\int_{-\infty}^\infty\d\tau\ &\Big(\vph^2(\tau) + 4\vph^2(\zeta_2(Z)-\zeta_1(Z)+\tau) 
					+ \vph^2(\zeta_3(Z)-\zeta_1(Z)+\tau)\\
			&+ 4\cos(\phi_{2,1}(Z))\vph(\tau)\vph(\zeta_2(Z)-\zeta_1(Z)+\tau)\\
			&+ 2\cos(\phi_{3,1}(Z))\vph(\tau)\vph(\zeta_3(Z)-\zeta_1(Z)+\tau)\\
			&+ 4\cos(\phi_{3,2}(Z))\vph(\zeta_2(Z)-\zeta_1(Z)+\tau)\vph(\zeta_3(Z)-\zeta_1(Z)+\tau)\Big)\ ,
	\end{split}
\end{align}
wobei wir hier nur reelle Einhüllende $\vph(\zeta)$ zugelassen haben.
Die ersten drei Summanden in der Klammer ergeben nach der Integration die Zahl sechs, d.h. wir erhalten
\begin{align}\label{e3:FPBspintFluss}
	P_{1x}(Z) = \frac38 &+ \frac14\cos(\phi_{2,1}(Z)) I_{2,1}(Z)\\
		&+ \frac18\cos(\phi_{3,1}(Z)) I_{3,1}(Z)\\
		&+ \frac14\cos(\phi_{3,2}(Z)) I_{3,2}(Z)\ ,
\end{align}
wobei wir die Integralfunktion
\begin{align}\label{e3:Ueberlapp}
	I_{\alpha,\beta}(Z) := \int_{-\infty}^\infty\d\tau\ \vph(\tau)\vph(\zeta_\alpha(Z)-\zeta_\beta(Z)+\tau)
\end{align}
verwendet haben. Der Wert von $I_{\alpha,\beta}(Z)$ ist proportional zum Überlapp der Wellenpakete der Komponenten mit
den Indizes $\alpha$ und $\beta$. Dieser Überlapp wird dann an genau der Stelle $Z$ auf der Strahlachse am größten, 
wo die Differenz $\zeta_\alpha(Z)-\zeta_\beta(Z)$ verschwindet, also wenn beide Wellenpakete exakt gleichzeitig am
Detektor eintreffen.
\FloatBarrier

\begin{figure}[hp]
\centering
\subfloat[]{\label{f3:FPFlussBsp.a}\includegraphics[width=12cm]{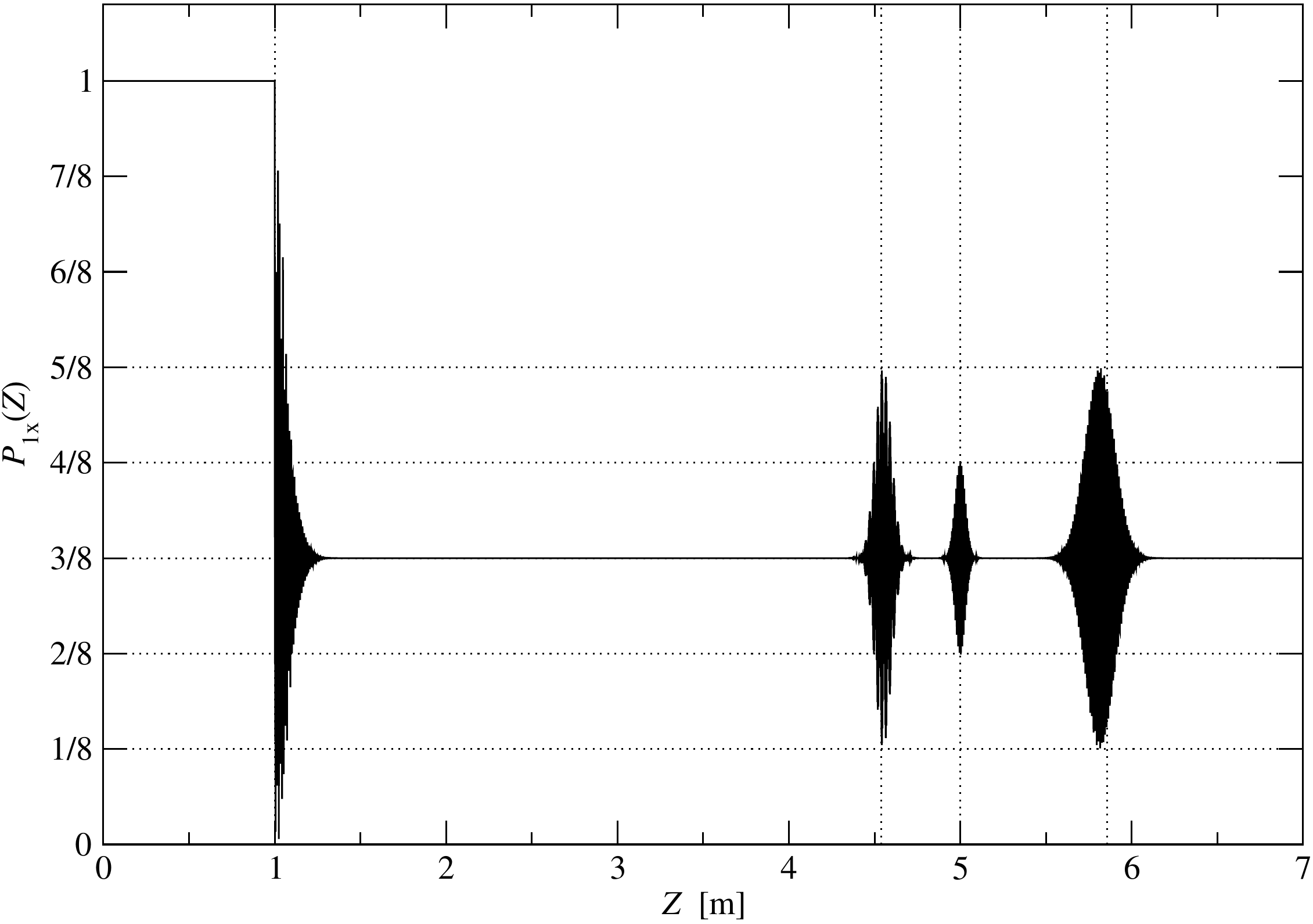}}\\[2mm] 
\subfloat[]{\label{f3:FPFlussBsp.b}\includegraphics[width=12cm]{FP1-Demo}}
\caption[Die berechnete Wahrscheinlichkeit $P\subt{1x}(Z)$ und die Schwerpunktsbewegungen der
Teilwellenpakete.]{
Abb. \subref*{f3:FPFlussBsp.a} zeigt den numerisch berechneten integrierte Fluss $P_{1x}(Z)$ der Atome,
die sich im Zustand $\ket{1_x}$ befinden in Abhängigkeit des Ortes $Z$
auf der Strahlachse. Das in \subref*{f3:FPFlussBsp.b} dargestellte Signal steht in direktem Zusammenhang
mit der Bewegung der Schwerpunkte. Man beachte, dass auch hier
für das zweite Wellenpaket ein Potential verwendet wurde, das um den Faktor
30 gegenüber dem realistischen Potential skaliert ist. Verwendet wurden weiter
Wellenpakete mit einer Breite von $\sigma=10^{-7}\u m$ und mit
mittlerer Geschwindigkeit $\bar v=2880\u {m/s}$.}
\label{f3:FPFlussBsp}
\end{figure}

\begin{figure}[hp]
  \centering
  \subfloat[]{\label{f3:FPFlussBsp2.a}\includegraphics[width=12cm]{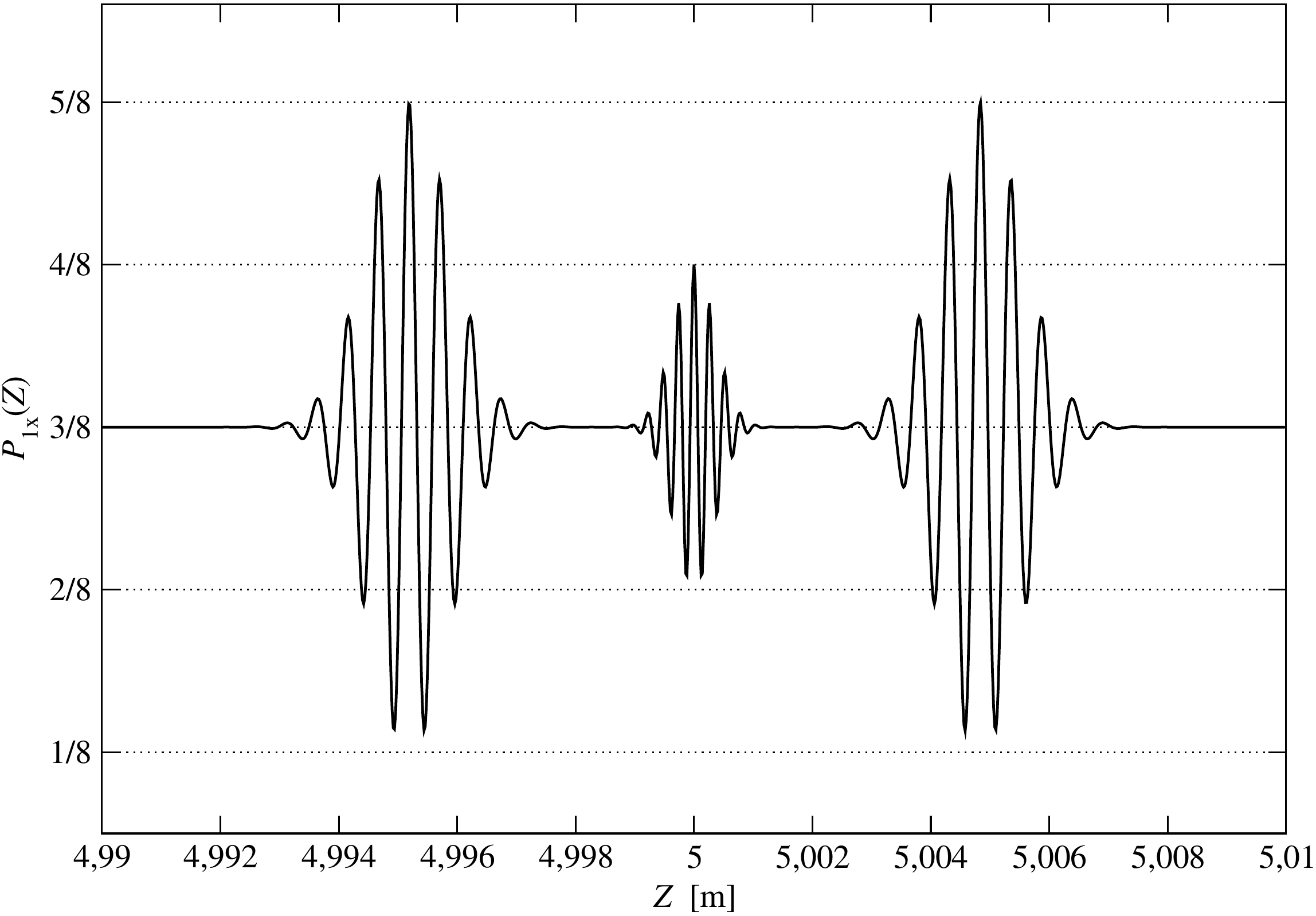}}
\subfloat[]{\label{f3:FPFlussBsp2.b}\includegraphics[width=12cm]{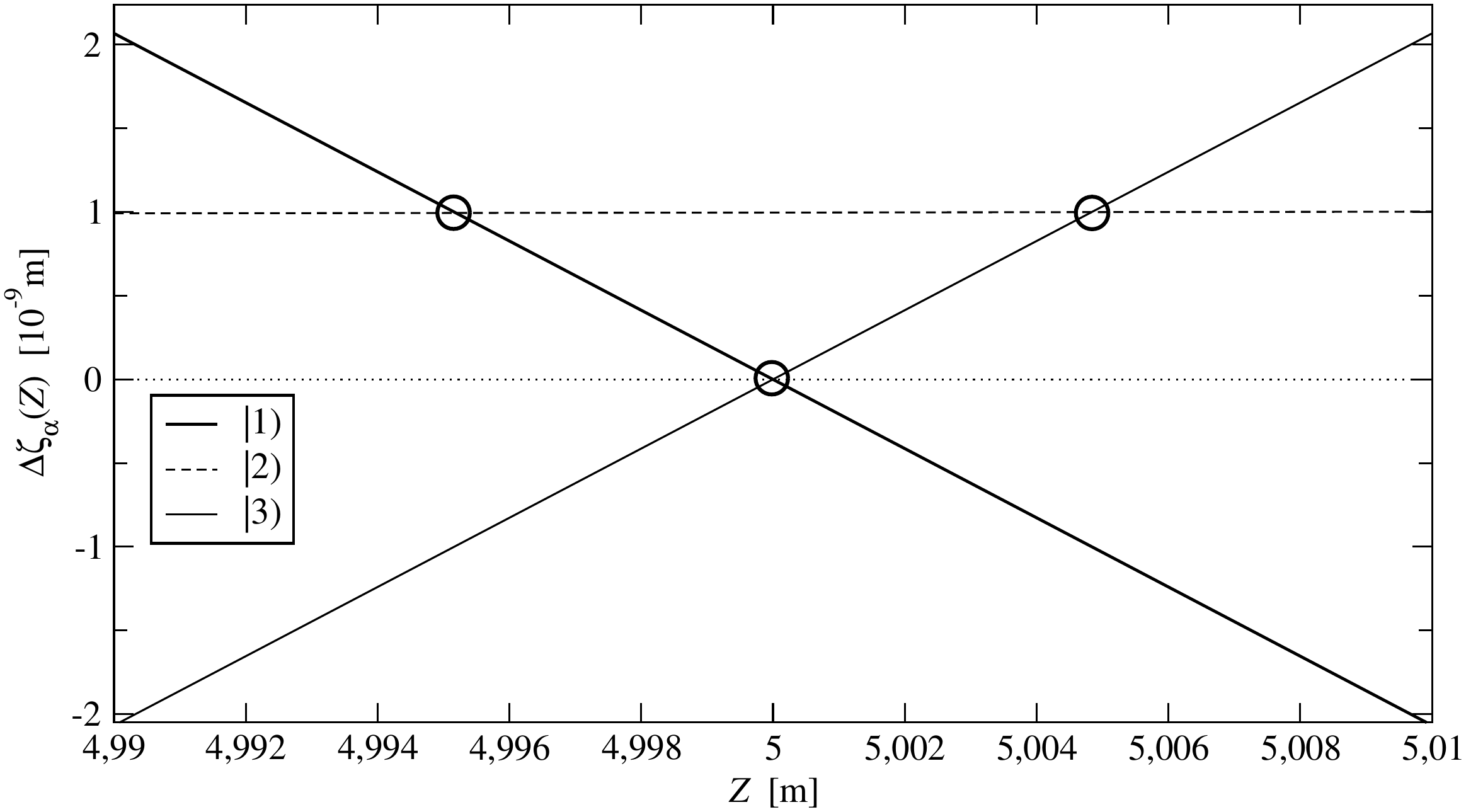}}
  \caption[Theoretisches Spinechosignal für Wasserstoff im Grundzustand (ohne
  Skalierung).]{
    Ein Beispiel für das Fahrplanmodell bei realistischen
    Potentialen. Die Potentiale nach Gl. (\ref{e3:EW-B-H1.1+3}) entsprechen nun einem Magnetfeld von $\mc
    B=0.5\u {mT}$, das von Wasserstoff mit einem Impuls bzw. einer Wellenzahl von  
$\bar k=11463\u {eV}\,\hat=\,5,8\ten{10}/\u m$ und einer de-Broglie-Wellenlänge von $\bar\lambda = 1,08\ten{-10}\u m$
    durchquert wird. Dies entspricht einer Geschwindigkeit der Atome von 
		$v = 3661\u {m/s}$. Die Breite der Wellenpakete beträgt
    $\sigma\approx 10^{-10}\u m$, was nicht mehr so gut die Bedingung (\ref{e2:kbar.k.approx.2})
    erfüllt. In \subref*{f3:FPFlussBsp2.a} kann man nun die korrekten Verhältnisse der Breiten
    und der Wellenlängen der einzelnen Signale zueinander erkennen.
    In \subref*{f3:FPFlussBsp2.b} sieht man, dass die Kreuzungspunkte nun viel dichter beisammen
    liegen, als bei Abb. \ref{f3:FPFlussBsp}.}
  \label{f3:FPFlussBsp2}
\end{figure}

In Abb. \subref*{f3:FPFlussBsp.a} ist die numerisch berechnete Ortsabhängigkeit des
integrierten Flusses $P_{1x}(Z)$ aus Gl. (\ref{e3:FPBspintFluss}) dargestellt. Es wurden die gleichen
Parameter wie für den in Abb. \ref{f3:FPPotBsp} gezeigten Fahrplan verwendet, sowie das in am Anfang
von Abschnitt \ref{s3:Fahrplanmodell} (siehe Abb. \ref{f3:Gausspaket}) eingeführte, 
um Null zentrierte Gaußsche Wellenpaket
\begin{align}\label{e3:Gausspaket}
 	\vph(\zeta) = \frac{1}{\sqrt{\sigma\sqrt\pi}}\exp\klg{-\frac{\zeta^2}{2\sigma^2}}\ ,
\end{align}
hier mit einer Breite $\sigma = 10^{-7}\u{m}$. Man erkennt in Abb. \subref*{f3:FPFlussBsp.a} folgendes:
\begin{enumerate}[(i)]
\item Vor dem Potential bewegen sich alle Wellenpakete gleich, d.h. der
  Polarisationszustand $\ket{1_x}$ des Atoms bleibt erhalten. Mit Beginn der Potentiale laufen die
  Zustände auseinander und es findet eine Oszillation der relativen Phase
  statt (die Spinrotation). Der Fluss pendelt sich schließlich beim konstanten Wert $3/8$ ein. 
	Hier gibt es keinen Überlapp zwischen den Wellenpaketen. Im weiteren Verlauf auf der
  Strahlachse treten schließlich drei Spinecho-Signale auf, verursacht durch
  die Oszillation der relativen Phase je zweier Wellenpakete und sichtbar gemacht
  durch deren Überlapp. 
\item Die Amplituden der unterschiedlichen Signale
  entsprechen den Kosinus-Vorfaktoren aus (\ref{e3:FPBspintFluss}).
\item Die Breite der Signale hängt vom Schnittwinkel der Schwerpunktsbewegung
  der beteiligten Wellenpakete aus Abb. \ref{f3:FPFlussBsp} ab. Ein kleiner
  Winkel bedeutet, dass der (zeitliche) Abstand (des Eintreffens) der Schwerpunkte 
	nur langsam mit dem betrachteten Ort auf der Strahlachse variiert, d.h. der Überlapp der
  Wellenpakete ist über einen längeren Bereich auf der Strahlachse groß.
  Bei einem großen Winkel ist der Überlapp nur sehr nahe am Kreuzungspunkt
  groß genug, um das Echo-Signal sichtbar zu machen.
\item Die Frequenz der Oszillation hängt von der Potentialdifferenz der
  beteiligten Wellenpakete ab. Dies ist in Abb. \ref{f3:FPFlussBsp} zwar nicht zu
  erkennen, wird aber weiter unten an einem geeigneteren Beispiel (siehe
  Abb. \ref{f3:FPFlussBsp2}) noch gezeigt. 
\end{enumerate}

Nun können wir auch besser die Skalierung von $V_2(z)$ um den Faktor 30
begründen: Hätten wir sie nicht vorgenommen, so
würden die Kreuzungspunkte wesentlich dichter beisammen liegen
und die drei Signale würden miteinander verschmelzen. Effektiv hätte man
dann an den Kreuzungspunkten den Ausgangszustand (fast) wieder hergestellt.
Eine Auflösung der drei Signale wäre dann nur bei sehr viel schmaleren
Wellenpaketen möglich.

Schließlich sei gesagt, dass Abb. \ref{f3:FPFlussBsp} der Abb. 3.12 auf S. 56
in \cite{DissAR} vom Prinzip her entspricht. Dort wurden allerdings korrekte
Energieverhältnisse verwendet, die Abbildung dann jedoch teilweise
reskaliert, so dass die Verhältnisse der Schwerpunktsbewegungen zueinander,
sowie die Lage der Kreuzungspunkte und die Form der Spinechosignale,
täuschen\footnote{M. DeKieviet, private Diskussion.}.

Der Vollständigkeit halber wollen wir nun aber noch ein Beispiel für
das Fahrplanmodell bei realistischen Energieverhältnissen angeben. Wir haben 
für ein magnetisches Feld von $\mc B=0.5\u {mT}$ bei veränderten
Strahlenergien (siehe Bildunterschrift des Signals) 
die Berechnungen erneut durchgeführt und das Ergebnis in
Abb. \ref{f3:FPFlussBsp2} dargestellt.

Aufgrund des geringeren Potentials für den Zustand $\rket2$ liegen die
Kreuzungspunkte nun sehr dicht beieinander. Die Abstände der Wellenpaketen
zwischen den Kreuzungspunkten sind daher sehr klein. Durch das insgesamt
sehr geringe Potential unterscheiden sich die Schwerpunktsgeschwindigkeiten 
nur gering, was dazu führt, dass die Pakete über eine längere Strecke hinweg 
einen signifikanten Überlapp haben.

Um zu erreichen, dass stets höchstens zwei Wellenpakete einen signifikanten Überlapp
haben, mussten wir für das Beispiel eine sehr geringe Breite der Wellenpakete von
$\sigma=10^{-10}\u m$ wählen, was bei dem verwendeten Wert für $\bar k$ etwa eine 
Größenordnung größer als $1/(2\bar k)$ ist. Die von Gl. (\ref{e2:kbar.k.approx.2}) geforderte
Bedingung $\sigma\gg 1/(2\bar k)$ ist hier mit $\sigma\approx 12/(2\bar k)$ immer noch
recht gut erfüllt.

Man sieht in Abb. \subref*{f3:FPFlussBsp2.a}, dass die äußeren Signale, bedingt durch die identischen Schnittwinkel
in Abb. \subref*{f3:FPFlussBsp2.b}, die gleiche Breite haben.
Da das Potential für $\rket2$ sehr nahe bei Null liegt und somit die Differenz zu den Potentialen von
$\rket1$ und $\rket3$ praktisch identisch ist, haben diese Signale auch die gleiche
Wellenlänge. Doppelt so groß ist der Unterschied zwischen den Potentialen von $\rket1$
und $\rket3$, was die halb so große Wellenlänge des mittleren Signals erklärt.
Diese beiden Zustände haben in Abb. \subref*{f3:FPFlussBsp2.b} auch den doppelten Schnittwinkel verglichen mit
den äußeren Kreuzungspunkten, daher ist das mittlere Signal auch nur halb
so breit wie die äußeren Signale.

\FloatBarrier

\subsection{Abschließende Diskussion}\label{s3:Conclusions}

Wie in der Einleitung zu diesem Abschnitt auf Seite \pageref{s3:Fahrplan} behauptet, haben wir hier 
tatsächlich gezeigt, dass wir in der Lage sind, sowohl das Fahrplanmodell 
als auch ein einfaches lABSE-Signal mit den bisherigen Methoden zu berechnen. Das von uns berechnete
Signal aus Abb.\ref{f3:FPFlussBsp2} mit realistischen Parametern kann direkt verglichen werden mit
dem theoretisch berechneten Signal aus \cite{DissAR}, Abb. 3.8, S. 50\footnote{
Auch hier wollen wir auf die rätselhafte, aber rein zufällige Übereinstimmung in den Nummern der Abbildungen 3.8 hinweisen.
}. 
Ebenso kann der Demonstrations-Fahrplan
aus Abb. \ref{f3:FPFlussBsp} direkt verglichen werden mit dem qualitativen Beispiel aus \cite{DissAR},
Abb. 3.12, S. 56 (Der Fahrplan ist dort allerdings relativ zu unserem Beispiel vertikal gespiegelt dargestellt).

Unsere Theorie erlaubt uns nun, die Grenzen des bisherigen Fahrplanmodells zu definieren.
Die WKB-Wellenpakete, die hier verwendet wurden, gelten nur unter Verwendung der Näherungen (W1)-(W3)
von Seite \pageref{WKB.WKB}, d.h. langsam variierende Potentiale (W1), die im Vergleich zur kinetischen Energie
sehr gering sind (W3), sowie sehr breite Wellenpakete (W2). Einzig (W3) dürfte für alle lABSE-Experimente, die wir
in der vorliegenden Arbeit betrachten wollen, eine gute Näherung sein. Die Breite der Wellenpakete im Ortsraum 
kann aber durchaus sehr schmal sein, wie im realistischen Beispiel gezeigt, so dass die Näherung (W2) nicht mehr
gut erfüllt ist und wir Dispersion berücksichtigen müssten. Noch deutlicher könnte die Näherung (W1) verletzt
sein, nämlich wenn wir nicht-adiabatische Verhältnisse betrachten. Insbesondere bei der
Betrachtung von P-verletzenden Rotationen der Polarisation von Atomen 
in elektrischen Feldern, wie sie z.B. in \cite{BoBrNa95}
vorgeschlagen wurden, benötigt man die nicht-adiabatische Erzeugung neuer Wellenpakete, die zu atomaren Zuständen
gehören, die gerade nicht im Anfangszustand vertreten waren. Es stellt sich dann die Frage, wie die Wellenpakete
der verschiedenen atomare Zustände im nicht-adiabatischen Fall miteinander wechselwirken.
Desweiteren benötigen wir gerade für die Untersuchung P-verletzender Effekte in Atomen metastabile Zustände 
\cite{BeNa83, BoBrNa95, DissTG, BrGaNa99, DiplTB}. Bei einer Beschreibung des Atoms als orts- und zeitabhängiges
Wellenpaket ist es nicht sofort klar, wie dieser Zerfall formal zu implementieren ist.

Wir brauchen also eine grundlegendere, exakte Theorie des lABSE. Im nächsten Kapitel werden wir daher eine
Methode zur prinzipiellen Berechnung der exakten Lösung der eindimensionalen Schrödinger-Gleichung
entwickeln. Dies erlaubt uns dann, einen quantitativen Vergleich zwischen der WKB-Lösung und der Charakteristiken-Lösung
anzustellen. In Kapitel \ref{s5:Zerfall} werden wir diese Methode dann zur Behandlung komplexer
skalare Potentiale verwenden. Hier werden wir sehen, wie der Zerfall eines instabilen Teilchen, dass durch das
Wellenpaket repräsentiert wird, formal zu berücksichtigen ist.

Die Übertragung der Methode auf matrixwertige Potentiale in Kapitel \ref{s6:Formalismus} bringt uns dann endlich
an das angestrebte Ziel der vorliegenden Arbeit, nämlich die vollständige theoretische Beschreibung eines lABSE-Experiments
mit statischen elektrischen und magnetischen Feldern.

\chapter{Die exakte Lösung der Schrödinger-Gleichung mit skalarem Potential}\label{s4:Formalismus}

\section{Ein kurzer Rückblick auf die Charakteristiken- und die WKB-Lösung}\label{s4:Vorbetrachtung}

In Abschnitt \ref{s2:Charakteristiken} haben wir eine Näherungslösung
der Schrödinger-Gleichung mit Hilfe der Methode der Charakteristiken
gefunden. Mit dem Ansatz
\begin{align}\label{e4:Ansatz.C}
	\Psi(z,t) = \exp\klg{-\I\frac{\bar k^2}{2m}t+\I\bar k z}A(z,t)
\end{align}
erhielten wir dort für den Amplitudenteil der Wellenfunktion im Falle eines skalaren Potentials $V(z)$
\begin{align}\label{e4:Lsg.C.2}
  A(z,t) = \exp\klg{-\I\frac{m}{\bar k}\int_{z-\tfrac{\bar k}m
        (t-t_0)}^{z} \d z'\ V(z')}
    \tilde\vph\klr{z-\frac{\bar k}{m}(t-t_0)}\ .
\end{align}

In Abschnitt \ref{s2:WKB} fanden wir mit Hilfe der WKB-Näherung für den Ansatz
\begin{align}\label{e4:Ansatz.WKB}
	\Psi(z,t) = \exp\klg{-\I\frac{\bar k^2}{2m}t + \I\int_{z_0}^z\d z'\ \sqrt{\bar k^2-2m V(z')}}A(z,t)
\end{align}
die Lösung
\begin{align}\label{e4:Lsg.WKB}
  \Psi(z,t) = \exp\klg{-\I\frac{\bar k^2}{2m}t 
      + \I\int^z_{z_0}\d z'\ \sqrt{\bar k^2 - 2mV(z')}}
    \vph(\zeta(z)-\tau(t))
\end{align}
mit den in (\ref{e2:Def.Zeta.Tau}) definierten Koordinatenfunktionen
\begin{align}\label{e4:ZetaTau}
\zeta(z):=\int_{z_0}^z\d z'\ \frac{\bar k}{\sqrt{\helpkbar\bar k^2-2mV(z')}}\ ,\qquad\tau(t) := \frac{\bar k}mt\ .
\end{align}

Beide Lösungen sind vom gleichen Typ, nämlich ein Produkt aus Phasenfaktor und Amplitudenfunktion,
\begin{align}\label{e4:AnsatzWF}
  \Psi(z,t) = \e^{\I\phi(z,t)}A(z,t)\ ,
\end{align}
mit jedoch unterschiedlichen Phasenwinkeln $\phi(z,t)$.
Ebenso sind beide Wellenfunktionen Lösungen der Schrödinger-Gleichung bei Vernachlässigung
der zweiten Ortsableitungen, d.h. unter der Näherung
\begin{align}\label{e4:Approximation}
  \del_z^2 \phi(z,t) \approx 0,\qquad \del_z^2A(z,t) \approx 0\ .
\end{align}
Der Charakteristiken-Lösung sieht man direkt an, dass $\del_z^2\phi(z,t) = 0$ identisch erfüllt ist,
während dies für die WKB-Lösung nur näherungsweise (im Rahmen der WKB-Näherung) gilt.
Es handelt sich bei dem Charakteristiken-Phasenwinkel\footnote{Hiermit wollen wir im
  folgenden den Phasenwinkel bezeichnen, den wir bei der Lösung mit der
  Methode der Charakteristiken verwendet haben.} damit wohl um eine Übersimplifikation,
die im WKB-Fall nicht gemacht wurde und die sich in entscheidender Weise auf das
Ergebnis auswirkt. Immerhin konnte die WKB-Lösung in Bezug auf die Schwerpunktsbewegung 
physikalisch sehr gut interpretiert werden, während dies offensichtlich bei der Charakteristiken-Lösung nicht
der Fall war (siehe Diskussion am Ende von Abschnitt \ref{s2:WKB-Eigenschaften}).

Die Methode zur Berechnung der exakten Lösung der Schrödinger-Gleichung in Form einer Reihenentwicklung,
die wir im nächsten Abschnitt entwickeln wollen, wird es uns erlauben, beide Näherungslösungen quantitativ
zu vergleichen. 

\section{Die Lösung der Schrödinger-Gleichung mit skalarem Potential}\label{s4:SG.Lsg}

Wir werden nun eine Lösung der Schrödinger-Gleichung mit skalarem Potential für einen
Ansatz der Form (\ref{e4:AnsatzWF}) berechnen. Wir werden zunächst die Schrödinger-Gleichung
für vorgegebenen Phasenwinkel $\phi(z,t)$ in geeigneter Weise umschreiben und danach
mit Hilfe geeigneter Green-Funktionen in eine Integralgleichung umwandeln, die wir dann iterativ lösen können.

\subsection{Umschreiben der Schrödinger-Gleichung}\label{s4:rewriteSG}

Das Einsetzen des Ansatzes (\ref{e4:AnsatzWF}),
\begin{align*}
\Psi(z,t) = \e^{\I\phi(z,t)}A(z,t)\ ,
\end{align*}
in die Schrödinger-Gleichung
\begin{align}\label{e4:SG}
(\del_z^2 - 2mV(z) + 2m\I\del_t)\Psi(z,t) = 0
\end{align}
liefert (siehe Gl. (\ref{e2:s.SG.2}))
\begin{align}\label{e4:SG.Ansatz}
\begin{split}
0 &= (\I \del_z^2\phi(z,t) - (\del_z\phi(z,t))^2 - 2mV(z) - 2m\del_t\phi(z,t))A(z,t)\\ 
&+ 2\I(\del_z\phi(z,t))(\del_z A(z,t)) + \del_z^2A(z,t) + 2m\I\del_t A(z,t)\ .
\end{split}
\end{align}
Wir können nun für $\phi(z,t)$ jeweils den WKB- oder den Charakteristiken-Phasenwinkel wählen, um eine
Gleichung für die Amplitudenfunktion $A(z,t)$ zu erhalten.

\subsubsection{Verwendung des WKB-Phasenwinkels}

Wählen wir zunächst den WKB-Phasenwinkel\footnote{
Wir bezeichnen den in Gl. (\ref{e2:SWKB}) definierten Anteil $S_k(z)$ des WKB-Phasenwinkels im Folgenden 
nur noch mit $S(z)$ und verstehen darunter den WKB-Anteil für die mittlere Wellenzahl, $S_{\bar k}(z)$.
}, also nach (\ref{e2:PW.WKB})
\begin{align}\label{e4:PW.WKB}
  \phi(z,t) = \phi\subt{WKB}(z,t) = -\frac{\bar k^2}{2m}t + S(z) 
    = -\frac{\bar k^2}{2m}t + \int_{z_0}^z\d z'\ k(z')\ .
\end{align}
Hier haben wir die lokale Wellenzahl
\begin{align}\label{e4:lokWZ}
  k(z) := \del_zS(z) = \sqrt{\bar k^2 - 2mV(z)}
\end{align}
eingeführt. Die Schrödinger-Gleichung (\ref{e4:SG.Ansatz}) lautet dann nach Berechnung aller Ableitungen
von $\phi(z,t)$:
\begin{align}
   \klr{\del_z^2 + 2\I k(z)\del_z - k^2(z) - \I \frac{m \del_z V(z)}{k(z)} 
     -2mV(z) + \bar k^2 + 2\I m\del_t}A(z,t) = 0\ .
\end{align}
Nach Definition der lokalen Wellenzahl gilt $\bar k^2 - k^2(z) = 2mV(z)$, also folgt
\begin{align}\label{e4:SG.WKB.1}
   \klr{\del_z^2 + 2\I k(z)\del_z - \I \frac{m \del_z V(z)}{k(z)} + 2\I m\del_t}A(z,t) = 0\ .
\end{align}
Nun gehen wir über zu den neuen Koordinaten $\zeta(z)$ und $\tau(t)$ aus (\ref{e4:ZetaTau}), denen wir im Zusammenhang
mit der WKB-Lösung bereits begegnet sind, d.h.
\begin{align*}
\zeta(z)=\int_{z_0}^z\d z'\ \frac{\bar k}{\sqrt{\helpkbar\bar k^2-2mV(z')}}\ ,\qquad \tau(t) = \frac{\bar
k}mt\ .
\end{align*}
Es gilt
\begin{align}
  \del_z = \ddp{}z = \frac{\bar k}{k(z)}\ddp{}\zeta = \frac{\bar k}{k(z)}\del_\zeta
\end{align}
und
\begin{align}
  \del_t = \frac{\bar k}m\del_\tau\ .
\end{align}
Mit der Amplitudenfunktion in den neuen Koordinaten $\zeta$ und $\tau$,
\begin{align}
B(\zeta(z),\tau(t)) := A(z,t)\qquad\text{(bzw. $B(\zeta,\tau) = A(z(\zeta),t(\tau))$)}
\end{align} 
können wir die Schrödinger-Gleichung (\ref{e4:SG.WKB.1}) umschreiben und erhalten
\begin{align}\label{e4:WKB.SG.tmp}
   2\I\bar k\klr{\del_\zeta + \del_\tau}B(\zeta,\tau) 
  + \klr{\del_z^2 - \I\frac{m\del_z V(z)}{k(z)}}B(\zeta,\tau) = 0\ .
\end{align}
Die zweite Ortsableitung der Amplitude berechnet sich zu
\begin{align}
  \begin{split}
    \del_z^2 B(\zeta,\tau) &= \del_z\frac{\bar k}{k(z)}\del_\zeta B(\zeta,\tau)\\ 
    &= -\frac{\bar k (\del_z k(z))}{k^2(z)}\del_\zeta B(\zeta,\tau) 
       + \frac{\bar k^2}{k^2(z)}\del_\zeta^2 B(\zeta,\tau)\\
    &= \frac{\bar k m (\del_z V(z))}{k^3(z)}\del_\zeta B(\zeta,\tau) + \frac{\bar
      k^2}{k^2(z)}\del_\zeta^2B(\zeta,\tau)\ ,
  \end{split}
\end{align}
womit sich (\ref{e4:WKB.SG.tmp}) schließlich in die Form
\begin{align}\label{e4:SG.WKB.2}
  \klr{\del_\tau + \del_\zeta}B(\zeta,\tau) = (\hat L\subt{WKB}B)(\zeta,\tau)
\end{align}
bringen lässt. Dabei ist $\hat L\subt{WKB}$ der für die Wahl des
WKB-Phasenwinkels charakteristische Differentialoperator
\begin{align}\label{e4:DefL.WKB}
  \begin{split}
    \hat L\subt{WKB} &= \frac{\I}{2\bar k}\kle{- \I\klr{\frac m{k(z)}\del_zV(z)}
      + \frac{m\bar k}{(k(z))^3}(\del_z V(z))\del_\zeta 
+ \frac{\bar k^2}{(k(z))^2}\del_\zeta^2}\ .
  \end{split}
\end{align}

\subsubsection{Verwendung des Charakteristiken-Phasenwinkels}\label{s4:C->SG}

Wir führen nun dieselbe Rechnung für den Charakteristiken-Phasenwinkel
\begin{align}\label{e4:PW.C}
  \phi(z,t) = \phi\subt{C}(z,t)= - \frac{\bar k^2}{2m}t + \bar k z
\end{align}
durch. Die Schrödinger-Gleichung (\ref{e4:SG.Ansatz}) lautet dann nach Berechnung aller Ableitungen
von $\phi(z,t)$
\begin{align}
  \klr{2\I\bar k\del_z + \del_z^2 - 2mV(z) + 2\I m\del_t}A(z,t) = 0\ .
\end{align}
Auch hier führen wir die Variable $\tau$ ein, lassen die Ortskoordinate $z$ aber unberührt.
Mit dem für den Charakteristiken-Phasenwinkel typischen Differentialoperator
\begin{align}\label{e4:DefL.C}
\hat L\subt{C} = \frac{\I}{2\bar k}\klr{\del_z^2 - 2mV(z)}
\end{align}
lautet die Schrödinger-Gleichung für die Funktion
\begin{align}
B(z,\tau(t)) = A(z,t)
\end{align}
dann schließlich
\begin{align}\label{e4:SG.C}
  \klr{\del_z + \del_\tau}B(z,\tau) = (\hat L\subt{C} B)(z,\tau)\ .
\end{align}

\subsection{Aufstellen und Lösen der Integralgleichung}\label{s4:Loesung}

Die beiden zu lösenden Schrödinger-Gleichungen (\ref{e4:SG.WKB.2}), (\ref{e4:SG.C})
sind sowohl im WKB- als auch im Charakteristiken-Fall vom gleichen Typ, nämlich
\begin{align}\label{e4:SG.typ}
  (\del_\zeta + \del_\tau)B(\zeta,\tau) = (\hat LB)(\zeta,\tau)\ .
\end{align}
Im Charakteristiken-Fall ist lediglich $\zeta=z$ zu setzen, im WKB-Fall ist 
$\zeta$ wie in (\ref{e4:ZetaTau}) definiert. In beiden Fällen ist
für $\hat L$ der entsprechende Differentialoperator $\hat L\subt{C}$ bzw. $\hat L\subt{WKB}$
zu verwenden.

Um (\ref{e4:SG.typ}) in eine Integralgleichung umschreiben zu können, benötigen wir eine
Greensche Funktion des Operators auf der linken Seite der Gleichung, für die
\begin{align}\label{e4:GreenDGL}
  (\del_\zeta + \del_\tau)G(\zeta,\tau) = \delta(\zeta)\delta(\tau)
\end{align}
erfüllt sein muss. Man kann sich leicht davon überzeugen, dass
\begin{align}\label{e4:GreenFkt}
  G(\zeta,\tau) = \Theta(\zeta+\tau)\delta(\zeta-\tau)
\end{align}
eine mögliche Lösung der Gl. (\ref{e4:GreenDGL}) ist.
Diese Funktion ist in der $(\zeta,\tau)$-Ebene nur auf der Winkelhalbierenden ($\zeta=\tau$)
im I. Quadranten von Null verschieden (also für $\zeta,\tau\geq 0$) und entspricht somit
einer retardierten Green-Funktion.

Die Gl. (\ref{e4:SG.typ}) kann damit in eine Integral-Gleichung umgewandelt werden:
\begin{align}
  B(\zeta,\tau) = \tilde B_0(\zeta-\tau) + \int_{-\infty}^\infty\d \zeta'\int_{-\infty}^\infty\d
\tau'
\ G(\zeta-\zeta',\tau-\tau')(\hat L B)(\zeta',\tau')\ .
\end{align}
Hierbei ist $\tilde B_0(\zeta-\tau)$ eine beliebige Funktion die nur von der Kombination
$\zeta-\tau$ der beiden Koordinaten abhängt. Jede solche Funktion ist automatisch
eine Lösung der homogenen Gleichung 
\begin{align}
(\del_\zeta+\del\tau)\tilde B_0(\zeta-\tau)=0\ .
\end{align}
Durch Einsetzen der Green-Funktion (\ref{e4:GreenFkt}) ergibt sich
\begin{align}\label{e4:Lsg.formal}
  B(\zeta,\tau) = \tilde B_0(\zeta-\tau) + \int_{-\infty}^\zeta\d \zeta'\ (\hat L
B)(\zeta',\tau')\big\vert_{\tau'=\tau-\zeta+\zeta'}\ .
\end{align}
In dem Integral ist zunächst der Operator $\hat L$ auf die Funktion $B(\zeta',\tau')$
anzuwenden, danach ist $\tau'=\tau-\zeta+\zeta'$ zu setzen und schließlich ist
die Integration auszuführen. Bei (\ref{e4:Lsg.formal}) handelt es sich um die formal exakte
Lösung der Gleichung (\ref{e4:SG.typ}), wobei man durch iteratives Einsetzen von $B(\zeta,\tau)$ im
Integral eine Reihenentwicklung für die Lösung bekommt. Betrachtet man das Integral jedoch genauer, so stellt
man fest, dass es divergent ist, was wir am folgenden Beispiel klar machen wollen. Nehmen wir z.B.
den Operator $\hat L\subt{WKB}$ aus (\ref{e4:DefL.WKB}) und betrachten den dort enthaltenen Beitrag
$(\bar k^2/k^2(z))\del_\zeta^2$ (neben dem konstanten Faktor $\I/2\bar k$). Dieser führt im
Integrand von (\ref{e4:Lsg.formal}) in erster Ordnung zu 
\begin{align}
  \int_{-\infty}^\zeta\d\zeta'\ \frac{\bar k^2}{k^2(z(\zeta'))}\tilde
B_0''(\zeta'-\tau')\big\vert_{\tau'=\tau-\zeta+\zeta'}
  = \tilde B_0''(\zeta-\tau)\int_{-\infty}^\zeta\d\zeta'\ \frac{\bar k^2}{k^2(z(\zeta'))}\ .
\end{align}
Da auf der rechten Seite der Gleichung der Integrand bei den sehr kleinen Potentialen, 
die wir hier betrachten, für alle $z$ nahe bei Eins liegt, divergiert das uneigentliche Integral 
und somit das gesamte Integral aus (\ref{e4:Lsg.formal}).

Dieses Problem kann man durch geschickte Wahl der homogenen Lösung $\tilde B_0(\zeta-\tau)$
aus der Welt schaffen, z.B.
\begin{align}\label{e4:homogene.Lsg}
\tilde B_0(\zeta-\tau) = \vph(\zeta-\tau) - \int_{-\infty}^{\zeta-\tau}\d \zeta'\ (\hat
LB)(\zeta',\tau')\big\vert_{\tau'=\tau-\zeta+\zeta'}\ .
\end{align}
Durch die Wahl der Integralobergrenze $\zeta-\tau$ stellen wir sicher, dass (\ref{e4:homogene.Lsg})
immer noch eine homogene Lösung der Gl. (\ref{e4:SG.typ}) ist. Wir beseitigen also die Divergenz
im Integral in (\ref{e4:Lsg.formal}) durch eine divergente homogene Lösung.

Es verbleibt in (\ref{e4:Lsg.formal}) noch
\begin{align}\label{e4:Loesung.rekursiv}
  B(\zeta,\tau) &= \vph(\zeta-\tau) 
    + \int_{\zeta-\tau}^\zeta\d \zeta'\ (\hat L B)(\zeta',\tau')\big\vert_{\tau'=\tau-\zeta+\zeta'}\ .
\end{align}
Wir können in diesem Zusammenhang den Integraloperator $\hat K$ definieren als
\begin{align}\label{e4:Def.K}
\boxed{(\hat K f)(\zeta,\tau) 
= \int_{\zeta-\tau}^\zeta\d \zeta'\ (\hat L f)(\zeta',\tau')\big\vert_{\tau'=\tau-\zeta+\zeta'}}\ ,
\end{align}
wobei $f = f(\zeta,\tau)$ eine beliebige Funktion von $\zeta$ und $\tau$ sein soll. Wir können die Lösung
(\ref{e4:Loesung.rekursiv}) also in abstrakter Form schreiben als
\begin{align}
  B = \vph + \hat K B
\end{align}
und können diese Gleichung nun algebraisch umformen zu
\begin{align}
B = \frac{1}{1-\hat K}\vph
\end{align}
und dies in eine geometrische Reihe entwickeln, 
\begin{align}
B = \sum_{i=0}^{\infty} \hat K^i\vph\ .
\end{align}
Damit haben wir die Amplitudenfunktion $B(\zeta,\tau)$ als Reihenentwicklung
nach der homogenen Lösung $\vph(\zeta-\tau)$ ausgedrückt,
\begin{align}\label{e4:Loesung}
\boxed{B(\zeta,\tau) = \sum_{i=0}^\infty (\hat K^i\vph)(\zeta,\tau)}\ .
\end{align}

Die algebraische Umstellung der rekursiv definierten Lösung (\ref{e4:Loesung.rekursiv}) und die
formale Entwicklung des Operators $(1-\hat K)^{-1}$ in eine geometrische Reihe entspricht genau
dem iterativen Einsetzen der Lösung (\ref{e4:Loesung.rekursiv}) in sich selbst (im Integraloperator).
Dies ist ein wohlbekanntes Lösungsverfahren und entspricht z.B. genau dem Verfahren, das in der Quantenmechanik
bei der Lösung des Streuproblems angewendet wird (Bornsche Reihe, vergleiche \cite{Cohen}, Bd. 2, Abschnitt 8.2.3 f., 
ab Seite 112).

In der Praxis wird man selten über die erste Ordnung hinaus entwickeln, insbesondere da in diesem Fall
der Integraloperator in höheren Potenzen sehr schwierig zu handhaben ist. Besonders deutlich werden wir
das bei der Lösung der matrixwertigen Schrödinger-Gleichung in Kapitel \ref{s6:Formalismus} sehen.
Konvergenzfragen im Bezug auf die Reihenentwicklung behandeln wir in der vorliegenden Arbeit nicht.

\subsection{Variablentransformation in der WKB-Lösung}\label{s4:Variablentrafo}

Letztendlich ist man an der Amplitudenfunktion $A(z,t) = B(\zeta(z),\tau(t))$ interessiert.
Wir müssen dazu den Integraloperator $\hat K$ in Abhängigkeit von der Ortskoordinate $z$ und
der Zeit $t$ schreiben. Bei der Verwendung der Charakteristiken-Phase ist dies kein Problem,
da dann anstelle von $\zeta$ ohnehin schon $z$ verwendet wird und $\tau(t) = \frac{\bar k}m t$
einfach eingesetzt werden kann.

Bei der WKB-Lösung muss man das Integral über $\zeta'$ in ein Integral über $z'$ umwandeln.
Dazu verwenden wir die Definition von $\zeta(z)$ aus Gl. (\ref{e4:ZetaTau}) und erhalten
\begin{align}\label{e4:Sub.Zeta.Z}
\d \zeta = \frac{\bar k}{k(z)}\d z\ .
\end{align}
Für die Integralgrenzen benötigen wir die Umkehrfunktion von $\zeta(z)$, die wir mit $\mc Z(\zeta)$
kennzeichnen wollen und für die
\begin{align}\label{e4:Umkehrfunktion}
\mc Z(\zeta(z)) = z
\end{align}
erfüllt sein muss. Für die Integraluntergrenze $\zeta-\tau$ im Operator $\hat K$ führen wir eine neue Funktion
ein, um die Schreibweise so kompakt wie möglich zu halten. Es sei daher
\begin{align}\label{e4:Def.u}
u(z,t) := \mc Z(\zeta(z)-\tau(t))\ .
\end{align}
Der Integraloperator $\hat K$ aus (\ref{e4:Def.K}) lautet dann
\begin{align}\label{e4:Def.K.WKB}
\boxed{(\hat K f)(z,t) 
= \int_{u(z,t)}^z\d z'\ \frac{\bar k}{k(z')}(\hat L f)(z',\tau')\big\vert_{\tau'=\tau(t)-\zeta(z)+\zeta(z')}}\ ,
\end{align}
wobei es sich in der Praxis anbietet, die Variable $\tau$ bei der Anwendung des Operator $\hat L$ im Integranden
zunächst beizubehalten und erst am Ende der Rechnung durch $\tau(t) = \frac{\bar k}m t$ zu ersetzen.

Wir wollen nun noch einige Anmerkungen zu der Funktion $u(z,t)$ machen, die uns im Folgenden noch häufig begegnen
wird. Zunächst ist es i.A. nicht möglich, die Umkehrfunktion zu $\zeta(z)$ analytisch anzugeben, da das Potential
$V(z)$ zum einen sehr kompliziert sein kann und zum anderen in der Praxis auch nur numerisch berechnet wird.
Deshalb ist es sinnvoll, die Integration über $z$ anstelle von $\zeta$ durchzuführen, da der Integrand hauptsächlich 
direkt von $z$ abhängt. Lediglich die Einhüllende $\vph(\zeta(z)-\tau(t))$ hängt von $\zeta(z)$
ab. Die Berechnung der Funktion $\zeta(z)$ ist aber wesentlich einfacher möglich als die Berechnung ihrer
Umkehrfunktion. 

Durch die Variablentransformation im Integral benötigen wir die Umkehrfunktion $u(z,t)$ 
nur noch an einer einzigen Stelle in der Formel, nämlich in der Integraluntergrenze. Hier kann $u(z,t)$
sehr gut physikalisch interpretiert werden, wie wir nun zeigen wollen.

In Abschnitt \ref{s3:Fahrplanmodell} haben wir bei der Diskussion des Fahrplanmodells die Zeit $T_\alpha(Z)$
definiert, die das Atom im Zustand $\rket{\alpha(Z)}$ benötigt, um im Potential $V_\alpha(Z)$ vom Ort $Z_0$
zum Ort $Z$ zu gelangen, siehe Gl. (\ref{e3:Delta.T}). Hier wollen wir die Zeit berechnen, die der Schwerpunkt
des Wellenpakets braucht, um im Potential $V(z)$ vom Ort $u(z,t)$ zum Ort $z$ zu gelangen. Analog zu (\ref{e3:Delta.T})
ist diese gegeben durch
\begin{align}
T(z,u(z,t)) = \frac{M}{\bar k}\int_{u(z,t)}^z\d z'\ \frac{\bar k}{k(z')}\ .
\end{align}
In diesem Integral können wir $z\to \zeta(z)$ substituieren und erhalten
\begin{align}
T(z,u(z,t)) = \frac{M}{\bar k}\int_{\zeta(u(z,t))}^{\zeta(z)}\d \zeta'\ 1\  = \frac{M}{\bar k}\klr{\zeta(z) - \zeta(u(z,t))}\ .
\end{align}
Wegen
\begin{align}
\zeta(u(z,t)) = \zeta(\mc Z(\zeta(z)-\tau(t))) = \zeta(z) - \tau(t)
\end{align}
folgt also das einfache Ergebnis
\begin{align}\label{e4:Interpretation.u}
T(z,u(z,t)) = \frac{M}{\bar k}\tau(t) = t\ .
\end{align}
Somit ist $u(z,t)$ gerade der Ort, an dem ein Teilchen zur Zeit $t=0$ starten muss, um zur Zeit $t$ am Ort
$z$ einzutreffen, nachdem es das Potential $V(z)$ im entsprechenden Intervall durchquert hat. Hierin drückt sich 
wieder der semiklassische Charakter der Entwicklung der Wellenfunktion aus. Die Wellenfunktion am Ort $z$ zur 
Zeit $t$ erhält nur Korrekturbeiträge, die (im klassischen Sinne) die Zeit hatten, zum betrachteten Ort zu gelangen.

Zum Ende dieses Abschnitts wollen wir noch die Taylor-Entwicklung für $u(z,t)$ für
sehr kleine Zeiten angeben. Mit
\begin{align}\label{e4:Dz:Dzeta}
\dd{\zeta(z)}z = \frac{\bar k}{k(z)}\qquad\Longleftrightarrow\qquad
\dd{\mc Z(\zeta)}\zeta = \frac{k(\mc Z(\zeta))}{\bar k}\ .
\end{align}
folgt nämlich für die Taylorentwicklung der Umkehrfunktion $\mc Z(\zeta-\tau)$
um $\zeta$
\begin{align}\label{e4:Taylor.z(zeta-tau)}
\begin{split}
\mc Z(\zeta-\tau) &= \sum_{n=0}^\infty \frac1{n!}\dd{^n\mc Z(\zeta)}{\zeta^n}(-\tau)^n\\
&= \kle{z + 
\sum_{n=1}^\infty \frac1{n!}\klr{\klr{\frac{k(z)}{
\bar k}\dd{}{z}}^{n-1}\frac{k(z)}{\bar k}}(-\tau)^n}_{z=\mc Z(\zeta)}\\
&= \kle{z - \frac{k(z)}{\bar k}\tau + \frac12\klr{\frac{k(z)}{
\bar k}\dd{}z\frac{k(z)}{\bar k}}\tau^2 \mp \ldots}_{z=\mc Z(\zeta)}\ .
\end{split}
\end{align}
Wenden wir dieses Ergebnis auf die Funktion $u(z,t)$ an, so folgt
\begin{align}\label{e4:u.taylor}
u(z,t) = \mc Z(\zeta(z)-\tau(t)) = z - \frac{k(z)}{m}t - \frac{(\del_zV(z))}{2m}t^2 + \OO(t^3)\ .
\end{align}
Für beliebig große Zeiten werden die Korrekturterme auch bei schwach veränderlichen Potentialen beliebig
groß. Häufig ist man nur an der Wellenfunktion in einem relativ kurzen Zeitraum interessiert\footnote{
Z.B. der Zeitraum, an dem das Wellenpaket an einem Detektor am Ort $Z_D$ eintrifft. Siehe dazu auch Abschnitt
\ref{s6:Signal}.}. Dann kann man analog die Entwicklung
\begin{align}\label{e4:u.taylor.2}
u(z,t+\Delta t) &= u(z,t) - \frac{k(\mc Z(\zeta(z)-\tau(t)))}{m}\Delta t  + \OO((\Delta t)^2)
\end{align}
betrachten.

\subsection{Anfangsbedingungen und Diskussion der Lösung}\label{s4:Anfangsbedingungen}

Die homogene Lösung in Gl. (\ref{e4:Loesung.rekursiv}) wurde nicht ohne Grund mit dem Buchstaben $\vph$
bezeichnet. Sie entspricht nämlich der Einhüllenden $\vph(\zeta-\tau)$ der WKB-Lösung, bzw.
im Falle der Verwendung der Charakteristiken-Phase der Einhüllenden $\tilde\vph(z-\tfrac{\bar k}mt)$.
Die Übereinstimmung ist exakt im Limes $t\to 0$ und damit für $\tau\to0$, also am Anfang der Zeitentwicklung, wenn
die Integralgrenzen im Integraloperator $\hat K$ identisch sind. Es folgt nach Gl. (\ref{e4:Loesung.rekursiv})
bei Verwendung des WKB-Phasenfaktors
\begin{align}\label{e4:AB.B}
B(\zeta(z),\tau=0) = \vph(\zeta(z))\ .
\end{align}
Die vollständige Wellenfunktion inklusive Phasenfaktor lautet dann gemäß (\ref{e4:AnsatzWF})
\begin{align}\label{e4:AB.Psi.WKB}
\Psi(z,t=0) = \e^{\I\phi\subt{WKB}(z,0)}\vph(\zeta(z))\ .
\end{align}
Geht man vom Charakteristiken-Phasenwinkel aus, so ist
\begin{align}
\Psi(z,t=0) = \e^{\I\phi\subt{C}(z,0)}\tilde\vph(z-z_0)\ .
\end{align}
Mit diesem Resultat rechtfertigen wir die im Folgenden gelegentlich verwendete Bezeichnung von $\vph$ bzw. $\tilde\vph$
als Anfangswellenpaket (auch wenn zum gesamten Anfangswellenpaket strenggenommen noch der jeweilige Phasenfaktor
gehört).

Zum Abschluss dieses Abschnitts wollen wir noch eine wichtige Bemerkung machen,
auf die wir im Verlaufe dieses Kapitels noch zurückgreifen werden. Die hier betrachteten
Operatoren $\hat L\subt{WKB}$ und $\hat L\subt{C}$ aus Gl. (\ref{e4:DefL.WKB}) und (\ref{e4:DefL.C})
enthalten beide einen Faktor proportional zu $\bar k^{-1}$. Auf der experimentell durch die
lABSE-Apparatur vorgegebenen Längenskala, die auch den Integrationsbereich des Integraloperators vorgibt und im
Bereich von einigen Metern liegt\footnote{M. DeKieviet, private Kommunikation.}, stellt die inverse mittlere Wellenzahl $\bar k^{-1}$, die nach den Betrachtungen der Beispiele 
in Abschnitt \ref{s3:Fahrplan} in der Größenordnung $\OO(10^{-10}\u m)$ liegt, eine extrem kleine Größe dar.
Wir können die Entwicklung nach dem Integraloperator (der dimensionslos ist), also als Entwicklung nach
Potenzen der inversen Wellenzahl auffassen.

Wir können nun die oben dargestellte Methode für die Berechnung der Amplitudenfunktion bei vorgegebenem Phasenwinkel
einmal mit dem Charakteristiken-Phasenwinkel und einmal mit dem WKB-Phasenwinkel durchführen. Die auf diese
Weise berechneten Wellenfunktion (also das Produkt von Phasenfaktor und Amplitudenfunktion) müssen dann in 
jeder Ordnung von $\bar k^{-1}$ übereinstimmen.

Desweiteren kann man bei fester Ordnung im Integraloperator $\hat K$ in beiden Fällen untersuchen, bis zu
welcher Ordnung in $\bar k^{-1}$ Terme in der jeweiligen Näherungslösung enthalten sind. Da jeder Integraloperator
einen Faktor $1/\bar k$ enthält, enthält die $n$-te Ordnung der Entwicklung in $\hat K$ sicher alle Beiträge
bis zur $n$-ten Ordnung in $1/\bar k$. Wie wir in Abschnitt \ref{s4:Vergleich} sehen werden, gibt ein Vergleich 
der Ordnungen in $\hat K$ Aufschluss über die Qualität des Ansatzes für den Phasenwinkel.

Wir werden also im Folgenden stets unterscheiden zwischen der Ordnung der Entwicklung im Integraloperator
$\hat K$ und der Ordnung der Entwicklung in der inversen mittleren Wellenzahl $\bar k^{-1}$.

\section{Anwendung des Formalismus}\label{s4:Anwendung}

\subsection{Verwendung des WKB-Phasenwinkels}\label{s4:WKB}
Bei Verwendung des WKB-Phasenwinkels müssen wir im Integraloperator $\hat K$ aus (\ref{e4:Def.K.WKB})
den Operator $\hat L\subt{WKB}$ aus Gl. (\ref{e4:DefL.WKB}) auf das Anfangswellenpaket $\vph(\zeta-\tau)$ anwenden.
Wir schreiben
\begin{align}\label{e4:LWKB.Short}
\hat L\subt{WKB} = \ms A(z) + \ms B(z)\del_\zeta + \ms C(z)\del_\zeta^2\ ,
\end{align}
wobei gemäß (\ref{e4:DefL.WKB})
\begin{subequations}
\begin{align}\label{e4:LWKB.A}
\ms A(z) &= \frac{1}{2\bar k}\klr{\frac m{k(z)}\del_zV(z)} = -\frac1{2\bar k}\del_zk(z)\ ,\\ \label{e4:LWKB.B}
\ms B(z) &= \frac{\I}{2\bar k}\frac{m\bar k}{(k(z))^3}(\del_z V(z)) = \frac{\I}{2\bar k}\del_z\frac{\bar k}{k(z)}\ ,\\ \label{e4:LWKB.C}
\ms C(z) &= \frac{\I}{2\bar k}\frac{\bar k^2}{(k(z))^2}
\end{align}
\end{subequations}
gilt. Da diese Funktionen nicht zeitabhängig sind, ist die Berechnung der Ableitungen der Anfangswellenpakete
und die anschließende Substitution $\tau'\to\tau-\zeta+\zeta'$ leicht durchzuführen. Es gilt
\begin{align}\label{e4:vph.in.K}
\kle{\del_{\zeta'}^{n}\vph(\zeta'-\tau')}_{\tau'\to\tau-\zeta+\zeta'} = \vph^{(n)}(\zeta-\tau)\ ,
\qquad (n\in \mb N_0)\ ,
\end{align}
also können wir im Operator $\hat K$ die Ableitungen der Wellenpakete vor das Integral ziehen erhalten dann
\begin{align}\label{e4:K.auf.vph}
\begin{split}
(\hat K\subt{WKB}\vph)(z,t) &= \ms U(z,u(z,t))\vph(\zeta(z)-\tau(t))\\
&+ \ms V(z,u(z,t))\vph'(\zeta(z)-\tau(t))\\
&+ \ms W(z,u(z,t))\vph''(\zeta(z)-\tau(t))
\end{split}
\end{align}
mit
\begin{subequations}
\begin{align}\label{e4:KWKB.U}
\ms U(z,u(z,t)) &= -\frac{1}{2\bar k}\int_{u(z,t)}^z\d z'\ \frac{\bar k}{k(z')}\del_{z'}k(z') 
= \ln\sqrt{\frac{k(u(z,t))}{k(z)}}\ ,\\ \label{e4:KWKB.V}
\ms V(z,u(z,t)) &= \frac{\I}{2\bar k}\int_{u(z,t)}^z\d z'\ \frac{\bar k}{k(z')}\del_{z'}\frac{\bar k}{k(z')}
=  \frac{\I}{4\bar k}\kle{\frac{\bar k^2}{k^2(z')}}^{z}_{u(z,t)}
\ ,\\ \label{e4:KWKB.W}
\ms W(z,u(z,t)) &= \frac{\I}{2\bar k}\int_{u(z,t)}^z\d z'\ \klr{\frac{\bar k}{k(z')}}^3\ .
\end{align}
\end{subequations}

In erster Ordnung $\hat K$ erhalten wir also bei Verwendung des WKB-Phasenwinkels die Wellenfunktion
\begin{align}\label{e4:1.Ord.WKB}
\begin{split}
\Psi\subt{WKB}^{(1)}(z,t) &= \e^{\I\phi\subt{WKB}(z,t)}A^{(1)}\subt{WKB}(z,t)\\
&= \e^{\I\phi\subt{WKB}(z,t)}\klg{\vph(\zeta(z)-\tau(t)) + (\hat K\subt{WKB}\vph)(z,t) }\\
&= e^{\I\phi\subt{WKB}(z,t)}\Bigg\{
\kle{1 + \ln\sqrt{\frac{k(u(z,t))}{k(z)}}\ }\vph(\zeta(z)-\tau(t))\\
&\hspace{22mm} +\frac{\I}{4\bar k}\kle{\frac{\bar k^2}{k^2(z')}}^{z}_{u(z,t)}\vph'(\zeta(z)-\tau(t))\\
&\hspace{22mm} +\klr{\frac{\I}{2\bar k}\int_{u(z,t)}^z\d z'\ \klr{\frac{\bar k}{k(z')}}^3}\vph''(\zeta(z)-\tau(t))
\Bigg\}\ .
\end{split}
\end{align}

\subsection{Verwendung des Charakteristiken-Phasenwinkels}\label{s4:Charakteristiken}

\subsubsection{Erste Ordnung}

Im Charakteristiken-Fall haben wir die Koordinaten $z,\tau$ anstelle von $\zeta,\tau$ zu verwenden,
sowie das Anfangswellenpaket $\tilde\vph(z-\tau)$. Der in (\ref{e4:DefL.C}) definierte Operator 
$\hat L\subt{C}$ führt zu
\begin{align}
  \begin{split}
    (\hat L\subt{C}\tilde\vph)(z,\tau) &= \frac{\I}{2\bar k}\tilde\vph''(z-\tau) - \frac{\I m}{\bar
k}V(z)\tilde\vph(z-\tau)
  \end{split}
\end{align}
und somit zu
\begin{align}
  \begin{split}
    (\hat K\subt{C}\tilde\vph)(z,t)
&= \tilde\vph''(z-\tau)\frac{\I}{2\bar k}\tau - \tilde\vph(z-\tau(t))\frac{\I m}{\bar
k}\int_{z-\tau(t)}^{z}\d z'\ V(z')\ .
  \end{split}
\end{align}
Also ist in erster Ordnung
\begin{align}\label{e4:1.Ord.C}
\begin{split}
\Psi\subt{C}^{(1)}(z,t) &= \e^{\I\phi\subt{C}(z,t)}A^{(1)}\subt{C}(z,t)\\
&= \e^{\I\phi\subt{C}(z,t)}\klg{\tilde\vph(z-\tau(t)) + (\hat K\subt{C}\tilde\vph)(z,t) }\\
&= e^{\I\phi\subt{C}(z,t)}\Bigg\{
\kle{1-\frac{\I m}{\bar k}\int_{z-\tau(t)}^z\d z'\ V(z')} \tilde\vph(z-\tau(t))\\
&\hspace{18mm} +\frac{\I}{2\bar k}\tau(t)\tilde\vph''(z-\tau(t))
\Bigg\}\ .
\end{split}
\end{align}

\subsubsection{Zweite Ordnung}

Für die Berechnung der zweiten Ordnung müssen wir die Wirkung des Integraloperators $\hat K\subt{C}$
auf $(\hat K_C\tilde\vph)(z,t)$ kennen. Zunächst berechnen wir die Wirkung des Differentialoperators
$\hat L\subt{C}$ und erhalten
\begin{align}
  \begin{split}
    \klr{\hat L\subt{C}\hat K\subt{C}\tilde\vph}(z,\tau) &= \tilde\vph(z-\tau)\kle{-\frac{m^2}{\bar
k^2}V(z)\int_{z-\tau}^{z}\d z'\ V(z')
      + \frac m{2\bar k^2}\kle{V'(z)-V'(z-\tau)}}\\
    &+\tilde\vph'(z-\tau)\frac{m}{\bar k^2}\kle{V(z)-V(z-\tau)}\\
    &+\tilde\vph''(z-\tau)\frac{m}{2\bar k^2}\kle{\tau V(z) + \int_{z-\tau}^{z}\d z'\ V(z')}\\
    &-\tilde\vph^{(4)}(z-\tau)\frac{1}{4\bar k^2}\tau
  \end{split}\raisetag{3cm}
\end{align}
und hieraus nach einiger Rechnung schließlich
\begin{align}\label{e4:2.Ord.C}
  \begin{split}
    B^{(2)}\subt{C}(z,\tau) &= \tilde\vph(z-\tau)\Bigg[1-\frac{\I m}{\bar k}\int_{z-\tau}^z\d z'\ 
V(z') - \frac{m^2}{\bar k^2}\int_{z-\tau}^z\d z'\ V(z')\int_{z-\tau}^{z'} \d z''\ V(z'')\\
      &\hspace{2.3cm} - \frac m{2\bar k^2}\klr{\tau V'(z-\tau) - V(z) + V(z-\tau)}\Bigg]\\
      &+\tilde\vph'(z-\tau)\frac{m}{\bar k^2}\kle{\int_{z-\tau}^z\d z'\ V(z') - \tau V(z-\tau)}\\
      &+\tilde\vph''(z-\tau)\kle{\frac{\I}{2\bar k}\tau + \frac{m}{2\bar k^2}\tau\int_{z-\tau}^z\d
z'\ V(z')}\\
      &-\tilde\vph^{(4)}(z-\tau)\frac{1}{8\bar k^2}\tau^2\ .
  \end{split}\raisetag{3.5cm}
\end{align}
Alle Beiträge proportional zu $\bar k^{-2}$ sind dabei in der zweiten Ordnung entstanden. Die Gesamtwellenfunktion
ergibt sich hieraus einfach durch 
\begin{align}
\Psi\subt{C}^{(2)}(z,t) = \e^{\I\phi\subt{C}(z,t)}A_C^{(2)}(z,t) = \e^{\I\phi\subt{C}(z,t)}B_C^{(2)}(z,\tau(t))\ .
\end{align}

\subsection{Vergleich der beiden Entwicklungen}\label{s4:Vergleich}

Wie am Ende von Abschnitt \ref{s4:Anfangsbedingungen} bereits angesprochen wurde, gibt es zwei
verschiedene Möglichkeiten des Vergleichs. Entweder man vergleicht die Beiträge in der Entwicklung 
zu vorgegebener Ordnung des Operators $\hat K$ oder man vergleicht
die Beiträge zu einer festen Ordnung in $1/\bar k$. Beide Varianten bringen
unterschiedliche Erkenntnisse.

Der Koeffizientenvergleich der Potenzen von $1/\bar k$
ermöglicht die Überprüfung der berechneten Entwicklungen für den WKB-Fall und den
Charakteristiken-Fall, denn beide müssen in jeder Ordnung von $1/\bar k$
übereinstimmen. Desweiteren liefert die
Entwicklung nach Potenzen von $1/\bar k$ für eine feste Ordnung im
Integraloperator $\hat K$ einen Hinweis auf die Qualität des Ansatzes.
Wurde der Ansatz gut gewählt, sollten bereits bei niedrigen Ordnungen in
$\hat K$ Beiträge von höheren Ordnungen in $1/\bar k$ zu finden sein.
Ebenso sollte der bessere Ansatz bei fester Ordnung in $\hat K$
zu betragsmäßig kleineren Korrekturen führen.

\subsubsection{Vergleich bis zur Ordnung $1/\bar k$}

Wir wenden uns zunächst den Entwicklungen in $1/\bar k$ zu. Wir betrachten
die jeweils erste Ordnung in $\hat K$, d.h. die Gln. (\ref{e4:1.Ord.WKB}) und
(\ref{e4:1.Ord.C}). In $A^{(1)}\subt C(z,t)$ aus (\ref{e4:1.Ord.C}), zur Erinnerung
\begin{align}\label{e4:1.Ord.C.copy}
A^{(1)}\subt C(z,t) &= \tilde\vph(z-\tau(t))\kle{1-\frac{\I m}{\bar k}\int_{z-\tau(t)}^z\d z'\ V(z')}
+ \tilde\vph''(z-\tau(t))\frac{\I}{2\bar k}\tau(t)\ ,
\end{align}
gibt es zwei Korrekturbeiträge, die beide erster Ordnung in $\bar k^{-1}$ sind. Diese vergleichen
wir nun mit den Beiträgen erster Ordnung in $A^{(1)}\subt{WKB}(z,t)$ aus (\ref{e4:1.Ord.WKB}), zur
Erinnerung
\begin{align}\label{e4:1.Ord.WKB.copy}
	\begin{split}
		A^{(1)}\subt{WKB}(z,t) 
&= \vph(\zeta(z)-\tau(t))\kle{1 + \ln\sqrt{\frac{k(u(z,t))}{k(z)}}\ }\\[1mm]
&+ \vph'(\zeta(z)-\tau(t))\frac{\I}{4\bar k}\klr{\frac{\bar k^2}{k^2(z)}
		 -\frac{\bar k^2}{k^2(u(z,t))}}\\[1mm]
&+ \vph''(\zeta(z)-\tau(t))\frac{\I}{2\bar k}\int_{u(z,t)}^z\d z'\ \frac{\bar k^3}{k^3(z')}\ .
	\end{split}
\end{align}
Hierzu müssen wir $A^{(1)}\subt{WKB}(z,t)$ weiter nach Potenzen
von $1/\bar k$ entwickeln, und zwar nur bis zur ersten Ordnung, um den
gewünschten Vergleich ziehen zu können. Der Übersichtlichkeit halber
schreiben wir die beiden Lösungen gemeinsame Variable $\tau(t)$ im Folgenden
kurz $\tau$. Die genaue Zeitabhängigkeit von $\tau(t)$ spielt für die folgenden
Betrachtungen keine Rolle.

Wir beginnen mit der Entwicklung der Argumente und der Wellenfunktion $\vph(\zeta(z)-\tau)$. Nach
Definition von $\zeta(z)$ in (\ref{e4:ZetaTau}) gilt
\begin{align}
\zeta(z) &=  \int_{z_0}^z\d z'\ \frac{\bar k}{k(z)}
=  \int_{z_0}^z\d z'\ \frac{\bar k}{\sqrt{\helpkbar\bar k^2 - 2mV(z')}}
=  \int_{z_0}^z\d z'\ \frac{1}{\sqrt{\helpkbar 1 - 2mV(z')/\bar k^2}}\ .
\end{align}
Hieran erkennt man, dass nach Entwicklung der Wurzel
\begin{align}
\zeta(z) = z - z_0 + \OO(\bar k^{-2})
\end{align}
folgt, also keine linearen Beiträge in $1/\bar k$ auftreten. In erster Ordnung in $1/\bar k$ kann das 
WKB-Anfangswellenpaket also geschrieben werden als
\begin{align}
  \vph(\zeta(z)-\tau) \approx \vph((z-z_0)-\tau)\ .
\end{align}
Nun betrachten wir die Koeffizienten der verschiedenen Ableitungen der
Wellenfunktion in (\ref{e4:1.Ord.WKB.copy}).
Wir beginnen bei dem Logarithmus-Term, der eine Korrektur zweiter Ordnung in $\bar k^{-1}$ ist. 
Dies sieht man, indem man das Argument des Logarithmus nach kleinen Potentialen entwickelt. Es gilt
mit $u(z,t)$ aus (\ref{e4:Def.u})
\begin{align*}
  \frac{k(z)}{k(u(z,t))} = \sqrt{\frac{1-2mV(z)/\bar k^2}{1-2mV(u(z,t))/\bar k^2}} 
  = \sqrt{1-\frac{2m(V(z)-V(u(z,t)))}{\bar k^2} + \OO(\bar k^{-4})}\ .
\end{align*}
Hiermit folgt
\begin{align}\label{e4:1.Korr:1.Ord.WKB}
\ln\sqrt{\frac{k(u(z,t))}{k(z)}} = -\frac12\ln\klr{\frac{k(z)}{k(u(z,t))}} = \frac{m(V(z)-V(u(z,t)))}{2\bar k^2} + \OO(\bar k^{-4})\ .
\end{align}
Der Koeffizient der ersten Ableitung von $\vph$ wird in der Entwicklung nach
Potenzen von $1/\bar k$ zu
\begin{align}\label{e4:2.Korr:1.Ord.WKB}
  \frac{\I}{4\bar k}\klr{\frac{\bar k^2}{k^2(z)}-\frac{\bar k^2}{k^2(u(z,t))}}
  = \frac{\I}{4\bar k}\klr{-\frac{m(V(z)-V(u(z,t)))}{\bar k^2}+\OO(\bar k^{-4})} = \OO(\bar k^{-3})
\end{align}
und der Koeffizient von $\vph''$ lautet unter Verwendung von (\ref{e4:u.taylor}), d.h.
$u(z,t) = z - \tau + \OO(\bar k^{-1})$ schließlich
\begin{align}\label{e4:3.Korr:1.Ord.WKB}
  \frac{\I}{2\bar k}\int_{u(z,t)}^z\d z'\ \frac{\bar k^3}{k^3(z')} = \frac{\I}{2\bar k}\tau +
\OO(\bar k^{-2})\ .
\end{align}
Wie man sieht, dominiert dieser Term die beiden anderen, da er der einzige Korrekturterm mit der
Ordnung $\OO(1/\bar k)$ ist. Insgesamt kann $A^{(1)}\subt{WKB}(z,t)$ in erster Ordnung in 
$1/\bar k$  damit geschrieben werden als 
\begin{align}\label{e4:1.Ord.WKB->1.Ord.kbar}
  A^{(1)}\subt{WKB}(z,t) = \vph((z-z_0)-\tau) +
  \vph''((z-z_0)-\tau)\frac{\I}{2\bar k}\tau + \OO(\bar k^{-2})\ .
\end{align}
Um diese Gleichung mit der ersten Ordnung in der Entwicklung der
Charakteristiken-Lösung vergleichen zu können, müssen wir die
Phasenfaktoren noch berücksichtigen. Wir entwickeln den WKB-Phasenwinkel
(\ref{e4:PW.WKB}) nach $1/\bar k$:
\begin{align}
  \begin{split}
    \phi\subt{WKB}(z,t) &= -\frac{\bar k^2}{2m}t + \int_{z_0}^z\d z'\ 
    \sqrt{\bar k^2-2m V(z')}\\
    &= -\frac{\bar k^2}{2m}t + \bar k(z-z_0) -
    \frac{m}{\bar k}\int_{z_0}^z\d z'\ V(z') + \OO(\bar k^{-2})\\
    &= \phi\subt{C}(z-z_0,t) - \frac{m}{\bar k}\int_{z_0}^z\d z'\ V(z') + \OO(\bar k^{-2})
  \end{split}
\end{align}
mit dem Charakteristiken-Phasenwinkel $\phi\subt{C}(z,t)$ aus
(\ref{e4:PW.C}). Entwickeln wir nun den gesamten WKB-Phasenfaktor, so folgt
\begin{align}
  \e^{\I\phi\subt{WKB}(z,t)} = \e^{\I\phi\subt{C}(z-z_0,t)}
  \klr{1 - \frac{\I m}{\bar k}\int_{z_0}^z\d z'\ V(z') + \OO(\bar k^{-2})}\ .
\end{align}
Nun können wir die Wellenfunktionen in erster Ordnung $1/\bar k$ insgesamt
vergleichen. Wir haben
\begin{align}\label{e4:Psi.C.1.Ord.kbar}
  \begin{split}
\Psi^{(1)}\subt{C}(z,t) &= \e^{\I\phi\subt{C}(z,t)}A^{(1)}\subt{C}(z,t)\\
    &=\exp\klg{-\I\frac{\bar k^2}{2m}t + \I\bar k z}
\Bigg(\tilde\vph(z-\tau(t))\kle{1-\frac{\I m}{\bar k}\int_{z-\tau(t)}^z\d z'\ 
      V(z')}\\
    &\hspace{40mm}+ \tilde\vph''(z-\tau(t))\frac\I{2\bar k}\tau(t)\Bigg)
\end{split}
\end{align}
und analog
\begin{align}\label{e4:Psi.WKB.1.Ord.kbar}
  \begin{split}
    \Psi^{(1)}\subt{WKB}(z,t) &= \e^{\I\phi\subt{WKB}(z,t)}A^{(1)}\subt{WKB}(z,t)\\
    &= \exp\klg{-\I\frac{\bar k^2}{2m}t + \I\bar k(z-z_0)}\\ 
    &\times \Bigg(\vph((z-z_0)-\tau(t))\kle{1-\frac{\I m}{\bar
        k}\int_{z_0}^z\d z'\ 
      V(z')}\\
    &\qquad+ \vph''((z-z_0)-\tau(t))\frac\I{2\bar k}\tau(t) + \OO(\bar
    k^{-2})\Bigg)\ .
  \end{split}
\end{align}
Der einzige sichtbare Unterschied zwischen den Gln. (\ref{e4:Psi.C.1.Ord.kbar}) und
(\ref{e4:Psi.WKB.1.Ord.kbar}) sind die verschiedenen Einhüllenden $\vph$ und $\tilde\vph$.
Bei gleichen Anfangsbedingungen gilt hier der Zusammenhang
\begin{align}
	\Psi(z,t=0) = \e^{\I\phi\subt C(z,t=0)}\tilde\vph(z) \overset!= \e^{\I\phi\subt{WKB}(z,t=0)}\vph(\zeta(z))
\end{align}
festgelegt. Nach Einsetzen der Phasenwinkel aus (\ref{e4:PW.C}) und (\ref{e4:PW.WKB}) folgt
\begin{align}
	\tilde\vph(z) = \exp\klg{-\I\bar k z + \I\int_{z_0}^z\d z'\ k(z')}\vph(\zeta(z))
\end{align}
und für Zeiten $t>0$, d.h. $z\to z-\tau$,
\begin{align}\label{e4:AWP.C<->WKB}
	\tilde\vph(z-\tau) = \exp\klg{-\I\bar k(z-\tau) + \I\int_{z_0}^{z-\tau}\d z'\ k(z')}\vph(\zeta(z-\tau))\ .
\end{align}
Entwickeln wir in dieser Gleichung mit den zuvor verwendeten Methoden wieder alles bis zur Ordnung $1/\bar k$,
erhalten wir
\begin{align}
	\begin{split}
		\tilde\vph(z-\tau) &= \exp\klg{-\I\bar k z_0 - \I\frac{m}{\bar k}\int_{z_0}^{z-\tau}\d z'\ V(z')}
			\vph\klr{(z-z_0)-\tau} + \OO(\bar k^{-2})\\
			&= \e^{-\I\bar k z_0}\klr{1 - \I\frac{m}{\bar k}\int_{z_0}^{z-\tau}\d z'\ V(z')}
			\vph\klr{(z-z_0)-\tau} + \OO(\bar k^{-2})\ .
	\end{split}
\end{align}
Nach zweifachem Ableiten nach $(z-\tau)$ erhalten wir
\begin{align}
  \tilde\vph''(z-\tau) = \e^{-\I\bar kz_0}
    \vph''((z-z_0)-\tau) + \OO(\bar k^{-1})\ .
\end{align}
Bei der zweiten Ableitung brauchen wir nur Terme nullter Ordnung in $1/\bar k$
zu berücksichtigen, da in (\ref{e4:Psi.C.1.Ord.kbar}) bereits ein Koeffizient erster
Ordnung in $1/\bar k$ vorhanden ist. Setzt man die Ausdrücke der letzten
beiden Gleichungen nun explizit in (\ref{e4:Psi.C.1.Ord.kbar}) ein, so erhält man sofort
das zu (\ref{e4:Psi.WKB.1.Ord.kbar}) identische Ergebnis, d.h. es
gilt tatsächlich
\begin{align}
  \Psi\subt{C}^{(1)}(z,t) = \Psi\subt{WKB}^{(1)}(z,t)\qquad\text{(in $\OO(\bar k^{-1})$)}\ .
\end{align}

Analog würde man einen Vergleich der Entwicklung bis zur Ordnung $\OO(\bar
k^{-2})$ angehen. Hierzu müsste man zunächst die zweite Ordnung in $\hat K\subt{WKB}$
für den WKB-Fall, $A^{(2)}\subt{WKB}(z,t)$, berechnen, 
da nur dann sichergestellt ist, das alle
Beiträge der zweiten Ordnung in $1/\bar k$ berücksichtigt werden.

\subsubsection{Vergleich der Entwicklungen in $\hat K$}

Nachdem wir uns nun hinreichend davon überzeugt haben, dass sowohl die 
Charakteristiken- wie auch die WKB-Entwicklung zu identischen Wellenfunktionen
führen, wollen wir nun weiter die Qualität dieser beiden Entwicklungen
vergleichen. 

Zunächst betrachten wir erneut den Zusammenhang zwischen der Entwicklung nach Potenzen
des Integraloperators $\hat K$ und der Entwicklung nach Potenzen von $1/\bar k$
und stellen dabei fest, dass hier der WKB-Fall klar im Vorteil ist. Bereits in
erster Ordnung in $\hat K\subt{WKB}$ erhält man dort Beiträge beliebig hoher Ordnung
in $1/\bar k$, aufsummiert in den einzelnen Korrekturbeiträgen, siehe
Gln. (\ref{e4:1.Korr:1.Ord.WKB}) bis (\ref{e4:3.Korr:1.Ord.WKB}). In der Charakteristiken-Entwicklung
hat man in erster Ordnung in $\hat K\subt{C}$ tatsächlich nur Korrekturen
erster Ordnung in $1/\bar k$.

Nun werfen wir einen Blick auf die zweite Ordnung der Entwicklung in $\hat
K\subt{C}$ der Charakteristiken-Lösung. In Gl. (\ref{e4:2.Ord.C}) erkennt man in
der ersten Zeile von $B^{(2)}\subt{C}(z,\tau)$ die Entwicklung eines Phasenfaktors
\begin{align}\label{e4:C.PF.Reihe}
\begin{split}
  &\phantom= \exp\klg{-\frac{\I m}{\bar k}\int_{z-\tau}^z\d z'\ V(z')}\\
  &\approx 1-\frac{\I m}{\bar k}\int_{z-\tau}^z\d z'\ V(z') 
  - \frac1{2!}\frac{m^2}{\bar k^2}\int_{z-\tau}^z\d z'\ V(z')\int_{z-\tau}^{z} \d z''\ V(z'')\\
  &= 1-\frac{\I m}{\bar k}\int_{z-\tau}^z\d z'\ V(z') 
  - \frac{m^2}{\bar k^2}\int_{z-\tau}^z\d z'\ V(z')\int_{z-\tau}^{z'} \d z''\ V(z'')\ .
\end{split}
\end{align}
Dieser Phasenfaktor entspricht dem bereits in Gl. (\ref{e4:Lsg.C.2}) enthaltenen
Phasenfaktor der Charakteristiken-Lösung. Während diese aber nur eine Näherungslösung
war, liefert die Entwicklung auf einfache Weise Korrekturterme höherer Ordnung.
Der Phasenfaktor (\ref{e4:C.PF.Reihe}) stellt aber auch, zusammen
mit dem Anteil $\e^{\I\bar k z}$ des Charakteristiken-Phasenfaktors aus (\ref{e4:Ansatz.C}), 
die Entwicklung eines WKB-ähnlichen Phasenfaktors
\begin{align}\label{e4:C.PF.Reihe.2}
  \exp\klg{+\I\bar k(z-\tau)+\I\int_{z-\tau}^z\d z'\ \sqrt{\bar k^2-2m V(z')}}
\end{align}
für sehr kleine Potentiale und kleine $\tau$ dar. Dies ist eine mathematische Bestätigung für die 
bereits aus physikalischen Argumenten gewonnene Einsicht, dass der WKB-Phasenfaktor
geeigneter für die Beschreibung einer im semiklassischen Sinne
realistischen Wellenfunktion ist. 

Wie man in Gl. (\ref{e4:C.PF.Reihe}) weiter sehen kann, sind im dort angegebenen Phasenfaktor
Terme aufsummiert, die sich typischerweise wie Potenzen von $m\tau V/\bar k$ verhalten.
Diese Terme werden trotz kleiner Potentiale bei großen Zeiten beliebig groß.
Nach der eben geführten Diskussion sind diese Terme im WKB-Phasenfaktor bereits
aufsummiert. Wählt man die Charakteristiken-Phase, bekommt man dagegen nur
die (beliebig großen) Einzelterme in jeder Ordnung der Entwicklung.
Dies ist ein weiteres Argument für den WKB-Phasenfaktor als Ausgangspunkt
für den Entwicklungs-Formalismus.

Schließlich gibt es im Zusammenhang mit dem grundlegenderen Phasenfaktor (\ref{e4:C.PF.Reihe.2}), der in
der Charakteristiken-Lösung zweiter Ordnung in entwickelter Form offenbar auftritt, noch eine weitere, 
bemerkenswerte Tatsache.
Korrigieren wir den ursprünglichen Charakteristiken-Phasenfaktor mit dem Phasenwinkel $\phi\subt{C}(z,t)$ aus
Gl. (\ref{e4:PW.C}) nämlich gerade um (\ref{e4:C.PF.Reihe.2}),
d.h. setzen wir
\begin{align}
	\Psi\subt C(z,t) \approx \exp\klg{-\I\frac{\bar k^2}{2m}t + \I\bar k(z-\tau(t))
		+ \I\int_{z-\tau(t)}^z\d z'\ \sqrt{\bar k^2-2m V(z')}}\tilde\vph(z-\tau(t))
\end{align}
und verwenden weiter den Zusammenhang (\ref{e4:AWP.C<->WKB}) zwischen $\tilde\vph$
und $\vph$, so erhalten wir gerade
\begin{align}\label{e4:Lsg.C.3}
	\Psi\subt C(z,t) \approx \exp\klg{-\I\frac{\bar k^2}{2m}t 
		+ \I\int_{z_0}^z\d z'\ \sqrt{\bar k^2-2m V(z')}}\vph(\zeta(z-\tau(t)))\ .
\end{align}
Dies entspricht beinahe der WKB-Näherungslösung (\ref{e4:Lsg.WKB}), dort ist lediglich
anstelle von $\zeta(z-\tau(t))$ der Ausdruck $\zeta(z)-\tau(t)$ im Argument
von $\vph$ zu finden. Man kann sich davon überzeugen, dass
\begin{align}
	\zeta(z-\tau) = \zeta(z)-\tau(t) + \OO(\tfrac{\tau m V}{\bar k})
\end{align}
gilt. Im Argument der WKB-Näherungslösung,
das ja die Bewegung eines Teilchens im klassischen Sinne richtig beschreibt,
sind also wieder alle Potenzen der bereits zuvor diskutierten Größe $\tau m V/\bar k$ aufsummiert, nicht aber
im Argument der Charakteristiken-Lösung, die mit dem WKB-Anfangswellenpaket über (\ref{e4:AWP.C<->WKB})
zusammenhängt.

Kommen wir zum Abschluss dieses Abschnitts noch zu einem letzten Argument, dass
für die Verwendung des WKB-Phasenwinkels bei der Berechnung der exakten Lösung
spricht.

Im Koeffizienten zu $\tilde\vph(z-\tau)$ der Charakteristiken-Entwicklung zweiter
Ordnung in $\hat K_C$ findet sich u.a. ein Term
\begin{align*}
  \frac{m(V(z)-V(z-\tau))}{2\bar k^2}\ .
\end{align*}
Dieser Beitrag entspricht dem $1/\bar k^2$-Beitrag des entwickelten Logarithmus-Terms aus der
WKB-Entwicklung, Gl. (\ref{e4:1.Korr:1.Ord.WKB}), allerdings in der weiteren Näherung
$u(z,t) = \mc Z(\zeta(z))-\tau(t))\approx z-\tau(t)$ (siehe Gl. (\ref{e4:u.taylor})).
Wir haben also in der Charakteristiken-Entwicklung in zweiter Ordnung eine Korrektur
erhalten, die eine Näherung einer Korrektur ist, die in der WKB-Entwicklung bereits in
aufsummierter Form in erster Ordnung enthalten ist.

Insgesamt zeigen die Ergebnisse dieses Kapitels, dass die Verwendung des WKB-Phasenwinkels
zu physikalisch sinnvolleren Resultaten führt als die Verwendung des Charakteristiken-Phasenwinkels.
Bereits die nullte Ordnung der Entwicklung (\ref{e4:Loesung}) gibt eine gute Näherung
mit korrekter Schwerpunktsbewegung des Wellenpakets und potentialabhängigem Phasenfaktor.

\subsection{Abschätzung der Gültigkeit der WKB-Lösung}\label{s4:WKB.Gueltigkeit}

Bevor wir noch einmal die bisherigen Ergebnisse zusammenfassen,
wollen wir noch untersuchen, in welchem experimentellen Rahmen die
WKB-Näherungslösung eine gute Approximation darstellt. 

Betrachten wir zunächst die Entwicklung der WKB-Lösung nach Gl. (\ref{e4:1.Ord.WKB.copy}).
Damit das ursprüngliche WKB-Wellenpaket eine gute Approximation darstellt,
muss der dominierende Anteil im Korrekturterm, siehe Gl. (\ref{e4:1.Ord.WKB->1.Ord.kbar}), verschwindend gering sein, 
d.h.
\begin{align}\label{e4:ValidityCond}
\abs{\frac{\tau}{2\bar k}\frac{\vph''(\zeta-\tau)}{\vph(\zeta-\tau)}} \ll 1\ .
\end{align}
Nehmen wir eine Gauß-Funktion
\begin{align}
\vph(\zeta-\tau) \propto \e^{-(\zeta-\tau)^2/2\sigma^2}
\end{align}
für die Einhüllende des Wellenpakets an, so folgt
\begin{align}
\frac{\vph''(\zeta-\tau)}{\vph(\zeta-\tau)} = \frac{(\zeta-\tau)^2}{\sigma^4}-\frac{1}{\sigma^2}
\end{align}
und mit (\ref{e4:ValidityCond}) also
\begin{align}
\abs{\frac{\tau}{2\bar k}\frac{\vph''(\zeta-\tau)}{\vph(\zeta-\tau)}}
= \frac{\tau}{2\bar k}\abs{\frac{(\zeta-\tau)^2}{\sigma^4}-\frac{1}{\sigma^2}}
= \frac{\tau}{2\bar k}\abs{\frac{\sigma^2-(\zeta-\tau)^2}{\sigma^4}}
\ll 1\ .
\end{align}
Von Interesse sind dabei nur Werte $\abs{\zeta-\tau}^2\leq \sigma^2$ des Arguments
des Gauß-Pakets. Maximal wird der Betrag von $\vph(\zeta-\tau)$ in diesem Bereich 
für $\zeta=\tau$, also am Maximum des Wellenpakets. Damit erhalten wir die Forderung
\begin{align}\label{e4:Fuzzi}
\epsilon := \frac{\tau}{2\bar k}\frac1{\sigma^2}\ll 1
\end{align}
und somit 
\begin{align}\label{e4:GaussBreiteMin}
\boxed{\sigma\gg \sqrt{\frac{\tau}{2\bar k}} =: \sigma\subt{min}}\ .
\end{align}

Der maximale Wert von $\tau$ entspricht der Länge der Apparatur, die im Bereich von unter zehn Metern
liegt\footnote{M. DeKieviet, private Diskussion.}. Wählen wir z.B. eine Apparatur-Länge von
\begin{align}
\tau_A = 5\u m\ ,
\end{align}
so folgen für typische Strahlgeschwindigkeiten die in Tabelle \ref{t4:Sigma.Bounds} gezeigten unteren Schranken
für die Breite der Wellenpakete.
\begin{table}[!htp]
\centering
\begin{tabular}{|c|c||c|c||c|c|}\hline
$\bar v\ [\u{m/s}]$ & $T\ [\u K]$ & $\bar k\ [10^{10}\u{m^{-1}}]$ & $\bar \lambda\ [10^{-10}\u{m}]$ & $\sigma\subt{min}\ [\u m]$ & $\sigma\subt{min}/\bar\lambda$\\ \hline \hline
1800 & 77 & $2,9$ & $2,197$ & $9,3\ten{-6}$ & $\approx 42000$\\ 
3500 & 300 & $5,6$ & $1,130$ & $1,5\ten{-5}$ & $\approx 132000$\\ \hline
\end{tabular}
\caption[Untere Schranken für die Breite der Wellenpakete für die Gültigkeit der WKB-Näherungslösung.]{Typische experimentelle Strahlgeschwindigkeiten und Temperaturen (M. DeKieviet, private Kommunikation) und die 
daraus für Wasserstoff folgenden mittleren Wellenzahlen, Wellenlängen und die unteren Schranken für die Breite des Wellenpakets gemäß (\ref{e4:GaussBreiteMin}).}
\label{t4:Sigma.Bounds}
\end{table}
Ein Blick auf die Werte in der letzten Spalte der Tabelle macht nachdenklich.
Haben wir in Gl. (\ref{e2:kbar.k.approx.2}) noch behauptet, das WKB-Wellenpaket wäre unter der 
Voraussetzung
\begin{align}
\sigma\gg \frac{1}{2\bar k} = \frac{\bar\lambda}{4\pi}
\end{align}
eine gute Näherung, so bekommen wir hier als untere Schranke einen Ausdruck, der proportional zur 
Wurzel dieses Wertes ist, was im realistischen Fall (wo $\tau = \OO(1\u m)$ ist) etwa 
dem $10^4$ bis $10^5$-fachen entspricht. Das Problem mit der damaligen Abschätzung ist,
dass sie unter der impliziten Annahme entstand, die WKB-Wellen, die wir dort überlagert haben, wären
exakte Lösungen der Schrödinger-Gleichung. Dies ist aber mitnichten der Fall, denn wir haben in
Abschnitt \ref{s2:WKB} ja gerade die zweiten Ortsableitungen des Phasenwinkels und der Amplitudenfunktion 
vernachlässigt und außerdem die Amplitude der ebenen WKB-Wellen zu Eins gesetzt.

Nachdem wir nun in der Lage sind, die exakte Wellenfunktion mit dem WKB-Wellenpaket als nullte Ordnung
zu entwickeln, zeigt uns die Korrektur erster Ordnung, dass der Term proportional
zur zweiten Ableitung nach $\zeta$ den größten Beitrag liefert. Dieser Term ist aber gerade aus 
der Transformation der zweiten Ortsableitung $\del_z^2$ der Amplitudenfunktion
in die zweite Ableitung nach $\zeta = \zeta(z)$ entstanden (siehe Abschnitt \ref{s4:rewriteSG}),
also genau aus dem Beitrag, den wir in Abschnitt \ref{s2:WKB} in der Schrödinger-Gleichung von vornherein
vernachlässigt haben.

Die einzig verlässliche Bedingung für die Gültigkeit der WKB-Näherungslösung ist also die aus Gl. (\ref{e4:GaussBreiteMin})

Betrachten wir nun ein freies Gaußpaket mit Dispersion. Eine anschauliche Rechnung hierzu findet sich in
\cite{Cohen}, Bd. 1, Abschnitt 1.11, S. 51 ff.. Dort wird in Gl. (1.170) ein Ausdruck für die Breite $\Delta x(t)$
des Wellenpakets zur Zeit $t$ angegeben, was in unserer Notation $\sigma(t)/\sqrt2$ entspricht. 
Wir erhalten mit Gl. (1.170) aus \cite{Cohen} ($\hbar = 1,\ a^2 = 2\sigma^2$) 
\begin{align}
\sigma(t) = \sigma\sqrt{1 + \frac{t^2}{m^2\sigma^4}} = \sigma\klr{1 + \frac{t^2}{2m^2\sigma^4} 
+ \OO\klr{\frac{t^4}{m^4\sigma^8}}}\ .
\end{align}
Die Breite des Wellenpakets ändert sich also genau dann nicht signifikant mit der Zeit, wenn
\begin{align}
\abs{\frac{t^2}{2m^2\sigma^4}} \ll 1
\end{align}
bzw.
\begin{align}\label{e4:Dispersion}
\sigma \gg \sqrt{\frac{t}{\sqrt2m}} = 2^{\frac14}\sqrt{\frac{\bar kt}{2\bar km}} 
= 2^{\frac14}\sqrt{\frac{\tau}{2\bar k}}
\overref{(\ref{e4:GaussBreiteMin})}= 2^{\frac14}\sigma\subt{min} \approx 1,18\sigma\subt{min}\ .
\end{align}
Der Fehler, den wir bei Verwendung des WKB-Wellenpakets (also der nullten Ordnung der Entwicklung) machen,
wird somit hauptsächlich durch die Dispersion der Wellenfunktion bestimmt, wie Gl. (\ref{e4:Dispersion})
in eindrucksvoller Weise zeigt.

\subsection{Ein Verfahren zur Berücksichtigung der Dispersion des Wellenpakets}\label{s4:Dispersion}

Beim aufmerksamen Leser mag zum jetzigen Zeitpunkt 
der Eindruck entstanden sein, dass gerade bei sehr schmalen Wellenpaketen
die Aussicht auf eine sehr genaue Beschreibung der Wellenfunktion in Anbetracht der großen Korrekturbeiträge
und der schwierigen Berechnung der Integraloperatoren in höheren Ordnungen nicht sehr gut ist.

Tatsächlich werden beim lABSE Wellenpakete mit einer Breite von wenigen \AA{}ngstr\o{}m bis Nanometern benötigt, um 
bei den üblicherweise verwendeten magnetischen Feldstärken brauchbare Spinechosignale zu liefern (siehe Abschnitt
\ref{s3:Fahrplan}). Solch ein schmales Wellenpaket im Ortsraum hätte aber eine entsprechend breite Verteilung im 
Impulsraum und daher eine starke Dispersion. Die Abschätzung (\ref{e4:Fuzzi}) ist für ein solches Wellenpaket
unter Umständen nicht mehr gültig, d.h. der eigentliche Entwicklungsparameter $\epsilon$ der Reihenentwicklung 
nach dem Integraloperator $\hat K$ ist dann nicht mehr klein gegenüber Eins. Es ist fraglich, ob
die Entwicklung nach Potenzen von $\hat K$ dann überhaupt noch gültig oder konvergent ist.
 Einen Ausweg aus dieser Situation stellt die
Superposition sehr breiter Wellenpakete ohne Dispersion zu einem sehr schmalen Wellenpaket mit Dispersion dar.
Wenn wir annehmen, dass für die breiten Wellenpakete die Berechnung der ersten Ordnung der Reihenentwicklung 
eine sehr gute Approximation an die exakte Lösung der Schrödinger-Gleichung liefert, sollte die Superposition
der Entwicklungen zu einem schmalen Wellenpaket eine gute Approximation des eigentlich gesuchten 
Wellenpakets liefern. Wir wollen im Folgenden am Beispiel eines Gaußschen Wellenpakets die mathematischen
Grundlagen dieser Vorgehensweise geben.

Zunächst verweisen wir auf Anhang \ref{sC:WFT}, wo wir einige Grundlagen über die sogenannte Gabor-Transformation
\cite{Gab46} zusammengestellt haben, die auch als {\em gefensterte} Fourier-Transformation bezeichnet wird.
Anstatt ein Wellenpaket aus unendlich ausgedehnten, ebenen Wellen zu superponieren (wie bei der gewöhnlichen
Fourier-Transformation), überlagert man hier Wellen (mit fester Wellenzahl) die durch eine um einen Ort $x_0$ zentrierte
Einhüllende (Fensterfunktion) abgeschnitten werden. 
Der tiefere Sinn hinter dieser Idee ist der, dass nur solche Basisfunktionen
für die Superposition benötigt werden, die auch in dem Ortsbereich definiert sind, wo die zu superponierende Funktion
von Null verschieden ist. Ein Nachteil der Gabor-Transformation ist die in ihrer Breite festgelegte 
Fensterfunktion. Will man eine Funktion superponieren, die im Ortsraum erheblich schmaler als die Fensterfunktion
ist, so hat man ähnliche Verhältnisse wie bei der Fourier-Transformation. In diesem Fall ist die Gabor-Transformation
ungeeignet und stellt sogar eine Verkomplizierung des Problems dar, da man eine eindimensionale Funktion in einen
zweidimensionalen Phasenraum (Wellenzahl $k$ und Zentrum $x_0$ der Fensterfunktion) 
transformiert hat. Eine Weiterentwicklung
der Gabor-Transformation ist daher die sogenannte Wavelet-Transformation, wo die Basisfunktionen eine festgelegte
Anzahl von Oszillationen enthalten. Bei kleinerer Wellenlänge, also höherer Ortsauflösung, hat man also ein
sehr viel schmaleres Fenster. Wavelets werden heute überall in den mathematischen und technischen Wissenschaften
angewendet, insbesondere in der Bild- und Signalverarbeitung. Eine gut verständliche, einfach gehaltene Einführung in
dieses sehr umfangreiche Teilgebiet der Mathematik gibt z.B. \cite{Kai94}.

In unserem Fall bietet sich eine vereinfachte Version der Gabor Transformation an. Wir betrachten ein schmales,
Gaußsches Wellenpaket mit der Breite $2\sigma$, das zur Zeit $t=0$ um einen Ort $z=0$ zentriert sei. In Anlehnung
an Gl. (\ref{eC:GT}) aus Anhang \ref{sC:WFT} machen wir den Ansatz für ein Gaußsches Wellenpaket (siehe Gl.
(\ref{e2:GaussWP}))
\begin{align}\label{e4:GT}
\begin{split}
\Psi(z,t=0) &= \frac{1}{\sqrt{\sigma\sqrt\pi}}\exp\klg{\I\bar k z}\exp\klg{-\frac{z^2}{2\sigma^2}}\\
&\overset!= \int \d\bar k'\ \tilde\Psi(\bar k'-\bar k)
\exp\klg{-\frac{z^2}{2\varkappa^2}}\exp\klg{\I\bar k'z}\ .
\end{split}
\end{align}
Die Breite $\varkappa$ der zu superponierenden Wellenpakete sei dabei so gewählt, dass die Abschätzung
(\ref{e4:Fuzzi}),
\begin{align}
\frac{\tau}{2\bar k \varkappa^2} \ll 1\ ,
\end{align}
sehr gut erfüllt ist und ferner $\varkappa\gg\sigma$ gilt. 
Wir gehen hier davon aus, dass ein evtl. vorhandenes Potential $V(z)$ über die ganze
Ausdehnung der sehr breiten Basisfunktionen als Null angesehen werden kann\footnote{Ansonsten hätten wir in
(\ref{e4:GT}) den WKB-Phasenfaktor anstelle von $\exp\klg{\I\bar k'z}$ verwenden müssen.}.

Im Gegensatz zur Gabor Transformation überlagern wir hier nur solche Basisfunktionen, deren Fensterfunktion
um Null zentriert ist. In dem speziellen Fall, den wir hier betrachten und der für die Zwecke dieser Arbeit
völlig ausreichend ist, können wir die Transformierte $\tilde\Psi(\bar k'-\bar k)$ dennoch exakt berechnen.
Auf der rechten Seite von Gleichung (\ref{e4:GT}) können wir die Fensterfunktion aus dem Integral über
$\bar k'$ herausziehen und erhalten
\begin{align}\label{e4:FT}
\int\d\bar k'\ \tilde\Psi(\bar k'-\bar k)\exp\klg{\I\bar k'z} = \frac{1}{\sqrt{\sigma\sqrt\pi}}\exp\klg{\I\bar kz}\exp\klg{-\frac{z^2}{2 \sigma_\varkappa^2}}
\end{align}
mit
\begin{align}
\sigma_\varkappa = \sqrt{\frac{\varkappa^2\sigma^2}{\varkappa^2-\sigma^2} }\ ,\qquad(\varkappa\gg\sigma)\ .
\end{align}
Auf der linken Seite von Gl. (\ref{e4:FT}) steht eine einfache Fourier-Transformation. Wir können also
$\tilde\Psi(\bar k'-\bar k)$ durch Fourier-Rücktransformation berechnen und erhalten
\begin{align}\label{e4:Transformierte}
\begin{split}
\tilde\Psi(\bar k'-\bar k) &= \frac1{2\pi}\frac{1}{\sqrt{\sigma\sqrt\pi}}\int\d z\ \exp\klg{-\frac{z^2}{2 \sigma_\varkappa^2}}
\exp\klg{-\I(\bar k'-\bar k)z}\\
&= \mc N(\sigma,\sigma_\varkappa)\exp\klg{-\frac{(\bar k'-\bar k)^2}{2\sigma_\varkappa^{-2}}}\ ,\\
\mc N(\sigma,\sigma_\varkappa) &= \frac{\sigma_\varkappa}{\sqrt{2\pi}}\frac{1}{\sqrt{\sigma\sqrt\pi}}
= \sqrt{\frac{1}{2\pi^{3/2}}\frac{\varkappa^2\sigma}{\varkappa^2-\sigma^2}}\ .
\end{split}
\end{align}
Dieses Ergebnis ist im Grenzfall $\sigma_\varkappa\to\sigma$, d.h. $\varkappa^2\to\infty$, 
identisch mit der Fourier-Transformierten einer Gaußfunktion aus Gl. (\ref{e2:Gauss.FT}).

Um nun die Berechnung höherer Ordnungen in der Entwicklung der exakten Lösung der Schrödinger-Gleichung
zu vermeiden und dennoch die Dispersion zu berücksichtigen, muss man (in nullter Ordnung
der Entwicklung der exakten Lösung) Wellenpakete
\begin{align}\label{e4:Base}
\Psi^{(0)}(\bar k',\varkappa,z,t) := \e^{\I\phi(\bar k',z,t)}\vph(\varkappa,\zeta(\bar k',z) - \tau(\bar k',t))
\end{align}
untersuchen, wobei $\phi(\bar k',z,t)$, $\zeta(\bar k',z)$ und $\tau(\bar k',t)$ aus den bisher
verwendeten Funktionen durch Ersetzen von $\bar k\to\bar k'$ entstehen. Die Einhüllende ist dabei gegeben durch
\begin{align}\label{e4:BaseEnvelope}
\vph(\varkappa,\zeta) := \exp\klg{-\frac{\zeta^2}{2\varkappa^2}}\ ,
\end{align}
ihre Normierung wurde in die Transformierte aus Gl. (\ref{e4:Transformierte}) verschoben.
Für jede Wellenfunktion nullter Ordnung aus Gl. (\ref{e4:Base}) mit \glqq mittlerer\grqq~ Wellenzahl $\bar k'$ 
muss die Anfangsbedingung
\begin{align}\label{e4:BaseAB}
\Psi(\bar k',\varkappa,z,t=0) = \exp\klg{\I\bar k'z}\exp\klg{-\frac{z^2}{2\varkappa^2}}
\end{align}
erfüllt sein, damit die Superposition (\ref{e4:GT}) mit der Transformierten (\ref{e4:Transformierte}) 
das gewünschte, schmale Wellenpaket mit Breite $\sigma$ und mittlerer Wellenzahl $\bar k$ ergibt.

Aus der Einhüllenden (\ref{e4:BaseEnvelope}) kann man nun mit Hilfe der Entwicklung nach Potenzen
des Integraloperator $\hat K$ aus (\ref{e4:Def.K.WKB}) die Korrektur erster Ordnung berechnen, d.h.
\begin{align}
A^{(1)}(\bar k',\varkappa,,z,t) = \vph(\varkappa,\zeta(\bar k',z) - \tau(\bar k',t))
+ (\hat K\vph)(\bar k',\varkappa,z,t)\ ,
\end{align}
wobei der Dispersionsterm, der im letzten Abschnitt diskutiert wurde, bei geeigneter Wahl der Breite
$\varkappa$ bereits vernachlässigbar ist. Man kann nun die Wellenfunktion erster Ordnung aufstellen,
\begin{align}\label{e4:Base.1.O}
\Psi^{(1)}(\bar k',\varkappa,z,t) = \e^{\I\phi(\bar k',z,t)}A^{(1)}(\bar k',\varkappa,,z,t)\ ,
\end{align}
die immer noch die Anfangsbedingung (\ref{e4:BaseAB}) erfüllt. Wie in Gl. (\ref{e4:GT}) kann damit
unter Verwendung der Transformierten (\ref{e4:Transformierte}) das Wellenpaket
\begin{align}\label{e4:1.O.Dispersion}
\Psi^{(1)}(\varkappa,z,t) = \int\d\bar k'\ \tilde\Psi(\bar k'-\bar k)\Psi^{(1)}(\bar k',\varkappa,z,t)
\end{align}
superponiert werden, dass auch für Zeiten $t>0$ gilt.

Da (\ref{e4:1.O.Dispersion}) bei ausreichend großem $\varkappa$ die Dispersion des Wellenpakets korrekt
beschreibt, sind hier nur noch die Beiträge proportional zur Einhüllenden und ihrer ersten Ableitung
in der Korrektur erster Ordnung relevant. Ist $\varkappa$ ausreichend groß, verbleibt nur noch der
zur Einhüllenden proportionale Beitrag.
Bei entsprechend gut gewähltem Phasenwinkel wird (\ref{e4:1.O.Dispersion}) in diesem Fall
die exakte Wellenfunktion sehr gut approximieren.

Die tatsächliche Berechnung der Superposition (\ref{e4:1.O.Dispersion}) kann i.A. nur numerisch durchgeführt
werden und unterliegt den damit verbundenen Schwierigkeiten\footnote{Man kann numerisch
nicht für unendlich viele Wellenzahlen $\bar k'$ die erste Ordnung der Entwicklung berechnen 
und diese dann alle superponieren.}.
Eine detaillierte Untersuchung dieses Aspekts muss aus Zeitgründen 
aber leider auf zukünftige Arbeiten verschoben werden.

\FloatBarrier

\newpage
\section{Zusammenfassung und Schlussfolgerungen}\label{s4:Conclusions}

Wir wollen am Ende dieses Kapitels die wesentlichen Resultate noch einmal in übersichtlicher Form festhalten.

\begin{enumerate}[(1)]
\item\label{i4:Formalismus} Der in diesem Kapitel entwickelte Formalismus erlaubt die im Prinzip
exakte Berechnung des Amplitudenanteils der Wellenfunktion bei vorgegebenem
Phasenfaktor.
\item\label{i4:Prinzip} Es handelt sich dabei um eine Entwicklung nach Potenzen eines
Integraloperators $\hat K\subt{WKB}$ bzw. $\hat K\subt{C}$, siehe Gl.
(\ref{e4:Def.K.WKB}) mit den Operatoren $\hat L\subt{WKB}$ aus (\ref{e4:DefL.WKB})
bzw. $\hat L\subt{C}$ aus (\ref{e4:DefL.C}).
\item\label{i4:Alternativ} Die Lösung (\ref{e4:Loesung}) kann auch als Entwicklung nach der inversen mittleren Wellenzahl
$1/\bar k$ aufgefasst werden.
\item\label{i4:kbar} Die Entwicklung bis zur $n$-ten Ordnung in 
$\hat K\subt{WKB}$ bzw. $\hat K\subt C$ enthält {\em mindestens} alle Beiträge bis zur Ordnung
$\bar k^{-n}$. Beiträge höherer Ordnung dagegen {\em können} enthalten sein. Es hat sich gezeigt,
dass dies bei der Charakteristiken-Entwicklung nicht der Fall ist, wohl aber bei der WKB-Entwicklung.
\item\label{i4:WKB} Die WKB-Lösung (\ref{e4:Lsg.WKB}) ist zur Beschreibung
  eines semiklassischen Wellenpakets besser geeignet als die
  Charakteristiken-Lösung (\ref{e4:Lsg.C.2}), weil ...
  \begin{enumerate}[a)]
\item\label{i4:WKB.SP}  ... sie die Bewegung des Schwerpunkts eines Wellenpakets durch ein
    Potential im klassischen Sinne richtig beschreibt, während der
    Schwerpunkt bei der Charakteristiken-Lösung auch im Potential mit
    konstanter Geschwindigkeit propagiert.
\item\label{i4:WKB.Korrekturen} ... die WKB-Entwicklung bereits in erster Ordnung in $\hat K\subt{WKB}$
    Beiträge zu beliebig hohen Ordnung in $1/\bar k$ enthält, die
    Charakteristiken-Entwicklung aber nur Beiträge bis zur
    Ordnung $1/\bar k$, siehe Punkt (\ref{i4:kbar}).
\item\label{i4:WKB.C} ... in der WKB-Näherungslösung bzw. der ersten Ordnung der Entwicklung in $\hat K\subt{WKB}$
    bereits viele Terme aus der Charakteristiken-Entwicklung aufsummiert sind, die proportional zu $\tau$
    sind und somit für große Zeiten beliebig groß werden können.
  \end{enumerate}
\item\label{i4:Vergleich} Die Entwicklungen in $\hat K\subt{WKB}$ und $\hat K\subt C$ müssen letztendlich
  die gleiche Wellenfunktion liefern, was überprüft werden kann, indem man
  einen Koeffizientenvergleich in $1/\bar k$ macht. Dies wurde für die erste Ordnung
  in $1/\bar k$ explizit durchgeführt. 
\item\label{i4:BingoWKB} Die Aufsummation des Phasenfaktors (\ref{e4:C.PF.Reihe}),
  der in der Charakteristiken-Entwicklung deutlich zu erkennen ist
  und der auch dem Phasenfaktor der Charakteristiken-Lösung (\ref{e4:Lsg.C.2})
  exakt entspricht, stellt offenbar eine Entwicklung des Phasenfaktors
  (\ref{e4:C.PF.Reihe.2}) für kleine Potentiale und kleine Zeiten dar. 

Zusammen mit dem zeitabhängigen Anteil des Charakteristiken-Phasenfaktors ergibt 
der aufsummierte Phasenfaktor (\ref{e4:C.PF.Reihe.2}) unter Verwendung des
Zusammenhangs zwischen den Anfangswellenpaketen $\tilde\vph$ und $\vph$ aus (\ref{e4:AWP.C<->WKB})
eine Näherung der WKB-Lösung in nullter Ordnung der Entwicklung, siehe Gl. (\ref{e4:Lsg.C.3}).

Dies bestätigt in eindrucksvoller Weise, dass die Wahl des WKB-Phasenfaktors
als Ausgangspunkt für die Anwendung des Formalismus zur Berechnung des Amplitudenteils der Wellenfunktion
nahe liegt.
\item Den größten Beitrag zur Korrektur liefert die in nullter Ordnung vernachlässigte Dispersion der
Wellenfunktion, siehe Abschnitt \ref{s4:WKB.Gueltigkeit}. Ein Maß für die Größe der Korrekturbeitrag
liefert der mit der Dispersion zusammenhängende Parameter $\epsilon$ aus in Gl. (\ref{e4:Fuzzi}). 
Er ist der eigentliche Entwicklungsparameter der Lösung (\ref{e4:Loesung}) für WKB-ähnliche Phasenfaktoren
und hängt neben der Länge $\tau$ auch von der Breite $\sigma$ des Wellenpakets ab. 
Nur wenn $\epsilon\ll 1$ ist, kann die Dispersion des Wellenpakets 
und somit Beiträge höherer Ordnung vernachlässigt werden.
\item Um die für lABSE benötigten, sehr schmalen Wellenpakete korrekt mit dem Entwicklungsformalismus
beschreiben zu können, müssen die in Abschnitt \ref{s4:Dispersion} behandelten 
Superpositionen sehr breiter Wellenpakete betrachtet werden, für die
die Dispersion jeweils vernachlässigbar ist. Für diese Basiswellenpakete stellt die erste Ordnung der
Entwicklung bei ausreichend großer Breite im Ortsraum eine sehr gute Approximation an die
exakte Lösung der Schrödinger-Gleichung dar. Die Superposition der Entwicklungen der Basiswellenpakete
ergibt dann die Entwicklung des gewünschten schmalen Wellenpakets inklusive Dispersion.
\end{enumerate}

Im Prinzip war es spätestens am Ende von Kapitel \ref{s3:Anwendungen} klar, dass die WKB-Näherungslösung
der Charakteristiken-Näherungslösung vorzuziehen ist, denn aufgrund der realistischeren Schwerpunktsbeschreibung
der WKB-Lösung waren wir bereits zu diesem Zeitpunkt in der Lage, das Fahrplanmodell theoretisch zu beschreiben
und einfache lABSE-Interferenzsignale zu berechnen.

Nun allerdings haben wir, durch die Entwicklung des Formalismus zur Berechnung der exakten Lösung (\ref{e4:Loesung})
der Schrödinger-Gleichung, auch die mathematische Kontrolle über die Korrekturen zur Näherungslösung. Desweiteren
waren wir in der Lage, die WKB- und die Charakteristiken-Lösung quantitativ zu vergleichen und besser zu verstehen.

Bereits in der ersten Ordnung des Formalismus kann man außerdem Rückschlüsse auf Korrekturen des Phasenwinkels
ziehen. Wir werden dies auch im nächsten Kapitel, wenn wir komplexe skalare Potential zulassen, in eindrucksvoller
Weise demonstrieren.

Als Ausgangspunkt für die Entwicklung nehmen wir, motiviert von den bisherigen Ergebnissen, stets den WKB-Phasenwinkel
oder Ergänzungen davon. Wir lassen daher im Folgenden den überflüssigen Index WKB bei Phasenwinkeln und
Amplitudenfunktionen fort.

\chapter{Implementation komplexer, skalarer Potentiale}\label{s5:Zerfall}

\section{Problemstellung}\label{s5:1.Ordnung}

Das Ziel der vorliegenden Arbeit ist die Entwicklung der Theorie zur
Beschreibung von lABSE-Experimenten zur Messung von paritätsverletzenden Effekten in Atomen.
Wie in der Einleitung (Kap. \ref{s1:Einleitung}) und auch in Abschnitt \ref{s2:LABSE} bereits
erwähnt, müssen wir dafür in der Lage sein, metastabile Atome zu beschreiben und wollen uns 
nun überlegen, wie wir dies erreichen können.

Wir erinnern uns dazu an die Zeitentwicklung des Eigenzustands $\ket{\Psi_E(t)}$ 
zur Energie $E$ eines stationären Hamiltonoperators $H$ im Schrödinger-Bild,
\begin{align}
\ket{\Psi_E(t)} = \exp\klg{-\I E (t-t_0)}\ket{\Psi_E(t_0)}\ .
\end{align}
Soll der Zustand $\ket{\Psi_E(t)}$ ein zerfallendes Teilchen repräsentieren,
so kann man dies durch Einführung komplexer Potentiale bzw. Energieeigenwerte erreichen, d.h.
man setzt
\begin{align}\label{e5:komplexe.Energie}
  E = V - \frac\I2\Gamma\ ,\qquad \Gamma>0\ ,
\end{align}
und erhält dann einen exponentiellen Abfall der Norm des Zustands gemäß
\begin{align}
\abs{\Psi_E(t)}^2 = \bracket{\Psi_E(t)}{\Psi_E(t)} = \e^{-\Gamma (t-t_0)}\ .
\end{align}
Eine grundlegende Methode zur Beschreibung instabiler Atome liefert der Wigner-Weisskopf-Formalismus,
siehe Anhang \ref{sA:WWF}, den Wigner und Weisskopf \cite{WeWi30} bereits 1930 zur Berechnung der natürlichen Linienbreite
von atomaren Übergängen entwickelt haben. Wir werden aber von diesem Formalismus erst im nächsten 
Kapitel \ref{s6:Formalismus}, wenn wir das metastabile Atom betrachten werden, wirklich Gebrauch machen
und uns in diesem Kapitel mit der oben motivierten Einführung eines komplexen skalaren Potentials zur
Beschreibung der Wellenfunktion eines zerfallenden Teilchens begnügen.

\section{Anwendung des Formalismus auf ein komplexes, skalares Potential}\label{s5:Anwendung}

\subsection{Berechnung des Integraloperators}

Wir betrachten also die Schrödinger-Gleichung
\begin{align}\label{e5:SG}
  \Big(\del_z^2 - 2m[V(z)-\tfrac\I2\Gamma(z)]+2m\I\del_t\Big)\Psi(z,t) = 0
\end{align}
und wenden zur Lösung den in Kapitel \ref{s4:Formalismus} entwickelten Formalismus an.
Dabei verwenden wir als Ausgangspunkt den von Gl. (\ref{e5:komplexe.Energie}) motivierten, WKB-ähnlichen
Phasenwinkel
\begin{align}\label{e5:PW}
\phi(z,t) = -\frac{\bar k^2}{2m}t + \int_{z_0}^z\d z'\ \sqrt{\bar k^2 - 2mV(z')} + \frac\I2\bar\Gamma t\ .
\end{align}
Wir weisen darauf hin, dass im WKB-Anteil dieses Phasenwinkels eine reelle, lokale Wellenzahl
\begin{align}\label{e5:lok.WZ}
k(z) = \sqrt{\bar k^2 - 2mV(z)}
\end{align}
verwendet wurde. Auf diese Weise bleibt die Koordinate $\zeta(z)$ ebenfalls reell und somit eindeutig umkehrbar
und wir können die übliche Definition des Anteils
\begin{align}\label{e5:SWKB}
S(z) = \int_{z_0}^z\d z'\ k(z')\ ,
\end{align}
im Phasenwinkel beibehalten.
Die Zerfallsbreite $\bar\Gamma$ aus Gl. (\ref{e5:PW}) sei eine über das Potential gemittelte Zerfallsbreite.
Sollte die Wahl des Phasenwinkels nicht optimal sein, werden wir dies an den Korrekturtermen ablesen können und
sind dann auch später noch in der Lage, einen besseren Phasenwinkel als Ausgangspunkt zu verwenden. Der Formalismus
zur Berechnung der exakten Lösung der Schrödinger-Gleichung hat in diesem Sinne eine eingebaute Selbstkorrektur.

Mit dem obigen Phasenwinkel und dem Ansatz (\ref{e4:AnsatzWF}), zur
Erinnerung
\begin{align*}
   \Psi(z,t)=\e^{\I\phi(z,t)}A(z,t)\ ,
\end{align*}
erhalten wir die benötigten Ableitungen für die Schrödinger-Gleichung:
\begin{subequations}
  \begin{align}\hspace{-3mm}
    2m\I\del_t\Psi(z,t) &= \e^{\I\phi(z,t)}\klr{2m\I\del_t + \bar k^2-\I m\bar\Gamma}A(z,t)\ ,\\
    \del_z\Psi(z,t)     &= \e^{\I\phi(z,t)}\klr{\del_z + \I\del_z S(z)}A(z,t)\ ,\\
    \del_z^2\Psi(z,t)   &= \e^{\I\phi(z,t)}\klr{\del_z^2 + 2\I k(z)\del_z
      -\bar k^2 + 2mV(z) + \I\del_zk(z)}A(z,t)\ .
  \end{align}
\end{subequations}

Damit lautet (\ref{e5:SG})
\begin{align}\label{e5:SG2}
   \klr{\del_z^2 + 2\I k(z)\del_z + \I \del_zk(z) 
     + 2\I m\del_t + \I m\kle{\Gamma(z)-\bar\Gamma}}A(z,t) = 0\ .
\end{align}
Der einzige Unterschied zu der SG (\ref{e4:SG.WKB.1}) für reelles Potential
ist der letzte, zusätzliche Summand in der Klammer. Wir können damit,
nach Einführung der neuen Koordinaten $\zeta(z)$ und $\tau(t)$ aus 
(\ref{e4:ZetaTau}), die Schrödinger-Gleichung sofort in der Form
\begin{align}\label{e5:SG3}
  \klr{\del_\zeta+\del_\tau}B(\zeta,\tau) = (\hat LB)(\zeta,\tau)
\end{align}
schreiben, wobei mit $\hat L\subt{WKB}$ aus (\ref{e4:DefL.WKB}) folgt, dass
\begin{align}\label{e5:DefL}
    \hat L &= \hat L\subt{WKB}-\frac{m}{2\bar k}\kle{\Gamma(z)-\bar\Gamma}
            = \ms A(z) + \ms B(z)\del_\zeta + \ms C(z)\del_\zeta^2
\end{align}
ist und
\begin{subequations}
\begin{align}\label{e5:L.A}
\ms A(z) &= -\frac1{2\bar k}\Big(\del_zk(z)  + m(\Gamma(z)-\bar\Gamma)\Big)\ ,\\ \label{e5:L.B}
\ms B(z) &= \frac\I{2\bar k}\klr{\del_z\frac{\bar k}{k(z)}}\ ,\\ \label{e5:L.C}
\ms C(z) &= \frac\I{2\bar k}\klr{\frac{\bar k}{k(z)}}^2\ .
\end{align}
\end{subequations}
Vergleicht man dies direkt mit den entsprechenden Ergebnissen für reelles Potential,
Gln. (\ref{e4:LWKB.A}) bis (\ref{e4:LWKB.C}), so erkennt man, dass hier lediglich $\ms A(z)$
einen zusätzlichen Beitrag $-m(\Gamma(z)-\bar\Gamma)/(2\bar k)$ bekommt. Wir können also
die Ergebnisse aus Abschnitt \ref{s4:WKB} fast direkt übernehmen und erhalten mit dem zusätzlichen Beitrag
\begin{align}\label{e5:K.auf.vph}
\begin{split}
(\hat K\vph)(z,t) &= \ms U(z,u(z,t))\vph(\zeta(z)-\tau(t))\\
&+ \ms V(z,u(z,t))\vph'(\zeta(z)-\tau(t))\\
&+ \ms W(z,u(z,t))\vph''(\zeta(z)-\tau(t))\ ,
\end{split}
\end{align}
\begin{subequations}
\begin{align}\label{e5:K.U}
\begin{split}
\ms U(z,u(z,t)) &= -\frac{1}{2\bar k}\int_{u(z,t)}^z\d z'\ \frac{\bar k}{k(z')}\Big(\del_{z'}k(z')  + m(\Gamma(z')-\bar\Gamma)\Big)\\
&= \ln\sqrt{\frac{k(u(z,t))}{k(z)}} 
- \frac{1}{2}\int_{u(z,t)}^z\d z'\ \frac{m\Gamma(z')}{k(z')} 
+ \frac{1}{2}\bar\Gamma t\ ,
\end{split}\\ \label{e5:K.V}
\ms V(z,u(z,t)) &= \frac{\I}{4\bar k}\kle{\frac{\bar k^2}{k^2(z')}}^{z}_{u(z,t)}
\ ,\\ \label{e5:K.W}
\ms W(z,u(z,t)) &= \frac{\I}{2\bar k}\int_{u(z,t)}^z\d z'\ \klr{\frac{\bar k}{k(z')}}^3\ .
\end{align}
\end{subequations}

Bei der Berechnung von $\ms U(z,u(z,t))$ haben wir die Relation
\begin{align}
\int_{u(z,t)}^z\d z'\ \frac{\bar k}{k(z')} = \int_{\zeta-\tau}^\zeta\d \zeta'
= \tau = \frac{\bar k}mt
\end{align}
verwendet um den Beitrag $+ \bar\Gamma t/2$ zu erhalten.

\subsection{Berechnung der Entwicklung bis zur zweiten Ordnung}\label{s5:2.Ordnung}

Wir haben nun bereits alle Ergebnisse, um die Wellenfunktion $\Psi^{(1)}(z,t)$ bis zur ersten
Ordnung anzugeben. Mit
\begin{align}
\Psi^{(1)}(z,t) = \e^{\I\phi(z,t)}A^{(1)}(z,t)
\end{align}
und $\phi(z,t)$ aus Gl. (\ref{e5:PW}) folgt nämlich
\begin{align}\label{e5:1.Ordnung}
\begin{split}
A^{(1)}(z,t) &= \vph(\zeta(z)-\tau(t))\Bigg[1 + \ln\sqrt{\frac{k(u(z,t))}{k(z)}} 
- \frac{1}{2}\int_{u(z,t)}^z\d z'\ \frac{m\Gamma(z')}{k(z')} 
+ \frac{1}{2}\bar\Gamma t\Bigg]\\
&+ \vph'(\zeta(z)-\tau(t))\frac{\I}{4\bar k}\kle{\frac{\bar k^2}{k^2(z')}}^{z}_{u(z,t)}\\
&+\vph''(\zeta(z)-\tau(t))\frac{\I}{2\bar k}\int_{u(z,t)}^z\d z'\ \klr{\frac{\bar k}{k(z')}}^3
\end{split}
\end{align}

Wir wollen nun die Berechnung der zweiten Ordnung skizzieren. Es bietet sich hier an,
die Notation so kurz wie möglich zu halten, um den Überblick nicht zu verlieren.
Zunächst gehen wir wieder zu den Koordinaten $\zeta$ und $\tau$ über, dann ist
die Amplitudenfunktion in erster Ordnung gegeben durch
\begin{align}
\begin{split}
B^{(1)}(\zeta,\tau) &= \vph(\zeta-\tau)\kle{1 + \int_{\zeta-\tau}^\zeta\d\zeta_1\ \ms A(\mc Z(\zeta_1))}\\
&+ \vph'(\zeta-\tau)\int_{\zeta-\tau}^\zeta\d\zeta_1\ \ms B(\mc Z(\zeta_1))\\
&+ \vph''(\zeta-\tau)\int_{\zeta-\tau}^\zeta\d\zeta_1\ \ms C(\mc Z(\zeta_1))\ .
\end{split}
\end{align}
Diese Form folgt übrigens aus jedem WKB-artigen Phasenwinkel, für den die im $\hat L$-Operator
auftretenden Funktionen $\ms A(z)$, $\ms B(z)$ und $\ms C(z)$ nicht zeitabhängig sind. Insbesondere
gelten die folgenden Ergebnisse damit für den reellen WKB-Fall, den wir in Abschnitt \ref{s4:WKB} betrachtet haben.

Nun muss der Integraloperator $\hat K$ auf $(\hat K\vph)(\zeta,\tau) = B^{(1)}(\zeta,\tau) - \vph(\zeta-\tau)$ 
angewendet werden um
die Korrektur zweiter Ordnung, d.h. $(\hat K^2\vph)(\zeta,\tau)$ zu erhalten. Dazu berechnen wir zunächst
die Wirkung des Differentialoperators $\hat L$ auf $(\hat K\vph)(\zeta,\tau)$ und erhalten
{\small
\begin{align}\label{e5:L.auf.K}
  \begin{split}
&\phantom=\ (\hat L\hat K\vph)(\zeta,\tau)\\
    &= \vph(\zeta-\tau)\Bigg[{\ms A}(\zeta)\int_{\zeta-\tau}^\zeta\d\zeta_1\ {\ms A}(\zeta_1)
    + {\ms B}(\zeta)\kle{{\ms A}}^\zeta_{\zeta-\tau} +
{\ms C}(\zeta)\kle{{\ms A}'}^\zeta_{\zeta-\tau}\Bigg]\\
    &+\vph'(\zeta-\tau)\Bigg[
      {\ms B}(\zeta)\int_{\zeta-\tau}^\zeta\d\zeta_1\ {\ms A}(\zeta_1)
      +{\ms A}(\zeta)\int_{\zeta-\tau}^\zeta\d\zeta_1\ {\ms B}(\zeta_1) 
      + {\ms B}(\zeta)\kle{{\ms B}}^\zeta_{\zeta-\tau}\\
&\hspace{22mm}     + 2{\ms C}(\zeta)\kle{{\ms A}}^\zeta_{\zeta-\tau} 
      + {\ms C}(\zeta)\kle{{\ms B}'}^\zeta_{\zeta-\tau}
      \Bigg]\\
    &+\vph''(\zeta-\tau)\Bigg[
      {\ms C}(\zeta)\int_{\zeta-\tau}^\zeta\d\zeta_1\ {\ms A}(\zeta_1)
      + {\ms A}(\zeta)\int_{\zeta-\tau}^\zeta\d\zeta_1\ {\ms C}(\zeta_1)
      + {\ms B}(\zeta)\int_{\zeta-\tau}^\zeta\d\zeta_1\ {\ms B}(\zeta_1)\\
&\hspace{22mm}      + 2{\ms C}(\zeta)\kle{{\ms B}}^\zeta_{\zeta-\tau}
   + {\ms B}(\zeta)\kle{{\ms C}}^\zeta_{\zeta-\tau} 
      + {\ms C}(\zeta)\kle{{\ms C}'}^\zeta_{\zeta-\tau}
    \Bigg]\\
    &+\vph^{(3)}(\zeta-\tau)\Bigg[
      {\ms C}(\zeta)\int_{\zeta-\tau}^\zeta\d\zeta_1\ {\ms B}(\zeta_1)
      + {\ms B}(\zeta)\int_{\zeta-\tau}^\zeta\d\zeta_1\ {\ms C}(\zeta_1)
      + 2{\ms C}(\zeta)\kle{{\ms C}}^\zeta_{\zeta-\tau}
    \Bigg]\\
    &+\vph^{(4)}(\zeta-\tau)\Bigg[{\ms C}(\zeta)\int_{\zeta-\tau}^\zeta\d\zeta_1\ 
{\ms C}(\zeta_1)\Bigg]
  \end{split}\raisetag{1cm}
\end{align}}

Dabei sind Ausdrücke der Art $\ms A(\zeta)$ stets als $\ms A(\mc Z(\zeta))$ zu verstehen. Oft haben wir
auch Argumente aus Platzgründen ganz weggelassen. 
Um $(\hat K^2\vph)(\zeta,\tau)$ aus (\ref{e5:L.auf.K}) zu
erhalten, müssen wir zunächst die Umbenennungen $\zeta\to\zeta_2$ und $\tau\to\tau'$
durchzuführen. Hiernach muss die Substitution $\tau'\to\tau-\zeta+\zeta_2$ erfolgen 
und dann das Integral über $\zeta_2$ von $\zeta-\tau$ bis $\zeta$ über den erhaltenen Ausdruck
berechnet werden. Das Ergebnis ist dann der gewünschte Ausdruck $(\hat K^2\vph)(\zeta,\tau)$,
mit dem sich die gesamte Amplitudenfunktion in zweiter Ordnung aus Addition ergibt:
\begin{align}
B^{(2)}(\zeta,\tau) = B^{(1)}(\zeta,\tau) + (\hat K^2\vph)(\zeta,\tau)\ .
\end{align}

Führt man diese Rechnung durch, so folgt
{\small
\begin{align}\label{e5:2.Ordnung}\hspace{-10mm}
  \begin{split}
    &B^{(2)}(\zeta,\tau)
    = \vph(\zeta-\tau)\Bigg[1 + \int_{\zeta-\tau}^\zeta\d\zeta_1\ {\ms A}(\zeta_1)
      + \int_{\zeta-\tau}^\zeta\d\zeta_2\ {\ms A}(\zeta_2)\int_{\zeta-\tau}^{\zeta_2}\d\zeta_1\ 
{\ms A}(\zeta_1)\\
    &\hspace{38mm} 
      + \int_{\zeta-\tau}^\zeta\d\zeta_2\ {\ms B}(\zeta_2)\kle{{\ms A}}^{\zeta_2}_{\zeta-\tau} 
      + {\ms C}(\zeta)\kle{{\ms A}'}^\zeta_{\zeta-\tau}\Bigg]\\
    &+\vph'(\zeta-\tau)\Bigg[\int_{\zeta-\tau}^\zeta\d\zeta_1\ {\ms B}(\zeta_1)
      + \int_{\zeta-\tau}^\zeta\d\zeta_2\ {\ms B}(\zeta_2)\int_{\zeta-\tau}^{\zeta_2}\d\zeta_1\ 
{\ms A}(\zeta_1)
      + \int_{\zeta-\tau}^\zeta\d\zeta_2\ {\ms A}(\zeta_2)\int_{\zeta-\tau}^{\zeta_2}\d\zeta_1\ 
{\ms B}(\zeta_1)\\
    &\hspace{22mm}
      + \int_{\zeta-\tau}^\zeta\d\zeta_2\ {\ms B}(\zeta_2)\kle{{\ms B}}^{\zeta_2}_{\zeta-\tau}
      + 2\int_{\zeta-\tau}^\zeta\d\zeta_2\ {\ms C}(\zeta_2)\kle{{\ms A}}^{\zeta_2}_{\zeta-\tau} 
      + \int_{\zeta-\tau}^\zeta\d\zeta_2\ {\ms C}(\zeta_2)\kle{{\ms B}'}^{\zeta_2}_{\zeta-\tau}
      \Bigg]\\
    &+\vph''(\zeta-\tau)\Bigg[\int_{\zeta-\tau}^\zeta\d\zeta_1\ {\ms C}(\zeta_1)
      + \int_{\zeta-\tau}^\zeta\d\zeta_2\ {\ms C}(\zeta_2)\int_{\zeta-\tau}^{\zeta_2}\d\zeta_1\ 
{\ms A}(\zeta_1)
      + \int_{\zeta-\tau}^\zeta\d\zeta_2\ {\ms A}(\zeta_2)\int_{\zeta-\tau}^{\zeta_2}\d\zeta_1\ 
{\ms C}(\zeta_1)\\
    &\hspace{22mm}  
      + \int_{\zeta-\tau}^\zeta\d\zeta_2\ {\ms B}(\zeta_2)\int_{\zeta-\tau}^{\zeta_2}\d\zeta_1\ 
{\ms B}(\zeta_1)
      + 2\int_{\zeta-\tau}^\zeta\d\zeta_2\ {\ms C}(\zeta_2)\kle{{\ms B}}^{\zeta_2}_{\zeta-\tau}
      + \int_{\zeta-\tau}^\zeta\d\zeta_2\ {\ms B}(\zeta_2)\kle{{\ms C}}^{\zeta_2}_{\zeta-\tau}\\
    &\hspace{22mm}
      + \int_{\zeta-\tau}^\zeta\d\zeta_2\ {\ms C}(\zeta_2)\kle{{\ms C}'}^{\zeta_2}_{\zeta-\tau}
    \Bigg]\\
    &+\vph^{(3)}(\zeta-\tau)\Bigg[
      \int_{\zeta-\tau}^\zeta\d\zeta_2\ {\ms C}(\zeta_2)\int_{\zeta-\tau}^{\zeta_2}\d\zeta_1\ 
{\ms B}(\zeta_1)
      + \int_{\zeta-\tau}^\zeta\d\zeta_2\ {\ms B}(\zeta_2)\int_{\zeta-\tau}^{\zeta_2}\d\zeta_1\ 
{\ms C}(\zeta_1)\\
    &\hspace{22mm}
      + 2\int_{\zeta-\tau}^\zeta\d\zeta_2\ {\ms C}(\zeta_2)\kle{{\ms C}}^{\zeta_2}_{\zeta-\tau}
    \Bigg]\\
    &+\vph^{(4)}(\zeta-\tau)\Bigg[
      \int_{\zeta-\tau}^\zeta\d\zeta_2\ {\ms C}(\zeta_2)\int_{\zeta-\tau}^{\zeta_2}\d\zeta_1\ 
{\ms C}(\zeta_1)
    \Bigg]
  \end{split}\raisetag{1cm}
\end{align}}

Hier erkennt man auf den ersten Blick die Komplexität des Formalismus. Bereits in zweiter Ordnung
muss man zur numerischen Berechnung der Wellenfunktion an einem Ort $z$ zur Zeit $t$ einige
geschachtelte Integrationen durchzuführen. 

Was man in Gl. (\ref{e5:2.Ordnung}) leicht erkennen kann ist, dass die ersten drei Terme
in der eckigen Klammer in der ersten Zeile der Gleichung offenbar zu einem Exponentialfaktor aufsummiert werden können:
\begin{align}\label{e5:Phasenkorrektur}
\begin{split}
&\phantom=\ 1 + \int_{\zeta-\tau}^\zeta\d\zeta_1\ {\ms A}(\zeta_1)
+ \int_{\zeta-\tau}^\zeta\d\zeta_2\ {\ms A}(\zeta_2)\int_{\zeta-\tau}^{\zeta_2}\d\zeta_1\ {\ms A}(\zeta_1)\\
&= 1 + \frac{1}{1!}\int_{\zeta-\tau}^\zeta\d\zeta_1\ {\ms A}(\zeta_1)
+ \frac{1}{2!}\int_{\zeta-\tau}^\zeta\d\zeta_2\ {\ms A}(\zeta_2)\int_{\zeta-\tau}^{\zeta}\d\zeta_1\ {\ms A}(\zeta_1)\\
&\approx \exp\klg{\int_{\zeta-\tau}^\zeta\d\zeta_1\ {\ms A}(\zeta_1)}
\end{split}
\end{align}
Man beachte, dass der Formalismus im Falle einer matrixwertigen Funktion $\uop A(\zeta_1)$ sogar die Pfadordnung 
korrekt liefert. Der Ausdruck (\ref{e5:Phasenkorrektur}) stellt also eine Korrektur des ursprünglichen Phasenfaktors
dar, was die Fähigkeit des Formalismus zur Selbstkorrektur eindrucksvoll demonstriert. Um einen sehr guten Phasenfaktor
für den Formalismus zu erhalten, kann man eigentlich stets beim gewöhnlichen WKB-Phasenwinkel (\ref{e2:PW.WKB})
starten, den Operator $\hat L$ und somit $\ms A(z)$ berechnen und kann diese Prozedur mit dem um (\ref{e5:Phasenkorrektur})
korrigierten Phasenfaktor wiederholen. Nach der ersten Iteration wird $\ms A \to \ms A(z,u(z,t))$, also
eine Funktion von $z$ und von der Untergrenzenfunktion $u(z,t)$ aus Gl. (\ref{e4:Def.u}), d.h. implizit ist
$\ms A$ dann eine Funktion von $z$ und $\zeta(z) - \tau(t)$. Iteriert man erneut, so bleibt $\zeta(z)-\tau(t)$
durch die $\tau'$-Substitution im Integraloperator stets erhalten\footnote{Im
Integraloperator werden alle $\zeta'-\tau'$ zu $\zeta-\tau$. Der Teil der Funktion $\ms A$,
der von $\zeta(z)-\tau(t)$ abhängt, ist also nicht von der Integration des Operators $\hat K$
betroffen.} und man verbleibt bei einer Funktion
$\ms A(z,u(z,t)) = \ms A(z,\mc Z(\zeta(z)-\tau(t)))$ auch nach weiteren Iterationen.

Wir können die Korrektur des ursprünglichen Phasenwinkels $\phi(z,t)$ aus (\ref{e5:PW}) direkt in 
der ersten Zeile der ersten Ordnung der Entwicklung der Amplitudenfunktion, Gl. (\ref{e5:1.Ordnung}), ablesen und
erhalten
\begin{align}\label{e5:PW.neu}
\begin{split}
  \phi(z,t) &\to \phi(z,t) - \I\int_{\zeta-\tau}^\zeta\d\zeta_1\ {\ms A}(\zeta_1)\\
&= -\frac{\bar k^2}{2m}t + \int_{z_0}^z\d z'\ \sqrt{\bar k^2 - 2mV(z')} + \frac\I2\bar\Gamma t\\
&- \I\ln\sqrt{\frac{k(u(z,t))}{k(z)}} 
+ \frac{\I}{2}\int_{u(z,t)}^z\d z'\ \frac{m\Gamma(z')}{k(z')} 
- \frac{\I}{2}\bar\Gamma t
\end{split}
\end{align}
Man beachte, dass der Phasenfaktor die Exponentialfunktion des mit der imaginären Einheit $\I$ multiplizierten
Phasenwinkels ist. Wir müssen also das Integral über $\ms A$ noch mit $-\I$ multiplizieren, um 
die Korrektur des Phasenwinkels zu erhalten.

Wieder kann man in Gl. (\ref{e5:PW.neu}) die Wirkung der Selbstkorrektur des Formalismus erkennen.
Der ursprünglich angesetzte Anteil $+\frac{\I}{2}\bar\Gamma t$ im Phasenwinkel wird
durch den Korrekturterm $-\frac{\I}{2}\bar\Gamma t$ gerade aufgehoben und durch einen anderen,
offenbar geeigneteren Beitrag ersetzt, den wir z.B. in der Form
\begin{align}\label{e5:TotaleZerfallsbreite}
\frac{\I}{2}\bar\Gamma(z,u(z,t))t := \frac{\I}{2}\int_{u(z,t)}^z\d z'\ \frac{m\,\Gamma(z')}{k(z')}
\end{align}
schreiben können und der physikalisch sehr gut interpretiert werden kann. Der Teil
\begin{align}
\frac{m}{k(z')}\d z' = \d t
\end{align}
ist die infinitesimale Zeit, die das Teilchen aufgrund seiner lokalen Geschwindigkeit $v\subt{lok}(z') = k(z')/m$
am Ort $z'$ verbringt. Während dieser Zeit hat es die lokale Zerfallsrate $\Gamma(z')$, 
also ist nach der Definition von Gl. (\ref{e5:TotaleZerfallsbreite}) $\bar\Gamma(z,u(z,t))$ nichts anderes 
als die über das Intervall $[u(z,t),z]$ gemittelte Zerfallsrate. Dieses Ortsintervall entspricht
nach den Ergebnissen aus Abschnitt \ref{s4:Variablentrafo} der Strecke, die das Teilchen in der Zeit $t$
im Potential $V(z)$ zurückgelegt hat, wenn es sich jetzt (zur Zeit $t$) am Ort $z$ befindet.
Die ursprünglich angesetzte, orts- und zeitunabhängige Zerfallsrate $\bar\Gamma$ wird also effektiv durch die
zu jeder Zeit und an jedem Ort korrekte mittlere Zerfallsrate $\bar\Gamma(z,u(z,t))$ ersetzt.

Der zweite Korrekturbeitrag des Phasenwinkels,
\begin{align}\label{e5:Def.kappa}
-\I\ln \kappa(z,t) := -\I \ln\sqrt{\frac{k(u(z,t))}{k(z)}}\ ,
\end{align}
ist eigentlich eine Korrektur der Amplitude der Wellenfunktion. Sie stellt im Falle eines stabilen
Wellenpakets (d.h. wenn $\Gamma(z)\equiv 0$) die Normierung des Wellenpakets 
zu allen Zeiten sogar in der nullten Ordnung der Entwicklung nach $\hat K$ sicher. Dies wollen wir nun zeigen.

{\bf Beweis:}\quad
Zur Zeit $t=0$ ist die Untergrenzenfunktion $u(z,0) = \mc Z(\zeta(z)-\tau(0)) = z$ und der
Integraloperator $\hat K$ aus (\ref{e4:Def.K.WKB}) ist aufgrund identischer Integralgrenzen
gleich Null. Damit verbleibt nur die nullte Ordnung der Entwicklung, also die homogene Lösung
der Gl. (\ref{e5:SG3}) und somit (siehe auch Gl. (\ref{e4:AB.Psi.WKB}) in Abschnitt \ref{s4:Anfangsbedingungen})
die Wellenfunktion
\begin{align}
\Psi(z,t=0) = \e^{\I\phi(z,0)}\vph(\zeta(z))\ .
\end{align}
Der Phasenwinkel $\phi(z,t)$ wird für $t=0$ nach Gl. (\ref{e5:PW.neu}) identisch mit dem WKB-Phasenwinkel
(\ref{e2:PW.WKB}). Möchte man von einer auf Eins normierten Wellenfunktion ausgehen, so muss man
$\vph(\zeta)$ derart wählen, dass
\begin{align}\label{e5:Norm.AWP}
\int_{-\infty}^\infty\d z\ \abs{\Psi(z,t=0)}^2 = \int_{-\infty}^\infty\d z\ \abs{\vph(\zeta(z))}^2 = 1
\end{align}
erfüllt ist. 

Wählt man z.B. für $\vph(\zeta)$ ein um $\zeta(z)=0$ und somit um $z=z_0$ zentriertes Wellenpaket und 
ferner einen Anfangsort $z_0$, der so weit vor allen Feldern liegt, dass sich das Anfangswellenpaket
vollständig im potentialfreien Raum befindet, dann kann man für den Bereich, in dem $\vph^2(\zeta(z))$
signifikant von Null verschieden ist $\zeta(z) = z-z_0$ setzen ohne eine Fehler zu machen und man erhält
\begin{align}\label{e5:Norm.AWP.Speziell}
1 = \int_{-\infty}^\infty\d z\ \abs{\vph(z-z_0)}^2\ .
\end{align}

Die folgenden Betrachtungen gelten aber auch im allgemeinen Fall.
Für Zeiten $t>0$ folgt für ein stabiles Wellenpaket (d.h. $\Gamma(z)\equiv 0$) mit dem Phasenwinkel
$\phi(z,t)$ aus (\ref{e5:PW.neu}) die Norm
\begin{align}
\mc N(z) = \int_{-\infty}^\infty\d z\ \kappa^2(z,t)\abs{\vph(\zeta(z)-\tau(t))}^2
= \int_{-\infty}^\infty\d z\ \frac{k(u(z,t))}{k(z)}\abs{\vph(\zeta(z)-\tau(t))}^2\ .
\end{align}
Nun führen wir die Substitution
\begin{align}
u = u(z,t) \overref{(\ref{e4:Def.u})}= \mc Z(\zeta(z)-\tau(t))
\end{align}
durch, wobei wir die Zeit $t$ festhalten wollen. Demnach folgt
\begin{align}
\zeta(z)-\tau(t) = \zeta(u(z,t)) = \zeta(u)
\end{align}
und
\begin{align}
\d u = \ddp{u(z,t)}{z}\d z \overref{(\ref{e5:Du:Dz})}= \kappa^2(z,t)\d z = \frac{k(u(z,t))}{k(z)}\d z\ .
\end{align}
Wir haben hier ein wenig vorgegriffen und auf die Ableitung von $u(z,t)$ nach $z$, die wir in Gl. (\ref{e5:Du:Dz})
berechnen werden, verwiesen. Die Grenzen des Integrals bleiben auch bei der Integration über $u$ unverändert,
es folgt also
\begin{align}\label{e5:Norm.Stabil}
\mc N(z) = \int_{-\infty}^\infty\d u\abs{\vph(\zeta(u))}^2 \overref{(\ref{e5:Norm.AWP})}= 1\ ,
\end{align}
was nach Voraussetzung, Gl. (\ref{e5:Norm.AWP}), gleich Eins ist.\hfill{\bf q.e.d.}

\section{Berechnung der Entwicklung mit korrigiertem Phasenfaktor und Amplitude}\label{s5:1.Ord.neu}
 
Motiviert vom Ergebnis des letzten Abschnitts berechnen wir nun die Entwicklung der Wellenfunktion
mit Hilfe des neuen Phasenwinkels (\ref{e5:PW.neu}),
\begin{align}\label{e5:PWneu}
  \phi(z,t) = -\frac{\bar k^2}{2m}t + S(z) + \frac\I2\int_{u(z,t)}^{z}\d z'\ 
\frac{m\Gamma(z')}{k(z')} -\I \ln \kappa(z,t)\ .  
\end{align}
mit $\kappa(z,t)$ aus (\ref{e5:Def.kappa}), $u(z,t)$ aus (\ref{e4:Def.u}) und $S(z)$ aus
(\ref{e5:SWKB}).

Um die Lesbarkeit der Formeln zu erhöhen, werden wir ab jetzt meist auf die
Angabe des Arguments $(z,t)$ bei den Funktionen $u\equiv u(z,t)$ und
$\kappa\equiv\kappa(z,t)$ verzichten und ebenso bei deren Ableitungen.
Weiterhin führen wir die folgende abkürzende Schreibweise
\begin{align}\label{e5:DefKLE}
  \diffuz f := \kle{f(z')}^z_{u(z,t)} = f(z) - f(u(z,t))
\end{align}
für die Differenzen einer rein $z$-abhängigen Funktion an den Stellen $z$ und $u(z,t)$ ein.

Bei der Anwendung der $\hat L$-Operators werden wir diverse Ableitungen der Funktionen $u(z,t)$
und $\kappa(z,t)$ benötigen. Diese wollen wir zunächst zusammenstellen. Es gilt
\begin{align}\label{e5:Du:Dz}
  \del_z u = \dd{\mc Z(\zeta)}{\zeta}\Big\vert_{\zeta(z)-\tau(t)}\dd{\zeta(z)}z =
\frac{k(z(\zeta(z)-\tau(t)))}{k(z)}
    = \frac{k(u(z,t))}{k(z)} = \kappa^2
\end{align}
und
\begin{align}\label{e5:Du:Dt}
  \del_tu = -\dd{\mc Z(\zeta)}{\zeta}\Big\vert_{\zeta(z)-\tau(t)}\dd{\tau(t)}t = -\frac{k(u(z,t))}m\ .
\end{align}
Die Orts- und Zeitableitungen von $\kappa(z,t)$ berechnen sich zu
\begin{subequations}\label{e5:Dkappa:D-}
\begin{align}\label{e5:Dkappa:Dt}
  \del_t\kappa &= -\frac\kappa2\frac{\del_zk(z)\big\vert_{z=u}}{m}\ ,\\ \label{e5:Dkappa:Dz}
  \del_z\kappa &= -\frac\kappa2\frac{\diffuz{\del_{z'}k}}{k(z)}\ ,\\ \label{e5:Dkappa:Dz2}
  \del_z^2\kappa &= \kappa\klr{\frac{\diffuz{\del_{z'}k}}{2k(z)}}^2 +
\frac{\kappa}{2k^2(z)}\klr{\diffuz{\del_{z'}k}\del_zk(z)-\diffuz{(\del_{z'}^2k)k}} \ ,
\end{align}
\end{subequations}
dabei bedeutet nach der Definition aus Gl. (\ref{e5:DefKLE}) z.B.
\begin{align}
  \diffuz{\del_{z'}k} = \kle{\del_{z'}k(z')}^{z'=z}_{z'=u(z,t)} 
  = \del_{z'}k(z')\big\vert_{z'=z} - \del_{z'}k(z')\big\vert_{z'=u(z,t)}\ .
\end{align}

Mit dem Phasenwinkel (\ref{e5:PWneu}) und dem Ansatz (\ref{e4:AnsatzWF}),
\begin{align*}
  \Psi(z,t) = \e^{\I\phi(z,t)}A(z,t)\ ,
\end{align*}
lautet die Schrödinger-Gleichung (\ref{e5:SG}) nach Berechnung und Einsetzen aller
Ableitungen von $\Psi(z,t)$ schließlich
\begin{align}
\klr{\del_z^2 + Q(z,u(z,t))\del_z + R(z,u(z,t)) + 2\I m\del_t}A(z,t) = 0\ ,
\end{align}
wobei die hier eingeführten Hilfsfunktionen definiert sind als
\begin{align}\label{e5:DefRinSG}
Q(z,u(z,t)) &= 2\I\del_z\phi(z,t) = 2\I\klr{k(z) + \frac{\I\diffuz{\del_{z'}k+ m\Gamma}}{2k(z)}}
\end{align}
und
\begin{align}
  \begin{split}\label{e5:DefFinSG}
R(z,u(z,t)) &= \I\del_z^2\phi(z,t) - (\del_z\phi(z,t))^2 - 2m\del_t\phi(z,t) - 2mV(z) + \I m\Gamma(z)\\
     &= \frac{\Big(\del_zk(z)+\tfrac12\diffuz{\del_{z'}k+m\Gamma}\Big)\diffuz{\del_{z'}k+m\Gamma}
        -\diffuz{k((\del_{z'}^2k)+m\del_{z'}\Gamma)}}{2k^2(z)}\ .
  \end{split}
\end{align}

Wie bei der Verwendung von WKB-artigen Phasenfaktoren üblich, gehen wir auch hier zu den Koordinaten
$\zeta(z)$ und $\tau(t)$ aus (\ref{e4:ZetaTau}) über und erhalten dann wieder eine Schrödinger-Gleichung
in der Form (\ref{e4:SG.typ}),
\begin{align}\label{e5:SGneu}
  (\del_\zeta+\del_\tau)B(\zeta,\tau) = (\hat LB)(\zeta,\tau)\ ,
\end{align}
mit dem Differentialoperator
\begin{align}\label{e5:DefL2}
  \begin{split}
\hat L &= \frac{\I}{2\bar k}\kle{R(z,u(z,t)) + \klr{\klr{\del_z\frac{\bar k}{k(z)}}
        -\frac{\bar k\diffuz{\del_{z'}k+m\Gamma}}{k^2(z)}}\del_\zeta + \klr{\frac{\bar
k}{k(z)}}^2\del_\zeta^2}\\
&=: \ms A_1(z,u(z,t)) + \ms B_1(z,u(z,t))\del_\zeta + \ms C(z)\del_\zeta^2\ ,
  \end{split}
\end{align}
wobei wir in Analogie zu (\ref{e5:DefL}) wieder die Hilfsfunktionen $\ms A_1$ usw. definiert haben.
Der Index soll verdeutlichen, dass es sich um die erste mit dem Formalismus berechnete 
Korrektur der Funktion $\ms A(z)$ aus (\ref{e5:L.A}) bzw. $\ms B(z)$ aus (\ref{e5:L.B}) handelt.
Wie bereits in der Diskussion nach Gl. (\ref{e5:Phasenkorrektur}) bemerkt, hängen die Hilfsfunktionen
nach der Verwendung des korrigierten Phasenwinkels (\ref{e5:PWneu}) nicht nur von $z$, sondern auch
von $u(z,t)$ ab. Lediglich die Funktion $\ms C=\ms C(z)$ bleibt zeitunabhängig und wird dies auch in
allen weiteren Korrekturen bleiben, da sie aus der Variablentransformation $z\to \zeta(z)$ 
in der Schrödinger-Gleichung entsteht.

Der Integraloperator berechnet sich nun analog zu (\ref{e5:K.auf.vph}) aus
\begin{align}\label{e5:K.auf.vph.2}
\begin{split}
(\hat K\vph)(z,t) &= \ms U(z,u(z,t))\vph(\zeta(z)-\tau(t))\\
&+ \ms V(z,u(z,t))\vph'(\zeta(z)-\tau(t))\\
&+ \ms W(z,u(z,t))\vph''(\zeta(z)-\tau(t))\ ,
\end{split}
\end{align}
wobei hier z.B.
\begin{align}
\ms U(z,u(z,t)) &= \int_{u(z,t)}^z\d z'\ \frac{\bar k}{k(z')}
\ms A_1(z',u(z',t' = t - \tfrac{m}{\bar k}\zeta(z) - \tfrac{m}{\bar k}\zeta(z')))
\end{align}
ist. Nun gilt aber, wie in der Diskussion nach Gl. (\ref{e5:Phasenkorrektur}) bereits prophezeit,
\begin{align}\label{e5:u.invariant}
\begin{split}
u(z',t' = t - \tfrac{m}{\bar k}\zeta(z) - \tfrac{m}{\bar k}\zeta(z'))) &= 
\mc Z(\zeta(z') - \tau(t - \tfrac{m}{\bar k}\zeta(z) - \tfrac{m}{\bar k}\zeta(z'))))\\
&= \mc Z(\zeta(z') - \tfrac{\bar k}mt + \zeta(z) - \zeta(z'))))\\
&= \mc Z(\zeta(z) - \tau(t))\\
&= u(z,t)\ ,
\end{split}
\end{align}
d.h. die Abhängigkeit einer Funktion von $u(z,t)$ bleibt im Integraloperator unberührt erhalten.
Demzufolge ist
\begin{subequations}
\begin{align}\label{e5:K.U.2}
\ms U(z,u(z,t)) &= \kle{\int_{u}^z\d z'\ \frac{\bar k}{k(z')}\ms A_1(z',u)}_{u=u(z,t)}\ ,\\ \label{e5:K.V.2}
\ms V(z,u(z,t)) &= \kle{\int_{u}^z\d z'\ \frac{\bar k}{k(z')}\ms B_1(z',u)}_{u=u(z,t)}\ ,\\ \label{e5:K.W.2}
\ms W(z,u(z,t)) &= \kle{\int_{u}^z\d z'\ \frac{\bar k}{k(z')}\ms C(z')}_{u=u(z,t)}\ .
\end{align}
\end{subequations}
Mit diesen Funktionen können wir die Entwicklung der Amplitudenfunktion $A^{(1)}(z,t)$ in kurzer Form schreiben
als
\begin{align}\label{e5:1.Ord.neu}
\begin{split}
A^{(1)}(z,t) &= \vph(\zeta(z)-\tau(t)\Big[1+\ms U(z,u(z,t))\Big]\\
&+ \vph'(\zeta(z)-\tau(t))\ms V(z,u(z,t)) + \vph''(\zeta(z)-\tau(t))\ms W(z,u(z,t))
\end{split}
\end{align}

\section{Diskussion und Vergleich der beiden Entwicklungen}\label{s5:Vgl.alt.neu}

Wir vergleichen nun die Entwicklungen (\ref{e5:1.Ordnung}) und (\ref{e5:1.Ord.neu}) mit altem und neuen Phasenwinkel.

Nach (\ref{e5:L.A}) ist
\begin{align}
\ms A(z) = -\frac{\del_zk(z)}{2\bar k}+\frac{m(\Gamma(z)-\bar\Gamma)}{2\bar k}
\end{align}
und nach (\ref{e5:DefL2})
\begin{align}
\ms A_1(z,u(z,t)) = \frac\I{2\bar k}R(z,u(z,t))
\end{align}
mit $R(z,u(z,t))$ aus (\ref{e5:DefFinSG}). 
Wie in Abschnitt \ref{s4:Vergleich} wollen wir wieder einen Vergleich der Ordnung in $1/\bar k$ machen,
um zu sehen, welche der beiden Funktionen zu größeren Korrekturen der Wellenfunktion führen wird.

Man sieht leicht, dass $\ms A(\zeta) = \OO(1/\bar k)$ ist, 
denn der Anteil $\del_zk(z)/2\bar k = -mV(z)/(\bar k k(z))$ ist offensichtlich 
von der Ordnung $1/\bar k^2$ und es verbleibt dann nur noch
der zu $\Gamma(z)-\bar\Gamma$ proportionale Term $m(\Gamma(z)-\bar\Gamma)/2\bar k$.

Bei genauerer Untersuchung von $\ms A(z,u(z,t))$ unter Verwendung der Definition von $R(z,u(z,t))$ 
aus (\ref{e5:DefFinSG}) findet man dagegen heraus, dass der
der führende Term in Entwicklung nach Potenzen von $1/\bar k$
\begin{align}
\ms A_1(z,u(z,t)) \approx \frac{m\diffuz{k\del_{z'}\Gamma}}{(4\bar k k^2(z))} = \OO(1/\bar k^2)
\end{align}
lautet, d.h. $\ms A_1(z,u(z,t))$ führt zu Korrekturen, die
um eine Ordnung in $1/\bar k$ höher liegen als die durch $\ms A(z)$
verursachten Korrekturen.

Die Ordnung der Korrekturfunktionen $\ms B(\zeta)$ aus (\ref{e5:L.B}) und
$\ms B_1(z,u(z,t))$ aus (\ref{e5:DefL2}) ist gleich, 
da beide den $\del_z(\bar k/k(z))$-Term enthalten, der von der Ordnung $1/\bar k^2$
und es in $\ms B_1(z,u(z,t))$ keine Beiträge niedrigerer Ordnung gibt.

Gegenüber dem alten Phasenwinkel hat sich an dieser Stelle der Entwicklung nach Potenzen
des Integraloperators $\hat K$ also keine wesentliche Verbesserung ergeben. Auch zukünftige
Korrekturen (in weiteren Iterationsschritten) könnten dies nicht leisten, da der führende
Beitrag in $\ms B_1(z,u(z,t))$ aus der Koordinatentransformation $z\to \zeta(z)$ 
stammt und deshalb stets von der Ordnung $1/\bar k^2$ ist.

Aus demselben Grund bleibt der nach wie vor größte Korrekturbeitrag bei der Anwendung
des Formalismus zur Berechnung der exakten Wellenfunktion der Beitrag $\ms C(z)$, der
vor der zweiten Ableitung nach $\zeta$ steht und der aus der 
Vernachlässigung der Dispersion des Wellenpakets stammt. Er
ist eine Korrektur erster Ordnung in $\OO(1/\bar k)$, siehe Gl. (\ref{e4:3.Korr:1.Ord.WKB}), und  wurde
in Abschnitt \ref{s4:Vergleich} ausführlich diskutiert. 
Da die Korrektur des Phasenwinkels (bzw. der Amplitude über Logarithmus-Beiträge zum Phasenwinkel) offenbar
keine Auswirkungen auf diesen Beitrag der Entwicklung in $\hat K$ hat, ist die Fähigkeit
der Entwicklungsmethode zur formalen Korrektur des WKB-Phasenwinkels eigentlich auf den ersten Iterationsschritt
begrenzt. Man kann zwar auch höhere Ordnungen von $1/\bar k$ in den Phasenwinkel mit aufnehmen,
doch es stellt sich dann stets die Frage, ob die damit erzielte Verbesserung der Genauigkeit des Ansatzes
den wachsenden numerischen Rechenaufwand wert ist, den die Verwendung eines komplizierteren Phasenwinkels
mit sich bringt.

Zur Minimierung des zu $\ms C(z)$ proportionalen Beitrags in der
Entwicklung muss man wieder das in Abschnitt \ref{s4:Dispersion} erläuterte Verfahren anwenden, nachdem
man sich für einen geeigneten Phasenfaktor entschieden hat. Wählt man die für dieses Verfahren
benötigten Wellenpakete breit genug, so kann auch eine Vernachlässigung des Entwicklungsbeitrags,
der proportional zur ersten Ableitung nach $\zeta$ ist, in Betracht gezogen werden.

Nachdem wir nun einen geeigneten Phasenwinkel zur Beschreibung eines zerfallenden Wellenpakets gefunden
haben, wollen wir im nächsten Kapitel den Übergang zu allgemeinen, matrixwertigen Potentialen vollziehen.
Wir werden sehen, dass, obwohl dieser Schritt mit den bisher erzielten Resultaten vergleichsweise einfach ist, 
der Grad der Komplexität der physikalischen Beschreibung und der mathematische Aufwand enorm steigen wird.

\chapter{Atome mit mehreren inneren Zuständen in
  matrixwertigen Potentialen}\label{s6:Formalismus}

\section{Aufstellen der Schrödinger-Gleichung}\label{s6:SG}

Bisher haben wir eindimensionale Wellenpakete betrachtet, die nicht speziell auf die Beschreibung von
Atomen ausgelegt waren. Wie bereits im vorgreifenden Abschnitt \ref{s3:FahrplanBasics}
erläutert, kann das Atom als Superposition mehrerer, innerer Zustände beschrieben werden,
die sich durch gewisse Quantenzahlen unterscheiden. Jeder Zustand bekommt dann ein eigenes Wellenpaket,
siehe Gl. (\ref{e3:Initial}). Um in diesem Kapitel möglichst schnell zu Ergebnissen zu kommen
haben wir die Grundlagen zur Beschreibung wasserstoffartiger Atome in den Anhang \ref{sA:QMBasics}
verlegt. Wir wollen hier nur einen kurzen Überblick über diese Grundlagen geben.

Wasserstoffartige Atome lassen sich als quantenmechanisches Zweiteilchen-Problem behandeln, 
siehe Anhang \ref{sA:ZweiTeilchen}.
Die aufgrund der Kopplung der Drehimpulse von Elektron und Atomkern zu unterscheidenden
inneren Zustände des Atoms führen bei vorhandenen externen Einflüssen 
zu einem matrixwertigen Potential, siehe Gl. (\ref{eA:2PPotential}). 
Bei der Darstellung dieser Matrix unterscheiden wir streng
zwischen zwei Möglichkeiten. Eine mögliche Wahl der Basiszustände, die sich insbesondere für die
numerische Behandlung anbietet, sind die Zustände mit Gesamtdrehimpuls $\v F$, der sich aus
den gekoppelten Drehimpulsen des Elektrons (Spin $\v S$ und Bahndrehimpuls $\v L$ zu $\v J$) und dem
Kernspin $\v I$ zusammensetzt, siehe Gl. (\ref{eA:Basis}). Eine weitere mögliche Basis sind die lokalen Eigenzustände
der Potentialmatrix $\uop M(Z)$, die wir mit $\rket{\alpha(Z)}$ bezeichnen wollen.
Für diese gilt
\begin{align}\label{e6:EWG}
\uop M(Z)\rket{\alpha(Z)} &= E_\alpha(Z)\rket{\alpha(Z)}\ ,\qquad(\alpha=1,\ldots,N)\ ,
\end{align}
wobei $N$ die Dimension des durch die inneren atomaren Zustände aufgespannten Hilbert-Raums sei.
Wie in Anhang
\ref{sA:WWF} ausführlich diskutiert, kann man im Fall metastabiler Atome mit dem Wigner-Weisskopf-Formalismus \cite{WeWi30}
die Potentialmatrix im Unterraum der atomaren Zustände mit Hauptquantenzahl $n=2$ in der Form
\begin{align}\label{e6:MM}
\uop M(Z) = \unl V(Z) - \tfrac\I2\unl \Gamma(Z)
\end{align}
schreiben und spricht dann von einer sogenannten nichthermiteschen Massen- oder Energiematrix\footnote{
Der Begriff Massenmatrix wurde aus der Anwendung des Wigner-Weisskopf-Formalismus auf den Zerfall 
der K-Mesonen übernommen. Der Wigner-Weisskopf-Formalismus, den wir in Anhang \ref{sA:WWF} präsentieren,
richtet sich nach der sehr empfehlenswerten Darstellung aus \cite{Nac91}, Anhang I. Dort wird auch
der Zusammenhang mit dem Zerfall der K-Mesonen behandelt.}.
Die Eigenwerte $E_\alpha(Z)$ aus Gl. (\ref{e6:EWG}) werden dann komplex und man muss zwischen rechten
und linken Eigenzuständen unterscheiden. Für die linken Eigenzustände, die wir mit einer zusätzlichen Tilde
kennzeichnen wollen, gilt die Eigenwertgleichung
\begin{align}\label{e6:lEWG}
\lrbra{\alpha(Z)}\uop M(Z) &= E_\alpha(Z)\lrbra{\alpha(Z)}\ ,\qquad(\alpha=1,\ldots,N)\ .
\end{align}
Für die linken und rechten Eigenvektoren gilt die Orthonormalitätsrelation (\ref{eA:lrOrthonorm}),
die rechten Eigenvektoren können außerdem zusätzlich normiert werden. 
Die Eigenenergien kann man in Real- und Imaginärteil aufspalten, d.h.
\begin{align}\label{e6:cEW}
E_\alpha(Z) = \Real E_\alpha(Z) + \I \Imag E_\alpha(Z) =: V_\alpha(Z)-\tfrac\I2\Gamma_\alpha(Z)\ .
\end{align}
Die (lokalen) totalen Zerfallsbreiten $\Gamma_\alpha(Z)$ müssen analog zum eindimensionalen
Fall aus Gl. (\ref{e5:komplexe.Energie}) stets positiv sein, ansonsten würde die Norm des Eigenzustands
exponentiell anwachsen. Für weitere Informationen zur Beschreibung des Zerfalls eines metastabilen Atoms verweisen
wir auf Anhang \ref{sA:WWF}. Wir wollen betonen, dass der im Folgenden entwickelte Formalismus ohne weiteres
auch auf den Fall stabiler Atome angewendet werden kann, wo $\uop M$ eine hermitesche Matrix ist und
$\Gamma_\alpha(Z)\equiv 0$ für alle Indizes $\alpha$ gilt.

Wie am Ende von Anhang \ref{sA:ZweiTeilchen} diskutiert, lautet bei Verwendung der Massenmatrix $\uop M(Z)$
die Schrödinger-Gleichung (\ref{eA:MatrixSG})
\begin{align}\label{e6:SG1}
  \Big(\del_Z^2 - 2M\uop M(Z)+2M\I\del_t\Big)\ket{\Psi(Z,t)} = 0\ ,
\end{align}
wobei $M$ die Gesamtmasse des Atoms sei. Nun denken wir uns das lokalen Eigenwertproblem
(\ref{e6:EWG}) als gelöst und machen den bereits in (\ref{e3:Initial}) betrachteten,
lokalen Ansatz
\begin{align}\label{e6:Psi}
  \ket{\Psi(Z,t)} = \sum_\alpha\Psi_\alpha(Z,t)\rket{\alpha(Z)}
\end{align}
und berechnen damit die Ableitungen, die in der Gl. (\ref{e6:SG1}) vorkommen. Es gilt
\begin{align}
    2M\I\del_t\ket{\Psi(Z,t)} &= 2M\I\sum_\alpha(\del_t\Psi_\alpha(Z,t))\rket{\alpha(Z)}\ ,\\
    \begin{split}
      \del_Z^2\ket{\Psi(Z,t)} &=
      \sum_\alpha \Big[(\del_Z^2\Psi_\alpha(Z,t))\rket{\alpha(Z)} +
      \Psi_\alpha(Z,t)(\del_Z^2\rket{\alpha(Z)})\\ 
      &\hspace{46mm}+ 2(\del_Z\Psi_\alpha(Z,t))(\del_Z\rket{\alpha(Z)})\Big]\ ,
  \end{split}
\end{align}
und somit folgt nach Einsetzen in (\ref{e6:SG1})
\begin{align}
  \begin{split}
    &\ \sum_\alpha \kle{(\del_Z^2 - 2M E_\alpha(Z) 
      + 2M\I\del_t)\Psi_\alpha(Z,t)}\rket{\alpha(Z)}\\
   =&\ \sum_\alpha \Big[- 2(\del_Z\Psi_\alpha(Z,t))[\del_Z\rket{\alpha(Z)}]
      - \Psi_\alpha(Z,t)[\del_Z^2\rket{\alpha(Z)}] \Big]\ .
\end{split}
\end{align}
Wir projizieren nun von links mit einem linken Eigenvektor $\lrbra{\beta(Z)}$ und erhalten daraus
unter Verwendung der Orthonormalität (\ref{eA:lrOrthonorm}) der linken und rechten Eigenvektoren
\begin{align}\label{e6:SG2}
  \begin{split}
    &\ (\del_Z^2 - 2M E_\beta(Z) + 2M\I\del_t)\Psi_\beta(Z,t)\\
    =&\ \sum_\alpha \Big[- 2(\del_Z\Psi_\alpha(Z,t))
        \big(\lrbra{\beta(Z)}\del_Z\rket{\alpha(Z)}\big)
      - \Psi_\alpha(Z,t)\big(\lrbra{\beta(Z)}\del_Z^2\rket{\alpha(Z)}\big) \Big]\ .
\end{split}
\end{align}
Der nächste Schritt wäre die Wahl eines speziellen WKB-ähnlichen Phasenwinkels $\phi_\alpha(Z,t)$
und das Einsetzen des Ansatzes
\begin{align}\label{e6:Ansatz}
\Psi_\alpha(Z,t) = \e^{\I\phi_\alpha(Z,t)}A_\alpha(Z,t)\ .
\end{align}
Bevor wir $\phi_\alpha(Z,t)$ angeben können, müssen wir einen genaueren Blick auf die Gesamtenergie
der einzelnen Wellenpakete werfen. Betrachten wir dazu einmal den adiabatischen Grenzfall, d.h.
die Potentiale seinen derart gewählt, dass Mischungen der Zustände aufgrund der Ortsableitungen
auf der rechten Seite von (\ref{e6:SG2}) vernachlässigt werden können.
In diesem Fall ist die rechte Seite von (\ref{e6:SG2}) ungefähr Null und wir
haben es mit unabhängigen Gleichungen für jede Komponente $\beta$ zu tun, die
wir jeweils genauso behandeln können, wie die eindimensionale Schrödinger-Gleichung
in den vergangenen Kapiteln.

Die Schwierigkeit hier ist allerdings die Wahl der (mittleren) Gesamtenergie der Wellenpakete.
Wählen wir die gleiche Gesamtenergie für alle Wellenpakete, so setzt diese sich
im feldfreien Raum nach der Diskussion in Anhang (\ref{sA:ZweiTeilchen}) aus der kinetischen und
der inneren Energie zusammen. Zustände mit unterschiedlicher innerer Energie (z.B. im Termschema
des Atoms) hätten dann unterschiedliche Geschwindigkeiten im feldfreien Raum und die korrespondierenden
Wellenpakete in der WKB-Näherung würden auseinanderdriften. 
Durch Einführung unterschiedlicher Gesamtenergien kann dies verhindert werden und dies ist auch der
Ansatz, den wir hier machen wollen. 

Letztendlich hängt es von der experimentellen Präparation des Anfangszustands ab. Startet der
Experimentator mit einem Atom, dass mit nur einem einzigen inneren Zustand beschrieben werden kann,
so kann man durch nichtadiabatische Potentiale Superpositionen von inneren Zuständen erzeugen.
Die zugehörigen Wellenpakete hätten in diesem Fall alle die gleiche Gesamtenergie und somit unterschiedliche
Schwerpunktsgeschwindigkeiten.

Startet der Experimentator mit einem Atom im Grundzustand, generiert die angeregten Zustände dann durch
unterschiedliche Energiezufuhr und präpariert dann den gewünschten Anfangszustand (z.B. mit einer
Stern-Gerlach-Methode), so sollten die
Wellenpakete der einzelnen Zustände unterschiedliche Gesamtenergien haben, aber gleiche kinetische Energien, d.h.
\begin{align}\label{e6:gesEnergie}
E\supt{ges}_\beta(Z) = \frac{\bar k^2}{2M} + V_\beta(Z)\ .
\end{align}
Wir wollen diesen Fall hier untersuchen. Der Fall identischer Gesamtenergien, der
immer dann relevant wird, wenn der Anfangszustand über ein statisches Potential aus einem
einzelnen inneren atomaren Zustand präpariert wird, kann aus Zeitgründen in 
der vorliegenden Arbeit nicht mehr behandelt werden, sollte aber mit den hier aufgestellten Formeln
relativ leicht zugänglich sein. Für alle Experimente, an die wir (nicht nur im Rahmen
dieser Arbeit)  denken, wird ohnehin ein Anfangszustand von Wasserstoff oder Deuterium
aus metastabilen Zuständen mit gleichem Gesamtdrehimpuls ausreichend sein. Da diese Zustände
im feldfreien Fall entartet sind, sind beide Betrachtungsweisen in diesem Fall äquivalent.

Gehen wir also davon aus, dass sich im Anfangszustand des Atoms, der am Ort $Z_0$ im feldfreien Raum
vorliegen soll, nur Wellenpakete mit Gesamtenergien
\begin{align}\label{e6:gesEnergie.2}
E\supt{ges}_\alpha := \frac{\bar k^2}{2M} + V_\alpha(Z_0)\ .
\end{align}
befinden. Dann ist die lokale Wellenzahl am Ort $Z$ für den Zustand mit Index $\alpha$ gegeben durch
\begin{align}\label{e6:Defk}
k_\alpha(Z) = \sqrt{2M E\supt{ges}_\alpha - 2M V_\alpha(Z)} = \sqrt{\bar k^2 - 2M\Delta V_\alpha(Z)}
\end{align}
mit
\begin{align}\label{e6:DefDV}
\Delta V_\alpha(Z) &:= V_\alpha(Z) - V_\alpha(Z_0)\ .
\end{align}
Für den zeitabhängigen Anteil des Phasenfaktors definieren wir
\begin{align}\label{e6:Defkbar}
\bar k_\alpha &:= \sqrt{2M E\supt{ges}_\alpha} = \sqrt{\bar k^2 + 2M\,V_\alpha(Z_0)}
\end{align}

Mit diesen Definitionen ist im adiabatischen Grenzfall die ebene Welle
\begin{align}\label{e6:ebeneWellen}
  \psi_\beta(\bar k,Z,t) = \exp\klg{-\I \frac{\bar k_\beta^2}{2M}t+\I\int_{Z_0}^Z\d Z'\ k_\beta(Z')}
\end{align}
eine Lösung mit Wellenzahl $\bar k$ der Schrödinger-Gleichung (\ref{e6:SG2}) 
im Sinne der WKB-Näherung. Aus Überlagerung von ebenen Wellen dieses Typs mit Wellenzahlen $k$
würden wir in Analogie zu Abschnitt \ref{s2:WKB} Wellenpakete der Form
\begin{align}\label{e6:Lsg.WKB}
  \Psi_\beta(Z,t) \approx \exp\klg{-\I\frac{\bar k^2_\beta}{2M}t + \I\int_{Z_0}^Z\d Z'\ k_\beta(Z')}\vph(\zeta_\beta(Z)-\tau(t))
\end{align}
erhalten, wobei die Koordinaten $\zeta_\beta(Z)$ und $\tau(t)$ definiert sind als
\begin{align}
  \zeta_\beta(Z) = \int_{Z_0}^Z\d Z'\ \frac{\bar k}{k_\beta(Z')}\ ,\qquad \tau(t) = \frac{\bar k}{M}t\ ,
\end{align}
mit $k_\beta(Z)$ aus Gl. (\ref{e6:Defk}). 

Durch die Verwendung der eben definierten Größen ist, wie oben bereits behauptet,
sichergestellt, dass im feldfreien Bereich alle Komponenten-Wellenpakete eine identische, mittlere
Geschwindigkeit $\bar k/M$ haben. Es gilt ja im feldfreien Raum nach Definition
\begin{align}
  V_\beta(Z) \to V_\beta(Z_0)\quad\Rightarrow\quad \Delta V_\beta(Z) \to 0
  \quad\Rightarrow\quad k_\beta(Z) \to \bar k\ .
\end{align}
Die Schwerpunktsgeschwindigkeit des Wellenpakets mit Index $\beta$
berechnet sich - wie am Ende von Abschnitt \ref{s2:WKB-Eigenschaften} bereits diskutiert - aus
\begin{align}\label{e6:SPvelo}
    0 = \dd{}{t}\klr{\zeta_\beta(\hat Z_\beta(t))-\tau(t)}
    \qquad\Longrightarrow\qquad   \dd{\hat Z_\beta(t)}t&= \frac{k_\beta(\hat Z_\beta(t))}M
\end{align}
und somit folgt wegen $k_\beta(Z)\to \bar k$ im feldfreien Raum die Behauptung. 
Aus (\ref{e6:SPvelo}) folgt mit (\ref{e6:Defk}) für das betrachtete Komponenten-Wellenpaket weiterhin
eine Energieerhaltungsgleichung
\begin{align}
\frac12M\klr{\dd{\hat Z_\beta(t)}{t}}^2 + V_\beta(Z) = \frac{\bar k_\beta^2}{2M} = E_\beta\supt{ges}\ . 
\end{align}
Für die potentiellen Energien $V_\beta(Z)$,
die ja der Realteil des komplexen Eigenwerts $E_\beta(Z)$ der nichthermiteschen Massenmatrix
$\uop M$ ist, kann man durch die relative Definition (\ref{e6:Defk}) der lokalen Wellenzahl 
den Nullpunkt frei wählen, ohne dass sich die Schwerpunktsgeschwindigkeiten
ändern.

Kommen wir nun zurück zur Schrödinger-Gleichung (\ref{e6:SG2}), die wir ja in eine zu (\ref{e4:SG.typ})
analoge Form bringen müssen, um den Formalismus zur Berechnung der exakten Lösung auf den
Fall matrixwertiger Potential übertragen zu können.

In Kapitel \ref{s5:Zerfall} haben wir bereits in Gl. (\ref{e5:PWneu})
einen guten Phasenwinkel für den Ansatz (\ref{e6:Ansatz})
für die Wellenfunktion berechnet. Wir müssen ihn nur noch auf den hier betrachteten Fall 
übertragen, was aber nach den eben gemachten Vorbetrachtungen kein Problem mehr darstellt.
Ohnehin wäre die Wahl eines ungeeigneten Phasenwinkels nach den Ergebnissen aus Kapitel \ref{s5:Zerfall}
nicht allzu tragisch, da wir dann leicht aus den Korrekturen erster Ordnung auf einen geeigneteren Phasenwinkel
schließen könnten.
Wir wählen also den aus (\ref{e5:PWneu}) und der obigen Diskussion motivierten
Phasenwinkel
\begin{align}\label{e6:PW}
  \begin{split}
    \phi_\alpha(Z,t) &= -\frac{\bar k_\alpha^2}{2M}t + S_\alpha(Z)
       + \I\int_{u_\alpha(Z,t)}^Z\d Z'\ \lrmatelem{\alpha(Z')}{\del_{Z'}}{\alpha(Z')}\\
      &+ \frac\I2\int_{u_\alpha(Z,t)}^Z\d Z'\ \frac{M\Gamma_\alpha(Z')}{k_\alpha(Z')}
       - \I\,\ln\kappa_\alpha(Z,t)
  \end{split}
\end{align}
mit
\begin{align}
  \label{e6:DefF2amma}
  \Gamma_\alpha(Z)   &:= -2\Imag(E_\alpha(Z))\ ,\\
  \label{e6:DefSWKB}
  S_\alpha(Z)        &:= \int_{Z_0}^Z\d Z'\ k_\alpha(Z')\ ,\\ 
  \label{e6:DefKappa}
  \kappa_\alpha(Z,t) &:= \sqrt{\frac{k_\alpha(u_\alpha(Z,t))}{k_\alpha(Z)}}
\end{align}
und
\begin{align}\label{e6:DefU}
  u_\alpha(Z,t) := \mc Z_\alpha(\zeta_\alpha(Z)-\tfrac{\bar k}M t)\ ,
\end{align}
wobei $\mc Z_\alpha(\zeta)$ die Umkehrfunktion zu $\zeta_\alpha(Z)$ ist. Der im Vergleich zu (\ref{e5:PWneu}) 
neu hinzugekommene Anteil
\begin{align}\label{e6:PhiGeom}
  \phi_\alpha\supt{geom}(Z,t) = \I\int_{u_\alpha(Z,t)}^Z\d Z'\ \lrmatelem{\alpha(Z')}{\del_{Z'}}{\alpha(Z')}
\end{align}
ist motiviert durch das diagonale Matrixelement der Ortsableitung auf der rechten Seite
von Gl. (\ref{e6:SG2}). Die physikalische Interpretation dieses Anteils ist der einer geometrischen Phase,
die der Zustand $\rket{\alpha(Z)}$ auf seinem Weg durch das Potential aufsammelt. Wir werden uns
in Kapitel \ref{s7:Berry} intensiv mit diesem Anteil des Phasenwinkels beschäftigen, 
wenn wir P-verletzende geometrischen Phasen untersuchen.

Durch eine Rechnung, die analog zu der in Abschnitt \ref{s5:1.Ord.neu} verläuft,
erhält man aus (\ref{e6:SG2}) die Schrödinger-Gleichung
\begin{align}\hspace{-3mm}\label{e6:SGtmp}
  \begin{split}
    &\e^{\I\phi_\beta(Z,t)}\kle{2\I\bar k(\del_{\zeta_\beta}+\del_\tau) 
      - 2\I\bar k\hat L_\beta} B_\beta(\zeta_\beta,\tau)\\
    = \sum_{\alpha\neq\beta} &\e^{\I\phi_\alpha(Z,t)}\,\Big[-
      2(\I(\del_Z\phi_\alpha(Z,t))A_\alpha(Z,t)
      +\del_Z A_\alpha(Z,t))\lrbra{\beta(Z)}\del_Z\rket{\alpha(Z)}\\
      &\hspace*{15mm}-A_\alpha(Z,t)\,\lrbra{\beta(Z)}\del_Z^2\rket{\alpha(Z)} \Big]\ ,
  \end{split}
\end{align}
mit einem zu (\ref{e5:DefL2}) ähnlichen Operator
\begin{align}\label{e6:DefL}
  \begin{split}
\hat L_\beta &= \frac{\I}{2\bar k}\Bigg[R_\beta(Z,u_\beta(Z,t)) + \klr{\frac{\bar k}{k_\beta(Z)}}^2\del_{\zeta_\beta}^2\\ 
      &+ \klr{\klr{\del_Z\frac{\bar k}{k_\beta(Z)}}
        - \frac{\bar k\diffuza[Z]{\del_{Z'}k_\beta+M\Gamma_\beta}{u_\beta}
- 2\uop D^{(1)}_{\beta\beta}(Z')k_\beta(Z')\big\vert_{Z'=u_\beta(Z,t)}}{k_\beta^2(Z)}}\del_{\zeta_\beta}\Bigg]\ .
  \end{split}
\end{align}
Hier haben wir die wie folgt definierten Matrixdarstellungen $\uop D^{(1,2)}(Z)$ der ersten und zweiten Ortsableitung
verwendet:
\begin{align}\label{e6:D1Matrix}
  \uop D^{(1)}(Z) &:= \klr{\lrbra{\beta(Z)}{\ddp{}Z}\rket{\alpha(Z)}}\ ,\\ \label{e6:D2Matrix}
  \uop D^{(2)}(Z) &:= \klr{\lrbra{\beta(Z)}{\ddp{^2}{Z^2}}\rket{\alpha(Z)}}\ .
\end{align}
Desweiteren tritt in $\hat L_\beta$ noch die Funktion
\begin{align}\label{e6:DefF}
  \begin{split}
R_\beta(Z,u_\beta(Z,t)) &= \uop D^{(2)}_{\beta\beta}(Z) 
      - \frac{\uop D^{(1)}_{\beta\beta}(Z)\diffuza[Z]{\ms R_\beta}{u_\beta}}{k_\beta(Z)}\\
      &+ \frac{1}{2k_\beta^2(Z)}\Bigg[
              \Big(\del_Zk_\beta(Z)+\tfrac12\diffuza[Z]{\ms R_\beta}{u_\beta}\Big)\diffuza[Z]{\ms R_\beta}{u_\beta} 
            - \diffuza[Z]{k_\beta(\del_{Z'}\ms R_\beta)}{u_\beta}
         \Bigg]
  \end{split}
\end{align}
mit
\begin{align}\label{e6:DefF2}
  \ms R_\beta(Z) := \del_Zk_\beta(Z) + M\Gamma_\beta(Z) + 2\uop D^{(1)}_{\beta\beta}(Z)k_\beta(Z)
\end{align}
auf. Beim Vergleich von $R_\beta(Z,u_\beta(Z,t))$ mit seinem skalaren Analogon aus Gl. (\ref{e5:DefFinSG}) erkennt man,
das hier neben den Beiträgen aus (\ref{e5:DefFinSG}) weitere, zu den Diagonalelementen $\uop D^{(1,2)}_{\beta\beta}(Z)$
proportionale Terme auftreten, die erst durch die Verwendung eines matrixwertigen Potentials entstanden sind.
Das gleiche gilt für den Vergleich der Koeffizienten von $\del_\zeta$ aus (\ref{e5:DefL2}) und
von $\del_{\zeta_\beta}$ aus (\ref{e6:DefL}).

Nach Übergang auf der rechten Seite von (\ref{e6:SGtmp}) zu $B_\alpha(\zeta_\alpha(Z),\tau(t)) = A_\alpha(Z,t)$ und nach
Umformung folgt schließlich
\begin{align}\label{e6:MatrixSG}
  (\del_{\zeta_\beta}+\del_\tau)B_\beta(\zeta_\beta,\tau)  = 
    \sum_\alpha\uell_{\beta\alpha}B_\alpha(\zeta_\alpha,\tau)
\end{align}
mit einem Matrixoperator $\uell$, der die Komponenten
\begin{align}\label{e6:MatrixL}
\begin{split}
\uell_{\beta\alpha} &= \hat L_\beta\delta_{\beta\alpha}
    + (1-\delta_{\beta\alpha})\e^{\I\klr{\phi_\alpha(Z,t)-\phi_\beta(Z,t)}}\\
    &\times \frac\I{2\bar k}\Bigg[2\uop D^{(1)}_{\beta\alpha}(Z)\klr{\I(\del_Z\phi_\alpha(Z,t))
         + \frac{\bar k}{k_\alpha(Z)}\del_{\zeta_\alpha}}
    +\uop D^{(2)}_{\beta\alpha}(Z)\Bigg]
\end{split}
\end{align}
hat.

Wir sehen also, dass sich auch im Falle eines matrixwertigen Potentials eine Schrödinger-Gleichung
vom Typ (\ref{e4:SG.typ}) aufstellen lässt. Wie zu erwarten, müssen wir nun einen matrixwertigen Differentialoperator
$\uell$ betrachten, der zu Wechselwirkungen zwischen den einzelnen Komponenten-Wellenfunktionen führen wird.
Die einzige, wesentliche Neuerung, die sich hier gegenüber dem eindimensionalen Fall ergab, war die Einführung
des geometrischen Anteils (\ref{e6:PhiGeom}) im Phasenwinkel (\ref{e6:PW}).

Durch die klare Aufspaltung des Operators $\uell$ in einen diagonalen und einen nichtdiagonalen Anteil
kann man leicht den adiabatischen Grenzfall studieren, indem man den nichtdiagonalen Anteil vernachlässigt.
In diesem Fall gilt für jede Komponenten-Wellenfunktion $B_\beta(\zeta_\beta,\tau)$ eine eindimensionale
Schrödinger-Gleichung vom Typ (\ref{e4:SG.typ}) mit dem Operator $\hat L_\beta$ aus (\ref{e6:DefL}),
die mit den bisher entwickelten Methoden gelöst werden kann.

\section{Lösen der Schrödinger-Gleichung}\label{s6:Solution}

Der einzige Unterschied der linken Seite von (\ref{e6:MatrixSG}) zur linken Seite
der Schrödinger-Gleichung in der Form (\ref{e4:SG.typ}) ist der zusätzliche Index $\beta$.
Wir können also leicht die Greensche Funktion aus (\ref{e4:GreenFkt}) auf den hier vorliegenden
Fall übertragen und erhalten
\begin{align}\label{e6:GreenFkt}
  G_\beta(\zeta_\beta,\tau) := \Theta(\zeta_\beta+\tau)\delta(\zeta_\beta-\tau)\ .
\end{align}
Damit können wir (\ref{e6:MatrixSG}) in eine Integralgleichung umwandeln,
\begin{align}
  \begin{split}
    B_\beta(\zeta_\beta,\tau) &= \tilde B_{0,\beta}(\zeta_\beta-\tau)\\
    &+\sum_\alpha
    \int_{-\infty}^{\infty}\d\zeta_\beta'\int_{-\infty}^{\infty}\d\tau' \ 
    G_\beta(\zeta_\beta-\zeta'_\beta,\tau-\tau')
    (\uell_{\beta\alpha}B_\alpha)
    (\zeta_\alpha(\mc Z_\beta(\zeta_\beta')),\tau')\ .
  \end{split}
\end{align}
Hier erkennt man bereits ein wesentliches Merkmal des Matrixformalismus.
Um die Integration über $\zeta_\beta'$
einer Funktion durchzuführen, die von $\zeta_\alpha$
abhängt, muss zunächst die implizite Abhängigkeit dieser
Größen voneinander berücksichtigt werden. Alle $\zeta_\alpha$ hängen von {\em
  einer} Ortskoordinate $Z$ ab. Dieses $Z$ kann aus jeder beliebigen Koordinate
$\zeta_\beta'$ über die entsprechende Umkehrfunktion 
$\mc Z_\beta(\zeta)$ berechnet werden. 
Der Zusammenhang zwischen $\zeta_\beta'$ und $\zeta_\alpha$
ist also, wie im Integranden zu erkennen, über den Ausdruck 
$\zeta_\alpha(\mc Z_\beta(\zeta_\beta'))$ gegeben.

Mit den Methoden aus Kapitel \ref{s4:Formalismus} erhält man also schließlich
unter Verwendung der Green-Funktion aus (\ref{e6:GreenFkt}):
\begin{align}\label{e6:OperatorK}
  \begin{split}
    B_\beta(\zeta_\beta,\tau) &= \vph_\beta(\zeta_\beta-\tau)+
    \sum_\alpha\int_{\zeta_\beta-\tau}^{\zeta_\beta}\d\zeta'_\beta\ 
    \kle{(\uell_{\beta\alpha}B_\alpha)(\zeta_\alpha(\mc Z_\beta(\zeta_\beta')),
      \tau')}_{\tau'=\tau-\zeta_\beta+\zeta_\beta'}\\
    &=: \vph_\beta(\zeta_\beta-\tau) +
    \sum_\alpha(\uk_{\beta\alpha}B_\alpha)(\zeta_\alpha(\mc
    Z_\beta(\zeta_\beta)),\tau)\ .
  \end{split}
\end{align}
Hierbei haben wir den neuen, matrixwertigen Integraloperator $\uk$
definiert. Durch iteratives Einsetzen erhalten wir also
\begin{align}\label{e6:AmplitudeB}
B_\beta(\zeta_\beta,\tau) = \sum_{n,\alpha}\klr{\klr{\hat{\umc K}^n}_{\beta\alpha}\vph_\alpha}
  (\zeta_\alpha(\mc Z_\beta(\zeta_\beta)),\tau)  \qquad(\beta=1,\ldots,N)\ ,
\end{align}
bzw. nach Übergang zur Amplitudenfunktion $A_\beta(Z,t)$ in den alten Koordinaten
\begin{align}\label{e6:AmplitudeA}
A_\beta(Z,t) = \sum_{n,\alpha}\klr{\klr{\hat{\umc K}^n}_{\beta\alpha}\vph_\alpha}(Z,t)  \qquad(\beta=1,\ldots,N)\ .
\end{align}
Um die Verwendung von Umkehrfunktionen $\mc Z_\beta(\zeta)$ weitestgehend zu vermeiden,
ist es günstig, wieder zur eigentlichen Ortskoordinate $Z$ zurückzukehren. 
Der Integraloperator $\uk_{\beta\alpha}$ hat dann die Gestalt
\begin{align}\label{e6:DefK}
(\uk_{\beta\alpha}\vph_\alpha)(Z,t) = \int_{u_\beta(Z,t)}^Z\d Z'
\ \frac{\bar k}{k_\beta(Z')}(\uell_{\beta\alpha}\vph_\alpha)(Z',t')
\big\vert_{t' = t - M(\zeta_\beta(Z) - \zeta_\beta(Z'))/\bar k}
\end{align}

Wir wollen nun die nullte und die erste Ordnung formal berechnen, um einen
Eindruck von der Gestalt der Lösung zu bekommen. Hierzu schreiben wir den 
Operator $\uell$ aus (\ref{e6:MatrixL}) abstrakt in der Form
\begin{align}\label{e6:L.ABC}
  \uell_{\beta\alpha} &= \uop A_{\beta\alpha}(Z,t) +
  \uop B_{\beta\alpha}(Z,t)\del_{\zeta_\alpha} +
  \uop C_{\beta\alpha}(Z)\del_{\zeta_\alpha}^2\ .
\end{align}
Die matrixwertigen Funktionen $\uop A(Z,t)$, $\uop B(Z,t)$ und $\uop C(Z,t)$,
die aus (\ref{e6:MatrixL}) unter Verwendung von (\ref{e6:DefL}) folgen, 
lauten
\begin{subequations}
\begin{align}\label{e6:L.A}
  \begin{split}
\uop A_{\beta\alpha}(Z,t) &= \delta_{\beta\alpha}\,\frac{\I}{2\bar k}R_\beta(Z,u(Z,t))\\
      &+ (1-\delta_{\beta\alpha})\e^{\I(\phi_\alpha(Z,t)-\phi_\beta(Z,t))}
      \Bigg[
      \frac{\I}{2\bar k}\uop D^{(2)}_{\beta\alpha}(Z)
    - \frac{(\del_Z\phi_\alpha(Z,t))}{\bar k}\uop D^{(1)}_{\beta\alpha}(Z)
    \Bigg]\ ,
  \end{split}
\end{align}
\begin{align}\label{e6:L.B}
  \begin{split}
    \uop B_{\beta\alpha}(Z,t) &= \delta_{\beta\alpha}\,\frac{\I}{2\bar k}\klr{\klr{\del_Z\frac{\bar k}{k_\beta(Z)}}
        - \frac{\bar k\diffuza[Z]{k_\beta+m\Gamma_\beta}{u_\beta}
          - 2\uop D^{(1)}_{\beta\beta}(Z')k_\beta(Z')\big\vert_{Z'=u_\beta(Z,t)}}{k_\beta^2(Z)}}\\
       &+ (1-\delta_{\beta\alpha})\e^{\I(\phi_\alpha(Z,t)-\phi_\beta(Z,t))}
         \frac{\I}{2 k_\alpha(Z)}\uop D^{(1)}_{\beta\alpha}(Z)\ ,
  \end{split}
\end{align}
\begin{align}\label{e6:L.C}
  \begin{split}
    \uop C_{\beta\alpha}(Z) &= \delta_{\beta\alpha}\,\frac{\I}{2\bar k}
      \klr{\frac{\bar k}{k_\beta(Z)}}^2\ .
  \end{split}
\end{align}
\end{subequations}
Vergleichen wir dies mit den eindimensionalen Funktion $\ms A$, $\ms B$ und $\ms C$, die in Gl. (\ref{e5:DefL2})
definiert sind, so fällt auf, dass zwar $\uop C(Z)$ immer noch zeitunabhängig ist, aber
die Funktionen $\uop A(Z,t)$ und $\uop B(Z,t)$ nicht mehr ausschließlich über die Untergrenzenfunktion $u_\beta(Z,t)$
zeitabhängig sind. Durch den in den nichtdiagonalen Beiträgen enthaltenen Phasenfaktor 
\begin{align}
\uop P_{\beta\alpha}(Z,t) := \e^{\I(\phi_\alpha(Z,t)-\phi_\beta(Z,t))}
\end{align}
verbleibt zumindest im Fall verschiedener Gesamtenergien der Zustände mit Index $\alpha$ und $\beta$ in den
nichtdiagonalen Anteilen der Funktionen $\uop A(Z,t)$ und $\uop B(Z,t)$ eine weitere Zeitabhängigkeit in Form
einer dynamische Phase, die gerade der Differenz der beteiligten Gesamtenergien entspricht.

Berechnen wir nun die nullte und die erste Ordnung der Entwicklung der Wellenfunktion. In nullter Ordnung $\uk$
erhalten wir gerade den vollkommen adiabatischen Grenzfall
\begin{align}\label{e6:0.Ordnung}
  A^{(0)}_\beta(Z,t) = \vph_\beta(\zeta_\beta(Z)-\tau(t))\quad\Longrightarrow\quad
\Psi_\beta(Z,t) = \e^{\I\phi_\beta(Z,t)}\vph_\beta(\zeta_\beta(Z)-\tau(t))\ ,
\end{align}
Hier erkennt man, dass im atomaren Gesamtzustand (\ref{e6:Psi}),
\begin{align*}
\ket{\Psi(Z,t)} = \sum_\alpha \Psi_\alpha(Z,t)\rket{\alpha(Z)}\ ,
\end{align*}
jeder lokale, innere, atomare Zustand $\rket{\alpha(Z)}$
ein unabhängiges Wellenpaket mit sich führt,
dass der nullten Ordnung der Entwicklung im eindimensionalen Fall entspricht.
Jedes dieser Wellenpakete bewegt sich somit dispersionsfrei mit der klassischen
Schwerpunktsgeschwindigkeit durch das Potential $\Delta V_\alpha(Z)$, das vom Realteil
der lokalen Eigenenergie seines zugehörigen Eigenzustands $\rket{\alpha(Z)}$ 
der Massenmatrix $\uop M(Z)$ abhängt (siehe Gl. (\ref{e6:DefDV})).

Mit der Lösung (\ref{e6:0.Ordnung}) ist man also bereits in der Lage, adiabatische lABSE-Experimente
zu beschreiben, unterliegt dann aber auch den gleichen Näherungen und Beschränkungen, wie bei der Lösung der
skalaren Schrödinger-Gleichung, die wir in den vergangenen Kapiteln untersucht haben.

Der wesentliche Unterschied zu dem bereits in Abschnitt \ref{s3:Fahrplan}
durch einfache Überlegungen eingeführten atomaren Gesamtzustand (\ref{e3:PsiDetektor})
ist, dass hier der geometrische Anteil des Phasenfaktors bereits automatisch enthalten ist.

Nun wollen wir zur Berechnung der ersten Ordnung kommen. Hier erhalten wir
\begin{align}\label{e6:1.Ordnung}
  \begin{split}
    A^{(1)}_\beta(Z,t) &= \vph_\beta(\zeta_\beta(Z)-\tau(t))\\
    &+ \sum_\alpha\int_{u_\beta(Z,t)}^{Z}\d Z'\ 
\frac{\bar k}{k_\beta(Z')}\uop A_{\beta\alpha}(Z',t-M(\zeta_\beta(Z)-\zeta_\beta(Z'))/\bar k)\\
    	&\quad\times \vph_\alpha(\zeta_\alpha(Z')-\tau(t)+\zeta_\beta(Z)-\zeta_\beta(Z'))\\
    &+ \sum_\alpha\int_{u_\beta(Z,t)}^{Z}\d Z'\ 
\frac{\bar k}{k_\beta(Z')}\uop B_{\beta\alpha}(Z',t-M(\zeta_\beta(Z)-\zeta_\beta(Z'))/\bar k)\\
			&\quad\times \vph_\alpha'(\zeta_\alpha(Z')-\tau(t)+\zeta_\beta(Z)-\zeta_\beta(Z'))\\
    &+ \sum_\alpha\int_{u_\beta(Z,t)}^{Z}\d Z'\ 
\frac{\bar k}{k_\beta(Z')}\uop C_{\beta\alpha}(Z')\\
    	&\quad\times \vph_\alpha''(\zeta_\alpha(Z')-\tau(t)+\zeta_\beta(Z)-\zeta_\beta(Z'))\ .
  \end{split}
\end{align}
Hier sind nach wie vor die Ableitungen der Anfangswellenpakete $\vph_\alpha(\zeta)$ stets mit einem Strich gekennzeichnet,
d.h. es ist
\begin{align}\label{e6:DefStrich}
  \vph_\alpha'(\zeta) \equiv \del_\zeta\vph_\alpha(\zeta)\ ,\quad
  \vph_\alpha''(\zeta) \equiv \del^2_\zeta\vph_\alpha(\zeta)\ ,\quad
  \vph_\alpha^{(i)}(\zeta) \equiv \del^i_\zeta\vph_\alpha(\zeta)\ ,\quad (i\geq 3)\ .
\end{align}
Im Gegensatz dazu notieren wir die Amplitudenfunktionen $A_\alpha(Z,t)$ bzw. $B_\alpha(\zeta_\alpha,\tau)$,
die bis zur $n$-ten Ordnung entwickelt wurden, mit $A_\alpha^{(n)}(Z,t)$ bzw. $B_\alpha^{(n)}(\zeta_\alpha,\tau)$,
was nicht mit der $n$-ten Ableitung zu verwechseln ist!

Kommen wir nun zur Interpretation der ersten Ordnung (\ref{e6:1.Ordnung}), d.h. wir wollen die Korrekturbeiträge
nun im Einzelnen verstehen. Dazu beginnen wir mit einer Betrachtung der Argumente der Funktionen in
den Integranden. Wir definieren
\begin{align}\label{e6:DefDT}
  \Delta T_\beta(Z',Z) &:= \frac{M}{\bar k}\klr{\zeta_\beta(Z)-\zeta_\beta(Z')}\ ,
\end{align}
was der Zeit entspricht, die ein klassisches Teilchen im Potential $V_\beta(Z)$
benötigt, um die Strecke zwischen den Orten $Z'$ und $Z$ zurückzulegen. Weiter machen wir uns
klar, das $\vph_\alpha(\zeta_\alpha(Z)-\tau(t)) = A^{(0)}_\alpha(Z,t)$ der nullten Ordnung
der Amplitudenfunktion mit Index $\alpha$ entspricht.

Die erste Ordnung (\ref{e6:1.Ordnung}) kann damit geschrieben werden als
\begin{align}\label{e6:1.OrdnungA}
  \begin{split}
    A^{(1)}_\beta(Z,t) &= A_\beta^{(0)}(Z,t)\\
    &+ \sum_\alpha\int_{u_\beta(Z,t)}^{Z}\d Z'\ 
    \frac{\bar k}{k_\beta(Z')}\uop A_{\beta\alpha}(Z',t-\Delta T_\beta(Z',Z))\\
    	&\quad\times A_\alpha^{(0)}(Z',t-\Delta T_\beta(Z',Z))\\
    &+ \sum_\alpha\int_{u_\beta(Z,t)}^{Z}\d Z'\ 
    \frac{\bar k}{k_\beta(Z')}\uop B_{\beta\alpha}(Z',t-\Delta T_\beta(Z',Z))\\
			&\quad\times (\del_{\zeta_\alpha}A_\alpha^{(0)})(Z',t-\Delta T_\beta(Z',Z))\\
    &+ \sum_\alpha\int_{u_\beta(Z,t)}^{Z}\d Z'\ 
    \frac{\bar k}{k_\beta(Z')}\uop C_{\beta\alpha}(Z')\\
    	&\quad\times (\del^2_{\zeta_\alpha} A_\alpha^{(0)})(Z',t-\Delta T_\beta(Z',Z))\ .
  \end{split}
\end{align}
Die Korrekturen zum Anfangswellenpaket $A^{(0)}_\beta(Z,t)$ am Punkt $(Z,t)$ stammen
also von den Anfangswellenpaketen aller Komponenten $\alpha$ am Ort $Z'<Z$, zu einer
früheren Zeit $t-\Delta T_\beta(Z',Z)$. Es gilt aber z.B. gerade 
$A^{(0)}_\beta(Z,t)\equiv A^{(0)}_\beta(Z',t-\Delta T_\beta(Z',Z))$,
d.h. insbesondere die Selbstkorrekturbeiträge hängen alle nur von der betrachteten
Stelle des ungestörten Anfangswellenpakets ab. Bei den diagonalen Korrekturtermen
können wir also die Funktion $A^{(0)}_\beta(Z,t)$ und deren Ableitungen aus dem Integral
herausziehen und erhalten letztendlich die gleiche Entwicklung wie im eindimensionalen Fall.
Zusätzlich müssen natürlich noch nichtdiagonale Beiträge berücksichtigt werden.
Bei Wellenpaketen mit $\alpha\neq\beta$ tragen - je nach Potentialverlauf - auch andere Stellen 
der Amplituden in nullter Ordnung zur Korrektur von $A^{(0)}_\beta(Z,t)$ bei.
Hier ist eine weitere Vereinfachung wie bei den diagonalen Termen nicht mehr möglich.

Insgesamt sieht man an der Formel (\ref{e6:1.OrdnungA}) sehr schön, wie die Amplitudenfunktionen
aller inneren atomaren Zustände sich gegenseitig beeinflussen.
Tritt z.B. eine Komponente im Anfangszustand überhaupt nicht auf, d.h.
ist z.B. $A^{(0)}_\beta(Z,t)\equiv 0$, so gehen einerseits von dieser Komponente
(in erster Ordnung) keine Korrekturen aus, andererseits kann durch den Einfluss anderer
Komponenten eine von Null verschiedene Amplitudenfunktion in dieser
Komponente entstehen. All diese Effekte sind erst durch Einführung eines
Matrixpotentials entstanden, womit wir nun die Beschränkung auf adiabatische
Zustandsänderungen überwunden haben.

Der Hauptbeitrag der diagonalen (Selbst-)Korrekturen stammt aus der
Funktion $\uop C(Z)$. Dies war bereits bei den skalaren Potentialen der Fall
und zeigt, dass auch hier die Dispersion den größten Effekt bei der Korrektur
ausmacht. Hier kommt  wieder die Anwendung des in Abschnitt \ref{s4:Dispersion}
erläuterten Verfahrens in Frage, bei dem sehr breite Wellenpakete zu einem sehr schmalen Wellenpaket
superponiert werden. Für jedes Basiswellenpaket wird die Lösung in erster Ordnung berechnet,
wobei der Dispersionsterm $\uop C(Z)$, der vor der zweiten Ableitung des Wellenpakets steht, 
bei ausreichender Breite der Basiswellenpakete vernachlässigbar wird. 
Ebenso kann man auch die Vernachlässigung des Korrekturbeitrags $\uop B(Z,t)$ in Betracht ziehen, falls
bereits die erste Ableitung der Einhüllenden relativ zur Einhüllenden selbst vernachlässigbar ist.
In diesem Fall würde man lediglich den Korrekturbeitrag $\uop A(Z,t)$ berücksichtigen, in
dessen nichtdiagonalen Beiträgen die (nichtdiagonalen)
Matrixelemente der Ableitungsmatrizen $\uop D^{(1,2)}(Z)$ auftreten, die für die nichtadiabatische
Mischung der Zustände und somit auch der Wellenpakete verantwortlich sind. Diese Matrizen
können z.B. störungstheoretisch berechnet werden, siehe dazu die Anhänge \ref{sA:StoeRe}
und \ref{sA:OrtsablMat} sowie die Gln. (\ref{eA:DMatElem1}) und (\ref{eA:DMatElem2}),
aber in der Praxis wird man sie in der Regel numerisch als Differenzenquotienten
berechnen.

Insgesamt wird bereits die numerische Berechnung der ersten Ordnung sehr anspruchsvoll werden,
da sehr viele Zutaten in die Formel eingehen, über die dann letztendlich noch integriert
werden muss. Selbst bei Verwendung der Methode aus Abschnitt \ref{s4:Dispersion} hat man
trotz der Vernachlässigung einiger Korrekturbeiträge in der ersten Ordnung
durch die anschließende Superposition der Wellenpakete einen großen Rechenaufwand.

Aus Zeitgründen konnte bisher kein numerisches Beispiel zur Anwendung
der ersten Ordnung des Formalismus gerechnet werden. Dies wird aber in naher
Zukunft nachgeholt, wenn an anderer Stelle P-erhaltende und P-verletzende Rotationen der
Polarisation von Atomen in elektrischen Feldern untersucht werden. Mehr dazu im Ausblick, siehe
Abschnitt \ref{s8:Ausblick}.

Wir wollen diesen Abschnitt mit einer Anmerkung zur Berechnung der zweiten Ordnung
der Entwicklung abschließen. In der Definition des Operators $\uell$ aus Gl. (\ref{e6:L.ABC})
stehen Ableitungen nach den Koordinaten $\zeta_\alpha$. In erster Ordnung macht dies Sinn,
da der Operator dann nur auf die Einhüllenden $\vph_\alpha(\zeta_\alpha-\tau)$ wirkt. In
zweiter Ordnung allerdings wirkt $\uell$ auf $(\uk\vph_\alpha)(Z,t)$, was eine gemischte
Funktion von $Z$ und allen $\zeta_\beta(Z)$ $(\beta=1,\ldots,N)$ ist. 
Hier bietet es sich an, die Ableitungen nach $\zeta_\alpha$ in $\uell$ wieder in Ableitungen nach
$Z$ zu transformieren.

Man hat dann
\begin{align}
\del_{\zeta_\alpha} = \frac{k_\alpha(Z)}{\bar k}\del_Z\ ,\quad
\del_{\zeta_\alpha}^2 = \frac{k_\alpha(Z)(\del_Zk_\alpha(Z))}{\bar k^2}\del_Z + \klr{\frac{k_\alpha(Z)}{\bar k}}^2\del_Z^2
\end{align}
in Gl. (\ref{e6:L.ABC}) einzusetzen und erhält
\begin{align}\label{e6:L.ABC.2}
\uell_{\beta\alpha} = \uop A_{\beta\alpha}(Z,t) + \unl{\ms{\tilde B}}_{\beta\alpha}(Z,t)\del_Z 
+ \unl{\ms{\tilde C}}_{\beta\alpha}(Z)\del_Z^2
\end{align}
mit den neuen Funktionen
\begin{align}\label{e6:L.B.2}
\begin{split}
\unl{\ms{\tilde B}}_{\beta\alpha}(Z,t) &= \uop B_{\beta\alpha}(Z,t)\frac{k_\alpha(Z)}{\bar k} 
+ \uop C_{\beta\alpha}(Z)\frac{k_\alpha(Z)(\del_Zk_\alpha(Z))}{\bar k^2}\\
&= \delta_{\beta\alpha}\,\frac{\I}{2\bar k}\Bigg(k_\alpha(Z)\klr{\del_Z\frac{1}{k_\beta(Z)}}
+ \frac{k_\alpha(Z)(\del_Zk_\alpha(Z))}{k^2_\beta(Z)}
\\
&\hspace{5mm}- \frac{k_\alpha(Z)\diffuza[Z]{k_\beta+m\Gamma_\beta}{u_\beta}}{k_\beta^2(Z)}
+ \frac{2k_\alpha(Z)\kle{\uop D^{(1)}_{\beta\beta}(Z')k_\beta(Z')}_{Z'=u_\beta(Z,t)}}{\bar k\,k_\beta^2(Z)}\Bigg)\\
&+ (1-\delta_{\beta\alpha})\e^{\I(\phi_\alpha(Z,t)-\phi_\beta(Z,t))}
\frac{\I}{2\bar k}\uop D^{(1)}_{\beta\alpha}(Z)\ ,
\end{split}
\end{align}
und
\begin{align}\label{e6:L.C.2}
\unl{\ms{\tilde C}}_{\beta\alpha}(Z) &= \uop C_{\beta\alpha}(Z)\klr{\frac{k_\alpha(Z)}{\bar k}}^2
= \delta_{\beta\alpha}\frac{\I}{2\bar k}\klr{\frac{k_\alpha(Z)}{k_\beta(Z)}}^2\ .
\end{align}
Hier haben wir die Funktionen $\uop B_{\beta\alpha}(Z)$ und $\uop C_{\beta\alpha}(Z)$ aus (\ref{e6:L.B}) und (\ref{e6:L.C})
verwendet. Die Funktion $\uop A(Z,t)$ in $\uell$ kann unverändert von (\ref{e6:L.A}) übernommen werden.

\section{Das Rezept zur theoretischen Beschreibung eines lABSE-Experiments}\label{s6:Rezept}

In diesem Abschnitt wollen wir die Vorgehensweise zur theoretischen Beschreibung eines
allgemeinen lABSE-Experiments angeben und evtl. noch benötigte Werkzeuge bereitstellen.

Der prinzipielle Ablauf ist der Folgende. Zunächst wählt man die Anfangsbedingungen, die
wir im nächsten Abschnitt genauer betrachten wollen. Dann berechnet man für verschiedene
Feldkonfigurationen den (erweiterten) Fahrplan und schließlich für eine gegebene Feldkonfiguration
das mit dem Experiment vergleichbare Spinecho-Signal.

Das erweiterte Fahrplanmodell werden wir in Abschnitt \ref{s6:xFP} diskutieren
und die Vorgehensweise zur Berechnung des lABSE-Signals in Abschnitt \ref{s6:Signal} angeben.

\subsection{Implementierung der Anfangsbedingungen}\label{s6:InitialConditions}

Es bietet sich, allein schon aufgrund der leichteren Wahl der normierten Einhüllenden
$\vph(\zeta)$ (siehe die Diskussion vor Gl. (\ref{e5:Norm.AWP.Speziell}), S. \pageref{e5:Norm.AWP.Speziell}  
im Abschnitt \ref{s5:2.Ordnung}) wieder an, den Ort $Z_0$ vor allen Feldern zu wählen.

Als nächstes muss man den inneren Anfangszustand des Atoms festlegen, z.B. anhand einer vorgegebenen
Polarisation. Da man meistens eine Polarisation in eine eindeutige Raumrichtung wählt, kann man
den entsprechenden Drehimpulsoperator $\unl F_i$ $(i=1,2,3)$ in der Darstellung der
Gesamtdrehimpulsbasis $\ket j$ berechnen, die wir auch für die Darstellung der Massenmatrizen
in Anhang  \ref{sB:MundEWP} verwendet haben.

Nun kann man den Polarisationszustand mit vorgegebener Gesamtdrehimpulsquantenzahl $F$
und der Komponente $F_i$ in {\em eine} bestimmte Raumrichtung $i=1,2,3$ durch Lösen des Eigenwertproblems
\begin{align}
\begin{split}
\unl F_i\ket{\chi} &= \unl F_i\sum_j \chi_j\ket j \overset!= F_i\sum_j \chi_j\ket j\ ,\\
\vu F^2\ket{\chi} &= (\unl F_1^2 + \unl F_2^2 +\unl F_3^2)\ket{\chi}=  F(F+1)\ket{\chi}
\end{split}
\end{align}
festlegen. Im allgemeinen brauchen wir die Koeffizienten $\chi_j$ bzgl. der Basis der lokalen atomaren Eigenzustände
$\rket{\alpha(Z_0)}$ am Anfangsort $Z_0$. Diese bekommen wir durch Projektion auf den zugeordneten linken Eigenzustand
$\lrbra{\alpha(Z_0)}$,
\begin{align}
\hat\chi_\alpha = \lrbra{\alpha(Z_0)}\chi\rangle\ .
\end{align}
Der gesamte Polarisationszustand kann dann in der lokalen Basis angegeben werden als
\begin{align}\label{e6:chi}
\begin{split}
\ket{\chi} &= \unl\um\ket{\chi} = \sum_\alpha\Pro_\alpha(Z_0)\ket\chi = \sum_{\alpha,j} \chi_j\Pro_\alpha(Z_0)\ket{j}
= \sum_{\alpha,j} \chi_j\rket{\alpha(Z_0)}\lrbra{\alpha(Z_0)}{j\rangle}\\ 
&\overref!= \sum_\alpha \hat\chi_\alpha\rket{\alpha(Z_0)}\ ,
\end{split}
\end{align}
wobei wir die Vollständigkeitsrelation (\ref{eA:QProVollst}) und die Quasiprojektoren (\ref{eA:QPro}) aus
Anhang \ref{sA:ZweiTeilchen} verwendet haben.

Nun betrachten wir den gesamten atomaren Zustand zur Zeit $t=0$, bei der der
Integraloperator $\uk$ (\ref{e6:DefK}) verschwindet und die Wellenpakete der
nullten Ordnung der Entwicklung entsprechen, also gleich $\vph_\alpha(\zeta_\alpha(Z))$ sind.
Zusammen mit der Definition (\ref{e6:Psi}) des atomaren Gesamtzustands folgt dann
\begin{align}\label{e6:InitialState}
  \ket{\Psi(Z,t=0)} = \sum_\alpha\Psi_\alpha(Z,t=0)\rket{\alpha(Z)} 
    = \sum_\alpha \e^{\I\phi_\alpha(Z,t=0)}\vph_\alpha(\zeta_\alpha(Z))\rket{\alpha(Z)}
\end{align}
mit dem Anfangsphasenwinkel nach Gl. (\ref{e6:PW}),
\begin{align}
\phi_\alpha(Z,t=0) = S_\alpha(Z) = \int_{Z_0}^Z\d Z'\ \frac{\bar k}{k_\alpha(Z')}\ .
\end{align}
Die Anfangswellenpakete $\vph_\alpha(\zeta)$ sind festgelegt durch die Wahl
der Einhüllenden $\vph(\zeta)$ und der Koeffizienten $\hat\chi_\alpha$ des Anfangszustands des Atoms.
Wählen wir für alle Komponenten eine identische Einhüllende, so erhalten wir insgesamt
\begin{align}\label{e6:Envelopes}
  \vph_\alpha(\zeta) = \hat\chi_\alpha\vph(\zeta)\ .
\end{align}
Dass dies durchaus sinnvoll ist, haben wir bereits in Abschnitt \ref{s3:FahrplanBasics} diskutiert, denn
zur Zeit $t=0$ wählen wir den Anfangszustand des Atoms ja nach Gl. (\ref{e3:Initial.Atom}) in der Form
\begin{align}
\ket{\Psi(Z,t=0)} = \Psi(Z,0)\ket{\chi} = \e^{\I\phi(Z,0)}\vph(Z-Z_0)\ket{\chi}\ ,
\end{align}
wobei wir hier wieder davon ausgegangen sind, dass wir wie in Gl. (\ref{e5:Norm.AWP.Speziell}) $Z_0$ und die Breite
$\sigma$ des Anfangswellenpakets so wählen, dass
\begin{align}
\zeta_\alpha(Z) \approx Z-Z_0\quad\text{für}\quad \vph(\zeta_\alpha(Z)) \neq 0
\end{align}
erfüllt ist.

\subsection{Das erweiterte Fahrplanmodell}\label{s6:xFP}

Das in Abschnitt \ref{s3:Fahrplanmodell} beschriebene Fahrplanmodell gibt die Bewegung der Schwerpunkte
der einzelnen Wellenpakete nur im adiabatischen Grenzfall richtig wieder. Wir haben gesehen, dass diese
Bewegung dann völlig identisch mit der Bewegung klassischer Punktteilchens in unterschiedlichen Potentialen
$V_\alpha(Z)$ ist. 

Nun wollen wir das Fahrplanmodell dahingehend erweitern, dass auch die Bewegung neu erzeugter
Wellenpakete dargestellt werden kann, was aber zu grundsätzlichen Schwierigkeiten führt. Mit dem in diesem Kapitel
aufgestellten Formalismus ist die Schwerpunktsbewegung neu erzeugter Wellenpakete nicht direkt ablesbar. Wir wissen
zunächst weder, wo dieser Schwerpunkt liegt, noch, wie das neue Wellenpaket überhaupt aussieht. Erst eine numerische
Berechnung der höheren Ordnung der Entwicklung könnte diese Informationen liefern. Den Schwerpunkt könnte man dann
als zeitabhängigen Ortserwartungswert des neuen Wellenpakets numerisch berechnen und die Breite des Wellenpakets
aus der Varianz erhalten. Diese Rechnung wäre aber für das Aufstellen des Fahrplans viel zu aufwändig, denn der
Fahrplan soll ja gerade eine Orientierungshilfe sein, die man vor der Berechnung der atomaren Gesamtwellenfunktion
braucht. Der Fahrplan ist (in unseren Augen) einzig dazu gedacht, einen groben Überblick über ein evtl. zu machendes
Experiment zu liefern. Aus dem Fahrplan soll letztendlich die Entscheidung für die Feldkonfiguration und den
verwendeten Anfangszustand des Atoms fallen.

Wie können wir also mit einfachen Mitteln einen erweiterten Fahrplan erstellen, der die Bewegung neuer Wellenpakete zumindest
in grober Näherung visualisiert? Die Antwort auf diese Frage ist praktisch schon im zweiten Satz dieses Abschnitts gegeben:
Wir müssen uns zunächst von den Wellenpaketen lösen und uns ganz auf den klassischen Standpunkt stellen, d.h. von
punktförmigen Teilchen sprechen. Dann kann man sich vorstellen, dass sich ein Teilchen der \glqq Sorte\grqq~
$\alpha$ an einem Ort $Z$ im Potential in ein Teilchen der Sorte $\beta$ verwandelt, wobei aufgrund des
zeitunabhängigen Potentials die Gesamtenergie erhalten ist.

Man kann sich auch mathematisch davon überzeugen, dass dies unter gewissen Voraussetzungen für die Schwerpunkte
der Wellenpakete erfüllt ist, indem man z.B. ein einfaches Zwei-Zustands-System betrachtet und die erste
Ordnung der Entwicklung auf ebene WKB-Wellen anwendet. Superponiert man anschließend diese ebenen Wellen
unter geschickter Anwendung einiger Näherungen zu Wellenpaketen, erhält man hinter allen Potentialen 
eine Energieerhaltungsgleichung für den Schwerpunkt des im Potential erzeugten Wellenpakets, die die Gesamtenergie
des ursprünglichen Wellenpakets enthält. Wir wollen diese lange und sehr technische Rechnung hier nicht
vorführen, sondern uns mit der im letzten Absatz dargestellten, klassischen Sichtweise begnügen. Wir werden
ohnehin feststellen, dass das erweiterte Fahrplanmodell nur sehr eingeschränkt in der Lage ist, die physikalischen
Vorgänge darzustellen. 

Im einfachen Fahrplanmodell braucht ein Teilchen im Zustand $\rket{\alpha(Z)}$ die Zeit
\begin{align}\label{e6:Zeit}
T_\alpha(Z) = \frac{M}{\bar k}\zeta_\alpha(Z) = \int_{Z_0}^Z\d Z'\ \frac{M}{k_\alpha(Z')}
\end{align}
um vom Ort $Z_0$ zum Ort $Z$ unter dem Einfluss des Potentials $V_\alpha(Z)$ zu gelangen 
(vergleiche Gl. (\ref{e3:Delta.T}) auf Seite \pageref{e3:Delta.T}). Dieses Potential geht in der
lokalen Wellenzahl gemäß (\ref{e6:Defk}) ein,
\begin{align}
k_\alpha(Z) = \sqrt{\bar k^2_\alpha - 2M V_\alpha(Z)}\ ,
\end{align}
wobei nach Gl. (\ref{e6:Defkbar})
\begin{align}
\bar k^2_\alpha = 2M E_\alpha\supt{ges} = \bar k^2 + V_\alpha(Z_0)
\end{align}
die von der Gesamtenergie des Zustands $\rket{\alpha(Z_0)}$ am Ort $Z_0$ abhängige, totale Wellenzahl ist.

Geht nun das Teilchen im Zustand $\rket{\alpha(Z)}$ im stationären Potential $V_\alpha(Z)$ an der Stelle $Z$ in den Zustand 
$\rket{\beta(Z)}$ über, so wird sich seine Gesamtenergie nicht verändern. 
Das lokale Potential, dass das Teilchen im Zustand $\rket{\beta(Z)}$ aber spürt, ist $V_\beta(Z)$, somit
wird sich seine lokale Wellenzahl ändern zu
\begin{align}\label{e6:lokWZ.xFP}
k_{\alpha\to\beta}(Z) := \sqrt{\bar k^2_\alpha - 2M V_\beta(Z)} 
= \sqrt{\bar k^2 - 2M (V_\beta(Z) - V_\alpha(Z_0))}\ .
\end{align}
Hieraus folgt, dass selbst im feldfreien Raum, wo $V_\beta(Z) \to V_\beta(Z_0)$ ist, bei nichtentarteten Zuständen
$\rket{\alpha(Z_0)}$ und $\rket{\beta(Z_0)}$ die freie Wellenzahl des übergegangenen Teilchens von der freien Wellenzahl
$\bar k$ abweicht und somit auch seine freie (mittlere) Geschwindigkeit.

Ist das Teilchen am Ort $Z_0$ gestartet, am Ort $Z_1$ vom Zustand $\rket{\alpha(Z_1)}$ 
in den Zustand $\rket{\beta(Z_1)}$ übergegangen und
befindet sich nun am Ort $Z>Z_1$, so hat es die Gesamtzeit
\begin{align}
T_{\alpha\to\beta}(T_\alpha(Z_1),Z) := T_\alpha(Z_1) + \int_{Z_1}^Z\d Z'\ \frac{M}{k_{\alpha\to\beta}(Z')}
\end{align}
benötigt. Wir können dies nach Multiplikation mit $\bar v$ auch wieder mit den Koordinaten $\zeta_\alpha(Z)$
schreiben und erhalten 
\begin{align}
\zeta_{\alpha\to\beta}(\zeta_\alpha(Z_1),Z) 
:= \zeta_\alpha(Z_1) + \int_{Z_1}^Z\d Z'\ \frac{\bar k}{k_{\alpha\to\beta}(Z')}\ .
\end{align}
Man kann sich leicht denken, dass bei mehrfachen, aufeinanderfolgenden Übergängen, die Ausdrücke auf analoge Weise
erweitert werden und an Komplexität gewinnen werden. Eine übersichtliche und gleichzeitig informative 
Notation zu finden ist nicht einfach. Wir wollen den Wert für $\zeta$ am Ort $Z$ eines Teilchens,
das am Ort $Z_{\beta\to\gamma}$ bei einem Wert $\zeta_{\alpha\to\beta}$ 
vom Zustand $\rket{\beta(Z_{\beta\to\gamma})}$ in den 
Zustand $\rket{\gamma(Z_{\beta\to\gamma})}$ übergegangen ist und insgesamt aus einem Anfangszustand
$\rket{\alpha(Z_0)}$ entstanden ist, bezeichnen als
\begin{align}\label{e6:Zeta.xFP}
\zeta_{\alpha\to\gamma}(\zeta_{\alpha\to\beta},Z_{\beta\to\gamma},Z) =
\zeta_{\alpha\to\beta} + \int_{Z_{\beta\to\gamma}}^Z\d Z'\ \frac{\bar k}{k_{\alpha\to\gamma}(Z')}\ .
\end{align}
Es handelt sich dabei um eine rekursive Definition, da $\zeta_{\alpha\to \beta}$ selbst aus mehrfachen
aufeinanderfolgenden Übergängen aus dem Zustand $\rket{\alpha(Z_0)}$ entstanden sein kann. Die Verwendung
von Gl. (\ref{e6:Zeta.xFP}) erlaubt die Beschreibung der Position des Teilchens relativ zur Position eines
freien Teilchen gemäß
\begin{align}\label{e6:DZeta.xFP}
\Delta\zeta_{\alpha\to\gamma}(\zeta_{\alpha\to\beta},Z_{\beta\to\gamma},Z) :=
\zeta_{\alpha\to\gamma}(\zeta_{\alpha\to\beta},Z_{\beta\to\gamma},Z) - (Z-Z_0)\ .
\end{align}

Nachdem wir uns bisher Gedanken über Teilchenpositionen gemacht haben, sollten wir uns nun überlegen,
wie wir die obigen Gleichungen im Hinblick auf die Bewegung von Wellenpaketen interpretieren. Das grundlegende Problem,
nämlich die Unkenntnis über die Form des neuen Wellenpakets und die Lage dessen Schwerpunkts, verhindert die
genaue Kenntnis des Ortes $Z_{\alpha\to\beta}$ innerhalb eines Potentials, an dem wir den Übergang ansetzen sollen.

\begin{figure}[!htb]
\centering
\subfloat[Visualisierung der Breite und der Bewegung der erzeugten Wellenpakete]{\label{f6:xFP1}\includegraphics[width=6cm]{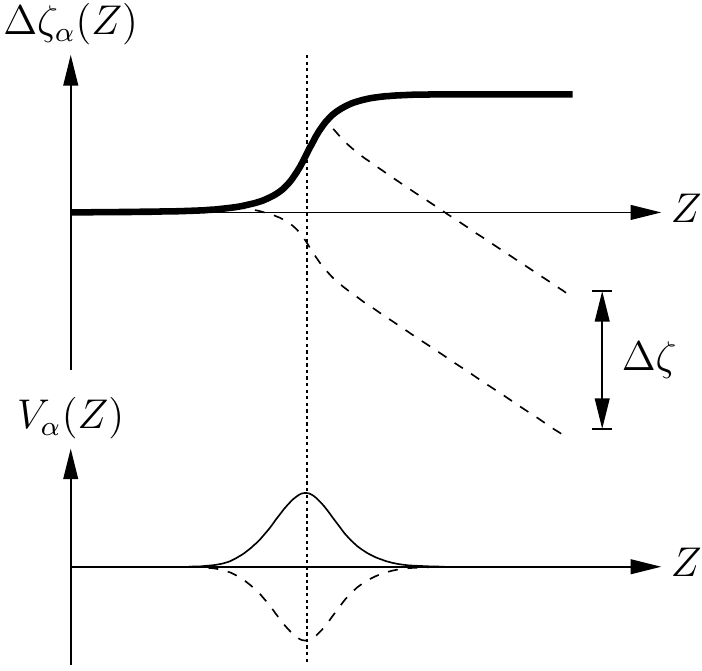}}\qquad
\subfloat[Visualisierung der Bewegung der erzeugten Wellenpakete.]{\label{f6:xFP2}
\includegraphics[width=6cm]{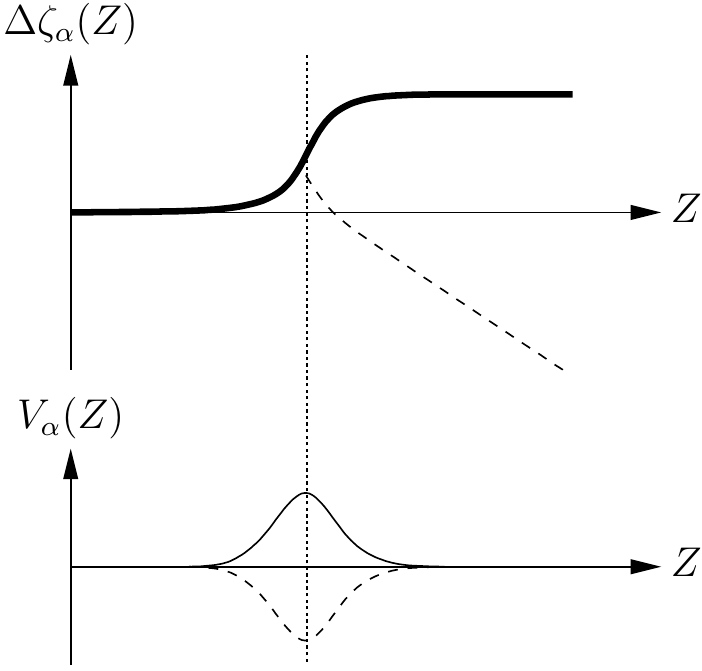}}\\[5mm]
\caption{Zwei Beispiele zur Anwendung des erweiterten Fahrplanmodells.}
\end{figure}
In Abb. \subref*{f6:xFP1} ist eine Möglichkeit der Darstellung des erweiterten Fahrplans schematisch gezeigt.
Hier betrachten wir zwei Zustände (durchgezogene und gestrichelte Linien), die entgegengesetzten Potentialen
(z.B. erzeugt durch ein magnetisches Feld) unterliegen. Anfangs starten wir mit einem einzigen Zustand. Die Bewegung
des Schwerpunkts des zugeordneten Wellenpakets im Potential ist durch die durchgezogene Linie im oberen Teil
der Abbildung eingezeichnet. Die Dicke der Linie soll in etwa die Breite bzw. die Unschärfe des Wellenpakets 
repräsentieren. Trifft das Wellenpaket nun auf das Potential, beginnt die Produktion des Wellenpakets des
zweiten Zustands, dargestellt durch die erste gestrichelten Linie in der oberen Abbildung.
Zum einen ist die
Geschwindigkeit (Steigung) dieser \glqq Vorderkante\grqq~ im potentialfreien Raum von der eines 
freien Wellenpakets verschieden, zum anderen sieht man auch die Auswirkungen des zugehörigen Potentials auf den
Verlauf der gestrichelten Linie. Die Produktion des neuen Wellenpakets hört erst auf, wenn das erste Wellenpaket
das Potential vollständig verlassen hat. Die zweite gestrichelte Linie symbolisiert somit die hintere Kante des
neuen Wellenpakets.

Durch die in Abb. \subref*{f6:xFP1} verwendete Darstellung bekommt man einen Eindruck von der Breite 
(die in der Abbildung mit $\Delta\zeta$ bezeichnet wurde) des neuen
Wellenpakets, die sowohl von der Breite des erzeugenden Wellenpakets als auch von der Breite des Potentials
und der Differenz der Geschwindigkeit zum freien Wellenpaket bestimmt wird. Der Nachteil dieser Art der
Darstellung ist die Unübersichtlichkeit, die zwangsläufig bei der Visualisierung mehrerer Übergänge entsteht.
In Abb. \subref*{f6:xFP2} ist nur eine mittlere Linie für das neue Wellenpaket eingezeichnet, die man dann
als Schwerpunkt bezeichnen kann\footnote{Die tatsächliche Lage des Schwerpunkts des neuen Wellenpakets ist
wie in der Einleitung zu diesem Abschnitt bereits erklärt nicht bekannt. Man kann aber erwarten, dass auch
das neu erzeugte Wellenpaket in nullter Näherung als dispersionsfreies Wellenpaket beschrieben werden kann,
so dass in diesem speziellen Fall jeder Punkt des Wellenpakets als Bezugspunkt angesehen werden kann.}.
Den Vorteil der größeren Übersichtlichkeit in Abb. \subref*{f6:xFP2} erkauft man sich allerdings
durch das Verwerfen der Information über die Breite der neuen Wellenpakete. 

\pagebreak
Man kann sich vorstellen, dass aus Kreuzungspunkten mit neu entstandenen Wellenpaketen aufgrund der
(bei ungünstig gewählten Potentialen) sehr großen Breite der Wellenpakete ganze Kreuzungsflächen werden.
Bei stark ausgeschmierten neuen Wellenpaketen kann kein ausgeprägtes Spinechosignal mehr gemessen werden.

\begin{floatingfigure}[r]{6cm}
\centering\vspace{-5mm}
\includegraphics[width=6cm]{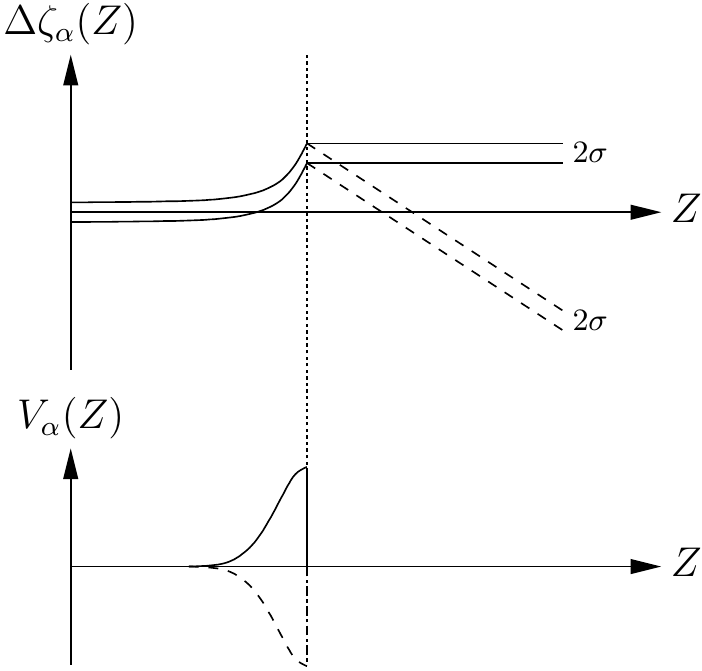}
\caption{Ein drittes Beispiel zur Anwendung des erweiterten Fahrplanmodells.}\label{f6:xFP3}
\end{floatingfigure}

Um ein scharfes Spinechosignal zu erhalten, muss man die Potentiale also derart wählen, dass die Breite
des erzeugenden Wellenpakets erhalten bleibt. In Abb. \ref{f6:xFP3} ist der effektivste Weg gezeigt,
dies zu erreichen, nämlich die Verwendung eines zunächst adiabatischen Potentials, dass dann einen
plötzlichen Sprung hat. Nur an der Potentialstufe wird das neue Wellenpaket erzeugt. Die Breite $2\sigma$ des
ursprünglichen Wellenpakets, die durch den Abstand der durchgezogenen Linie symbolisiert ist, bleibt so beim
neuen Wellenpaket direkt erhalten. In der Praxis könnte man eine leicht aufgeweichte Kante verwenden, um
das in erster Ordnung entstehende Wellenpaket numerisch zu berechnen.

\FloatBarrier
Wir fassen die Anleitung zum Erstellen des erweiterten Fahrplans zusammen:
\begin{enumerate}[(1)]
\item Zunächst entscheidet man, welche Zustände im Anfangszustand enthalten sein sollen.
\item Dann wählt man die äußeren elektrischen und magnetischen Felder und berechnet 
damit für das betrachtete Atom die Potentialverläufe für alle inneren atomaren Zustände.
\item Man beginnt dann mit der Erstellung des einfachen Fahrplans wie in Kap. \ref{s3:Fahrplanmodell}
beschrieben.
\item Nun hat man sich für eine der oberen Betrachtungsweisen des erweiterten Fahrplanmodells
zu entscheiden. Alternativ dazu kann man natürlich auch eine gemischte Darstellung verwenden.

Es empfiehlt sich hier, ein adiabatisches Potential mit wenigen Sprüngen zu verwenden und
die Darstellung aus Abb. \ref{f6:xFP3} zu verwenden. Auf diese Weise wird die Übersichtlichkeit
des Fahrplans nicht gefährdet und man bekommt keine allzu breiten Wellenpakete.

\item Man muss sich auch anhand der Massenmatrix $\uop M(Z)$ überlegen, welche Mischungen am stärksten auftreten
und dann nur diese berücksichtigen.
\end{enumerate}

Als Resultat ergeben sich die Kreuzungspunkte/-flächen der einzelnen Zustände. 
Mit dieser Verfahrensweise wird man also allein mit Hilfe des erweiterten Fahrplanmodells in der Lage
sein, sich schnell und einfach mögliche lABSE-Experimente auszudenken. Wir wollen noch einmal klar
und deutlich sagen, was das erweiterte Fahrplanmodell kann bzw. nicht kann:
\begin{itemize}
\item Es kann nur die Bewegungen der Schwerpunkte der Wellenpakete im adiabatischen Grenzfall korrekt wiedergeben.
\item Die Bewegung neu erzeugter Wellenpakete kann nur qualitativ angezeigt werden.
\item Das Fahrplanmodell kann weder die Lage der Schwerpunkte, noch die Form der neuen Wellenpakete liefern.
\item Zerfall oder die Größe der Amplituden der neu entstandenen Wellenpakete können nicht dargestellt werden.
\end{itemize}
Während der einfache Fahrplan ohne Beimischungen wenigstens im adiabatischen Grenzfall eine korrekte Beschreibung
der Propagation der Wellenpakete liefert, ist die Einbeziehung der Bewegung neu entstandener Wellenpakete im 
Rahmen des erweiterten Fahrplans zum größten Teil Handarbeit. Er zeigt einem nur, was man auch sehen will, da
man selbst entscheidet, welche Mischungen man an welcher Stelle einbaut. Nichtsdestotrotz sollte auch die
Linien der neuen Wellenpakete eine recht gute Näherung sein, solange man sich nicht zu weit vom adiabatischen
Grenzfall entfernt.

\newpage
\subsection{Berechnung des Spinechosignals}\label{s6:Signal}

Hat man sich einmal für einen Anfangszustand und eine Feldkonfiguration entschieden,
kann man unter Verwendung der Entwicklung für die exakte Lösung der Schrödinger-Gleichung
das lABSE-Signal berechnen. Wir wollen in diesem Abschnitt kurz erklären, wie dabei
die prinzipielle Vorgehensweise ist.

\subsubsection{Experimentelle Vorgehensweise zur Messung des lABSE-Signals}

\begin{figure}[!hp]
\centering
\includegraphics[width=12cm]{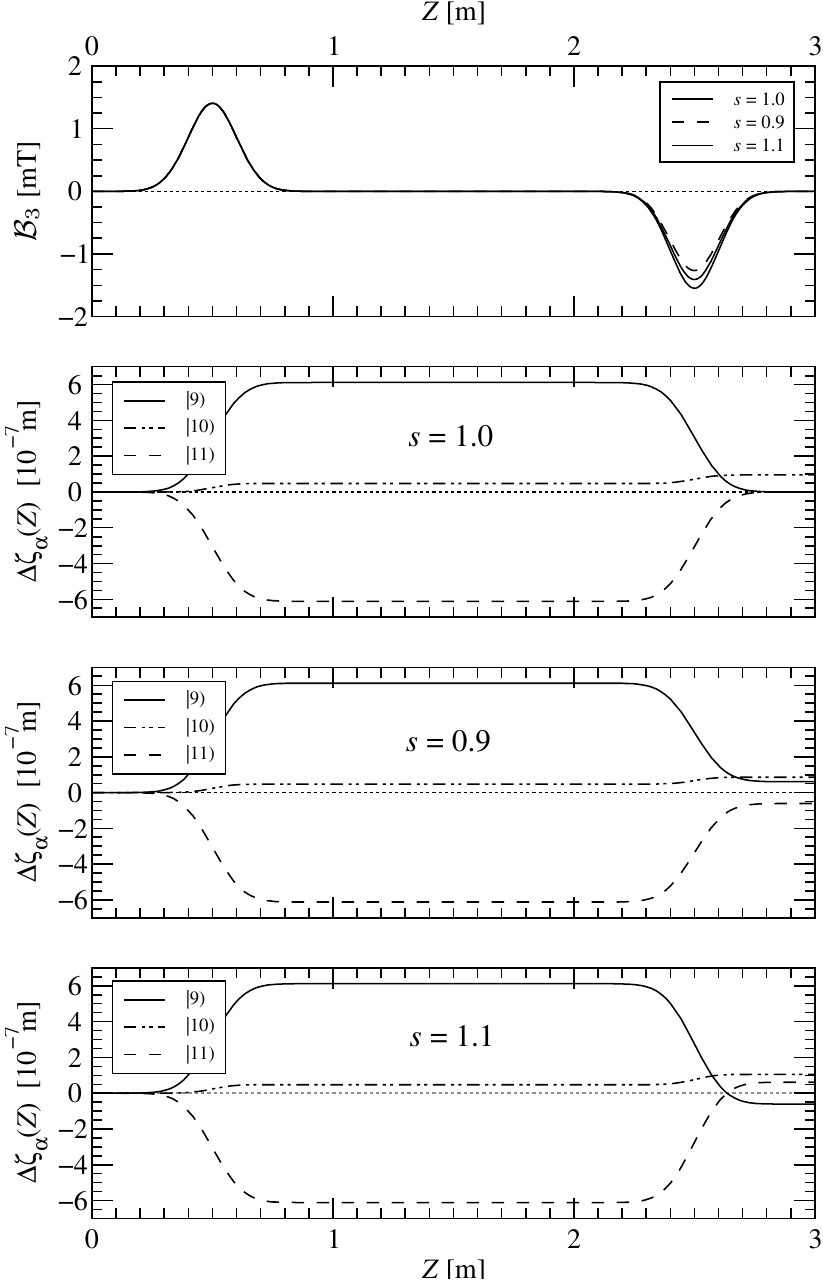}
\caption[Veranschaulichung der experimentellen Vorgehensweise beim lABSE.]{Veranschaulichung der experimentellen Vorgehensweise. Erläuterungen siehe Text.}\label{f6:FP}
\end{figure}
In Abschnitt \ref{s3:Fahrplan} haben wir noch den Standpunkt vertreten, dass
man den hypothetischen Detektor entlang der $Z$-Achse verschiebt, um durch die Kreuzungspunkte
(siehe z.B. Abb. \ref{f3:FPFlussBsp2} auf S. \pageref{f3:FPFlussBsp2}) zu fahren und
an jedem Ort den Fluss der Atome in einem bestimmten Zustand zu messen und aufzuintegrieren.
Normiert auf den Gesamtfluss der Atome durch den Detektor ergibt sich dann ein ortsabhängiges
Spinechosignal, das die Wahrscheinlichkeit angibt, dass sich ein Atom am Ort $Z$ im detektierten
Zustand befindet.

In der experimentellen Praxis geht man anders vor. Anstatt den Detektor tatsächlich zu verschieben
und die Messung an verschiedenen (relativ ungenau bekannten) Orten durchzuführen, platziert man
den Detektor an einem festen Ort $Z_D$. Dann variiert man durch eine sogenannte $\Delta\mc B$-Spule\footnote{
Hierbei handelt es sich um einige wenige Windungen, die zusätzlich um die zweite Spinecho-Spule
gewickelt werden. Durch Variation des Stroms variiert man das Feld der zusätzlichen Spule und somit
das gesamte zweite Spinechofeld.
}
(siehe \cite{DissAR}, Abschnitt 2.9.3, S. 32) das Magnetfeld in der zweiten Spinechospule, so dass sich die relative
Lage der Wellenpakete am Ende der zweiten Spule verändert. Dies kann man so lange tun, bis sich jeweils
zwei Wellenpakete überlappen, d.h. bis man einen Kreuzungspunkt am Ende des zweiten Magnetfelds vorliegen hat.
Durch die identischen Geschwindigkeiten im feldfreien Raum\footnote{Wie in Abschnitt \ref{s6:SG} diskutiert,
kann man dies durch im Allgmeinen unterschiedliche Gesamtenergien der Wellenpakete der atomaren Zustände erreichen.
Darüberhinaus ist bei einer Superposition entarteter, innerer atomarer Zustände im Anfangszustand sowohl die Gesamtenergie als auch die
kinetische Energie im feldfreien Raum identisch.} treffen beide Wellenpakete am Detektor ein,
ohne ihren Überlapp zu verändern. Durch Variation des $\Delta\mc B$-Feldes verändert man den Überlapp
am Detektor und fährt so das gesamte Spinechosignal durch.

In der Theorie wird die $\Delta\mc B$-Spule durch einen Parameter $s$ simuliert. In dem oberen Diagramm in
Abb. \ref{f6:FP} haben wir ein grundlegendes Spinechofeld dargestellt\footnote{ 
Beide Spinechofelder haben jeweils ein Magnetfeldintegral von $50\u{\mu T}\cdot\u m$ und weisen in $3$-Richtung.
}. Das zweite Spinechofeld ist stets antiparallel zum ersten Feld und mit diesem für $s=1$ betragsmäßig identisch.
In den unteren Diagrammen haben wir als Beispiel die adiabatischen Fahrpläne für die gegebenen Spinechofelder
und für metastabilen Wasserstoff angegeben. Die Nomenklatur ist dabei wie folgt,
\begin{align}
\rket{9} = \rket{2\hat S_{1/2},1,1,\mc B_3}\ ,\quad\rket{10} = \rket{2\hat S_{1/2},1,0,\mc B_3}\ ,\quad\rket{11} = \rket{2\hat S_{1/2},1,-1,\mc B_3}\ ,
\end{align}
siehe auch Anhang \ref{sB:H2}, Tabelle \ref{tB:H2States} auf S. \pageref{tB:H2States}. 
Die beiden Zahlen vor dem Magnetfeld $\mc B_3$ stehen für die Quantenzahlen des Gesamtdrehimpulses 
$F$ und dessen Projektion $F_3$ auf die $3$-Achse, 
die beide für das hier betrachtete Feld noch gute Quantenzahlen sind. Das Breit-Rabi-Diagramm für diese
Zustände kann Abb. \subref*{fB:BR-H2-2} auf Seite \pageref{fB:BR-H2} entnommen und gleicht von der
äußeren Gestalt her dem Breit-Rabi-Diagramm der Wasserstoff-Grundzustände.

Im Diagramm für $s=1$ in Abb. \ref{f6:FP} erkennt man, dass aufgrund der antisymmetrischen Felder und
der antisymmetrischen Energieabhängigkeit der Zustände $\rket{9}$ und $\rket{11}$ die Schwerpunkte der
zugehörigen Wellenpakete sich am Detektor genau treffen. Der Detektor befindet sich also genau am
sogenannten $9$-$11$-Kreuzungspunkt. Wählen wir nun $s=0.9$ erhalten wir ein leicht abgeschwächtes
zweites Spinechofeld und die Schwerpunkte der Zustände $\rket{9}$ und $\rket{11}$ treffen sich nicht mehr
am Ende des Feldes. Allerdings liegen nun die Schwerpunkte der Zustände $\rket{9}$ und $\rket{10}$ dichter beisammen
und man kann sich leicht vorstellen, dass bei einer weiteren Abschwächung des Feldes (also bei noch kleinerem Wert
für $s$) irgendwann der $9$-$10$-Kreuzungspunkt erreicht ist. Analog dazu findet man bei $s>1.1$ irgendwo den
$10$-$11$-Kreuzungspunkt, siehe unteres Diagramm in Abb. \ref{f6:FP}.

\begin{floatingfigure}[r]{9cm}
\centering\vspace{-5mm}
\includegraphics[width=9cm]{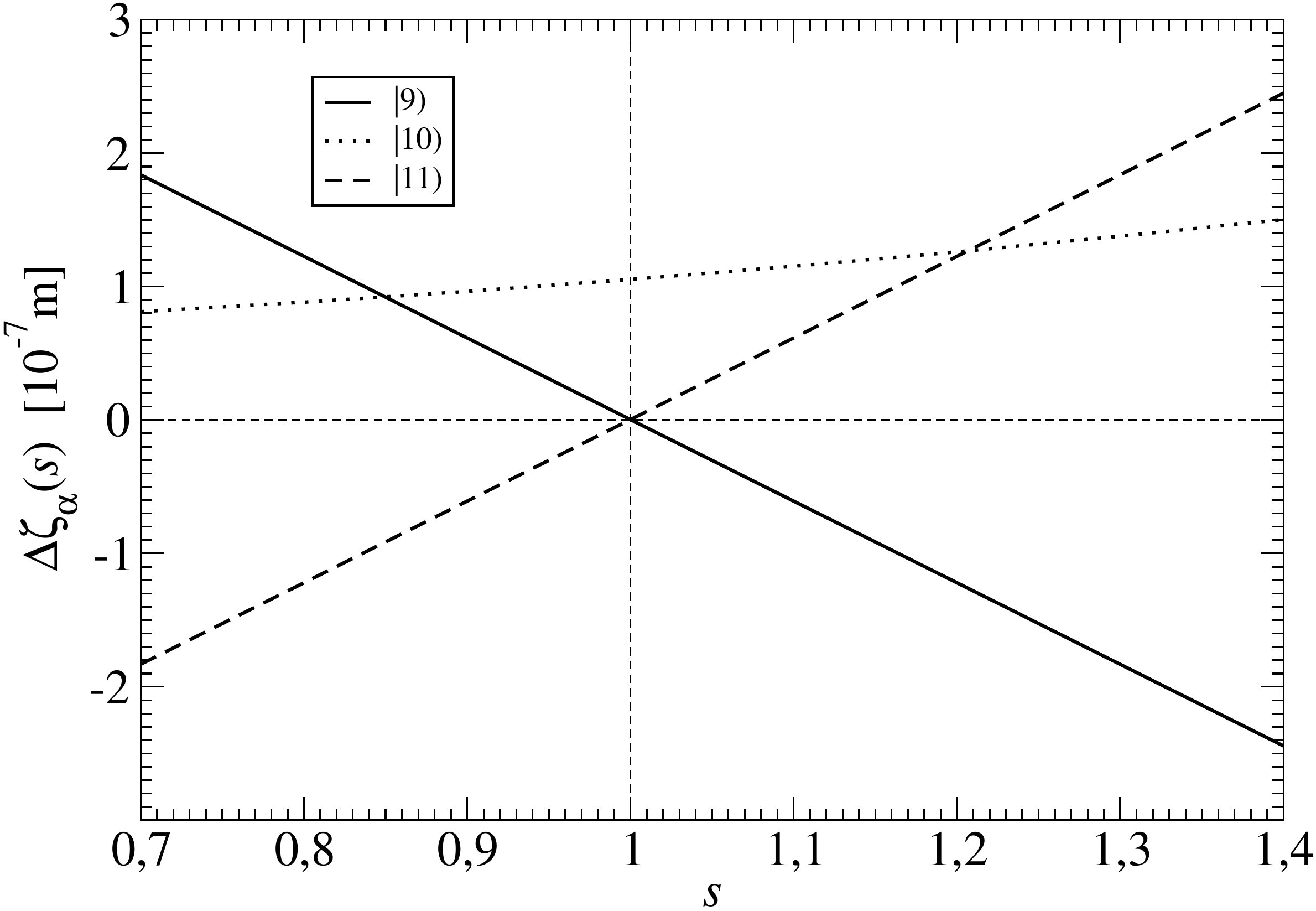}
\caption[Die Kreuzungspunkte in Abhängigkeit vom Skalierungsparameter $s$.]{Die Kreuzungspunkte in Abhängigkeit vom Skalierungsparameter $s$, der direkt mit dem
Spulenstrom $\Delta I$ der $\Delta\mc B$-Spule in Verbindung gebracht werden kann.}\label{f6:FPs}
\end{floatingfigure}
Man kann die Lage der Kreuzungspunkte am festen Ort $Z_D$ also in Abhängigkeit vom Parameter $s$ darstellen
und würde dann in dem hier betrachteten Beispiel das in Abb. \ref{f6:FPs} gezeigte Bild erhalten. Rein optisch
ähnelt dieses Bild stark dem in Abb. \ref{f3:FPFlussBsp2} auf S. \pageref{f3:FPFlussBsp2} gezeigten Fahrplan, 
der ortsabhängige Kreuzungspunkte für Wasserstoff im Grundzustand bei einem ähnlichen Magnetfeld ($s=1$)
darstellt. Wir würden daher in der $s$-Darstellung auch ein ähnliches Signal erhalten wie im ortsabhängigen Fall.
In experimentellen Arbeiten gibt es den Skalierungsparameter $s$ natürlich nicht. Er kann aber direkt mit
dem Strom $\Delta I$ in der $\Delta\mc B$-Spule in Verbindung gebracht werden, siehe hierzu \cite{DissAR},
Gl. (2.58) auf S. 32. Es gilt hiernach nämlich die Proportionalität
\begin{align}
\Delta\mc B \propto \Delta I
\end{align}
aber in unserem Fall sicher auch
\begin{align}
\Delta\mc B \propto (s-1)
\end{align}
also insgesamt $(s-1) \propto \Delta I$.

\subsubsection{Theoretische Vorgehensweise bei der Berechnung des lABSE-Signals}

Nachdem man mit der in Abschnitt \ref{s6:InitialConditions} beschriebenen Methode
die Einhüllende $\vph(\zeta)$ sowie die Koeffizienten $\hat\chi_\alpha$ und somit
die Anfangswellenpakete $\vph_\alpha(\zeta_\alpha(Z)-\tau(t))$ bestimmt hat,
kann man mit der Entwicklung, d.h. z.B. in erster Ordnung mit Hilfe von Gleichung
(\ref{e6:1.Ordnung}), den Gesamtzustand 
\begin{align}\label{e6:PsiDetektor}
\ket{\Psi(Z_D,s,t)} = \sum_\alpha \Psi_\alpha(Z_D,s,t)\rket{\alpha(Z_D)} 
= \sum_\alpha \e^{\I\phi_\alpha(Z_D,s,t)}A_\alpha(Z_D,s,t)\rket{\alpha(Z_D)}\ .
\end{align}
am Ort $Z_D$ des Detektors zu jeder Zeit $t$ und für jeden Wert des Skalierungsparameters $s$ 
berechnen. Wie in Abschnitt \ref{s6:Solution} erläutert, berücksichtigt man dabei die Dispersion
des Wellenpakets durch (numerische) Superposition der Entwicklungen für sehr breite
Wellenpakete.

Wir gehen wieder davon aus, dass $Z_D$ hinter allen angelegten Felder liege und können dann
\begin{align}\label{e6:Ende}
\rket{\alpha(Z_0)} = \rket{\alpha(Z_D)}\, \qquad(\alpha=1,\ldots,N)
\end{align}
voraussetzen (die geometrische Phase (\ref{e6:PhiGeom}) wurde ja explizit abgespalten).

Die Wellenfunktion, die der Detektor messen wird, hängt von seinem experimentellen Aufbau ab. 
Wir nehmen hier an, dass der Detektor auf einen gewissen atomaren Zustand 
\begin{align}\label{e6:chiDetektor}
  \ket{\chi'} := \sum_j \chi_j'\ket{j} = \sum_\alpha \hat\chi'_\alpha\rket{\alpha(Z_D)}\ ,
\end{align}
empfindlich sein soll. Dieser Zustand muss aus dem atomaren Gesamtzustand herausprojiziert
werden. Ist z.B. $\ket{\chi'} = \rket{\alpha(Z_D)}$ mit einem Index $\alpha=1,\ldots,N$,
so ist die Wahrscheinlichkeit für die Detektion dieses Zustands das Betragsquadrat des Koeffizienten 
dieses Zustands im atomaren Gesamtzustand $\ket{\Psi(Z_D,s,t)}$. Diesen können wir durch Projektion auf den linken
Eigenzustand $\lrbra{\alpha(Z_D)} = \lbra[X]{\chi'}$ aus $\ket{\Psi(Z_D,s,t)}$ herausfiltern.
Hieraus schließen wir, dass wir die detektierte Wellenfunktion 
für einen allgemeinen Zustand $\ket{\chi'}$ aus
\begin{align}\label{e6:Psi.pr}
\Psi\subt{pr}(Z_D,s,t) := \lbra[X]{\chi'}\Psi(Z_D,s,t)\rangle 
= \sum_\alpha\hat{\chi'}^*_\alpha\lrbra{\alpha(Z_D)}\Psi(Z_D,s,t)\rangle\ .
\end{align}
erhalten können. Man kann beim Berechnen dieses Ausdrucks dann die Orthogonalität der
linken und rechten Eigenvektoren ausnutzen, um die projizierte Wellenfunktion zu erhalten.
Wir wollen hier aber auch darauf hinweisen, dass die linken Eigenzustände der nichthermiteschen Massenmatrix
nicht als physikalische Zustände betrachtet werden können. Erwartungswerte von Operatoren beispielsweise 
sind nur bzgl. der rechten Eigenzustände definiert.

Mit der Wellenfunktion (\ref{e6:Psi.pr}) können wir die Wahrscheinlichkeitsdichte
\begin{align}\label{e6:rho.pr}
\rho\subt{pr}(s,t) := \abs{\Psi\subt{pr}(Z_D,s,t)}^2
\end{align}
und die Wahrscheinlichkeitsstromdichte
\begin{align}\label{e6:j.pr}
j\subt{pr}(s,t) := \Real\kle{\Psi^*\subt{pr}(Z,s,t)\frac{1}{\I M}\ddp{}{Z}\Psi\subt{pr}(Z,s,t)}_{Z=Z_D}
\end{align}
aufstellen, die in unserem (räumlich) eindimensionalen Fall identisch mit dem Fluss der Atome im
Zustand $\ket{\chi'}$ durch den Detektor ist, d.h.
\begin{align}\label{e6:Phi.pr}
\Phi\subt{pr}(s,t) \equiv j\subt{pr}(s,t)\ .
\end{align}
Hiermit folgt durch Integration über die Zeit (was äquivalent mit dem Zählen und anschließenden Normieren 
von detektierten Atomen ist) das gesuchte Spinechosignal
\begin{align}\label{e6:Probability}
P_{\chi'}(s) := \int_{t_0}^\infty\d t\ \Phi\subt{pr}(s,t)\ ,
\end{align}
das die vom Parameter $s$ abhängige Wahrscheinlichkeit darstellt, das Atom am Ort $Z_D$ im Zustand
$\ket{\chi'}$ zu finden. Die für ein Interferenzsignal typischen Oszillationen zeigen sich
dann bei Variation des Skalierungsparameters $s$, der wie bereits diskutiert 
den Überlapp der einzelnen Wellenpakete steuert.

Die Integration in (\ref{e6:Probability}) ist nur von dem Zeitpunkt $t_0$ an auszuführen, an dem 
der atomare Anfangszustand präpariert wurde. Gerade bei einem zerfallenden Wellenpaket macht eine Betrachtung
für Zeiten $t<t_0$ keinen Sinn, da der für den Zerfall verantwortliche Phasenfaktor dann ein exponentielles
Anwachsen der Amplitude verursachen würde und zu einer Norm des Wellenpakets, die größer als Eins ist.
Der Grund dafür liegt darin, dass die Untergrenzenfunktionen $u_\alpha(Z_D,t)$ für $t<0$ (wir nehmen hier o.B.d.A.
$t_0=0$ an) Werte größer als $Z_D$ liefern würde. Im für den Zerfall des Wellenpakets verantwortlichen Anteil
des Phasenwinkels, der eine Integration von $u_\alpha(Z_D,t)$ bis $Z_D$ über die lokale Zerfallsrate des Zustands
mit Index $\alpha$ ist, würde dann bei Vertauschung der Integralgrenzen ein zusätzliches Minuszeichen auftreten.

\chapter{Paritätsverletzende Berry-Phasen in Wasserstoff}\label{s7:Berry}

\section{Allgemeine Einführung in geometrische Phasen}\label{s7:Berry.Introduction}

Im Jahre 1984 führte M.V. Berry die nach ihm benannte geometrische Berry-Phase 
bei adiabatischen Zustandsänderungen in der Quantenmechanik ein \cite{Ber84}. 
Wir wollen in diesem Abschnitt die Grundlagen für die Behandlung
geometrischer Phasen angeben, wobei wir die Notation der Originalarbeit \cite{Ber84}
von Berry übernommen haben. 

Betrachten wir zunächst die Schrödinger-Gleichung eines rein zeitabhängigen Systems,
\begin{align}\label{e7:Berry.SG.1}
H(t)\ket{t} = \I\del_t\ket t\ .
\end{align}
Die Abhängigkeit des Hamiltonoperators von der Zeit sei nun implizit über die Zeitabhängigkeit verschiedener
Parameter gegeben, die wir hier zu einem $r$-dimensionalen Vektor $\v R(t) = \sum_{i=1}^r R_i(t)\v e_i$ zusammenfassen wollen.
Hiermit können wie den sogenannten Parameterraum, d.h. die Menge aller möglichen Konfigurationen des Parametervektors
$\v R(t)$, definieren. Betrachten wir einen sich adiabatisch ändernden Hamiltonoperator, so wird der Vektor $\v R(t)$
im Parameterraum entlang einer Kurve $\mc C$ laufen. Die Schrödinger-Gleichung lautet nun mit $H(t) = \hat H(\v R(t))$
\begin{align}\label{e7:Berry.SG.2}
\hat H(\v R(t))\ket{t} = \I\del_t\ket t\ .
\end{align}

Wir können nun für jeden Wert von $\v R$ das Eigenwertproblem des Hamiltonoperators lösen, d.h.
\begin{align}\label{e7:Berry.SG.stat}
\hat H(\v R)\ket{n(\v R)} = E_n(\v R)\ket{n(\v R)}\ ,\qquad(n=1,\ldots,N)\ .
\end{align}
Diese Eigenwertgleichung legt die Beziehung der relativen Phase der Zustände
$\ket{n(\v R)}$ für verschiedene $\v R$ nicht fest, wir können hierfür irgendeine differenzierbare,
lokale Phase (d.h. abhängig von $\v R$) wählen. Der Gesamtzustand $\ket t$ kann daher stets als Superposition
der lokalen Eigenvektoren von $\hat H(\v R)$ in der Form
\begin{align}\label{e7:Berry.State}
\ket t = \sum_n c_n\exp\klg{-\I\int_{t_0}^t\d t'\ E_n(\v R(t'))}\exp\klg{\I\gamma_n(t)}\ket{n(\v R(t))}
\end{align}
geschrieben werden. Hierbei wurde bereits die adiabatische Näherung verwendet, bei der die Koeffizienten $c_n$ zeitunabhängig
sind, d.h. ein Zustand \ket{t_0}, der in einem Eigenzustand $\ket{n(\v R(t_0))}$ präpariert ist, bleibt für alle
Zeiten $t$ in diesem (sich zeitlich verändernden) Zustand $\ket{n(\v R(t))}$.

Setzen wir nun $\ket t$ in die Schrödinger-Gleichung ein, so erhalten wir
\begin{align}
\begin{split}
&~\sum_n c_n \exp\klg{-\I\int_{t_0}^t\d t'\ E_n(\v R(t'))}\exp\klg{\I\gamma_n(t)}E_n(\v R(t))\ket{n(\v R(t))}\\
=&~\sum_n c_n \exp\klg{-\I\int_{t_0}^t\d t'\ E_n(\v R(t'))}\exp\klg{\I\gamma_n(t)}(E_n(\v R(t)) - \dot\gamma_n(t) + \I\del_t)\ket{n(\v R(t))}\ .
\end{split}
\end{align}
Die Beiträge proportional zu $E_n(\v R(t))$ können auf beiden Seiten subtrahiert werden. Multiplizieren wir von links mit
einem linken\footnote{Ist $\hat H(\v R(t))$ hermitesch, so ist der linke Bravektor natürlich gleich dem hermitesch
konjugierten, rechten Ketvektor. Da wir in der vorliegenden Arbeit aber auch nichthermitesche Matrixdarstellungen des
Hamiltonoperators berücksichtigen, wollen wir die Diskussion an dieser Stelle möglichst allgemein halten.
Wir weichen diesbezüglich von der Originalarbeit \cite{Ber84} ab.} 
Eigenzustand $\lbra{m(\v R(t))}$, so können wir die (hier angenommene) Orthogonalität der Eigenvektoren
von $\hat H(\v R(t))$ ausnutzen und erhalten schließlich
\begin{align}
\begin{split}
c_m\dot\gamma_m(t) &= \I\sum_m c_n \exp\klg{-\I\int_{t_0}^t\d t'\ (E_n(\v R(t'))-E_m(\v R(t')))}\\
&\times\exp\klg{\I(\gamma_n(t)-\gamma_m(t))}\lbra{m(\v R(t))}\del_t\ket{n(\v R(t))}\ .
\end{split}
\end{align}
Im adiabatischen Grenzfall gilt
\begin{align}
\lbra{m(\v R(t))}\del_t\ket{n(\v R(t))} \approx 0\ ,\quad(m\neq n)\ ,
\end{align}
da in diesem Fall nach Voraussetzung Übergänge zwischen den einzelnen Eigenzuständen aufgrund der zeitlichen Änderung
der äußeren Parameter vernachlässigen werden. Somit verbleibt
\begin{align}\label{e7:Berry.Phase.Dot}
\dot\gamma_n(t) = \I \lbra{n(\v R(t))}\del_t\ket{n(\v R(t))}
\end{align}
bzw.
\begin{align}\label{e7:Berry.Phase}
\gamma_n(t) = \I\int_{t_0}^t\d t'\ \lbra{n(\v R(t'))}\del_{t'}\ket{n(\v R(t'))} 
= \I\int_{\mc C(t_0,t)}\d\v R'\cdot\lbra{n(\v R')}\nab_{\v R'}\ket{n(\v R')}\ .
\end{align}
Dabei ist $\mc C(t_0,t)$ die Kurve im Parameterraum, die in der Zeit von $t_0$ bis $t$ von der Spitze des Parametervektors
$\v R(t')$ beschrieben wird. Ist der Weg bei einer Zeit $t=T$ geschlossen, d.h. $\v R(t_0) = \v R(T)$, so schreiben wir
einfach
\begin{align}\label{e7:Berry.Phase.Closed}
\gamma_n(\mc C) = \I\oint_{\mc C}\d\v R'\cdot\lbra{n(\v R')}\nab_{\v R'}\ket{n(\v R')}\ .
\end{align}
Ist der zugrundeliegende Hamiltonoperator hermitesch, also
\begin{align}
\lbra{n(\v R(t)))} = \ket{n(\v R(t))}\HC = \bra{n(\v R(t))}\ ,
\end{align}
so kann man aus der Normierung der Zustände leicht zeigen\footnote{
Es ist $1 = \bracket{n(\v R)}{n(\v R)}$, also $0 = \nab_{\v R}\bracket{n(\v R)}{n(\v R)} = 
\bra{n(\v R)}\overset{\leftharpoonup}\nab_{\v R}\ket{n(\v R)} + \bra{n(\v R)}\nab_{\v R}\ket{n(\v R)}$.
Damit folgt $\bra{n(\v R)}\overset{\leftharpoonup}\nab_{\v R}\ket{n(\v R)} = \bra{n(\v R)}\nab_{\v R}\ket{n(\v R)}^*
= -\bra{n(\v R)}\nab_{\v R}\ket{n(\v R)}$, also $\bra{n(\v R)}\nab_{\v R}\ket{n(\v R)}\in \I\mb R$.}, dass
\begin{align}\label{e7:Berry.Grad}
\bra{n(\v R')}\nab_{\v R'}\ket{n(\v R')} \in \I\mb R
\end{align}
gilt, also
\begin{align}
\gamma_n(t) \in \mb R\qquad (\text{falls}\ \hat H = \hat H\HC)\ .
\end{align}
Ist dagegen $\hat H \neq \hat H\HC$, so gilt zwar aufgrund der Normierung der rechten Eigenvektoren Gl. (\ref{e7:Berry.Grad})
immer noch, jedoch gilt mit der Normierung der linken und rechten Eigenvektoren auch
\begin{align}\label{e7:Berry.Grad.LR}
\lbra{n(\v R')}\nab_{\v R'}\ket{n(\v R')}^* = - \bra{n(\v R')}\nab_{\v R'}\lket{n(\v R')}
\end{align}
Im allgemeinen wird bei Verwendung nichthermitescher Hamiltonoperatoren die geometrische Phase
also einen Real- und einen Imaginärteil haben,
\begin{subequations}
\begin{align}\label{e7:Re.Berry}
\begin{split}
\Real \gamma_n(t) &= \tfrac12\klr{\gamma_n(t) + \gamma_n^*(t)}\\
&= \frac{\I}{2}\int_{\mc C(t_0,t)}\d\v R'\cdot\klr{
\lbra{n(\v R')}\nab_{\v R'}\ket{n(\v R')}-\lbra{n(\v R')}\nab_{\v R'}\ket{n(\v R')}^*}\ ,
\end{split}\\
\begin{split}\label{e7:Im.Berry}
\Imag \gamma_n(t) &= \tfrac1{2\I}\klr{\gamma_n(t) - \gamma_n^*(t)}\\
&= \frac{1}{2}\int_{\mc C(t_0,t)}\d\v R'\cdot\klr{
\lbra{n(\v R')}\nab_{\v R'}\ket{n(\v R')}+\lbra{n(\v R')}\nab_{\v R'}\ket{n(\v R')}^*}\ .
\end{split}
\end{align}
\end{subequations}

\section{Geometrische Phasen im Kontext der vorliegenden Arbeit}\label{s7:Berry.Context}

Bei der Lösung der matrixwertigen Schrödinger-Gleichung mit Hilfe des Entwicklungformalismus
in Kapitel \ref{s6:Formalismus} haben wir bereits den geometrischen Anteil (\ref{e6:PhiGeom}),
\begin{align}\label{e7:BP}
\gamma_\alpha(Z,t) := 
\Phi\supt{geom}_\alpha(Z,t) = \I\int_{u_\alpha(Z,t)}^Z\d Z'\ \lrbra{\alpha(Z')}\del_{Z'}\rket{\alpha(Z')}
\end{align}
im Phasenwinkel berücksichtigt. Wären wir bei der Umformung der Schrödinger-Gleichung in Abschnitt \ref{s6:SG}
zunächst von einem Phasenwinkel ohne geometrischen Anteil ausgegangen, so hätten wir in erster
Ordnung der Entwicklung (wie z.B. auch in Abschnitt \ref{s5:Anwendung}) sicher erkannt, dass
eine entsprechende Korrektur des Phasenwinkels nötig ist. Wir können also behaupten, 
dass bei dem in Kapitel \ref{s6:Formalismus} vorgestellten Formalismus zur Berechnung 
der exakten Lösung der Schrödinger-Gleichung die geometrische Phase automatisch auftritt.

Die in der vorliegenden Arbeit auftretende geometrische Phase (\ref{e7:BP}) ist direkt zu vergleichen mit
dem Ausdruck in Gl. (\ref{e7:Berry.Phase}). Wir haben hier allerdings die Orts- und Zeitabhängigkeit zu berücksichtigen.
In diesem Zusammenhang tritt an die Stelle der Zeitintegration die Integration über den Ort mit der üblichen,
zeitabhängigen Untergrenze $u_\alpha(Z,t)$ aus Gl. (\ref{e6:DefU}),
\begin{align}\label{e7:Def.u}
u_\alpha(Z,t) := \mc Z_\alpha(\zeta_\alpha(Z) - \tau(t))\ .
\end{align}
Setzen wir uns bei der Beobachtung des Wellenpakets an den Ort $Z_D$ des Detektors hinter allen äußeren Feldern,
so trägt zu einer Zeit $t$ nur das Ortsintervall $\kle{u_\alpha(Z_D,t),Z_D}$ zur geometrischen Phase bei, 
das in der Zeit $t$ bis zum Detektor vom Wellenpaket durchlaufen werden konnte.

Betrachten wir nun ausschließlich den adiabatischen Grenzfall, d.h. die nullte Ordnung der Entwicklung der
Wellenfunktion und somit die Wellenpakete (\ref{e6:0.Ordnung}),
\begin{align}
\Psi_\alpha(Z_D,t) = \e^{\I\phi_\alpha(Z_D,t)}\vph_\alpha(\zeta_\alpha(Z_D) - \tau(t))
\end{align}
für die einzelnen Komponenten des atomaren Gesamtzustands $\ket{\Psi(Z,t)}$. Wir wollen nun zeigen,
dass die in (\ref{e7:BP}) auftretende Integraluntergrenze bei der Berechnung eines Spinechosignals
unter einigen Annahmen durch die Konstante $Z_0$ ersetzt werden kann. Bei der Berechnung eines
solchen Signals ist nur der sehr kleine Zeitraum relevant, während dessen die Wellenpakete am Detektorort $Z_D$
eintreffen, also die $\vph_\alpha(\zeta_\alpha(Z_D) - \tau(t))$ signifikant von Null verschieden sind.

Der Schwerpunkt des Wellenpakets mit Index $\alpha$ wird sich nach Gl. (\ref{e6:Zeit}) zur Zeit 
\begin{align}
T_\alpha(Z_D) = \frac{M}{\bar k}\zeta_\alpha(Z_D) = \int_{Z_0}^{Z_D}\d Z\ \frac{M}{k_\alpha(Z)}
\end{align}
am Detektorort befinden. Wir können hier ohne große Einschränkungen fordern, dass die Feldkonfiguration
am Ort $Z_D$ des Detektors identisch mit der Feldkonfiguration in einem Bereich mit der Breite des 
Wellenpakets um den Anfangsort $Z_0$ sei. Wir stellen weiter fest, dass
\begin{align}
\begin{split}
u_\alpha(Z_D,T_\alpha(Z_D)) &= \mc Z_\alpha\klr{\zeta_\alpha(Z_D) - \frac{\bar k}{M}T_\alpha(Z_D)}\\
&= \mc Z_\alpha(\zeta_\alpha(Z_D) - \zeta_\alpha(Z_D)) = \mc Z_\alpha(0) = Z_0\ ,
\end{split}
\end{align}
gilt. Wir können also davon ausgehen, dass während der Zeit des Eintreffens des Wellenpakets am Ort des
Detektors der Wert der Untergrenzenfunktion $u_\alpha(Z_D,t)$ stets in dem Bereich um $Z_0$ liegt,
bei dem die gleiche Feldkonfiguration wie am Ort $Z_D$ herrscht. Der Wert der geometrischen Phase
wird sich dann während dieser Zeit nicht mehr verändern und wir können den vereinfachten Ausdruck
\begin{align}\label{e7:BP2}
\gamma_\alpha(\mc C) = \I\int_{Z_0}^{Z_D}\d Z \lrbra{\alpha(Z)}\del_{Z}\rket{\alpha(Z)}
\end{align}
betrachten, wobei $\mc C$ wieder die (geschlossene) Kurve im Parameterraum ist, die im Intervall von $Z_0$ bis
$Z_D$ durchlaufen wird.

Da wir nur lABSE-Experimente mit statischen elektrischen und magnetischen Feldern in Betracht ziehen wollen,
ist der Parameterraum in unserem Fall der sechsdimensionale Raum ($r=6$) der Feldstärken und der Parametervektor $\v R$
lässt sich in der Form
\begin{align}\label{e7:R}
  \v R = (\vmc E, \vmc B)
\end{align}
schreiben. Wir wollen nun in Analogie zur Vorgehensweise in der Originalarbeit von M.V. Berry \cite{Ber84},
die geometrische Phase (\ref{e7:BP2}) in ein Flächenintegral im Parameterraum umschreiben.
Wir benötigen dazu mathematische Hilfsmittel aus der Theorie der Differentialformen, nämlich die äußere Ableitung
\begin{align}\label{e7:ext.d}
d := \sum_{i=1}^r \d R_i\ddp{}{R_i}
\end{align}
und das äußere Produkt $\wedge$ (auch Keilprodukt\footnote{Engl.: {\em wedge product}.} genannt). Wir begnügen uns
hier mit der einfachen Anwendung der Rechenregeln für die äußere Ableitung und das Keilprodukt. Die mathematischen
Hintergründe können in den ersten Kapiteln des Buches über Differentialformen von H. Flanders \cite{Fla63}
nachgelesen werden.

Das Keilprodukt ist eine Verallgemeinerung des Kreuzprodukts, wobei es nicht auf Vektoren, sondern auf die sogenannten
1-Formen wirkt. Ein Beispiel für eine 1-Form sind die differentiellen Komponenten des Parametervektors $\d R_i$, die
z.B. in der Definition (\ref{e7:ext.d}) der äußeren Ableitung vorkommen. Die wesentliche Eigenschaft des Keilprodukts
ist die Antisymmetrie, d.h. man vereinbart
\begin{align}\label{e7:Keil.Antisymm}
\d R_i\wedge\d R_j = - \d R_j\wedge\d R_i\ ,
\end{align}
woraus man bereits die Eigenschaft
\begin{align}\label{e7:w0}
\d R_i\wedge\d R_j = 0\quad\text{für $i = j$}
\end{align}
ableiten kann. Eine weitere Konsequenz der Antisymmetrie ist das Verschwinden der zweifachen äußeren Ableitung einer
0-Form (d.h. einer gewöhnlichen Funktion des Parametervektors $\v R$). Es gilt nämlich das sogenannte Poincar\'e-Lemma
\begin{align}\label{e7:dd0}
\begin{split}
dd f(\v R) &= \sum_{i,j=1}^r \d R_i\wedge\d R_j\ddp{}{R_i}\ddp{}{R_j}f(\v R)\\
&= \sum_{i<j}\d R_i\wedge\d R_j\klr{\ddp{}{R_i}\ddp{}{R_j} - \ddp{}{R_j}\ddp{}{R_i}}f(\v R)\\
&= 0\ .
\end{split}
\end{align}
Hier wurde die Vertauschbarkeit der partiellen Ableitungen ausgenutzt. In der Terminologie der Theorie
der Differentialformen bezeichnet man also Funktionen wie $f(\v R)$ als 0-Formen und Ausdrücke, die
nur einfache differentielle Größen $\d R_i$ enthalten als 1-Form. Das $n$-fache Keilprodukt
von 1-Formen bezeichnet man als $n$-Form. Hiermit wollen wir diesen kleinen Ausflug in die Theorie der
Differentialformen abschließen und uns wieder der geometrischen Phase (\ref{e7:BP2}) zuwenden.

Wir betrachten die in Gl. (\ref{e6:D1Matrix}) definierten Matrixelemente der Ortsableitung,
\begin{align}
\uop D^{(1)}_{\alpha\beta}(Z) = \lrbra{\alpha(Z)}\del_Z\rket{\beta(Z)}\ .
\end{align}
Die geometrische Phase (\ref{e7:BP2}) ist das Integral über das Diagonalelement $\uop D^{(1)}_{\alpha\alpha}(Z)$
über eine mit $Z$ parametrisierte, geschlossene Kurve $\mc C$ im Parameterraum. Die Abhängigkeit der
Zustände von $Z$ geht nur implizit über die Abhängigkeit von den äußeren Feldern in die atomaren Zustände ein,
wir können also auch
\begin{align}\label{e7:Z.implicit}
\rket{\alpha(Z)} \equiv \rket{\alpha(\v R(Z))}\ ,\quad\lrbra{\alpha(Z)} \equiv \lrbra{\alpha(\v R(Z))}
\end{align}
schreiben. Das Matrixelement $\uop D^{(1)}_{\alpha\alpha}(Z)$ kann dann als
\begin{align}\label{e7:Z.to.R}
\uop D^{(1)}_{\alpha\alpha}(Z) = \lrbra{\alpha(\v R(Z))}\del_Z\rket{\alpha(\v R(Z))} 
= \sum_{i=1}^r\ \lrbra{\alpha(\v R)}\ddp{}{R_i}\rket{\alpha(\v R)}\big\vert_{\v R=\v R(Z)}\dd{R_i(Z)}{Z}
\end{align}
geschrieben werden und die geometrische Phase somit als
\begin{align}\label{e7:BP3}
\begin{split}
\gamma_\alpha(\mc C) &= \I\int_{Z_0}^{Z_D}\d Z\ \uop D_{\alpha\alpha}^{(1)}(Z)
= \I\oint_{\mc C}\d\v R\cdot\lrbra{\alpha(\v R)}\nab_{\v R}\rket{\alpha(\v R)}\\
&= \I\oint_{\mc C}\lrbracket{\alpha(\v R)}{d|\alpha(\v R)}\ .
\end{split}
\end{align}
Wir haben hier die kompakte Schreibweise mit dem Operator $d$ der äußeren Ableitung aus (\ref{e7:ext.d}) verwendet,
die wir auch im Folgenden beibehalten wollen. Es gilt dabei
\begin{align}
d\rket{\alpha(\v R)} = \sum_{i=1}^r\ddp{}{R_i}\rket{\alpha(\v R)}\d R_i\ .
\end{align}

Wir können nun den verallgemeinerten Stokes'schen Satz (\cite{Fla63}, Abschnitt 5.8, S. 64) anwenden und erhalten 
ein Flächenintegral über eine 2-Form über eine von der Kurve $\mc C$ umrandete Fläche $\mc F$ im Parameterraum:
\begin{align}\label{e7:BP.Stokes}
\gamma_{\alpha}(\mc C) = \I\oint_{\mc C = \del{\mc F}}\lrbracket{\alpha(\v R)}{d|\alpha(\v R)} 
= \I\int_{\mc F}d\lrbracket{\alpha(\v R)}{d|\alpha(\v R)}\ .
\end{align}
Mit den Rechenregeln für das Keilprodukt und die äußere Ableitung können wir den Integranden berechnen und erhalten
\begin{align}\label{e7:Integrand}
\begin{split}
d\lrbracket{\alpha(\v R)}{d|\alpha(\v R)} &= \klr{d\lrbra{\alpha(\v R)}}d\rket{\alpha(\v R)} 
+ \lrbracket{\alpha(\v R)}{dd|\alpha(\v R)}\\
&= \klr{d\lrbra{\alpha(\v R)}}d\rket{\alpha(\v R)} \\
&=\sum_{i,j}\lrbra{\alpha(\v R)}\klr{\overset{\leftharpoonup}\del_i\del_j}\rket{\alpha(\v R)}
\d R_i\wedge\d R_j\\
&=\sum_{i<j}\lrbra{\alpha(\v R)}\klr{\overset{\leftharpoonup}\del_i\del_j - \overset{\leftharpoonup}\del_j\del_i}
\rket{\alpha(\v R)}\d R_i\wedge\d R_j
\end{split}
\end{align}
Im ersten Schritt wurde das Poincar\'e-Lemma (\ref{e7:dd0}) angewendet und
im letzten Schritt die Antisymmetrie (\ref{e7:Keil.Antisymm}) des Keilprodukts ausgenutzt, 
um die Summe nur über $i<j$ laufen zu lassen. Wir haben hier außerdem die verkürzte Schreibweise
\begin{align}
\del_i = \ddp{}{R_i}
\end{align}
verwendet.

In Berrys Originalarbeit \cite{Ber84} 
wird an dieser Stelle eine Eins in der Form $\um=\sum_{m\neq n}\ket{m}\bra{m}$ eingeschoben.
Danach wird verwendet, dass bei Eigenzuständen $\ket{n}$ eines hermiteschen Hamiltonoperators die geometrische Phase 
wegen $\bracket{n(\v R)}{d|n(\v R)}\in\I\mb R$ reell ist. Wir haben bereits am Ende des letzten Abschnitts gezeigt, 
dass diese Argumentation bei den Eigenzuständen der nichthermiteschen Massenmatrix nicht mehr angewendet werden kann. 
In der Schreibweise mit der äußeren Ableitung lautet die zu (\ref{e7:Berry.Grad.LR}) analoge Gleichung
\begin{align}\label{e7:lb.da}
\lrbracket{\beta(\v R)}{d|\alpha(\v R)} = - \klr{d\lrbra{\beta(\v R)}}\rket{\alpha(\v R)}\ ,
\end{align}
sie folgt aus
\begin{align}\label{e7:d.lb.a}
d\lrbracket{\beta(\v R)}{\alpha(\v R)} = d\delta_{\beta\alpha} = 0\ .
\end{align}
Aus der Normierungsbedingung für rechte Eigenzustände, 
$\rbracket{\alpha(\v R)}{\alpha(\v R)} = 1$, erhält man zwar auch weiterhin
\begin{align}\label{e7:a.da}
\rbracket{\alpha(\v R)}{d|\alpha(\v R)} \in \I\mb R\ ,
\end{align}
diese Bedingung hilft uns hier jedoch nicht weiter, da wir für die Eigenzustände der nichthermiteschen Massenmatrix 
nur eine Vollständigkeitsrelation für Quasiprojektoren (\ref{eA:QProVollst}) zur Verfügung haben. Durch Einschieben
dieser Vollständigkeitsrelation erhalten wir aus (\ref{e7:Integrand}) unter Verwendung von (\ref{e7:d.lb.a})
\begin{align}
\begin{split}
d\lrbracket{\alpha(\v R)}{d|\alpha(\v R)} &= \klr{d\lrbra{\alpha(\v R)}}d\rket{\alpha(\v R)} 
= \sum_\beta \klr{d\lrbra{\alpha(\v R)}}\rket{\beta(\v R)}\lrbracket{\beta(\v R)}{d|\alpha(\v R)}\\
&= -\sum_\beta \lrbracket{\alpha(\v R)}{d|\beta(\v R)}\lrbracket{\beta(\v R)}{d|\alpha(\v R)}\ .
\end{split}
\end{align}
Der Beitrag zur Summe mit $\beta=\alpha$ verschwindet, denn es gilt
\begin{align}
\begin{split}
&\qquad\qquad\!\!\klr{d\lrbra{\alpha(\v R)}}\rket{\alpha(\v R)}\lrbracket{\alpha(\v R)}{d|\alpha(\v R)}\\
&\overref{(\ref{e7:d.lb.a})}= -\lrbracket{\alpha(\v R)}{d|\alpha(\v R)}\lrbracket{\alpha(\v R)}{d|\alpha(\v R)}\\
&\overref{(\ref{e7:ext.d})}=  -\sum_{i,j = 1}^r 
\lrbra{\alpha(\v R)}\del_i\rket{\alpha(\v R)}\lrbra{\alpha(\v R)}\del_j\rket{\alpha(\v R)}
\d R_i\wedge\d R_j\\
&\overref{(\ref{e7:Keil.Antisymm})}=  
-\sum_{i<j} \kle{\lrbra{\alpha(\v R)}\del_i\rket{\alpha(\v R)}\lrbra{\alpha(\v R)}\del_j\rket{\alpha(\v R)} 
- (i\leftrightarrow j)}\d R_i\wedge\d R_j\\
&\ \ \ =\ \ \  0
\end{split}
\end{align}
und somit lautet der Integrand von Gl. (\ref{e7:BP.Stokes})
\begin{align}\label{e7:Integrand2}
d\lrbracket{\alpha(\v R)}{d|\alpha(\v R)} 
&= -\sum_{\beta\neq\alpha} \lrbracket{\alpha(\v R)}{d|\beta(\v R)}\lrbracket{\beta(\v R)}{d|\alpha(\v R)}\ .
\end{align}

Nun werfen wir einen Blick auf die Eigenwergleichung für die Zustände $\rket{\alpha(\v R)}$,
\begin{align}\label{e7:EWG}
\uop M(\v R)\rket{\alpha(\v R)} = E_\alpha(\v R)\rket{\alpha(\v R)}
\end{align}
und gehen im folgenden stets von nichtentarteten Zuständen bei allen betrachteten Parametervektoren $\v R$ aus, 
d.h.
\begin{align}\label{e7:Nichtentartung}
\alpha\neq\beta\qquad\Longleftrightarrow\qquad E_\alpha(\v R)\neq E_\beta(\v R)\ .
\end{align}

Wenden wir hierauf die äußere Ableitung an, so folgt
\begin{align}\label{e7:d.EWG}
\begin{split}
d\Big(\uop M(\v R)\rket{\alpha(\v R)}\Big) &= \klr{d\uop M(\v R)}\rket{\alpha(\v R)} + \uop M(\v R)d\rket{\alpha(\v R)}\\
&= d\Big(E_\alpha(\v R)\rket{\alpha(\v R)}\Big)\\
&= \klr{d E_\alpha(\v R)}\rket{\alpha(\v R)} + E_\alpha(\v R)d\rket{\alpha(\v R)}\ .
\end{split}
\end{align}
Aus der Projektion von links mit $\lrbra{\alpha(\v R)}$ folgert man, dass die äußere Ableitung des Eigenwerts
$E_\alpha(\v R)$ gleich dem Diagonalelement der äußeren Ableitung der Massenmatrix ist:
\begin{align}\label{e7:dEa}
d E_\alpha(\v R) = \lrbra{\alpha(\v R)}\klr{d\uop M(\v R)}\rket{\alpha(\v R)}\ .
\end{align}
Aus der Projektion von links mit $\lrbra{\beta(\v R)}$ ($\beta\neq\alpha$) erhält man aus Gl. (\ref{e7:d.EWG})
\begin{align}\label{5:prop.mmod}
\lrbracket{\beta(\v R)}{d|\alpha(\v R)} = \frac{\lrbra{\beta(\v R)}\klr{d\uop M(\v R)}\rket{\alpha(\v R)}}{
E_\alpha(\v R)-E_\beta(\v R)}
\overref{(\ref{e7:lb.da})}= -\klr{d\lrbra{\beta(\v R)}}\rket{\alpha(\v R)}\ .
\end{align}
Hiermit können wir den Integranden (\ref{e7:Integrand2}) des Flächenintegrals der 
geometrischen Phase nach direkt umschreiben in
\begin{align}\label{e7:Integrand.final}
d\lrbracket{\alpha(\v R)}{d|\alpha(\v R)} =
\sum_{\beta\neq\alpha}\frac{
\lrbra{\alpha(\v R)}\klr{d\uop M(\v R)}\rket{\beta(\v R)}\lrbra{\beta(\v R)}\klr{d\uop M(\v R)}\rket{\alpha(\v R)}
}{\klr{E_\alpha(\v R)-E_\beta(\v R)}^2}\ .
\end{align}
Das Minuszeichen aus (\ref{e7:Integrand}) wurde hier durch die Vertauschung der Energieeigenwerte im Nenner von 
$\lrbracket{\alpha(\v R)}{d|\beta(\v R)}$ beseitigt. Mit diesem Ergebnis lautet die geometrische Phase (\ref{e7:BP.Stokes})
also unter der Voraussetzung (\ref{e7:Nichtentartung})
\begin{align}\label{e7:BP.final}
\gamma_\alpha(\mc C) &= \I\sum_{\beta\neq\alpha}\int_{\mc F}\frac{
\lrbra{\alpha(\v R)}\klr{d\uop M(\v R)}\rket{\beta(\v R)}\lrbra{\beta(\v R)}\klr{d\uop M(\v R)}\rket{\alpha(\v R)}
}{\klr{E_\alpha(\v R)-E_\beta(\v R)}^2}\ .
\end{align}
Diese Gleichung entspricht Gl. (10) in Berrys Originalarbeit \cite{Ber84}, die allerdings nur für einen dreidimensionalen
Parameterraum gilt. In der Diskussion dieser Gleichung macht Berry jedoch die Bemerkung, dass für einen höherdimensionalen
Parameterraum die Kreuzprodukte durch Keilprodukte zu ersetzen sind und die Gradienten durch äußere Ableitungen,
was dann genau auf eine zu (\ref{e7:BP.final}) äquivalente Form der Gl. (10) aus Berrys Arbeit führt.
Wir haben hier allerdings durch die Berücksichtigung nichthermitescher Hamiltonoperatoren ein noch allgemeineres
Resultat erhalten.

\section{Paritätsverletzende geometrische Phasen}\label{s7:PV}

\subsection{Paritätsverletzung in Atomen}\label{s7:APV}

Wie in der Einleitung (Kap. \ref{s1:Einleitung}) bereits erwähnt, wird Paritätsverletzung in Atomen
bereits seit 30 Jahren theoretisch und experimentell untersucht \cite{Bou74}. 
In dieser Arbeit, die eine Fortsetzung von \cite{BeNa83, BoBrNa95, DissTG, BrGaNa99, GaNa00, DiplTB} ist,
wollen wir im Gegensatz zu den üblicherweise betrachteten, schweren Atomen die leichtesten
Atome studieren, insbesondere Wasserstoff und Deuterium.

In \cite{BoBrNa95} und \cite{GaNa00} wurden Wege angegeben, wie man die P-verletzenden Effekte in 
leichten Atomen theoretisch so verstärken kann, dass sie prinzipiell auch experimentell nachweisbar
werden könnten. Wir haben die Hoffnung, dass die 
in dieser Arbeit untersuchte Atomstrahl-Spinecho-Methode \cite{ASSE1} verwendet werden kann, um
P-verletzende Effekte in Wasserstoff oder Deuterium zu messen. In diesem Zusammenhang besteht eine
enge Zusammenarbeit mit der Arbeitsgruppe von PD M. DeKieviet, PhD., dem wir an dieser Stelle
besonders für sein Engagement danken wollen.

Wir wollen hier nur einige Anmerkungen dazu machen, wie man Paritätsverletzung formal in der Massenmatrix $\uop M(Z)$
berücksichtigen kann und für genauere Informationen auf Anhang \ref{sB:M.Berechnung} und die dort zitierten Referenzen
verweisen. In der Diplomarbeit \cite{DiplTB} findet sich eine ausführliche Herleitung der Matrixelemente des
P-verletzenden Hamiltonoperators. Ausgangspunkt bildet dabei die effektive Lagrange-Dichte der vier-Fermion-Kopplung
zwischen Elektron und Quarks, die den neutralen, schwachen Strom enthält. Der aus diesem Teil der
Lagrange-Dichte folgende Hamiltonoperator kann mit der nichtrelativistischen Reduktion auf die in Atomen herrschenden 
Verhältnisse angepasst werden. Als Ergebnis erhält man einen P-verletzenden Hamiltonoperator
\begin{align}\label{e7:HPV}
\begin{split}
H\subt{PV} &= H^{(1)}\subt{PV} + H^{(2)}\subt{PV}\ ,\\
H^{(1)}\subt{PV} &= \frac{G}{4\sqrt2}\frac{Q_W^{(1)}}{m_e}\klg{\delta^3(\v x)(\v\sigma\cdot\v p) 
  + (\v\sigma\cdot\v p)\delta^3(\v x)}\ ,\\
H^{(2)}\subt{PV} &= \frac{G}{4\sqrt2}\frac{Q_W^{(2)}}{m_e}
\klg{\delta^3(\v x)(\v I\cdot\v p)(\v\sigma\cdot\v p) 
+ (\v\sigma\cdot\v p)(\v I\cdot\v p)\delta^3(\v x)}\ ,
\end{split}
\end{align}
wobei $\v x$ der Ortsoperator des Elektrons (relativ zum Ort des Atomkerns) ist, $\v p$ der Operator des Elektronimpulses,
$\v \sigma$ der Elektronspin und $\v I$ der Kernspin. Weiterhin treten die Fermi-Konstante $G$, die Elektronmasse
$m_e$ sowie die sogenannten schwachen Kernladungen $Q_W^{(1)}$ und $Q_W^{(2)}$ auf, die bereits in der Einleitung
diskutiert wurden.

Die Matrixelemente von $H\subt{PV}$, die sich bezüglich der Gesamtdrehimpulsbasis $\ket{2L_J,F,F_3}$ ergeben,
sind im Anhang \ref{sB:M.Berechnung} in Gl. (\ref{eB:HPV}) aufgeführt. Sie treten (natürlich) nur zwischen Zuständen
unterschiedlicher Parität auf, d.h. im Unterraum mit Hauptquantenzahl $n=2$ nur zwischen $S$- und $P$-Zuständen.
Da $H\subt{PV}$ als Skalarprodukt von Operatoren offenbar rotationsinvariant ist, gibt es weiterhin 
(im feldfreien Raum) nur Mischungen zwischen Zuständen mit gleichem Gesamtdrehimpuls $F$ und gleicher 
dritter Komponente $F_3$.

Die Matrixdarstellungen des Operators $H\subt{PV}$ wollen wir im Folgenden mit
\begin{align}
\Big(\bra{2L'_{J'},F',F_3'}H\PV\ket{2L_{J},F,F_3}\Big) = \delta_1\uop M^{(1)}\PV + \delta_2\uop M^{(2)}\PV
\end{align}
bezeichnen, wobei wir die in (\ref{eB:HPV}) auftretenden, P-verletzenden Parameter $\delta_1$ und $\delta_2$
(siehe Gl. (\ref{eB:PV.Parameter}) aus Anhang \ref{sB:Beitraege})
explizit ausgeklammert haben, da wir später eine Störungsentwicklung nach diesen sehr kleinen Parametern
machen wollen. In der nun folgenden Rechnung werden wir
stellvertretend für diese Summe von Matrizen die Schreibweise $\delta\uop M\PV$ verwenden.

Die gesamte Massenmatrix für das zu beschreibende (neutrale) Atom in einem statischen elektrischen und
magnetischen Feld, die sich aus den Rechnungen in Anhang \ref{sB:M.Berechnung}
ergibt\footnote{Eine sehr ausführliche Herleitung und Diskussion des Hamiltonoperators eines Atoms
in einem allgemeinen elektrischen und magnetischen Feld findet sich in \cite{DissTG}, Anhang A, S. 131ff.}, lautet also
\begin{align}\label{e7:MM}
\uop M(\v R(Z),\delta_1,\delta_2) 
= \uop M(\vmc E(Z),\vmc B(Z),\delta_1,\delta_2) = \uop M_0(\delta_1,\delta_2) 
- \vu D\cdot\vmc E(Z) - \vu\mu\cdot\vmc B(Z)\ ,
\end{align}
wobei der freie Teil $\uop M_0\delta_1,\delta_2)$ der Massenmatrix die P-verletzenden Beiträge enthält, d.h.
\begin{align}\label{e7:MM0}
\uop M_0(\delta_1,\delta_2) = \uop{\tilde M}_0 + \delta_1\uop M^{(1)}\PV + \delta_2\uop M^{(2)}\PV \sim \uop{\tilde M}_0 + \delta\uop M\PV\ .
\end{align}

\subsection{Geometrische Flussdichten im Raum der elektrischen und magnetischen Feldstärke}\label{s7:Fluxes}

Wir kehren nun zurück zur geometrischen Phase aus Gl. (\ref{e7:BP.final}). Dort wollen wir die äußere Ableitung
der Massenmatrix nun explizit einsetzen. In den folgenden Gleichungen haben wir die P-verletzenden Parameter
$\delta_{1,2}$ nicht explizit ausgeschrieben, da die hier berechneten Formeln allgemein gültig sind. 
Wir werden im nächsten Abschnitt die Massenmatrix und
ihre Eigenzustände durch ihre Störungsentwicklung nach $\delta_{1,2}$ ersetzen und dann sehen, dass die Formeln
dieses Abschnitts formal für den P-erhaltenden Beitrag nullter Ordnung übernommen werden können.
Wir weisen darauf hin, dass lokale Matrixelemente der Gestalt $\uop M_{\alpha\beta}(\v R)$ in diesem Abschnitt
stets bzgl. der Basis der lokalen Eigenzustände $\rket{\alpha(\v R)}$ der vollen Massenmatrix 
(inkl. der P-verletzenden Beiträge) zu verstehen sind. Im
nächsten Abschnitt werden wir dagegen vereinbaren, dass diese Matrixelemente nur bzgl. der
Eigenzustände des P-erhaltenden Anteils der Massenmatrix zu verstehen sind.

Mit $\uop M(\v R)$ aus (\ref{e7:MM}) und $\v R = (\vmc E,\vmc B)$ folgt
\begin{align}
d\uop M(\v R) = \sum_{i=1}^r \ddp{\uop M(\v R)}{R_i}\d R_i = -\vu D\cdot\d\vmc E - \vu\mu\cdot\d\vmc B\ .
\end{align}
Definieren wir nun also die lokalen Matrixelemente des Dipoloperators $\vu D$ und des Operators des magnetischen Moments
$\vu\mu$ als
\begin{align}
\vu D_{\beta\alpha}(\v R) := \lrbra{\beta(\v R)}\vu D\rket{\alpha(\v R)}\ ,\quad
\vu\mu_{\beta\alpha}(\v R) := \lrbra{\beta(\v R)}\vu\mu\rket{\alpha(\v R)}\ ,
\end{align}
so folgt für die geometrische Phase (\ref{e7:BP.final})
\begin{align}\label{e7:BP.D.Mu}\hspace{-8mm}
\gamma_\alpha(\mc C) &= \I\sum_{\beta\neq\alpha}\int_{\mc F}\frac{
\klr{\vu D_{\alpha\beta}(\v R)\cdot\d\vmc E + \vu\mu_{\alpha\beta}(\v R)\cdot\d\vmc B}\wedge
\klr{\vu D_{\beta\alpha}(\v R)\cdot\d\vmc E + \vu\mu_{\beta\alpha}(\v R)\cdot\d\vmc B}
}{\klr{E_\alpha(\v R)-E_\beta(\v R)}^2}\ .
\end{align}
Multiplizieren wir das Keilprodukt aus, so können wir die geometrische Phase in der Form
\begin{align}
\gamma_\alpha(\mc C) = \int_{\mc F}\klr{\mc I_\alpha^{(\mc E)}(\v R)+\mc I_\alpha^{(\mc B)}(\v R)
+\mc I_\alpha^{(\mc E,\mc B)}(\v R)}
\end{align}
schreiben, wobei die darin vorkommenden 2-Formen definiert sind als
\begin{align}\label{e7:I.E}
\begin{split}
\mc I^{(\mc E)}_\alpha(\v R) &= \I\sum_{\beta\neq\alpha}\frac{\klr{\vu D_{\alpha\beta}(\v R)\cdot\d\vmc E}
\wedge\klr{\vu D_{\beta\alpha}(\v R)\cdot\d\vmc E}}{\klr{E_\alpha(\v R)-E_\beta(\v R)}^2}\\
&= \I\sum_{\beta\neq\alpha}\sum_{i,j=1}^3 \frac{\unl D_{i,\alpha\beta}(\v R)\unl D_{j,\beta\alpha}(\v R)}{
\klr{E_\alpha(\v R)-E_\beta(\v R)}^2}\d\mc E_i\wedge\d\mc E_j\\
&= \frac\I2\sum_{\beta\neq\alpha}\sum_{i,j=1}^3\frac{\unl D_{i,\alpha\beta}(\v R)\unl D_{j,\beta\alpha}(\v R) - (i\leftrightarrow j)}{\klr{E_\alpha(\v R)-E_\beta(\v R)}^2}
\d\mc E_i\wedge\d\mc E_j\ ,
\end{split}\\[5mm] \label{e7:I.B}
\begin{split}
\mc I^{(\mc B)}_\alpha(\v R) &= \I\sum_{\beta\neq\alpha}\frac{\klr{\vu\mu_{\alpha\beta}(\v R)\cdot\d\vmc B}
\wedge\klr{\vu\mu_{\beta\alpha}(\v R)\cdot\d\vmc B}}{\klr{E_\alpha(\v R)-E_\beta(\v R)}^2}\\
&= \frac\I2\sum_{\beta\neq\alpha}\sum_{i,j=1}^3\frac{\unl\mu_{i,\alpha\beta}(\v R)\unl\mu_{j,\beta\alpha}(\v R) - (i\leftrightarrow j)}{\klr{E_\alpha(\v R)-E_\beta(\v R)}^2}
\d\mc B_i\wedge\d\mc B_j
\end{split}
\end{align}
und
\begin{align}\label{e7:I.E.B}
\begin{split}\hspace{-5mm}
\mc I^{(\mc E,\mc B)}_\alpha(\v R) &= \I\sum_{\beta\neq\alpha}\Bigg[
\frac{\klr{\vu D_{\alpha\beta}(\v R)\cdot\d\vmc E}\wedge\klr{\vu\mu_{\beta\alpha}(\v R)\cdot\d\vmc B}}{
\klr{E_\alpha(\v R)-E_\beta(\v R)}^2}\\
&\quad\qquad -\frac{\klr{\vu D_{\beta\alpha}(\v R)\cdot\d\vmc E}\wedge\klr{\vu\mu_{\alpha\beta}(\v R)\cdot\d\vmc B}}{
\klr{E_\alpha(\v R)-E_\beta(\v R)}^2}
\Bigg]\\
&= \I\sum_{\beta\neq\alpha}\sum_{i,j=1}^3
\frac{\unl D_{i,\alpha\beta}(\v R)\unl\mu_{j,\beta\alpha}(\v R) 
- \unl\mu_{j,\alpha\beta}(\v R)\unl D_{i,\beta\alpha}(\v R)}{
\klr{E_\alpha(\v R)-E_\beta(\v R)}^2}\d\mc E_i\wedge\d\mc B_j\ .
\end{split}
\end{align}
Die ersten beiden 2-Formen, $\mc I^{(\mc E)}_\alpha(\v R)$ und $\mc I^{(\mc B)}_\alpha(\v R)$, sind das Produkt
zweier antisymmetrischer Tensoren zweiter Stufe (mit Indizes $i$ und $j$), also vom Typ
\begin{align}
\mc I = \sum_{i,j=1}^3 Q_{ij}R_{ij}\ ,\quad Q_{ij} = - Q_{ji}\ ,\quad R_{ij}=-R_{ji}\ .
\end{align}
Wir können diese Antisymmetrie unter Verwendung des Levi-Civita-Tensors explizit machen, indem wir
\begin{align}
Q_{ij} = \sum_{k=1}^3 \vep_{ijk}\tilde Q_k\ ,\qquad
R_{ij} = \sum_{\ell=1}^3 \vep_{ij\ell}\tilde R_\ell\ ,
\end{align}
schreiben. Hiermit ergibt sich
\begin{align}\label{e7:I.Skalarprodukt}
\mc I = \sum_{i,j=1}^3\sum_{k,\ell=1}^3 
\vep_{ijk}\vep_{ij\ell}\tilde Q_k\tilde R_\ell = 2\sum_{k,\ell=1}^3\delta_{k\ell}\tilde Q_k\tilde R_\ell
= 2\tilde{\v Q}\cdot\tilde{\v R}\ .
\end{align}
Hierbei wurde die bekannte Relation $\sum_{i,j=1}^3 \vep_{ijk}\vep_{ij\ell} = 2\delta_{k\ell}$ verwendet.
Die Komponenten $\tilde Q_k$ und $\tilde R_\ell$ berechnen sich ebenfalls mit Hilfe dieser Relation zu
\begin{align}\label{e7:Def.Vektoren}
\tilde Q_k = \frac12\sum_{i,j=1}^3\vep_{ijk}Q_{ij}\ ,\quad
\tilde R_\ell = \frac12\sum_{i,j=1}^3\vep_{ij\ell}R_{ij}\qquad (k,\ell = 1,2,3)\ .
\end{align}
Aus der antisymmetrischen 2-Form $\d\mc E_i\wedge\d\mc E_j$ aus Gl. (\ref{e7:I.E}) können wir damit das gerichtete
Flächenelement
\begin{align}\label{e7:FE.E}
\d f_\ell^{(\mc E)} := \frac12\sum_{i,j=1}^3\vep_{ij\ell}\d\mc E_i\wedge\d\mc E_j\ ,\qquad(\ell = 1,2,3)
\end{align}
bilden und analog dazu natürlich auch
\begin{align}\label{e7:FE.B}
\d f_\ell^{(\mc B)} := \frac12\sum_{i,j=1}^3\vep_{ij\ell}\d\mc B_i\wedge\d\mc B_j\ ,\qquad(\ell = 1,2,3)\ .
\end{align}
Aus den antisymmetrischen Tensorkomponenten in den eckigen Klammern von Gl. (\ref{e7:I.E}) und (\ref{e7:I.B})
können wir Vektorfelder
\begin{align}\label{e7:Flux.E}
{\mc J}_{\alpha,\ell}^{(\mc E)}(\v R) := \frac{\I}{2}\sum_{i,j=1}^3\vep_{ij\ell}\klr{\sum_{\beta\neq\alpha}
\frac{\unl D_{i,\alpha\beta}(\v R)\unl D_{j,\beta\alpha}(\v R) - (i\leftrightarrow j)}{
\klr{E_\alpha(\v R)-E_\beta(\v R)}^2}}\ ,\qquad(\ell = 1,2,3)
\end{align}
und
\begin{align}\label{e7:Flux.B}
{\mc J}_{\alpha,\ell}^{(\mc B)}(\v R) := \frac{\I}{2}\sum_{i,j=1}^3\vep_{ij\ell}\klr{\sum_{\beta\neq\alpha}
\frac{\unl\mu_{i,\alpha\beta}(\v R)\unl\mu_{j,\beta\alpha}(\v R) - (i\leftrightarrow j)}{
\klr{E_\alpha(\v R)-E_\beta(\v R)}^2}}\ ,\qquad(\ell = 1,2,3)
\end{align}
definieren und erhalten dann gemäß (\ref{e7:I.Skalarprodukt})
\begin{align}
\mc I^{(\mc E)}_\alpha(\v R) = \v{\mc J}_{\alpha}^{(\mc E)}(\v R)\cdot\d\v f^{(\mc E)}\ ,\qquad
\mc I^{(\mc B)}_\alpha(\v R) = \v{\mc J}_{\alpha}^{(\mc B)}(\v R)\cdot\d\v f^{(\mc B)}\ .
\end{align}
Der Faktor zwei, der hier im Vergleich zu (\ref{e7:I.Skalarprodukt}) fehlt, 
wurde bereits bei der Definition der ${\mc J}_{\alpha,\ell}(\v R)$ berücksichtigt, wo
wir den Faktor $1/2$ aus (\ref{e7:Def.Vektoren}) unterschlagen haben.

Wir können diese Rechnung leider nicht auf den Beitrag $\mc I^{(\mc E,\mc B)}_\alpha(\v R)$ aus Gl. (\ref{e7:I.E.B})
übertragen, da dieser Ausdruck aufgrund der gemischten 2-Form $\d\mc E_i\wedge\d\mc B_j$ nicht als Produkt antisymmetrischer
Tensoren geschrieben werden kann. Die geometrische Phase lautet mit den bisherigen Ergebnissen also
\begin{align}\label{e7:BP.Flux}
\gamma_\alpha(\mc C) = \int_{\mc F}\v{\mc J}_{\alpha}^{(\mc E)}(\v R)\cdot\d\v f^{(\mc E)}
+ \int_{\mc F}\v{\mc J}_{\alpha}^{(\mc B)}(\v R)\cdot\d\v f^{(\mc B)}
+ \int_{\mc F}\mc I^{(\mc E,\mc B)}_\alpha(\v R)\ .
\end{align}

Besonders interessant an diesem Ergebnis sind die ersten beiden Beiträge. Hier haben wir offenbar zwei separate
Integrale über Flächen im Raum der elektrischen bzw. magnetischen Feldstärke. Die Integranden sind dreidimensionale
Vektorfelder, die wir als Flussdichten interpretieren können und die wir im folgenden kurz als
geometrische Flussdichten bezeichnen wollen. Bei konstantem elektrischen (magnetischen) Feld ist
die geometrische Phase also durch den Gesamtfluss durch die geschlossene Kurve $\mc C$ im Raum der
magnetischen (elektrischen) Feldstärke gegeben. Durch visuelle Darstellung der in (\ref{e7:Flux.E}) und
(\ref{e7:Flux.B}) definierten Flussdichten können wir also einerseits zu einem tieferen Verständnis der 
zugrundeliegenden Physik kommen und darüberhinaus auch eine Entscheidung über die genaue Lage der geschlossenen
Kurve $\mc C$ im Parameterraum treffen. Diese Möglichkeit der Visualisierung ist besonders interessant im Hinblick
auf die Aufspaltung der Flussdichten in einen P-erhaltenden und P-verletzenden Anteil, die wir im Folgenden untersuchen
wollen. Beide Beiträge zur geometrischen Phase können dann separat untersucht werden. 

Der gemischte Beitrag zur geometrischen Phase in Gl. (\ref{e7:BP.Flux}) kann nicht auf diese Weise visualisiert werden.
Hier kann man lediglich jeweils zwei Komponenten des elektrischen bzw. magnetischen Feldes als konstant betrachten und
in der verbleibenden Fläche im Parameterraum den Integranden untersuchen.

\subsection{Die P-erhaltenden und P-verletzenden geometrischen Flussdichten}\label{s7:PC.PV.Fluxes}

Wir knüpfen nun an die Diskussion aus Abschnitt \ref{s7:APV} an und betrachten die Massenmatrix
\begin{align}\label{e7:MMPV}
\uop M(\v R,\delta) 
= \uop M(\vmc E,\vmc B,\delta) = \uop{\tilde M}_0 - \vu D\cdot\vmc E - \vu\mu\cdot\vmc B + \delta\uop M\PV
\end{align}
mit der symbolischen Schreibweise für den P-verletzenden Anteil. Die Eigenzustände der vollen Massenmatrix
erfüllen die Gleichung
\begin{align}
\uop M(\vmc E,\vmc B,\delta)\rket{\alpha(\vmc E,\vmc B,\delta)} 
= E_\alpha(\vmc E,\vmc B,\delta)\rket{\alpha(\vmc E,\vmc B,\delta)}\ .
\end{align}
Nun wollen wir den P-verletzenden Anteil der zuvor berechneten geometrischen Flussdichten separat behandeln und müssen dazu
die Eigenzustände nach dem betragsmäßig kleinen Parameter $\delta$ in eine Störungsreihe entwickeln.
Wir verwenden dabei die Formeln aus Anhang \ref{sA:StoeRe} und entwickeln nur bis zur ersten Ordnung in
$\delta$. Wir erhalten dann
\begin{align}\label{e7:R.stoer}
\rket{\alpha(\vmc E,\vmc B,\delta)} = \rket{\alpha^{(0)}(\vmc E,\vmc B)} 
+ \delta\rket{\alpha^{(1)}(\vmc E,\vmc B)} + \OO(\delta^2)
\end{align}
und analog
\begin{align}\label{e7:L.stoer}
\lrbra{\alpha(\vmc E,\vmc B,\delta)} = \lrbra{\alpha^{(0)}(\vmc E,\vmc B)} 
+ \delta\lrbra{\alpha^{(1)}(\vmc E,\vmc B)} + \OO(\delta^2)
\end{align}
sowie
\begin{align}\label{e7:E.stoer}
E_\alpha(\vmc E,\vmc B,\delta) = E_\alpha^{(0)}(\vmc E,\vmc B) + \delta E_\alpha^{(1)}(\vmc E,\vmc B) + \OO(\delta^2)
\end{align}
und mit Gl. (\ref{eA:ProEntOrt}) aus Anhang \ref{sA:StoeRe} (für nichtentartete Zustände, siehe 
Voraussetzung (\ref{e7:Nichtentartung}))
\begin{align}
\rket{\alpha^{(1)}(\vmc E,\vmc B)} = \sum_{\gamma\neq\alpha}\rket{\gamma^{(0)}(\vmc E,\vmc B)}
\frac{\uop M\PVU_{\gamma\alpha}(\vmc E,\vmc B)}{E^{(0)}_\alpha(\vmc E,\vmc B)-E^{(0)}_\gamma(\vmc E,\vmc B)}\ ,\\
\lrbra{\alpha^{(1)}(\vmc E,\vmc B)} = \sum_{\gamma\neq\alpha}
\frac{\uop M\PVU_{\alpha\gamma}(\vmc E,\vmc B)}{E^{(0)}_\alpha(\vmc E,\vmc B)-E^{(0)}_\gamma(\vmc E,\vmc B)}
\lrbra{\gamma^{(0)}(\vmc E,\vmc B)}\ ,
\end{align}
wobei hier und auch im weiteren Verlauf dieses Kapitels die 
Matrixelemente in der lokalen Basis immer bzgl. der ungestörten Zustände zu
verstehen sind, also z.B.
\begin{align}
\uop M\PVU_{\alpha\beta}(\vmc E,\vmc B) := \lrbra{\alpha^{(0)}(\vmc E,\vmc B)}\uop M\PV\rket{\beta^{(0)}(\vmc E,\vmc B)}
\end{align}
und analog für $\vu D_{\alpha\beta}(\vmc E,\vmc B)$ und $\vu\mu_{\alpha\beta}(\vmc E,\vmc B)$.
Die erste Ordnung der Störungsentwicklung für den Energieeigenwert lautet nach Gl. (\ref{eA:Energieentwicklung})
\begin{align}
E_\alpha^{(1)}(\vmc E,\vmc B) = \lrbra{\alpha^{(0)}(\vmc E,\vmc B)}\uop M\PV\rket{\alpha^{(0)}(\vmc E,\vmc B)}
= \uop M\PVU_{\alpha\alpha}(\vmc E,\vmc B)\ .
\end{align}
In der Basis der Gesamtdrehimpulszustände $\ket{2L_J,F,F_3}$ hat die Matrix $\uop M\PV$ keine Diagonalelemente
und die Korrekturen der Energie sind quadratisch in $\delta$. Für Felder mit einer gewissen Symmetrie unter Spiegelungen,
kann man sogar allgemein zeigen, dass die Eigenenergien $E_\alpha(\vmc E,\vmc B,\delta)$ nur von
geraden Potenzen von $\delta$ abhängen (\cite{BoBrNa95}, Abschnitt 3.4). Für allgemeine Felder wird dies jedoch nicht
der Fall sein \cite{DissTG,BrGaNa99}. In unserem Fall wird die Vernachlässigung der ersten Ordnung aufgrund der schwachen Felder
und der Tatsache, dass wir hier lediglich quadratische Energiedifferenzen in den Nennern der geometrischen Flussdichten 
betrachten, dennoch eine sehr gute Näherung darstellen. Tatsächlich zeigt die numerische Berechnung der lokalen
Matrixdarstellung, dass selbst bei wesentlich höheren Feldern als wir sie hier betrachten, nämlich bei 
$\vmc B = \OO(1\u T)$ und $\vmc E = \OO(100\u{V/m})$, die Diagonalelemente der P-verletzenden Matrizen 
in der lokalen Darstellung immer noch viele Größenordnungen kleiner als die Nebendiagonalelemente sind.
Wir setzen im Folgenden also stets
\begin{align}
E_\alpha(\vmc E,\vmc B,\delta) \approx E_\alpha^{(0)}(\vmc E,\vmc B)
\end{align}
voraus.

\newcommand{\PCU}{\supt{PC}}

Um nun die geometrischen Flussdichten in P-erhaltenden Teil und P-verletzenden Teil aufspalten zu können, müssen wir
alle lokalen Matrixelemente der Operatoren für das Dipolmoment und das magnetische Moment nach $\delta$ entwickeln.
Wir betrachten exemplarisch für beide Operatoren einen Vektoroperator $\vu\eta$ und dessen Matrixelement
\begin{align}\label{e7:eta}
\lrbra{\alpha(\vmc E, \vmc B,\delta)}\unl\eta_i\rket{\beta(\vmc E,\vmc B,\delta)}
= \unl\eta\PCU_{i,\alpha\beta}(\vmc E, \vmc B) + \delta\unl\eta\PVU_{i,\alpha\beta}(\vmc E, \vmc B) + \OO(\delta^2)\ ,
\quad (i=1,2,3)\ ,
\end{align}
das wir in einen P-erhaltenden (Superskript PC für {\em parity conserving}) und einen P-verletzenden Anteil aufgespaltet
haben. Mit der Entwicklung der Eigenzustände folgt
\begin{align}\label{e7:eta.PC}
\unl\eta\PCU_{i,\alpha\beta}(\vmc E, \vmc B) \equiv \unl\eta_{i,\alpha\beta}(\vmc E, \vmc B) 
= \lrbra{\alpha^{(0)}(\vmc E, \vmc B)}\unl\eta_i\rket{\beta^{(0)}(\vmc E,\vmc B)}
\end{align}
und
\begin{align}\label{e7:eta.PV}
\begin{split}
\unl\eta\PVU_{i,\alpha\beta}(\vmc E, \vmc B) &= 
\lrbra{\alpha^{(1)}(\vmc E, \vmc B)}\unl\eta_i\rket{\beta^{(0)}(\vmc E,\vmc B)}
+ \lrbra{\alpha^{(0)}(\vmc E, \vmc B)}\unl\eta_i\rket{\beta^{(1)}(\vmc E,\vmc B)}\\
&= \sum_{\gamma\neq\alpha}\frac{
\uop M\PVU_{\alpha\gamma}(\vmc E, \vmc B)\unl\eta_{i,\gamma\beta}(\vmc E, \vmc B)}{
E^{(0)}_\alpha(\vmc E,\vmc B) - E^{(0)}_\gamma(\vmc E,\vmc B)}
+\sum_{\gamma\neq\beta}\frac{\unl\eta_{i,\alpha\gamma}(\vmc E, \vmc B)\uop M\PVU_{\gamma\beta}(\vmc E, \vmc B)
}{
E^{(0)}_\beta(\vmc E,\vmc B) - E^{(0)}_\gamma(\vmc E,\vmc B)}\ .
\end{split}
\end{align}
Wir haben hier bereits von der weiter oben getroffenen Vereinbarung Gebrauch gemacht, 
dass alle Matrixelemente, die in der Form $\unl\eta_{\alpha\beta}$ geschrieben werden, ab sofort
bzgl. der ungestörten, feldabhängigen Basiszuständen $\rket{\alpha^{(0)}(\vmc E,\vmc B)}$ zu verstehen sind.
Dies gilt natürlich nicht für die Matrixelemente des letzten Abschnitt \ref{s7:Fluxes}, die bzgl.
der Eigenzustände $\rket{\alpha(\vmc E,\vmc B,\delta)}$ der vollen Massenmatrix (\ref{e7:MMPV}) zu verstehen sind.
Nichtsdestotrotz sagt uns Gl. (\ref{e7:eta.PC}), dass zusammen mit der Vernachlässigung der P-verletzenden Beiträge
zu den Energiedifferenzen die Ergebnisse des letzten Abschnitts formal übernommen werden können, um die P-erhaltenden
geometrischen Flussdichten zu erhalten. Aus (\ref{e7:Flux.E}) und (\ref{e7:Flux.B}) folgt dann
\begin{subequations}\label{e7:PC.Fluxes}
\begin{align}\label{e7:PC.Flux.E}
{\mc J}_{\alpha,\ell}^{(\mc E,\,\rm PC)}(\vmc E,\vmc B) &= \frac{\I}{2}\sum_{i,j=1}^3\vep_{ij\ell}\klr{\sum_{\beta\neq\alpha}
\frac{\unl D_{i,\alpha\beta}(\vmc E,\vmc B)\unl D_{j,\beta\alpha}(\vmc E,\vmc B) - (i\leftrightarrow j)}{
\klr{E^{(0)}_\alpha(\vmc E,\vmc B)-E^{(0)}_\beta(\vmc E,\vmc B)}^2}}\ ,\\[5mm] \label{e7:PC.Flux.B}
{\mc J}_{\alpha,\ell}^{(\mc B,\,\rm PC)}(\vmc E,\vmc B) &= \frac{\I}{2}\sum_{i,j=1}^3\vep_{ij\ell}\klr{\sum_{\beta\neq\alpha}
\frac{\unl\mu_{i,\alpha\beta}(\vmc E,\vmc B)\unl\mu_{j,\beta\alpha}(\vmc E,\vmc B) - (i\leftrightarrow j)}{
\klr{E^{(0)}_\alpha(\vmc E,\vmc B)-E^{(0)}_\beta(\vmc E,\vmc B)}^2}}\ ,\\
\nonumber
(\ell &= 1,2,3)\ .
\end{align}
\end{subequations}
In Anhang \ref{sA:Parity} haben wir gezeigt, dass die so definierten Flussdichten zu Recht als P-erhaltend bezeichnet
werden können. Unter P-Transformation ändert sich das Vorzeichen des elektrischen Feldes $\vmc E$, während das
magnetische Feld $\vmc B$ in sich selbst übergeht. In Gl. (\ref{eA:PC.Fluxes}) ist gezeigt, dass bei diesem Übergang
die P-erhaltenden Flussdichten unverändert bleiben, d.h. es gilt
\begin{align}\label{e7:P.PC.Fluxes}
\v{\mc J}^{(\mc E,\,{\rm PC})}_{\alpha}(\vmc E,\vmc B) = \v{\mc J}^{(\mc E,\,{\rm PC})}_{\alpha}(-\vmc E,\vmc B)\ ,
\qquad
\v{\mc J}^{(\mc B,\,{\rm PC})}_{\alpha}(\vmc E,\vmc B) = \v{\mc J}^{(\mc B,\,{\rm PC})}_{\alpha}(-\vmc E,\vmc B)\ .
\end{align}

Um die P-verletzende Anteile zu berechnen, die in erster Ordnung in $\delta$ zu den Flussdichten beitragen, müssen wir
in den Matrixprodukten, die im Zähler der geometrischen Flussdichten stehen, formal jeweils ein Matrixelement durch das
P-verletzende Matrixelement aus Gl. (\ref{e7:eta.PV}) ersetzen. Wir erinnern uns außerdem daran, dass es zwei P-verletzende
Beiträge zur Massenmatrix gab, jeweils proportional zu $\delta_1$ und $\delta_2$. Also müssen wir zwei
P-verletzende geometrische Flussdichten unterscheiden. Das Endergebnis lautet dann für $\varkappa\in\klg{1,2}$
und $\ell\in\klg{1,2,3}$

{\footnotesize
\begin{subequations}
\begin{align}\label{e7:PV.Flux.E}
\begin{split}
&~{\mc J}_{\varkappa,\alpha,\ell}^{(\mc E,\,\rm PV)}(\vmc E,\vmc B)\\
=&~\frac{\I}{2}\sum_{i,j=1}^3\vep_{ij\ell}\klr{\sum_{\beta\neq\alpha}
\frac{\kle{\unl D^{(\varkappa,\,\rm PV)}_{i,\alpha\beta}(\vmc E,\vmc B)\unl D_{j,\beta\alpha}(\vmc E,\vmc B) 
+ \unl D_{i,\alpha\beta}(\vmc E,\vmc B)\unl D^{(\varkappa,\,\rm PV)}_{j,\beta\alpha}(\vmc E,\vmc B)} 
- (i\leftrightarrow j)}{
\klr{E^{(0)}_\alpha(\vmc E,\vmc B)-E^{(0)}_\beta(\vmc E,\vmc B)}^2}}\ ,
\end{split}\\[5mm] \label{e7:PV.Flux.B}
\begin{split}
&~{\mc J}_{\varkappa,\alpha,\ell}^{(\mc B,\,\rm PV)}(\vmc E,\vmc B)\\
=&~\frac{\I}{2}\sum_{i,j=1}^3\vep_{ij\ell}\klr{\sum_{\beta\neq\alpha}
\frac{\kle{\unl\mu^{(\varkappa,\,\rm PV)}_{i,\alpha\beta}(\vmc E,\vmc B)\unl\mu_{j,\beta\alpha}(\vmc E,\vmc B)
+ \unl\mu_{i,\alpha\beta}(\vmc E,\vmc B)\unl\mu^{(\varkappa,\,\rm PV)}_{j,\beta\alpha}(\vmc E,\vmc B)} - (i\leftrightarrow j)}{
\klr{E^{(0)}_\alpha(\vmc E,\vmc B)-E^{(0)}_\beta(\vmc E,\vmc B)}^2}}\ .
\end{split}
\end{align}
\end{subequations}}

wobei nach Gl. (\ref{e7:eta.PV})
{\small 
\begin{subequations}
\begin{align}\label{e7:D.PV}
\unl D^{(\varkappa,\,\rm PV)}_{i,\alpha\beta}(\vmc E,\vmc B) &= \sum_{\gamma\neq\alpha}\frac{
\uop M^{(\varkappa)}_{{\rm PV}, \alpha\gamma}(\vmc E,\vmc B)\unl D_{i,\gamma\beta}(\vmc E, \vmc B)}{
E^{(0)}_\alpha(\vmc E,\vmc B) - E^{(0)}_\gamma(\vmc E,\vmc B)}
+\sum_{\gamma\neq\beta}\frac{\unl D_{i,\alpha\gamma}(\vmc E, \vmc B)\uop M^{(\varkappa)}_{{\rm PV}, \gamma\beta}(\vmc E,\vmc B)
}{
E^{(0)}_\beta(\vmc E,\vmc B) - E^{(0)}_\gamma(\vmc E,\vmc B)}\ ,\\[5mm] \label{e7:Mu.PV}
\unl\mu^{(\varkappa,\,\rm PV)}_{i,\alpha\beta}(\vmc E,\vmc B) &= \sum_{\gamma\neq\alpha}\frac{
\uop M^{(\varkappa)}_{{\rm PV}, \alpha\gamma}(\vmc E,\vmc B)\unl\mu_{i,\gamma\beta}(\vmc E, \vmc B)}{
E^{(0)}_\alpha(\vmc E,\vmc B) - E^{(0)}_\gamma(\vmc E,\vmc B)}
+\sum_{\gamma\neq\beta}\frac{\unl\mu_{i,\alpha\gamma}(\vmc E, \vmc B)\uop M^{(\varkappa)}_{{\rm PV}, \gamma\beta}(\vmc E,\vmc B)
}{
E^{(0)}_\beta(\vmc E,\vmc B) - E^{(0)}_\gamma(\vmc E,\vmc B)}\ ,\\[5mm]\nonumber
(\varkappa &= 1,2)
\end{align}
\end{subequations}}

einzusetzen ist. Auch diese Flussdichten werden zu Recht als P-verletzend bezeichnet, denn wie in Gl. (\ref{eA:PV.Fluxes})
in Anhang \ref{sA:Parity} gezeigt ist, gilt
\begin{align}\label{e7:P.PV.Fluxes}
\begin{split}
\v{\mc J}^{(\mc E,\,{\rm PV})}_{\varkappa,\alpha}(\vmc E,\vmc B) 
&= -\v{\mc J}^{(\mc E,\,{\rm PV})}_{\varkappa,\alpha}(-\vmc E,\vmc B)\ ,\quad
\v{\mc J}^{(\mc B,\,{\rm PV})}_{\varkappa,\alpha}(\vmc E,\vmc B) 
= -\v{\mc J}^{(\mc B,\,{\rm PV})}_{\varkappa,\alpha}(-\vmc E,\vmc B)\ ,\\[5mm]
(\varkappa &= 1,2)\ .
\end{split}
\end{align}

Die geometrischen Flussdichten (\ref{e7:Flux.E}) und (\ref{e7:Flux.B}) lauten damit insgesamt bis zur Ordnung
$\OO(\delta)$

{\small
\begin{subequations}
\begin{align}\label{e7:Flux.E.final}
\v{\mc J}^{(\mc E)}_{\alpha}(\vmc E,\vmc B,\delta) &= \v{\mc J}^{(\mc E,\,{\rm PC})}_{\alpha}(\vmc E,\vmc B)
+ \delta_1 \v{\mc J}^{(\mc E,\,{\rm PV})}_{1,\alpha}(\vmc E,\vmc B)
+ \delta_2 \v{\mc J}^{(\mc E,\,{\rm PV})}_{2,\alpha}(\vmc E,\vmc B) + \OO(\delta^2)\ ,\\[5mm]\label{e7:Flux.B.final}
\v{\mc J}^{(\mc B)}_{\alpha}(\vmc E,\vmc B,\delta) &= \v{\mc J}^{(\mc B,\,{\rm PC})}_{\alpha}(\vmc E,\vmc B)
+ \delta_1 \v{\mc J}^{(\mc B,\,{\rm PV})}_{1,\alpha}(\vmc E,\vmc B)
+ \delta_2 \v{\mc J}^{(\mc B,\,{\rm PV})}_{2,\alpha}(\vmc E,\vmc B) + \OO(\delta^2)\ .
\end{align}
\end{subequations}}

Eingesetzt in (\ref{e7:BP.Flux}) ergeben sich nach der Integration über die Fläche hieraus die P-erhaltenden und
P-verletzenden Beiträge zur geometrischen Phase. Für konstantes magnetisches Feld ist
\begin{align}\label{e7:BP.const.B}
\begin{split}
\gamma_\alpha(\mc C) \xrightarrow{\vmc B=\const} \gamma^{(\mc E)}_\alpha(\mc C) &=  
\gamma^{(\mc E,\,{\rm PC})}_\alpha(\mc C)
+ \delta_1\gamma^{(\mc E,\,{\rm PV})}_{1,\alpha}(\mc C) + \delta_2\gamma^{(\mc E,\,{\rm PV})}_{2,\alpha}(\mc C)\ ,\\[5mm]
\gamma^{(\mc E,\,{\rm PC})}_\alpha(\mc C) &= 
\int_{\mc F}\v{\mc J}^{(\mc E,\,{\rm PC})}_{\alpha}(\vmc E,\vmc B)\cdot\d\v f^{(\mc E)}\ ,\\
\gamma^{(\mc E,\,{\rm PV})}_{\varkappa,\alpha}(\mc C) &= 
\int_{\mc F}\v{\mc J}^{(\mc E,\,{\rm PV})}_{\varkappa,\alpha}(\vmc E,\vmc B)\cdot\d\v f^{(\mc E)}\ ,\quad 
(\varkappa=1,2)
\end{split}
\end{align}
und analog folgt für konstantes elektrisches Feld 
\begin{align}\label{e7:BP.const.E}
\begin{split}
\gamma_\alpha(\mc C) \xrightarrow{\vmc E=\const} \gamma^{(\mc B)}_\alpha(\mc C) &=  
\gamma^{(\mc B,\,{\rm PC})}_\alpha(\mc C)
+ \delta_1\gamma^{(\mc B,\,{\rm PV})}_{1,\alpha}(\mc C) + \delta_2\gamma^{(\mc B,\,{\rm PV})}_{2,\alpha}(\mc C)\ ,\\[5mm]
\gamma^{(\mc B,\,{\rm PC})}_\alpha(\mc C) &= 
\int_{\mc F}\v{\mc J}^{(\mc B,\,{\rm PC})}_{\alpha}(\vmc E,\vmc B)\cdot\d\v f^{(\mc B)}\ ,\\
\gamma^{(\mc B,\,{\rm PV})}_{\varkappa,\alpha}(\mc C) &= 
\int_{\mc F}\v{\mc J}^{(\mc B,\,{\rm PV})}_{\varkappa,\alpha}(\vmc E,\vmc B)\cdot\d\v f^{(\mc B)}\ ,\quad 
(\varkappa=1,2)\ .
\end{split}
\end{align}
Wie bereits am Ende des letzten Abschnitts erwähnt wird man in der Praxis die P-erhaltenden und P-verletzenden Flussdichten
studieren, um einen geeigneten Integrationsweg $\mc C$ im Parameterraum festzulegen. Die numerische Berechnung der
geometrischen Phasen allerdings wird sich leichter durch die Verwendung der diagonalen Matrixelemente der Ortsableitungsmatrix
bewerkstelligen lassen. Diese müssen zuvor ebenfalls in einen P-erhaltenden und P-verletzenden Anteil aufgespalten
werden. Es folgt dann
\begin{align}\label{e7:D.full}
\begin{split}
\lrbra{\alpha(Z,\delta)}\del_Z\rket{\alpha(Z,\delta)} &= 
\lrbra{\alpha(\vmc E(Z),\vmc B(Z),\delta)}\del_Z\rket{\alpha(\vmc E(Z),\vmc B(Z),\delta)}\\
&= \uop D^{(1)}_{\alpha\alpha}(Z) + \delta_1\uop D^{(1,\,{\rm PV})}_{1,\alpha\alpha}(Z) 
+ \delta_2\uop D^{(1,\,{\rm PV})}_{2,\alpha\alpha}(Z)
\end{split}
\end{align}
mit dem ungestörten Matrixelement 
\begin{align}\label{e7:D.PC}
\uop D^{(1)}_{\alpha\alpha}(Z) = \lrbra{\alpha(Z)}\del_Z\rket{\alpha(Z)}
\end{align}
und den P-verletzenden Anteilen
\begin{align}\label{e7:D.PV.2}
\uop D^{(1,\,{\rm PV})}_{\varkappa,\alpha\alpha}(Z)
&= \sum_{\beta\neq\alpha}\frac{
\uop M^{(\varkappa)}_{{\rm PV},\alpha\beta}\uop D^{(1)}_{\beta\alpha}(Z) 
+\uop D^{(1)}_{\alpha\beta}(Z)\uop M^{(\varkappa)}_{{\rm PV},\beta\alpha}
}{E^{(0)}_\alpha(Z) - E^{(0)}_\beta(Z)}\ .
\end{align}
Wir sind hier wieder zur Schreibweise mit der $Z$-Abhängigkeit übergegangen, da die Felder numerisch in Abhängigkeit
von dieser Ortskoordinate vorliegen. Generell wird dann für jeden Ort $Z$ das Eigenwertproblem (ohne P-Verletzung) 
mit den an diesem Ort geltenden Werten für die Felder numerisch gelöst und damit alle benötigten Ausdrücke berechnet, 
wie z.B. die lokalen Darstellungen der Matrizen $\uop D^{(1)}(Z)$, $\uop M\PV^{(1,2)}(Z)$, $\vu D(Z)$ und $\vu\mu(Z)$. 
Unter Verwendung von (\ref{e7:D.full}) lautet also 
der numerisch zu berechnende Ausdruck für die geometrische Phase gemäß Gl. (\ref{e7:BP3})
\begin{align}\label{e7:BP.num}
\gamma_\alpha(\mc C) = \gamma\PCU_\alpha(\mc C) + \delta_1\gamma\PVU_{1,\alpha}(\mc C) 
+ \delta_2\gamma\PVU_{2,\alpha}(\mc C)\ ,
\end{align}
mit
\begin{subequations}
\begin{align}\label{e7:PC.BP.num}
\gamma\PCU_\alpha(\mc C) &= \I\int_{Z_0}^Z\d Z\ \uop D^{(1)}_{\alpha\alpha}(Z)\ ,\\ \label{e7:PV.BP.num}
\gamma\PVU_{\varkappa,\alpha}(\mc C)  &= \I\int_{Z_0}^Z\d Z\ \uop D^{(1,\,{\rm PV})}_{\varkappa,\alpha\alpha}(Z)\ ,
\qquad(\varkappa=1,2)\ .
\end{align}
\end{subequations}

\subsection{Numerische Resultate}\label{s7:Results}

Wie bereits mehrfach erwähnt bietet es sich an, zunächst die geometrischen Flussdichten zu studieren und dann geeignete Kurven
im Parameterraum zu wählen. Leider hat man bereits für metastabilen Wasserstoff eine überwältigende Menge an
möglichen Flussdichten, die man analysieren könnte. Zu den vier metastabilen $2S$-Zuständen von Wasserstoff kann man
jeweils eine P-erhaltende und eine P-verletzende Flussdichte berechnen und in Form eines dreidimensionalen
Vektordiagramms darstellen. Alle Flussdichten sind komplexwertig, liefern also je einen Plot für den Realteil und einen
weiteren für den Imaginärteil. Man muss außerdem stets die Flussdichten zu konstantem $\vmc E$-Feld und konstantem
$\vmc B$-Feld unterscheiden. Zu jeder Variante kann man die Flussdichte im entsprechenden Raum der Feldstärken plotten,
wobei jeder Wert des konstanten Feldes einen separaten Plot liefert. Um einen gewissen Bereich in beiden Feldstärken
abzudecken, kann man also selbst bei großen Schrittweiten in den Feldstärken leicht auf eine Zahl von über 1000
Vektordiagrammen kommen.

Zum Glück ist die Situation nicht so tragisch, wie der letzte Absatz vermuten lässt. Durch die numerische Berechnung
und Betrachtung verschiedener Vektordiagramme entsteht relativ schnell ein Eindruck von den zugrundeliegenden
Gesetzmäßigkeiten, so dass man sich letztendlich auf die exemplarische Betrachtung sehr weniger Plots
beschränken kann. Wir wollen dabei anmerken, dass die zeitliche Dauer der numerischen Berechnung der Plots
kein Problem darstellt: Mit einem handelsüblichen Rechner und einem in C++ geschriebenen Programm sind wir
in der Lage, etwa 2000 Plots (mit je einigen 100 Vektoren) in etwa 15 Minuten zu berechnen.

Wir wollen in diesem Abschnitt diese Gesetzmäßigkeiten an einem anschaulichen Beispiel erklären und 
für ganz spezielle Integrationswege auch die numerisch berechneten Werte der geometrischen Phasen angeben, 
um ein Gefühl für die Größenordnung des P-verletzenden Effekts zu bekommen. 
Die numerische Untersuchung der P-verletzenden geometrischen Phasen mit den in diesem
Kapitel aufgestellten Formeln steht gerade erst am Anfang und wird in der näheren Zukunft hoffentlich noch viele
Früchte tragen. Wir werden im Ausblick in Abschnitt \ref{s8:Ausblick} genauer darauf eingehen, 
welche Aspekte in kommenden Arbeiten noch zu studieren sind.

Wir betrachten nun die geometrische Flussdichte des metastabilen Wasserstoffzustands mit Index $\alpha=9$, 
$\rket{9,\vmc E,\vmc B,\delta} := \rket{2\hat S_{1/2},1,1,\vmc E,\vmc B,\delta}$ (siehe
Anhang \ref{sB:H2}, Tabelle \ref{tB:H2States}) bei konstantem Magnetfeld 
\begin{align}
\vmc B = (0,0,1\u{mT})
\end{align}
im Raum der elektrischen Feldstärke $\vmc E$. Wir wählen für die Berechnung der Vektordiagramme ein um
$\vmc E=0$ zentriertes Intervall $\mc E_i \in \kle{-10\u{V/cm},10\u{V/cm}}$, $i\in\klg{1,2,3}$ und verwenden
die in Anhang \ref{sB:H2}, Tabelle \ref{tB:mu1H2} ff. dargestellten Matrixdarstellungen für die Operatoren
des magnetischen Moments $\vu\mu$ und des Dipolmoments $\vu D$. Wir beschränken uns
auf die Darstellung der Realteile der P-erhaltenden und P-verletzenden Flussdichten, die sich qualitativ genauso wie
die Imaginärteile verhalten. Weiterhin betrachten wir exemplarisch nur die zu $\delta_1$ proportionalen, P-verletzenden
Beiträge.
\begin{figure}[!htp]
\centering
\includegraphics[width=7cm]{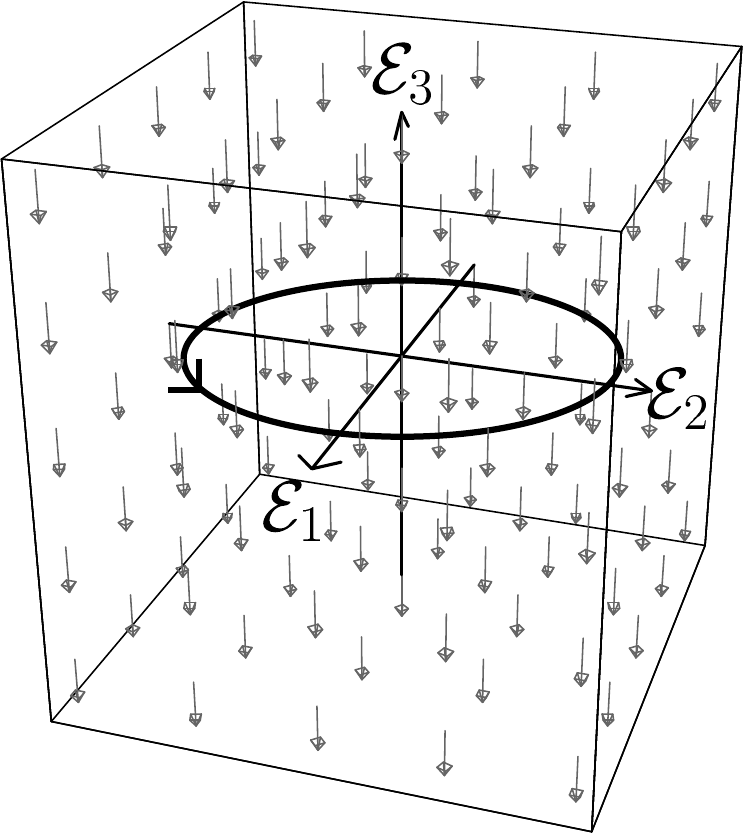}\\[5mm]
\includegraphics[width=7cm]{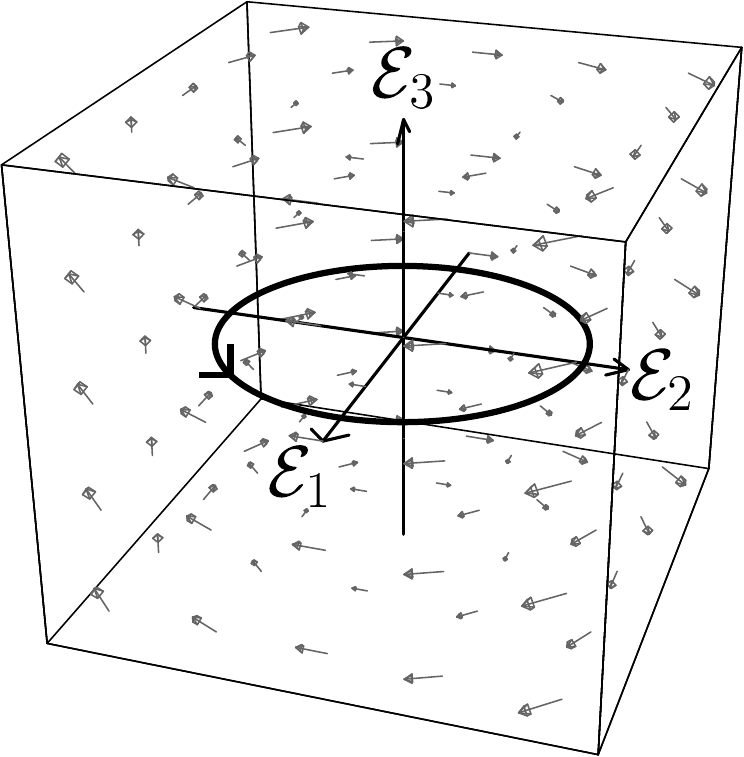}\qquad
\includegraphics[width=7cm]{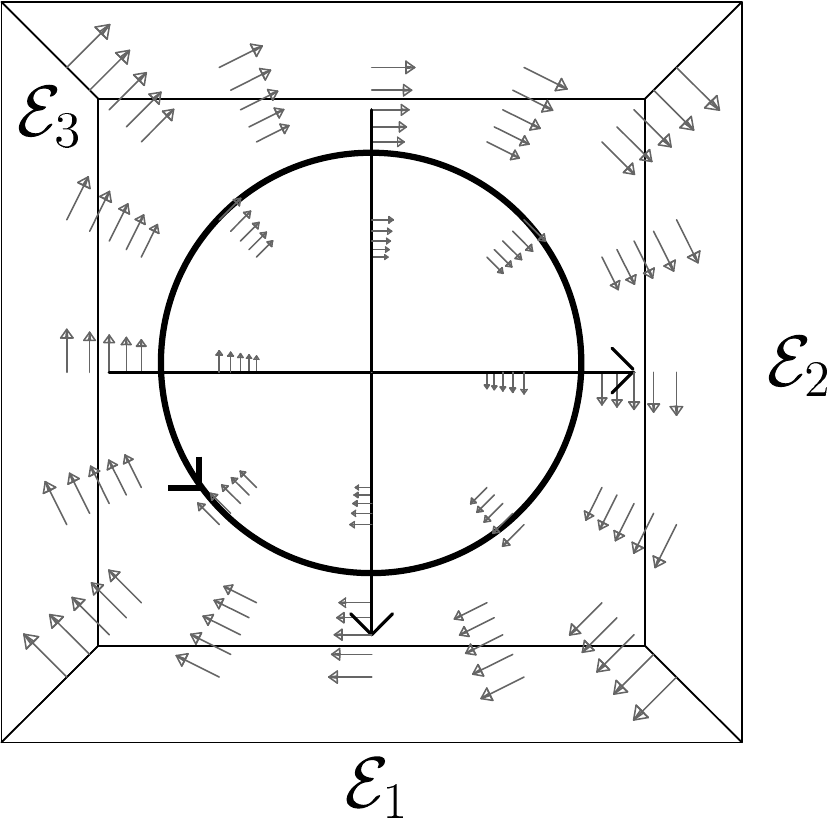}
\caption[Die Realteile der geometrischen Flussdichten bei konstantem magnetischen Feld.]{Die Realteile der geometrischen Flussdichten des Zustands $\rket{9,\vmc E,\vmc B,\delta} 
:= \rket{2\hat S_{1/2},1,1,\vmc E,\vmc B,\delta}$ bei $\vmc B = (0,0,1\u{mT})$ im Raum der elektrischen Feldstärke, 
$\mc E_i \in \kle{-10\u{V/cm},10\u{V/cm}}$, $i\in\klg{1,2,3}$. Oben ist der P-erhaltende Teil 
$\Real\v{\mc J}^{(\mc E,\,{\rm PC})}_{9}(\vmc E,\vmc B)$ dargestellt, unten der 
P-verletzende Anteil $\Real\v{\mc J}^{(\mc E,\,{\rm PV})}_{1,9}(\vmc E,\vmc B)$, der zum Parameter $\delta_1$
gehört.}\label{f7:Phi.PC.E}
\end{figure}

Betrachten wir zunächst den linken Plot in Abb. \ref{f7:Phi.PC.E}, in dem 
$\Real\v{\mc J}^{(\mc E,\,{\rm PC})}_{9}(\vmc E,\vmc B)$ dargestellt ist. Offenbar orientiert sich die Ausrichtung
aller Vektoren an der Ausrichtung des magnetischen Feldes. Der Einfluss des magnetischen Feldes auf die Eigenzustände 
ist viel größer als der Einfluss der sehr kleinen elektrischen Feldstärken, die wir
hier betrachtet haben. Die Richtung des Magnetfelds kann also als ausgezeichnete Richtung bezeichnet werden und
gibt somit die Quantisierungsachse vor. Wie klein die Auswirkung des elektrischen Feldes auf die Flussdichtevektoren ist,
kann man schon darin erkennen, dass alle im Diagramm vorkommenden Vektoren scheinbar gleich sind\footnote{
Numerisch exakt identisch sind diese Vektoren i.A. jedoch nicht. Die Änderung der Beträge der Energieeigenwerte
im Definitionsbereich von Abb. \ref{f7:Phi.PC.E} ist um einige Größenordnung geringer als die Energiedifferenzen,
die aufgrund des zugrunde liegenden Magnetfelds entstanden sind. In diesem Zusammenhang sei auch auf Anhang \ref{sB:H2.EW}
verwiesen, wo numerische Ergebnisse für die Energieeigenwerte von metastabilem Wasserstoff in elektrischen und 
magnetischen Feldern präsentiert werden.}. Insbesondere
sind die Flussdichtevektoren an zueinander P-transformierten Feldstärken identisch, wie in Anhang \ref{sA:Parity},
Gl. (\ref{eA:PC.Fluxes}) gezeigt. Da das magnetische Feld invariant unter P-Transformation ist, kann man in Abb.
\ref{f7:Phi.PC.E} direkt die Vektoren an punktgespiegelten Koordinaten miteinander vergleichen.

Numerische Rechnungen zeigen, dass der P-erhaltende Anteil der geometrischen Phase, der sich unter
Verwendung des im Diagramm eingezeichneten Integrationsweg $\mc C$ ergibt, etwa
\begin{align}\label{e7:PC.9}
\gamma_9^{{\rm PC}}(\mc C) = \OO(10^{-3})
\end{align}
ist. Berechnet man dagegen den P-verletzenden Anteil der geometrischen Phase mit diesem Pfad $\mc C$, so erhält man
\begin{align}
\gamma_{1,9}^{{\rm PV}}(\mc C) = 0\ .
\end{align}
Dieses Ergebnis kann man verstehen, wenn man das mittlere und das rechte Diagramm in Abb. \ref{f7:Phi.PC.E} betrachtet.
Der Integrationsweg $\mc C$ wurde von uns so gewählt, dass er ein optimales Ergebnis für die P-erhaltende Phase
liefert\footnote{Er ist senkrecht zu den Vektoren der P-erhaltenden Flussdichte.}. Es zeigt sich jedoch, dass
die P-verletzende Flussdichte gerade ein Wirbelfeld um die $\mc E_3$-Achse ist. Alle P-verletzenden Flussdichtevektoren liegen in
der gleichen Ebene, wie der Integrationsweg, also gibt es auch keinen P-verletzenden Fluss durch die von
$\mc C$ eingeschlossene Fläche.

Wie bei der P-erhaltenden Flussdichte sind auch die Vektoren der P-verletzenden Flussdichte offenbar (aus demselben Grund, siehe Fußnote)
betragsmäßig unabhängig vom elektrischen Feld im dargestellten Definitionsbereich. Die Richtung allerdings weist eine sehr 
starke Feldabhängigkeit auf. Sie drückt den P-verletzenden Charakter der Flussdichte aus, denn nach Gl. 
(\ref{eA:PV.Fluxes}) aus Anhang \ref{sA:Parity} folgt, dass die Vektoren der P-verletzenden Flussdichte 
an zueinander punktgespiegelten Koordinaten entgegengesetzt gleich sein müssen. Dies kann man in den beiden
Diagrammen in Abb. \ref{f7:Phi.PC.E} sehr gut erkennen.

Ein optimaler Integrationsweg für die P-verletzende Phase müsste also in einer Ebene liegen, die die $\mc E_3$-Achse
enthält und dürfte die $\mc E_3$-Achse nicht überschreiten. Wählt man z.B. einen Weg mit $\mc E_1 = 0$ und $\mc E_2>0$
im hier verwendeten Bereich für die elektrischen Feldstärken, so erhält man P-verletzende, geometrische Phasen
der Größenordnung
\begin{align}\label{e7:PV.9}
\gamma_{1,9}^{(\mc E,\,{\rm PV})}(\mc C') = \OO(10^{-4}\delta_1)\ .
\end{align}
Für den hier beschriebenen Weg $\mc C'$, der senkrecht zu $\mc C$ liegt, verschwindet wie zu erwarten der
P-erhaltende Anteil der geometrischen Phase, $\gamma_9^{(\mc E,\,{\rm PC})}(\mc C')$.

\begin{floatingfigure}[r]{7cm}
\centering\vspace{-5mm}
\includegraphics[width=7cm]{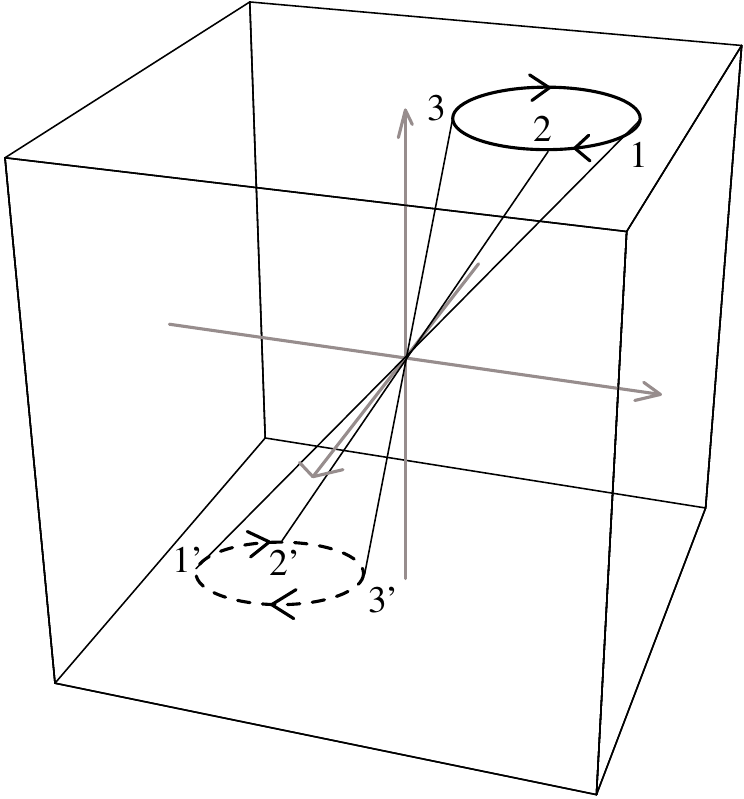}
\caption[Verhalten einer geschlossenen Kurve unter P-Transformation]{Verhalten einer geschlossenen Kurve unter 
P-Transformation. Diese Betrachtung gilt für alle Koordinatensysteme, die unter P-Transformation ihr Vorzeichen 
wechseln, also z.B. für Ortskoordinaten oder die elektrische Feldstärke, nicht aber für magnetische 
Feldstärken.\\}\label{f7:PV.Weg}
\end{floatingfigure}
Wir wollen nun diskutieren, wie man mit den Relationen (\ref{eA:PC.Fluxes}) und (\ref{eA:PV.Fluxes}) 
für die P-transformierten geometrischen Flussdichten zeigen kann, dass die zugehörigen geometrischen Phasen
ebenfalls P-erhaltend bzw. P-verletzend sind.
Allgemein erhält man die P-transformierte geometrische Phase durch Integration über den P-transformierten Weg. 
Betrachtet man einen Integrationsweg im Raum der elektrischen Feldstärke bei konstantem Magnetfeld, 
so ergibt sich aufgrund der P-Invarianz des Magnetfelds z.B. das in Abb. \ref{f7:PV.Weg} dargestellte Bild. 
Jede Koordinate des Integrationsweges muss punktgespiegelt werden, wodurch der gesamte Integrationsweg 
punktgespiegelt wird, der Umlaufsinn jedoch erhalten bleibt. Eine analoge Betrachtungsweise im Raum der
magnetischen Feldstärken bei konstantem elektrischen Feld gilt nicht. In diesem Fall müsste man den
gleichen Integrationsweg in einem anderen Diagramm für entgegengesetztes elektrisches Feld betrachten. Wir
werden weiter unten erläutern, wie man sich aus diesem Dilemma befreien kann.

Bei der P-erhaltenden geometrischen Phase $\gamma_\alpha^{(\mc E,\,{\rm PC})}(\mc C)$
hat man das Integral über die P-erhaltende Flussdichte zu bilden.
Transformiert man den Integrationsweg, so integriert man also bei gleichem Umlaufsinn über die Flussdichte
mit der gespiegelten elektrischen Feldstärke, der aber nach (\ref{eA:PC.Fluxes}) identisch ist mit der
Flussdichte an der ursprünglichen Stelle. Der P-erhaltende Anteil der geometrischen Phase bleibt also
invariant\footnote{Würde sich auch der Umlaufsinn des Integrationsweges unter P-Transformation umkehren,
so wäre dadurch ein Vorzeichenwechsel entstanden.} unter P-Transformation. 
Bei der P-verletzenden Phase tritt wegen Gl. (\ref{eA:PV.Fluxes}) nach der P-Transformation
der entgegengesetzt gleiche Fluss durch den P-transformierten Integrationsweg auf, 
also kehrt die P-verletzende geometrische Phase unter P-Transformation insgesamt ihr
Vorzeichen um. Zusammenfassend gilt also
\begin{align}\label{e7:BP.PC.PT}
\gamma_\alpha^{(\mc E,\,{\rm PC})}(\mc C) &\xrightarrow{\quad P\quad} 
\phantom-\gamma_\alpha^{(\mc E,\,{\rm PC})}(\mc C)\ ,\\ \label{e7:BP.PV.PT}
\begin{split}
\gamma_{\varkappa,\alpha}^{(\mc E,\,{\rm PV})}(\mc C) &\xrightarrow{\quad P\quad} 
-\gamma_{\varkappa,\alpha}^{(\mc E,\,{\rm PV})}(\mc C)\ ,\\[2mm]
&(\varkappa=1,2)\ .
\end{split}
\end{align}

Eine Idee, die man aufgrund der bisherigen Diskussion in zukünftigen Arbeit 
weiterverfolgen sollte, ist die Wahl eines Integrationsweges im Raum der elektrischen Feldstärken,
der die Form einer Acht hat, wobei der Mittelpunkt bei $\vmc E=0$ liegen sollte. Ein Beispiel für einen
solchen Weg ist in Abb. \ref{f7:Acht} dargestellt. Der Integrationsweg muss dabei in einer Ebene liegen und
die eine Schleife muss die Spiegelung der anderen Schleife am Ursprung sein (d.h. P-Transformation mit umgekehrtem
Umlaufsinn). Nach den bisherigen Ergebnissen sollte für einen solchen Integrationsweg die P-erhaltende geometrische
Phase stets verschwinden, während die P-verletzende geometrische Phase verdoppelt wird (sie ist identisch für jede 
Integrationsschleife). Durch die Verwendung dieser Art von Integrationswegen würde man eine rein P-verletzende
geometrische Phase erhalten.

\pagebreak
Kommen wir nun zu dem zur bisher betrachteten Situation entgegengesetzten Fall und
betrachten die geometrischen Phasen und Flussdichten für ein konstantes elektrisches Feld $\vmc E$
im Raum der magnetischen Feldstärke $\vmc B$. Dies erfordert ein gewisses Umdenken im Vergleich zu dem bisher diskutierten
Beispiel. Die Relationen aus (\ref{eA:PC.Fluxes}) und (\ref{eA:PV.Fluxes}) gelten für identisches $\vmc B$-Feld,
aber entgegengesetzt gleiches $\vmc E$-Feld. Ginge man genauso vor, wie in diesem Abschnitt, so müsste man also
stets identische Koordinaten in zwei separaten Vektordiagrammen im $\vmc B$-Raum miteinander vergleichen, 
die für entgegengesetzte elektrische Felder berechnet wurden. 

\begin{floatingfigure}[r]{7cm}
\centering\vspace{-5mm}
\includegraphics[width=7cm]{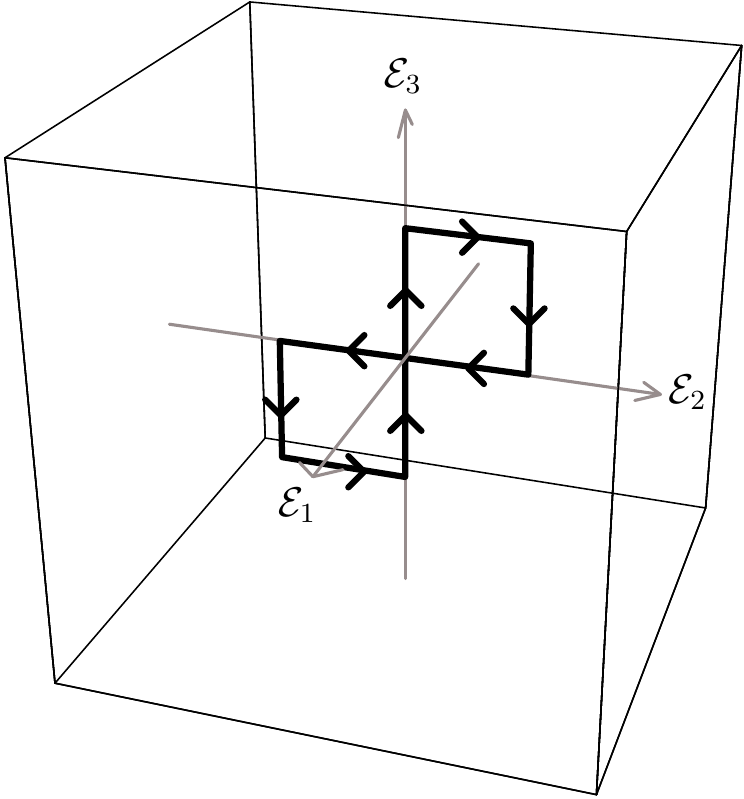}
\caption[Ein Integrationsweg für eine rein P-verletzende geometrische Phase]{Ein möglicher
Integrationsweg für eine rein P-verletzende geometrische Phase.\\[5mm]}\label{f7:Acht}
\end{floatingfigure}
Wollte man eine Vergleichsmöglichkeit innerhalb eines Diagramms für konstantes elektrisches Feld und variabler
magnetischer Feldstärke schaffen, so müsste man anstelle
der P-Transformation Spiegelungen an einer Ebene betrachten, die $\vmc E$ enthält. 
Diese Spiegelung $R$ kann dann als Kombination der Paritätstransformation und einer Rotation um die Normale $\v n$ der
Spiegelungsebene um $\pi$ darstellen. Der entsprechende Operator, der auf die inneren atomaren Zustände angewendet
werden kann, lautet dann
\begin{align}
R(\v n) = \e^{\I\pi(\v F\cdot\v n)}P\ ,
\end{align}
wobei $\v F$ der Operator des Gesamtdrehimpulses ist, siehe Anhang \ref{sA:ZweiTeilchen}. 
Mit den Gesamtdrehimpulszuständen aus Anhang \ref{sA:ZweiTeilchen} könnte man dann eine
Matrixdarstellung $\umc R$ des Spiegelungsoperators angeben und damit ähnliche Relationen wie in den Gleichungen
(\ref{eA:PC.Fluxes}) und (\ref{eA:PV.Fluxes}) aufstellen.
Eine weitere Schwierigkeit ist, dass auch die Zustände unter einer solchen Spiegelung im Allgemeinen in
andere Zustände übergehen, da z.B. die dritten Komponenten $F_3$ des Gesamtdrehimpulses sich unter Transformation
i.A. ändern werden. Ist dies der Fall, so muss man wiederum zwei Vektordiagramme (je einen für jeden Zustand)
miteinander vergleichen. Eine Möglichkeit, die Änderung des Zustands zu vermeiden, wäre
eine Spiegelung, die die dritte Komponente des Magnetfelds invariant lässt.
Es würden dann für konstantes elektrisches Feld $\vmc E$ die Flussdichten und geometrischen Phasen ein und desselben
der zueinander gespiegelten Magnetfelder $\vmc B$ und $\vmc B' = R\vmc B$ miteinander in Beziehung gesetzt. 
Anhand dieser Relationen könnte man innerhalb eines Vektordiagramms im Raum der magnetischen Feldstärke
bei eingezeichneter Spiegelungsebene wieder optisch die Flussdichtevektoren
miteinander vergleichen. Integrationswege in diesen Diagrammen wären dann zu vergleichen mit gespiegelten Integrationswegen,
die jedoch (im Gegensatz zu P-transformierten Wegen) einen umgekehrten Umlaufsinn hätten.

Die von uns gezeigte Existenz P-verletzender geometrischer Phasen bestätigt die Ergebnisse aus 
der Dissertation \cite{DissTG}, in der allerdings Atome in Ruhe in einem zeitabhängigen elektrischen
Feld betrachtet wurden. Ein Vergleich der Resultate aus \cite{DissTG} mit numerischen Werten, die
mit Hilfe der hier aufgestellten Formeln berechnet werden, ist eine interessante Aufgabe, die
in zukünftigen Arbeiten behandelt werden muss.

Bei der Betrachtung der Gln. (\ref{e7:PC.9}) und (\ref{e7:PV.9}) fällt auf, dass zumindest bei den
hier exemplarisch verwendeten Integrationspfaden der P-erhaltende Anteil und der P-verletzende Anteil der geometrischen
Phase um viele Größenordnungen auseinanderliegen\footnote{Man beachte, dass die P-verletzenden Parameter 
$\delta_{1,2}$ Werte im Bereich von $10^{-13}\,...\,10^{-12}$ annehmen, siehe dazu auch 
\cite{BoBrNa95} und \cite{DiplTB}.}.
Hier muss man weitere Maßnahmen zur Vergrößerung des Effekts in Angriff nehmen. 
Neben einem modularen, periodischen Versuchaufbau, bei dem die geometrische Phase einige tausend Mal 
aufgesammelt werden würde, wäre noch die Verwendung von beinahe entarteten Zuständen denkbar 
(z.B. durch ein noch kleineres, zugrundeliegendes Magnetfeld). Dann nämlich wird
die in den Formeln dieses Kapitels überall auftretende Energiedifferenz im Nenner zu einer Vergrößerung der
geometrischen Phasen führen. Eine weitere Idee wäre die Verwendung eines Versuchsaufbaus, der den P-erhaltenden
Anteil der geometrischen Phase verschwinden lässt, nicht aber den P-verletzenden Anteil, was zu einer
erhöhten Sensitivität auf die P-verletzende Phase führen würde.

\chapter{Zusammenfassung und Ausblick}\label{s8:Summary}

\section{Zusammenfassung}\label{s8:Zusammenfassung}

In dieser Arbeit haben wir eine Theorie zur Beschreibung longitudinaler Atomstrahl-Spinecho-Experimente
(lABSE, abgeleitet von {\em longitudinal Atomic Beam Spin Echo}) entwickelt.
Bei solchen Experimenten durchquert ein Strahl neutraler Atome eine Anordnung elektrischer und magnetischer
Felder. Eine wesentliches Merkmal der lABSE-Experimente ist das sogenannte Spinecho-Feld, das eine antiparallele
Anordnung zweier hintereinandergeschalteter Magnetfelder in Strahlrichtung ist. 

Die Atome sind anfangs in einer bestimmten Polarisation präpariert, die quantenmechanisch als
eine Superposition verschiedener atomarer Gesamtdrehimpulszustände dargestellt werden kann.
Eine Beschreibung der Atome im Ortsraum gelingt durch die Zuordnung von Wellenpaketen zu den einzelnen
atomaren Zuständen. Im ersten Magnetfeld spalten die Energieniveaus dieser Zustände auf 
unterschiedliche Weise auf, was zu einem Auseinanderdriften
der zugeordneten Wellenpakete führt. Im zweiten, antiparallelen Magnetfeld werden die einzelnen Wellenpakete
durch geschickte Wahl der Feldstärke wieder zum Überlapp gebracht, so dass man die Oszillation der relativen
Phase in dem sogenannten Spinecho-Interferenzsignal mit einem geeigneten Detektor messen kann. 
Fügt man neben den Spinechofeldern weitere elektrische und magnetische Felder zum experimentellen Aufbau hinzu, 
so wird sich das Spinecho-Signal verändern. Hieraus kann man Rückschlüsse ziehen auf die physikalischen Prozesse
im Atom, die beim Durchqueren der Feldanordnung stattgefunden haben.

In Kapitel \ref{s2:Vorbetrachtungen} bis Kapitel \ref{s6:Formalismus} erarbeiteten wir uns Schritt für Schritt
die vollständige Theorie zur Beschreibung dieser Art von Experimenten. Ausgehend von zwei sehr einfachen Ansätzen
berechneten wir in Kapitel \ref{s2:Vorbetrachtungen} zunächst Näherungslösungen der Schrödinger-Gleichung,
deren unterschiedliche Eigenschaften wir diskutierten. Einige Anwendungen der WKB-Näherungslösung, die die
besseren physikalischen Eigenschaften aufweist, folgten dann in Kapitel \ref{s3:Anwendungen}. 
In Kapitel \ref{s4:Formalismus} entwickelten wir eine Methode zur 
Berechnung der Lösung der Schrödinger-Gleichung mit skalarem Potential in
Form einer Reihenentwicklung. Diese Methode wendeten wir in Kapitel \ref{s5:Zerfall} zur Beschreibung eines
Wellenpakets in einem komplexen, skalaren Potential an, in dem das durch das Wellenpaket repräsentierte 
Teilchen mit der Zeit zerfällt.
In Kapitel \ref{s6:Formalismus} übertrugen wir dann den Formalismus auf den Fall nichthermitescher, matrixwertiger
Potentiale, die zur Beschreibung metastabiler Atome notwendig sind. In Kapitel \ref{s7:Berry} wendeten wir die Ergebnisse
aus Kapitel \ref{s6:Formalismus} auf die Untersuchung paritätsverletzender geometrischer Phasen an.

Wir wollen nun auf die einzelnen Schritte im Detail eingehen und die wesentlichen Ergebnisse festhalten.
In Kapitel \ref{s2:Vorbetrachtungen} wählten wir als Ansatz für die Lösung der Schrödinger-Gleichung ein
Produkt aus Phasenfaktor und Amplitudenfunktion und vernachlässigten weiterhin alle zweiten Ortsableitungen
von Phasenwinkel und Amplitudenfunktion. Legten wir den Phasenfaktor durch eine einfache, ebene Welle fest, 
so erhielten wir als Näherungslösung ein Wellenpaket mit potentialunabhängiger Schwerpunktsbewegung. Wählten wir hingegen
den Phasenfaktor der WKB-Näherung, so ergab sich ein Wellenpaket mit einer im klassischen Sinne korrekten
Schwerpunktsbewegung, bei der die erhaltene Gesamtenergie stets die Summe aus lokaler kinetischer und potentieller
Energie war.

Mit dieser WKB-Näherungslösung konnten wir in Kapitel \ref{s3:Anwendungen} bereits das sogenannte 
Fahrplanmodell im adiabatischen Grenzfall theoretisch reproduzieren und die zugehörigen Spin\-echosignale berechnen. 
Im Fahrplanmodell wird bei vorgegebener Feldkonfiguration der relative, zeitliche Abstand des Eintreffens 
der Schwerpunkte der atomaren Teilwellenpakete in Abhängigkeit von der Position auf der Strahlachse dargestellt. 
Dies ermöglicht die ungefähre Vorhersage der Lage von sogenannten Kreuzungspunkten, 
an denen sich jeweils zwei Wellenpakete maximal überlappen und die Oszillation der relativen Phase 
als Spinechosignal gemessen werden kann.

Motiviert von diesem Erfolg entwickelten wir in Kapitel \ref{s4:Formalismus} einen Formalismus zur Berechnung
der exakten Lösung der skalaren Schrödinger-Gleichung in Form einer Reihenentwicklung nach einem Integraloperator
$\hat K$. Die Anwendung des Formalismus basiert dabei stets auf dem Ansatz für die Wellenfunktion als Produkt eines
Phasenfaktors und einer Amplitudenfunktion. Ausgehend vom WKB-Phasenfaktor 
berechnet man den Integraloperator $\hat K$, der die Entwicklung des Amplitudenanteils
der Wellenfunktion bestimmt, wobei die nullte Ordnung dieser Entwicklung der aus Kapitel \ref{s2:Vorbetrachtungen}
bekannten WKB-Näherungslösung entsprach. Der Formalismus erlaubt dabei eine gewisse Freiheit bei der
Wahl des Phasenfaktors, der neben dem WKB-Phasenwinkel noch weitere Beiträge enthalten kann. Hiervon
hängt die Gestalt des Integraloperators $\hat K$ und die Qualität der Entwicklung ab.

Die Entwicklung der Lösung der Schrödinger-Gleichung erlaubte uns in Abschnitt \ref{s4:Anwendung} einen 
quantitativen Vergleich der Näherungslösungen aus
Kapitel \ref{s2:Vorbetrachtungen} und darüberhinaus ein tieferes Verständnis der zugehörigen Korrekturen, 
die sich aus der Reihenentwicklung bis zur ersten Ordnung ergaben.
Es stellte sich heraus, dass der Hauptbeitrag der Korrektur der WKB-Näherungslösung
aus der Vernachlässigung der Dispersion des Wellenpakets stammte. Um die Reihenentwicklung auch auf sehr
schmale Wellenpakete mit ausgeprägter Dispersion anwenden zu können, entwickelten wir in Abschnitt
\ref{s4:Dispersion} ein Verfahren, bei dem ein schmales Wellenpaket als Superposition sehr breiter Wellenpakete
geschrieben wird, für die die Entwicklung bereits in erster Ordnung eine sehr gute Approximation der
exakten Lösung der Schrödinger-Gleichung liefert.

Die Anwendung des Formalismus auf komplexe, skalare Potentiale in Kapitel \ref{s5:Zerfall} demonstrierte eindrucksvoll
seine Fähigkeit zur Selbstkorrektur des Phasenwinkels. Ausgehend von einem einfachen Zusatz im WKB-Phasenwinkel,
der den Zerfall des Wellenpakets beschreibt, berechneten wir die erste und die zweite Ordnung der Entwicklung der
exakten Lösung der Schrödinger-Gleichung und konnten zwei Korrekturbeiträge zum Phasenwinkel identifizieren.
Der eine Korrekturbeitrag ersetzte den zuvor angesetzten Zerfallsanteil gerade durch einen verbesserten Ausdruck,
den wir in Abschnitt \ref{s5:2.Ordnung} auch sehr gut physikalisch interpretieren konnten. Der zweite Korrekturbeitrag
konnte als Logarithmus einer Funktion geschrieben werden, die somit eine Korrektur der Amplitude des Wellenpakets
darstellte. Wir haben gezeigt, dass diese Amplitudenkorrektur die Normierung des Wellenpakets bereits in
nullter Ordnung der Entwicklung zu jeder Zeit sicherstellt (in reellen Potentialen ohne Zerfall des Wellenpakets).

In Kapitel \ref{s6:Formalismus} übertrugen wir den Formalismus zur Berechnung der Reihenentwicklung der
exakten Lösung der Schrödinger-Gleichung mit dem in Kapitel \ref{s5:Zerfall} berechneten Phasenwinkel
auf den Fall nichthermitescher, matrixwertiger Potentiale, die für die Beschreibung metastabiler Atome
mit mehreren inneren Zuständen benötigt werden. Dabei sei angemerkt, dass der dort entwickelte Matrixformalismus
leicht auf andere Fälle übertragen werden kann, bei denen ein stationäres Matrixpotential (hermitesch oder
nichthermitesch) zugrundeliegt.

Im Gegensatz zum skalaren Fall erhielten wir in Kapitel \ref{s6:Formalismus} 
die Entwicklung der exakten Lösungen $\ket{\Psi(Z,t)}$
nach Potenzen eines matrixwertigen Integraloperators $\uk$. Als Basis der
Darstellung der Operatoren und des verallgemeinerten Spinors $\ket{\Psi(Z,t)}$, der das Atom insgesamt beschreibt, 
verwendeten wir die lokalen Eigenzustände $\rket{\alpha(Z)}$ ($\alpha=1,\ldots,N$) der Potentialmatrix $\uop M(Z)$.

Die nullte Ordnung der Entwicklung der Amplitudenfunktionen jeder Spinorkomponente von $\ket{\Psi(Z,t)}$ 
nach Potenzen von $\uk$ liefert den adiabatischen Grenzfall, bei dem die Wellenpakete
der verschiedenen atomaren Zustände unabhängig voneinander durch individuelle Potentiale propagieren,
die sich als Realteil der lokalen Eigenwerte der Potentialmatrix $\uop M(Z)$ ergeben. Die adiabatischen
Wellenpakete jeder Spinorkomponente sind dabei bis auf einen sehr wichtigen Zusatz zum Phasenwinkel identisch
mit der WKB-Näherungslösung der skalaren Schrödinger-Gleichung. Der zusätzliche Beitrag im Phasenwinkel
entspricht der geometrischen Phase, die der atomare Zustand, zu dem das Wellenpaket gehört, bei seinem Weg durch
die Feldkonfiguration erhält. Da die Wellenpakete in der adiabatischen Näherung wie im skalaren Fall zunächst 
dispersionsfrei sind, bietet sich das in Abschnitt \ref{s4:Dispersion} diskutierte Verfahren zur Berücksichtigung
der Dispersion auch für die Lösung der matrixwertigen Schrödinger-Gleichung an.

In erster Ordnung der Entwicklung treten Mischungen der Zustände in den Lösungen der
Amplitudenfunktionen jedes Teilwellenpakets auf. Diese sind proportional zu den 
nichtdiagonalen, lokalen Matrixelementen der in der Schrödinger-Gleichung enthaltenen
Operatoren für die erste und zweite Ortsableitung. Bei Änderung der Massenmatrix $\uop M(Z)$ mit
dem Ort $Z$, ändert sich die Zusammensetzung ihrer Eigenzustände $\rket{\alpha(Z)}$ und
die Matrixelemente der Ableitungsoperatoren, z.B. $\lrbra{\beta(Z)}\dd{}{Z}\rket{\alpha(Z)}$
($\alpha,\beta = 1,\ldots,N$) werden im Allgemeinen ungleich Null. 
Auf diese Weise können Wellenpakete von atomaren Zuständen entstehen,
die nicht im Anfangszustand vertreten waren. Auch in der ersten Ordnung der Entwicklung
der Lösung der matrixwertigen Lösung stellte der Dispersionsterm den größten Korrekturbeitrag
dar, weshalb insbesondere in diesem Fall Verfahren zur Berücksichtigung der Dispersion der Wellenpakete 
angewendet werden muss.

Eine genauere Betrachtung der ersten Ordnung der Entwicklung für matrixwertige Potentiale erlaubte 
uns die Einführung eines erweiterten Fahrplanmodells, das das adiabatische Fahrplanmodell
um die grafische Darstellung der Bewegung neu entstandener Wellenpakete in den äußeren Felder erweitert.
In diesem Zusammenhang haben wir in Abschnitt \ref{s6:xFP} verschiedene Möglichkeiten
der Visualisierung diskutiert, dabei jedoch festgestellt, dass die Erweiterung des Fahrplanmodell 
nicht als exakte Methode zur Vorhersage der Lage von Kreuzungspunkten zu verstehen ist, sondern vielmehr als ein
Werkzeug zur übersichtlichen Darstellung der wesentlichen bzw. interessierenden 
Prozesse beim lABSE angesehen werden sollte.

In Kapitel \ref{s7:Berry} haben wir die Verbindung zwischen der Theorie zur Beschreibung
von lABSE-Experimenten und der Paritätsverletzung in Atomen am Beispiel der geometrischen Phasen
hergestellt, die auch im adiabatischen Grenzfall auftreten können.
Bewegt sich das Atom, während es die Feldanordnung durchquert, entlang einer geschlossenen Kurve
in dem als Parameterraum bezeichneten sechsdimensionalen Raum der elektrischen und magnetischen Feldstärken, so können die
atomaren Eigenzustände unter gewissen Umständen eine geometrische Phase erhalten.

Wir haben in Kapitel \ref{s7:Berry} am Beispiel des metastabilen Wasserstoffs mit Hauptquantenzahl
$n=2$ gezeigt, wie diese geometrische Phase in einen P-erhaltenden und einen P-verletzenden
Anteil aufgespalten werden kann. Weiterhin haben wir für den Fall, dass das magnetische (elektrische)
Feld konstant ist, geometrische Flussdichten im Raum der elektrischen (magnetischen) Feldstärke
definiert. Die geometrischen Phasen konnten dann als geometrischer Gesamtfluss durch eine vorgegebene
geschlossene Kurve im Raum der elektrischen (magnetischen) Feldstärke interpretiert werden.
Darüberhinaus erleichtert die grafische Darstellung der Flussdichten in einem dreidimensionalen Vektordiagramm
die Wahl einer Integrationskurve zur Berechnung der geometrischen Phasen.

Anhand der grafischen Darstellung der numerisch berechneten P-erhaltenden 
und P-verletzenden geometrischen Flussdichten in einem speziellen Fall konnten
wir den P-erhaltenden
bzw. P-verletzenden Charakter der Flussdichten veranschaulichen. Wir haben dazu
die geometrischen Flussdichten eines bestimmten metastabilen Zustands von Wasserstoff
im Raum der elektrischen Feldstärke bei konstantem Magnetfeld betrachtet.
Anhand dieses Beispiels haben wir ferner
das Verhalten der Integrationswege unter P-Transformation diskutiert, sowie die daraus
im Zusammenhang mit dem Verhalten der Flussdichten folgenden Transformationseigenschaften 
der geometrischen Phasen. Schließlich konnten wir aus diesem Beispiel direkt
die Existenz einer P-verletzenden geometrischen Phase für einen bestimmten Integrationsweg
zeigen. Wir konnten darüberhinaus einen Vorschlag für einen geeigneten Typ von Integrationswegen
machen, die die Form einer Acht im Parameterraum haben und eine rein P-verletzende geometrische Phase liefern.

Wir haben schließlich festgestellt, dass eine Betrachtung der Flussdichten für konstantes
elektrisches Feld im Raum der magnetischen Feldstärken ein gewisses Umdenken erfordert, da
die P-Transformation die elektrische Feldstärke umkehrt, das Magnetfeld aber invariant lässt.
Will man P-transformierte Flussdichten, Integrationswege und geometrische Phasen miteinander
vergleichen, muss man in diesem Fall also je zwei Vektordiagramme betrachten. Als Alternative
haben wir die Verwendung von Spiegelungen vorgeschlagen, die als Produkt der P-Transformation
und einer geeigneten Drehung dargestellt werden können.

Viele interessante Fragen ergeben sich aus unseren bisherigen Überlegungen und motivieren 
vertiefende und weiterführende Studien. 
Wir wollen einige davon im nun folgenden Abschnitt präsentieren.

\section{Ausblick}\label{s8:Ausblick}

Die in \cite{DiplTB} diskutierten P-verletzenden und P-erhaltenden Polarisationsrotationen 
von Deuterium in elektrischen Feldern 
sind interessante Anwendungen des in dieser Arbeit aufgestellten Formalismus. In diesem Zusammenhang
ist ein vertiefendes Studium der numerischen Methoden zur Berechnung der ersten Ordnung des
Matrixformalismus unerlässlich, auch in Verbindung mit der in Abschnitt \ref{s4:Dispersion}
diskutierten Methode zur Berücksichtigung der Dispersion der Wellenpakete. 
Hier kann der erweiterte Fahrplan zur Anwendung kommen. 
Die in \cite{DiplTB} behandelten P-erhaltenden Polarisationsrotationen
von Deuterium, die einen wesentlichen größeren Effekt als die P-verletzenden Drehungen darstellen, 
können als Test einer lABSE-Apparatur dienen und einen Vergleich von Theorie und Experiment ermöglichen. 
Hierauf aufbauend bietet sich eine Untersuchung der P-verletzenden Polarisationsrotationen 
von Wasserstoff in elektrischen Feldern mit der in \cite{BoBrNa95} vorgeschlagenen Vorgehensweise
zur resonanten Verstärkung an.

Ein tiefergehendes Studium der P-verletzenden geometrischen Phasen ist ein weiteres, interessantes
und vielversprechendes Gebiet. Die Betrachtung periodischer Feldkonfigurationen, bei denen die geometrischen Phasen im 
Idealfall viele tausend mal addiert werden, ist für die Untersuchung P-verletzender geometrischer Phasen
unerlässlich. Die Untersuchung der geometrischen Flussdichten muss auf
den Fall konstanter elektrischer und variabler magnetischer Feldstärken ausgedehnt werden.
Hier stößt man auch auf Entartungen der Zustände, die mit den in dieser Arbeit entwickelten
Methoden nicht behandelt werden können. In diesem Zusammenhang muss man die sogenannten
nichtabelschen, geometrischen Phasen studieren, die völlig neue Aspekte in Bezug auf die
P-Verletzung in Atomen bringen könnten. Die hier betrachteten, abelschen geometrischen Phasen
müssen allerdings eingehender in der Nähe von Punkten im Parameterraum betrachtet werden,
an denen Entartungen auftreten. Durch das Auftreten der Energiedifferenzen in den Nennern der Formeln für die Berechnung
der geometrischen Phasen ist hier eine Verstärkung der P-verletzenden Beiträge zu erwarten.

Eine Untersuchung der geometrischen Phasen bei sowohl variabler elektrischer als auch magnetischer
Feldstärke steht noch aus. Eine Definition geometrischer Flussdichten ist in diesem Fall zwar
nicht mehr möglich, jedoch kann man zweidimensionale Ebenen im Parameterraum betrachten, bei
denen jeweils eine Komponente des elektrischen und magnetischen Feldes variiert wird.

Die Verstärkung P-verletzender Effekte bei fast entarteten Zuständen wurde in einem anderen Zusammenhang
bereits in \cite{GaNa00} ausgenutzt, wo die sogenannten Floquet-Eigenwerte (siehe z.B. \cite{Cyc87}) 
eines vereinfachten, zeitlich periodischen, zweidimensionalen Hamiltonoperators für atomares Dysprosium 
betrachtet wurden. Eine Übertragung auf die in dieser Arbeit untersuchten geometrischen Phasen ist 
ein weiteres interessantes Projekt.

Die in dieser Arbeit entwickelte Theorie hat also das Potential, viele interessante Methoden
zur Messung P-verletzender Effekte im Rahmen eines lABSE-Experiments theoretisch zu untersuchen.
Die enge Zusammenarbeit mit der Arbeitsgruppe von Prof. D. Dubbers und PD M. DeKieviet, PhD.,
die das Atomstrahl-Spinecho-Experiment betreuen, bietet darüberhinaus ein ideales Umfeld für
zielgerichtete, theoretische Untersuchung mit der Hoffnung auf einen experimentellen 
Nachweis P-verletzender Effekte in leichten Atomen in den nächsten Jahren. 
Wir wollen uns dafür an dieser Stelle herzlich bedanken.


\FloatBarrier


\appendix
\renewcommand{\chaptermark}[1]{\markboth{#1}{}}
\chapter{Quantenmechanische Grundlagen}\label{sA:QMBasics}

In diesem Anhang wollen wir die quantenmechanischen Grundlagen zur Beschreibung metastabiler
Atome erarbeiten. Hierzu gehört zunächst eine Behandlung des Zerfalls der inneren atomaren Zustände
im Rahmen der Wigner-Weisskopf-Methode. Danach studieren wir die quantenmechanische Behandlung
des Zwei-Teilchen-Problems, um die Propagation eines wasserstoffähnlichen Atoms, das aus einem Kern und
einem Elektron besteht, richtig beschreiben zu können. Im darauf folgenden Abschnitt
stellen wir die Störungsrechnung mit nichthermiteschen Massenmatrizen vor, was uns in
Abschnitt \ref{sA:OrtsablMat} die Berechnung der Ableitungsmatrizen $\uop D^{(1,2)}(z)$ aus 
Kapitel \ref{s6:Formalismus} erlaubt. Im letzten Abschnitt \ref{sA:Parity} geben wir einige nützliche
Relationen im Zusammenhang mit der Paritätstransformation an, die wir in Kapitel \ref{s7:Berry} benötigen.

Für die Notation vereinbaren wir hier, wie in der gesamten, vorliegenden Arbeit, dass Matrizen
generell mit einem Unterstrich gekennzeichnet (z.B. $\uop M$) werden. Innere atomare Zustände schreiben
wir als Ket-/Bra-Vektoren mit runden Klammern (z.B. $\rket{\alpha}$). Den verallgemeinerten Spinor
$\ket{\Psi(z,t)}$ der Gesamtwellenfunktion eines Atoms und andere Zustände schreiben wir 
wie gewohnt mit spitzen Klammern.

\section{Die Wigner-Weisskopf-Methode}\label{sA:WWF}

In diesem Abschnitt wollen wir lediglich eine Zusammenfassung der von V. Weisskopf und F. Wigner
entwickelten Methode \cite{WeWi30} geben.
Eine (gegenüber der Originalarbeit \cite{WeWi30} zeitgemäßere) Herleitung 
der im folgenden aufgeführten Ergebnisse findet sich in \cite{Nac91}, Anhang I,
wo die Methode auf den Zerfall von Kaonen angewendet wurde.

Allgemein kommt die Wigner-Weisskopf-Methode dann zur Anwendung, wenn ein quantenmechanisches
System aus einer diskreten Menge von Zuständen in ein Kontinuum von Zuständen zerfällt.
Betrachten wir metastabile Atome mit Hauptquantenzahl $n=2$ und Kernspin $I$, dann wollen wir
mit $N_I$ die daraus resultierende Zahl von atomaren Zuständen $\rket{2,\alpha}$
in diesem $(n=2)$-Unterraum bezeichnen. Dabei ist $\alpha$ ein geeignet gewählter Index zur
Durchnummerierung der Zustände. Die Zustände $\rket{2,\alpha}$ sind diskret und seien weiterhin
nichtentartet.

Das metastabile Atom, das zu Anfang als Superposition der $\rket{2,\alpha}$ beschrieben werden kann,
wird nun mit der Zeit unter Aussendung von Photonen in Zustände aus dem $(n=1)$-Unterraum $\rket{1,\beta}$ zerfallen.
Wir erhalten also mit der Zeit eine Beimischung von Produktzuständen
\begin{align}
  \ket{1,\beta;\gamma,k} \equiv \rket{1,\beta}\otimes\ket{\gamma,k}\ ,
\end{align}
wobei die $\ket{\gamma,k}$, $k=1,2,3,\ldots$ eine geeignete Basis des Fock-Raums der Photonen darstellen.
Der Fock-Raum der Photonen ist dabei kontinuierlich, da Photonen beliebiger Impulse darin enthalten sind, 
womit die Menge der Produktzustände insgesamt auch kontinuierlich ist, so dass die Voraussetzung für die Anwendung der Wigner-Weisskopf-Methode gegeben sind\footnote{Natürlich sind die atomaren Energieniveaus in der Theorie durch
scharf definierte Energiedifferenzen voneinander getrennt, aber durch die endliche Lebensdauer der angeregten
Zustände tritt eine gewisse Unschärfe der Energiedifferenzen auf, so dass die ausgesendeten Photonen ein
kontinuierliches Spektrum aufweisen.}.

Insgesamt wollen wir die zeitliche Entwicklung des gesamten atomaren Zustands $\ket{t}$ durch Lösen der
Schrödinger-Gleichung
\begin{align}
  H\ket t = \I\dd{}t\ket t
\end{align}
gewinnen. Der Zustand $\ket t$ lässt sich dabei schreiben als ein verallgemeinerter Spinor
\begin{align}
  \ket t = \begin{pmatrix}
              \Psi(t)\\ \Phi_1(t)\\ \Phi_2(t)\\ \vdots
           \end{pmatrix}
\end{align}
mit den Komponenten-Spinoren
\begin{align}
  \Psi(t) = \begin{pmatrix}
              (2,1|t\rangle\\ (2,2|t\rangle\\ (2,3|t\rangle\\ \vdots
            \end{pmatrix}\qquad
  \Phi_k(t) = \begin{pmatrix}
                \bracket{1,1;\gamma,k}{t}\\ \bracket{1,2;\gamma,k}{t}\\ \bracket{1,3;\gamma,k}{t}\\ \vdots
              \end{pmatrix}\ .
\end{align}
Wie man sieht, lebt der Spinor $\Psi(t)$ im $(n=2)$-Unterraum der atomaren Zustände und der Spinor
$\Phi(t) = (\Phi_k(t))$ im Produkt-Raum der $(n=1)$-Zustände mit dem Unterraum aller Photon-Zustände 
$\ket{\gamma,k}$, $k=1,2,3,\ldots$.

Den Hamilton-Operator 
\begin{align}\label{eA:WWH}
  H = \hat H + H\subt{rad}
\end{align}
zerlegen wir nun in einen stabilen Anteil $\hat H$ und einen kleinen, 
störenden Anteil $H\subt{rad}$, der für den radiativen Zerfall verantwortlich ist. Wir vernachlässigen hier
Übergänge aufgrund von $H\subt{rad}$ zwischen Zuständen mit gleicher Hauptquantenzahl $n$. Da die Energiedifferenzen
im Vergleich zu $(n=1)$-Zuständen sehr gering sind, haben solche Übergänge sehr große Lebensdauern.
Die relevanten Matrixelemente von $H\subt{rad}$ bezeichnen wir mit
\begin{align}
  \uop H_k := \klr{\bra{1,\beta;\gamma,k}{H\subt{rad}}\rket{2,\alpha}}\ .
\end{align}

Im stabilen Anteil $\hat H$ des Hamiltonoperators befinden sich Beiträge des Coulomb-Potentials des Kerns, in dem 
das Elektron sich befindet, sowie Beiträge der schwachen Wechselwirkung und der äußeren elektrischen und
magnetischen Felder. Wir werden auf diese Beiträge an anderer Stelle noch genauer eingehen. Hier wollen wir 
nur festhalten, dass wir Mischungen zwischen Zuständen verschiedener Hauptquantenzahlen aufgrund des Anteils
$\hat H$ des Hamiltonoperators vernachlässigen.

Wählen wir den Anfangszustand so, dass er nur Komponenten im $(n=2)$-Unterraum besitzt, d.h.
\begin{align}
  \ket{t=0} = \begin{pmatrix}
                \Psi(0)\\ 0\\ 0\\ \vdots
              \end{pmatrix}\ ,
\end{align}
so liefert die Wigner-Weisskopf-Methode die Zeitentwicklung des verallgemeinerten Spinors im $(n=2)$-Unterraum,
\begin{align}\label{eA:n=2Lsg}
  \Psi(t) = \e^{-\I\uop M t}\Psi(0)\ ,\qquad(t\geq 0)\ .
\end{align}
Dabei ist $\uop M$ die sogenannte nichthermitesche Massenmatrix\ , die in einen hermiteschen Anteil $\unl E$
und einen antihermiteschen Anteil $-\tfrac\I2\unl\Gamma$ zerlegt werden kann, wobei
\begin{align}\label{eA:Massenmatrix}
  \uop M &= \unl E - \tfrac\I2\unl\Gamma\ ,\\
  \unl E &= \big(\rmatelem{2,\alpha}{\hat H}{2,\alpha'}\big)\ ,\\
  \unl\Gamma &= 2\pi\sum_k\uop H_k\HC\delta(E_1+E_{\gamma,k}-E_2)\uop H_k\ .
\end{align}
Die so definierten, hermiteschen Matrizen $\unl E$ und $\unl\Gamma$ leben im Unterraum der
$(n=2)$-Zustände. In der Diracschen Delta-Funktion sind $E_{1,2}$ jeweils die Energie-Schwerpunkte der
Zustände mit den Hauptquantenzahlen $n=1,2$.

An Gl. (\ref{eA:n=2Lsg}) erkennt man, dass die nichthermitesche Massenmatrix $\uop M$ wie ein zeitunabhängiger
Hamiltonoperator im stabilen Fall wirkt, d.h. die Lösung (\ref{eA:n=2Lsg}) entspricht der Lösung einer
matrixwertigen Schrödinger-Gleichung
\begin{align}\label{eA:effSG}
  \uop M\Psi(t) = \I\dd{}t\Psi(t)\ .
\end{align}
Dass $\uop M$ nichthermitesch ist, zieht einige Besonderheiten nach sich.
Zunächst stellen wir fest, dass $\uop M$ nur im $(n=2)$-Unterraum lebt. Dort hat man die Eigenwertgleichung
\begin{align}\label{eA:rEG}
  \uop M\rket{\alpha} = E_\alpha\rket{\alpha}
\end{align}
mit komplexen Energie-Eigenwerten
\begin{align}\label{eA:EEW}
  E_\alpha = V_\alpha - \tfrac\I2\Gamma_\alpha\ .
\end{align}
Der Realteil $V_\alpha$ entspricht der Energie des Zustand $\rket{\alpha}$ (wir verzichten im folgenden auf die
Angabe der Hauptquantenzahl bei der Dirac-Notation, da wir uns von nun an stets auf den $(n=2)$-Unterraum beziehen,
solange es nicht explizit anders erwähnt wird). Der Anteil $\Gamma_\alpha$ im Imaginärteil des Eigenwerts
muss positiv sein und führt dann nach Gl. (\ref{eA:n=2Lsg}) zu einem exponentiellen Abklingen des Beitrags des atomaren
Zustands $\rket\alpha$ im Gesamtspinor $\Psi(t)$.

Die Nichthermitezität von $\uop M$ macht die Einführung von linken Eigenvektoren notwendig, die wir
mit einer zusätzlichen Tilde kennzeichnen und die die Eigenwertgleichung
\begin{align}\label{eA:lEG}
  \rbra{\tilde\alpha}\uop M = \rbra{\tilde{\alpha}}E_\alpha
\end{align}
mit den gleichen Eigenwerten $E_\alpha$ wie in Gl. (\ref{eA:rEG}) erfüllen. Für die linken Ket-Vektoren
gilt dann eine Eigenwertgleichung mit komplex konjugierten Eigenwerten und hermitesch konjugierter Massenmatrix:
\begin{align}
  \uop M\HC\rket{\tilde\alpha} = E_\alpha^*\rket{\tilde\alpha}\ .
\end{align}
Gehen wir davon aus, dass wir bei einem äußeren (elektrischen oder magnetischen) Feld die 3-Achse in stets in Feldrichtung
wählen können, so kann man die Massenmatrix $\uop M$ gleichzeitig mit der Matrix der dritten Komponente des
Gesamtdrehimpulses $\unl F_3$ diagonalisieren. Man erhält dann gemeinsame linke und rechte Eigenzustände
und es gilt
\begin{align}\label{eA:MF3kommu}
  \kle{\uop M,\unl F_3}      &= 0
\end{align}
und
\begin{subequations}\label{eA:MF3Basis}
\begin{align}
  \uop M\rket{\alpha,F_3}    &= E_{\alpha,F_3}\rket{\alpha,F_3}\ ,\\
  \unl F_3\rket{\alpha,F_3}  &= F_3\rket{\alpha,F_3}\ ,\\
  \lrbra{\alpha,F_3}\uop M   &= \lrbra{\alpha,F_3}E_{\alpha,F_3}\ ,\\
  \lrbra{\alpha,F_3}\unl F_3 &= \lrbra{\alpha,F_3}F_3\ .
\end{align}
\end{subequations}
Sind die Energieeigenwerte weiterhin nichtentartet im Unterraum zu festem $F_3$, d.h. gilt
\begin{align}\label{eA:NichtEntartet}
  E_{\alpha',F_3} \neq E_{\alpha,F_3}\qquad(\alpha'\neq\alpha)\ ,
\end{align}
dann bilden die in (\ref{eA:MF3Basis}) definierten Zustände eine Basis des $(n=2)$-Unterraum,
d.h. wir können die Orthonormierung
\begin{align}\label{eA:lrOrthonorm}
  \lrbracket{\alpha',F_3'}{\alpha,F_3} = \delta_{\alpha',\alpha}\delta_{F_3',F_3}
\end{align}
wählen, die sich als günstig für die Definition von Quasiprojektoren
\begin{subequations}\label{eA:QProAll}
\begin{align}\label{eA:QPro}
  \Pro_{\alpha,F_3} &= \rket{\alpha,F_3}\lrbra{\alpha,F_3}\\ \label{eA:QProOrth}
  \Pro_{\alpha',F_3'}\Pro_{\alpha,F_3} &= \delta_{\alpha',\alpha}\delta_{F_3',F_3}\Pro_{\alpha,F_3}\\ \label{eA:SpQPro}
  \Spur(\Pro_{\alpha,F_3}) &= 1\\ \label{eA:QProVollst}
  \sum_{\alpha,F_3}\Pro_{\alpha,F_3} &= \unl\um
\end{align}
\end{subequations}
erweist. Darüberhinaus ist es für die Normierung von Erwartungswerten angebracht, zusätzlich die Normierung
der rechten Eigenvektoren zu fordern, d.h.
\begin{align}\label{eA:rNormierung}
  \rbracket{\alpha,F_3}{\alpha,F_3} = 1\ .
\end{align}
Mit den Quasiprojektoren kann man wegen (\ref{eA:QProVollst}) 
die Matrixdarstellungen $\uop M$ und $\unl F_3$ mit den Eigenwerten umschreiben in
\begin{align}\label{eA:MQPro}
  \uop M   &= \sum_{\alpha,F_3}E_{\alpha,F_3}\Pro_{\alpha,F_3}\ ,\\ \label{eA:F3QPro}
  \unl F_3 &= \sum_{\alpha,F_3}F_3\Pro_{\alpha,F_3}\ .
\end{align}

\section{Quantenmechanik des Zwei-Teilchen-Problems}\label{sA:ZweiTeilchen}

In diesem Abschnitt wollen wir die Grundlagen zur Beschreibung eines wasserstoffähnlichen Atoms als
Produktzustand von atomaren Zuständen und Wellenpaketen im Ortsraum erarbeiten.
Wir betrachten das Atom als Zwei-Teilchen-System, d.h. wir betrachten den Atomkern
als punktförmiges, geladenes Teilchen mit Kernspin $I$.

Wir beginnen mit möglichst allgemeinen Voraussetzungen und werden gegen Ende des
Abschnitts ein wenig spezieller. Der Hamiltonoperator kann unter Verwendung der 
(zunächst dreidimensionalen) Schwerpunktskoordinate $\v R$
und der Relativkoordinate (des Elektrons) $\v r$ allgemein geschrieben werden als
\begin{align}\label{eA:2PH}
  \begin{split}
    H &= H\subt{ext} + H\subt{int} + V(\v r,\v R)\\
      &= \frac{\v P^2}{2M} + V\subt{ext}(\v R) + \frac{\v p^2}{2m} + V\subt{int}(\v r) + V(\v r,\v R)\ .
  \end{split}
\end{align}
Dabei sei $\v P$ der Schwerpunktsimpuls, $M$ die Gesamtmasse, $\v p$ der Relativimpuls
und $m$ die reduzierte Masse des Elektrons. $V\subt{ext}(\v R)$ sei ein auf das
Atom als ganzes wirkendes Potential, $V\subt{int}$ sei der Operator für die
Wechselwirkung zwischen Elektron und Atomkern und $V(\v r,\v R)$ sei ein
Potential, das Schwerpunkts- und Relativbewegung miteinander verknüpft\footnote{Ein Beispiel
hierfür wäre das Potential des elektrischen Dipolmoments $\v D = -e\v r$ des Atoms 
in einem von $\v R$ abhängigen, äußeren elektrischen Feld $\vmc E(\v R)$, also $V(\v r,\v R) = -\v D\cdot\vmc E(\v R)$}.

Zunächst denken wir uns das Problem der Relativbewegung gelöst, d.h. es sei unter Verwendung
von $\v p = \frac{1}\I\nab_{\v r}$
\begin{align}\label{eA:intEG}\hspace*{-3mm}
  H\subt{int}\vph_j(\v r) = \klr{-\frac1{2m}\v \nabla_{\v r}^2+V\subt{int}(\v r)}\vph_j(\v r) 
    = E^{(0)}_j(\v r)\vph_j(\v r)\ ,\quad(j=1,\ldots,N)\ .
\end{align}
Die $\vph_j(\v r)$ könnten z.B. die Wasserstoff-Wellenfunktionen $\psi_{n,L,L_3}(\v r)$ sein, die
mit dem Index $j$ geeignet durchnummeriert seien. Wir wollen uns im folgenden nur auf den diskreten
Teil des Spektrums beschränken und die Streulösungen, die ein kontinuierliches Eigenwertspektrum
aufweisen, aus der Diskussion ausklammern.

An dieser Stelle bietet sich als Einschub eine Diskussion über innere atomare Eigenzustände an.
Eine genauere Betrachtung des Atoms beinhaltet nämlich auch die Einbeziehung der 
Spin-Bahn-Kopplung (Feinstruktur) sowie der Wechselwirkung
der magnetischen Momente des Elektrons und der Atomkerns (Hyperfeinstruktur). In diesem Fall genügt aber
eine Bezeichnung der Zustände mit den Quantenzahlen $n,L,L_3$ nicht mehr, sondern
die inneren atomaren Zustände sind dann Linearkombinationen der Gesamtdrehimpulszustände 
$\ket{nL_J,S,I,F,F_3}$, in denen Kernspin $\v I$
und Gesamtdrehimpuls $\v J$ des Elektrons zum Gesamtdrehimpuls $\v F = \v I + \v J$ gekoppelt sind.
Der Gesamtdrehimpuls des Elektrons seinerseits ist die Summe aus Bahndrehimpuls und Spin, d.h.
$\v J = \v L + \v S$. Die Zustände $\ket{nL_J,S,I,F,F_3}$ sind dann Eigenzustände der Drehimpulsoperatoren
$\v L^2, \v S^2, \v J^2, \v I^2, \v F^2$ und $F_3 = \v e_3\cdot\v F$. Da Elektronspin und Kernspin
feste Größen für jedes Atom sind, lassen wir die zugehörigen Quantenzahlen in der Dirac-Schreibweise
weg, d.h. wir haben
\begin{align}\label{eA:Basis}
  \ket{nL_J,F,F_3}\quad=\quad\text{\parbox{10cm}{\small Gesamtdrehimpulszustände von Kernspin, Bahndrehimpuls und Spin des Elektrons.}}
\end{align}
Wie bereits erwähnt sind die inneren atomaren Zustände i.A. Linearkombinationen der eben eingeführten reinen
Gesamtdrehimpulszustände, da z.B. die Hyperfeinstruktur zu einer Mischung von Zuständen verschiedener
Elektrongesamtdrehimpulse $J$ führt. Diese inneren atomaren Zustände wollen wir mit einem zusätzlichen
Hut über der Quantenzahl $L$ kennzeichnen und mit einer runden, statt einer spitzen Klammer schreiben, d.h.
\begin{align}\label{eA:Atomzustand}
  \rket{n\hat L_J,F,F_3}\quad=\quad\text{\small innerer, atomarer Zustand im feldfreien Fall.}
\end{align}
Wenn möglich, kürzen wir die Gesamtdrehimpulszustände und die inneren atomaren, feldfreien Zustände
durch die Schreibweisen $\ket{j}$ und $\rket{j}$, $j=1,\ldots,N$, ab. Dabei ist $j$ einfach ein
Index, der alle möglichen Zustände geeignet durchnummeriert. Oft verwenden wir auch den Index $\alpha$
anstelle von $j$ in den inneren Zuständen $\rket{\alpha}$, um den Unterschied in der
Kurzschreibweise noch deutlicher zu machen.
Zum Abschluss dieses kurzen Einschubs wollen wir noch anmerken, dass die inneren atomaren Zustände
bei vorhanden äußeren Feldern in der Form
\begin{align}\label{eA:Zustand.im.Feld}
  \rket{n\hat L_J,F,F_3,\v{\mc E},\v{\mc B}} = \text{\small innerer atomarer Zustand im elektromagnetischen Feld}\ ,
\end{align}
bzw. kurz als $\rket{\alpha,\v{\mc E},\v{\mc B}}$ oder, um die Ortsabhängigkeit
zu verdeutlichen, als $\rket{\alpha(\v R)}$ geschrieben werden. Wir wollen betonen, dass für allgemeine äußere
Felder $\vmc E$ und $\vmc B$ die in der Notation auftretenden Symbole $L$, $J$, $F$ und $F_3$ keine guten Quantenzahlen
mehr sein müssen. Die Notation ist so zu verstehen, dass der Zustand $\rket{n\hat L_J,F,F_3,\v{\mc E},\v{\mc B}}$
adiabatisch aus dem atomaren Zustand $\rket{n\hat L_J,F,F_3}$ für verschwindende äußere Felder entsteht.
Wir wollen weiterhin vereinbaren, dass diese adiabatische Änderung stets entlang 
einer geraden Verbindungslinie der beiden Punkte $(\vmc E=0,\vmc B=0)$ und $(\vmc E,\vmc B)$
stattfindet.

Nun zurück zur Diskussion des Zwei-Teilchen-Problems.
Die Eigenfunktionen $\vph_j(\v r)$ aus Gl. (\ref{eA:intEG})
seien orthonormiert, d.h. es gelte
\begin{align}\label{eA:intON}
  \int_{\mathbb R^3}\d^3r\ \vph^*_{j'}(\v r)\vph_j(\v r) = \delta_{j',j}\ .
\end{align}
Eine Vollständigkeitsrelation gilt für die diskreten Lösungen und die Kontinuumslösungen,
die wir mit einem kontinuierlichen Index $a$ bezeichnen wollen. Sie lautet dann
\begin{align}\label{eA:Vollstaendigkeit}
\sum_j\vph_j^*(\v r')\vph_j(\v r) + \int_{a}\d a\ \vph_a^*(\v r')\vph_a(\v r) = \delta(\v r'-\v r)\ .
\end{align}
Die Gesamtwellenfunktion des Hamiltonoperators (\ref{eA:2PH}) erfüllt die Schrödinger-Gleichung
\begin{align}\label{eA:2PSG}
  H\Psi(\v R,\v r, t) = \I\ddp{}t\Psi(\v R,\v r,t)\ .
\end{align}
Betrachten wir nur gebundene Lösungen des Hamiltonoperators können wir mit der Vollständigkeitsrelation
(\ref{eA:Vollstaendigkeit}) die Gesamtwellenfunktion als Superposition der gebundenen Lösungen
$\vph_j(\v r)$ mit diskretem Index $j$ schreiben, d.h.
\begin{align}\label{eA:2PAnsatz}
  \Psi(\v R,\v r,t) = \sum_j\psi_j(\v R,t)\vph_j(\v r)
\end{align}
mit
\begin{align}\label{eA:2PPart}
\begin{split}
\psi_j(\v R,t) &= \int_{\mb R^3}\d^3r\ \vph^*_j(\v r)\Psi(\v R,\v r,t)\ ,\\
\psi_a(\v R,t) &= \int_{\mb R^3}\d^3r\ \vph^*_a(\v r)\Psi(\v R,\v r,t) = 0\quad (\text{für alle $a$})\ .
\end{split}
\end{align}
Stellen wir die inneren Eigenfunktionen $\vph_j(\v r)$ bzw. ihre Ket-Vektoren
als Vektoren dar, so können wir aus den Schwerpunktsanteilen $\psi_j(\v R,t)$
einen verallgemeinerten Spinor
\begin{align}\label{eA:2PSpinor}
  \ket{\Psi(\v R,t)} = \big(\psi_j(\v R,t)\big) = \begin{pmatrix}
                                                         \psi_1(\v R,t)\\ \vdots\\ \psi_N(\v R,t) 
                                                       \end{pmatrix}
\end{align}
bilden. Wir wollen nun zu einer matrixwertigen Schrödinger-Gleichung für den verallgemeinerten
Spinor (\ref{eA:2PSpinor}) gelangen. Dazu schreiben wir zunächst die Gesamtwellenfunktion in
der Schrödinger-Gleichung (\ref{eA:2PSG}) in der Form (\ref{eA:2PAnsatz}) und erhalten mit
der Eigenwertgleichung (\ref{eA:intEG})
\begin{align}
  \begin{split}
    H\Psi(\v R,\v r,t) &= \sum_j \klr{-\frac{1}{2M}\v \nabla_{\v R}^2 + V\subt{ext}(\v R) + E^{(0)}_j + V(\v r,\v R)}
      \psi_j(\v R,t)\vph_j(\v r)\\
        &= \sum_j\vph_j(\v r)\klr{\I\ddp{}t\psi_j(\v R,t)}\ .
  \end{split}
\end{align}
Nun projizieren wir von links auf die Eigenfunktion $\vph^*_k(\v r)$ und erhalten
\begin{align}
  \sum_j \kle{(-\frac{1}{2M}\v \nabla_{\v R}^2 + V\subt{ext}(\v R)+E^{(0)}_j)\delta_{k,j} + \unl V_{kj}(\v R)}\psi_j(\v R,t)
  = \sum_j \delta_{k,j}\I\ddp{}t\psi_j(\v R,t)
\end{align}
mit der Matrix
\begin{align}\label{eA:2PPotential}
  \unl V(\v R) = \klr{\,\int_{\mb R^3}\d^3r\ \vph^*_k(\v r)V(\v r,\v R)\vph_j(\v r)\,}\ .
\end{align}
Definieren wir noch die Matrizen
\begin{align}\label{eA:intAnteilPot}
  \unl E_0 &= \big(\delta_{k,j}E^{(0)}_j\big)\ ,\\ \label{eA:RestPotential}
  \unl V\subt{ges}(\v R) &= V\subt{ext}(\v R)\unl\um  + \unl V(\v R)\ ,
\end{align}
so gelangen wir zu einer Schrödinger-Gleichung für den in (\ref{eA:2PSpinor}) definierten,
verallgemeinerten Spinor $\ket{\Psi(\v R,t)}$:
\begin{align}\label{eA:MatrixSG}
  \boxed{\klr{-\frac{1}{2M}\v \nabla^2 + \unl E_0 + \unl V\subt{ges}(\v R)}\ket{\Psi(\v R,t)} = \I\ddp{}{t}\ket{\Psi(\v R,t)}}\ .
\end{align}
Der Vollständigkeit halber sollte am Laplace-Operator noch eine Einheitsmatrix im $(N\times N)$-dimensionalen
Raum der inneren atomaren Zustände stehen, also $\v \nabla^2\equiv \unl\um\v \nabla^2$. Es versteht sich weiterhin
von selbst, das der Laplace-Operator nur auf die Schwerpunkts-Koordinate $\v R$ wirkt.

Die in (\ref{eA:Basis}) eingeführten Gesamtdrehimpulszustände $\ket{j}$ bieten sich als Basis für die 
Darstellung aller Matrizen an. Man beachte, dass die Matrix $\unl E_0$ zwar diagonal in der Basis der
freien inneren atomaren Zustände $\rket{\alpha}$ diagonal ist, i.A. aber nichtdiagonal in
der Gesamtdrehimpulsbasis $\ket{j}$. Die Verwendung der Gesamtdrehimpulsbasis bietet sich aber
schon allein aufgrund der Möglichkeit der Zerlegung in elementare Drehimpulseigenzustände unter
Verwendung der Clebsch-Gordan-Koeffizienten an. Wir werden dies bei der expliziten Berechnung der
Massenmatrizen in Kapitel \ref{sB:MundEWP} sehen.

Wir führen nun die lokalen Eigenzustände
\begin{align}\label{eA:lokEZ}
  \rket{\alpha(\v R)} = \sum_j c_{\alpha,j}(\v R)\ket j
\end{align}
der Potentialmatrix
\begin{align}\label{eA:PotMat}
  \unl V(\v R) = \unl E_0 + \unl V\subt{ges}(\v R)
\end{align}
aus der Schrödinger-Gleichung (\ref{eA:MatrixSG}) ein. 
Im Falle eines metastabilen Atoms kommt zum freien Anteil $\unl E_0$ die Matrix
$-\tfrac\I2\unl\Gamma$ der Zerfallsraten hinzu und die Potentialmatrix $\unl V(\v R)$
geht über in die nichthermitesche Massenmatrix
\begin{align}\label{eA:nhMM}
  \uop M(\v R) = \unl V(\v R) - \tfrac\I2\unl\Gamma\ .
\end{align}
In diesem Fall muss man - wie in Abschnitt \ref{sA:WWF} diskutiert - rechte und linke Eigenvektoren von
$\uop M(\v R)$ unterscheiden und man erhält für die linken Bra-Vektoren die zu (\ref{eA:lokEZ})
analoge Zerlegung
\begin{align}\label{eA:lokEZ.lbra}
  \lrbra{\alpha(\v R)} = \sum_j \tilde c_{\alpha,j}^*(\v R)\bra j
\end{align}
mit i.A. von den $c_{\alpha,j}(\v R)$ verschiedenen Koeffizienten $\tilde c_{\alpha,j}(\v R)$.

Im Moment wollen wir uns jedoch der Einfachheit halber auf stabile Atome beschränken
und haben dann an jedem Ort $\v R$ die Eigenwertgleichung
\begin{align}\label{eA:lokEWG}
  \unl V(\v R)\rket{\alpha(\v R)} = V_\alpha(\v R)\rket{\alpha(\v R)}\ ,\qquad(\alpha=1,\ldots,N)\ ,
\end{align}
die wir als gelöst ansehen wollen. Wir gehen davon aus, dass keine Entartungen in den Eigenwerten
auftreten. Für den Zustand des Atoms können wir dann den Ansatz
\begin{align}\label{eA:lokDZ}
  \ket{\Psi(\v R,t)} = \sum_\alpha\Psi_\alpha(\v R,t)\rket{\alpha(\v R)} 
    = \sum_{\alpha,j}\Psi_\alpha(\v R,t)c_{\alpha,j}(\v R)\ket j
\end{align}
machen. Die ursprünglichen Wellenpakete $\psi_j(\v R,t)$ in der Darstellung der Gesamtdrehimpulszustände
sind jetzt Linearkombination der Wellenpakete $\Psi_\alpha(\v R,t)$ in der neuen, lokalen Basis von
Eigenzuständen. Es gilt nach Projektion auf den Zustand $\ket j$
\begin{align}\label{eA:conWP}
  \psi_j(\v R,t) = \bracket{j}{\Psi(\v R,t)} = \sum_\alpha\Psi_\alpha(\v R,t)c_{\alpha,j}(\v R)\ .
\end{align}
Setzen wir nun den Zustand $\ket{\Psi(\v R,t)}$ in der lokalen Darstellung (\ref{eA:lokDZ})
in die Schrödinger-Gleichung (\ref{eA:2PSG}) ein, so folgt
\begin{align}
  \sum_\alpha\klr{-\frac{1}{2M}\v \nabla^2 + V_\alpha(\v R)}\Psi_\alpha(\v R,t)\rket{\alpha(\v R)}
  = \sum_\alpha\I\ddp{}{t}\Psi_\alpha(\v R,t)\rket{\alpha(\v R)}\ .
\end{align}
Nehmen wir nun an, es gäbe keine äußeren Potentiale, dann hängen die Zustände $\rket{\alpha(\v R)}$
nicht mehr von $\v R$ ab und es folgt nach Projektion auf einen Zustand $\rket\beta$
\begin{align}\label{eA:freeSG}
  \klr{-\frac{1}{2M}\v \nabla^2 + V_\beta}\Psi_\beta(\v R,t)
  = \I\ddp{}{t}\Psi_\beta(\v R,t)\ .
\end{align}
Die Lösungen dieser Schrödinger-Gleichung sind einfach die ebenen Wellen, 
die zu einem Wellenpaket superponiert werden können, d.h.
\begin{align}\label{eA:freeLsg}
  \Psi_\beta(\v R,t) = \int\d^3k\ \tilde\Psi_\beta(\v k)\exp\klg{-\I\klr{\frac{\v k^2}{2M} + V_\beta}t + \I\v k\cdot\v R}\ .
\end{align}
Jede ebene Welle hat dabei eine Gesamtenergie
\begin{align}\label{eA:gesEnergie}
  E_\beta(\v k) = \frac{\v k^2}{2M} + V_\beta\ ,
\end{align}
die die Summe aus kinetischer Energie ${\v k^2}/{2M}$ und innerer Energie $V_\beta$ 
des (freien) atomaren Zustands $\rket\beta$ ist.

Die in diesem Abschnitt erarbeiteten Ergebnisse stellen die Grundlage zur Betrachtung der metastabilen Atome
mit mehreren inneren Zuständen in matrixwertigen Potentialen in Kapitel \ref{s6:Formalismus} dar.

\section{Störungsrechnung für nichthermitesche Massenmatrizen}\label{sA:StoeRe}

Wir übertragen den Formalismus der Störungsrechnung auf den Fall
nichthermitescher Massenmatrizen. Dazu gehen wir aus von
einer Matrix
\begin{align}\label{eA:S1}
  \uop M(\lambda) = \uop M_0 +\lambda \uop M_1\ ,
\end{align}
wobei $\uop M_0$ und $\uop M_1$ beide nichthermitesch sein können und $\lambda$
ein reeller Parameter ist. Wir nehmen an, dass das Eigenwertproblem für $\uop
M_0$ gelöst sei und die Eigenwerte nicht entartet sind\footnote{Diese Annahme wird uns auch
in den folgenden Abschnitten begleiten, sowie in Kapitel \ref{s7:Berry}, wo wir viele der im Folgenden
erhaltenen Ergebnisse anwenden.}, d.h. es gelte
\begin{align}\label{eA:S2}
  E^0_\alpha \neq E^0_\beta \quad\text{für}\quad \alpha\neq\beta
\end{align}
und
\begin{align}
  \uop M_0\rket{\alpha^0} =
  E^0_\alpha\rket{\alpha^0},\qquad\lrbra[X]{\alpha^0}\uop M_0 =
  \lrbra[X]{\alpha^0}E_\alpha^0
\end{align}
mit den ungestörten rechten Eigenvektoren $\rket{\alpha^0}$ und linken
Eigenvektoren $\rket{\tilde{\alpha^0}}$.  Wie bereits in Anhang \ref{sA:WWF}
erklärt, können wir für die linken und rechten Eigenvektoren die
Orthonormalitätsrelation
\begin{align}
  \rbracket{\tilde{\alpha^0}}{\beta^0} = \delta_{\alpha\beta}
\end{align}
mit $1\leq \alpha,\beta\leq N$ fordern und damit Quasiprojektoren
\begin{align}\label{eA:SQPro}
    \Pro_\alpha^0 &:= \rket{\alpha^0}\lrbra[X]{\alpha^0}\ ,\qquad(\alpha=1,\ldots,N)\ ,
\end{align}
definieren, die die Eigenschaften
\begin{align}\label{eA:QProProp}
\begin{split}
    \sum_\alpha \Pro_\alpha^0 &= \um\ ,\\
    \Pro_\alpha^0\Pro_\beta^0 &= \Pro_\alpha^0\delta_{\alpha\beta}
  \end{split}
\end{align}
besitzen. Die Eigenwerte von $\uop M(\lambda)$ berechnen sich aus der charakteristischen
Gleichung
\begin{align}
  \det\kle{\uop M(\lambda) - E\cdot\um} = 0\ .
\end{align}
Für genügend kleine $\lambda$ können wir auch die gestörten
Eigenwerte als nicht entartet annehmen:
\begin{align}
  E_\alpha(\lambda) \neq
  E_\beta(\lambda)\quad\text{für}\quad\alpha\neq\beta\ .
\end{align}
Dies impliziert (siehe \cite{BoBrNa95}) die Existenz von linken und rechten
Eigenvektoren und von  Quasiprojektoren für die gestörte Matrix $\uop
M(\lambda)$:
\begin{align}\label{eA:S3}
  \rket{\alpha(\lambda)},\quad\lrbra{\alpha(\lambda)},\quad\mbm
  P_\alpha(\lambda)\ .
\end{align}
In \cite{BoBrNa95}, Anhang C, wurde weiter gezeigt, dass die in (\ref{eA:S3})
dargestellten Größen differenzierbar nach $\lambda$ (in einer geeigneten
Umgebung von Null) sind. Es zeigt sich, dass die korrekt normierten
Eigenvektoren gegeben sind als
\begin{align}\label{eA:Zustandsentwicklung}
  \begin{split}
    \rket{\alpha(\lambda)} &= \Pro_\alpha(\lambda)\rket{\alpha^0}\ ,\\
    \lrbra{\alpha(\lambda)} &= \kle{\Spur(\Pro_\alpha(\lambda)\mbm
      P_\alpha(0))}^{-1}\lrbra[X]{\alpha^0}\Pro_\alpha(\lambda)\ .
  \end{split}
\end{align}

Wir wollen nun die Störungsentwicklung der Quasiprojektoren und somit der
Zustände berechnen und gehen dabei wie im Falle hermitescher Matrizen vor,
wobei die Methode aus \cite{Messiah}, Kap. 16.3, verwendet wird.  Wir
definieren die Resolvente
\begin{align}
  G_0(w) := \frac{1}{w-\uop M_0}\ ,\quad (w\in\mb C)\ ,
\end{align}
die man wegen
\begin{align}
  G_0(w)\Pro_\alpha^0 = \frac{\Pro_\alpha^0}{w-E_\alpha^0}
\end{align}
und (\ref{eA:SQPro}) schreiben kann als
\begin{align}
  G_0(w) =\sum_\alpha\frac{\Pro_\alpha^0}{w-E_\alpha^0}\ .
\end{align}
Der Quasiprojektor $\Pro_\alpha^0$ ist also das Residuum des einfachen Poles von
$G_0(w)$ an der Stelle $E_\alpha^0$ und kann daher mit der Cauchyschen
Integralformel geschrieben werden als
\begin{align}
  \Pro_\alpha^0 = \frac1{2\pi\I}\oint_{C_\alpha}G_0(w)\d w\ ,
\end{align}
wobei $C_\alpha$ ein in der komplexen Ebene geschlossener Weg sei, der
entgegen dem Uhrzeigersinn läuft und ausschließlich die Singularität
bei $E_\alpha^0$ einschließt. Da wir für hinreichend kleine $\lambda$
die Eigenwerte $E_\alpha(\lambda)$ ebenfalls als diskret und nicht entartet
annehmen können, gilt die analoge Relation für den gestörten
Quasiprojektor $\Pro_\alpha(\lambda)$:
\begin{align}\label{eA:SQProCF}
  \Pro_\alpha(\lambda) = \frac1{2\pi\I}\oint_{C_\alpha}G(w)\d w\ ,
\end{align}
mit
\begin{align}
  G(w) := \frac1{w-\uop M_0 - \lambda\uop M_1}\ .
\end{align}
Wir entwickeln nun die Resolvente $G(w)$ nach $\lambda$. Es gilt
\begin{align*}
  G(w) &= \frac1{w-\uop M_0-\lambda\uop M_1}
  = \frac1{w-\uop M_0}\kle{(w-\uop M_0-\lambda\uop M_1) + \lambda\uop M_1}\frac1{w-\uop
M_0-\lambda\uop M_1}\\
  &= \frac1{w-\uop M_0} + \frac1{w-\uop M_0}\lambda\uop M_1\frac1{w-\uop M_0-\lambda\uop M_1}\\
  &= G_0(w)\kle{1+\lambda\uop M_1G(w)}\ .
\end{align*}
Durch iteratives Einsetzen von $G(w)$ ergibt sich
\begin{align}\label{eA:Gz1}
  G(w) = \sum_{n=0}^\infty G_0(w)\kle{\lambda\uop M_1 G_0(w)}^n =
  \sum_{n=0}^\infty (w-\uop M_0)^{-1}\kle{\lambda\uop M_1(w-\uop M_0)^{-1}}^n
\end{align}
und durch Einschieben der Vollständigkeitsrelation (\ref{eA:SQPro}) für die ungestörten
Quasiprojektoren erhalten wir weiter
\begin{align}\label{eA:Gz2}
  G(w) = \sum_{n=0}^{\infty} \klg{\sum_{\beta}\frac{\mbm
      P_\beta^0}{w-E_\beta^0}\kle{\sum_{\gamma}\lambda\uop M_1\frac{\mbm
        P_\gamma^0}{w-E_\gamma^0}}^n}\ .
\end{align}
Nun berechnen wir aus (\ref{eA:SQProCF}) die Störungsentwicklung des
Quasiprojektors $\Pro_\alpha(\lambda)$. Bis zur zweiten Ordnung in $\lambda$ haben wir
mit (\ref{eA:Gz2}):
\begin{align}\label{eA:SQProEntw1}
  \begin{split}
    \Pro_\alpha(\lambda) &= \sum_\beta\mbm
    P_\beta^0\,\frac1{2\pi\I}\oint_{C_\alpha}\frac{1}{w-E_\beta^0}\d w +
    \sum_{\beta,\gamma}\Pro_\beta^0\lambda\uop M_1\Pro_\gamma^0
    \,\frac1{2\pi\I}\oint_{C_\alpha}\frac{1}{(w-E_\beta^0)(w-E_\gamma^0)}\d w\\
    &+\sum_{\beta,\gamma,\kappa}\Pro_\beta^0\lambda\uop M_1\mbm
    P_\gamma^0\lambda\uop M_1\Pro_\kappa^0
    \,\frac1{2\pi\I}\oint_{C_\alpha}\frac{1}{(w-E_\beta^0)(w-E_\gamma^0)(w-E_\kappa^0)}\d w+
\OO(\lambda^3)\\
  \end{split}
\end{align}
Zur Berechnung der Integrale verwendet man die Residuenformel (siehe z.B.
Theorem 6.3 in \cite{FrBu00}), die in unserem
Fall in der Form
\begin{align}\label{eA:Residuenformel}
  \begin{split}
    \frac1{2\pi\I}\oint_{C_\alpha}f(w)\d w &= \Res(f(w);E_\alpha^0)\ ,\\
    \Res(f(w);E_\alpha^0) &=
    \frac{\kle{(\del_w)^{k-1}(w-E_\alpha^0)^kf(w)}}{(k-1)!}\Bigg\vert_{w=E_\alpha^0}
  \end{split}
\end{align}
geschrieben werden kann, 
wobei $f(w)$ jeweils aus den Integralen in (\ref{eA:SQProEntw1}) zu übernehmen
ist und angenommen wurde, dass der Faktor $(w-E_\alpha^0)^{-1}$ in der $k$-ten
Potenz vorkommt, d.h. $f(w)$ einen Pol $k$-ter Ordnung in $E_\alpha^0$
aufweist. In der Ordnung $\lambda^0$ erhalten wir also den ungestörten Quasiprojektor
$\Pro_\alpha^0$, da $C_\alpha$ ja nur die Singularität bei $E_\alpha^0$
einschließen soll. Bei höheren Ordnungen muss man, je nachdem wieviele
der Summationsindizes $\beta,\gamma,\kappa,\ldots$ gleich $\alpha$ sind, das
Residuum des Poles bei $E_\alpha^0$ mit (\ref{eA:Residuenformel})
berechnen. Sind alle Indizes gleich $\alpha$, so ergibt das Integral also 
gerade Null.

Bis zur zweiten Ordnung in $\lambda$ ergibt sich schließlich
\begin{align}\label{eA:Projektorentwicklung}
  \begin{split}
    &\Pro_\alpha(\lambda) = \Pro_\alpha^0 + \sum_{\beta\neq\alpha}
    \frac{\kle{\Pro_\alpha^0\lambda\uop M_1\Pro_\beta^0+ \Pro_\beta^0\lambda\uop
M_1\Pro_\alpha^0}}{E_\alpha^0-E_\beta^0}\\
    &\ +\sum_{\beta,\gamma\neq\alpha}\frac{\kle{\Pro_\alpha^0\lambda\uop
        M_1\Pro_\beta^0\lambda\uop M_1 \Pro_\gamma^0+\mbm
        P_\beta^0\lambda\uop M_1\Pro_\gamma^0\lambda\uop M_1 \mbm
        P_\alpha^0+\Pro_\gamma^0\lambda\uop M_1\Pro_\alpha^0\lambda\uop M_1
        \Pro_\beta^0}}{(E_\alpha^0-E_\gamma^0)(E_\alpha^0-E_\beta^0)}\\
    &\ -\sum_{\beta\neq \alpha}\frac{ \big[\Pro_\alpha^0\lambda\uop M_1\mbm
      P_\alpha^0\lambda\uop M_1\Pro_\beta^0 + \Pro_\alpha^0\lambda\uop
      M_1\Pro_\beta^0\lambda\uop M_1\Pro_\alpha^0
      + \Pro_\beta^0\lambda\uop M_1\Pro_\alpha^0\lambda\uop
M_1\Pro_\alpha^0\big]}{(E_\alpha^0-E_\beta^0)^2}\\
    &\ + \OO(\lambda^3)\raisetag{27mm}
  \end{split}
\end{align}
Die beiden letzten Summen gehören dabei zur Ordnung $\lambda^2$. Die Norm
berechnet sich dann zu
\begin{align}\label{eA:Norm}
  \Spur[\Pro_\alpha(\lambda)\Pro_\alpha^0] &= 1 -
  \sum_{\beta\neq\alpha}\frac1{(E_\alpha^0-E_\beta^0)^2}\Spur[\lambda\uop
  M_1\Pro_\beta^0\lambda\uop M_1\Pro_\alpha^0] + \OO(\lambda^3)
\end{align}
Mit den letzten beiden Gleichungen und zusammen mit
(\ref{eA:Zustandsentwicklung}) können wir die Entwicklung der linken und
rechten Eigenvektoren bei gegebener Matrix $\uop M(\lambda)$ durchführen.
Die Entwicklung der Eigenwerte erhält man aus der Gleichung für den
Erwartungswert der Matrix $\uop M(\lambda)$:
\begin{align}\label{eA:Energieentwicklung}
  \begin{split}
    E_\alpha(\lambda) &= \Spur[\uop M(\lambda)\Pro_\alpha(\lambda)]
    = \Spur[(\uop M_0 + \lambda\uop M_1)\Pro_\alpha(\lambda)]\\
    &= E_\alpha^0 + \Spur[\lambda\uop M_1\Pro_\alpha^0] +
    \sum_{\beta\neq\alpha}\frac1{E_\alpha^0-E_\beta^0} \Spur[\lambda\uop
    M_1\Pro_\beta^0\lambda\uop M_1\Pro_\alpha^0] + \OO(\lambda^3)
  \end{split}
\end{align}

\section{Störungstheoretische Berechnung der Ableitungsmatrizen}\label{sA:OrtsablMat}

In diesem Kapitel wollen wir (eindimensional) ortsabhängige Matrizen und Eigenvektoren
betrachten, d.h. es soll
\begin{align}\label{eA:EWGort}
  \uop M(z)\rket{\alpha(z)} = E_\alpha(z)\rket{\alpha(z)}
\end{align}
für alle Orte $z\in\mb R$ gelten. Wir interessieren uns nun für die
Eigenwertgleichung an einem infinitesimal benachbarten Ort $z'=z+\delta z$.
Auch dort muss die Eigenwertgleichung
\begin{align}\label{eA:EWGort2}
  \uop M(z+\delta z)\rket{\alpha(z+\delta z)} = E_\alpha(z+\delta z)\rket{\alpha(z+\delta z)}  
\end{align}
gelten. Nach Taylorentwicklung aller $z$-abhängigen Größen erhält man daraus
\begin{align}\label{eA:EWGexp}
  \begin{split}
    &\ \Big(\uop M(z) + \delta z\klr{\del_z\uop M(z)} 
      + \tfrac1{2!}(\delta z)^2\klr{\del_z^2\uop M(z)}\Big)\\
    \times &\ \Big(\rket{\alpha(z)} + \delta z\del_z\rket{\alpha(z)}
      + \tfrac1{2!}(\delta z)^2\del_z^2\rket{\alpha(z)}\Big) + \OO((\delta z)^3)\\
    =&\ \Big(E_\alpha(z) + \delta z\klr{\del_z E_\alpha(z)}
      + \tfrac1{2!}(\delta z)^2\klr{\del_z^2 E_\alpha(z)}\Big)\\
    \times &\ \Big(\rket{\alpha(z)} + \delta z\del_z\rket{\alpha(z)}
      + \tfrac1{2!}(\delta z)^2\del_z^2\rket{\alpha(z)}\Big) + \OO((\delta z)^3)\ .
  \end{split}
\end{align}
Wie man sieht, läuft dies auf eine Störungsentwicklung hinaus.
Der reelle Parameter $\lambda$ entspricht nun $\delta z$, desweiteren
identifiziert man
\begin{align}\label{eA:Identifikationen}
  \rket{\alpha^n}\,\hat=\,\kle{\tfrac1{n!}(\del_z)^n\rket{\alpha(z)}},\quad
  E_\alpha^n \,\hat=\, \kle{\tfrac1{n!}(\del_z)^nE_\alpha(z)}\ ,\quad
  \uop M_n \,\hat=\, \kle{\tfrac1{n!}(\del_z)^n\uop M(z)}\ .
\end{align}
Der einzige Unterschied zum letzten Abschnitt ist, dass nun
die Störung der Matrix $\uop M(z)$
nicht mehr linear in $\delta z$ ist, 
sondern Beiträge von beliebig hoher Ordnung
in $\delta z$ enthält, d.h. es gilt mit $\uop M_n$ aus
(\ref{eA:Identifikationen})
\begin{align}\label{eA:MStoerung.z}
  \uop M(z+\delta z) = \uop M(z) + \sum_{n=1}^\infty (\delta z)^n\uop M_n\ .
\end{align}
In den Formeln des letzten Abschnitts entspricht dies einer Entwicklung der
gestörten Massenmatrix nach dem Parameter $\lambda$,
\begin{align}\label{eA:MStoerung.lambda}
  \uop M(\lambda) = \uop M_0 + \sum_{n=1}^\infty\lambda^n\uop M_n\ ,
\end{align}
wobei die $\uop M_n$ in dieser Gleichung natürlich
nichts mit den Ortsableitungen aus (\ref{eA:Identifikationen}) zu tun haben.
Die Herleitung aus dem letzten Abschnitt kann nun im Prinzip übernommen werden.
Auf diese Weise gelangt man schließlich zu einer Störungsentwicklung des
Quasiprojektors $\Pro_\alpha(\lambda)$, in der folgende Ersetzung des
ursprünglichen Störanteils $\lambda\uop M_1$ vorgenommen wurde:
\begin{align}\label{eA:Ersetzung}
\lambda\uop M_1\ \longrightarrow\  \sum_{n=1}^\infty \lambda^n\uop M_n\ .
\end{align}
Die Entwicklung (\ref{eA:Projektorentwicklung}) wird dann nach Ausmultiplikation zu
\begin{align}\label{eA:AllgProEnt}\hspace{-5mm}
  \begin{split}
      \Pro_\alpha&(\lambda) = \Pro_\alpha^0 + \sum_{\beta\neq\alpha}
      \frac{\kle{\Pro_\alpha^0\lambda\uop M_1\Pro_\beta^0
        + \Pro_\beta^0\lambda\uop M_1\Pro_\alpha^0}}{E_\alpha^0-E_\beta^0}\\
    &+ \sum_{\beta\neq\alpha}
      \frac{\kle{\Pro_\alpha^0\lambda^2\uop M_2\Pro_\beta^0
        + \Pro_\beta^0\lambda^2\uop M_2\Pro_\alpha^0}}{E_\alpha^0-E_\beta^0}\\
    &+ \sum_{\beta,\gamma\neq\alpha}\frac{\kle{\Pro_\alpha^0\lambda\uop
        M_1\Pro_\beta^0\lambda\uop M_1 \Pro_\gamma^0 +
        \Pro_\beta^0\lambda\uop M_1\Pro_\gamma^0\lambda\uop M_1
        \Pro_\alpha^0+\Pro_\gamma^0\lambda\uop M_1\Pro_\alpha^0\lambda\uop M_1
        \Pro_\beta^0}}{(E_\alpha^0-E_\gamma^0)(E_\alpha^0-E_\beta^0)}\\
    &- \sum_{\beta\neq \alpha}\frac{ \big[\Pro_\alpha^0\lambda\uop M_1
        \Pro_\alpha^0\lambda\uop M_1\Pro_\beta^0 + \Pro_\alpha^0\lambda\uop
        M_1\Pro_\beta^0\lambda\uop M_1\Pro_\alpha^0
        + \Pro_\beta^0\lambda\uop M_1\Pro_\alpha^0\lambda\uop M_1\Pro_\alpha^0\big]}{(E_\alpha^0-E_\beta^0)^2}\\
    &+ \OO(\lambda^3)\ .
  \end{split}
\end{align}
Nur der Beitrag aus der zweiten Zeile, der zweiter Ordnung in $\lambda$ ist, ist
neu hinzugekommen. Er stammt aus der (ursprünglichen) ersten Ordnung
von (\ref{eA:Projektorentwicklung}). Alle weiteren, neuen Beiträge sind
mindestens von der Ordnung $\lambda^3$ und treten deshalb nicht in obiger
Formel auf.

Nun wollen wir diese allgemeinen Ergebnisse wieder auf den hier vorliegenden Fall über\-tragen.
Wir identifizieren wieder $\delta z$ als den Störparameter und erhalten dann
eine gestörte Massenmatrix der Form (\ref{eA:MStoerung.z}), wobei nun die $\uop M_n$
aus (\ref{eA:Identifikationen}) zu übernehmen sind. Wir wollen die Entwicklung der gestörten Quasiprojektoren
\begin{align}
  \Pro_\alpha(z+\delta z) = \rket{\alpha(z+\delta z)}\lrbra{\alpha(z+\delta z)}\ ,\quad(\alpha=1,\ldots,N)\ ,
\end{align}
nach den ungestörten Quasiprojektoren
\begin{align}
  \Pro_\alpha(z) = \rket{\alpha(z)}\lrbra{\alpha(z)}\ ,\quad(\alpha=1,\ldots,N)\ ,
\end{align}
erhalten, um damit die Entwicklung der Zustände $\rket{\alpha(z+\delta z)}$
durch die Projektion
\begin{align}\label{eA:Pro.Diff}
  \rket{\alpha(z+\delta z)} = \Pro_\alpha(z+\delta z)\rket{\alpha(z)}
    = \sum_{n=0}^\infty \tfrac1{n!}(\delta_z)^n\del_z^n\rket{\alpha(z)}
\end{align}
zu berechnen. Aus der Ordnung $(\delta z)^n$ der Entwicklung des
Quasiprojektors erhalten wir dann gemäß Gl. (\ref{eA:Pro.Diff})
einen Ausdruck für die $n$-te Ableitung des Zustand $\rket{\alpha(z)}$
an der Stelle $z$. Dies wird dann schließlich die Berechnung der Ableitungsmatrizen
\begin{align}\label{eA:dMat}
  \uop D^{(n)}_{\beta\alpha}(z) = \lrmatelem{\beta(z)}{\del_z^n}{\alpha(z)}\ 
  ,\qquad (n\in\mb{N})\ ,
\end{align}
ermöglichen. Wir erhalten also aus (\ref{eA:AllgProEnt}) mit den Ersetzungen
\begin{align}\label{eA:Ersetzungen2}
\begin{split}
  \Pro_\alpha(\lambda)\ &\longrightarrow\ \Pro_\alpha(z+\delta z)\ ,\\
  \Pro_\alpha^0\ &\longrightarrow\ \Pro_\alpha(z)\ ,\\
  E_\alpha^0\ &\longrightarrow\ E_\alpha(z)\ ,\\
  \lambda\ &\longrightarrow\ \delta z\ ,\\
  \uop M_n\ &\longrightarrow\ \tfrac1{n!}\del_z^n\uop M(z)
\end{split}
\end{align}
die Formel
{\small
\begin{align}\label{eA:ProEntOrt}
  \begin{split}
    \Pro_\alpha(z+\delta z) &= \hspace{1cm}\Pro_\alpha(z)\\[2mm]
        &+ (\delta z)\ \klr{\sum_{\beta\neq\alpha}
      \frac{\kle{\Pro_\alpha(z)(\del_z\uop M(z))\Pro_\beta(z)
          + \Pro_\beta(z)(\del_z\uop M(z))\Pro_\alpha(z)}}{E_\alpha(z)-E_\beta(z)}}\\[2mm]
        &+ (\delta z)^2\Bigg(\sum_{\beta\neq\alpha}
      \frac{\kle{\Pro_\alpha(z)(\del_z^2\uop M(z))\Pro_\beta(z)
        + \Pro_\beta(z)(\del_z^2\uop M(z))\Pro_\alpha(z)}}{2!(E_\alpha(z)-E_\beta(z))}\\[1mm]
    &\hspace{11mm}-\hspace{2mm} \sum_{\beta\neq \alpha}\Bigg[
      \frac{\Pro_\alpha(z)(\del_z\uop M(z))\Pro_\alpha(z)(\del_z\uop 
        M(z))\Pro_\beta(z)}{(E_\alpha(z)-E_\beta(z))^2}\\
    &\hspace{23mm}+ 
      \frac{\Pro_\alpha(z)(\del_z\uop M(z))\Pro_\beta(z)(\del_z\uop
        M(z))\Pro_\alpha(z)}{(E_\alpha(z)-E_\beta(z))^2}\\
    &\hspace{23mm}+ 
      \frac{\Pro_\beta(z)(\del_z\uop M(z))\Pro_\alpha(z)(\del_z\uop 
        M(z))\Pro_\alpha(z)}{(E_\alpha(z)-E_\beta(z))^2}
      \Bigg]\\[1mm]
    &\hspace{11mm}+ \sum_{\beta,\gamma\neq\alpha}\Bigg[
        \frac{\Pro_\alpha(z)
        (\del_z\uop M(z))\Pro_\beta(z)(\del_z\uop M(z))\Pro_\gamma(z)}
{(E_\alpha(z)-E_\gamma(z))(E_\alpha(z)-E_\beta(z))}\\
    &\hspace{23mm}+ \frac{\Pro_\beta(z)(\del_z\uop M(z))\Pro_\gamma(z)(\del_z\uop M(z))
        \Pro_\alpha(z)}{(E_\alpha(z)-E_\gamma(z))(E_\alpha(z)-E_\beta(z))}\\
    &\hspace{23mm}+ \frac{\Pro_\gamma(z)(\del_z\uop M(z))\Pro_\alpha(z)(\del_z\uop M(z))
        \Pro_\beta(z)}{(E_\alpha(z)-E_\gamma(z))(E_\alpha(z)-E_\beta(z))}
      \Bigg]\Bigg)\\[0mm]
    &+ \OO((\delta z)^3)\ .
  \end{split}
\end{align}}

\vspace{-5mm}
Wir lesen also gemäß (\ref{eA:Pro.Diff}) die Ableitungen des Zustands $\rket{\alpha(z)}$
wie folgt ab:
\begin{align}\label{eA:dKetOrt}
  \begin{split}
    \del_z\rket{\alpha(z)}   &= \sum_{\beta\neq\alpha}\rket{\beta(z)}
      \frac{\lrmatelem{\beta(z)}{(\del_z\uop M(z))}{\alpha(z)}}{E_\alpha(z)-E_\beta(z)}
  \end{split}
\end{align}
und
\begin{align}\label{eA:ddKetOrt}
  \begin{split}
    \del_z^2\rket{\alpha(z)} &= \sum_{\beta\neq\alpha}\rket{\beta(z)}
      \Bigg[\frac{\lrmatelem{\beta(z)}{(\del_z^2\uop M(z))}{\alpha(z)}}
{E_\alpha(z)-E_\beta(z)}\\
      &\hspace{20mm}-2!\frac{\lrmatelem{\beta(z)}{(\del_z\uop M(z))}{\alpha(z)}
          \lrmatelem{\alpha(z)}{(\del_z\uop M(z))}{\alpha(z)}}
{(E_\alpha(z)-E_\beta(z))^2}\\
      &\hspace{20mm}+ 2!\sum_{\gamma\neq\alpha}
         \frac{\lrmatelem{\beta(z)}{(\del_z\uop M(z))}{\gamma(z)}
           \lrmatelem{\gamma(z)}{(\del_z\uop M(z))}{\alpha(z)}}
{(E_\alpha(z)-E_\gamma(z))(E_\alpha(z)-E_\beta(z))}\Bigg]\\[3mm]
      &- \phantom{\sum_{\beta\neq\alpha}}\,\rket{\alpha(z)}\Bigg[2!\sum_{\beta\neq\alpha}
         \frac{\lrmatelem{\alpha(z)}{(\del_z\uop M(z))}{\beta(z)}
           \lrmatelem{\beta(z)}{(\del_z\uop M(z))}{\alpha(z)}}
{(E_\alpha(z)-E_\beta(z))^2}\Bigg]\ .
     \end{split}
\end{align}
Man erkennt, dass $\del_z^2\rket{\alpha(z)}$ im Gegensatz zu
$\del_z\rket{\alpha(z)}$ auch eine Komponente in $\rket{\alpha(z)}$-Richtung
hat. Man kann sich leicht davon überzeugen, dass dies dazu führt, 
dass die Norm des gestörten Zustands $\rket{\alpha(z+\delta z)}$
dem bereits früher in (\ref{eA:Norm}) berechneten Ausdruck
entspricht (natürlich nach den Ersetzungen aus (\ref{eA:Ersetzungen2})).

Die Matrixelemente der Ortsableitungsmatrizen (\ref{eA:dMat}) lauten also
mit (\ref{eA:dKetOrt})
\begin{subequations}\label{eA:DMatElem1}
\begin{align}
\begin{split}
    \uop D^{(1)}_{\beta\alpha}(z) &= \lrmatelem{\beta(z)}{\del_z}{\alpha(z)}\\[2mm]
    &\!\!\overset{\beta\neq\alpha}= 
      \frac{\lrmatelem{\beta(z)}{(\del_z\uop M(z))}{\alpha(z)}}
{E_\alpha(z)-E_\beta(z)}\ ,
\end{split}
\end{align}
\begin{align}
    \uop D^{(1)}_{\alpha\alpha}(z) &= 0
\end{align}
\end{subequations}
und mit (\ref{eA:ddKetOrt})
\begin{subequations}\label{eA:DMatElem2}
\begin{align}
\begin{split}
\uop D^{(2)}_{\beta\alpha}(z) &= \lrmatelem{\beta(z)}{\del_z^2}{\alpha(z)}\\[2mm]
    &\!\!\overset{\beta\neq\alpha}=\!\! 
      \phantom{2!}\frac{\lrmatelem{\beta(z)}{(\del_z^2\uop M(z))}{\alpha(z)}}
{E_\alpha(z)-E_\beta(z)}\\
    &-2!\frac{\lrmatelem{\beta(z)}{(\del_z\uop M(z))}{\alpha(z)}
          \lrmatelem{\alpha(z)}{(\del_z\uop M(z))}{\alpha(z)}}
{(E_\alpha(z)-E_\beta(z))^2}\\
    &+ 2!\sum_{\gamma\neq\alpha}
         \frac{\lrmatelem{\beta(z)}{(\del_z\uop M(z))}{\gamma(z)}
           \lrmatelem{\gamma(z)}{(\del_z\uop M(z))}{\alpha(z)}}
{(E_\alpha(z)-E_\gamma(z))(E_\alpha(z)-E_\beta(z))}\ ,
\end{split}
\end{align}
\begin{align}
\begin{split}
\uop D^{(2)}_{\alpha\alpha}(z) &=
         2!\sum_{\beta\neq\alpha}
         \frac{\lrmatelem{\alpha(z)}{(\del_z\uop M(z))}{\beta(z)}
           \lrmatelem{\beta(z)}{(\del_z\uop M(z))}{\alpha(z)}}
{(E_\alpha(z)-E_\beta(z))^2}\ .
\end{split}
\end{align}
\end{subequations}
Dabei ist zu beachten, dass die Nebendiagonalelemente und die Diagonalelemente
jeweils separat angegeben wurden. Um die obigen Matrixelemente berechnen zu
können, muss man sich also zunächst die Massenmatrix $\uop M(z)$ beschaffen. In
deren Matrixelementen treten dann die $z$-abhängigen elektrischen und
magnetischen Felder explizit auf, die vorgegeben sein müssen. 
Diese Matrixelemente sind dann die einzigen, die auch in den Ableitungen 
der Massenmatrix ungleich Null sind. Schließlich muss man noch die
ortsabhängigen Eigenvektoren $\rket{\alpha(z)}$ zu $\uop M(z)$ kennen.

Eine weitere nützliche Schreibweise der obigen
Matrixelemente können wir unter Verwendung der Resultate aus
(\ref{eA:DMatElem1}) erhalten:
\begin{align}\label{eA:DMatElem2a}
  \begin{split}
    \uop D^{(2)}_{\beta\alpha}(z)\, &\!\!\overset{\beta\neq\alpha}= 
      \frac{\lrmatelem{\beta(z)}{(\del_z^2\uop M(z))}{\alpha(z)}}
{E_\alpha(z)-E_\beta(z)}\\
    &- 2\uop D^{(1)}_{\beta\alpha}(z)
       \frac{\kle{\lrmatelem{\alpha(z)}{(\del_z\uop M(z))}{\alpha(z)}
           -\lrmatelem{\beta(z)}{(\del_z\uop M(z))}{\beta(z)}}}
{E_\alpha(z)-E_\beta(z)}\\
    &+ 2\sum_{\substack{\gamma\neq\alpha,\\ \gamma\neq\beta}}
        \uop D^{(1)}_{\beta\gamma}(z)\uop D^{(1)}_{\gamma\alpha}(z)
        \frac{E_\beta(z)-E_\gamma(z)}{E_\alpha(z)-E_\gamma(z)}\\[4mm]
    \uop D^{(2)}_{\alpha\alpha}(z) &= -2\sum_{\beta\neq\alpha}
    \uop D^{(1)}_{\alpha\beta}(z)\uop D^{(1)}_{\beta\alpha}(z)\ .
  \end{split}
\end{align}

An dieser Stelle wollen wir ausdrücklich darauf hinweisen, dass die hier vorgestellte Methode zur Berechnung der 
Matrixelemente für die Betrachtung geometrischer Phasen (siehe Kapitel \ref{s7:Berry}) ungeeignet ist. Durch die
Verwendung störungstheoretischer Methoden enthält die erste Ortsableitung eines Eigenzustands $\rket{\alpha(z)}$ nur 
Beiträge von Eigenzuständen mit Indizes $\beta\neq\alpha$. Für geometrische Phasen (im adiabatischen Grenzfall)
muss aber gerade $\lrbra{\alpha(z)}\del_z\rket{\alpha(z)}\neq 0$ gelten.

Die hier verwendete Methode legt also automatisch für jedes $Z$ eine spezielle, lokale Phase der Eigenzustände fest.
Wählt man nun eine andere lokale Phase\footnote{Die Eigenzustände der Massenmatrix sind nur bis auf eine lokale Phase 
bestimmt. Die Orthonormalität der linken und rechten Eigenzustände bleibt erhalten, solange 
$\rket{\alpha(z)} = \exp\klg{\I\phi_\alpha(z)}\rket{\alpha'(z)}$ und 
$\lrbra{\alpha(z)} = \exp\klg{-\I\phi_\alpha(z)}\lrbra{\alpha'(z)}$ gewählt wird, wobei $\phi_\alpha(z)$ eine beliebige,
komplexe Funktion sein darf. Fordert man allerdings die Normierung der rechten Eigenzustände, so wie wir es in der
vorliegenden Arbeit stets getan haben, muss $\phi_\alpha(z)$ reell sein.
Wir müssen berücksichtigen, dass der geometrische Phasenwinkel hier im Allgemeinen
nicht reell ist, da wir Eigenzustände einer nichthermiteschen Matrix betrachten, siehe dazu auch Anhang \ref{s7:Berry}.
} 
d.h.
\begin{align}
  \rket{\alpha(z)} = \e^{\I\phi_\alpha(z)}\rket{\alpha'(z)}\ ,
\end{align}
so folgt
\begin{align}
  0 = \lrbra{\alpha(z)}\del_z\rket{\alpha(z)} = \I \del_z \phi_\alpha(z) + \lrbra{\alpha'(z)}\del_z\rket{\alpha'(z)}
\end{align}
und die geometrische Phase der Zustände $\rket{\alpha'(z)}$ wäre für $\del_z \phi_\alpha(z) \neq 0$ i.A. nicht mehr 
identisch Null.

Mit der hier vorgestellten Methode sammelt sich die geometrische Phase in den Zuständen an. Hat man ein adiabatisches
Potential, das sich an den Orten $z_0$ und $z_D$ gleicht, so ist mit der hier verwendeten Methode nicht mehr sichergestellt,
dass $\rket{\alpha(z_0)} = \rket{\alpha(z_D)}$. Der Unterschied ist gerade die geometrische Phase, d.h. es gilt
$\rket{\alpha(z_D)} = \exp\klg{\I\gamma_\alpha(z_D)}\rket{\alpha(z_0)}$, wobei $\gamma_\alpha(z_D)$ der geometrische 
Phasenwinkel ist.

In praktischen Rechnungen wird das Eigenwertproblem der Massenmatrix an jedem Ort numerisch gelöst. 
Ist $\uop M(z_0) = \uop M(z_D)$, so gilt sicher $\rket{\alpha(z_0)} = \rket{\alpha(z_D)}$,
d.h. in numerischen Rechnungen sollte der geometrische Phasenfaktor leicht zu extrahieren sein,
denn dann ist i.A. auch $\uop D_{\alpha\alpha}^{(1)}(z)\neq 0$. 

\section{Die Paritätstransformation der geometrischen Flussdichten}\label{sA:Parity}

In Abschnitt \ref{s7:Results} diskutieren wir P-erhaltende und P-verletzende geometrische Flussdichten.
Um wirklich zu verstehen, warum diese Flussdichten sich unter P-Transformation entsprechend verhalten,
müssen wir wissen, wie die P-Transformation auf die ungestörte (d.h. $\delta_1=\delta_2=0$) 
Massenmatrix $\uop M(\vmc E,\vmc B)$, die P-verletzenden Matrizen $\uop M\PV^{(1,2)}$ und die ungestörten 
Eigenzustände $\rket{\alpha^{(0)}(\vmc E,\vmc B)}$ wirkt. Wir halten uns an die in Abschnitt \ref{s7:Results} 
verwendete Notation.

Die P-Transformation (auch räumliche Inversion genannt) ist die Punktspiegelung der räumlichen Koordinaten,
\begin{align}\label{eA:Def.P}
P\ :\quad\v x\ \longrightarrow -\v x\ .
\end{align}
Bei der Anwendung der P-Transformation auf vektorwertige Größen unterscheidet man zwischen polaren und axialen Vektoren.
Während polare Vektoren ihr Vorzeichen unter P-Transformation ändern, sind axiale Vektoren invariant unter P-Transformation.
Der Ortsvektor $\v x$ und der Impuls $\v p$ sind z.B. polare Vektoren, Drehimpulse dagegen sind axiale Vektoren.
Das elektrische Feld $\vmc E$ wechselt als polarer Vektor sein Vorzeichen unter P-Transformation, das magnetische Feld
$\vmc B$ dagegen bleibt als axialer Vektor invariant (siehe hierzu auch \cite{Jac99}, Kap. 6.10, Abschnitt B, S. 269 f. und
Tabelle 6.1, S. 271).

Die Transformationseigenschaften der einfachen Vektoren übertragen sich direkt auf die in der Quantenmechanik zugeordneten 
Operatoren, die wir uns im Unterraum der atomaren Zustände mit Hauptquantenzahl $n=2$ in der Gesamtdrehimpulsbasis
$\ket{j}$ (siehe Gl. (\ref{eA:Basis}) in Anhang \ref{sA:ZweiTeilchen}) als Matrizen dargestellt denken können. 
Die Darstellung der Paritätstransformation in dieser Basis wollen wir mit $\umc P$ bezeichen. Es gilt dann
\begin{align}
 \umc P\,\vu x\,\umc P^\dag = - \vu x
\end{align}
und somit auch für das Dipolmoment $\vu D$,
\begin{align}\label{eA:P.D.P}
\umc P\,\vu D\,\umc P^\dag = -e\umc P\,\vu x\,\umc P^\dag = +e \vu x = -\vu D\ .
\end{align}
Das magnetische Moment des Atoms (bzw. des Elektrons, da der Beitrag des Kerns vernachlässigbar klein ist), bleibt
dagegen invariant unter P-Transformation, da es nur von Drehimpulsoperatoren abhängt. 
Es gilt (siehe auch Gl. (\ref{eB:e.magn.Moment}))
\begin{align}\label{eA:P.Mu.P}
\umc P\,\vu\mu\,\umc P^\dag = -\mu_B \umc P\klr{\v L + g\v S}\umc P^\dag = -\mu_B\klr{\v L + g\v S} = \vu\mu\ .
\end{align}
Betrachten wir nun den P-erhaltenden Anteil der Massenmatrix aus Gl. (\ref{e7:MMPV})\ ,
\begin{align}\label{eA:MMPC}
\uop M(\vmc E,\vmc B) = \uop{\tilde M}_0 - \vu D\cdot\vmc E - \vu\mu\cdot\vmc B\ .
\end{align}
Wenden wir hierauf die P-Transformation an, müssen wir dabei nur die Wirkung auf die freie, P-erhaltende
Massenmatrix $\uop{\tilde M}_0$, sowie auf die Matrizen $\vu D$ und $\vu\mu$ für das Dipolmoment und das magnetische
Moment berücksichtigen. Die Feldstärken bleiben von der P-Transformation, die eine Matrixmultiplikation ist,
unberührt. Der Anteil $\uop{\tilde M}_0$ bleibt invariant unter P-Transformation, denn er repräsentiert
das freie Atom ohne P-verletzende Beiträge. Das Coulomb-Potential, in dem das Elektron sich befindet, ist kugelsymmetrisch
und hängt somit nur vom Betrag des Ortsvektors des Elektrons ab. Weitere (relativistische) Korrekturen wie
Fein- und Hyperfeinstruktur hängen neben Drehimpulsoperatoren ebenfalls nur vom Betrag des Ortsvektors ab.
Eine ausführliche Abhandlung über die quantenmechanischen Beschreibung des Wasserstoffatoms findet sich z.B. in 
\cite{BeSa57}.

Mit Gl. (\ref{eA:P.D.P}) und (\ref{eA:P.Mu.P}) erhalten wir also
\begin{align}\label{eA:P.M.P}
\umc P\,\uop M(\vmc E,\vmc B)\,\umc P^{\dag} = \uop{\tilde M}_0 + \vu D\cdot\vmc E - \vu\mu\cdot\vmc B
= \uop M(-\vmc E,\vmc B)\ .
\end{align}
Eine Anwendung der P-Transformation auf die Eigenzustände von $\uop M(\vmc E,\vmc B)$ wird demzufolge einen
Eigenzustand zu den P-Transformierten Feldstärken $(-\vmc E,\vmc B)$ liefern. Im allgemeinen wird aber noch
eine Phasenfaktor aufgrund der intrinsischen Parität des atomaren Zustands hinzukommen. Um diese Phasenfaktoren
in der Diskussion der geometrischen Flussdichten in Abschnitt \ref{s7:Results} zu vermeiden, bietet sich die
Betrachtung der P-Transformation der (ebenfalls ungestörten) Quasiprojektoren $\Pro_\alpha(\vmc E,\vmc B)$ an.
Diese kann man mit Hilfe der Resolvente (vgl. \cite{BoBrNa95}, Abschnitt 3.3)
\begin{align}\label{eA:Resolvente}
\frac{1}{\uop M(\vmc E,\vmc B) - \xi\unl\um} 
= \sum_\alpha \frac{\Pro^{(0)}_\alpha(\vmc E,\vmc B)}{E^{(0)}_\alpha(\vmc E,\vmc B) - \xi}\ ,
\qquad (\xi\in\mb C)
\end{align}
erhalten. Wir haben hier die Vollständigkeitsrelation für die Quasiprojektoren 
$\Pro^{(0)}_\alpha(\vmc E,\vmc B) = \rket{\alpha^{(0)}(\vmc E,\vmc B)}\lrbra{\alpha^{(0)}(\vmc E,\vmc B)}$ aus
Gl. (\ref{eA:QProVollst}) verwendet und die Schreibweise $E^{(0)}_\alpha(\vmc E,\vmc B)$ für die Energieeigenwerte
der P-erhaltenden Massenmatrix (\ref{eA:MMPC}) aus Kap. \ref{s7:Berry} übernommen. Wenden wir die P-Transformation auf die linke Seite dieser Gleichung an, so
folgt
\begin{align}
\umc P\frac{1}{\uop M(\vmc E,\vmc B) - \xi\unl\um}\umc P^\dag
= \frac{1}{\uop M(-\vmc E,\vmc B) - \xi\unl\um} 
= \sum_\alpha \frac{\Pro^{(0)}_\alpha(-\vmc E,\vmc B)}{E^{(0)}_\alpha(-\vmc E,\vmc B) - \xi}\ .
\end{align}
Dies muss aber gleich der P-Transformieren rechten Seite von Gl. (\ref{eA:Resolvente}) sein, d.h.
\begin{align}
\sum_\alpha \frac{\Pro^{(0)}_\alpha(-\vmc E,\vmc B)}{E^{(0)}_\alpha(-\vmc E,\vmc B) - \xi} 
= \sum_\alpha \frac{\umc P\,\Pro^{(0)}_\alpha(\vmc E,\vmc B)\,\umc P^\dag}{E^{(0)}_\alpha(\vmc E,\vmc B) - \xi}\ .
\end{align}
Bei angenommener Nichtentartung der Eigenwerte $E_\alpha(\vmc E,\vmc B)$ kann man nun die Polstruktur in 
der Variable $\xi$ betrachten, die auf beiden Seiten der Gleichung natürlich identisch sein muss. Bei
geeigneter Durchnummerierung in $\alpha$ für elektrische Felder $\vmc E$ und $-\vmc E$ erhält man also
\begin{align}\label{eA:Stark}
E^{(0)}_\alpha(\vmc E,\vmc B) &= E^{(0)}_\alpha(-\vmc E,\vmc B)\ ,\\ \label{eA:P.Pro.P}
\umc P\,\Pro^{(0)}_\alpha(\vmc E,\vmc B)\,\umc P^\dag &= \Pro^{(0)}_\alpha(-\vmc E,\vmc B)\ ,\\ \nonumber
(\alpha &= 1\,\ldots, N)\ .
\end{align}
Die Gleichung (\ref{eA:Stark}) drückt dabei nichts anderes aus als die Abwesenheit eines linearen Stark-Effekts,
die (auch für den Fall metastabiler Atome) für ein statisches elektrisches Feld schon seit langer
Zeit bekannt ist (siehe z.B. \cite{Bou77}, Abschnitt 4.1 und die dort aufgeführten Referenzen).

Nun untersuchen wir die Wirkung auf die (auf $\delta_{1,2}$ normierten) Matrixdarstellungen
\begin{align}\label{eA:MPV}
\uop M\PV^{(\varkappa)} (\vmc E,\vmc B)
= \Big(\lrbra{\alpha^{(0)}(\vmc E,\vmc B)}H\PV^{(\varkappa)}\rket{\beta^{(0)}(\vmc E,\vmc B)}/\delta_\varkappa\Big)\ ,
\quad (\varkappa=1,2)\ ,
\end{align}
der P-verletzenden Hamiltonoperatoren $H\PV^{(1,2)}$, 
deren Matrixelemente in der Darstellung der Gesamtdrehimpulszustände
in Anhang \ref{sB:Beitraege}, Gl. (\ref{eB:HPV}) angegeben sind.

Wie man \cite{BoBrNa95}, Gl. (2.11) und (2.12) entnehmen kann, sind die Operatoren $H^{(1,2)}\PV$
nach der sogenannten nichtrelativistischen Reduktion gegeben durch
\begin{align}\label{eA:HPV}
H^{(1)}\PV &= C\PV^{(1)}\klg{\delta^3(\v x)(\v\sigma\cdot\v p) + (\v\sigma\cdot\v p)\delta^3(\v x)}\ ,\\
H^{(2)}\PV &= C\PV^{(2)}\klg{\delta^3(\v x)(\v I\cdot\v\sigma)(\v\sigma\cdot\v p) 
+ (\v\sigma\cdot\v p)(\v I\cdot\v\sigma)\delta^3(\v x)}\ ,
\end{align}
mit zwei Konstanten $C\PV^{(1,2)}\in\mb R$, die für die folgende Untersuchung nicht weiter interessant sind.
In $H^{(1)}\PV$ steht ein Produkt des axialen Vektoroperators $\v\sigma$ für den Elektronspin mit dem polaren
Vektoroperator des Impulses $\v p$. Unter P-Transformation wird $H^{(1)}\PV$ also sicher sein Vorzeichen
ändern. In $H^{(2)}\PV$ kommt noch das P-invariante Produkt von Kernspin und Elektronspin hinzu, also haben wir
\begin{align}\label{eA:P.HPV.P}
P H^{(\varkappa)}\PV P^\dag = - H^{(\varkappa)}\PV\ ,\qquad(\varkappa=1,2)\ .
\end{align}
Schreiben wir $\uop M^{(\varkappa)}\PV(\vmc E,\vmc B)$ mit Hilfe der Quasiprojektoren als
\begin{align}
\uop M\PV^{(\varkappa)} (\vmc E,\vmc B) 
= \sum_{\alpha,\beta} \Pro^{(0)}_\alpha(\vmc E,\vmc B)\unl H^{(\varkappa)}\PV\Pro^{(0)}_\beta(\vmc E,\vmc B)\ ,
\end{align}
wobei $\unl H^{(\varkappa)}\PV$ die P-verletzenden Matrizen in der $\ket j$-Darstellung seien, so folgt
mit (\ref{eA:P.Pro.P}) und (\ref{eA:P.HPV.P}) 
\begin{align}\label{eA:P.MPV.lok.P}
\begin{split}
\umc P\,\uop M\PV^{(\varkappa)}(\vmc E,\vmc B)\umc P^\dag
&= \sum_{\alpha,\beta} \umc P\,\Pro^{(0)}_\alpha(\vmc E,\vmc B)\umc P^\dag
\umc P\,\unl H^{(\varkappa)}\PV\umc P^\dag\umc P\,\Pro^{(0)}_\beta(\vmc E,\vmc B)\umc P^\dag\\
&= -\sum_{\alpha,\beta} \Pro^{(0)}_\alpha(-\vmc E,\vmc B)\unl H^{(\varkappa)}\PV\Pro^{(0)}_\beta(-\vmc E,\vmc B)\\
&= -\uop M\PV^{(\varkappa)}(-\vmc E,\vmc B)\ .
\end{split}
\end{align}
Mit der gleichen Methode\footnote{Durch Ersetzen von $\unl H^{(\varkappa)}\PV$
durch $\vu D$ bzw. $\vu\mu$ in (\ref{eA:P.MPV.lok.P}) und Anwendung der Relationen (\ref{eA:P.D.P}),
(\ref{eA:P.Mu.P}) sowie (\ref{eA:P.Pro.P}).} kann man sich davon überzeugen, dass für die lokalen Matrixdarstellung 
des Dipolmoments und des magn. Moments gilt
\begin{align}\label{eA:P.D.lok.P}
\umc P\,\vu D(\vmc E,\vmc B)\umc P^\dag = -\vu D(-\vmc E,\vmc B)\ ,
\end{align}
bzw.
\begin{align}\label{eA:P.Mu.lok.P}
\umc P\,\vu\mu(\vmc E,\vmc B)\umc P^\dag = +\vu\mu(-\vmc E,\vmc B)\ .
\end{align}

In der P-erhaltenden geometrischen Flussdichte ${\mc J}^{(\mc E,\,{\rm PC})}_{\alpha,\ell}(\vmc E,\vmc B)$
aus Gl. (\ref{e7:PC.Flux.E}) genügt es, den repräsentativen Beitrag
\begin{align}\label{eA:PCF.Term}
\begin{split}
&~\sum_{\beta\neq\alpha}\frac{\unl D_{i,\alpha\beta}(\vmc E,\vmc B)\unl D_{j,\beta\alpha}(\vmc E,\vmc B)}{
\klr{E^{(0)}_\alpha(\vmc E,\vmc B)-E^{(0)}_\beta(\vmc E,\vmc B)}^2}\\
=&~\Spur\kle{\unl D_i\klr{
\sum_{\beta\neq\alpha}\frac{\Pro^{(0)}_\beta(\vmc E,\vmc B)}{\klr{E^{(0)}_\alpha(\vmc E,\vmc B)-E^{(0)}_\beta(\vmc E,\vmc B)}^2}
}\unl D_j\Pro^{(0)}_\alpha(\vmc E,\vmc B)}
\end{split}
\end{align}
zu betrachten. Durch geschicktes Einschieben von Einheitsmatrizen in der Form $\unl\um = \umc P^\dag\umc P$ zwischen den
Dipolmomenten und den Projektoren, erhalten wir unter Verwendung der obigen Relationen
\begin{align}
\begin{split}
&~\sum_{\beta\neq\alpha}\frac{\unl D_{i,\alpha\beta}(\vmc E,\vmc B)\unl D_{j,\beta\alpha}(\vmc E,\vmc B)}{
\klr{E^{(0)}_\alpha(\vmc E,\vmc B)-E^{(0)}_\beta(\vmc E,\vmc B)}^2}\\
=&~\Spur\kle{\unl D_i\klr{
\sum_{\beta\neq\alpha}\frac{\Pro^{(0)}_\beta(-\vmc E,\vmc B)}{
\klr{E^{(0)}_\alpha(-\vmc E,\vmc B)-E^{(0)}_\beta(-\vmc E,\vmc B)}^2}
}\unl D_j\Pro^{(0)}_\alpha(-\vmc E,\vmc B)}\\
=&~\sum_{\beta\neq\alpha}\frac{\unl D_{i,\alpha\beta}(-\vmc E,\vmc B)\unl D_{j,\beta\alpha}(-\vmc E,\vmc B)}{
\klr{E^{(0)}_\alpha(-\vmc E,\vmc B)-E^{(0)}_\beta(-\vmc E,\vmc B)}^2}\ .
\end{split}
\end{align}
Die beiden Vorzeichen, die bei der P-Transformation der Dipolmomente entstehen, heben sich gegenseitig auf. Um das Vorzeichen
des elektrischen Feldes in den Argumenten der Energien umzukehren, haben wir Gl. (\ref{eA:Stark}) verwendet.
Die Argumentation für die Flussdichte $\v{\mc J}^{(\mc B,\,{\rm PC})}_{\alpha}(\vmc E,\vmc B)$ verläuft vollkommen analog.
Es sind lediglich die Dipolmomente durch die magnetischen Momente zu ersetzen, die ja ohnehin unter P-Transformation
invariant sind.

Insgesamt folgt hieraus also die Invarianz der P-erhaltenden geometrischen Flussdichten unter P-Transformation\footnote{Die 
Bezeichnung als P-erhaltende geometrische Flussdichten ist somit gerechtfertigt.},
\begin{align}\label{eA:PC.Fluxes}
\v{\mc J}^{(\mc E,\,{\rm PC})}_{\alpha}(\vmc E,\vmc B) = \v{\mc J}^{(\mc E,\,{\rm PC})}_{\alpha}(-\vmc E,\vmc B)\ ,
\qquad
\v{\mc J}^{(\mc B,\,{\rm PC})}_{\alpha}(\vmc E,\vmc B) = \v{\mc J}^{(\mc B,\,{\rm PC})}_{\alpha}(-\vmc E,\vmc B)\ .
\end{align}
In den P-verletzenden Flussdichten $\v{\mc J}_{\varkappa,\alpha}^{(\mc E,\,{\rm PV})}(\vmc E,\vmc B)$ ($\varkappa=1,2$)
wird jeweils ein Dipolmoment durch den P-verletzenden Ausdruck $D^{(\varkappa,\,{\rm PV})}_{i,\alpha\beta}(\vmc E,\vmc B)$
aus Gl. (\ref{e7:D.PV}) ersetzt, der wiederrum eine Produkt von Dipolmatrixelementen und Matrixelementen der
P-verletzenden Matrix $\uop M\PV^{(\varkappa)}$ ist. Letztendlich müssen wir das Transformationsverhalten von
Termen der Form
\begin{align}\label{eA:PVF.Term}
\begin{split}
&~\sum_{\beta,\gamma\neq\alpha}\frac{
\uop M^{(\varkappa)}_{{\rm PV},\alpha\gamma}(\vmc E,\vmc B)\unl D_{i,\gamma\beta}(\vmc E,\vmc B)
\unl D_{j,\beta\alpha}(\vmc E,\vmc B)}{
\klr{E^{(0)}_\gamma(\vmc E,\vmc B)-E^{(0)}_\alpha(\vmc E,\vmc B)}\klr{E^{(0)}_\alpha(\vmc E,\vmc B)-E^{(0)}_\beta(\vmc E,\vmc B)}^2}\\
=&~\Spur\Bigg[\uop M^{(\varkappa)}\PV\klr{
\sum_{\gamma\neq\alpha}\frac{\Pro^{(0)}_\beta(\vmc E,\vmc B)}{\klr{E^{(0)}_\alpha(\vmc E,\vmc B)-E^{(0)}_\beta(\vmc E,\vmc B)}^2}
}\\
&\qquad\times \unl D_i\klr{
\sum_{\beta\neq\alpha}\frac{\Pro^{(0)}_\beta(\vmc E,\vmc B)}{\klr{E^{(0)}_\alpha(\vmc E,\vmc B)-E^{(0)}_\beta(\vmc E,\vmc B)}^2}
}\unl D_j\Pro^{(0)}_\alpha(\vmc E,\vmc B)\Bigg]
\end{split}
\end{align}
studieren. Der wesentliche Unterschied zu dem in Gl. (\ref{eA:PCF.Term}) ist die zusätzliche P-verletzende Matrix
$\uop M^{(\varkappa)}\PV$, die beim Einschieben der P-Transformationen ein zusätzliches Minuszeichen liefern wird.
Während sich die Vorzeichenwechsel in den Dipolmomenten gegenseitig aufheben (bzw. die magn. Momente im Fall von
$\v{\mc J}_{\varkappa,\alpha}^{(\mc B,\,{\rm PV})}(\vmc E,\vmc B)$ invariant bleiben), entsteht so also ein
globales Vorzeichen beim Übergang von $\vmc E\to -\vmc E$ und es folgt schließlich
\begin{align}\label{eA:PV.Fluxes}
\begin{split}
\v{\mc J}^{(\mc E,\,{\rm PV})}_{\varkappa,\alpha}(\vmc E,\vmc B) 
&= -\v{\mc J}^{(\mc E,\,{\rm PV})}_{\varkappa,\alpha}(-\vmc E,\vmc B)\ ,\quad
\v{\mc J}^{(\mc B,\,{\rm PV})}_{\varkappa,\alpha}(\vmc E,\vmc B) 
= -\v{\mc J}^{(\mc B,\,{\rm PV})}_{\varkappa,\alpha}(-\vmc E,\vmc B)\ ,\\[5mm]
(\varkappa &= 1,2)\ .
\end{split}
\end{align}

\chapter{Grundlagen für die Beschreibung von metastabilem Wasserstoff und Deuterium}\label{sB:MundEWP}

In diesem Anhang wollen wir die nichthermiteschen Massenmatrizen für
Wasserstoff und Deuterium in den Unterräumen mit den Hauptquantenzahlen
$n=1,2$ explizit angeben und deren Berechnung skizzieren.
Wir verweisen dabei häufig auf \cite{DiplTB}, wo die Berechnungen für
Deuterium im $(n=2)$-Unterraum ausführlich dargestellt werden.

Weiterhin werden wir die Eigenwertprobleme für Wasserstoff und Deuterium
numerisch lösen, auch in Abhängigkeit von einem elektrischen bzw. magnetischen Feld in 3-Richtung.
Beide Felder werden getrennt voneinander betrachtet.

\section{Allgemeine Vorbetrachtungen}\label{sB:Vorspiel}

\subsection{Beiträge zur Massenmatrix}\label{sB:Beitraege}

In der vorliegenden Arbeit betrachten wir nur wasserstoffartige Atome, d.h. Atome mit nur einem
Elektron. Weiterhin betrachten wir den Atomkern näherungsweise als punktförmiges Objekt.
Wir haben es somit mit einem quantenmechanischen Zwei-Teilchen-Problem zu tun, wie wir es bereits in
Abschnitt \ref{sA:ZweiTeilchen} diskutiert haben. Den Zerfall der angeregten Atome, die wir ja nur im
Unterraum der Zustände mit Hauptquantenzahl $n=2$ betrachten wollen, implementieren wir mit der
Wigner-Weisskopf-Methode, die bereits in Abschnitt \ref{sA:WWF} eingeführt wurde.

Wir interessieren uns in diesem Abschnitt nur für die Beschreibung des atomaren Elektrons, nachdem
die Schwerpunktsbewegung abgespalten wurde, d.h. wir wollen die Matrixelemente der Massenmatrix
\begin{align}\label{eB:MaMa}
  \uop M(\v R) = \unl E_0 - \tfrac\I2\unl\Gamma + \unl V\subt{ges}(\v R) =: \uop M_0 + \unl V\subt{ges}(\v R)
\end{align}
berechnen, die von Gl. (\ref{eA:nhMM}) motiviert wird. Wir arbeiten hier mit der Basis der
Gesamtdrehimpulszustände $\ket{2L_J,F,F_3}$ aus Gl. (\ref{eA:Basis}). Der radiative Anteil $H\subt{rad}$
des internen atomaren Hamiltonoperators $H\subt{int}$, siehe Gl. (\ref{eA:WWH}),
\begin{align}\label{eB:Hint}
  H\subt{int} = \hat H + H\subt{rad}
\end{align}
ist bereits in der Matrix $\unl\Gamma$ berücksichtigt, siehe dazu Abschnitt \ref{sA:WWF}. 
Wir verwenden für $\unl\Gamma$ stets eine diagonale
Matrix mit entsprechenden Einträgen $\Gamma_L$, $L=S,P$, die aus experimentellen Zerfallsraten gewonnen
werden. Der stabile Anteil $\hat H$, der in der Matrix $\unl E_0$ enthalten ist, setzt sich zusammen aus
\begin{align}\label{eB:Hstable}
  \hat H = H_0 + H\subt{PV}\ ,
\end{align}
wobei $H_0$ bereits relativistische Korrekturen enthalten soll. Wir verwenden für die Bestimmung der Matrixelemente
von $H_0$ experimentelle Literaturwerte für die Lamb-Shift, die Feinstrukturaufspaltung, die Hyperfeinaufspaltung
und die Zerfallsbreiten. Genauere Informationen dazu geben wir in den
nächsten Abschnitten. 
Der P-verletzende Hamiltonoperator $H\subt{PV}$ enthält den Beitrag
der schwachen Wechselwirkung zwischen Elektron und Atomkern in Form des Austauschs von Z-Bosonen.

Der ortsabhängige Anteil $\unl V\subt{ges}(\v R)$ der Massenmatrix berechnet sich aus den angelegten
äußeren Feldern. Gemäß Definition (\ref{eA:RestPotential}) kommt noch der Beitrag eines auf das gesamte
Atom wirkenden äußeren Potentials $V\subt{ext}(\v R)\unl\um$ hinzu, den wir hier allerdings nicht brauchen,
da wir nur neutrale, wasserstoffartige Atome (Wasserstoff und Deuterium) betrachten wollen.
Für unsere Zwecke genügt also
\begin{align}\label{eB:V.ext}
  \unl V\subt{ges}(\v R) \equiv \unl V(\v R) 
  = \Big(\bra{j}e\v r\cdot\v{\mc E}(\v R) + \mu_B(\v L + g\v S)\cdot\v{\mc B}(\v R)\ket{k}\Big)\ ,
\end{align}
wobei $\v r$ der Ortsoperator, $\v L$ der Operator des Bahndrehimpulses und $\v S$ der Operator des Spins
des Elektrons ist. Weiter ist $e$ die (positive) Elementarladung, $g$ der Land\'e-Faktor des Elektrons
und $\mu_B=e/(2m_e)$ das Bohrsche Magneton. 
Im einzelnen lauten die Hamiltonoperatoren des elektrischen und magnetischen Felds
für das Elektron des neutralen Atoms
\begin{align}\label{eB:H.E}
  H_{\v{\mc E}}(\v R) = -\v D\cdot\v{\mc E}(\v R) = + e\v r\cdot\v{\mc E}(\v R)
\end{align}
und
\begin{align}\label{eB:H.B}
  H_{\v{\mc B}}(\v R) = -\v\mu\cdot\v{\mc B}(\v R) = + \mu_B(\v L + g\v S)\cdot\v{\mc B}(\v R)\ .
\end{align}
Die Vorzeichen kehren sich hier jeweils um aufgrund der negativen Ladung und des negativen magnetischen Moments
des Elektrons.

\subsection{Vorgehensweise und Formeln zur Berechnung der Massenmatrix}\label{sB:M.Berechnung}

Die Vorgehensweise bei der Berechnung der einzelnen Beiträge zur Massenmatrix ist stets die gleiche:
Man zerlegt die Zustände $\ket{j}$ der Gesamtdrehimpulsbasis, siehe Gl. (\ref{eA:Basis}), 
unter Verwendung der Clebsch-Gordan-Koeffizienten
(CGK) in Produkte der einzelnen Spinoren für Elektronspin und Kernspin und der nichtrelativistischen Basiszustände
$\ket{n,L,L_3}$ des Atoms:
\begin{align}\label{eB:Basis.Zerlegung}
  \hspace{-5mm}\ket{nL_J,F,F_3} = \sum_{I_3,S_3,L_3,J_3}
  \bracket{L,L_3;\tfrac12,S_3}{J,J_3}\bracket{J,J_3;I,I_3}{F,F_3}\ket{n,L,L_3}\ket{\tfrac12,S_3}\ket{I,I_3}\ .
\end{align}

Zunächst wollen wir die Matrix $\unl E_0$, also den hermiteschen Anteil der freien Massenmatrix $\uop M_0$
berechnen. Hierzu verwenden wir die Energien aus dem Termschema des jeweils betrachteten Atoms
(siehe nächster Abschnitt) und sind so in der Lage, die Diagonalelemente von $\unl E_0$
in Abhängigkeit der Lamb-Shift $\LambShift$, der Feinstrukturaufspaltung $\FineStructure$ und der Hyperfeinstrukturaufspaltung $\HyperFineSplitting$ des Grundzustandes auszudrücken. Es verbleiben noch die nichtdiagonalen
Beiträge des Hyperfeinstruktur-Hamiltonoperators $H\subt{Hfs}$ und die Beiträge des P-verletzenden
Hamiltonoperators $H\subt{PV}$.

Die Berechnung der Matrixelemente von $H\subt{Hfs}$ ist
ausführlich in \cite{DiplTB}, Kap. 3.3.1, S. 20, diskutiert. Ein Standardwerk
über die quantenmechanische Behandlung von Atomen mit einem oder zwei Elektronen,
in dem neben vielen anderen Dingen auch die Hyperfeinstruktur von Wasserstoff
detailliert behandelt wird, ist das Buch von Bethe und Salpeter \cite{BeSa57}.
Für $n=1$ ist
$H\subt{Hfs}$ diagonal, für $n=2$ tritt eine Mischung innerhalb der
$2P$-Zustände auf. Mit der in \cite{DiplTB} hergeleiteten Formel
\begin{align}\label{eB:Hfs}
  \begin{split}
    &\matelem{2P_{J'},F',F_3'}{H\subt{Hfs}}{2P_J,F,F_3} 
    = C\subt{Hfs}(J',J,I,F)\delta_{F',F}\delta_{F_3',F_3}\ ,\\
    &C\subt{Hfs}(J',J,I,F) 
    = \frac1{20}\frac e{m_e}\mu_K g_I\verw{\frac1{4\pi r^3}}_{n=2,L=1}\\   
&\quad\times \Bigg((-1)^{F+I-\tfrac12}\sqrt{6(2I+1)(I+1)I(2J'+1)(2J+1)}\kle{37-6J'(J'+1)-6J(J+1)}\\
    &\quad\times \wigjb{J}{1}{J'}{1}{\tfrac12}{1}\wigjb{J}{1}{J'} IFI +
8\kle{F(F+1)-I(I+1)-J(J+1)}\delta_{J',J}\Bigg)\raisetag{5mm}
  \end{split}
\end{align}
lässt sich das Verhältnis zwischen den (experimentell gemessenen)
Diagonalelementen und den Nebendiagonalelementen berechnen. Hierzu
benötigt man allerdings entsprechende Formeln für die sog. Wignerschen
$6j$-Symbole, die in (\ref{eB:Hfs}) an den geschweiften Klammern
zu erkennen sind.

Die Berechnung der Matrixelemente von $H\subt{PV}$ ist ebenfalls
ausführlich in \cite{DiplTB} dargestellt. Dort wurde für $n=2$
gezeigt, dass
\begin{align}\label{eB:HPV}
  \begin{split}
     \matelem{2S_{1/2},F',F_3'}{H\subt{PV}^{(1)}}{2P_{1/2},F,F_3} 
     &= -\I\delta_1\,\LambShift\,\delta_{F',F}\delta_{F_3',F_3}\ ,\\    
     \matelem{2S_{1/2},F',F_3'}{H\subt{PV}^{(2)}}{2P_{1/2},F,F_3} 
     &= -\I\delta_2\,\LambShift\\
     &\quad\times \kle{F(F+1)-I(I+1)-\tfrac34}\delta_{F',F}\delta_{F_3',F_3}
  \end{split}
\end{align}
gilt, wobei die paritätsverletzenden Parameter $\delta_{1,2}$
gegeben sind durch
\begin{align}\label{eB:PV.Parameter}
  \delta_i := -\frac{\sqrt3 G}{64\pi\sqrt2}\frac{Q_W^{(i)}}{m_e}\frac1{r_B^4
  \LambShift}\ ,\qquad(i=1,2)\ .
\end{align}
Hier tauchen neben der Elektronmasse $m_e$, dem Bohrschen Radius $r_B$
und der Lamb-Verschiebung $\LambShift=E_{2S_{1/2}}-E_{2P_{1/2}}$ noch die
Fermi-Konstante
\begin{align}\label{eB:GFermi}
  G = 1.1663\ten{-5}\u {GeV}^{-2}
\end{align}
und die schwachen Ladungen
\begin{align}\label{eB:Qweak}
  \begin{split}
    Q_W^{(1)}(Z,N) &= \phantom-(1-4\sin^2\vth_W)Z-N\ ,\\
    Q_W^{(2)}(Z,N) &= -(1-4\sin^2\vth_W)\kle{G_A^{(u)}(Z,N)-G_A^{(d)}(Z,N)-G_A^{(s)}(Z,N)}
  \end{split} 
\end{align}
auf. Beiträge der schwereren Quarks $c,b,t$ wurden vernachlässigt. Die
$G_A^{(q)}$ sind die axialen Formfaktoren der Quarks mit Flavour $q$, die
aus experimentellen Daten berechnet werden können. Für weitere Informationen dazu sei
auf \cite{BoBrNa95}, Anhang A verwiesen, sowie auf dort angegebene Referenzen.

Nun wollen wir die Matrixelemente der Hamiltonoperatoren (\ref{eB:H.E}) und (\ref{eB:H.B})
für die äußeren Felder berechnen. Für $H_{\v{\mc E}}(\v R)$ benötigen wir die Matrixelemente
des Ortsoperators $\v r$ des Elektrons, die wir natürlich in der Ortsdarstellung berechnen
wollen. Hierzu machen wir Gebrauch von der Ortsdarstellung der Wasserstoff-Wellenfunktionen
\begin{align}\label{eB:H-Wellenfkt}
  \bracket{\v r}{n,L,L_3} = \psi_{n,L,L_3}(\v r) = R_{n,L}(r)Y_{L,L_3}(\theta,\phi)
\end{align}
mit der Radialwellenfunktion $R_{n,L}(r)$ und den Kugelflächenfunktionen $Y_{L,L_3}(\theta,\phi)$.
Die Radialwellenfunktionen für $n=1,2$ können aus \cite{Cohen}, Bd. 2, S. 25, Gl. (7.123) entnommen
werden, die Kugelflächenfunktionen z.B. aus dem Buch über Drehimpulse in der Quantenmechanik von
Edmonds \cite{Edm64}, Anhang 2, Tab. 1, S. 148. Die Konvention aus \cite{Cohen} für die Radialwellenfunktion $R_{n,L}(r)$
ist, dass $R_{n,L}(r) \geq 0$ für $r\to 0$. Die für uns interessanten Radialwellenfunktionen lauten
dann nach \cite{Cohen}, Bd. 2, Gl. (7.123), S. 25:
\begin{align}\hspace{-5mm}
\begin{split}
R_{1,0}(r) &= \frac{2}{r_B^{3/2}}\e^{-r/r_B}\ ,\\
R_{2,0}(r) &= \frac{2}{(2r_B)^{3/2}}\klr{1-\frac{r}{2r_B}}\e^{-r/(2r_B)}\ ,
\quad
R_{2,1}(r) = \frac{1}{(2r_B)^{3/2}}\frac{r}{\sqrt3\,r_B}\e^{-r/(2r_B)}\ .
\end{split}
\end{align}

Die Matrixelemente von $\v r$ sind nur zwischen Zuständen verschiedener Parität, also verschiedener
Bahndrehimpulse ungleich Null. Somit gibt es für $n=1$ keine von Null verschiedenen Matrixelemente
und für $n=2$ nur zwischen $S$- und $P$-Zuständen. Hierfür erhält man
\begin{align}\label{eB:r}
  \begin{split}
    &\ \matelem{2S_{1/2},F',F'_3}{\v r}{2P_J,F,F_3}\\
    =&\ -3r_B\sum_{I_3,J_3,J_3',L_3}\v e_{L_3}\bracket{\tfrac12,J_3';I,I_3}{F',F'_3}
    \bracket{J,J_3;I,I_3}{F,F_3}\bracket{1,L_3;\tfrac12,J'_3}{J,J_3}\ .
  \end{split}
\end{align}
In der letzten Gleichung ist $r_B$ der Bohrsche Radius und
$\v e_{L_3}$ einer der folgenden sphärischen Basisvektoren
\begin{align}\label{eB:SphericalBasis}
  \v e_0 = \v e_3,\qquad\v e_±= \mp\tfrac1{\sqrt2}\klr{\v e_1±\I\v e_2}\ .
\end{align}
Somit ist $\v r$ im wesentlichen eine Summe über Produkte von CGK, was
sich am einfachsten mit einschlägigen Computeralgebra-Programmen berechnen
lässt. 

Für die Berechnung der Matrix des Operators $H_{\v{\mc B}}(\v R)$ benötigen wir
die Matrix des Operators des magnetischen Moments $\v\mu$ des Elektrons,
\begin{align}\label{eB:e.magn.Moment}
  \v\mu = -\mu_B(\v L + g\v S)\ ,
\end{align}
siehe Gl. (\ref{eB:H.B}).
Die dritte Komponente des magnetischen Moments kann man sofort anschreiben. Man erhält
\begin{align}\label{eB:mu3}\hspace{-2mm}
  \begin{split}
    &\ \matelem{n'L'_{J'},F',F'_3}{\mu_3}{nL_J,F,F_3}
    =-\mu_B\delta_{F'_3,F_3}\delta_{n',n}\delta_{L',L}\sum_{I_3,J_3,L_3,S_3}(L_3+g\,S_3)\\
    \times &\ \bracket{J',J_3;I,I_3}{F',F_3}\bracket{L',L_3;\tfrac12,S_3}{J',J_3}
    \bracket{L,L_3;\tfrac12,S_3}{J,J_3}\bracket{J,J_3;I,I_3}{F,F_3}\ .
  \end{split}
\end{align}
Die weiteren Komponenten berechnet man mit Hilfe der Leiteroperatoren für Bahndrehimpuls und
Elektronspin. Es ist (\cite{Edm64}, Seite 22, Gl. (2.3.1)) für einen allgemeinen Drehimpulsoperator $\v J$
\begin{align}\label{eB:LeiterOps}
  J_\pm := J_1 \pm \I J_2\ ,\qquad J_1 = \frac12\klr{J_+ + J_-}\ ,\quad J_2 = \frac1{2\I}\klr{J_+ - J_-}
\end{align}
und es zeigt sich, das für die Matrixelemente der Operatoren $J_\pm$ (\cite{Edm64}, Gl. (2.3.16)f.)
\begin{align}\label{eB:LeiterMatElem}
  \matelem{J',J'_3}{J_\pm}{J,J_3} = \sqrt{(J\mp J_3)(J\pm J_3 + 1)}\delta_{J',J}\delta_{J_3',J_3\pm 1}
    =: C_\pm(J,J_3)\delta_{J',J}\delta_{J_3',J_3\pm 1}
\end{align}
gilt, wobei wir als Abkürzung die Funktionen $C_\pm(J,J_3)$ definiert haben. 
Für die Matrixelemente der Operatoren $J_{1,2}$ folgt also
\begin{subequations}
\begin{align}\label{eB:J1MatElem}
  \begin{split}
    \matelem{J',J'_3}{J_1}{J,J_3} = \frac12\delta_{J',J}\Big(
        &C_+(J,J_3)\delta_{J_3',J_3 + 1} + C_-(J,J_3)\delta_{J_3',J_3 - 1}\Big)\ ,
  \end{split}\\
  \begin{split}\label{eB:J2MatElem}
    \matelem{J',J'_3}{J_2}{J,J_3} = \frac1{2\I}\delta_{J',J}\Big(
        &C_+(J,J_3)\delta_{J_3',J_3 + 1} - C_-(J,J_3)\delta_{J_3',J_3 - 1}\Big)\ .
  \end{split}
\end{align}
\end{subequations}
Somit folgt
\begin{align}\label{e9:mu1}
  \begin{split}
    &~\matelem{n'L'_{J'},F',F'_3}{\mu_1}{nL_J,F,F_3}\\ 
   =&~-\mu_B\matelem{n'L'_{J'},F',F'_3}{L_1 + g S_1}{nL_J,F,F_3}\\
   =&~-\frac12\mu_B\Bigg(\matelem{n'L'_{J'},F',F'_3}{L_+ + L_-}{nL_J,F,F_3}\\
    &\quad\qquad+ g\matelem{n'L'_{J'},F',F'_3}{S_+ + S_-}{nL_J,F,F_3}\Bigg)
  \end{split}
\end{align}
Die einzelnen Matrixelemente berechnen sich zu
\begin{align}\label{e9:L_pm}
  \begin{split}
    &~\matelem{n'L'_{J'},F',F'_3}{L_\pm}{nL_J,F,F_3} = \delta_{n',n}\delta_{L',L}\delta_{F_3',F_3\pm 1}\sum_{I_3,S_3,L_3,J_3}C_\pm(L,L_3)\\
    \times&~\bracket{J',J_3\pm1;I,I_3}{F',F_3\pm1}\bracket{L',L_3\pm1;\tfrac12,S_3}{J',J_3\pm1}\\
    \times&~\bracket{L,L_3;\tfrac12,S_3}{J,J_3}\bracket{J,J_3;I,I_3}{F,F_3}
  \end{split}
\end{align}
und
\begin{align}\label{e9:S_pm}
  \begin{split}
    &~\matelem{n'L'_{J'},F',F'_3}{S_\pm}{nL_J,F,F_3} = \delta_{n',n}\delta_{L',L}\delta_{F_3',F_3\pm 1}\sum_{I_3,S_3,L_3,J_3}C_\pm(\tfrac12,S_3)\\
    \times&~\bracket{J',J_3\pm1;I,I_3}{F',F_3\pm1}\bracket{L',L_3;\tfrac12,S_3\pm1}{J',J_3\pm1}\\
    \times&~\bracket{L,L_3;\tfrac12,S_3}{J,J_3}\bracket{J,J_3;I,I_3}{F,F_3}\ .
  \end{split}
\end{align}
Wir ersparen uns an dieser Stelle die Angabe des Endresultats nach Einsetzen von (\ref{e9:L_pm}) und
(\ref{e9:S_pm}) in (\ref{e9:mu1}). Die Matrixdarstellung der zweiten Komponente des magnetischen Moments
folgt analog aus
\begin{align}\label{e9:mu2}
  \begin{split}
    &~\matelem{n'L'_{J'},F',F'_3}{\mu_2}{nL_J,F,F_3}\\ 
   =&~-\mu_B\matelem{n'L'_{J'},F',F'_3}{L_2 + g S_2}{nL_J,F,F_3}\\
   =&~-\frac1{2\I}\mu_B\Bigg(\matelem{n'L'_{J'},F',F'_3}{L_+ - L_-}{nL_J,F,F_3}\\
    &\quad\qquad+ g\matelem{n'L'_{J'},F',F'_3}{S_+ - S_-}{nL_J,F,F_3}\Bigg)\ .
  \end{split}
\end{align}

\section{Wasserstoff}\label{sB:H}

\subsection{Die Massenmatrix im Unterraum mit Hauptquantenzahl \texorpdfstring{$n=1$}{}}\label{sB:H1}

Die Betrachtung des $(n=1)$-Unterraums von Wasserstoff bietet sich als Einführung
für den $(n=2)$-Unterraum an. Hier haben wir die denkbar einfachste
Situation: Es gibt keine Metastabilität\footnote{Wir vernachlässigen hier den Übergang zwischen
den durch die Hyperfeinstruktur getrennten Niveaus. Die zugehörige Lebensdauer der höherliegenden
Energieniveaus liegt im Bereich von 10 Millionen Jahren, bzw. es ist $\Gamma\approx 10^{-15}\u s^{-1}$.}
($\Gamma\approx 0$)
und somit ist die Massenmatrix $\uop M$ hermitesch (und sollte deshalb als
Hamiltonoperator bezeichnet werden). Weiterhin gibt es keine
nichtdiagonalen Beiträge vom Hyperfeinstruktur-Hamiltonian und keinerlei Beiträge
des paritätsverletzenden Hamiltonoperators und des Dipoloperators,
da diese nur Zustände mit unterschiedlicher Parität mischen.
Wir benötigen hier also nur die Matrixdarstellung des 
Magnetfeld-Hamiltonoperators $\unl H_{\mc B}$.

Der $(n=1)$-Unterraum für Wasserstoff hat die Dimension $N = 4$.
Der Bahndrehimpuls $L$ aller Zustände ist Null, der Kernspin ist
$I=\tfrac12$, womit nur die Gesamtdrehimpulse $F=0,1$ möglich sind.
Die Zustände mit $F=1$ sind (ohne Magnetfeld) dreifach entartet
(symmetrisches Triplett), der Zustand zu $F=0$ ist ein Singulett-Zustand.

\begin{figure}[htbp]
  \centering
  \includegraphics{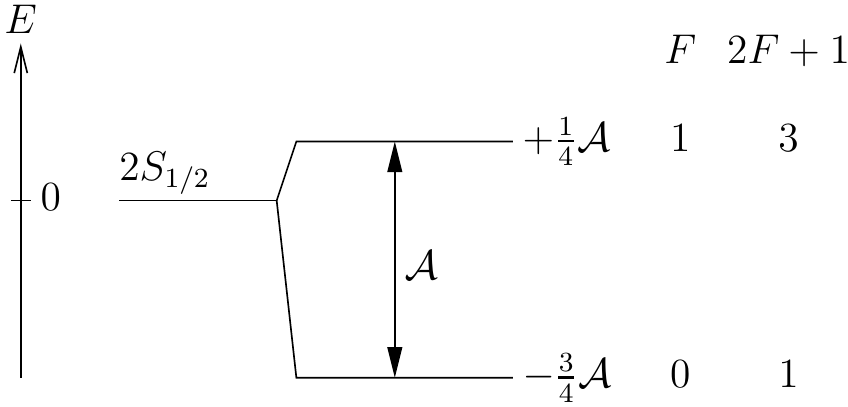}
  \caption{Termschema des Wasserstoff-Grundzustands.}
  \label{fB:TermschemaH1}
\end{figure}
Das zugehörige Termschema ist in Abb. \ref{fB:TermschemaH1} dargestellt,
wobei wir den Schwerpunkt der Energien als Nullpunkt gewählt haben.
Die Zustände zu verschiedenen $F$ sind durch die Hyperfeinaufspaltung\footnote{Die zitierten Quellen
für die in dieser Arbeit verwendeten experimentell gemessenen Aufspaltungen der Energieniveaus von Wasserstoff
und Deuterium sind oft sehr alt, aber immer noch aktuell. Einen guten Überblick über die experimentellen
Werte auf dem Stand von 1990 bietet das Buch von Kinoshita \cite{Kin90}, Kap. 13 und 14, 
dass auch von neueren Arbeiten noch zitiert wird, siehe z.B.
\cite{Hae04}, wo eine neue Messung der 2S-Hyperfeinaufspaltung vorgestellt wird.}
\begin{align}\label{eB:HfsH1S}
  \HyperFineSplitting/h = 1420.405751767\u {MHz}\qquad(\text{aus \cite{Hel70}})
\end{align}
voneinander getrennt ($h=$Planksches Wirkungsquantum). In der Basis (\ref{eA:Basis})
der Gesamtdrehimpulszustände, die wir absteigend zunächst nach $F_3$ und dann nach
$F$ sortieren (siehe z.B. Tabelle \ref{tB:mu3H1}), 
lautet die feldfreie Massenmatrix $\uop M_0$ also
\begin{align}\label{eB:H1.M0}
  \uop M_0 = \diag\klr{\tfrac14\HyperFineSplitting,\tfrac14\HyperFineSplitting,
                      -\tfrac34\HyperFineSplitting,\tfrac14\HyperFineSplitting}\ .
\end{align}
Die auf das Bohrsche Magneton normierte Matrix $\unl\mu_3/\mu_B$ ist in
Tabelle \ref{tB:mu3H1} dargestellt, die nun durch das magnetische Feld 
$\mc B(Z)$ ortsabhängig gewordene Massenmatrix 
$\uop M(Z) = \uop M_0 - \unl\mu_3\mc B(Z)$ findet sich in Tabelle \ref{tB:MH1}.
\begin{table}[hb!]\centering\small
  \begin{tabular}{r||c||c|c||c|}
    &
    $1S_{1/2}$ &
    $1S_{1/2}$ &
    $2S_{1/2}$ &
    $1S_{1/2}$ \\
    &
    $1,1$ &
    $1,0$ &
    $0,0$ &
    $1,-1$ \\ \hlx{vhhv}
    $1S_{1/2},1,1$   & $-\tfrac g2$ & 0 & 0 & 0\\ \hlx{vhhv}
    $1S_{1/2},1,0$   & 0 & 0 & $-\tfrac g2$ & 0\\ \hlx{vhv}
    $1S_{1/2},0,0$   & 0 & $-\tfrac g2$ & 0 & 0\\ \hlx{vhhv}
    $1S_{1/2},1,-1$  & 0 & 0 & 0 & $\tfrac g2$\\ \hlx{vh}
  \end{tabular}
  \caption[Die Matrix der dritten Komponente des magn. Moments für
    Wasserstoff im Grundzustand]{Die Matrix der dritten Komponente des magn. Moments
    \unl{$\mu$}$_3/\mu_B$, normiert auf das Bohrsche Magneton. Es ist $g=2.0023$
  der Landé-Faktor des Elektrons}\label{tB:mu3H1}
\end{table}
\begin{table}[htb!]\centering\small
  \begin{tabular}{r||c||c|c||c|}
    &
    $1S_{1/2}$ &
    $1S_{1/2}$ &
    $2S_{1/2}$ &
    $1S_{1/2}$ \\
    &
    $1,1$ &
    $1,0$ &
    $0,0$ &
    $1,-1$ \\ \hlx{vhhv}
    $1S_{1/2},1,1$   & $\tfrac14\HyperFineSplitting+\tfrac12g\mu_B\mc B(Z)$ & 0 & 0 & 0\\ \hlx{vhhv}
    $1S_{1/2},1,0$   & 0 & $\tfrac14\HyperFineSplitting$ & $\tfrac12g\mu_B\mc B(Z)$ & 0\\ \hlx{vhv}
    $1S_{1/2},0,0$   & 0 & $\tfrac12g\mu_B\mc B(Z)$ & $-\tfrac34\HyperFineSplitting$ & 0\\ \hlx{vhhv}
    $1S_{1/2},1,-1$  & 0 & 0 & 0 & $\tfrac14\HyperFineSplitting-\tfrac12g\mu_B\mc B(Z)$\\ \hlx{vh}
  \end{tabular}
  \caption[Die Massenmatrix für Wasserstoff im Grundzustand]{Die Massenmatrix \unl{$\ms
M$}$(z)$.}\label{tB:MH1}
\end{table}

\begin{table}[hb!]
  \centering
  \begin{tabular}[c]{|c||c|c||l|l|}\hlx{hv}
    $\alpha$ & $F$ & $F_3$ & $E_\alpha(Z)$ & $\rket{\alpha(Z)}=$\\ \hlx{vhhv}
    1 & 1 & 1 & $\tfrac14\HyperFineSplitting(1+4\hat b(Z))$ & $\ket{1S_{1/2},1,1}$\\
    \hlx{vhv}
    2 & 1 & 0 & $-\tfrac14\HyperFineSplitting(1-2\sqrt{h(Z)})$ & 
    $\mc N_+(Z)\klr{c_+(Z)\ket{1S_{1/2},1,0} + \ket{1S_{1/2},0,0}}$\\ \hlx{vhv}
    3 & 1 & -1 & $\tfrac14\HyperFineSplitting(1-4\hat b(Z))$ &
    $\rket{1S_{1/2},1,-1}$\\ \hlx{vhhv}
    4 & 0 & 0 & $-\tfrac14\HyperFineSplitting(1+2\sqrt{h(Z)})$ &
    $\mc N_-(Z)\klr{c_-(Z)\ket{1S_{1/2},1,0} + \ket{1S_{1/2},0,0}}$\\ \hlx{vh}
  \end{tabular}
  \caption[Eigenwerte und Eigenvektoren der Massenmatrix]{Eigenwerte und Eigenvektoren der
Massenmatrix \unl{$\ms M$}$(Z)$. Man
    beachte, dass $F$ für die $\rket{\alpha(Z)}$ i.A. keine gute Quantenzahl mehr ist.}
  \label{tB:MEWEV}
\end{table}
\FloatBarrier

Die Berechnung der Eigenvektoren und Eigenwerte gestaltet sich in diesem Fall
sehr einfach. In Tabelle \ref{tB:MEWEV} sind die Ergebnisse dargestellt,
dabei ist zu beachten, dass die Reihenfolge der Vektoren etwas anders gewählt
wurde als bisher. Die Nummerierung entspricht nun der in \cite{DissAR}
verwendeten Konvention. 

In Tabelle \ref{tB:MEWEV} wurden die folgenden Abkürzungen verwendet:
\begin{align}\label{eB:bH1}
  \hat b(Z)     &:= \tfrac12 g\mu_B\mc B(Z)/\HyperFineSplitting\ ,\\ \label{eB:hH1}
  h(Z) &:= 1+4\hat b^2(Z)\ ,\\ \label{eB:cH1}
  c_±(Z) &:= \frac{1±\sqrt{h(Z)}}{2 \hat b(Z)}\ ,\\ \label{eB:NH1}
  \mc N_±(Z) &:= \sqrt{\frac{2\hat b^2(Z)}{h(Z) ±\sqrt{h(Z)}}}\ .
\end{align}
\begin{floatingfigure}[r]{8cm}
\centering
  \includegraphics[width=8cm]{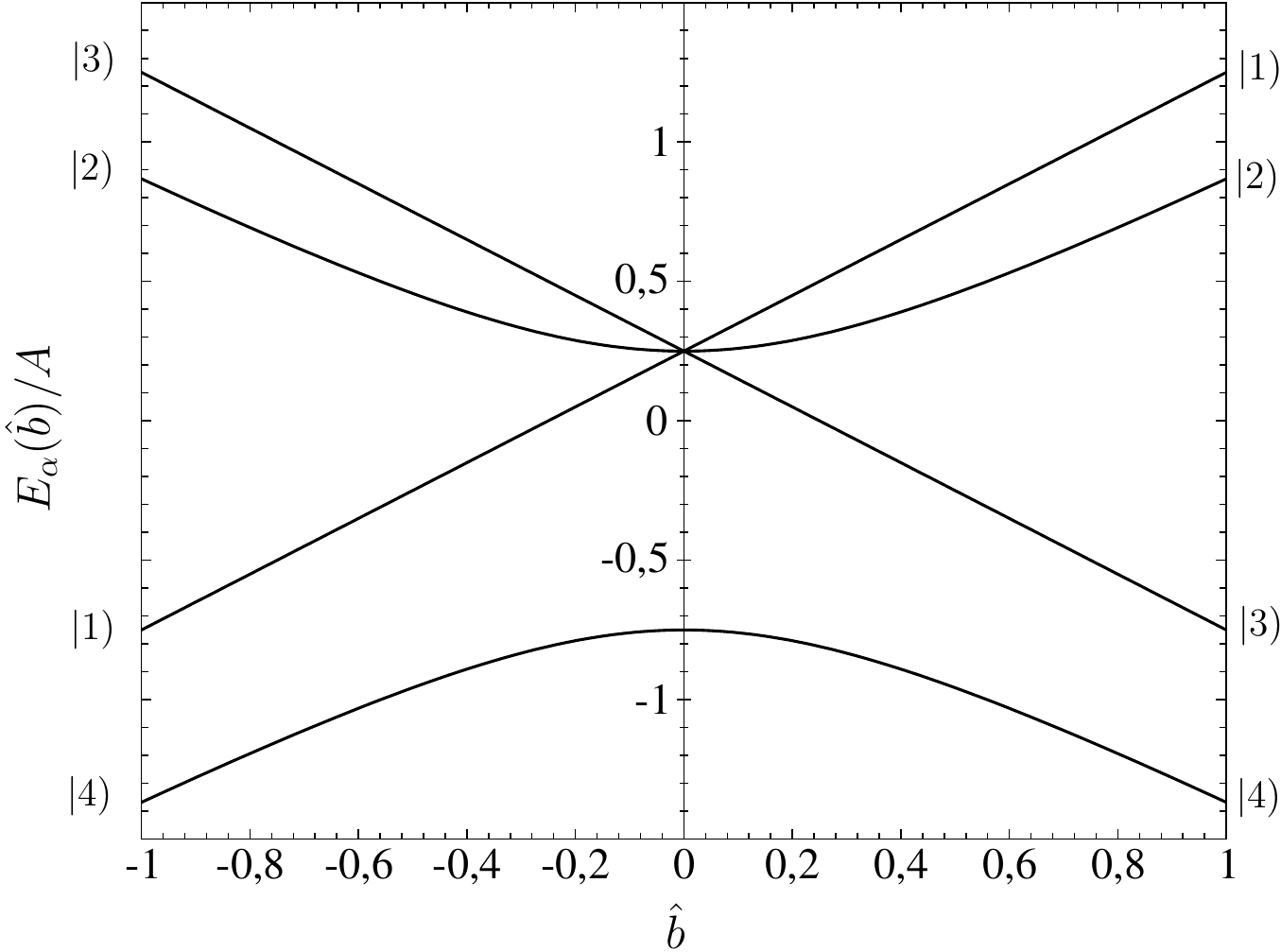}
\caption{Breit-Rabi-Diagramm für Wasserstoff im $(n=1)$-Unterraum}
\label{fB:BR-H1}
\end{floatingfigure}
Mit den berechneten Eigenenergien im Magnetfeld ergibt sich das in Abb. \ref{fB:BR-H1}
dargestellte Breit-Rabi-Diagramm. Man erkennt den Übergang zwischen dem Zeeman-Bereich,
wo sich die Energien der Zustände mit $\alpha=2,4$ quadratisch mit $\hat b$ ändern, und dem
Paschen-Back-Bereich, wo sich die Energien aller Zustände linear mit $\hat b$ ändern.
An der Stelle $\hat b \approx 0.5$, wo etwa die Grenze dieser Bereich liegt, definiert man
das sog. kritische Magnetfeld, was hier einen Wert 
\begin{align}\label{eB:BkritH1}
\mc B\subt{krit.} = \frac{\HyperFineSplitting}{g\mu_B} \approx 50\u{mT}
\end{align}
hat.

Nun berechnen wir Ableitungsmatrizen $\uop D^{(1,2)}(Z)$, die in
(\ref{e6:D1Matrix}), (\ref{e6:D2Matrix}) definiert wurden. Es bietet sich dabei die
in Anhang \ref{sA:OrtsablMat} dargestellte Vorgehensweise an. Hierzu
benötigen wir jedoch zunächst die Matrixdarstellung von $\del_Z\uop M(Z)$ bezüglich 
der lokalen Eigenvektoren $\rket{\alpha(Z)}$. Eine Unterscheidung zwischen
linken und rechten Eigenvektoren gibt es dabei in diesem Fall nicht.
Die Darstellung von $(\del_Z\uop M(Z))$ in der Basis $\klg{\ket{1S_{1/2},F,F_3}}$ 
der Gesamtdrehimpulszustände aus Gl. {\ref{eA:Basis}}
ergibt sich sofort aus Tabelle \ref{tB:MH1}, sie kann in Tabelle \ref{tB:dMH1}
abgelesen werden (für $k=1$).

Die Berechnung von $\uop D^{(1)}_{\beta\alpha}(Z)$ erfolgt nun durch
zweifaches Einschieben der Basis $\klg{\ket{1S_{1/2},F,F_3}}$ in den Ausdruck
(\ref{e6:D1Matrix}):
\begin{align}
  \begin{split}
    \uop D^{(1)}_{\beta\alpha}(Z)\, &\!\!\overset{\beta\neq\alpha}= \sum_{F,F',F_3,F_3'}
    \rbra{\beta(Z)}1S_{1/2},F',F'_3\rangle\langle 1S_{1/2},F,F_3\rket{\alpha(Z)}\\
    &\times \frac{\matelem{1S_{1/2},F',F'_3}{(\del_Z\uop M(Z))}{1S_{1/2},F,F_3}}
{E_\alpha(Z)-E_\beta(Z)}
  \end{split}
\end{align}
Mit den Definitionen der Zustände $\rket{\alpha(Z)}$ aus Tabelle
\ref{tB:MEWEV} ergibt sich nach einiger Rechnung das in Tabelle
\ref{tB:D1H1} dargestellt Ergebnis.

Desweiteren erhält man mit
der Formel (\ref{eA:DMatElem2a}), $\uop D^{(1)}(Z)$ aus
Tabelle \ref{tB:D1H1} und mit Tabelle \ref{tB:dMH1} ($k=2$),
$\uop D^{(2)}(Z)$ wie in Tabelle \ref{tB:D2H1} dargestellt.
\FloatBarrier

\begin{table}[htb]\centering\small
  \begin{tabular}{r||c||c|c||c|}
    &
    $1S_{1/2}$ &
    $1S_{1/2}$ &
    $2S_{1/2}$ &
    $1S_{1/2}$ \\
    &
    $1,1$ &
    $1,0$ &
    $0,0$ &
    $1,-1$ \\ \hlx{vhhv}
    $1S_{1/2},1,1$   & $\tfrac12g\mu_B\mc B^{(k)}(Z)$ & 0 & 0 & 0\\ \hlx{vhhv}
    $1S_{1/2},1,0$   & 0 & 0 & $\tfrac12g\mu_B\mc B^{(k)}(Z)$ & 0\\ \hlx{vhv}
    $1S_{1/2},0,0$   & 0 & $\tfrac12g\mu_B\mc B^{(k)}(Z)$ & 0 & 0\\ \hlx{vhhv}
    $1S_{1/2},1,-1$  & 0 & 0 & 0 & $\tfrac12g\mu_B\mc B^{(k)}(Z)$\\ \hlx{vh}
  \end{tabular}
  \caption[Die $k$-fach abgeleitete Massenmatrix $\del_Z^k$\unl{$\ms
    M$}$(Z)$]{
    Die $k$-fach abgeleitete Massenmatrix $\del_Z^k$\unl{$\ms M$}$(Z)$. $\mc B^{(k)}(Z)$ steht
    für die $k$-te Ableitung des Magnetfelds nach $Z$.}\label{tB:dMH1}
\end{table}
\begin{table}[!htbp]
  \centering
  \begin{tabular}[c]{|c||c|c|c|c|}\hlx{hv}
    $\beta\backslash\alpha$& 1 & 2 & 3 & 4\\ \hlx{vhhv}
    1 & 0 & 0 & 0 & 0 \\ \hlx{vhv}
    2 & 0 & 0 & 0 & $-\frac{\hat b'(Z)}{h(Z)}\frac{\abs{\hat b(Z)}}{\hat b(Z)}$\\ \hlx{vhv}
    3 & 0 & 0 & 0 & 0 \\ \hlx{vhv}
    4 & 0 & $\frac{\hat b'(Z)}{h(Z)}\frac{\abs{\hat b(Z)}}{\hat b(Z)}$ & 0 & 0 \\ \hlx{vh} 
  \end{tabular}
  \caption[Die Matrix \unl{$\ms D$}$^{(1)}(Z)$]{Die Matrix \unl{$\ms D$}$^{(1)}(Z)$ in der
Darstellung der
    Basis der lokalen Eigenzustände von \unl{$\ms M$}$(Z)$.}
  \label{tB:D1H1}
\end{table}
\begin{table}[!htbp]
  \centering
  \begin{tabular}[c]{|c||c|c|c|c|}\hlx{hv}
    $\beta\backslash\alpha$& 1 & 2 & 3 & 4\\ \hlx{vhhv}
    1 & 0 & 0 & 0 & 0 \\ \hlx{vhv}
    2 & 0 & $2\klr{\frac{\hat b'(Z)}{h(Z)}}^2$ & 0 & 
    $\klr{8\hat b(Z)\klr{\frac{\hat b'(Z)}{h(Z)}}^2-\frac{\hat
        b''(Z)}{h(Z)}}\frac{\abs{\hat b(Z)}}{\hat b(Z)}$\\ \hlx{vhv}
    3 & 0 & 0 & 0 & 0 \\ \hlx{vhv}
    4 & 0 & $-\klr{8\hat b(Z)\klr{\frac{\hat b'(Z)}{h(Z)}}^2-\frac{\hat
        b''(Z)}{h(Z)}}\frac{\abs{\hat b(Z)}}{\hat b(Z)}$ 
    & 0 & $2\klr{\frac{\hat b'(Z)}{h(Z)}}^2$ \\ \hlx{vh} 
  \end{tabular}
  \caption[Die Matrix \unl{$\ms D$}$^{(2)}(Z)$]{Die Matrix \unl{$\ms D$}$^{(2)}(Z)$ in der
Darstellung der
    Basis der lokalen Eigenzustände von \unl{$\ms M$}$(Z)$.}
  \label{tB:D2H1}
\end{table}
\FloatBarrier

\subsection{Die Massenmatrix im Unterraum mit Hauptquantenzahl \texorpdfstring{$n=2$}{}}\label{sB:H2}

Der $(n=2)$-Unterraum für Wasserstoff ist 16-dimensional. Der maximale
Gesamtdrehimpuls wird im Zustand $2P_{3/2}$ mit $F=2$ erreicht.
Die $2P$-Zustände mit unterschiedlichem $J$ sind durch die Feinstruktur
$\FineStructure$, die $2S_{1/2}$- und die
$2P_{1/2}$ sind durch die Lamb-Verschiebung $\LambShift$ voneinander getrennt.
Die experimentellen Werte für diese Energien lauten
\begin{align}\label{eB:LambH2}
  \LambShift/h &=    1057.8514(19)\u {MHz}\qquad\text{(aus \cite{PaSoYa85})}\ ,\\ \label{eB:FSH2}
  \FineStructure/h &= 10969.127(95)\u {MHz}\qquad\text{(aus \cite{Eri77})}\ .&
\end{align}
Die Hyperfeinaufspaltungen der Zustände mit den Quantenzahlen
$2S_{1/2}$, $2P_{1/2}$ und $2P_{3/2}$ können aus der
Hyperfeinaufspaltung $\HyperFineSplitting$ des Grundzustands berechnet werden.
Die hierfür verwendete Formel, die aus den Ergebnissen aus
\cite{BeSa57}, S. 110, abgeleitet werden kann, lautet
\begin{align}\label{eB:HfsSkalierung}
  \HyperFineSplitting(I,nL_J) &= \frac{3}{2n^3(2L+1)(J+1)}\frac{f(I,1/2)}{f(I,J)}\HyperFineSplitting\ ,\quad
  f(I,J) =
  \begin{cases}
    I+\tfrac12 & J\leq I\\
    \frac{I(J+\tfrac12)}J & J\geq I
  \end{cases}\ .
\end{align}
Für Wasserstoff mit Kernspin $I=\tfrac12$ ergeben sich damit die in
Tab. \ref{tB:HfsH2} angegebenen Werte. 
\begin{table}[!htbp]
  \centering
  \begin{tabular}{|l||c|c|c|c|}\hlx{hv}
    $nL_J$ & $1S_{1/2}$ & $2S_{1/2}$ & $2P_{1/2}$ & $2P_{3/2}$\\ \hlx{vhv}
    $\HyperFineSplitting(\tfrac12,nL_J)$ & $\HyperFineSplitting=1420.4\u {MHz}\,h$ 
  & $\frac18\HyperFineSplitting$ & $\frac1{24}\HyperFineSplitting$ & $\frac1{60}\HyperFineSplitting$\\
\hlx{vh}
  \end{tabular}
  \caption{Skalierung der Hyperfeinaufspaltung für Wasserstoff mit $n\leq 2$.}
  \label{tB:HfsH2}
\end{table}

Die $(n=2)$-Zustände zerfallen auf unterschiedliche Weise in die
$(n=1)$-Grundzustände: Die $2P_J$-Zustände gehen hauptsächlich
durch spontane Emission eines Photons (E1-Übergang)
in die $1S_{1/2}$-Zustände über, die $2S_{1/2}$-Zustände
dagegen sind metastabil und gehen hauptsächlich durch den
selteneren zwei-Photon-Zerfall (2E1-Übergang)
in den Grundzustand über. Die mittleren
Lebensdauern $\tau_{S,P}$, bzw. die totalen Zerfallsbreiten
$\Gamma_{S,P} = \tau^{-1}_{S,P}$, sind daher für beide Zustandsgruppen stark unterschiedlich
und lauten:
{\small
\begin{align}\label{eB:DR-H2-P}
  &\hspace{-5mm}\Gamma_P/\hbar = \tau_P^{-1} = \Gamma(2P\to
  1S+\gamma)(1+\OO(\alpha\subt{em}))
  = (1.6\ten{-9}\u s)^{-1}\quad (\text{aus \cite{BeSa57}})\ ,\\ \label{eB:DR-H2-S}
  &\hspace{-5mm}\Gamma_S/\hbar = \tau_S^{-1} = \Gamma(2S\to 1S+2\gamma)(1+\OO(\alpha\subt{em}))
  = (0.1215\u s)^{-1}\qquad (\text{aus \cite{Dra86}})\ .
\end{align}}

\begin{figure}[!hb]
  \centering
  \includegraphics[width=15cm]{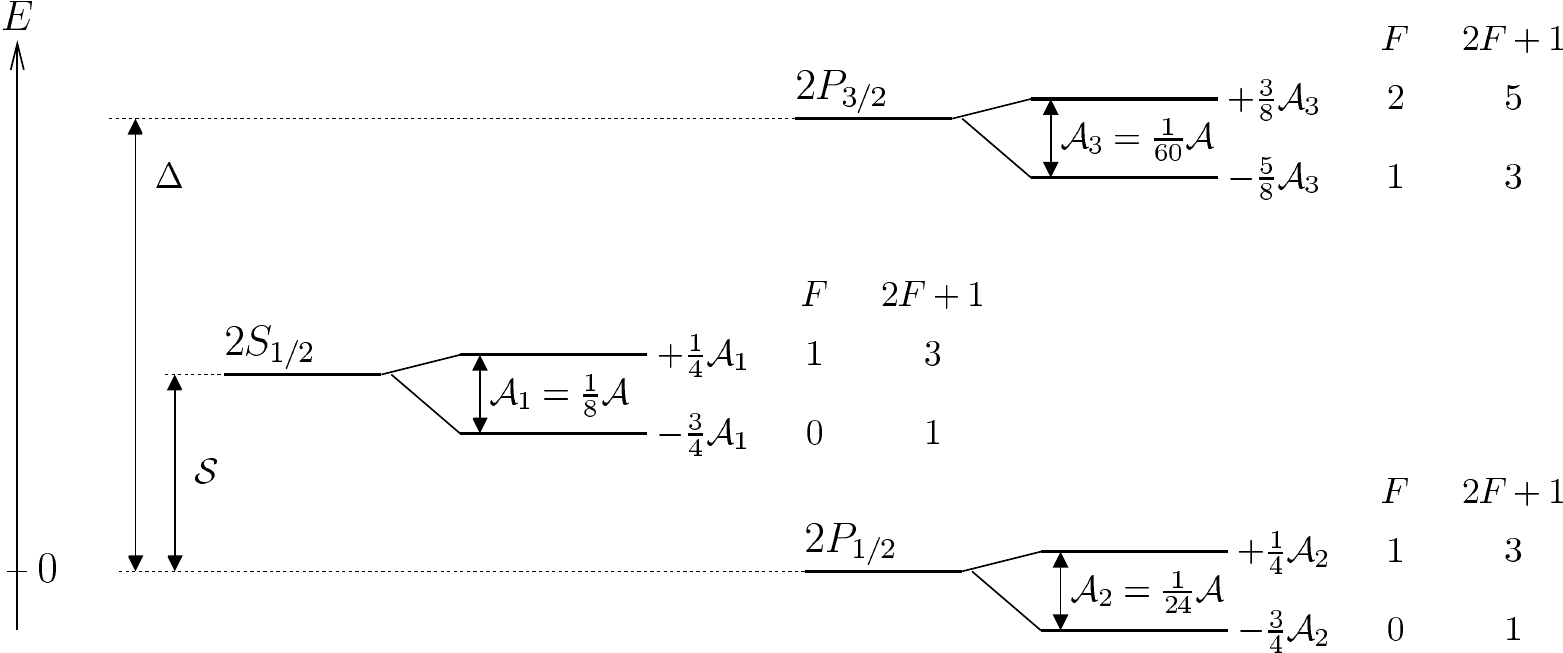}
  \caption{Termschema der $(n=2)$-Zustände von Wasserstoff (nicht maßstabsgetreu).}
  \label{fB:TermschemaH2}
\end{figure}

Insgesamt ergibt sich damit das in Abb. \ref{fB:TermschemaH2} dargestellte Termschema. Die 
freie Massenmatrix $\uop M_0$ ist in den Tabellen \subref*{tB:M0H2a} bis
\subref*{tB:M0H2c} angegeben. Die Tabellen \ref{tB:mu1H2}, \ref{tB:mu2H2} und \ref{tB:mu3H2}
enthalten die Matrixdarstellungen der drei Komponenten des Operators des (auf das Bohrsche Magneton $\mu_B$ normierten) magnetischen Moments
$\vu\mu = (\unl\mu_1,\unl\mu_2,\unl\mu_3)$. Während die Matrix $\unl\mu_3$ blockdiagonal
bzgl. der dritten Komponente $F_3$ des Gesamtdrehimpulses ist, sind in $\unl\mu_1$ und
$\unl\mu_2$ die nebendiagonalen Blöcke besetzt. Eine Übersicht über die Gestalt der Matrizen 
findet sich in den Tabellen \subref*{tB:mu1overview} und \subref*{tB:mu2overview}.
Nach demselben Schema geben wir in den Tabellen \ref{tB:Dip1H2} und \ref{tB:Dip2H2}
die Matrixdarstellungen der (auf $e r_B$ normierten) ersten beiden Komponenten des Dipoloperators 
$\vu D = (\unl D_1,\unl D_2,\unl D_3)$ an. Die dritte Komponente $\unl D_3$ ist in 
Tabelle \ref{tB:Dip3H2} dargestellt.

Die Lösung des Eigenwertproblems der Massenmatrix und damit auch
die Angabe der lokal definierten Ableitungsmatrizen $\uop D^{(1,2)}(Z)$
ist hier im Gegensatz zum Unterraum mit Hauptquantenzahl $n=1$ nur noch numerisch möglich.

\begin{table}[!hb]
  \centering
  \small
  \subfloat[]{
    \begin{tabular}{l||>{\centering\baselineskip=18pt }m{20mm}||*{4}{>{\centering\baselineskip=18pt }m{20mm}|}}
      & $2P_{3/2}$ & $2P_{3/2}$ & $2P_{3/2}$ & $2P_{1/2}$ & $2S_{1/2}$\\ \hlx{v}
      & $2,2$ & $2,1$ & $1,1$ & $1,1$ & $1,1$\\ \hlx{vhhv}
      $2P_{3/2},2,2$ & \mycell{18mm}{$\FineStructure+\frac{\HyperFineSplitting}{160}$\\$-\tfrac{\I }{2}\Gamma_P$}
      & 0 & 0 & 0 & 0\\ \hlx{vhhv}
      $2P_{3/2},2,1$ & 0 & 
      \mycell{18mm}{${\FineStructure}+\frac{\HyperFineSplitting}{160}$\\$-\frac{\I}{2}\,{\Gamma_P}$}
      & 0 & 0 & 0\\ \hlx{vhv}
      $2P_{3/2},1,1$ & 0 & 0 & 
      \mycell{18mm}{${\FineStructure}-\frac{\HyperFineSplitting}{96}$\\$-\frac{\I}{2}\,{\Gamma_P}$} & 
      $-\frac{\HyperFineSplitting}{192\,{\sqrt{2}}}$ & 0\\ \hlx{vhv}
      $2P_{1/2},1,1$ & 0 & 0 & 
      $-\frac{\HyperFineSplitting}{192\,{\sqrt{2}}}$ & 
      $\frac{\HyperFineSplitting}{96}-\frac\I2{\Gamma_P}$ & 
      \mycell{18mm}{$\I{\delta_1}{\LambShift}$\\$+\frac\I2\delta_2 \LambShift$} \\ \hlx{vhv}
      $2S_{1/2},1,1$ & 0 & 0 & 0 & 
      \mycell{18mm}{$-\I{\delta_1}{\LambShift}$\\$-\frac\I2 {\delta_2}{\LambShift}$} &
      \mycell{18mm}{${\LambShift}+\frac{\HyperFineSplitting}{32}$\\$-\frac{\I}{2}\,{\Gamma_S}$}\\ \hlx{vh}
    \end{tabular}
\label{tB:M0H2a}}
  \caption[Die freie, nichthermitesche Massenmatrix für
  Wasserstoff im $(n=2)$-Unterraum]{Die freie, nichthermitesche Massenmatrix \underline{$\ms M_0$}\ 
für
  Wasserstoff im $(n=2)$-Unterraum ($F_3=2,1$).}\label{tB:M0H2}
\end{table}
\begin{table}[!htp]
  \ContinuedFloat
  \centering
  \small
  \subfloat[]{
    \begin{tabular}{l||*{6}{>{\centering\baselineskip=18pt }m{18mm}|}}
      & $2P_{3/2}$ & $2P_{3/2}$ & $2P_{1/2}$ & $2S_{1/2}$ & $2P_{1/2}$ & $2P_{3/2}$\\ \hlx{v}
      & $2,0$ & $1,0$ & $1,0$ & $1,0$ & $0,0$ & $0,0$\\ \hlx{vhhv}
      $2P_{3/2},2,0$ & 
      \mycell{18mm}{${\FineStructure}+\frac{\HyperFineSplitting}{160}$\\$-\frac{\I}{2}{\Gamma_P}$} & 0 & 0 & 0
      & 0 & 0\\  \hlx{vhv}
      $2P_{3/2},1,0$ & 0 & 
      \mycell{18mm}{${\FineStructure}-\frac{\HyperFineSplitting}{96}$\\$-\frac{\I}{2}{\Gamma_P}$} & $-\frac{\HyperFineSplitting}{192{\sqrt{2}}}$ & 0 & 0 & 0 \\ \hlx{vhv}
      $2P_{1/2},1,0$ & 0 & 
      $-\frac{\HyperFineSplitting}{192{\sqrt{2}}}$ & $\frac{\HyperFineSplitting}{96} - \frac\I2{\Gamma_P}$ & 
      \mycell{18mm}{$\I\delta_1{\LambShift}$\\$+\frac\I2\delta_2{\LambShift}$} & 0 & 0 \\ \hlx{vhv}
      $2S_{1/2},1,0$ & 0 & 0 & 
      \mycell{18mm}{$-\I{\delta_1}{\LambShift}$\\$-\frac\I2{\delta_2}{\LambShift}$} & 
      \mycell{18mm}{${\LambShift}+\frac{\HyperFineSplitting}{32}$\\$-\frac{\I}{2}{\Gamma_S}$} & 0 & 0 \\\hlx{vhv}
      $2P_{1/2},0,0$ & 0 & 0 & 0 & 0 & 
      $-\frac{\HyperFineSplitting}{32} - \frac\I2{\Gamma_P}$ & 
      \mycell{18mm}{$\I{\delta_1}{\LambShift}$\\$-\frac{3}{2}\I{\delta_2}{\LambShift}$} \\ \hlx{vhv}
      $2S_{1/2},0,0$ & 0 & 0 & 0 & 0 & 
      \mycell{18mm}{$-\I{\delta_1}{\LambShift}$\\$-\frac{3}{2}\I{\delta_2}{\LambShift}$} & 
      \mycell{18mm}{${\LambShift}-\frac{3\HyperFineSplitting}{32}$\\$-\frac{\I}{2}{\Gamma_S}$} \\ \hlx{vh}
    \end{tabular}
\label{tB:M0H2b}}\\
  \subfloat[]{

    \begin{tabular}{l||*{4}{>{\centering\baselineskip=18pt }m{20mm}|}|>{\centering\baselineskip=18pt }m{20mm}|}
      & $2P_{3/2}$ & $2P_{3/2}$ & $2P_{1/2}$ & $2S_{1/2}$ & $2P_{3/2}$\\ \hlx{v}
      & $2,-1$ & $1,-1$ & $1,-1$ & $1,-1$ & $2,-2$\\ \hlx{vhhv}
      $2P_{3/2},2,-1$ & 
      \mycell{18mm}{${\FineStructure}+\frac{\HyperFineSplitting}{160}$\\$-\frac{\I}{2}{\Gamma_P}$} 
      & 0 & 0 & 0 & 0 \\  \hlx{vhv}
      $2P_{3/2},1,-1$ & 0 & 
      \mycell{18mm}{${\FineStructure}-\frac{\HyperFineSplitting}{96}$\\$-\frac{\I}{2}{\Gamma_P}$} & 
      $-\frac{\HyperFineSplitting}{192{\sqrt{2}}}$ & 0 & 0 \\ \hlx{vhv}
      $2P_{1/2},1,-1$ & 0 & 
      $-\frac{\HyperFineSplitting}{192{\sqrt{2}}}$ & 
      $\frac{\HyperFineSplitting}{96} - \frac\I2{\Gamma_P}$ & 
      \mycell{18mm}{$\I{\delta_1}{\LambShift}$\\$+\frac\I2{\delta_2}{\LambShift}$} & 0 \\  \hlx{vhv}
      $2S_{1/2},1,-1$ & 0 & 0 & 
      \mycell{18mm}{$-\I{\delta_1}{\LambShift}$\\$-\frac\I2{\delta_2}{\LambShift}$} & 
      \mycell{18mm}{${\LambShift}+\frac{\HyperFineSplitting}{32}$\\$-\frac{\I}{2}{\Gamma_S}$} & 0 \\  \hlx{vhhv}
      $2P_{3/2},2,-2$ & 0 & 0 & 0 & 0 & 
      \mycell{18mm}{${\FineStructure}+\frac{\HyperFineSplitting}{160}$\\$-\frac{\I}{2}{\Gamma_P}$} \\ \hlx{vh}
    \end{tabular}

\label{tB:M0H2c}}
  \caption[]{Die freie, nichthermitesche Massenmatrix \underline{$\ms M_0$}\ für
  Wasserstoff im $(n=2)$-Unterraum ($F_3=0,-1,-2$).}
\end{table}
\FloatBarrier

\begin{table}[!hp]
\centering
\small\vspace{1cm}
\subfloat[Die Zuordnung der Tabellennummern zu den einzelnen Blöcken der Matrix $\unl\mu_1$.
Es werden nur die Blöcke der oberen Hälfte der Matrix angegeben. Die unteren Blöcke folgen durch hermitesche Konjugation.]{
\begin{tabular}{c||c|c|c|c|c|}
$F_3'\backslash F_3$ & $2$ & $1$ & $0$ & $-1$ & $-2$\\ \hlx{vhhv}
$2$  & 0 & \subref*{tB:mu1H2b} & 0 & 0 & 0 \\ \hlx{vhv}
$1$  & (\subref*{tB:mu1H2b})$^\dag$ & 0 & \subref*{tB:mu1H2c} & 0 & 0 \\ \hlx{vhv}
$0$  & 0 & (\subref*{tB:mu1H2c})$^\dag$ & 0 & \subref*{tB:mu1H2d} & 0 \\ \hlx{vhv}
$-1$ & 0 & 0 & (\subref*{tB:mu1H2d})$^\dag$ & 0 & \subref*{tB:mu1H2e} \\ \hlx{vhv}
$-2$ & 0 & 0 & 0 & (\subref*{tB:mu1H2e})$^\dag$ & 0 \\ \hlx{vhv}
\end{tabular}\label{tB:mu1overview}}\\[15mm]
\subfloat[]{

    \begin{tabular}{l||*{4}{>{\centering\baselineskip=18pt }m{16mm}|}}
      & $2P_{3/2}$ & $2P_{3/2}$ & $2P_{1/2}$ & $2S_{1/2}$\\ \hlx{v}
      & $2,1$ & $1,1$ & $1,1$ & $1,1$\\ \hlx{vhhv}
      $2P_{3/2},2,2$ & 
      $-\frac{g+2}{4}$ & 
      $\frac{g+2}{4 \sqrt{3}}$ & 
      $-\frac{g-1}{\sqrt{6}}$ & 0\\ \hlx{vh}
    \end{tabular}\label{tB:mu1H2b}}\\[10mm]
\subfloat[]{

    \begin{tabular}{l||*{6}{>{\centering\baselineskip=18pt }m{16mm}|}}
      & $2P_{3/2}$ & $2P_{3/2}$ & $2P_{1/2}$ & $2S_{1/2}$ & $2P_{1/2}$ & $2S_{1/2}$\\ \hlx{v}
      & $2,0$ & $1,0$ & $1,0$ & $1,0$ & $0,0$ & $0,0$\\ \hlx{vhhv}
      $2P_{3/2},2,1$ & 
      $-\frac{g+2}{4} \sqrt{\frac{3}{2}}$ & 
      $\frac{g+2}{4 \sqrt{6}}$ & 
      $-\frac{g-1}{2 \sqrt{3}}$ & 0 & 0 & 0\\ \hlx{vhv}
      $2P_{3/2},1,1$ & 
      $-\frac{g+2}{12 \sqrt{2}}$ & 
      $-\frac{5 (g+2)}{12 \sqrt{2}}$ & 
      $\frac{1-g}{6}$ & 0 & 
      $\frac{1-g}{3}$ & 0\\ \hlx{vhv}
      $2P_{1/2},1,1$ & 
      $\frac{g-1}{6}$ & 
      $\frac{1-g}{6}$ & 
      $\frac{g-4}{6 \sqrt{2}}$ & 0 & 
      $-\frac{g-4}{6 \sqrt{2}}$ & 0\\ \hlx{vhv}
      $2S_{1/2},1,1$ & 0 & 0 & 0 & 
      $-\frac{g}{2 \sqrt{2}}$ & 0 & 
      $\frac{g}{2 \sqrt{2}}$\\ \hlx{vh}
    \end{tabular}\label{tB:mu1H2c}}\\[5mm]
\caption[Die Matrix \unl{$\mu$}$_1$ der ersten Komponente des magnetischen Moments
von Wasserstoff im $(n=2)$-Unterraum]{
Die auf $\mu_B$ normierte Matrix \unl{$\mu$}$_1$ der ersten Komponente des
magnetischen Moments von Wasserstoff im $(n=2)$-Unterraum. (Fortsetzung auf der nächsten Seite)}
\label{tB:mu1H2}
\end{table}
\FloatBarrier

\begin{table}[!ht]
\ContinuedFloat
\centering
\small
\subfloat[]{

    \begin{tabular}{l||*{4}{>{\centering\baselineskip=18pt }m{16mm}|}}
      & $2P_{3/2}$ & $2P_{3/2}$ & $2P_{1/2}$ & $2S_{1/2}$\\ \hlx{v}
      & $2,-1$ & $1,-1$ & $1,-1$ & $1,-1$\\ \hlx{vhhv}
      $2P_{3/2},2,0$ & 
      $-\frac{g+2}{4}\sqrt{\frac{3}{2}}$ & 
      $\frac{g+2}{12 \sqrt{2}}$ & 
      $\frac{1-g}{6}$ & 0\\ \hlx{vhv}
      $2P_{3/2},1,0$ & 
      $-\frac{g+2}{4 \sqrt{6}}$ & 
      $-\frac{5 (g+2)}{12 \sqrt{2}}$ & 
      $\frac{1-g}{6}$ & 0\\ \hlx{vhv}
      $2P_{1/2},1,0$ & 
      $\frac{g-1}{2 \sqrt{3}}$ & 
      $\frac{1-g}{6}$ & 
      $\frac{g-4}{6 \sqrt{2}}$ & 0\\ \hlx{vhv}
      $2S_{1/2},1,0$ & 0 & 0 & 0 & 
      $-\frac{g}{2 \sqrt{2}}$\\ \hlx{vhv}
      $2P_{1/2},0,0$ & 0 & 
      $\frac{g-1}{3}$ & 
      $\frac{g-4}{6 \sqrt{2}}$ & 0\\ \hlx{vhv}
      $2S_{1/2},0,0$ & 0 & 0 & 0 & 
      $-\frac{g}{2 \sqrt{2}}$\\ \hlx{vh}
    \end{tabular}\label{tB:mu1H2d}}\qquad
\subfloat[]{

    \begin{tabular}{l||*{1}{>{\centering\baselineskip=18pt }m{13mm}|}}
      & $2P_{3/2}$\\ \hlx{v}
      & $2,-2$\\ \hlx{vhhv}
      $2P_{3/2},2,-1$ & 
      $-\frac{g+2}{4}$\\ \hlx{vhv}
      $2P_{3/2},1,-1$ & 
      $-\frac{g+2}{4 \sqrt{3}}$\\ \hlx{vhv}
      $2P_{1/2},1,-1$ & 
      $\frac{g-1}{\sqrt{6}}$\\ \hlx{vhv}
      $2S_{1/2},1,-1$ & 0\\ \hlx{vh}
    \end{tabular}\label{tB:mu1H2e}}
\caption[]{
(Fortsetzung) Die auf $\mu_B$ normierte Matrix \unl{$\mu$}$_1$ der ersten Komponente des
magnetischen Moments von Wasserstoff im $(n=2)$-Unterraum.}
\end{table}

\begin{table}[!hb]
\centering
\small\vspace{1cm}
\subfloat[Die Zuordnung der Tabellennummern zu den einzelnen Blöcken der Matrix $\unl\mu_2$.
Es werden nur die Blöcke der oberen Hälfte der Matrix angegeben. Die unteren Blöcke folgen durch hermitesche Konjugation.]{
\begin{tabular}{c||c|c|c|c|c|}
$F_3'\backslash F_3$ & $2$ & $1$ & $0$ & $-1$ & $-2$\\ \hlx{vhhv}
$2$  & 0 & \subref*{tB:mu2H2b} & 0 & 0 & 0 \\ \hlx{vhv}
$1$  & (\subref*{tB:mu2H2b})$^\dag$ & 0 & \subref*{tB:mu2H2c} & 0 & 0 \\ \hlx{vhv}
$0$  & 0 & (\subref*{tB:mu2H2c})$^\dag$ & 0 & \subref*{tB:mu2H2d} & 0 \\ \hlx{vhv}
$-1$ & 0 & 0 & (\subref*{tB:mu2H2d})$^\dag$ & 0 & \subref*{tB:mu2H2e} \\ \hlx{vhv}
$-2$ & 0 & 0 & 0 & (\subref*{tB:mu2H2e})$^\dag$ & 0 \\ \hlx{vhv}
\end{tabular}\label{tB:mu2overview}}\\[5mm]
\caption[Die Matrix \unl{$\mu$}$_2$ der zweiten Komponente des magnetischen Moments 
von Wasserstoff im $(n=2)$-Unterraum]{
Die auf $\mu_B$ normierte Matrix \unl{$\mu$}$_2$ der zweiten Komponente des
magnetischen Moments von Wasserstoff im $(n=2)$-Unterraum. (Fortsetzung auf der nächsten Seite)}
\label{tB:mu2H2}
\end{table}
\FloatBarrier

\begin{table}[!hp]
\ContinuedFloat
\centering
\small
\subfloat[]{

    \begin{tabular}{l||*{4}{>{\centering\baselineskip=18pt }m{13mm}|}}
      & $2P_{3/2}$ & $2P_{3/2}$ & $2P_{1/2}$ & $2S_{1/2}$\\ \hlx{v}
      & $2,1$ & $1,1$ & $1,1$ & $1,1$\\ \hlx{vhhv}
      $2P_{3/2},2,2$ & 
$\frac{\I (g+2)}{4}$ & 
      $-\frac{\I (g+2)}{4 \sqrt{3}}$ & 
      $\frac{\I (g-1)}{\sqrt{6}}$ & 0\\ \hlx{vh}
    \end{tabular}\label{tB:mu2H2b}}\\[10mm]
\subfloat[]{

    \begin{tabular}{l||*{6}{>{\centering\baselineskip=18pt }m{16mm}|}}
      & $2P_{3/2}$ & $2P_{3/2}$ & $2P_{1/2}$ & $2S_{1/2}$ & $2P_{1/2}$ & $2S_{1/2}$\\ \hlx{v}
      & $2,0$ & $1,0$ & $1,0$ & $1,0$ & $0,0$ & $0,0$\\ \hlx{vhhv}
      $2P_{3/2},2,1$ & 
      $\frac{\I(g+2)}{4} \sqrt{\frac{3}{2}}$ & 
      $-\frac{\I (g+2)}{4 \sqrt{6}}$ & 
      $\frac{\I (g-1)}{2 \sqrt{3}}$ & 0 & 0 & 0\\ \hlx{vhv}
      $2P_{3/2},1,1$ & 
      $\frac{\I (g+2)}{12 \sqrt{2}}$ & 
      $\frac{5 \I (g+2)}{12 \sqrt{2}}$ & 
$\frac{\I (g-1)}{6}$ & 0 & 
$\frac{\I (g-1)}{3}$ & 0\\ \hlx{vhv}
      $2P_{1/2},1,1$ & 
$-\frac{\I (g-1)}{6}$ & 
$\frac{\I (g-1)}{6}$ & 
      $-\frac{\I (g-4)}{6 \sqrt{2}}$ & 0 & 
      $\frac{\I (g-4)}{6 \sqrt{2}}$ & 0\\ \hlx{vhv}
      $2S_{1/2},1,1$ & 0 & 0 & 0 & 
      $\frac{\I g}{2 \sqrt{2}}$ & 0 & 
      $-\frac{\I g}{2 \sqrt{2}}$\\ \hlx{vh}
    \end{tabular}\label{tB:mu2H2c}}\\[10mm]
\subfloat[]{

    \begin{tabular}{l||*{4}{>{\centering\baselineskip=18pt }m{16mm}|}}
      & $2P_{3/2}$ & $2P_{3/2}$ & $2P_{1/2}$ & $2S_{1/2}$\\ \hlx{v}
      & $2,-1$ & $1,-1$ & $1,-1$ & $1,-1$\\ \hlx{vhhv}
      $2P_{3/2},2,0$ & 
      $\frac{\I(g+2)}{4}\sqrt{\frac{3}{2}}$ & 
      $-\frac{\I (g+2)}{12 \sqrt{2}}$ & 
      $\frac{\I(g-1)}{6}$ & 0\\ \hlx{vhv}
      $2P_{3/2},1,0$ & 
      $\frac{\I(g+2)}{4 \sqrt{6}}$ & 
      $\frac{5\I(g+2)}{12 \sqrt{2}}$ & 
      $\frac{\I(g-1)}{6}$ & 0\\ \hlx{vhv}
      $2P_{1/2},1,0$ & 
      $-\frac{\I (g-1)}{2 \sqrt{3}}$ & 
      $\frac{\I(g-1)}{6}$ & 
      $-\frac{\I (g-4)}{6 \sqrt{2}}$ & 0\\ \hlx{vhv}
      $2S_{1/2},1,0$ & 0 & 0 & 0 & 
      $\frac{\I g}{2 \sqrt{2}}$\\ \hlx{vhv}
      $2P_{1/2},0,0$ & 0 & 
      $-\frac{\I(g-1)}{3}$ & 
      $-\frac{\I (g-4)}{6 \sqrt{2}}$ & 0\\ \hlx{vhv}
      $2S_{1/2},0,0$ & 0 & 0 & 0 & 
      $\frac{\I g}{2 \sqrt{2}}$\\ \hlx{vh}
    \end{tabular}\label{tB:mu2H2d}}\qquad
\subfloat[]{

    \begin{tabular}{l||*{1}{>{\centering\baselineskip=18pt }m{13mm}|}}
      & $2P_{3/2}$\\ \hlx{v}
      & $2,-2$\\ \hlx{vhhv}
      $2P_{3/2},2,-1$ & 
      $\frac{\I(g+2)}{4}$\\ \hlx{vhv}
      $2P_{3/2},1,-1$ & 
      $\frac{\I (g+2)}{4 \sqrt{3}}$\\ \hlx{vhv}
      $2P_{1/2},1,-1$ & 
      $-\frac{\I (g-1)}{\sqrt{6}}$\\ \hlx{vhv}
      $2S_{1/2},1,-1$ & 0\\ \hlx{vh}
    \end{tabular}\label{tB:mu2H2e}}\\[5mm]
\caption[]{
(Fortsetzung) Die auf $\mu_B$ normierte Matrix \unl{$\mu$}$_2$ der zweiten Komponente des
magnetischen Moments von Wasserstoff im $(n=2)$-Unterraum.}
\end{table}
\FloatBarrier

\begin{table}[!hp]
  \centering
  \small
  \subfloat[]{
    \begin{tabular}{l||>{\centering\baselineskip=18pt }m{16mm}||*{4}{>{\centering\baselineskip=18pt }m{16mm}|}}
      & $2P_{3/2}$ & $2P_{3/2}$ & $2P_{3/2}$ & $2P_{1/2}$ & $2S_{1/2}$\\ \hlx{v}
      & $2,2$ & $2,1$ & $1,1$ & $1,1$ & $1,1$\\ \hlx{vhhv}
      $2P_{3/2},2,2$ & 
      $-\frac{2 + g}{2}$ & 0 & 0 & 0 & 0\\ \hlx{vhhv}
      $2P_{3/2},2,1$ & 0 & 
      $-\frac{2 + g}{4}$ & 
      $-\frac{2 + g}{4{\sqrt{3}}}$ & 
      $\frac{1-g}{{\sqrt{6}}}$ & 
      0\\ \hlx{vhv}
      $2P_{3/2},1,1$ & 0 & 
      $-\frac{2 + g}{4{\sqrt{3}}}$ & 
      $-\frac{5(2 + g)}{12}$ & 
      $-\frac{1-g}{3{\sqrt{2}}}$ & 
      0\\ \hlx{vhv}
      $2P_{1/2},1,1$ & 0 & 
      $\frac{1 - g}{{\sqrt{6}}}$ & 
      $-\frac{1 - g}{3{\sqrt{2}}}$ & 
      $-\frac{4 - g}{6}$ & 
      0\\ \hlx{vhv}
      $2S_{1/2},1,1$ & 0 & 0 & 0 & 0 & 
      $-\frac{g}{2}$\\ \hlx{vh}
    \end{tabular}\label{tB:mu3H2a}}\\
  \subfloat[]{
    \begin{tabular}{l||*{6}{>{\centering\baselineskip=18pt }m{16mm}|}}
      & $2P_{3/2}$ & $2P_{3/2}$ & $2P_{1/2}$ & $2S_{1/2}$ & $2P_{1/2}$ & $2S_{1/2}$\\ \hlx{v}
      & $2,0$ & $1,0$ & $1,0$ & $1,0$ & $0,0$ & $0,0$\\ \hlx{vhhv}
      $2P_{3/2},2,0$ & 0 & 
      $-\frac{2 + g}{6}$ & 
      $\frac{\sqrt2(1 - g)}{3}$ & 0 & 0 & 0\\ \hlx{vhv}
      $2P_{3/2},1,0$ & 
      $-\frac{2 + g}{6}$ & 0 & 0 & 0 & 
      $\frac{{\sqrt{2}}(1 - g)}{3}$ & 0\\ \hlx{vhv}
      $2P_{1/2},1,0$ & 
      $\frac{{\sqrt{2}}(1 - g)}{3}$ & 0 & 0 & 0 & 
      $-\frac{4 - g}{6}$ & 0\\ \hlx{vhv}
      $2S_{1/2},1,0$ & 0 & 0 & 0 & 0 & 0 & 
      $-\frac{g}{2}$\\ \hlx{vhv}
      $2P_{1/2},0,0$ & 0 & 
      $\frac{{\sqrt{2}}(1 - g)}{3}$ & 
      $-\frac{4 - g}{6}$ & 0 & 0 & 0\\ \hlx{vhv}
      $2S_{1/2},0,0$ & 
      0 & 0 & 0 & 
      $-\frac{g}{2}$ & 
      0 & 0\\ \hlx{vh}
    \end{tabular}\label{tB:mu3H2b}}\\
  \subfloat[]{
    \begin{tabular}{l||*{4}{>{\centering\baselineskip=18pt }m{16mm}|}|>{\centering\baselineskip=18pt }m{16mm}|}
      & $2P_{3/2}$ & $2P_{3/2}$ & $2P_{1/2}$ & $2S_{1/2}$ & $2P_{3/2}$\\ \hlx{v}
      & $2,-1$ & $1,-1$ & $1,-1$ & $1,-1$ & $2,-2$\\ \hlx{vhhv}
      $2P_{3/2},2,-1$ & 
      $\frac{2 + g}{4}$ & 
      $-\frac{2 + g}{4{\sqrt{3}}}$ & 
      $\frac{1 - g}{{\sqrt{6}}}$ & 
      0 & 0\\ \hlx{vhv}
      $2P_{3/2},1,-1$ & 
      $-\frac{2 + g}{4{\sqrt{3}}}$ & 
      $\frac{5(2 + g)}{12}$ & 
      $\frac{1 - g}{3{\sqrt{2}}}$ & 
      0 & 0\\ \hlx{vhv}
      $2P_{1/2},1,-1$ & 
      $\frac{1 - g}{{\sqrt{6}}}$ & 
      $\frac{1 - g}{3{\sqrt{2}}}$ & 
      $\frac{4 - g}{6}$ & 
      0 & 0\\ \hlx{vhv}
      $2S_{1/2},1,-1$ & 
      0 & 0 & 0 & 
      $\frac{g}{2}$ & 0\\ \hlx{vhhv}
      $2P_{3/2},2,-2$ & 0 & 0 & 0 & 0 & 
      $\frac{2 + g}{2}$\\ \hlx{vh}
    \end{tabular}\label{tB:mu3H2c}}
  \caption[Die Matrix der dritten Komponente des magnetischen Moments 
  von Wasserstoff im $(n=2)$-Unterraum]{
    Die auf $\mu_B$ normierte Matrix \unl{$\mu$}$_3$ der dritten Komponente des
  magnetischen Moments von Wasserstoff im $(n=2)$-Unterraum.}
  \label{tB:mu3H2}
\end{table}
\FloatBarrier

\begin{table}[!hp]
\centering
\small\vspace{1cm}
\subfloat[Die Zuordnung der Tabellennummern zu den einzelnen Blöcken der Matrix $\unl D_1$.
Es werden nur die Blöcke der oberen Hälfte der Matrix angegeben. Die unteren Blöcke folgen durch hermitesche Konjugation.]{
\begin{tabular}{c||c|c|c|c|c|}
$F_3'\backslash F_3$ & $2$ & $1$ & $0$ & $-1$ & $-2$\\ \hlx{vhhv}
$2$  & 0 & \subref*{tB:Dip1H2b} & 0 & 0 & 0 \\ \hlx{vhv}
$1$  & (\subref*{tB:Dip1H2b})$^\dag$ & 0 & \subref*{tB:Dip1H2c} & 0 & 0 \\ \hlx{vhv}
$0$  & 0 & (\subref*{tB:Dip1H2c})$^\dag$ & 0 & \subref*{tB:Dip1H2d} & 0 \\ \hlx{vhv}
$-1$ & 0 & 0 & (\subref*{tB:Dip1H2d})$^\dag$ & 0 & \subref*{tB:Dip1H2e} \\ \hlx{vhv}
$-2$ & 0 & 0 & 0 & (\subref*{tB:Dip1H2e})$^\dag$ & 0 \\ \hlx{vhv}
\end{tabular}\label{tB:Dip1overview}}\\[15mm]
\subfloat[]{

    \begin{tabular}{l||*{4}{>{\centering\baselineskip=18pt }m{13mm}|}}
      & $2P_{3/2}$ & $2P_{3/2}$ & $2P_{1/2}$ & $2S_{1/2}$\\ \hlx{v}
      & $2,1$ & $1,1$ & $1,1$ & $1,1$\\ \hlx{vhhv}
      $2P_{3/2},2,2$ & 0 & 0 & 0 & 
      $-\frac{3}{\sqrt{2}}$\\ \hlx{vh}
    \end{tabular}\label{tB:Dip1H2b}}\\[10mm]
\subfloat[]{

    \begin{tabular}{l||*{6}{>{\centering\baselineskip=18pt }m{13mm}|}}
      & $2P_{3/2}$ & $2P_{3/2}$ & $2P_{1/2}$ & $2S_{1/2}$ & $2P_{1/2}$ & $2S_{1/2}$\\ \hlx{v}
      & $2,0$ & $1,0$ & $1,0$ & $1,0$ & $0,0$ & $0,0$\\ \hlx{vhhv}
      $2P_{3/2},2,1$ & 0 & 0 & 0 & 
      $-\frac{3}{2}$ & 0 & 0\\ \hlx{vhv}
      $2P_{3/2},1,1$ & 0 & 0 & 0 & 
      $-\frac{\sqrt{3}}{2}$ & 0 & 
      $-\sqrt{3}$\\ \hlx{vhv}
      $2P_{1/2},1,1$ & 0 & 0 & 0 & 
      $-\sqrt{\frac{3}{2}}$ & 0 & 
      $\sqrt{\frac{3}{2}}$\\ \hlx{vhv}
      $2S_{1/2},1,1$ & 
      $\frac{\sqrt{3}}{2}$ & 
      $-\frac{\sqrt{3}}{2}$ & 
      $-\sqrt{\frac{3}{2}}$ & 0 & 
      $\sqrt{\frac{3}{2}}$ & 0\\ \hlx{vh}
    \end{tabular}\label{tB:Dip1H2c}}\\[5mm]
\caption[Die Matrix \unl{$D$}$_1$ der ersten Komponente des Dipolmoments
von Wasserstoff im $(n=2)$-Unterraum]{
Die auf $e r_B$ normierte Matrix \unl{$D$}$_1$ der ersten Komponente des
Dipolmoments von Wasserstoff im $(n=2)$-Unterraum. (Fortsetzung auf der nächsten Seite)}
\label{tB:Dip1H2}
\end{table}
\FloatBarrier

\begin{table}[!ht]
\ContinuedFloat
\centering
\small
\subfloat[]{

    \begin{tabular}{l||*{4}{>{\centering\baselineskip=18pt }m{13mm}|}}
      & $2P_{3/2}$ & $2P_{3/2}$ & $2P_{1/2}$ & $2S_{1/2}$\\ \hlx{v}
      & $2,-1$ & $1,-1$ & $1,-1$ & $1,-1$\\ \hlx{vhhv}
      $2P_{3/2},2,0$ & 0 & 0 & 0 & 
      $-\frac{\sqrt{3}}{2}$\\ \hlx{vhv}
      $2P_{3/2},1,0$ & 0 & 0 & 0 & 
      $-\frac{\sqrt{3}}{2}$\\ \hlx{vhv}
      $2P_{1/2},1,0$ & 0 & 0 & 0 & 
      $-\sqrt{\frac{3}{2}}$\\ \hlx{vhv}
      $2S_{1/2},1,0$ & 
      $\frac{3}{2}$ & 
      $-\frac{\sqrt{3}}{2}$ & 
      $-\sqrt{\frac{3}{2}}$ & 0\\ \hlx{vhv}
      $2P_{1/2},0,0$ & 0 & 0 & 0 & 
      $-\sqrt{\frac{3}{2}}$\\ \hlx{vhv}
      $2S_{1/2},0,0$ & 0 & 
      $\sqrt{3}$ & 
      $-\sqrt{\frac{3}{2}}$ & 0\\ \hlx{vh}
    \end{tabular}\label{tB:Dip1H2d}}\qquad
\subfloat[]{

    \begin{tabular}{l||*{1}{>{\centering\baselineskip=18pt }m{13mm}|}}
      & $2P_{3/2}$\\ \hlx{v}
      & $2,-2$\\ \hlx{vhhv}
      $2P_{3/2},2,-1$ & 0\\ \hlx{vhv}
      $2P_{3/2},1,-1$ & 0\\ \hlx{vhv}
      $2P_{1/2},1,-1$ & 0\\ \hlx{vhv}
      $2S_{1/2},1,-1$ & 
      $\frac{3}{\sqrt{2}}$\\ \hlx{vh}
    \end{tabular}\label{tB:Dip1H2e}}
\caption[]{
(Fortsetzung) Die auf $e r_B$ normierte Matrix \unl{$D$}$_1$ der ersten Komponente des
Dipolmoments von Wasserstoff im $(n=2)$-Unterraum.}
\end{table}

\begin{table}[!hb]
\centering
\small\vspace{1cm}
\subfloat[Die Zuordnung der Tabellennummern zu den einzelnen Blöcken der Matrix $\unl D_2$.
Es werden nur die Blöcke der oberen Hälfte der Matrix angegeben. Die unteren Blöcke folgen durch hermitesche Konjugation.]{
\begin{tabular}{c||c|c|c|c|c|}
$F_3'\backslash F_3$ & $2$ & $1$ & $0$ & $-1$ & $-2$\\ \hlx{vhhv}
$2$  & 0 & \subref*{tB:Dip2H2b} & 0 & 0 & 0 \\ \hlx{vhv}
$1$  & (\subref*{tB:Dip2H2b})$^\dag$ & 0 & \subref*{tB:Dip2H2c} & 0 & 0 \\ \hlx{vhv}
$0$  & 0 & (\subref*{tB:Dip2H2c})$^\dag$ & 0 & \subref*{tB:Dip2H2d} & 0 \\ \hlx{vhv}
$-1$ & 0 & 0 & (\subref*{tB:Dip2H2d})$^\dag$ & 0 & \subref*{tB:Dip2H2e} \\ \hlx{vhv}
$-2$ & 0 & 0 & 0 & (\subref*{tB:Dip2H2e})$^\dag$ & 0 \\ \hlx{vhv}
\end{tabular}\label{tB:Dip2overview}}\\[5mm]
\caption[Die Matrix \unl{$D$}$_2$ der zweiten Komponente des Dipolmoments 
von Wasserstoff im $(n=2)$-Unterraum]{
Die auf $e r_B$ normierte Matrix \unl{$D$}$_2$ der zweiten Komponente des
Dipolmoments von Wasserstoff im $(n=2)$-Unterraum. (Fortsetzung auf der nächsten Seite)}
\label{tB:Dip2H2}
\end{table}
\FloatBarrier

\begin{table}[!hp]
\ContinuedFloat
\centering
\small
\subfloat[]{

    \begin{tabular}{l||*{4}{>{\centering\baselineskip=18pt }m{13mm}|}}
      & $2P_{3/2}$ & $2P_{3/2}$ & $2P_{1/2}$ & $2S_{1/2}$\\ \hlx{v}
      & $2,1$ & $1,1$ & $1,1$ & $1,1$\\ \hlx{vhhv}
      $2P_{3/2},2,2$ & 0 & 0 & 0 & 
      $\frac{3 \I}{\sqrt{2}}$\\ \hlx{vh}
    \end{tabular}\label{tB:Dip2H2b}}\\[10mm]
\subfloat[]{

    \begin{tabular}{l||*{6}{>{\centering\baselineskip=18pt }m{13mm}|}}
      & $2P_{3/2}$ & $2P_{3/2}$ & $2P_{1/2}$ & $2S_{1/2}$ & $2P_{1/2}$ & $2S_{1/2}$\\ \hlx{v}
      & $2,0$ & $1,0$ & $1,0$ & $1,0$ & $0,0$ & $0,0$\\ \hlx{vhhv}
      $2P_{3/2},2,1$ & 0 & 0 & 0 & 
      $\frac{3 \I}{2}$ & 0 & 0\\ \hlx{vhv}
      $2P_{3/2},1,1$ & 0 & 0 & 0 & 
      $\frac{\I \sqrt{3}}{2}$ & 0 & 
      $i \sqrt{3}$\\ \hlx{vhv}
      $2P_{1/2},1,1$ & 0 & 0 & 0 & 
      $\I \sqrt{\frac{3}{2}}$ & 0 & 
      $-\I \sqrt{\frac{3}{2}}$\\ \hlx{vhv}
      $2S_{1/2},1,1$ & 
      $-\frac{\I \sqrt{3}}{2}$ & 
      $\frac{\I \sqrt{3}}{2}$ & 
      $\I \sqrt{\frac{3}{2}}$ & 0 & 
      $-\I \sqrt{\frac{3}{2}}$ & 0\\ \hlx{vh}
    \end{tabular}\label{tB:Dip2H2c}}\\[10mm]
\subfloat[]{

    \begin{tabular}{l||*{4}{>{\centering\baselineskip=18pt }m{13mm}|}}
      & $2P_{3/2}$ & $2P_{3/2}$ & $2P_{1/2}$ & $2S_{1/2}$\\ \hlx{v}
      & $2,-1$ & $1,-1$ & $1,-1$ & $1,-1$\\ \hlx{vhhv}
      $2P_{3/2},2,0$ & 0 & 0 & 0 & 
      $\frac{\I \sqrt{3}}{2}$\\ \hlx{vhv}
      $2P_{3/2},1,0$ & 0 & 0 & 0 & 
      $\frac{\I \sqrt{3}}{2}$\\ \hlx{vhv}
      $2P_{1/2},1,0$ & 0 & 0 & 0 & 
      $\I \sqrt{\frac{3}{2}}$\\ \hlx{vhv}
      $2S_{1/2},1,0$ & 
      $-\frac{3 \I}{2}$ & 
      $\frac{\I \sqrt{3}}{2}$ & 
      $\I \sqrt{\frac{3}{2}}$ & 0\\ \hlx{vhv}
      $2P_{1/2},0,0$ & 0 & 0 & 0 & 
      $\I \sqrt{\frac{3}{2}}$\\ \hlx{vhv}
      $2S_{1/2},0,0$ & 0 & 
      $-\I \sqrt{3}$ & 
      $\I \sqrt{\frac{3}{2}}$ & 0\\ \hlx{vh}
    \end{tabular}\label{tB:Dip2H2d}}\qquad
\subfloat[]{

    \begin{tabular}{l||*{1}{>{\centering\baselineskip=18pt }m{13mm}|}}
      & $2P_{3/2}$\\ \hlx{v}
      & $2,-2$\\ \hlx{vhhv}
      $2P_{3/2},2,-1$ & 0\\ \hlx{vhv}
      $2P_{3/2},1,-1$ & 0\\ \hlx{vhv}
      $2P_{1/2},1,-1$ & 0\\ \hlx{vhv}
      $2S_{1/2},1,-1$ & 
      $-\frac{3 \I}{\sqrt{2}}$\\ \hlx{vh}
    \end{tabular}\label{tB:Dip2H2e}}\\[5mm]
\caption[]{
(Fortsetzung) Die auf $e r_B$ normierte Matrix \unl{$D$}$_2$ der zweiten Komponente des
Dipolmoments von Wasserstoff im $(n=2)$-Unterraum.}
\end{table}
\FloatBarrier

\begin{table}[!hp]
  \centering
  \small
  \subfloat[]{

    \begin{tabular}{l||>{\centering\baselineskip=18pt }m{13mm}||*{4}{>{\centering\baselineskip=18pt }m{13mm}|}}
      & $2P_{3/2}$ & $2P_{3/2}$ & $2P_{3/2}$ & $2P_{1/2}$ & $2S_{1/2}$\\ \hlx{v}
      & $2,2$ & $2,1$ & $1,1$ & $1,1$ & $1,1$\\ \hlx{vhhv}
      $2P_{3/2},2,2$ & 0 & 0 & 0 & 0 & 0\\ \hlx{vhhv}
      $2P_{3/2},2,1$ & 0 & 0 & 0 & 0 & 
      $\frac{3}{{\sqrt{2}}}$\\ \hlx{vhv}
      $2P_{3/2},1,1$ & 0 & 0 & 0 & 0 & 
      $-{\sqrt{\frac{3}{2}}}$\\ \hlx{vhv}
      $2P_{1/2},1,1$ & 0 & 0 & 0 & 0 & 
      $-{\sqrt{3}}$\\ \hlx{vhv}
      $2S_{1/2},1,1$ & 0 & 
      $\frac{3}{{\sqrt{2}}}$ & 
      $-{\sqrt{\frac{3}{2}}}$ & 
      $-{\sqrt{3}}$ & 0\\ \hlx{vh}
    \end{tabular}

\label{tB:Dip3H2a}}
  \vspace{-2mm}
  \subfloat[]{

    \begin{tabular}{l||*{6}{>{\centering\baselineskip=18pt }m{13mm}|}}
      & $2P_{3/2}$ & $2P_{3/2}$ & $2P_{1/2}$ & $2S_{1/2}$ & $2P_{1/2}$ & $2S_{1/2}$\\ \hlx{v}
      & $2,0$ & $1,0$ & $1,0$ & $1,0$ & $0,0$ & $0,0$\\ \hlx{vhhv}
      $2P_{3/2},2,0$ & 0 & 0 & 0 & 
      ${\sqrt{6}}$ & 0 & 0\\ \hlx{vhv}
      $2P_{3/2},1,0$ & 0 & 0 & 0 & 0 & 0 & 
      ${\sqrt{6}}$\\ \hlx{vhv}
      $2P_{1/2},1,0$ & 0 & 0 & 0 & 0 & 0 & 
      $-{\sqrt{3}}$\\ \hlx{vhv}
      $2S_{1/2},1,0$ & 
      ${\sqrt{6}}$ & 0 & 0 & 0 & 
      $-{\sqrt{3}}$ & 0\\ \hlx{vhv}
      $2P_{1/2},0,0$ & 0 & 0 & 0 & 
      $-{\sqrt{3}}$ & 0 & 0\\ \hlx{vhv}
      $2S_{1/2},0,0$ & 0 & 
      ${\sqrt{6}}$ & 
      $-{\sqrt{3}}$ & 0 & 0 & 0\\ \hlx{vh}
    \end{tabular}

\label{tB:Dip3H2b}}
  \vspace{-2mm}
  \subfloat[]{

    \begin{tabular}{l||*{4}{>{\centering\baselineskip=18pt }m{13mm}|}|>{\centering\baselineskip=18pt }m{13mm}|}
      & $2P_{3/2}$ & $2P_{3/2}$ & $2P_{1/2}$ & $2S_{1/2}$ & $2P_{3/2}$\\ \hlx{v}
      & $2,-1$ & $1,-1$ & $1,-1$ & $1,-1$ & $2,-2$\\ \hlx{vhhv}
      $2P_{3/2},2,-1$ & 0 & 0 & 0 & 
      $\frac{3}{{\sqrt{2}}}$ & 0\\ \hlx{vhv}
      $2P_{3/2},1,-1$ & 0 & 0 & 0 & 
      ${\sqrt{\frac{3}{2}}}$ & 0\\ \hlx{vhv}
      $2P_{1/2},1,-1$ & 0 & 0 & 0 & 
      ${\sqrt{3}}$ & 0\\ \hlx{vhv}
      $2S_{1/2},1,-1$ & 
      $\frac{3}{{\sqrt{2}}}$ & 
      ${\sqrt{\frac{3}{2}}}$ & 
      ${\sqrt{3}}$ & 0 & 0\\ \hlx{vhhv}
      $2P_{3/2},2,-2$ & 0 & 0 & 0 & 0 & 0\\ \hlx{vh}
    \end{tabular}

\label{tB:Dip3H2c}}
  \caption[Die Matrix der dritten Komponente des Dipoloperators von
  Wasserstoff im $(n=2)$-Unterraum]{
    Die auf $e r_B$ normierte Matrix \unl{$D$}$_3$ der dritten Komponente des
  Dipoloperators von Wasserstoff im $(n=2)$-Unterraum.}
  \label{tB:Dip3H2}
\end{table}
\FloatBarrier

\subsection{Numerische Berechnung der Eigenwerte in elektrischen und magnetischen Feldern}\label{sB:H2.EW}

Mit den expliziten Matrizen aus dem letzten Abschnitt ist es uns nun möglich, die Eigenwerte bei
vorhandenen äußeren Feldern numerisch zu berechnen. Wir betrachten exemplarisch in 3-Richtung orientierte,
statische Felder $\mc E=\v{\mc E}\cdot\v e_3$ und $\mc B=\v{\mc B}\cdot\v e_3$.

\begin{figure}[!hp]
  \centering
  \subfloat[]{
    \includegraphics[width=15cm]{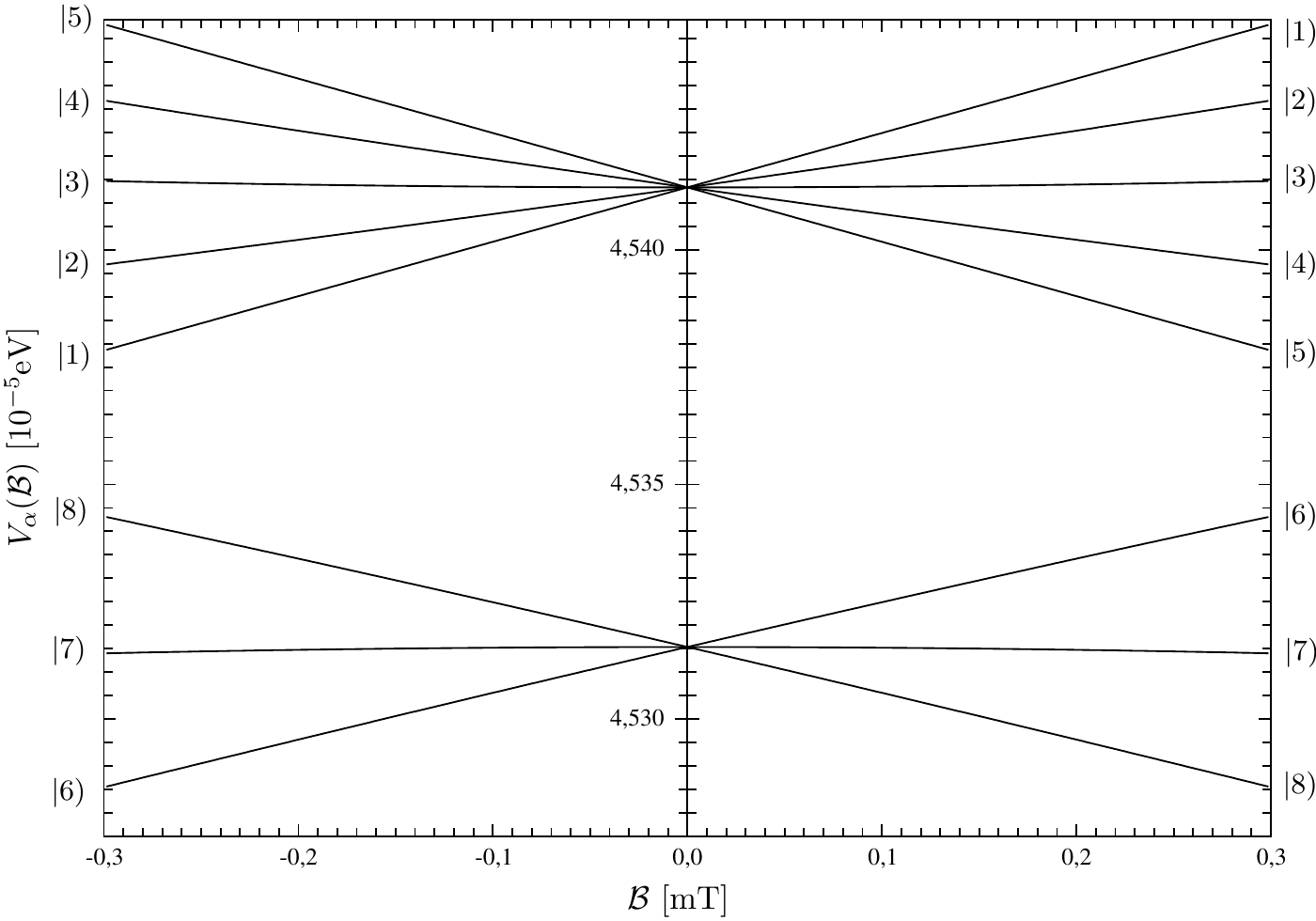}\label{fB:BR-H2-1}
}\\[5mm]
  \subfloat[]{
    \includegraphics[width=7.6cm]{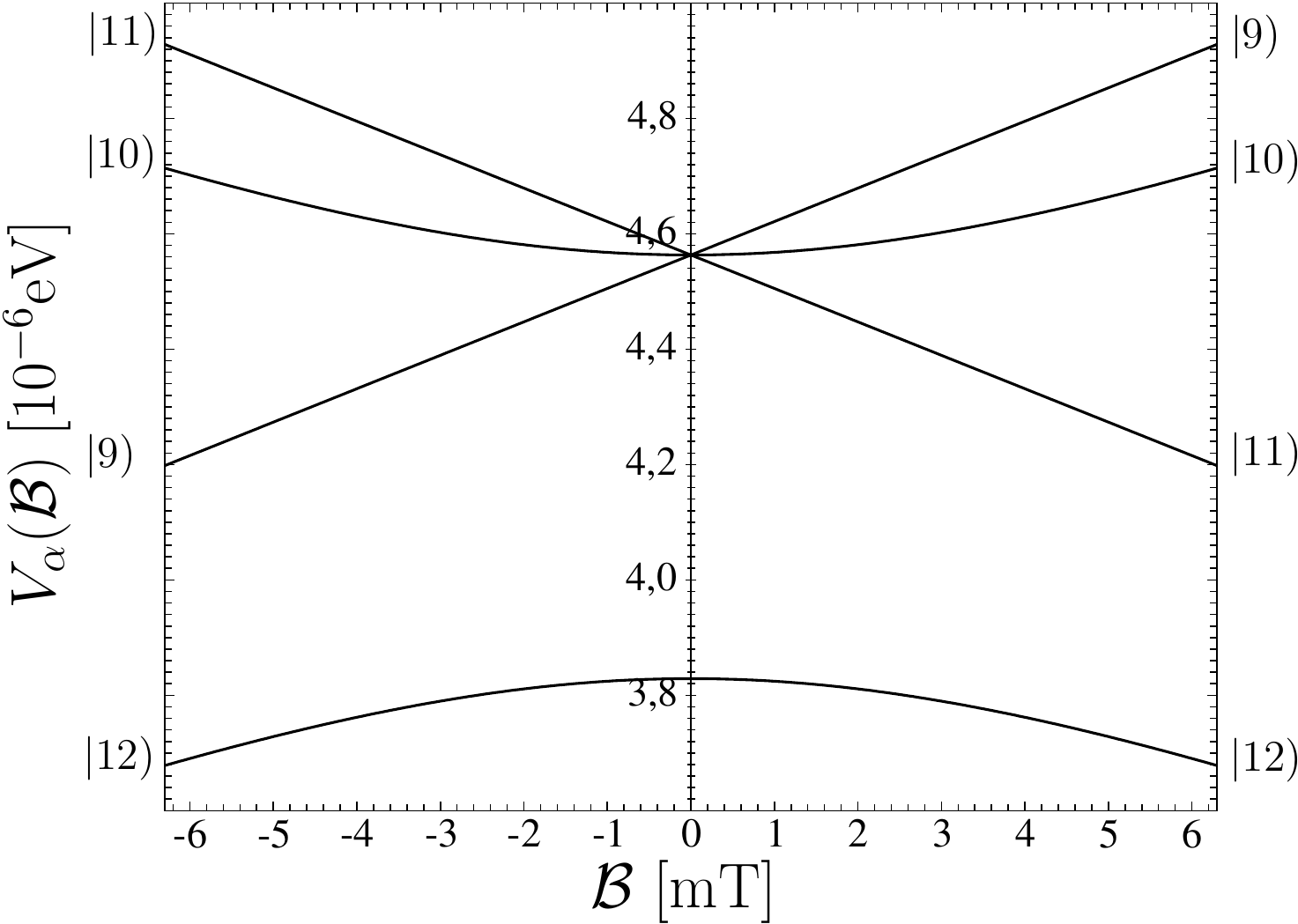}\label{fB:BR-H2-2}
}
  \subfloat[]{
    \includegraphics[width=7.6cm]{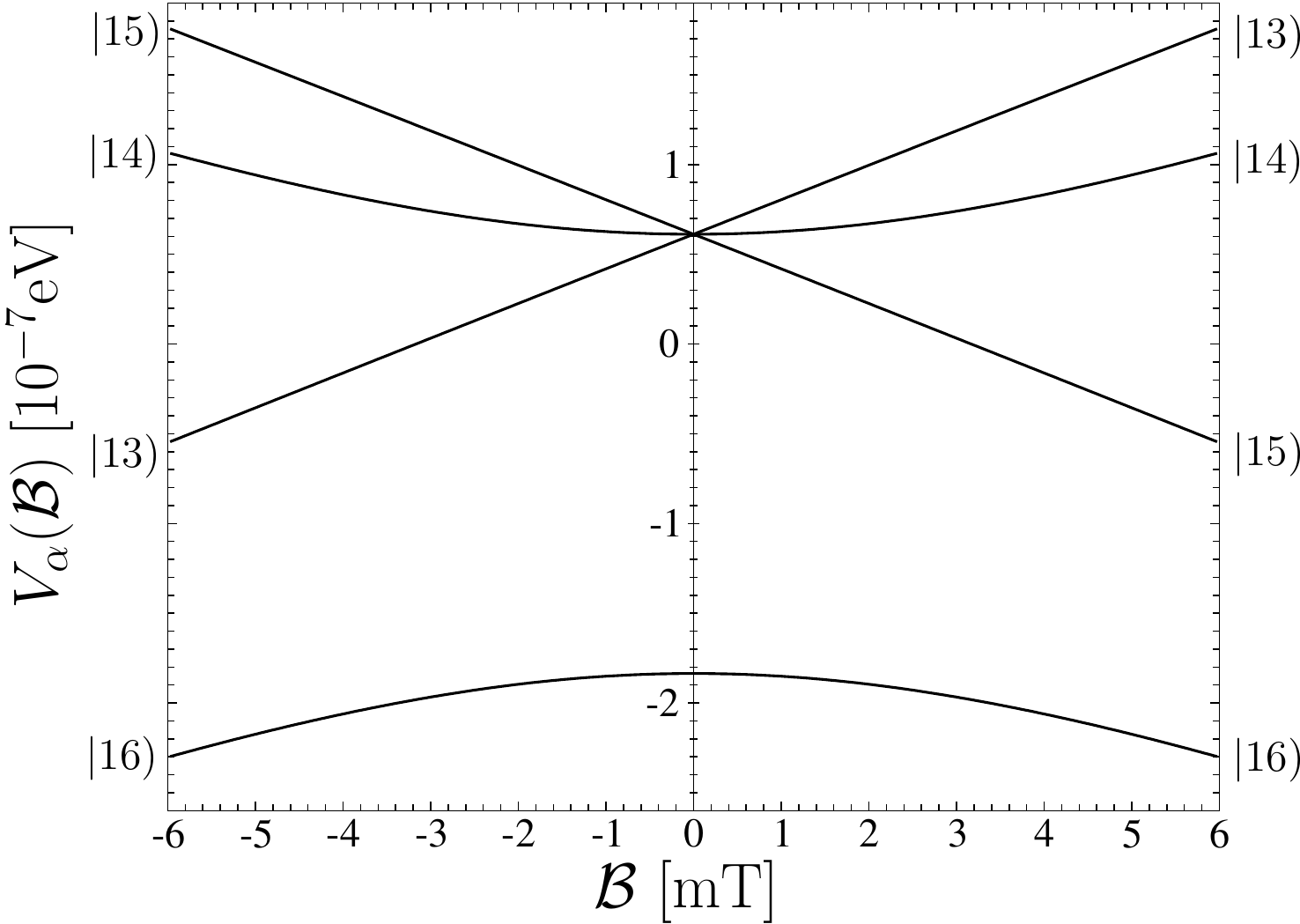}\label{fB:BR-H2-3}
}
  \caption[Breit-Rabi-Diagramme für Wasserstoff im $(n=2)$-Unterraum]{
    Breit-Rabi-Diagramme für Wasserstoff im $(n=2)$-Unterraum. Die Bedeutung
    der mit $\rket\alpha$, $(\alpha=1,…,16)$, bezeichneten Zustände
    kann Tabelle \ref{tB:H2States} entnommen werden.
}
  \label{fB:BR-H2}
\end{figure}

In einem rein magnetischen Feld tritt keine Mischung von Zuständen unterschiedlicher Bahndrehimpulse auf,
also bleiben die Lebensdauern der $2P$- und $2S$-Zustände konstant. Die Realteile der Eigenwerte liefern
uns dagegen die Breit-Rabi-Diagramme, Abb. \ref{fB:BR-H2}. Die Reihenfolge der Energien für ein kleines,
positives Magnetfeld in 3-Richtung entspricht für Wasserstoff der Reihenfolge der Durchnummerierung
aus Tabelle \ref{tB:H2States}\footnote{Dies wird für Deuterium nicht mehr der Fall sein, wie in Abschnitt
\ref{sB:D2.EW} gezeigt wird.}. Im Falle allgemeiner elektrischer und magnetischer Felder sind in der Notation
aus Tab. \ref{tB:H2States} die Buchstaben $\mc E$ und $\mc B$ durch die Vektoren $\vmc E$ und
$\vmc B$ zu ersetzen.
\begin{table}[htbp]
  \centering\small
  \begin{tabular}{|Mc|Ml||Mc|Ml|}\hlx{hv}
    \alpha & \rket{2\hat L_J,F,F_3,\mc E,\mc B} & \alpha & \rket{2\hat L_J,F,F_3,\mc E,\mc B}\\ \hlx{vhhv} 
    1 & \rket{2\hat P_{3/2},2,2,\mc E,\mc B} & 9  & \rket{2\hat S_{1/2},1,1,\mc E,\mc B}\\ \hlx{vhv}
    2 & \rket{2\hat P_{3/2},2,1,\mc E,\mc B} & 10 & \rket{2\hat S_{1/2},1,0,\mc E,\mc B}\\ \hlx{vhv}
    3 & \rket{2\hat P_{3/2},2,0,\mc E,\mc B}  & 11 & \rket{2\hat S_{1/2},1,-1,\mc E,\mc B}\\ \hlx{vhv}
    4 & \rket{2\hat P_{3/2},2,-1,\mc E,\mc B}  & 12 & \rket{2\hat S_{1/2},0,0,\mc E,\mc B}\\ \hlx{vhv}
    5 & \rket{2\hat P_{3/2},2,-2,\mc E,\mc B} & 13 & \rket{2\hat P_{1/2},1,1,\mc E,\mc B}\\ \hlx{vhv}
    6 & \rket{2\hat P_{3/2},1,1,\mc E,\mc B}  & 14 & \rket{2\hat P_{1/2},1,0,\mc E,\mc B}\\ \hlx{vhv}
    7 & \rket{2\hat P_{3/2},1,0,\mc E,\mc B}  & 15 & \rket{2\hat P_{1/2},1,-1,\mc E,\mc B}\\ \hlx{vhv}
    8 & \rket{2\hat P_{3/2},1,-1,\mc E,\mc B}  & 16 & \rket{2\hat P_{1/2},0,0,\mc E,\mc B}\\ \hlx{vh}
  \end{tabular}
  \caption[Zuordnung der verschiedenen Notationen für Wasserstoff im
    $(n=2)$-Unterraum]{
      Zuordnung der verschiedenen Notationen für Wasserstoff im
    $(n=2)$-Unterraum. Die Reihenfolge der
    Zustände wurde absteigend nach den ungestörten Energien und 
    innerhalb eines Multipletts absteigend nach $F$ und $F_3$ gewählt.}
  \label{tB:H2States}
\end{table}

\begin{figure}[!ht]
  \centering
  \includegraphics[width=15cm]{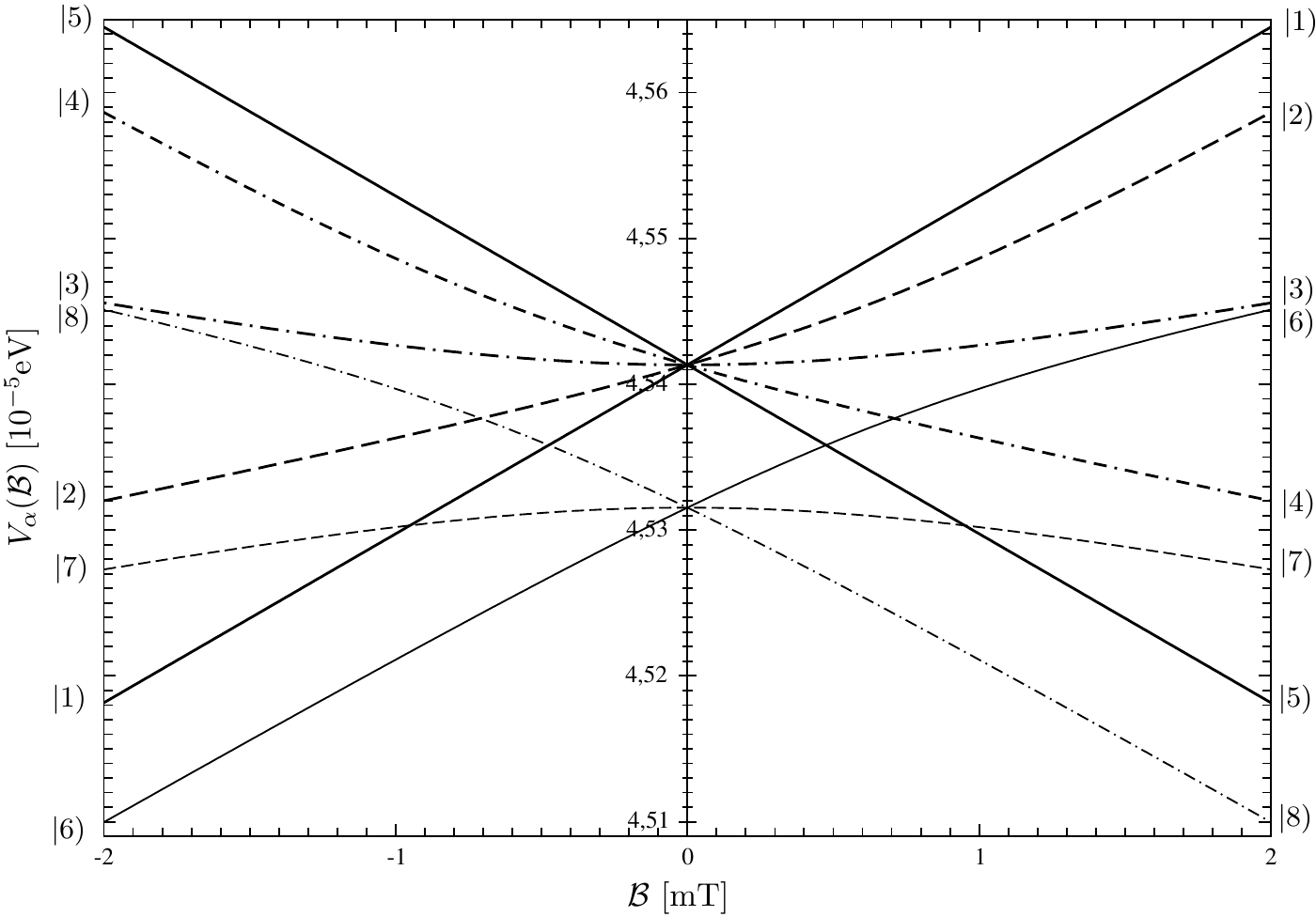}
  \caption[Breit-Rabi-Diagramm für die $2P_{3/2}$-Zustände von Wasserstoff]{
    Breit-Rabi-Diagramm für die $2P_{3/2}$-Zustände von Wasserstoff. Gegenüber
    Abb. \subref*{fB:BR-H2-1} ist hier ein größeres Magnetfeldintervall dargestellt. }
  \label{fB:BR-H2-1x}
\end{figure}
Das Breit-Rabi-Diagramm der $2P_{3/2}$-Zustände ist in Abb. \ref{fB:BR-H2-1x}
noch einmal bis zu einem etwas größeren Magnetfeld dargestellt. Man sieht dort,
dass bereits ab $|\mc B|\approx 0.5\u{mT}$ eine Überschneidung der Energien
verschiedener Zustände auftritt.

Die kritischen Magnetfelder, die aus den Breit-Rabi-Diagrammen
entnommen werden können, liegen bei etwa 
\begin{align}
\begin{split}
  \mc B\subt{krit.}(2S_{1/2}) &\approx \mc B\subt{krit.}(2P_{1/2}) \approx 6\u{mT}\ ,\\
  \mc B\subt{krit.}(2P_{3/2}) &\approx 2\u{mT}\ ,
\end{split}
\end{align}
wobei die Angabe für $\mc B\subt{krit.}(2P_{3/2})$ sehr vage ist, da der Übergang zwischen
Zeeman- und Paschen-Back-Bereich für die einzelnen Zustände sehr unterschiedlich ist.
Die Breit-Rabi-Diagramme \subref*{fB:BR-H2-2} und \subref*{fB:BR-H2-3} 
der $2S_{1/2}$- und $2P_{1/2}$-Zustände entsprechen dem Diagramm
der $1S_{1/2}$-Zustände aus Abb. \ref{fB:BR-H1}. Das Diagramm der $2P_{3/2}$-Zustände
dagegen zeigt einen deutlich komplexeren Verlauf der Energien dieser Zustände im Magnetfeld.

\begin{figure}[!hp]
  \centering
  \subfloat[]{
    \includegraphics[width=15cm]{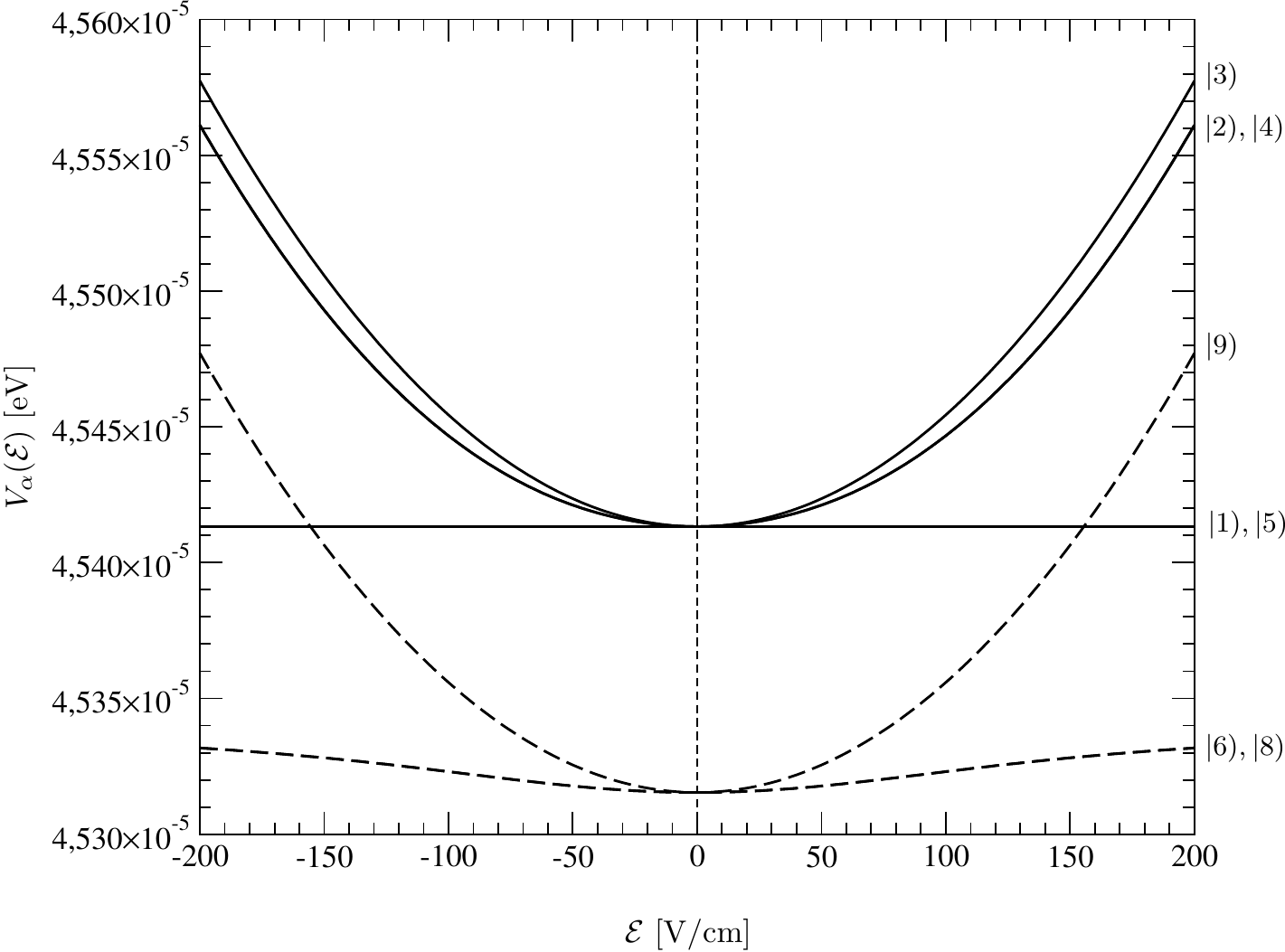}\label{fB:EW-E-Real-H2-1}
}\\[5mm]
  \subfloat[]{
    \includegraphics[width=7.6cm]{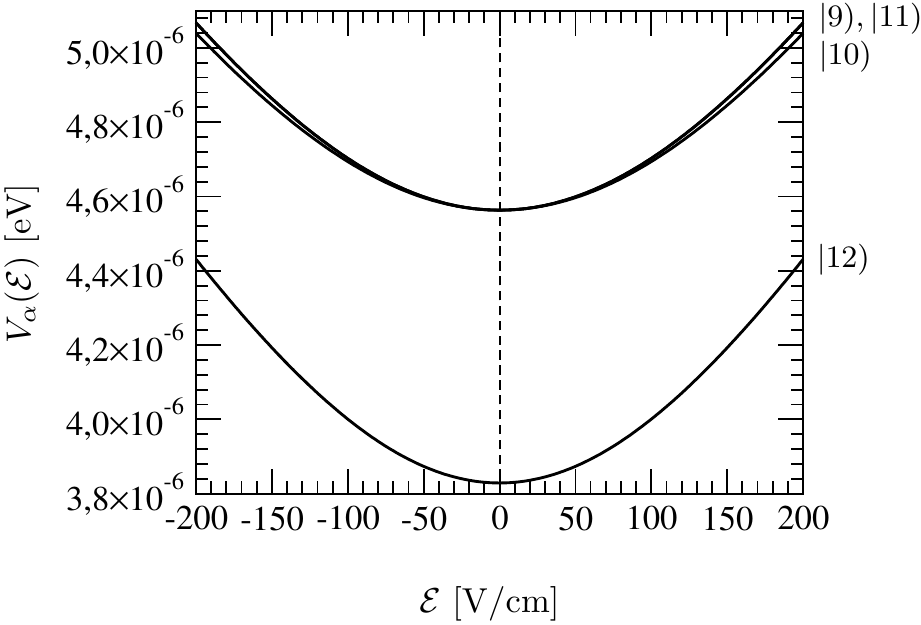}\label{fB:EW-E-Real-H2-2}
}
  \subfloat[]{
    \includegraphics[width=7.6cm]{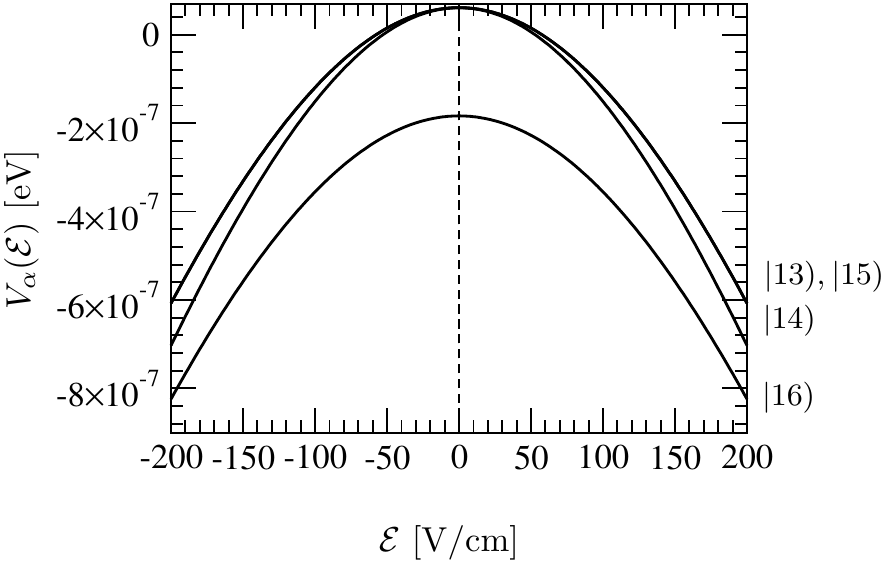}\label{fB:EW-E-Real-H2-3}
}
  \caption[Abhängigkeit der atomaren Energieniveaus vom elektrischen Feld]{
    Abhängigkeit der atomaren Energieniveaus (Realteile der Eigenwerte der Massenmatrix)
    vom elektrischen Feld $\mc E=\v e_3\cdot\v{\mc E}$. Die zu den Diagrammen gehörenden
    Zustände sind die \subref{fB:EW-E-Real-H2-1} $2P_{3/2}$-, 
    \subref{fB:EW-E-Real-H2-2} $2S_{1/2}$- und die \subref{fB:EW-E-Real-H2-3} $2P_{1/2}$-Zustände.
}
  \label{fB:EW-E-Real-H2}
\end{figure}
\begin{figure}[!hp]
  \centering
  \subfloat[]{\label{fB:EW-E-Imag-H2-1}
    \includegraphics[width=7.6cm]{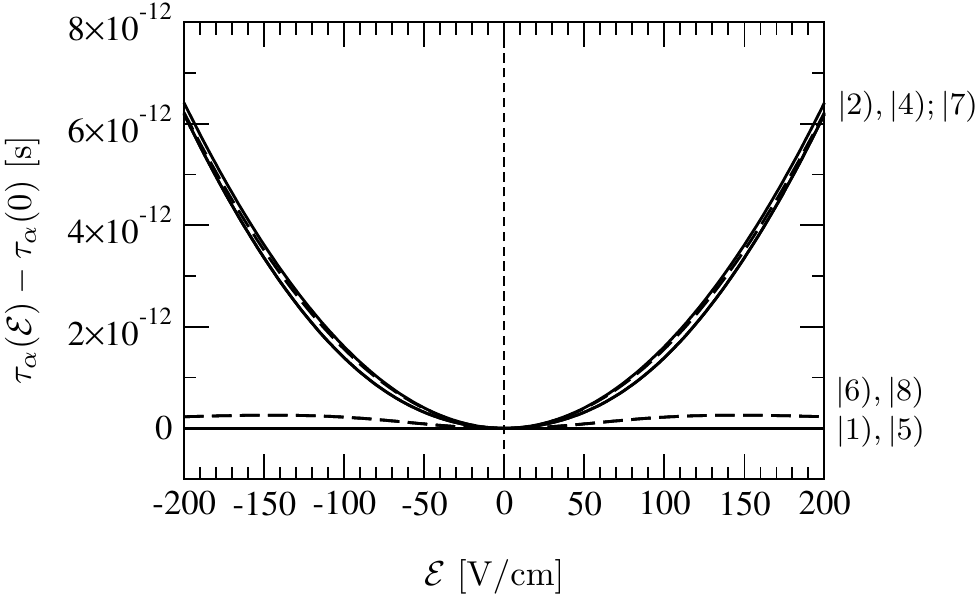}
}
  \subfloat[]{\label{fB:EW-E-Imag-H2-2}
    \includegraphics[width=7.6cm]{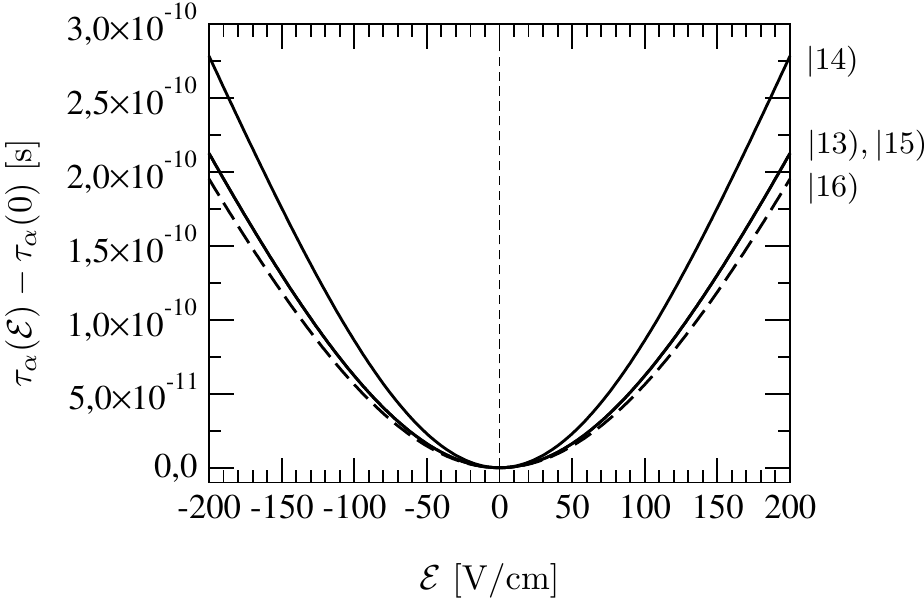}
}\\
  \subfloat[]{\label{fB:EW-E-Imag-H2-3}
    \includegraphics[width=15cm]{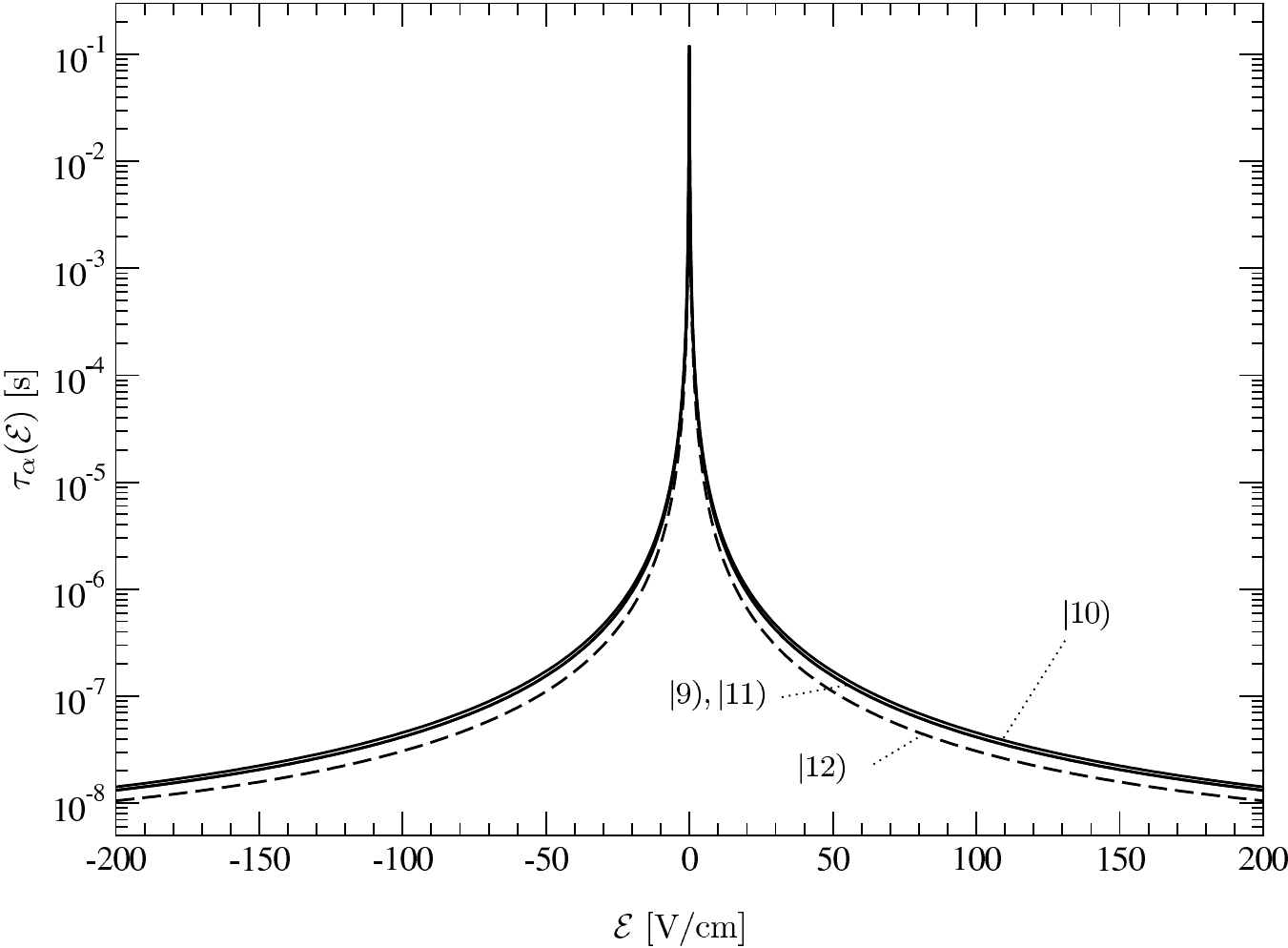}
}
  \caption[Abhängigkeit der mittleren Lebensdauern vom elektrischen Feld]{
    Abhängigkeit der mittleren Lebensdauern ($\tau_{S,P} = \Gamma^{-1}_{S,P}$)
    vom elektrischen Feld $\mc E=\v e_3\cdot\v{\mc E}$. Die Diagramme \subref*{fB:EW-E-Imag-H2-1} 
    und \subref*{fB:EW-E-Imag-H2-2} zeigen die relative Änderung der Lebensdauern 
    der $2P$-Zustände, Diagramm \subref*{fB:EW-E-Imag-H2-3} zeigt in
    logarithmischer Auftragung die absolute mittlere Lebensdauer der $2S$-Zustände.
}
  \label{fB:EW-E-Imag-H2}
\end{figure}

Nun kommen wir zur numerischen Berechnung der Verschiebung der atomaren Energieniveaus in einem äußeren
konstanten und homogenen elektrischen Feld (Stark-Effekt). Als Grundlage hierfür dient die im letzten Abschnitt
diskutierte Matrix des Dipoloperators, Tab. \ref{tB:Dip3H2}. Da der Dipoloperator zu einer Mischung
der $2P$- und $2S$-Zustände führt, erwarten wir eine Veränderung der Lebensdauern der Zustände, die
ja im Zusammenhang mit den Imaginärteilen der komplexen Energieeigenwerte der Massenmatrix stehen.

Betrachten wir zunächst die Diagramme für die Abhängigkeit der Realteile der Eigenwerte vom elektrischen
Feld $\mc E$ in 3-Richtung, siehe Abb. \subref*{fB:EW-E-Real-H2-1}-\subref*{fB:EW-E-Real-H2-3}. 
Man sieht für (fast) alle Zustände eine
in guter Näherung quadratischen Verlauf der Energien für kleine ($|\mc E| \lesssim 50\u V$) Felder, 
für größere Feldstärken tragen höhere, gerade Potenzen von $\mc E$ bei. 
Beiträge von Zuständen mit anderen Hauptquantenzahlen sind hier vernachlässigt worden.

In Abb. \ref{fB:EW-E-Imag-H2} sind die mittleren Lebensdauern (inverse Zerfallsraten) der einzelnen Zustände
dargestellt. Während sich die Lebensdauern der $2P$-Zustände aufgrund des Einflusses der $2S$-Zustände
relativ gering verlängern, fällt die Lebensdauer der $2S$-Zustände rapide ab. 

Wir wollen dies ein wenig besser verstehen und betrachten dazu exemplarisch den atomaren Zustand
$\rket{2\hat S_{1/2},1,1}$. Wir vernachlässigen bei der folgenden Diskussion die P-verletzenden Beiträge
zur Massenmatrix und die nichtdiagonalen Beiträge des Hyperfeinstruktur-Hamiltonoperators. Die inneren atomaren
Zustände sind dann näherungsweise durch die Gesamtdrehimpulszustände gegeben.

In einem schwachen elektrischen Feld $\mc E$ in 3-Richtung wird der feldfreie
Zustand $\rket{2\hat S_{1/2},1,1}\approx \ket{2S_{1/2},1,1}$ gestört und erhält z.B. einen Beitrag des Zustands $\rket{2\hat P_{1/2},1,1}\approx \ket{2P_{1/2},1,1}$, d.h. es gilt (siehe Tabelle \subref*{tB:Dip3H2a}, 
S. \pageref{tB:Dip3H2a}) nach störungstheoretischen Überlegungen
\begin{align}\label{eB:2S.in.E}
  \rket{2\hat S_{1/2},1,1,\mc E} \approx \rket{2\hat S_{1/2},1,1} 
  + \frac{\sqrt3 er_B\mc E}{\Delta E}\rket{2\hat P_{1/2},1,1}\ ,
\end{align}
mit der komplexen Energiedifferenz
\begin{align}\label{eB:Delta.E.2S.2P}
  \Delta E \approx \mc S -\tfrac\I2(\Gamma_S-\Gamma_P)\ .
\end{align}
Dabei ist $\mc S = E_{2S_{1/2}}-E_{2P_{1/2}}$ die Energie-Differenz der Schwerpunkte der $2S_{1/2}$- und
$2P_{1/2}$-Zustände, also die Lamb-Shift. Die Hyperfeinaufspaltung wurde in der Energiedifferenz außer Acht
gelassen, da wir hier nur grobe Abschätzungen vornehmen wollen. Desweiteren wurden in (\ref{eB:2S.in.E})
alle Beiträge der $2P_{3/2}$-Zustände aufgrund der etwa zehn mal größeren Energiedifferenz vernachlässigt.

Der zu $\rket{2\hat S_{1/2},1,1,\mc E}$ gehörende Energieeigenwert bekommt erst in zweiter Ordnung 
Störungstheorie einen Beitrag des elektrischen Felds, d.h. es gilt näherungsweise
\begin{align}
\begin{split}
  E(2\hat S_{1/2},1,1,\mc E) &\approx E(2\hat S_{1/2},1,1) + \frac{(\sqrt3 e r_B\mc E)^2}{\Delta E}\\
  &\approx \mc S -\tfrac\I2\Gamma_S  + \frac{(\sqrt3 e r_B\mc E)^2}{|\Delta E|^2}\Delta E^*\ .
\end{split}
\end{align}
Setzt man $\Delta E^*$ gemäß Gl. (\ref{eB:Delta.E.2S.2P}) hier ein, so erhält man
\begin{align}
  E(2\hat S_{1/2},1,1,\mc E) &\approx \mc S\kle{1 + \klr{\frac{\sqrt 3e r_B\mc E}{|\Delta E|}}^2} 
  - \frac\I2\Gamma_S\kle{1 + \frac{\Gamma_P-\Gamma_S}{\Gamma_S}\klr{\frac{\sqrt3 e r_B\mc E}{|\Delta E|}}^2}\ .
\end{align}
Wir sehen also, dass selbst bei einer sehr kleinen Verschiebung des Realteils der Energie, d.h. für
\begin{align}
  \klr{\frac{\sqrt 3e r_B\mc E}{|\Delta E|}}^2 \ll 1
\end{align}
eine sehr große Änderung der Zerfallsrate
\begin{align}\label{eB:DR-H2-S.in.E}
  \Gamma(2\hat S_{1/2},1,1,\mc E)\approx 
  \Gamma_S\kle{1 + \frac{\Gamma_P-\Gamma_S}{\Gamma_S}\klr{\frac{\sqrt3 e r_B\mc E}{|\Delta E|}}^2}
\end{align}
möglich ist, da für die freien Zerfallsraten $\Gamma_{S,P}$ nach den Gln. (\ref{eB:DR-H2-S}),
({\ref{eB:DR-H2-P}}) $\Gamma_P/\Gamma_S \approx 10^7$ und somit auch
\begin{align}
  \frac{\Gamma_P-\Gamma_S}{\Gamma_S} \approx 10^7
\end{align}
gilt. Selbst bei einem schwachen elektrischen Feld tritt also eine Vervielfachung 
der Zerfallsrate und eine damit
verbundene, starke Verkürzung der Lebensdauer der metastabilen $2S$-Zustände auf.

Bei einer analogen Rechnung für den Zustand $\rket{2\hat P_{1/2},1,1,\mc E}$ würde man dagegen
die Zerfallsrate
\begin{align}\label{eB:DR-H2-P.in.E}
  \Gamma(2\hat P_{1/2},1,1,\mc E)\approx 
  \Gamma_P\kle{1 - \frac{\Gamma_P-\Gamma_S}{\Gamma_P}\klr{\frac{\sqrt3 e r_B\mc E}{|\Delta E|}}^2}
\end{align}
erhalten. Hier tritt der positive Faktor
\begin{align}
  \frac{\Gamma_P-\Gamma_S}{\Gamma_P} \approx 1
\end{align}
auf, der in (\ref{eB:DR-H2-P.in.E}) mit einem negativen Vorzeichen behaftet ist. Eine kleine Änderung
des Realteils der Energie ist hier also auch mit einer kleinen Änderung der Zerfallsrate verbunden.
Das negative Vorzeichen erklärt die relative Verlängerung der Lebensdauer der $2P$-Zustände.

\newpage
\section{Deuterium}\label{sB:D}

\subsection{Die Massenmatrix im Unterraum mit Hauptquantenzahl \texorpdfstring{$n=2$}{}}\label{sB:D2}

Deuterium hat einen Kernspin von $I=1$. Somit treten im
24-dimensionalen $(n=2)$-Unterraum Gesamtdrehimpulse im Bereich $\tfrac12\leq F\leq\tfrac52$
auf. Mit der Formel (\ref{eB:HfsSkalierung}) folgt für
die Hyperfeinaufspaltung der $(n=2)$-Zustände insgesamt
\begin{table}[!htbp]
  \centering\small
  \begin{tabular}{|l||c|c|c|c|}\hlx{hv}
    $nL_J$ & $1S_{1/2}$ (aus \cite{WiRa72})& $2S_{1/2}$ & $2P_{1/2}$ & $2P_{3/2}$\\ \hlx{vhv}
    $\HyperFineSplitting(I=1,nL_J)$ & $\HyperFineSplitting=327.384\u {MHz}\,h$ & $\frac18\HyperFineSplitting$ & $\frac1{24}\HyperFineSplitting$ & $\frac1{45}\HyperFineSplitting$\\ \hlx{vh}
  \end{tabular}
  \caption{Skalierung der Hyperfeinaufspaltung für Deuterium mit $n\leq 2$.}
  \label{tB:HfsD2}
\end{table}

\begin{figure}[!ht]
  \centering
  \includegraphics[width=15cm]{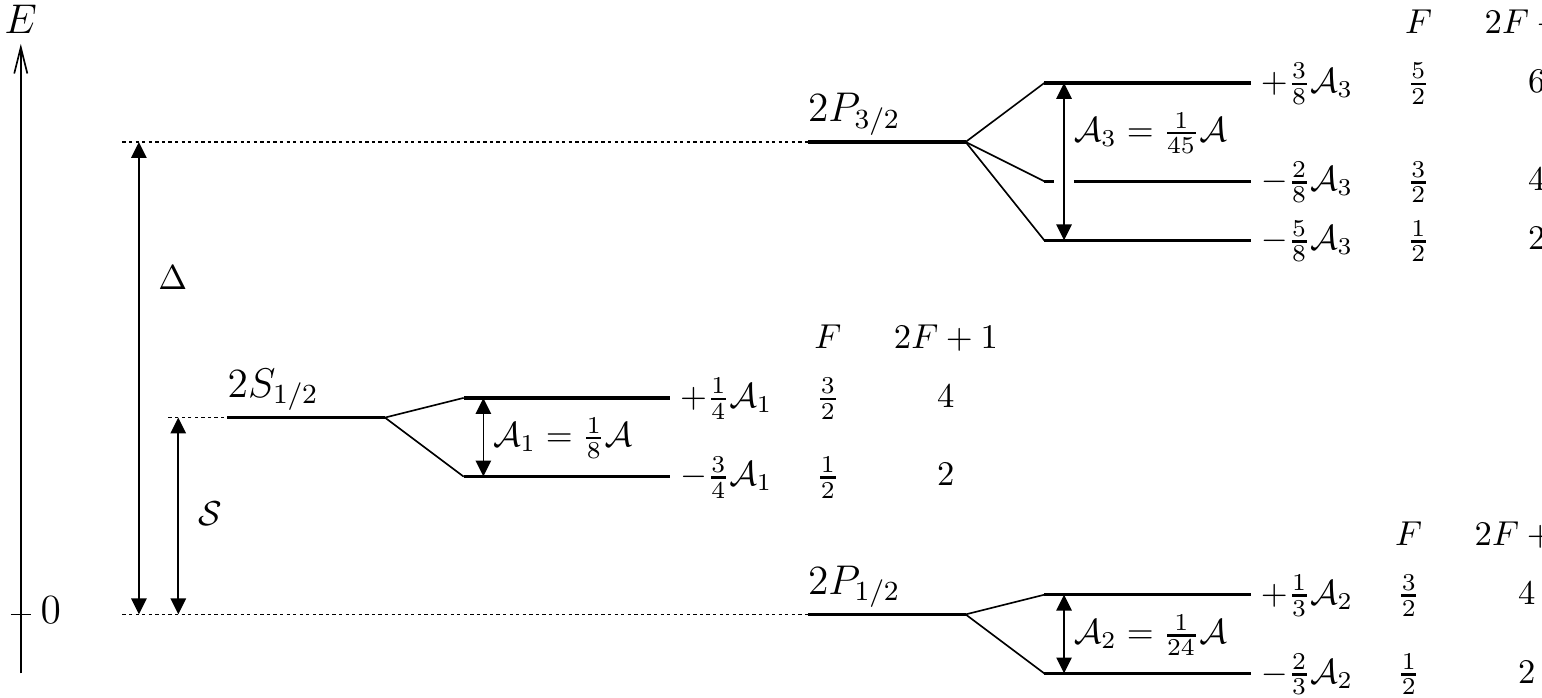}
  \caption{Termschema der $(n=2)$-Zustände von Deuterium (nicht maßstabsgetreu).}
  \label{fB:TermschemaD2}
\end{figure}

Das Termschema der $(n=2)$-Zustände von Deuterium ist in
Abb. \ref{fB:TermschemaD2} dargestellt. Die
experimentellen Werte (aus \cite{Eri77}) für die Lamb-Shift und die Feinstrukturaufspaltung
betragen hier aber
\begin{align}\label{eB:LambD2}
  \LambShift/h &=  1059.282(64)\u {MHz} ,&\\ \label{eB:FSD2}
  \FineStructure/h &= 10971.90(12)\u {MHz}\ .&
\end{align}
\FloatBarrier
Für die Zerfallszeiten werden die gleichen Werte wie für Wasserstoff
verwendet, siehe Gln. (\ref{eB:DR-H2-P}), (\ref{eB:DR-H2-S}). 
Die Matrizen $\uop M_0$, $\vu\mu/\mu_B$ und $\vu D/(e r_B)$ sind auf den 
folgenden Seiten in den Tabellen \ref{tB:M0D2} bis \ref{tB:Dip3D2} dargestellt.

\begin{table}[!hp]
  \centering
  \footnotesize
  \subfloat[]{

    \begin{tabular}{l||>{\centering\baselineskip=15pt }m{22mm}||*{4}{>{\centering\baselineskip=15pt }m{22mm}|}}
      & $2P_{3/2}$ & $2P_{3/2}$ & $2P_{3/2}$ & $2P_{1/2}$ & $2S_{1/2}$\\ \hlx{v}
      & $\tfrac{5}{2},\tfrac{5}{2}$ & $\tfrac{5}{2},\tfrac{3}{2}$ & $\tfrac{3}{2},\tfrac{3}{2}$ & $\tfrac{3}{2},\tfrac{3}{2}$ & $\tfrac{3}{2},\tfrac{3}{2}$\\ \hlx{vhhv}
      $2P_{3/2},\tfrac{5}{2},\tfrac{5}{2}$ & 
      ${\FineStructure}+\frac{\HyperFineSplitting}{120} - \frac{\I}{2}{\Gamma_P}$ & 0 & 0 & 0 & 0\\ \hlx{vhhv}
      $2P_{3/2},\tfrac{5}{2},\tfrac{3}{2}$ & 0 & 
      ${\FineStructure}+\frac{\HyperFineSplitting}{120} - \frac{\I}{2}{\Gamma_P}$ & 0 & 0 & 0\\ \hlx{vhv}
      $2P_{3/2},\tfrac{3}{2},\tfrac{3}{2}$ & 0 & 0 & 
      \mycell{22mm}{${\FineStructure}-\frac{\HyperFineSplitting}{180}$\\$ - \frac{\I}{2}{\Gamma_P}$} & 
      $-\frac{{\sqrt{5}}\HyperFineSplitting}{576}$ & 0\\ \hlx{vhv}
      $2P_{1/2},\tfrac{3}{2},\tfrac{3}{2}$ & 0 & 0 & 
      $-\frac{{\sqrt{5}}\HyperFineSplitting}{576}$ & 
      $\frac{\HyperFineSplitting}{72} - \frac\I2{\Gamma_P}$ & 
      $\I({\delta_1} + {\delta_2}){\LambShift}$\\ \hlx{vhv}
      $2S_{1/2},\tfrac{3}{2},\tfrac{3}{2}$ & 0 & 0 & 0 & 
      $-\I({\delta_1} + {\delta_2}) {\LambShift}$ & 
      ${\LambShift}+\frac{\HyperFineSplitting}{24} - \frac{\I}{2}{\Gamma_S}$\\ \hlx{vh}
    \end{tabular}

\label{tB:M0D2a}}\\[5mm]
  \subfloat[]{

    \begin{tabular}{l||*{7}{>{\centering\baselineskip=15pt }m{15mm}|}}
      & $2P_{3/2}$ & $2P_{3/2}$ & $2P_{1/2}$ & $2S_{1/2}$ & $2P_{3/2}$ & $2P_{1/2}$ & $2S_{1/2}$\\ \hlx{v}
      & $\tfrac{5}{2},\tfrac{1}{2}$ & $\tfrac{3}{2},\tfrac{1}{2}$ & $\tfrac{3}{2},\tfrac{1}{2}$ & $\tfrac{3}{2},\tfrac{1}{2}$ & $\tfrac{1}{2},\tfrac{1}{2}$ & $\tfrac{1}{2},\tfrac{1}{2}$ & $\tfrac{1}{2},\tfrac{1}{2}$\\ \hlx{vhhv}
      $2P_{3/2},\tfrac{5}{2},\tfrac{1}{2}$ & 
      \mycell{15mm}{${\FineStructure}+\frac{\HyperFineSplitting}{120}$\\$ - \frac{\I}{2}{\Gamma_P}$} & 0 & 0 & 0 & 0 & 0 & 0\\ \hlx{vhv}
      $2P_{3/2},\tfrac{3}{2},\tfrac{1}{2}$ & 0 & 
      \mycell{15mm}{${\FineStructure}-\frac{\HyperFineSplitting}{180}$\\$ - \frac{\I}{2}{\Gamma_P}$} & 
      $-\frac{\sqrt{5}\HyperFineSplitting}{576}$ & 0 & 0 & 0 & 0\\ \hlx{vhv}
      $2P_{1/2},\tfrac{3}{2},\tfrac{1}{2}$ & 0 & 
      $-\frac{{\sqrt{5}}\HyperFineSplitting}{576}$ & 
      $\frac{\HyperFineSplitting}{72} - \frac\I2{\Gamma_P}$ & 
      \mycell{15mm}{$\I{\delta_1}\LambShift$\\$ + \I{\delta_2}{\LambShift}$} & 0 & 0 & 0\\ \hlx{vhv}
      $2S_{1/2},\tfrac{3}{2},\tfrac{1}{2}$ & 0 & 0 & 
      \mycell{15mm}{$-\I{\delta_1}\LambShift$\\$ - \I{\delta_2}{\LambShift}$} & 
      \mycell{15mm}{${\LambShift}+\frac{\HyperFineSplitting}{24}$\\$ - \frac{\I}{2}{\Gamma_S}$} & 0 & 0 & 0\\ \hlx{vhv}
      $2P_{3/2},\tfrac{1}{2},\tfrac{1}{2}$ & 0 & 0 & 0 & 0 & 
      \mycell{15mm}{${\FineStructure}-\frac{\HyperFineSplitting}{72}$\\$ - \frac{\I}{2}{\Gamma_P}$} & 
      $-\frac{\HyperFineSplitting}{288{\sqrt{2}}}$ & 0\\ \hlx{vhv}
      $2P_{1/2},\tfrac{1}{2},\tfrac{1}{2}$ & 0 & 0 & 0 & 0 & 
      $-\frac{\HyperFineSplitting}{288{\sqrt{2}}}$ & 
      \mycell{15mm}{$-\frac{\HyperFineSplitting}{36}$\\$ - \frac\I2{\Gamma_P}$} & 
      \mycell{15mm}{$\I{\delta_1}\LambShift$\\$ - 2\I{\delta_2} {\LambShift}$}\\
\end{tabular}

\label{tB:M0D2b}}
  \caption[Die freie, nichthermitesche Massenmatrix für
  Deuterium im $(n=2)$-Unterraum]{Die freie, nichthermitesche Massenmatrix \underline{$\ms M_0$}\ 
für
  Deuterium im $(n=2)$-Unterraum ($F_3=\tfrac52,\tfrac32,\tfrac12$).}\label{tB:M0D2}
\end{table}
\begin{table}[!hp]
  \ContinuedFloat
  \centering
  \footnotesize
  \subfloat[]{

    \begin{tabular}{l||*{7}{>{\centering\baselineskip=15pt }m{15mm}|}}
      & $2P_{3/2}$ & $2P_{3/2}$ & $2P_{1/2}$ & $2S_{1/2}$ & $2P_{3/2}$ & $2P_{1/2}$ & $2S_{1/2}$\\ \hlx{v}
      & $\tfrac{5}{2},-\tfrac{1}{2}$ & $\tfrac{3}{2},-\tfrac{1}{2}$ & $\tfrac{3}{2},-\tfrac{1}{2}$ & $\tfrac{3}{2},-\tfrac{1}{2}$ & $\tfrac{1}{2},-\tfrac{1}{2}$ & $\tfrac{1}{2},-\tfrac{1}{2}$ & $\tfrac{1}{2},-\tfrac{1}{2}$\\ \hlx{vhhv}
      $2P_{3/2},\tfrac{5}{2},-\tfrac{1}{2}$ & 
      \mycell{15mm}{${\FineStructure}+\frac{\HyperFineSplitting}{120}$\\$ - \frac{\I}{2}{\Gamma_P}$} 
      & 0 & 0 & 0 & 0 & 0 & 0\\ \hlx{vhv}
      $2P_{3/2},\tfrac{3}{2},-\tfrac{1}{2}$ & 0 & 
      \mycell{15mm}{${\FineStructure}-\frac{\HyperFineSplitting}{180}$\\$ - \frac{\I}{2}{\Gamma_P}$} & 
      $-\frac{{\sqrt{5}}\HyperFineSplitting}{576}$ & 0 & 0 & 0 & 0\\ \hlx{vhv}
      $2P_{1/2},\tfrac{3}{2},-\tfrac{1}{2}$ & 0 & 
      $-\frac{{\sqrt{5}}\HyperFineSplitting}{576}$ & 
      $\frac{\HyperFineSplitting}{72} - \frac\I2{\Gamma_P}$ & 
      \mycell{15mm}{$\I{\delta_1}\LambShift$\\$ + \I{\delta_2} {\LambShift}$} & 0 & 0 & 0\\ \hlx{vhv}
      $2S_{1/2},\tfrac{3}{2},-\tfrac{1}{2}$ & 0 & 0 & 
      \mycell{15mm}{$-\I {\delta_1}\LambShift$\\$ - \I{\delta_2} {\LambShift}$} & 
      \mycell{15mm}{${\LambShift}+\frac{\HyperFineSplitting}{24}$\\$ - \frac{\I}{2}{\Gamma_S}$} 
      & 0 & 0 & 0\\ \hlx{vhv}
      $2P_{3/2},\tfrac{1}{2},-\tfrac{1}{2}$ & 0 & 0 & 0 & 0 & 
      \mycell{15mm}{${\FineStructure}-\frac{\HyperFineSplitting}{72}$\\$ - \frac{\I}{2}{\Gamma_P}$} & 
      $-\frac{\HyperFineSplitting}{288{\sqrt{2}}}$ & 0\\ \hlx{vhv}
      $2P_{1/2},\tfrac{1}{2},-\tfrac{1}{2}$ & 0 & 0 & 0 & 0 & 
      $-\frac{\HyperFineSplitting}{288{\sqrt{2}}}$ & 
      \mycell{15mm}{$-\frac{\HyperFineSplitting}{36}$\\$ - \frac\I2{\Gamma_P}$} & 
      \mycell{15mm}{$\I{\delta_1}\LambShift$\\$ - 2\I{\delta_2} {\LambShift}$}\\
\end{tabular}

\label{tB:M0D2c}}\\[5mm]
  \subfloat[]{

    \begin{tabular}{l||*{4}{>{\centering\baselineskip=15pt }m{22mm}|}|>{\centering\baselineskip=15pt }m{22mm}|}
      & $2P_{3/2}$ & $2P_{3/2}$ & $2P_{1/2}$ & $2S_{1/2}$ & $2P_{3/2}$\\ \hlx{v}
      & $\tfrac{5}{2},-\tfrac{3}{2}$ & $\tfrac{3}{2},-\tfrac{3}{2}$ & $\tfrac{3}{2},-\tfrac{3}{2}$ & $\tfrac{3}{2},-\tfrac{3}{2}$ & $\tfrac{5}{2},-\tfrac{5}{2}$\\ \hlx{vhhv}
      $2P_{3/2},\tfrac{5}{2},-\tfrac{3}{2}$ & 
      ${\FineStructure}+\frac{\HyperFineSplitting}{120} - \frac{\I}{2}{\Gamma_P}$ & 0 & 0 & 0 & 0\\ \hlx{vhv}
      $2P_{3/2},\tfrac{3}{2},-\tfrac{3}{2}$ & 0 & 
      \mycell{22mm}{${\FineStructure}-\frac{\HyperFineSplitting}{180}$\\$ - \frac{\I}{2}{\Gamma_P}$} & 
      $-\frac{{\sqrt{5}}\HyperFineSplitting}{576}$ & 0 & 0\\ \hlx{vhv}
      $2P_{1/2},\tfrac{3}{2},-\tfrac{3}{2}$ & 0 & 
      $-\frac{{\sqrt{5}}\HyperFineSplitting}{576}$ & 
      $\frac{\HyperFineSplitting}{72} - \frac\I2 {\Gamma_P}$ & 
      $\I({\delta_1} + {\delta_2}) {\LambShift}$ & 0\\ \hlx{vhv}
      $2S_{1/2},\tfrac{3}{2},-\tfrac{3}{2}$ & 0 & 0 & 
      $-\I({\delta_1} + {\delta_2}) {\LambShift}$ & 
      ${\LambShift}+\frac{\HyperFineSplitting}{24} - \frac{\I}{2}{\Gamma_S}$ & 0\\ \hlx{vhhv}
      $2P_{3/2},\tfrac{5}{2},-\tfrac{5}{2}$ & 0 & 0 & 0 & 0 & 
      ${\FineStructure}+\frac{\HyperFineSplitting}{120} - \frac{\I}{2}{\Gamma_P}$\\ \hlx{vh}
    \end{tabular}

\label{tB:M0D2d}}
  \caption[]{Die freie, nichthermitesche Massenmatrix \underline{$\ms M_0$}\ für
  Deuterium im $(n=2)$-Unterraum ($F_3=-\tfrac12,-\tfrac32,-\tfrac52$).}
\end{table}

\begin{table}[!hp]
\centering
\footnotesize\vspace{1cm}
\subfloat[Die Zuordnung der Tabellennummern zu den einzelnen Blöcken der Matrix $\unl\mu_1$.
Es werden nur die Blöcke der oberen Hälfte der Matrix angegeben. Die unteren Blöcke folgen durch hermitesche Konjugation.]{
\begin{tabular}{c||c|c|c|c|c|c|}
$F_3'\backslash F_3$ & $5/2$ & $3/2$ & $1/2$ & $-1/2$ & $-3/2$ & $-5/2$\\ \hlx{vhhv}
$5/2$  & 0 & \subref*{tB:mu1D2b} & 0 & 0 & 0 & 0\\ \hlx{vhv}
$3/2$  & (\subref*{tB:mu1D2b})$^\dag$ & 0 & \subref*{tB:mu1D2c} & 0 & 0 & 0\\ \hlx{vhv}
$1/2$  & 0 & (\subref*{tB:mu1D2c})$^\dag$ & 0 & \subref*{tB:mu1D2d} & 0 & 0\\ \hlx{vhv}
$-1/2$ & 0 & 0 & (\subref*{tB:mu1D2d})$^\dag$ & 0 & \subref*{tB:mu1D2e} & 0\\ \hlx{vhv}
$-3/2$ & 0 & 0 & 0 & (\subref*{tB:mu1D2e})$^\dag$ & 0 & \subref*{tB:mu1D2f}\\ \hlx{vhv}
$-5/2$ & 0 & 0 & 0 & 0 & (\subref*{tB:mu1D2f})$^\dag$ & 0 \\ \hlx{vhv}
\end{tabular}\label{tB:mu1D2overview}}\\[15mm]
\subfloat[]{

    \begin{tabular}{l||*{4}{>{\centering\baselineskip=18pt }m{13mm}|}}
      & $2P_{3/2}$ & $2P_{3/2}$ & $2P_{1/2}$ & $2S_{1/2}$\\ \hlx{v}
      & $\tfrac{5}{2},\tfrac{3}{2}$ & $\tfrac{3}{2},\tfrac{3}{2}$ & $\tfrac{3}{2},\tfrac{3}{2}$ & $\tfrac{3}{2},\tfrac{3}{2}$\\ \hlx{vhhv}
      $2P_{3/2},\tfrac{5}{2},\tfrac{5}{2}$ & 
      $-\frac{g+2}{2 \sqrt{5}}$ & 
      $\frac{g+2}{\sqrt{30}}$ & 
      $-\frac{g-1}{\sqrt{6}}$ & 0\\ \hlx{vh}
    \end{tabular}\label{tB:mu1D2b}}\\[10mm]
\subfloat[]{

    \begin{tabular}{l||*{7}{>{\centering\baselineskip=18pt }m{15mm}|}}
      & $2P_{3/2}$ & $2P_{3/2}$ & $2P_{1/2}$ & $2S_{1/2}$ & $2P_{3/2}$ & $2P_{1/2}$ & $2S_{1/2}$\\ \hlx{v}
      & $\tfrac{5}{2},\tfrac{1}{2}$ & $\tfrac{3}{2},\tfrac{1}{2}$ & $\tfrac{3}{2},\tfrac{1}{2}$ & $\tfrac{3}{2},\tfrac{1}{2}$ & $\tfrac{1}{2},\tfrac{1}{2}$ & $\tfrac{1}{2},\tfrac{1}{2}$ & $\tfrac{1}{2},\tfrac{1}{2}$\\ \hlx{vhhv}
      $2P_{3/2},\tfrac{5}{2},\tfrac{3}{2}$ & 
$-\frac{\sqrt{2}(g+2)}{5}$ & 
      $\frac{g+2}{5 \sqrt{2}}$ & 
      $-\frac{g-1}{\sqrt{10}}$ & 0 & 0 & 0 & 0\\ \hlx{vhv}
      $2P_{3/2},\tfrac{3}{2},\tfrac{3}{2}$ & 
      $-\frac{g+2}{10 \sqrt{3}}$ & 
      $-\frac{11 (g+2)}{30 \sqrt{3}}$ & 
      $-\frac{2 (g-1)}{3 \sqrt{15}}$ & 0 & 
$\frac{\sqrt5(g+2)}{6\sqrt3}$ & 
$-\frac{\sqrt5(g-1)}{3\sqrt6}$ & 0\\ \hlx{vhv}
      $2P_{1/2},\tfrac{3}{2},\tfrac{3}{2}$ & 
      $\frac{g-1}{2 \sqrt{15}}$ & 
      $-\frac{2 (g-1)}{3 \sqrt{15}}$ & 
      $\frac{g-4}{6 \sqrt{3}}$ & 0 & 
      $\frac{g-1}{6 \sqrt{3}}$ & 
      $-\frac{g-4}{3 \sqrt{6}}$ & 0\\ \hlx{vhv}
      $2S_{1/2},\tfrac{3}{2},\tfrac{3}{2}$ & 0 & 0 & 0 & 
      $-\frac{g}{2 \sqrt{3}}$ & 0 & 0 & 
      $\frac{g}{\sqrt{6}}$\\ \hlx{vh}
    \end{tabular}\label{tB:mu1D2c}}\\[10mm]
\caption[Die Matrix \unl{$\mu$}$_1$ der ersten Komponente des magnetischen Moments
von Deuterium im $(n=2)$-Unterraum]{
Die auf $\mu_B$ normierte Matrix \unl{$\mu$}$_1$ der ersten Komponente des
magnetischen Moments von Deuterium im $(n=2)$-Unterraum. (Fortsetzung auf der nächsten Seite)}
\label{tB:mu1D2}
\end{table}
\FloatBarrier

\begin{table}[!hp]
\ContinuedFloat
\centering\vspace{1cm}
\footnotesize
\subfloat[]{

    \begin{tabular}{l||*{7}{>{\centering\baselineskip=18pt }m{16mm}|}}
      & $2P_{3/2}$ & $2P_{3/2}$ & $2P_{1/2}$ & $2S_{1/2}$ & $2P_{3/2}$ & $2P_{1/2}$ & $2S_{1/2}$\\ \hlx{v}
      & $\tfrac{5}{2},-\tfrac{1}{2}$ & $\tfrac{3}{2},-\tfrac{1}{2}$ & $\tfrac{3}{2},-\tfrac{1}{2}$ & $\tfrac{3}{2},-\tfrac{1}{2}$ & $\tfrac{1}{2},-\tfrac{1}{2}$ & $\tfrac{1}{2},-\tfrac{1}{2}$ & $\tfrac{1}{2},-\tfrac{1}{2}$\\ \hlx{vhhv}
      $2P_{3/2},\tfrac{5}{2},\tfrac{1}{2}$ & 
$-\frac{3(g+2)}{10}$ & 
      $\frac{g+2}{10}$ & 
      $-\frac{g-1}{2 \sqrt{5}}$ & 0 & 0 & 0 & 0\\ \hlx{vhv}
      $2P_{3/2},\tfrac{3}{2},\tfrac{1}{2}$ & 
      $-\frac{g+2}{10}$ & 
$-\frac{11(g+2)}{45}$ & 
      $-\frac{4 (g-1)}{9 \sqrt{5}}$ & 0 & 
$\frac{(g+2)}{18} \sqrt{5} $ & 
$-\frac{\sqrt5(g-1)}{9\sqrt2}$ & 0\\ \hlx{vhv}
      $2P_{1/2},\tfrac{3}{2},\tfrac{1}{2}$ & 
      $\frac{g-1}{2 \sqrt{5}}$ & 
      $-\frac{4 (g-1)}{9 \sqrt{5}}$ & 
      $\frac{g-4}{9}$ & 0 & 
      $\frac{g-1}{18}$ & 
      $-\frac{g-4}{9 \sqrt{2}}$ & 0\\ \hlx{vhv}
      $2S_{1/2},\tfrac{3}{2},\tfrac{1}{2}$ & 0 & 0 & 0 & 
      $-\frac{g}{3}$ & 0 & 0 & 
      $\frac{g}{3 \sqrt{2}}$\\ \hlx{vhv}
      $2P_{3/2},\tfrac{1}{2},\tfrac{1}{2}$ & 0 & 
$-\frac{\sqrt5(g+2)}{18}$ & 
      $\frac{1-g}{18}$ & 0 & 
$-\frac{5(g+2)}{18}$ & 
$-\frac{2\sqrt2(g-1)}{9}$ & 0\\ \hlx{vhv}
      $2P_{1/2},\tfrac{1}{2},\tfrac{1}{2}$ & 0 & 
$\frac{\sqrt5(g-1)}{9\sqrt2}$ & 
      $\frac{g-4}{9 \sqrt{2}}$ & 0 & 
$-\frac{2\sqrt2(g-1)}{9}$ & 
      $\frac{4-g}{18}$ & 0\\ \hlx{vhv}
      $2S_{1/2},\tfrac{1}{2},\tfrac{1}{2}$ & 0 & 0 & 0 & 
      $-\frac{g}{3 \sqrt{2}}$ & 0 & 0 & 
      $\frac{g}{6}$\\ \hlx{vh}
    \end{tabular}\label{tB:mu1D2d}}\\[10mm]
\subfloat[]{

    \begin{tabular}{l||*{4}{>{\centering\baselineskip=18pt }m{16mm}|}}
      & $2P_{3/2}$ & $2P_{3/2}$ & $2P_{1/2}$ & $2S_{1/2}$\\ \hlx{v}
      & $\tfrac{5}{2},-\tfrac{3}{2}$ & $\tfrac{3}{2},-\tfrac{3}{2}$ & $\tfrac{3}{2},-\tfrac{3}{2}$ & $\tfrac{3}{2},-\tfrac{3}{2}$\\ \hlx{vhhv}
      $2P_{3/2},\tfrac{5}{2},-\tfrac{1}{2}$ & 
$-\frac{\sqrt{2} (g+2)}{5})$ & 
      $\frac{g+2}{10 \sqrt{3}}$ & 
      $-\frac{g-1}{2 \sqrt{15}}$ & 0\\ \hlx{vhv}
      $2P_{3/2},\tfrac{3}{2},-\tfrac{1}{2}$ & 
      $-\frac{g+2}{5 \sqrt{2}}$ & 
      $-\frac{11 (g+2)}{30 \sqrt{3}}$ & 
      $-\frac{2 (g-1)}{3 \sqrt{15}}$ & 0\\ \hlx{vhv}
      $2P_{1/2},\tfrac{3}{2},-\tfrac{1}{2}$ & 
      $\frac{g-1}{\sqrt{10}}$ & 
      $-\frac{2 (g-1)}{3 \sqrt{15}}$ & 
      $\frac{g-4}{6 \sqrt{3}}$ & 0\\ \hlx{vhv}
      $2S_{1/2},\tfrac{3}{2},-\tfrac{1}{2}$ & 0 & 0 & 0 & 
      $-\frac{g}{2 \sqrt{3}}$\\ \hlx{vhv}
      $2P_{3/2},\tfrac{1}{2},-\tfrac{1}{2}$ & 0 & 
$-\frac{\sqrt5(g+2)}{6\sqrt3}$ & 
      $-\frac{g-1}{6 \sqrt{3}}$ & 0\\ \hlx{vhv}
      $2P_{1/2},\tfrac{1}{2},-\tfrac{1}{2}$ & 0 & 
$\frac{\sqrt5(g-1)}{3\sqrt6}$ & 
      $\frac{g-4}{3 \sqrt{6}}$ & 0\\ \hlx{vhv}
      $2S_{1/2},\tfrac{1}{2},-\tfrac{1}{2}$ & 0 & 0 & 0 & 
      $-\frac{g}{\sqrt{6}}$\\ \hlx{vh}
    \end{tabular}\label{tB:mu1D2e}}\qquad\quad
\subfloat[]{

    \begin{tabular}{l||*{1}{>{\centering\baselineskip=18pt }m{13mm}|}}
      & $2P_{3/2}$\\ \hlx{v}
      & $\tfrac{5}{2},-\tfrac{5}{2}$\\ \hlx{vhhv}
      $2P_{3/2},\tfrac{5}{2},-\tfrac{3}{2}$ & 
      $-\frac{g+2}{2 \sqrt{5}}$\\ \hlx{vhv}
      $2P_{3/2},\tfrac{3}{2},-\tfrac{3}{2}$ & 
      $-\frac{g+2}{\sqrt{30}}$\\ \hlx{vhv}
      $2P_{1/2},\tfrac{3}{2},-\tfrac{3}{2}$ & 
      $\frac{g-1}{\sqrt{6}}$\\ \hlx{vhv}
      $2S_{1/2},\tfrac{3}{2},-\tfrac{3}{2}$ & 0\\ \hlx{vh}
    \end{tabular}\label{tB:mu1D2f}}\\[10mm]
\caption[]{
(Fortsetzung) Die auf $\mu_B$ normierte Matrix \unl{$\mu$}$_1$ der ersten Komponente des
magnetischen Moments von Deuterium im $(n=2)$-Unterraum.}
\end{table}
\FloatBarrier

\begin{table}[!hp]
\centering
\footnotesize\vspace{1cm}
\subfloat[Die Zuordnung der Tabellennummern zu den einzelnen Blöcken der Matrix $\unl\mu_2$.
Es werden nur die Blöcke der oberen Hälfte der Matrix angegeben. Die unteren Blöcke folgen durch hermitesche Konjugation.]{
\begin{tabular}{c||c|c|c|c|c|c|}
$F_3'\backslash F_3$ & $5/2$ & $3/2$ & $1/2$ & $-1/2$ & $-3/2$ & $-5/2$\\ \hlx{vhhv}
$5/2$  & 0 & \subref*{tB:mu2D2b} & 0 & 0 & 0 & 0\\ \hlx{vhv}
$3/2$  & (\subref*{tB:mu2D2b})$^\dag$ & 0 & \subref*{tB:mu2D2c} & 0 & 0 & 0\\ \hlx{vhv}
$1/2$  & 0 & (\subref*{tB:mu2D2c})$^\dag$ & 0 & \subref*{tB:mu2D2d} & 0 & 0\\ \hlx{vhv}
$-1/2$ & 0 & 0 & (\subref*{tB:mu2D2d})$^\dag$ & 0 & \subref*{tB:mu2D2e} & 0\\ \hlx{vhv}
$-3/2$ & 0 & 0 & 0 & (\subref*{tB:mu2D2e})$^\dag$ & 0 & \subref*{tB:mu2D2f}\\ \hlx{vhv}
$-5/2$ & 0 & 0 & 0 & 0 & (\subref*{tB:mu2D2f})$^\dag$ & 0 \\ \hlx{vhv}
\end{tabular}\label{tB:mu2D2overview}}\\[15mm]
\subfloat[]{

    \begin{tabular}{l||*{4}{>{\centering\baselineskip=18pt }m{13mm}|}}
      & $2P_{3/2}$ & $2P_{3/2}$ & $2P_{1/2}$ & $2S_{1/2}$\\ \hlx{v}
      & $\tfrac{5}{2},\tfrac{3}{2}$ & $\tfrac{3}{2},\tfrac{3}{2}$ & $\tfrac{3}{2},\tfrac{3}{2}$ & $\tfrac{3}{2},\tfrac{3}{2}$\\ \hlx{vhhv}
      $2P_{3/2},\tfrac{5}{2},\tfrac{5}{2}$ & 
      $\frac{\I (g+2)}{2 \sqrt{5}}$ & 
      $-\frac{\I (g+2)}{\sqrt{30}}$ & 
      $\frac{\I (g-1)}{\sqrt{6}}$ & 0\\ \hlx{vh}
    \end{tabular}\label{tB:mu2D2b}}\\[10mm]
\subfloat[]{

    \begin{tabular}{l||*{7}{>{\centering\baselineskip=18pt }m{16mm}|}}
      & $2P_{3/2}$ & $2P_{3/2}$ & $2P_{1/2}$ & $2S_{1/2}$ & $2P_{3/2}$ & $2P_{1/2}$ & $2S_{1/2}$\\ \hlx{v}
      & $\tfrac{5}{2},\tfrac{1}{2}$ & $\tfrac{3}{2},\tfrac{1}{2}$ & $\tfrac{3}{2},\tfrac{1}{2}$ & $\tfrac{3}{2},\tfrac{1}{2}$ & $\tfrac{1}{2},\tfrac{1}{2}$ & $\tfrac{1}{2},\tfrac{1}{2}$ & $\tfrac{1}{2},\tfrac{1}{2}$\\ \hlx{vhhv}
      $2P_{3/2},\tfrac{5}{2},\tfrac{3}{2}$ & 
$\frac{\I \sqrt{2} (g+2)}{5} $ & 
      $-\frac{\I (g+2)}{5 \sqrt{2}}$ & 
      $\frac{\I (g-1)}{\sqrt{10}}$ & 0 & 0 & 0 & 0\\ \hlx{vhv}
      $2P_{3/2},\tfrac{3}{2},\tfrac{3}{2}$ & 
      $\frac{\I (g+2)}{10 \sqrt{3}}$ & 
      $\frac{11 \I (g+2)}{30 \sqrt{3}}$ & 
      $\frac{2 \I (g-1)}{3 \sqrt{15}}$ & 0 & 
      $-\frac{\I\sqrt5(g+2)}{6\sqrt3}$ & 
      $\frac{\I\sqrt5(g-1)}{3\sqrt6}$ & 0\\ \hlx{vhv}
      $2P_{1/2},\tfrac{3}{2},\tfrac{3}{2}$ & 
      $-\frac{\I (g-1)}{2 \sqrt{15}}$ & 
      $\frac{2 \I (g-1)}{3 \sqrt{15}}$ & 
      $-\frac{\I (g-4)}{6 \sqrt{3}}$ & 0 & 
      $-\frac{\I (g-1)}{6 \sqrt{3}}$ & 
      $\frac{\I (g-4)}{3 \sqrt{6}}$ & 0\\ \hlx{vhv}
      $2S_{1/2},\tfrac{3}{2},\tfrac{3}{2}$ & 0 & 0 & 0 & 
      $\frac{\I g}{2 \sqrt{3}}$ & 0 & 0 & 
      $-\frac{\I g}{\sqrt{6}}$\\ \hlx{vh}
    \end{tabular}\label{tB:mu2D2c}}\\[10mm]
\caption[Die Matrix \unl{$\mu$}$_2$ der zweiten Komponente des magnetischen Moments
von Deuterium im $(n=2)$-Unterraum]{
Die auf $\mu_B$ normierte Matrix \unl{$\mu$}$_2$ der zweiten Komponente des
magnetischen Moments von Deuterium im $(n=2)$-Unterraum. (Fortsetzung auf der nächsten Seite)}
\label{tB:mu2D2}
\end{table}
\FloatBarrier

\begin{table}[!hp]
\ContinuedFloat
\centering\vspace{1cm}
\footnotesize
\subfloat[]{

    \begin{tabular}{l||*{7}{>{\centering\baselineskip=18pt }m{16mm}|}}
      & $2P_{3/2}$ & $2P_{3/2}$ & $2P_{1/2}$ & $2S_{1/2}$ & $2P_{3/2}$ & $2P_{1/2}$ & $2S_{1/2}$\\ \hlx{v}
& $\tfrac{5}{2},-\tfrac{1}{2}$ & $\tfrac{3}{2},-\tfrac{1}{2}$ & $\tfrac{3}{2},-\tfrac{1}{2}$ & $\tfrac{3}{2},-\tfrac{1}{2}$ & $\tfrac{1}{2},-\tfrac{1}{2}$ & $\tfrac{1}{2},-\tfrac{1}{2}$ & $\tfrac{1}{2},-\tfrac{1}{2}$\\ \hlx{vhhv}
      $2P_{3/2},\tfrac{5}{2},\tfrac{1}{2}$ & 
$\frac{3\I (g+2)}{10}$ & 
$-\frac{\I (g+2)}{10}$ & 
      $\frac{\I (g-1)}{2 \sqrt{5}}$ & 0 & 0 & 0 & 0\\ \hlx{vhv}
      $2P_{3/2},\tfrac{3}{2},\tfrac{1}{2}$ & 
$\frac{\I (g+2)}{10}$ & 
$\frac{11\I (g+2)}{45}$ & 
      $\frac{4 \I (g-1)}{9 \sqrt{5}}$ & 0 & 
$-\frac{\I \sqrt{5} (g+2)}{18}$ & 
      $\frac{\I\sqrt5(g-1)}{9\sqrt2}$ & 0\\ \hlx{vhv}
      $2P_{1/2},\tfrac{3}{2},\tfrac{1}{2}$ & 
      $-\frac{\I (g-1)}{2 \sqrt{5}}$ & 
      $\frac{4 \I (g-1)}{9 \sqrt{5}}$ & 
$-\frac{\I (g-4)}{9}$ & 0 & 
$-\frac{\I (g-1)}{18}$ & 
      $\frac{\I (g-4)}{9 \sqrt{2}}$ & 0\\ \hlx{vhv}
      $2S_{1/2},\tfrac{3}{2},\tfrac{1}{2}$ & 0 & 0 & 0 & 
      $\frac{\I g}{3}$ & 0 & 0 & 
      $-\frac{\I g}{3 \sqrt{2}}$\\ \hlx{vhv}
      $2P_{3/2},\tfrac{1}{2},\tfrac{1}{2}$ & 0 & 
$\frac{\I \sqrt{5} (g+2)}{18}$ & 
$\frac{\I (g-1)}{18}$ & 0 & 
$\frac{5\I (g+2)}{18}$ & 
$\frac{2\I \sqrt{2} (g-1)}{9}$ & 0\\ \hlx{vhv}
      $2P_{1/2},\tfrac{1}{2},\tfrac{1}{2}$ & 0 & 
      $-\frac{\I\sqrt5(g-1)}{9\sqrt2}$ & 
      $-\frac{\I (g-4)}{9 \sqrt{2}}$ & 0 & 
$\frac{2 \I \sqrt{2} (g-1)}{9}$ & 
$\frac{\I (g-4)}{18}$ & 0\\ \hlx{vhv}
      $2S_{1/2},\tfrac{1}{2},\tfrac{1}{2}$ & 0 & 0 & 0 & 
      $\frac{\I g}{3 \sqrt{2}}$ & 0 & 0 & 
      $-\frac{\I g}{6}$\\ \hlx{vh}
    \end{tabular}\label{tB:mu2D2d}}\\[10mm]
\subfloat[]{

    \begin{tabular}{l||*{4}{>{\centering\baselineskip=18pt }m{16mm}|}}
      & $2P_{3/2}$ & $2P_{3/2}$ & $2P_{1/2}$ & $2S_{1/2}$\\ \hlx{v}
      & $\tfrac{5}{2},-\tfrac{3}{2}$ & $\tfrac{3}{2},-\tfrac{3}{2}$ & $\tfrac{3}{2},-\tfrac{3}{2}$ & $\tfrac{3}{2},-\tfrac{3}{2}$\\ \hlx{vhhv}
      $2P_{3/2},\tfrac{5}{2},-\tfrac{1}{2}$ & 
$\frac{\I \sqrt{2} (g+2)}{5}$ & 
      $-\frac{\I (g+2)}{10 \sqrt{3}}$ & 
      $\frac{\I (g-1)}{2 \sqrt{15}}$ & 0\\ \hlx{vhv}
      $2P_{3/2},\tfrac{3}{2},-\tfrac{1}{2}$ & 
      $\frac{\I (g+2)}{5 \sqrt{2}}$ & 
      $\frac{11 \I (g+2)}{30 \sqrt{3}}$ & 
      $\frac{2 \I (g-1)}{3 \sqrt{15}}$ & 0\\ \hlx{vhv}
      $2P_{1/2},\tfrac{3}{2},-\tfrac{1}{2}$ & 
      $-\frac{\I (g-1)}{\sqrt{10}}$ & 
      $\frac{2 \I (g-1)}{3 \sqrt{15}}$ & 
      $-\frac{\I (g-4)}{6 \sqrt{3}}$ & 0\\ \hlx{vhv}
      $2S_{1/2},\tfrac{3}{2},-\tfrac{1}{2}$ & 0 & 0 & 0 & 
      $\frac{\I g}{2 \sqrt{3}}$\\ \hlx{vhv}
      $2P_{3/2},\tfrac{1}{2},-\tfrac{1}{2}$ & 0 & 
      $\frac{\I\sqrt5(g+2)}{6\sqrt3}$ & 
      $\frac{\I (g-1)}{6 \sqrt{3}}$ & 0\\ \hlx{vhv}
      $2P_{1/2},\tfrac{1}{2},-\tfrac{1}{2}$ & 0 & 
      $-\frac{\I\sqrt5(g-1)}{3\sqrt6}$ & 
      $-\frac{\I (g-4)}{3 \sqrt{6}}$ & 0\\ \hlx{vhv}
      $2S_{1/2},\tfrac{1}{2},-\tfrac{1}{2}$ & 0 & 0 & 0 & 
      $\frac{\I g}{\sqrt{6}}$\\ \hlx{vh}
    \end{tabular}\label{tB:mu2D2e}}\qquad\quad
\subfloat[]{

    \begin{tabular}{l||*{1}{>{\centering\baselineskip=18pt }m{13mm}|}}
      & $2P_{3/2}$\\ \hlx{v}
      & $\tfrac{5}{2},-\tfrac{5}{2}$\\ \hlx{vhhv}
      $2P_{3/2},\tfrac{5}{2},-\tfrac{3}{2}$ & 
      $\frac{\I (g+2)}{2 \sqrt{5}}$\\ \hlx{vhv}
      $2P_{3/2},\tfrac{3}{2},-\tfrac{3}{2}$ & 
      $\frac{\I (g+2)}{\sqrt{30}}$\\ \hlx{vhv}
      $2P_{1/2},\tfrac{3}{2},-\tfrac{3}{2}$ & 
      $-\frac{\I (g-1)}{\sqrt{6}}$\\ \hlx{vhv}
      $2S_{1/2},\tfrac{3}{2},-\tfrac{3}{2}$ & 0\\ \hlx{vh}
    \end{tabular}\label{tB:mu2D2f}}\\[10mm]
\caption[]{
(Fortsetzung) Die auf $\mu_B$ normierte Matrix \unl{$\mu$}$_2$ der zweiten Komponente des
magnetischen Moments von Deuterium im $(n=2)$-Unterraum.}
\end{table}
\FloatBarrier

\begin{table}[!hp]
\centering
\footnotesize
\subfloat[]{

    \begin{tabular}{l||>{\centering\baselineskip=15pt }m{18mm}||*{4}{>{\centering\baselineskip=15pt }m{18mm}|}}
      & $2P_{3/2}$ & $2P_{3/2}$ & $2P_{3/2}$ & $2P_{1/2}$ & $2S_{1/2}$\\ \hlx{v}
      & $\tfrac{5}{2},\tfrac{5}{2}$ & $\tfrac{5}{2},\tfrac{3}{2}$ & $\tfrac{3}{2},\tfrac{3}{2}$ & $\tfrac{3}{2},\tfrac{3}{2}$ & $\tfrac{3}{2},\tfrac{3}{2}$\\ \hlx{vhhv}
      $2P_{3/2},\tfrac{5}{2},\tfrac{5}{2}$ & 
      $-\frac{2 + g}{2}$ & 0 & 0 & 0 & 0\\ \hlx{vhhv}
      $2P_{3/2},\tfrac{5}{2},\tfrac{3}{2}$ & 0 & 
      $-\frac{3(2 + g)}{10}$ & 
      $-\frac{\sqrt{2}(2 + g)}{5\sqrt3}$ & 
      $-\frac{\sqrt2(1-g)}{\sqrt{15}} $ & 0\\ \hlx{vhv}
      $2P_{3/2},\tfrac{3}{2},\tfrac{3}{2}$ & 0 & 
      $-\frac{\sqrt{2}(2 + g)}{5\sqrt3}$ & 
      $-\frac{11\left( 2 + g \right) }{30}$ & 
      $\frac{2(1 - g)}{3{\sqrt{5}}}$ & 0\\ \hlx{vhv}
      $2P_{1/2},\tfrac{3}{2},\tfrac{3}{2}$ & 0 & 
      $-\frac{\sqrt2(1-g)}{\sqrt{15}}$ & 
      $\frac{2(1 - g)}{3{\sqrt{5}}}$ & 
      $-\frac{4 - g}{6}$ & 0\\ \hlx{vhv}
      $2S_{1/2},\tfrac{3}{2},\tfrac{3}{2}$ & 0 & 0 & 0 & 0 & 
      $-\frac{g}{2}$\\ \hlx{vh}
    \end{tabular}

\label{tB:mu3D2a}}\\[1cm]
\subfloat[]{\hspace{-4mm}

    \begin{tabular}{l||*{7}{>{\centering\baselineskip=15pt }m{16mm}|}}
      & $2P_{3/2}$ & $2P_{3/2}$ & $2P_{1/2}$ & $2S_{1/2}$ & $2P_{3/2}$ & $2P_{1/2}$ & $2S_{1/2}$\\ \hlx{v}
      & $\tfrac{5}{2},\tfrac{1}{2}$ & $\tfrac{3}{2},\tfrac{1}{2}$ & $\tfrac{3}{2},\tfrac{1}{2}$ & $\tfrac{3}{2},\tfrac{1}{2}$ & $\tfrac{1}{2},\tfrac{1}{2}$ & $\tfrac{1}{2},\tfrac{1}{2}$ & $\tfrac{1}{2},\tfrac{1}{2}$\\ \hlx{vhhv}
      $2P_{3/2},\tfrac{5}{2},\tfrac{1}{2}$ & 
      $-\frac{2 + g}{10}$ & 
      $-\frac{2 + g}{5}$ & 
      $-\frac{1 - g}{{\sqrt{5}}}$ & 0 & 0 & 0 & 0\\ \hlx{vhv}
      $2P_{3/2},\tfrac{3}{2},\tfrac{1}{2}$ & 
      $-\frac{2 + g}{5}$ & 
      $-\frac{11(2 + g )}{90}$ & 
      $\frac{2(1 - g)}{9{\sqrt{5}}}$ & 0 & 
      $-\frac{{\sqrt{5}}(2 + g )}{9}$ & 
      $-\frac{{\sqrt{10}}(1 - g )}{9}$ & 0\\ \hlx{vhv}
      $2P_{1/2},\tfrac{3}{2},\tfrac{1}{2}$ & 
      $-\frac{1 - g}{{\sqrt{5}}}$ & 
      $\frac{2(1 - g)}{9{\sqrt{5}}}$ & 
      $-\frac{4 - g}{18}$ & 0 & 
      $\frac{1 - g}{9}$ & 
      $-\frac{{\sqrt{2}}(4 - g)}{9}$ & 0\\ \hlx{vhv}
      $2S_{1/2},\tfrac{3}{2},\tfrac{1}{2}$ & 0 & 0 & 0 & 
      $-\frac{g}{6}$ & 0 & 0 & 
      $-\frac{{\sqrt{2}}g}{3}$\\ \hlx{vhv}
      $2P_{3/2},\tfrac{1}{2},\tfrac{1}{2}$ & 0 & 
      $-\frac{{\sqrt{5}}(2 + g )}{9}$ & 
      $\frac{1 - g}{9}$ & 0 & 
      $-\frac{5(2 + g )}{18}$ & 
      $\frac{2{\sqrt{2}}(1 - g)}{9}$ & 0\\ \hlx{vhv}
      $2P_{1/2},\tfrac{1}{2},\tfrac{1}{2}$ & 0 & 
      $-\frac{{\sqrt{10}}(1 - g )}{9}$ & 
      $-\frac{{\sqrt{2}}(4 - g)}{9}$ & 0 & 
      $\frac{2{\sqrt{2}}(1 - g)}{9}$ & 
      $\frac{4 - g}{18}$ & 0\\ \hlx{vhv}
      $2S_{1/2},\tfrac{1}{2},\tfrac{1}{2}$ & 
      0 & 0 & 0 & 
      $-\frac{{\sqrt{2}}g}{3}$ & 0 & 0 & 
      $\frac{g}{6}$\\ \hlx{vh}
    \end{tabular}

\label{tB:mu3D2b}}
\caption[Die Matrix der dritten Komponente des
magnetischen Moments von Deuterium im $(n=2)$-Unterraum]{
Die auf $\mu_B$ normierte Matrix \unl{$\mu$}$_3$ der dritten Komponente des
magnetischen Moments von Deuterium im $(n=2)$-Unterraum 
($F_3=\tfrac52,\tfrac32,\tfrac12$).}\label{tB:mu3D2}
\end{table}
\begin{table}[!hp]
\ContinuedFloat
\centering
\footnotesize
\subfloat[]{\hspace{-4mm}

    \begin{tabular}{l||*{7}{>{\centering\baselineskip=15pt }m{16mm}|}}
      & $2P_{3/2}$ & $2P_{3/2}$ & $2P_{1/2}$ & $2S_{1/2}$ & $2P_{3/2}$ & $2P_{1/2}$ & $2S_{1/2}$\\ \hlx{v}
      & $\tfrac{5}{2},-\tfrac{1}{2}$ & $\tfrac{3}{2},-\tfrac{1}{2}$ & $\tfrac{3}{2},-\tfrac{1}{2}$ & $\tfrac{3}{2},-\tfrac{1}{2}$ & $\tfrac{1}{2},-\tfrac{1}{2}$ & $\tfrac{1}{2},-\tfrac{1}{2}$ & $\tfrac{1}{2},-\tfrac{1}{2}$\\ \hlx{vhhv}
      $2P_{3/2},\tfrac{5}{2},-\tfrac{1}{2}$ & 
      $\frac{2 + g}{10}$ & 
      $-\frac{2 + g}{5}$ & 
      $-\frac{1 - g}{{\sqrt{5}}}$ & 
      0 & 0 & 0 & 0\\ \hlx{vhv}
      $2P_{3/2},\tfrac{3}{2},-\tfrac{1}{2}$ & 
      $-\frac{2 + g}{5}$ & 
      $\frac{11(2 + g )}{90}$ & 
      $-\frac{2(1 - g )}{9{\sqrt{5}}}$ & 0 & 
      $-\frac{{\sqrt{5}}(2 + g )}{9}$ & 
      $-\frac{{\sqrt{10}}(1 - g )}{9}$ & 0\\ \hlx{vhv}
      $2P_{1/2},\tfrac{3}{2},-\tfrac{1}{2}$ & 
      $-\frac{1 - g}{{\sqrt{5}}}$ & 
      $-\frac{2(1 - g)}{9{\sqrt{5}}}$ & 
      $\frac{4 - g}{18}$ & 0 & 
      $\frac{1 - g}{9}$ & 
      $-\frac{{\sqrt{2}}(4 - g)}{9}$ & 0\\ \hlx{vhv}
      $2S_{1/2},\tfrac{3}{2},-\tfrac{1}{2}$ & 0 & 0 & 0 & 
      $\frac{g}{6}$ & 0 & 0 & 
      $-\frac{{\sqrt{2}}g}{3}$\\ \hlx{vhv}
      $2P_{3/2},\tfrac{1}{2},-\tfrac{1}{2}$ & 0 & 
      $-\frac{{\sqrt{5}}(2 + g)}{9}$ & 
      $\frac{1 - g}{9}$ & 0 & 
      $\frac{5(2 + g )}{18}$ & 
      $-\frac{2{\sqrt{2}}(1 - g )}{9}$ & 0\\ \hlx{vhv}
      $2P_{1/2},\tfrac{1}{2},-\tfrac{1}{2}$ & 0 & 
      $-\frac{{\sqrt{10}}(1 - g )}{9}$ & 
      $-\frac{{\sqrt{2}}(4 - g )}{9}$ & 0 & 
      $-\frac{2{\sqrt{2}}(1 - g)}{9}$ & 
      $-\frac{4 - g}{18}$ & 0\\ \hlx{vhv}
      $2S_{1/2},\tfrac{1}{2},-\tfrac{1}{2}$ & 0 & 0 & 0 & 
      $-\frac{{\sqrt{2}}g}{3}$ & 0 & 0 & 
      $-\frac{g}{6}$\\ \hlx{vh}
    \end{tabular}

\label{tB:mu3D2c}}\\[1cm]
\subfloat[]{

    \begin{tabular}{l||*{4}{>{\centering\baselineskip=15pt }m{16mm}|}|>{\centering\baselineskip=15pt }m{16mm}|}
      & $2P_{3/2}$ & $2P_{3/2}$ & $2P_{1/2}$ & $2S_{1/2}$ & $2P_{3/2}$\\ \hlx{v}
      & $\tfrac{5}{2},-\tfrac{3}{2}$ & $\tfrac{3}{2},-\tfrac{3}{2}$ & $\tfrac{3}{2},-\tfrac{3}{2}$ & $\tfrac{3}{2},-\tfrac{3}{2}$ & $\tfrac{5}{2},-\tfrac{5}{2}$\\ \hlx{vhhv}
      $2P_{3/2},\tfrac{5}{2},-\tfrac{3}{2}$ & 
      $\frac{3(2 + g)}{10}$ & 
      $-\frac{\sqrt2(2 + g)}{5\sqrt3}$ & 
      $-\frac{\sqrt2(1-g)}{\sqrt{15}}$ & 0 & 0\\ \hlx{vhv}
      $2P_{3/2},\tfrac{3}{2},-\tfrac{3}{2}$ & 
      $-\frac{{\sqrt{\frac{2}{3}}}(2 + g)}{5}$ & 
      $\frac{11(2 + g)}{30}$ & 
      $-\frac{2(1 - g)}{3{\sqrt{5}}}$ & 0 & 0\\ \hlx{vhv}
      $2P_{1/2},\tfrac{3}{2},-\tfrac{3}{2}$ & 
      $-\frac{\sqrt2(1-g)}{\sqrt{15}}$ & 
      $-\frac{2(1 - g)}{3{\sqrt{5}}}$ & 
      $\frac{4 - g}{6}$ & 0 & 0\\ \hlx{vhv}
      $2S_{1/2},\tfrac{3}{2},-\tfrac{3}{2}$ & 0 & 0 & 0 & 
      $\frac{g}{2}$ & 0\\ \hlx{vhv}
      $2P_{3/2},\tfrac{5}{2},-\tfrac{5}{2}$ & 0 & 0 & 0 & 0 & 
      $\frac{2 + g}{2}$\\ \hlx{vh}
    \end{tabular}

\label{tB:mu3D2d}}
\caption[]{Die auf $\mu_B$ normierte Matrix \unl{$\mu$}$_3$ der dritten Komponente des
magnetischen Moments von Deuterium im $(n=2)$-Unterraum
($F_3=-\tfrac12,-\tfrac32,-\tfrac52$).\\[5mm]}
\end{table}

\begin{table}[!hp]
\centering
\footnotesize\vspace{1cm}
\subfloat[Die Zuordnung der Tabellennummern zu den einzelnen Blöcken der Matrix $\unl D_1$.
Es werden nur die Blöcke der oberen Hälfte der Matrix angegeben. Die unteren Blöcke folgen durch hermitesche Konjugation.]{
\begin{tabular}{c||c|c|c|c|c|c|}
$F_3'\backslash F_3$ & $5/2$ & $3/2$ & $1/2$ & $-1/2$ & $-3/2$ & $-5/2$\\ \hlx{vhhv}
$5/2$  & 0 & \subref*{tB:Dip1D2b} & 0 & 0 & 0 & 0\\ \hlx{vhv}
$3/2$  & (\subref*{tB:Dip1D2b})$^\dag$ & 0 & \subref*{tB:Dip1D2c} & 0 & 0 & 0\\ \hlx{vhv}
$1/2$  & 0 & (\subref*{tB:Dip1D2c})$^\dag$ & 0 & \subref*{tB:Dip1D2d} & 0 & 0\\ \hlx{vhv}
$-1/2$ & 0 & 0 & (\subref*{tB:Dip1D2d})$^\dag$ & 0 & \subref*{tB:Dip1D2e} & 0\\ \hlx{vhv}
$-3/2$ & 0 & 0 & 0 & (\subref*{tB:Dip1D2e})$^\dag$ & 0 & \subref*{tB:Dip1D2f}\\ \hlx{vhv}
$-5/2$ & 0 & 0 & 0 & 0 & (\subref*{tB:Dip1D2f})$^\dag$ & 0 \\ \hlx{vhv}
\end{tabular}\label{tB:Dip1D2overview}}\\[15mm]
\subfloat[]{

    \begin{tabular}{l||*{4}{>{\centering\baselineskip=18pt }m{13mm}|}}
      & $2P_{3/2}$ & $2P_{3/2}$ & $2P_{1/2}$ & $2S_{1/2}$\\ \hlx{v}
      & $\tfrac{5}{2},\tfrac{3}{2}$ & $\tfrac{3}{2},\tfrac{3}{2}$ & $\tfrac{3}{2},\tfrac{3}{2}$ & $\tfrac{3}{2},\tfrac{3}{2}$\\ \hlx{vhhv}
      $2P_{3/2},\tfrac{5}{2},\tfrac{5}{2}$ & 0 & 0 & 0 & 
      $-\frac{3}{\sqrt{2}}$\\ \hlx{vh}
    \end{tabular}\label{tB:Dip1D2b}}\\[10mm]
\subfloat[]{

    \begin{tabular}{l||*{7}{>{\centering\baselineskip=18pt }m{13mm}|}}
      & $2P_{3/2}$ & $2P_{3/2}$ & $2P_{1/2}$ & $2S_{1/2}$ & $2P_{3/2}$ & $2P_{1/2}$ & $2S_{1/2}$\\ \hlx{v}
      & $\tfrac{5}{2},\tfrac{1}{2}$ & $\tfrac{3}{2},\tfrac{1}{2}$ & $\tfrac{3}{2},\tfrac{1}{2}$ & $\tfrac{3}{2},\tfrac{1}{2}$ & $\tfrac{1}{2},\tfrac{1}{2}$ & $\tfrac{1}{2},\tfrac{1}{2}$ & $\tfrac{1}{2},\tfrac{1}{2}$\\ \hlx{vhhv}
      $2P_{3/2},\tfrac{5}{2},\tfrac{3}{2}$ & 0 & 0 & 0 & 
      $-3 \sqrt{\frac{3}{10}}$ & 0 & 0 & 0\\ \hlx{vhv}
      $2P_{3/2},\tfrac{3}{2},\tfrac{3}{2}$ & 0 & 0 & 0 & 
      $-\frac{2}{\sqrt{5}}$ & 0 & 0 & 
      $-\sqrt{\frac{5}{2}}$\\ \hlx{vhv}
      $2P_{1/2},\tfrac{3}{2},\tfrac{3}{2}$ & 0 & 0 & 0 & 
      $-1$ & 0 & 0 & 
      $\sqrt{2}$\\ \hlx{vhv}
      $2S_{1/2},\tfrac{3}{2},\tfrac{3}{2}$ & 
      $\frac{3}{2 \sqrt{5}}$ & 
      $-\frac{2}{\sqrt{5}}$ & 
      $-1$ & 0 & 
      $\frac{1}{2}$ & 
      $\sqrt{2}$ & 0\\ \hlx{vh}
    \end{tabular}\label{tB:Dip1D2c}}\\[10mm]
\caption[Die Matrix \unl{$D$}$_1$ der ersten Komponente des Dipolmoments
von Deuterium im $(n=2)$-Unterraum]{
Die auf $e r\subt B$ normierte Matrix \unl{$D$}$_1$ der ersten Komponente des
Dipolmoments von Deuterium im $(n=2)$-Unterraum. (Fortsetzung auf der nächsten Seite)}
\label{tB:Dip1D2}
\end{table}
\FloatBarrier

\begin{table}[!hp]
\ContinuedFloat
\centering\vspace{1cm}
\footnotesize
\subfloat[]{\input{Dip1D2d}\label{tB:Dip1D2d}}\\[10mm]
\subfloat[]{\input{Dip1D2e}\label{tB:Dip1D2e}}\qquad\quad
\subfloat[]{

    \begin{tabular}{l||*{1}{>{\centering\baselineskip=18pt }m{13mm}|}}
      & $2P_{3/2}$\\ \hlx{v}
      & $\tfrac{5}{2},-\tfrac{5}{2}$\\ \hlx{vhhv}
      $2P_{3/2},\tfrac{5}{2},-\tfrac{3}{2}$ & 0\\ \hlx{vhv}
      $2P_{3/2},\tfrac{3}{2},-\tfrac{3}{2}$ & 0\\ \hlx{vhv}
      $2P_{1/2},\tfrac{3}{2},-\tfrac{3}{2}$ & 0\\ \hlx{vhv}
      $2S_{1/2},\tfrac{3}{2},-\tfrac{3}{2}$ & 
      $\frac{3}{\sqrt{2}}$\\ \hlx{vh}
    \end{tabular}\label{tB:Dip1D2f}}\\[10mm]
\caption[]{
(Fortsetzung) Die auf $e r\subt B$ normierte Matrix \unl{$D$}$_1$ der ersten Komponente des
Dipolmoments von Deuterium im $(n=2)$-Unterraum.}
\end{table}
\FloatBarrier

\begin{table}[!hp]
\centering
\footnotesize\vspace{1cm}
\subfloat[Die Zuordnung der Tabellennummern zu den einzelnen Blöcken der Matrix $\unl D_2$.
Es werden nur die Blöcke der oberen Hälfte der Matrix angegeben. Die unteren Blöcke folgen durch hermitesche Konjugation.]{
\begin{tabular}{c||c|c|c|c|c|c|}
$F_3'\backslash F_3$ & $5/2$ & $3/2$ & $1/2$ & $-1/2$ & $-3/2$ & $-5/2$\\ \hlx{vhhv}
$5/2$  & 0 & \subref*{tB:Dip2D2b} & 0 & 0 & 0 & 0\\ \hlx{vhv}
$3/2$  & (\subref*{tB:Dip2D2b})$^\dag$ & 0 & \subref*{tB:Dip2D2c} & 0 & 0 & 0\\ \hlx{vhv}
$1/2$  & 0 & (\subref*{tB:Dip2D2c})$^\dag$ & 0 & \subref*{tB:Dip2D2d} & 0 & 0\\ \hlx{vhv}
$-1/2$ & 0 & 0 & (\subref*{tB:Dip2D2d})$^\dag$ & 0 & \subref*{tB:Dip2D2e} & 0\\ \hlx{vhv}
$-3/2$ & 0 & 0 & 0 & (\subref*{tB:Dip2D2e})$^\dag$ & 0 & \subref*{tB:Dip2D2f}\\ \hlx{vhv}
$-5/2$ & 0 & 0 & 0 & 0 & (\subref*{tB:Dip2D2f})$^\dag$ & 0 \\ \hlx{vhv}
\end{tabular}\label{tB:Dip2D2overview}}\\[15mm]
\subfloat[]{\input{Dip2D2b}\label{tB:Dip2D2b}}\\[10mm]
\subfloat[]{\input{Dip2D2c}\label{tB:Dip2D2c}}\\[10mm]
\caption[Die Matrix \unl{$D$}$_2$ der zweiten Komponente des Dipolmoments
von Deuterium im $(n=2)$-Unterraum]{
Die auf $e r\subt B$ normierte Matrix \unl{$D$}$_2$ der zweiten Komponente des
Dipolmoments von Deuterium im $(n=2)$-Unterraum. (Fortsetzung auf der nächsten Seite)}
\label{tB:Dip2D2}
\end{table}
\FloatBarrier

\begin{table}[!hp]
\ContinuedFloat
\centering\vspace{1cm}
\footnotesize
\subfloat[]{\input{Dip2D2d}\label{tB:Dip2D2d}}\\[10mm]
\subfloat[]{\input{Dip2D2e}\label{tB:Dip2D2e}}\qquad\quad
\subfloat[]{\input{Dip2D2f}\label{tB:Dip2D2f}}\\[10mm]
\caption[]{
(Fortsetzung) Die auf $e r\subt B$ normierte Matrix \unl{$D$}$_2$ der zweiten Komponente des
Dipolmoments von Deuterium im $(n=2)$-Unterraum.}
\end{table}
\FloatBarrier

\begin{table}[!htp]
  \centering
  \footnotesize
  \subfloat[]{\input{Dip3D2a}\label{tB:Dip3D2a}}\\[1cm]
  \subfloat[]{\input{Dip3D2b}\label{tB:Dip3D2b}}
  \caption[Die Matrix der dritten Komponente des
     Dipoloperators von Deuterium im $(n=2)$-Unterraum]{
     Die auf $e r_B$ normierte Matrix \unl{$D$}$_3$ der dritten Komponente des
     Dipolmoments von Deuterium im $(n=2)$-Unterraum ($F_3=\tfrac52,\tfrac32$,$\tfrac12$).}\label{tB:Dip3D2}
\end{table}
\begin{table}[!hp]  
  \ContinuedFloat
  \centering
  \footnotesize
  \subfloat[]{\input{Dip3D2c}\label{tB:Dip3D2c}}\\[1cm]
  \subfloat[]{\input{Dip3D2d}\label{tB:Dip3D2d}}
  \caption[]{Die auf $e r_B$ normierte Matrix \unl{$D$}$_3$ der dritten Komponente des
  Dipolmoments von Deuterium im $(n=2)$-Unterraum ($F_3=-\tfrac12,-\tfrac32,-\tfrac52$).}
\end{table}
\FloatBarrier

\subsection{Numerische Berechnung der Eigenwerte in elektrischen und magnetischen Feldern}\label{sB:D2.EW}

In den nun folgenden Diagrammen, wie auch in der gesamten vorliegenden Arbeit wollen wir die in Tabelle 
\ref{tB:D2States} dargestellte Nummerierung der Deuterium-Zustände vereinbaren. Auch hier betrachten wir
exemplarisch in 3-Richtung orientierte Felder $\mc E = \vmc E\cdot\v e_3$, $\mc B = \vmc B\cdot\v e_3$.
Bei allgemeinen Feldern ist wie gehabt in der Notation aus Tab. \ref{tB:D2States}
$(\mc E,\mc B)\to(\vmc E,\vmc B)$ zu ersetzen.

\begin{figure}[!hb]
\centering
\subfloat[]{\label{fB:BR-D2-1}\includegraphics[width=15cm]{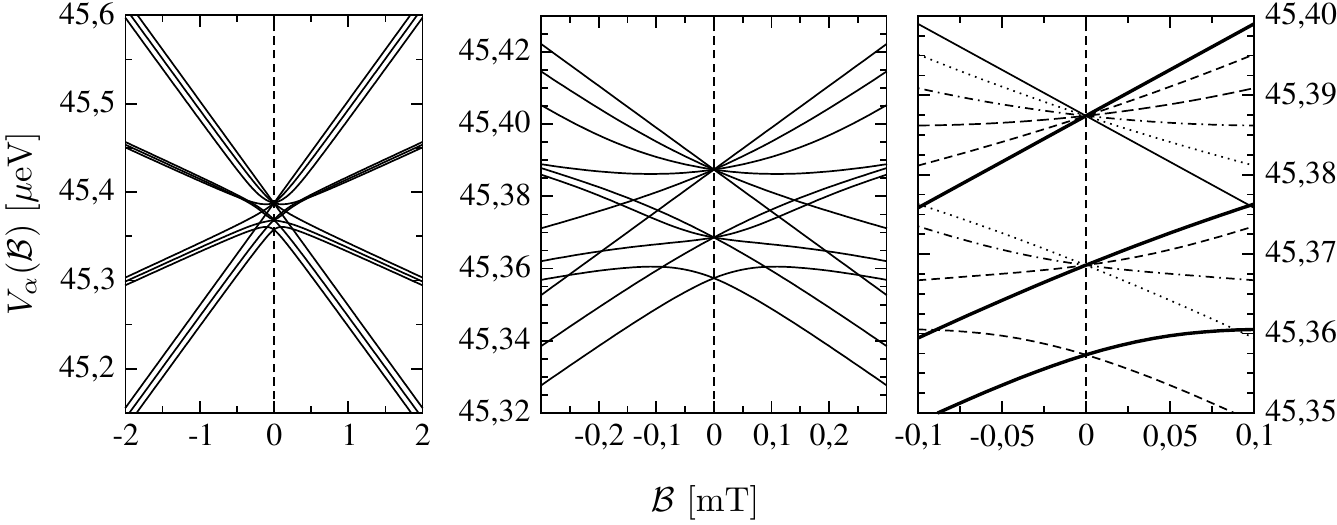}}\\[5mm]
\subfloat[]{\label{fB:BR-D2-2}\includegraphics[width=7cm]{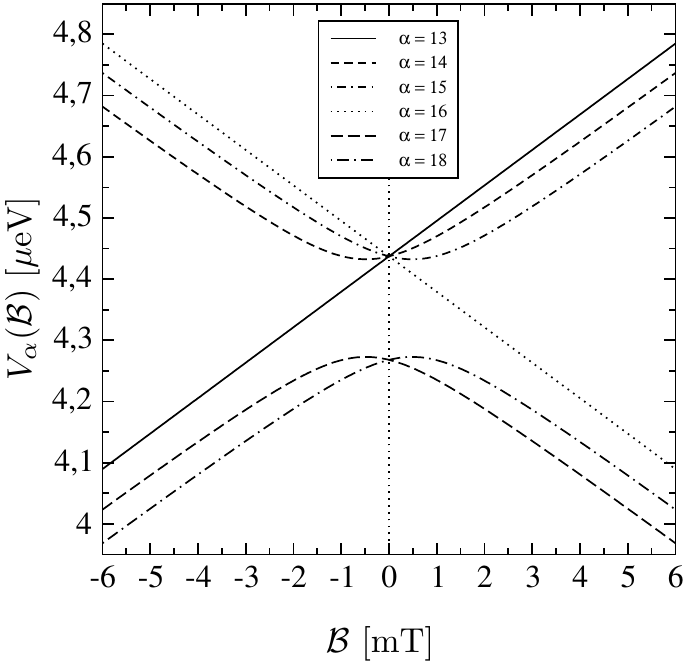}}\hspace{1cm}
\subfloat[]{\label{fB:BR-D2-3}\includegraphics[width=7cm]{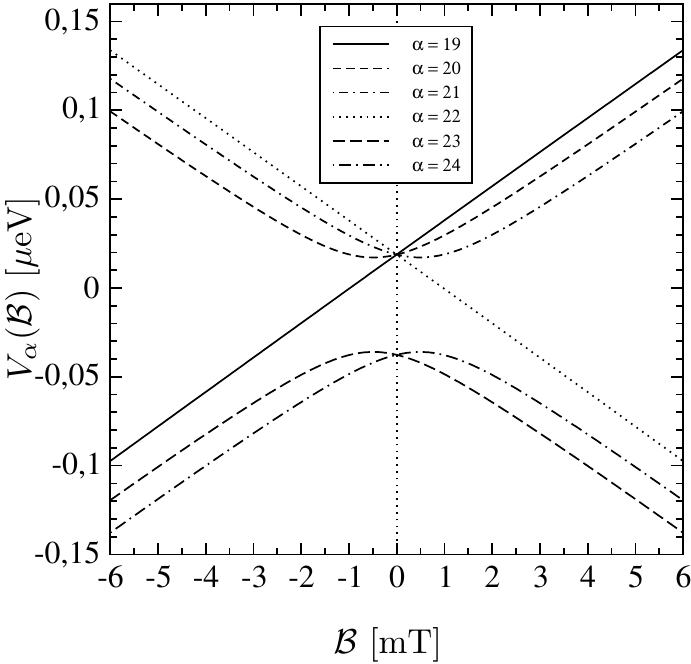}}
\caption[Breit-Rabi-Diagramme der $(n=2)$-Zustände von Deuterium]{
  Breit-Rabi-Diagramme der $(n=2)$-Zustände von Deuterium. Abb. \subref*{fB:BR-D2-1} zeigt
  einen Überblick über die Verschiebung der Energieniveaus der $2P_{3/2}$-Zustände
  im magnetischen Feld. Die Abb. \subref*{fB:BR-D2-2} und \subref*{fB:BR-D2-3} zeigen 
  die $2S_{1/2}$- und $2P_{1/2}$-Zustände. Zur Nummerierung siehe Tabelle \ref{tB:D2States}.
}
\label{fB:BR-D2}
\end{figure}

Die Reihenfolge der Durchnummerierung orientiert sich an der Reihenfolge beim Wasserstoff (d.h. innerhalb 
der Multipletts absteigend nach $F_3$). Wie man in Abb. \subref*{fB:BR-D2-2} jedoch erkennt, entspricht
dies nicht mehr der absteigenden Reihenfolge der Energieniveaus für kleines, in positive 3-Richtung
orientiertes Magnetfeld. 
\begin{table}[!hb]
  \centering\small
  \begin{tabular}{|Mc|Ml||Mc|Ml|}\hlx{hv}
    \alpha & \rket{2\hat L_J,F,F_3,\mc E,\mc B}         & \alpha & \rket{2\hat L_J,F,F_3,\mc E,\mc B}\\ \hlx{vhhv} 
    1  & \rket{2\hat P_{3/2},\tfxxx,\tfxxx,\mc E,\mc B} & 13 & \rket{2\hat S_{1/2},\tfxx,\tfxx,\mc E,\mc B}\\ \hlx{vhv}
    2  & \rket{2\hat P_{3/2},\tfxxx,\tfxx,\mc E,\mc B}  & 14 & \rket{2\hat S_{1/2},\tfxx,\tfx,\mc E,\mc B}\\ \hlx{vhv}
    3  & \rket{2\hat P_{3/2},\tfxxx,\tfx,\mc E,\mc B}   & 15 & \rket{2\hat S_{1/2},\tfxx,-\tfx,\mc E,\mc B}\\ \hlx{vhv}
    4  & \rket{2\hat P_{3/2},\tfxxx,-\tfx,\mc E,\mc B}  & 16 & \rket{2\hat S_{1/2},\tfxx,-\tfxx,\mc E,\mc B}\\ \hlx{vhv}
    5  & \rket{2\hat P_{3/2},\tfxxx,-\tfxx,\mc E,\mc B} & 17 & \rket{2\hat S_{1/2},\tfx,\tfx,\mc E,\mc B}\\ \hlx{vhv}
    6  & \rket{2\hat P_{3/2},\tfxxx,-\tfxxx,\mc E,\mc B}& 18 & \rket{2\hat S_{1/2},\tfx,-\tfx,\mc E,\mc B}\\ \hlx{vhv}
    7  & \rket{2\hat P_{3/2},\tfxx,\tfxx,\mc E,\mc B}   & 19 & \rket{2\hat P_{1/2},\tfxx,\tfxx,\mc E,\mc B}\\ \hlx{vhv}
    8  & \rket{2\hat P_{3/2},\tfxx,\tfx,\mc E,\mc B}    & 20 & \rket{2\hat P_{1/2},\tfxx,\tfx,\mc E,\mc B}\\ \hlx{vhv}
    9  & \rket{2\hat P_{3/2},\tfxx,-\tfx,\mc E,\mc B}   & 21 & \rket{2\hat P_{1/2},\tfxx,-\tfx,\mc E,\mc B}\\ \hlx{vhv}
    10 & \rket{2\hat P_{3/2},\tfxx,-\tfxx,\mc E,\mc B}  & 22 & \rket{2\hat P_{1/2},\tfxx,-\tfxx,\mc E,\mc B}\\ \hlx{vh}
    11 & \rket{2\hat P_{3/2},\tfx,\tfx,\mc E,\mc B}     & 23 & \rket{2\hat P_{1/2},\tfx,\tfx,\mc E,\mc B}\\ \hlx{vhv}
    12 & \rket{2\hat P_{3/2},\tfx,-\tfx,\mc E,\mc B}    & 24 & \rket{2\hat P_{1/2},\tfx,-\tfx,\mc E,\mc B}\\ \hlx{vh}
  \end{tabular}
  \caption[Zuordnung der verschiedenen Notationen für Deuterium im
    $(n=2)$-Unterraum]{
      Zuordnung der verschiedenen Notationen für Deuterium im
    $(n=2)$-Unterraum. Die Reihenfolge der
    Zustände wurde absteigend nach den ungestörten Energien und 
    innerhalb eines Multipletts absteigend nach $F$ und $F_3$ gewählt.}
  \label{tB:D2States}
\end{table}

Die Breit-Rabi-Diagramme der $(n=2)$-Zustände von Deuterium können Abb. \ref{fB:BR-D2}
entnommen werden. Dabei zeigt \subref*{fB:BR-D2-1} einen Überblick über das Verhalten
der $2P_{3/2}$-Energieniveaus im Magnetfeld (wir haben uns eine Zuordnung der Zustände
zu den einzelnen Linien erspart, um die Übersichtlichkeit der Diagramme nicht
noch weiter zu verringern). Anders ist dies
bei den Diagrammen \subref*{fB:BR-D2-2} und \subref*{fB:BR-D2-3}, die das Verhalten der
$2S_{1/2}$- und $2P_{3/2}$-Energieniveaus wiedergeben. Beide Diagramme gleichen sich optisch
und in der Reihenfolge der Zustände. Man sieht hier, dass im Gegensatz zu den Breit-Rabi-Diagrammen
von Wasserstoff (siehe Abb. \ref{fB:BR-H2}) die Reihenfolge der Nummerierung nicht mehr
der absteigenden Reihenfolge der Energieniveaus für kleines magnetisches Feld entspricht.
Weiterhin erkennt man hier - wie auch beim Wasserstoff - die Symmetrie der Diagramme
bzgl. des Übergangs $\mc B\to -\mc B$, $F_3\to-F_3$.
\FloatBarrier

Die Verschiebung der Energieniveaus der einzelnen Zustände bei Anlegen eines elektrischen Felds
in 3-Richtung von Deuterium ähnelt dem Verhalten der Wasserstoff-Zustände und 
kann Abb. \ref{fB:Stark-D2} entnommen werden. 

\begin{figure}[p]
\centering
\subfloat[]{\label{fB:LT-D2-1}\includegraphics[width=10cm]{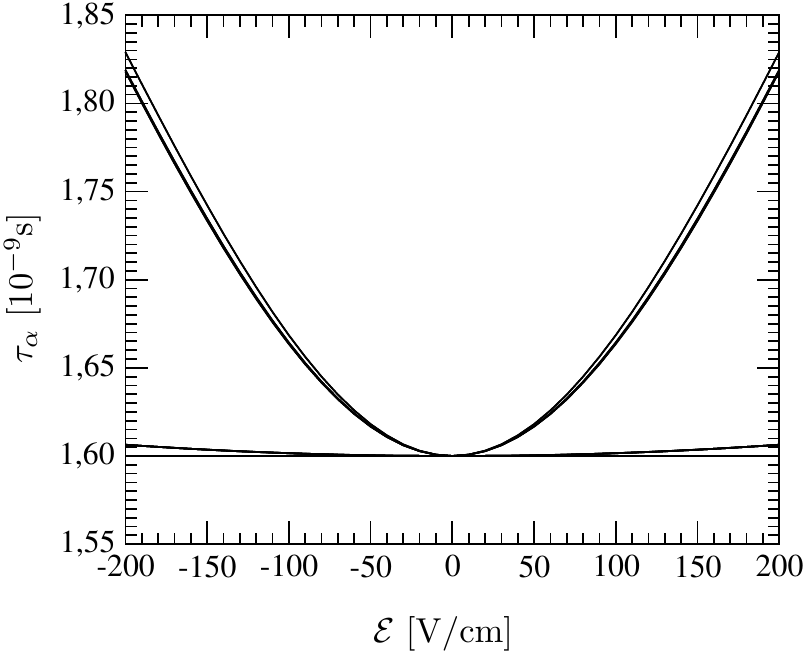}}\\[5mm]
\subfloat[]{\label{fB:LT-D2-2}\includegraphics[width=15cm]{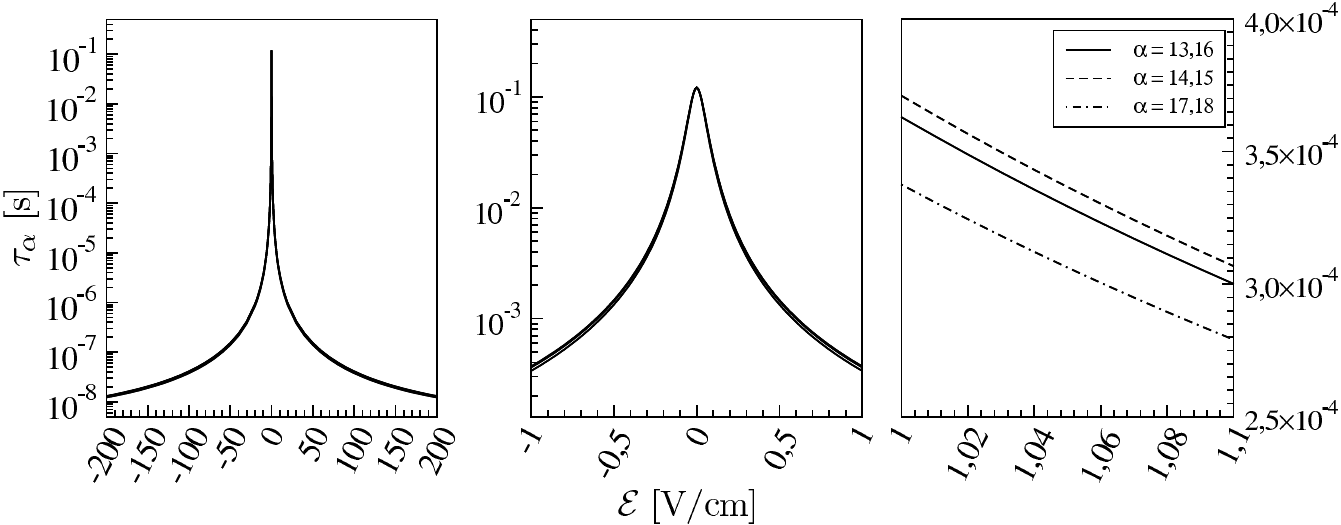}}
\caption[Lebensdauern der $(n=2)$-Zustände von Deuterium im elektrischen Feld]{
  Lebensdauern der $(n=2)$-Zustände von Deuterium im elektrischen Feld $\mc E = \v e_3\cdot\v{\mc E}$. 
  Diagramm \subref*{fB:LT-D2-1} zeigt die Abhängigkeit der $2P$-Zustände, \subref*{fB:LT-D2-2}
  die $2S_{1/2}$-Zustände. Zur Nummerierung siehe Tabelle \ref{tB:D2States}.
}
\label{fB:LT-D2}
\end{figure}

Auch die Abhängigkeit der Lebensdauern $\tau_\alpha(\mc E)=\Gamma_\alpha^{-1}(\mc E)$ 
vom elektrischen Feld, dargestellt in Abb. \ref{fB:LT-D2}, ist ähnlich wie beim Wasserstoff, siehe Abb.
\ref{fB:EW-E-Imag-H2}. Der Anstieg der Lebensdauern der $2P$-Zustände um bis zu $10^{-10}\u s$ 
entspricht von der Größenordnung her auch den Ergebnissen für Wasserstoff. Es wurde in Abb. 
\subref*{fB:LT-D2-1} wieder auf eine Zuordnung der Linien zu den einzelnen $2P$-Zuständen verzichtet.

Diagramm \subref*{fB:LT-D2-2} zeigt in logarithmischer Auftragung (links)
die Feld-Abhängigkeit der Lebensdauern der $2S_{1/2}$-Zustände. Es zeigt sich kein wesentlicher Unterschied
zum Wasserstoff, siehe Abb. \subref*{fB:EW-E-Imag-H2-3}. Weiterhin sind in Abb. \subref*{fB:LT-D2-2}
(Mitte und rechts) vergrößerte Ausschnitte der Lebensdauer-Abhängigkeit der $2S$-Zustände gezeigt.

Abschließend können wir feststellen, dass zwar die Breit-Rabi-Diagramme für Deuterium wesentlich
komplexer als die des Wasserstoffs sind, die Abhängigkeit der Eigenwerte vom elektrischen Feld
aber keine großen Unterschiede zeigt.
\begin{figure}[b!]
\centering
\subfloat[]{\label{fB:Stark-D2-1}\includegraphics[width=12cm]{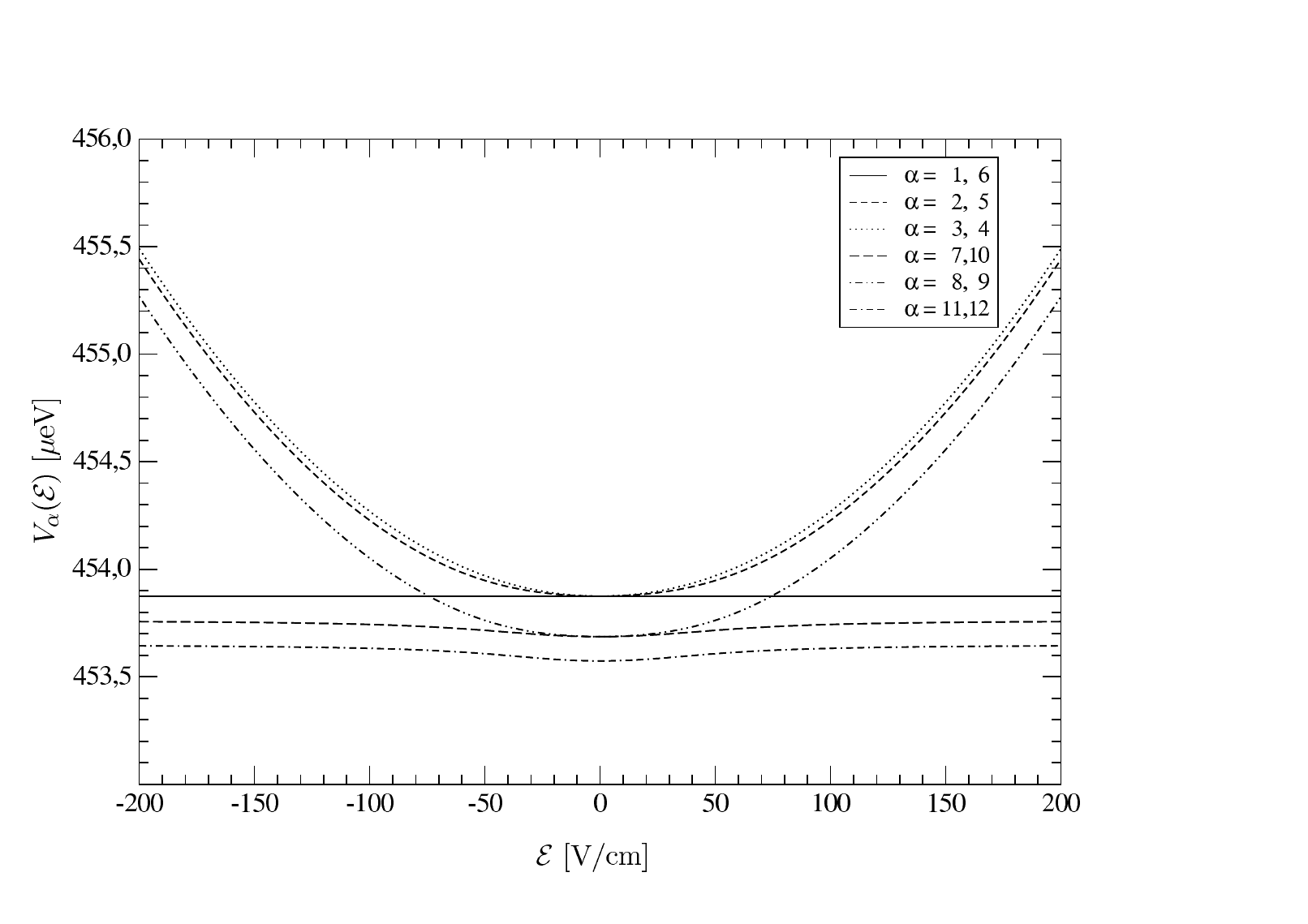}}
\caption[Energieniveaus der $(n=2)$-Zustände von Deuterium im elektrischen Feld]{
  Energieeigenwerte (Realteile) der $2P_{3/2}$-Zustände von Deuterium im elektrischen 
  Feld $\mc E = \v e_3\cdot\v{\mc E}$. Zur Nummerierung siehe Tabelle \ref{tB:D2States}.
}
\label{fB:Stark-D2}
\end{figure}
\begin{figure}[p]
\ContinuedFloat
\centering
\subfloat[]{\label{fB:Stark-D2-2}\includegraphics[width=12cm]{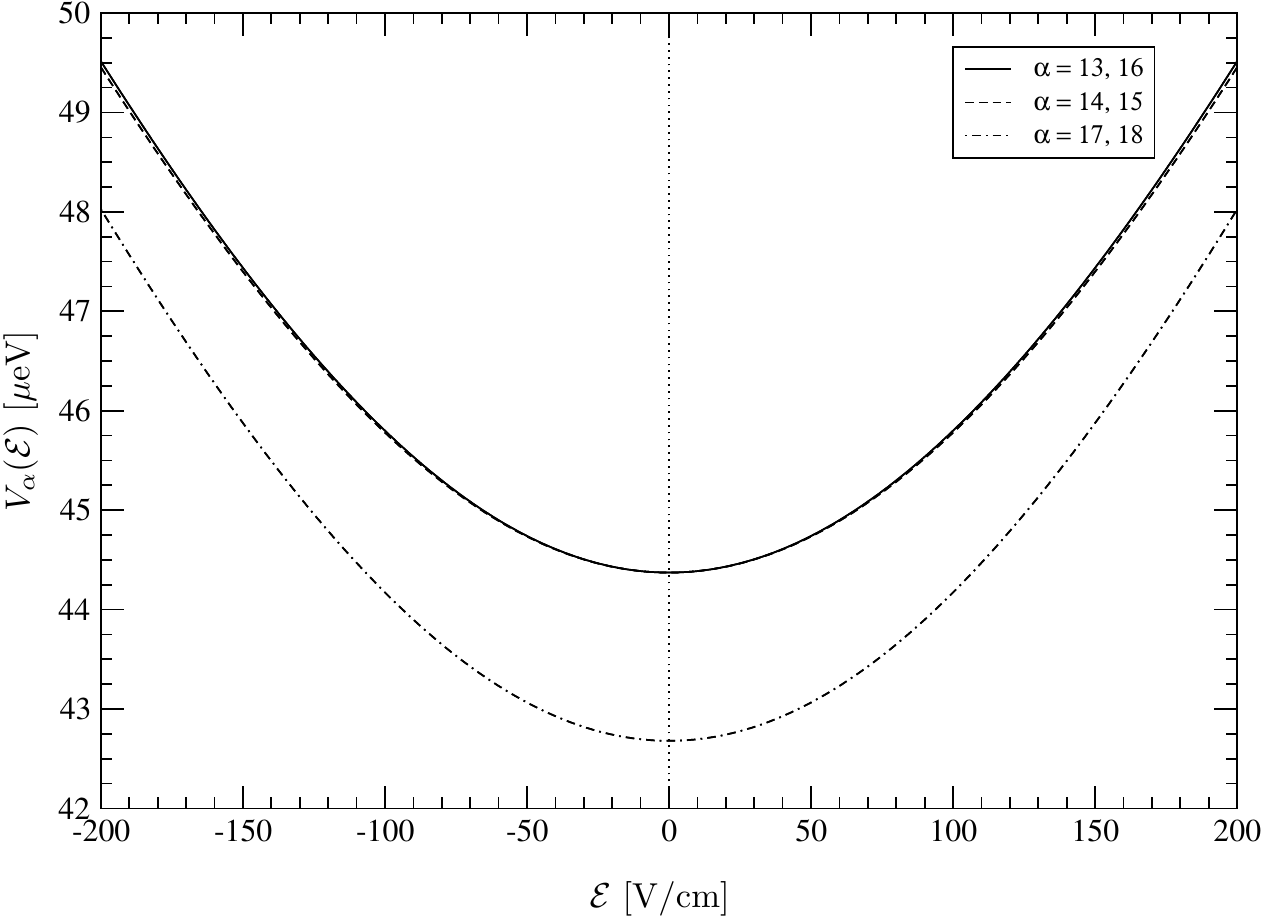}}\\[5mm]
\subfloat[]{\label{fB:Stark-D2-3}\includegraphics[width=12cm]{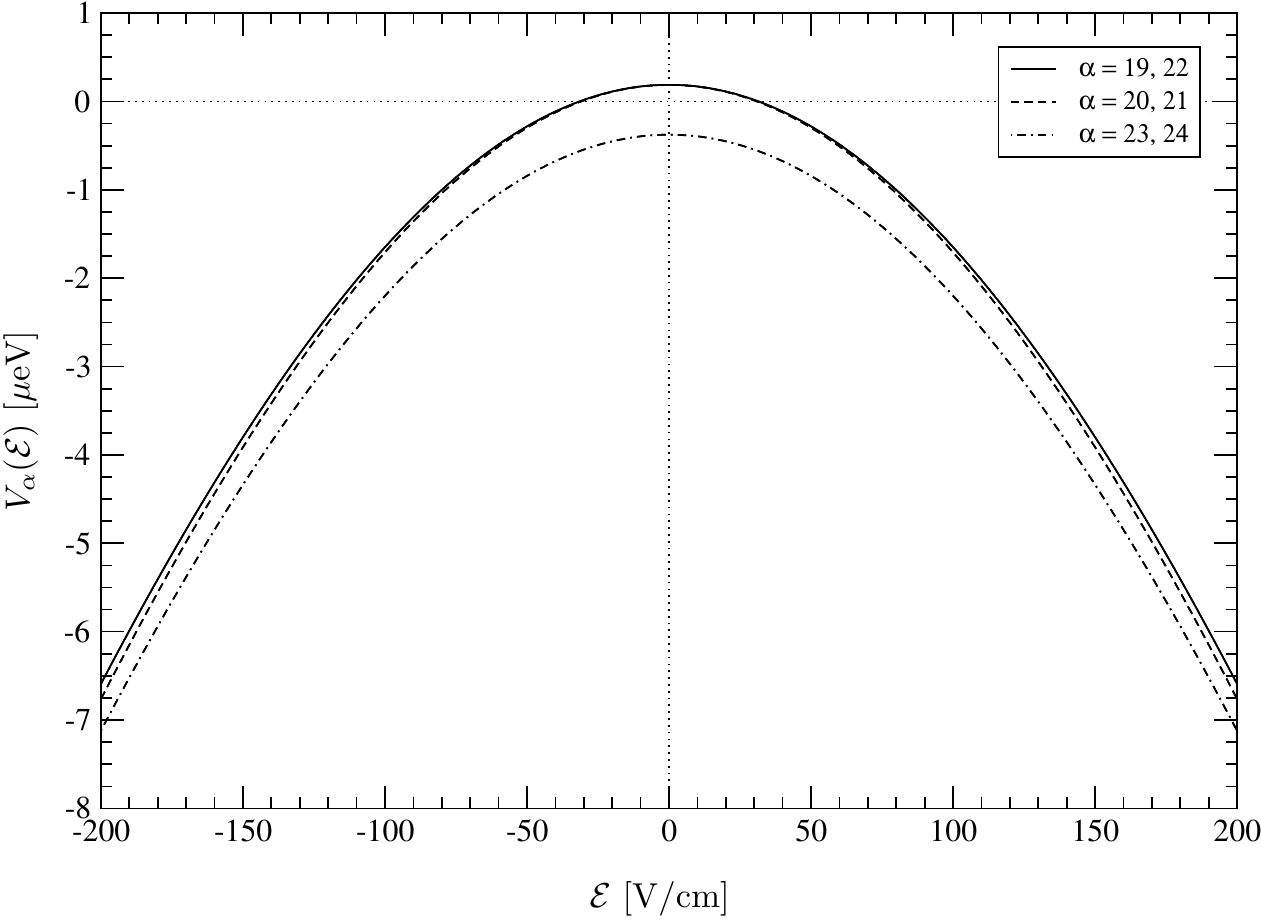}}
\caption[]{
  Energieniveaus der $2S_{1/2}$- \subref{fB:Stark-D2-2} und 
  $2P_{1/2}$-Zustände \subref{fB:Stark-D2-3} von Deuterium im elektrischen 
  Feld $\mc E = \v e_3\cdot\v{\mc E}$. Zur Nummerierung siehe Tabelle \ref{tB:D2States}.
}
\end{figure}

\chapter{Die Gabor Transformation}\label{sC:WFT}

Die Fourier Transformation zerlegt eine quadratintegrable Funktion $\psi(x)$ in ebene Wellen $\e^{\I k x}$.
Die Amplituden-Funktion $\tilde\psi(k)$, die als Fourier-Transformierte bezeichnet wird, lautet
\begin{align}\label{eC:FT}
	\tilde\psi(k) = \frac1{2\pi}\int\d x\ \psi(x)\e^{-\I k x}\ ,
\end{align}
während die inverse Fourier-Transformation gegeben ist durch
\begin{align}\label{eC:iFT}
	\psi(x) = \int\d k\ \tilde\psi(k)\e^{\I k x}\ .
\end{align}

Die Fourier Transformation ist nicht in der Lage, eine Phasenraum-Beschreibung der Funktion $\psi(x)$ zu geben,
da die Basis-Funktionen zwar einen scharfen Impuls $k$, jedoch - als ebene Wellen - keine Lokalisierung im Ortsraum haben.
In praktischen Fällen ist man oft am Frequenzspektrum eines Signals zu einer bestimmten Zeit interessiert, bzw.
am Impulsspektrum einer Funktion an einem bestimmen Ort. Zu diesem Zweck wurde 1946 von Gabor \cite{Gab46}
die nach ihm benannte Transformation eingeführt, die auch als {\em gefensterte} Fourier Transformation
(Windowed Fourier Transform) oder {\em Short Time Fourier Transform (STFT)} bekannt ist.

\begin{floatingfigure}{7cm}
\includegraphics[width=7cm]{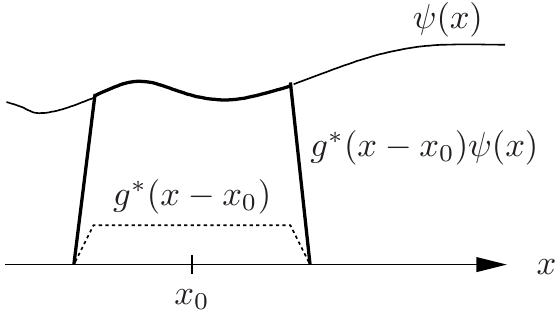}
\caption{Veranschaulichung der Gabor Transformation}\label{fC:GT}
\end{floatingfigure} 
Die Gabor Transformation führt eine um Null zentrierte Fensterfunktion $g(x)$ ein, für die gilt
\begin{align}\label{eC:Window}
	g(x) \begin{cases}
					\neq 0\qquad &\abs x\lesssim L/2\\
					= 0   &\abs x\gtrsim L/2
	     \end{cases}
\end{align} 
d.h. $g(x)$ hat etwa die Breite $L$. Die Gabor Transformation schneidet nun also einen um ein $x_0$
lokalisierten Teil der Funktion $\psi(x)$ durch Multiplikation mit der 
(komplex konjugierten\footnote{aus Konventionsgründen})
Fensterfunktion $g^*(x-x_0)$ aus und berechnet die (inverse) 
Fourier Transformation (siehe Abb. \ref{fC:GT}):
\begin{align}\label{eC:GT}
	\boxed{\hat\psi(k,x_0) := \int\d x\ \psi(x)g^*(x-x_0)\e^{-\I k x}}\ .
\end{align}
Dies definiert die Gabor-Transformierte, die nun eine Beschreibung der Funktion $\psi(x)$
um den Phasenraumpunkt $(k,x_0)$ liefert. Hat das Fenster etwa die Breite $L$ im Ortsraum,
so hat die Gabor-Transformierte aufgrund der Unschärferelation im Impulsraums etwa eine
Breie $1/L$. 

Die Umkehrtransformation zu (\ref{eC:GT}), mit der man die ursprüngliche Funktion aus der
Gabor-Transformierten rekonstruieren kann, lässt sich leicht berechnen. Hierzu kehrt man
die inverse Fourier Transformation in (\ref{eC:GT}) zunächst um und erhält
\begin{align}
	g^*(x-x_0)\psi(x) = \frac1{2\pi}\int\d k\ \hat\psi(k,x_0)\e^{\I k x}\ .
\end{align}
Da die Fenster-Funktion Null werden kann, können wir sie an dieser Stelle nicht dividieren.
Stattdessen multiplizieren wir mit $g(x-x_0)$ und integrieren über $x_0$:
\begin{align}
	\klr{\int\d x_0\ \abs{g(x-x_0)}^2}\psi(x) = \frac1{2\pi}\int\d k\int\d x_0\ \hat\psi(k,x_0)g(x-x_0)\e^{\I k x}\ .
\end{align}
Falls nun $g(x)$ quadratintegrabel ist und
\begin{align}\label{eC:Cg}
	\boxed{0 < C_g := \int\d x\ \abs{g(x)}^2 < \infty}
\end{align}
erfüllt ist, können wir durch $C_g$ dividieren und erhalten die Rekursionsformel
\begin{align}\label{eC:iGT}
	\boxed{
		\psi(x) = \frac1{C_g\,2\pi}\int_{-\infty}^\infty\d k\int_{-\infty}^\infty\d x_0\ \hat\psi(k,x_0)g(x-x_0)\e^{\I k x}
}\ .
\end{align}

Eine weitere nützliche Betrachtungsweise der Gabor Transformation (\ref{eC:GT}) ist die eines
Skalarprodukts im Raum der quadratintegrablen Funktionen $L^2(\mathbb R)$, die von einer Variable
abhängen. Dabei kann man die Funktionen
\begin{align}\label{eC:Notes}
	g_{k,x_0}(x) := g(x-x_0)\e^{\I k x}
\end{align}
als neue Basisfunktionen des $L^2(\mathbb R)$ betrachten. Gl. (\ref{eC:GT}) lautet dann
\begin{align}
	\hat\psi(k,x_0) = \langle g_{k,x_0},\psi\rangle_{L^2(\mathbb R)}\ .
\end{align}

Wir wollen nun einige Eigenschaften der Gabor Transformation auflisten, ohne diese näher zu begründen
Hierzu sei auf die Literatur \cite{Kai94, FT98} verwiesen.
\begin{enumerate}[(i)]
	\item Die Basisfunktionen $g_{k,x_0}(x)$ werden oft auch als {\em Noten} bezeichnet, da sie eine Länge (die Breite
		des Fensters) und eine Frequenz (bzw. Wellenzahl) haben.
	\item Das Auflösungsvermögen der einzelnen Basisfunktionen unterliegt einigen Beschränkungen aufgrund der festen Breite
		der $g_{k,x_0}(x)$. im Ortsraum, für große $k$ liegt eine schlechte Ortsauflösung vor, d.h. 
		die Basisfunktion $g_{k,x_0}(x)$ ist dann {\em unterlokalisiert} im Ort.
		Für kleine $k$ eine schlechte Skalenauflösung, d.h. die Basisfunktion $g_{k,x_0}(x)$ ist dann 
{\em überlokalisiert} im Ort (siehe \cite{FT98}).
	\item Die Basisfunktionen $\{g_{k,x_0}(x); k,x_0\in\mathbb R\}$ sind {\em übervollständig}, da sie nicht
		orthogonal sind, d.h. es gilt zwar
		\begin{align}
\int\d k\int\d x_0\ g^*_{k,x_0}(x)g_{k,x_0}(x') = \frac{C_g}{2\pi}\delta(x-x')\ ,
		\end{align}
		aber auch im Allgemeinen
		\begin{align}
			\int\d x\ g_{k',x_0'}(x)g_{k,x_0}(x) \neq C\delta(k-k')\delta(x-x')\ .
		\end{align}
	\item Es gilt das Analogon zur Parsevalschen Identität $\langle f,g\rangle = \langle \tilde f,\tilde g\rangle$
		der Fourier Transformation:
		\begin{align}\label{eC:ParsevalGT}
			\int\d x\ \psi_1^*(x)\psi_2(x) = \int\d k\int\d x_0\ \hat\psi_1^*(k,x_0)\hat\psi_2(k,x_0)
		\end{align}
	\item Hieraus folgt: Ist $\psi(x)$ quadratintegrabel, dann ist auch $\hat\psi(k,x_0)$ quadratintegrabel
		im Phasenraum.
	\item Nicht jede im Phasenraum quadratintegrable Funktion ist auch eine Gabor-Transformierte, da eine
		Gabor-Transformierte die Unschärferelation respektieren müssen.
\end{enumerate}

An dieser Stelle sei angemerkt, dass die Idee der Gabor Transformation, eine Funktion im Phasenraum lokalisiert
darzustellen, durch die Einführung von {\em Wavelets} ({\em Wellchen}) weiterentwickelt wurde. Es existieren
hierzu u.a. in der Mathematik, Physik und Informatik unzählbare Anwendungen. Für weitere Information sei
auf \cite{Kai94} und darin aufgeführte Referenzen verwiesen.

\begin{figure}
	\centerline{\includegraphics[width=5cm]{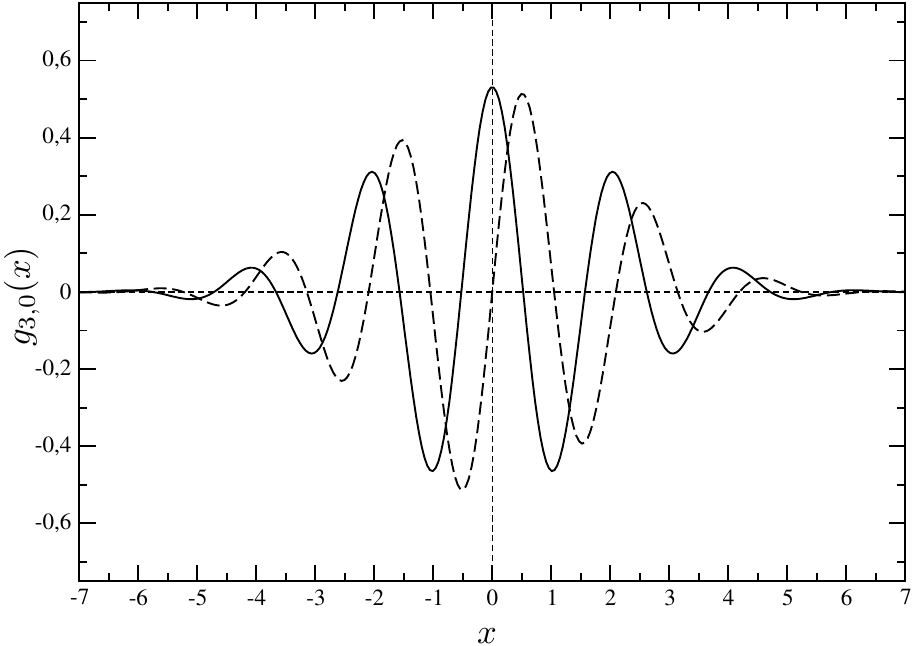}\quad\includegraphics[width=5cm]{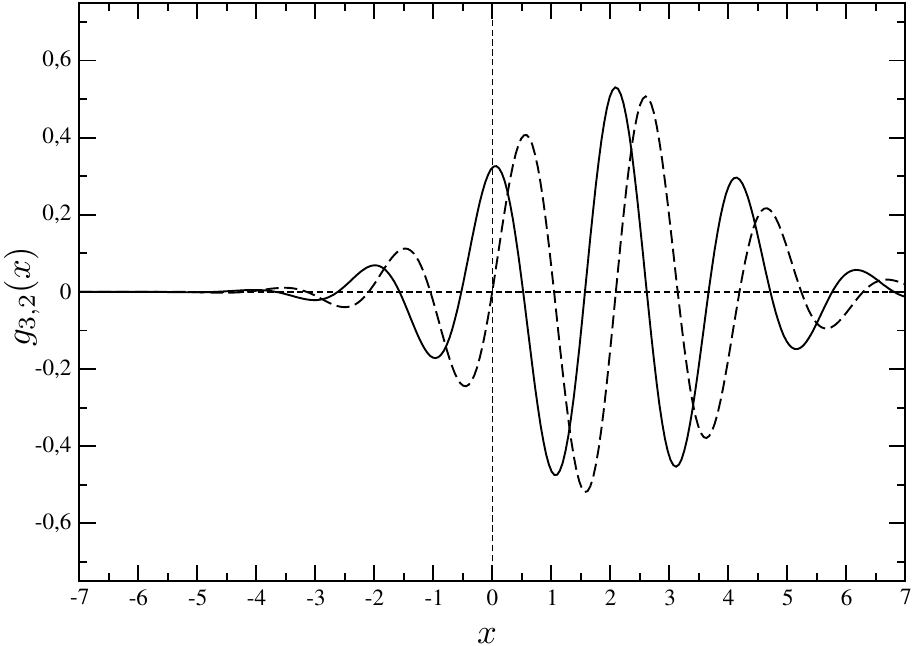}\quad
		\includegraphics[width=5cm]{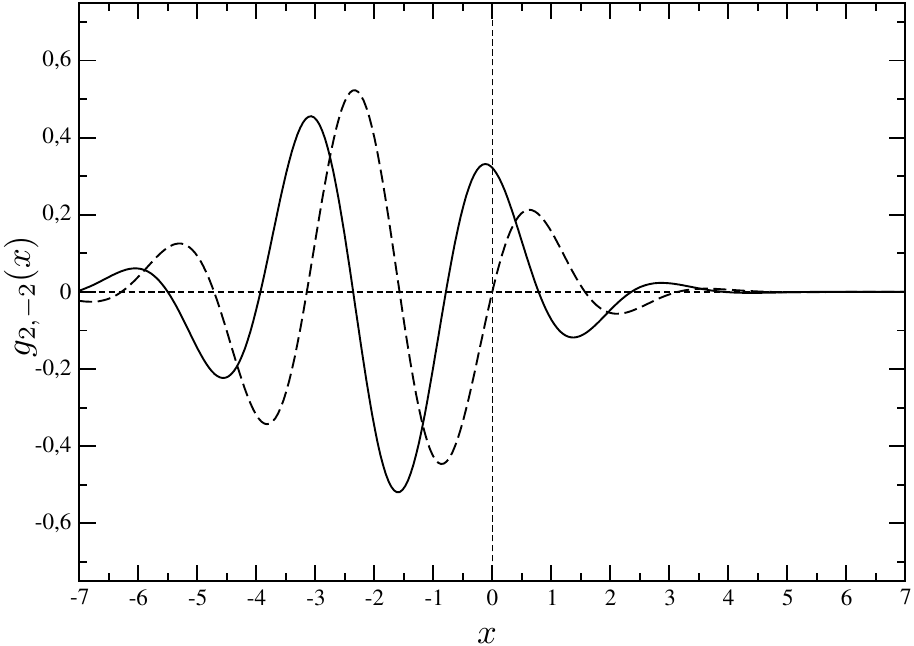}}
	\caption[Drei repräsentative \glqq Noten\grqq]{Drei repräsentative \glqq Noten\grqq~$g_{k,x_0}(x)$. 
		Von links nach rechts: $g_{3,0}(x)$, $g_{3,2}(x)$ und $g_{2,-2}(x)$. Der Realteil ist durchgezogen,
		der Imaginärteil gestrichelt dargestellt.}\label{fC:Noten}
\end{figure}

\begin{figure}
	\centerline{\includegraphics[width=6cm]{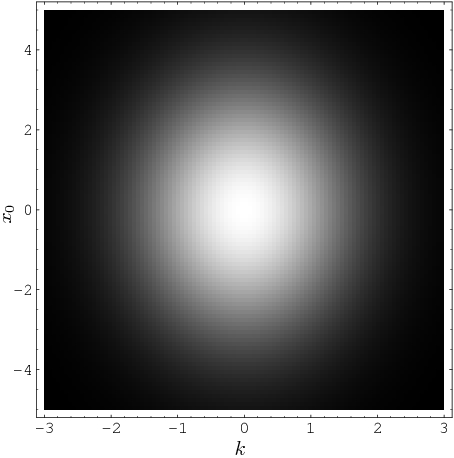}\qquad\raisebox{5mm}{\includegraphics[width=7cm]{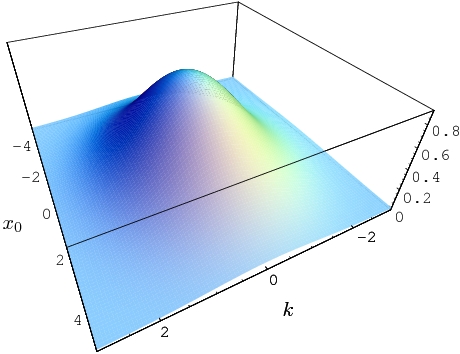}}}
	\centerline{\includegraphics[width=6cm]{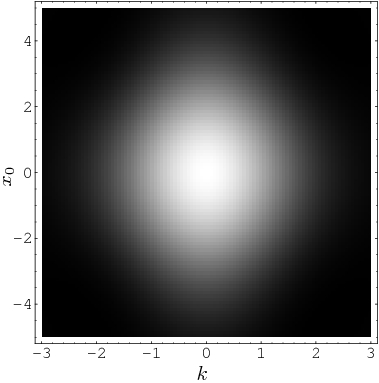}\qquad\raisebox{5mm}{\includegraphics[width=7cm]{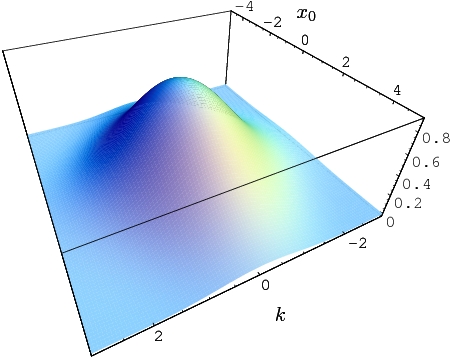}}}
	\centerline{\includegraphics[width=6cm]{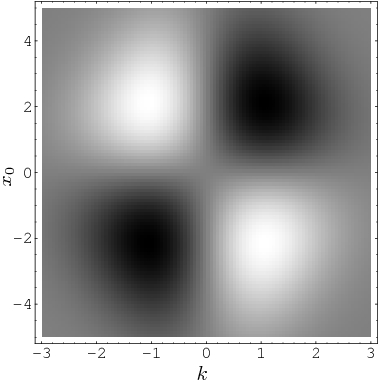}\qquad\raisebox{5mm}{\includegraphics[width=7cm]{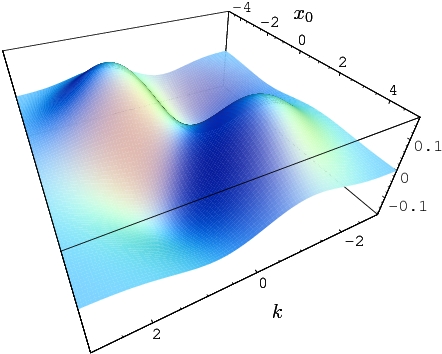}}}
	\caption[Beispiel: Gabor-Transformierte]{Darstellung der Gabor Transformierten $\hat\vph(k,x_0)$ 
		aus (\ref{eC:GaussGT}) für $\sigma=1$ und $\sigma_g=2$. Von oben nach unten: Betrag, Real- und
		Imaginärteil}\label{fC:GaussGT}
\end{figure}

Wir wollen dieses Kapitel mit einem Beispiel abschließen. Betrachten wir eine um Null zentrierte Gauß-Funktion
der Breite $\sigma$,
\begin{align}\label{eC:Gauss}
	\vph(x) = \frac{1}{\sqrt{\sigma\sqrt{\pi}}}\exp\klg{-\frac{x^2}{2\sigma^2}}\ ,
\end{align}
und wählen weiter ein Gauß-Fenster der Breite $\sigma_g$, so dass
\begin{align}\label{eC:GaussFenster}
	g_{k,x_0}(x) = \frac{1}{\sqrt{\sigma_g\sqrt{\pi}}}\exp\klg{-\frac{(x-x_0)^2}{2\sigma_g^2}+\I k x}\ ,
\end{align}
dann lautet die Gabor-Transformierte
\begin{align}\label{eC:GaussGT}
	\hat\vph(k,x_0) = \sqrt{\frac{2\sigma_g\sigma}{\sigma^2+\sigma_g^2}}
		\exp\klg{-\frac{x_0^2+2\I k x_0\sigma^2 + k^2\sigma_g^2\sigma^2}{2(\sigma_g^2+\sigma^2)}}\ .
\end{align}
Der Plot der Gabor-Transformierten ist in Abb. \ref{fC:GaussGT} für $\sigma=1$ und $\sigma_g=2$ dargestellt, 
drei Beispiele für Basisfunktionen $g_{k,x_0}(x)$ finden sich in Abb. \ref{fC:Noten}.
Wählt man andere Konfigurationen von $\sigma$ und $\sigma_g$, so kann man eine sehr breite Verteilung
in $x_0$ oder $k$ erreichen. Wir wollen nur einige Spezialfälle betrachten:
\begin{itemize}
	\item Eine breite Verteilung in $x_0$ erhält man, wenn man $(\sigma^2+\sigma_g^2)$ sehr groß wählt, so dass
		die Exponentialfunktion für $k=0$ und $x_0\to\infty$ gegen Null geht. Wählt man $(\sigma^2+\sigma_g^2)$
		dagegen sehr klein, so ist die Verteilung in $x_0$ nur schmal.
	\item Eine breite Verteilung in $k$ (für $x_0=0$) erhält man, wenn man
		$(\sigma^2+\sigma_g^2)$ sehr groß wählt, aber gleichzeitig das Produkt $\sigma_g^2\sigma^2$ klein gegenüber
		der Breite der Exponentialfunktion. Dies erreicht man z.B. durch $\sigma/\sigma_g\ll 1$ und durch
		$(k\subt{max}\sigma_g\sigma)^2\approx (\sigma^2+\sigma_g^2)$, wobei $k\subt{max}$ ein Maß für die
		gewünschte Breite der Gabor-Transformierten in $k$ ist.
\end{itemize}

Wir wollen zum Abschluss dieses Anhangs noch eine Bemerkung im Zusammenhang mit der in dieser Arbeit behandelten
Theorie machen. In Abschnitt \ref{s4:Dispersion} haben wir eine vereinfachte Version der Gabor-Transformation
angewendet, um ein schmales Gaußpaket als Superposition sehr breiter Gaußpakete zu schreiben. Dabei haben wir
nur Fensterfunktionen, die um $x_0=0$ zentriert waren, zugelassen. Im allgemeinen kann aber auch die Superposition
verschobener Basiswellenpakete mit verschiedenen $x_0$ interessant für die in dieser Arbeit vorgestellte Theorie
sein. Man muss hierbei immer abwägen, ob die Einführung einer weiteren Integrationsvariable $x_0$ und der damit
verbundene numerische Aufwand durch ein evtl. verbessertes Endresultat aufgewogen wird. Eine weitere Option ist
auch die Anwendung der oben bereits erwähnten Wavelet-Transformation, 
siehe z.B. \cite{Kai94}. Wavelets sind Basisfunktionen mit
einer festen Anzahl von Schwingungen und variabler Breite. Bei sehr kleinen Wellenlängen, also hoher Ortsauflösung,
hat man nur ein sehr schmales Wavelet, während auf großen Skalen (langen Wellenlängen) und somit niedriger
Ortsauflösung ein sehr breites Wavelet als Basisfunktion verwendet wird. Das Problem der Überlokalisierung
oder Unterlokalisierung, das bei der Gabor-Transformation auftreten kann, wird durch die Verwendung von
Wavelets umgangen.

\vfill
\cleardoublepage
\stepcounter{chapter}
\fancyhead[LE,LO,CE,CO,RE,RO]{}
\fancyhead[LO,RE]{Literaturverzeichnis}
\fancyhead[RO,LE]{\bfseries\thepage}

\cleardoublepage
\parskip3pt
\parindent1ex
\refstepcounter{dummy}
\addcontentsline{toc}{chapter}{Abbildungsverzeichnis}
\fancyhead[ER]{\slshape Abbildungsverzeichnis}
\fancyhead[OL]{\slshape Abbildungsverzeichnis}
\listoffigures
\cleardoublepage
\refstepcounter{dummy}
\addcontentsline{toc}{chapter}{Tabellenverzeichnis}
\fancyhead[ER]{\slshape Tabellenverzeichnis}
\fancyhead[OL]{\slshape Tabellenverzeichnis}
\listoftables

\cleardoublepage
\refstepcounter{dummy}
\addcontentsline{toc}{chapter}{Literatur}
\fancyhead[ER]{\slshape Literaturverzeichnis}
\fancyhead[OL]{\slshape Literaturverzeichnis}

\bibliography{mybibg}

\chapter*{Danksagung}
\refstepcounter{dummy}
\addcontentsline{toc}{chapter}{Danksagung}
\fancyhead[ER]{\slshape Danksagung}
\fancyhead[OL]{\slshape Danksagung}
\parskip5pt
\parindent2ex

Meinem Betreuer Prof. Dr. Otto Nachtmann möchte ich von ganzem Herzen danken, dass er
mich nach meiner Diplomarbeit, die ich zuvor in seiner Arbeitsgruppe durchgeführt habe,
auch als Doktorand angenommen hat. Sein Vertrauen in mich war und ist auch heute noch
eine zusätzliche Motivation für mich. Seine ruhige und zuversichtliche Art führte
zu der entspannten Arbeitsatmosphäre, in der sich ein junger Geist erst frei entfalten 
kann. Stets konnte ich mich auf seine Unterstützung und bewundernswerte 
fachliche Kompetenz beim Lösen von Problemen verlassen, sei es nun bei den Formeln dieser Arbeit oder bei der langfristigen
Sicherung meiner Finanzierung. Herr Nachtmann hat sich dabei immer auch für das Wohlergehen
meiner Familie interessiert und eingesetzt, wofür ich ihm besonders dankbar bin.

Herrn Priv.-Doz. Maarten DeKieviet, PhD., möchte ich für seinen bereitwilligen Einsatz als
Zweitgutachter danken, darüberhinaus aber auch für viele interessante und fruchtbare Diskussionen
über die Möglichkeiten der Atomstrahl-Spinecho-Methode. Seine lockere und ungezwungene Art
machten die Zusammenarbeit zwischen Theorie und Experiment zu einer Freude. 

Herr Priv.-Doz. Dr. Thomas Gasenzer danke ich für zahlreiche Diskussionen über die Theorie
der Paritätsverletzung in Atomen, er war für mich ein Mentor und Freund.
Besonders danken möchte ich ihm für seine wertvollen Tipps bei der Erstellung eines Vortrags
für die DPG-Frühjahrstagung, sowie für das Lesen und konstruktive Kritisieren der Zusammenfassung
dieser Arbeit.

Dieter Greiner danke ich für viele Diskussionen über unser gemeinsames Arbeitsgebiet. Er stellte mir
oft die Fragen, die ich mir selbst schon längst hätte stellen müssen. Bei der Betreuung seiner Diplomarbeit
musste ich über viele, lange zurückliegende Dinge noch einmal in einem neuen Licht nachdenken, was
mit oft geholfen hat, sie klarer zu sehen. Darüberhinaus bedanke ich mich für das Korrekturlesen
der Einleitung und Zusammenfassung dieser Arbeit.

Vielen weiteren Personen möchte ich danken für die schöne gemeinsame Zeit, die interessanten oder
auch einfach nur netten Gespräche, die zu der schönen Atmosphäre im Institut beigetragen haben:
Thorsten Zöller, Andreas von Manteuffel, Juliane Behrend, Thomas Bittig, Sebastian Scheffler,
Wolfgang Unger, Urs Bergmann und allen, die ich an dieser Stelle vergessen habe, die sich aber dennoch
angesprochen fühlen.

Der Deutschen Forschungsgemeinschaft danke ich für die finanzielle
Förderung meiner Arbeit, die mich und meine Familie ernährt.

Ich danke Kaffee, weil ich es ohne ihn
nicht geschafft hätte, wach zu bleiben oder mich zu konzentrieren. 
Ich danke außerdem dem Internet und meinen Kollegen vom Sim-Racing-Team \glqq${}^\circ$HoT\grqq, die für
Zerstreuung in meiner Freizeit gesorgt haben. 
Den Komponisten James Horner, James Newton Howard, Hans Zimmer und Alan Silvestri danke ich für ihre wunderschöne
Filmmusik, die mir beim stundenlangen Schreiben geholfen hat.

Meinen Eltern danke ich für alles, was sie für mich getan haben. Ihr habt mir immer die nötigen
Freiheiten gelassen, mich in Liebe erzogen und mir ein Gefühl für das gegeben, 
was im Leben wichtig ist, was richtig und was falsch ist. 
Ihr habt mich bis zum heutigen Tage gefördert und unterstützt, finanziell wie
moralisch. Mein besonderer Dank gilt meinem kürzlich verschiedenen Großvater Wilhelm Hesse, 
der immer für mich da war. Seine Liebenswürdigkeit, Aufrichtigkeit, Ehrlichkeit und Treue werden mir
stets in Erinnerung bleiben und mir ein Vorbild sein.

Liebste Dajana, Dir möchte ich für alles danken, was Du bist und was Du mir gibst. 
Ohne Deinen Rückhalt und Dein positives Wesen, Deine Liebe und Dein Vertrauen in mich
hätte ich das hier niemals geschafft.

Liebe Elena, Dir danke ich dafür, dass es Dich gibt. Danke, dass Du mich immer wieder verblüffst, dass Du
mir den Spiegel vorhältst, wie das nur ein Kind kann und dass mich Dein Lächeln stets daran erinnert, was im
Leben wirklich zählt.\\[1cm]

\begin{flushright}
\em Dossenheim, Mai 2006
\end{flushright}

\end{document}